\documentclass[graybox,envcountchap,envcountsame, sectrefs]{svmono}

\usepackage{mathptmx}
\usepackage{helvet}
\usepackage{courier}
\usepackage{type1cm}         

\usepackage{makeidx}         
\usepackage{graphicx}        
\usepackage{multicol}        
\usepackage[bottom]{footmisc}

\makeindex             

\begin{document}

\author{N.S. Gonchar}
\title{Mathematical foundations of information economics}
\subtitle{-- Monograph --}
\maketitle

\frontmatter

%
%

\preface

The state of economic theory and accumulated facts from the different branches of the economic science require to analyze the concept of the  description of economy systems. The economic reality generates the problems the solution of that is only possible by a new paradigm of the description of economy system. The classical mathematical economics
is based on a  notion of the rational consumer choice\index{rational consumer choice} generated by a certain  preference relation on some set of goods a  consumer wanted   and the concept of maximization of the firm profit\index{maximization of the firm profit}. The sense of   the notion of the rational consumer choice  is that it is determined by   a certain utility  function,\index{utility  function} defining
the choice of a  consumer by maximization of it on a certain budget set of goods.\index{ budget set of goods}
Moreover,  choices of consumers are independent. In the reality  choices of consumers are not independent because they depend on the firms supply.\index{firms supply}

Except the firms  supply, the   consumer choice\index{consumer choice} is also determined by  information about the state of the economy system that the consumer has and respectively evaluates   at the moment of the choice. In turn, the firms supply is made on the basis  of needs of the consumers\index{needs of the consumers} and their buying power.
By  information about the state of the economy system we understand a certain information about the equilibrium
price vector  and   productive processes realized  in the economy system under the equilibrium  price vector.\index{equilibrium  price vector}

 The new concept of the description  of the economy system is  to construct the stochastic model of the  economy system\index{stochastic model of the  economy system} based on the principle that the  firms supply  is primary and the  consumers  choice is secondary. Consumers make their choice  having information about the state of the economy system that  is taken into account by them under the choice. Firms make decisions relative to  productive processes  on the basis of  information about the real needs of consumers\index{real needs of consumers} and this principle is called the agreement of the supply  structure with  the choice structure.\index{agreement of the supply  structure with  the choice structure}

To construct such  a theory, it is necessary  to formulate  adequate to the  reality the theory of  consumers choice and  decisions making  by firms. Thus, the main foundations of the stochastic model of economy system are the notions of  consumers choice and   decisions making by firms\index{consumers choice and   decisions making by firms} relative to  productive processes. Under uncertainty these two notions have the stochastic nature. Besides, the theory of consumers choice\index{consumers choice} must take into account the structure of supply and mutual dependence of the consumers choice.

So, the sense of the new paradigm of the description of the economic phenomena is to construct   models of the economy systems,  describing adequately   consumers choice  and   decisions making  by firms\index{decisions making  by firms} relative to  productive processes, that give possibility to use them for finding the conditions of the stable growth of  real economy systems.

To construct the stochastic model  of  economy system, it is necessary:\\
1) to formulate a new concept of consumers choice and decisions  making  by firms relative to  productive processes in that it is taken into account that:\\
a) consumers do their own choice having a certain information  about the state of the economy system at the moment of the choice  and this information is respectively  taken into account by them  under the choice;\\
b) under uncertainty conditions   we describe consumers choice  by mathematical objects  having the probability nature and  choices of consumers are mutually dependent in the probability sense;\\
c)   we describe  decisions  making by firms\index{decisions  making by firms} relative to productive processes, taking into consideration uncertainty in actions of competitors,  by objects of  the probability nature   that include
mutual dependence of decisions  making   by firms.\index{mutual dependence of decisions  making   by firms}\\
d) as a result, in  such a model of economy system, firms produce  the final product\index{ final product} that economic agents  consume.

In the monograph, the original results, developing essentially ideas of the author, that were  appeared   initially in the papers of the author and  were  after summarized in the monographs  \cite{55, 71},  are presented .

In this monograph, the following results are  first obtained:

the axioms of consumers choice and    decisions making by firms\index{axioms of consumers choice and    decisions making by firms} relative to productive processes are formulated that take into consideration the stochastic character of the consumers choice and their mutual dependence, the stochastic character of  decisions  making by  firms and their mutual dependence;

a  notion of the comprehensive description of a  consumer choice \index{comprehensive description of a  consumer choice }is introduced that contains  information
about the state  of the economy system and the possible evaluation of this
information by the consumer under the choice;

  new mathematical objects such as  random fields of consumers choice\index{random fields of consumers choice} and   decisions making by firms\index{random fields of decisions making by firms}  describing  consumers choice and   decisions making  by firms are constructed;

 the existence of the  common probability space  is proved on which  the existence of  random fields of consumers choice and  decisions making  by firms   satisfying  axioms of consumers choice and   decisions making  by firms  is established;

a notion of productive economic process\index{productive economic process} is introduced that allows to construct model economy systems in that the final product is produced in the process of the  economy operation and
 to single out model economy systems that, for example, operate only  in profit regime or certain
firms  have   preferences for their formation in the period of the economy operation.
With the help of  certain productive economic processes and on the basis of the same firms, whose production is described by the same technological maps, one can construct model  economy  systems with  different  strategies of development;

 important notions of income  pre-functions\index{income  pre-functions  of consumers}  and  income  functions of consumers\index{income  functions of consumers} under uncertainty are introduced that are  objects determining  the theory of economic equilibrium;

on the basis of a new
notion of conditionally independent random fields,\index{conditionally independent random fields}
the  constructive mathematical theory of random fields of consumer choices\index{random fields of consumer choices} is built. In the economy science,  a new notion of  random fields of evaluation of information by consumers\index{random fields of evaluation of information by consumers}  determining the structure of consumers choice   is introduced; the  random field of evaluation of information by a consumer  is  evaluation of a certain available information about the state of economy system that influences  his choice;  subjective evaluation of information by a consumer
and  incomplete information about the state of economy system\index{incomplete information about the state of economy system} that the  consumer has and uncertainty in actions of competitors relative to their strategy of behavior is a reason for that  the evaluation of the available  information by the consumer is rather random;

a general dependence of random fields of consumers choice\index{ random fields of consumers choice} on   random fields of   decisions making\index{random fields of   decisions making} by firms and  random fields of evaluation of information by consumers is established;

examples of the specification of random fields of consumers choice and  decisions making by firms satisfying axioms formulated are presented and the correspondence between the classical description of a consumer choice and proposed one in this monograph is established;
possible enclosures of the classical description and proposed  one are realized under correspondent interpretation of  the consumer utility function;

illustrations of the stochastic models of economy systems  are given;

a special class of  convex down technological mappings\index{class of  convex down technological mappings}  is introduced and  the  belonging of it to the CTM (compact technological mappings) class in a wide sense  is proved;

for the   special  technological map the     Lemma on the existence of  a continuous strategy of firm behaviour\index{Lemma on the existence of  a continuous strategy of firm behavior}  being arbitrarily close by profit  to the optimal strategy of firm behaviour\index{optimal strategy of firm behavior}
that is,  generally speaking, discontinuous one
 is proved;

a fundamental Theorem on the  existence of a continuous strategy of firm behaviour\index{ Theorem on the  existence of a continuous strategy of firm behavior} being arbitrarily close by profit  to the optimal strategy  of firm behavior is proved  for technological maps from the CTM class in a wide sense,\index{technological maps from the CTM class in a wide sense}  optimal strategies of that are generally speaking discontinuous;

the structure of convex down  and  Kakutani continuous technological maps\index{Kakutani continuous technological maps}  is found. The Theorem about   continuation of the  two   Kakutani continuous and convex down technological maps  on the  minimal convex span of the expenditure sets of these technological maps is proved. As a result of the proved Theorem,  the statement about continuation of a  technological map from the CTM class in a wide sense\index{ technological map from the CTM class in a wide sense} to a convex down technological map from the CTM class\index{technological map from the CTM class}  is proved. It is  very important in the case of default of the firm described by a convex down technological map from  the CTM class in a wide sense;

the conditions of the  absence of arbitrage\index{absence of arbitrage} in the constructed model of economy system are found both under aggregative   and unaggrega\-tive description of economy systems;

algorithms of finding   equilibrium states in the  proposed models of economy   under uncertainty  conditions are constructed;

a notion of equilibrium state based on minimization  of the unused capital  function\index{unused capital  function} is introduced; on this basis   notions of the  minimal and maximal degrees  of  possibility of the financial crisis\index{minimal and maximal degrees  of  possibility of the financial crisis} are given;

under  agreement conditions of  the supply structure  with the choice  structure,  the mathematical theory of the control over  economy systems, whose technologies of firms  production
 are described by  convex down technological maps from the  CTM class  is constructed. For this purpose a notion of the economically compatible  productive processes of firms\index{economically compatible  productive processes of firms} is introduced.
On the basis of this  notion   an algorithm of construction of economically admissible taxation systems\index{economically admissible taxation systems} under which all firms in equilibrium state operate without default  in  the regime given by an  productive economic process\index{ productive economic process}  is constructed;

 in the exchange models with constant elasticity\index{exchange models with constant elasticity} of demand,  the conditions are found under which   the consumption of  goods takes place  in accordance with the structure of demand in the equilibrium state. An  algorithm of the construction of such equilibrium states is proposed;

for  models of economy systems with proportional consumption\index{economy systems with proportional consumption}    described  aggregately,   conditions for a  structure matrix of production,\index{structure matrix of production}  a matrix of  unproductive consumption,\index{matrix of  unproductive consumption}  levels of consumption\index{levels of consumption} and taxation, and an output vector are established  under which every industry of  economy is profitable in the state of economic equilibrium and consumers satisfy their own needs in accordance with the demand structure.
A constructive algorithm of finding  of the economic equilibrium state  that provides distribution of  produced goods  in accordance with the demand structure  is given and   the uniqueness  of that state up to constant  factor is proved;

the structure of  dependence of the output vector\index{ output vector} on the levels of taxation of industries,\index{levels of taxation of industries} the levels of  consumption,\index{levels of  consumption} the structure of production and consumption\index{structure of production and consumption} is found under which every industry is profitable and the  distribution of produced goods takes place  in accordance with the  demand structure;

 based on the principle of  openness of  economy systems,\index{principle of  openness of  economy systems} the  theory of economic transformations is constructed. A class of economic transformations\index{class of economic transformations} over characteristics of an economy system
 with proportional consumption, that do not change the equilibrium price vector, within the   non-aggregate description,  is investigated. Among the most important  transformations are those  that are connected with
the improvement of  production technologies,\index{improvement of  production technologies} increasing of  consumption and levels of the production, the change of taxation system;\index{taxation system}

to a non-aggregate  description of the economy system  an aggregate description in  cost parameters is put into correspondence and the set of  relations are found that allows  to construct  the structure matrix of unproductive consumption\index{structure matrix of unproductive consumption} by aggregate data that gives the possibility to an aggregate description  to put into correspondence
  non-aggregate one. The last gives the possibility to formulate  principles of   the economy of the social agreement\index{economy of the social agreement}. The theory of economic transformations is a good framework for the investigation of real economy systems to discover deformations that do not assist the permanent economic development;

under the  aggregate description of an  economy system with proportional consumption,\index{aggregate description of an  economy system with proportional consumption}    its reaction   on the influence of monopoly prices\index{monopoly prices} and the changes of the external economic relations\index{external economic relations }   is found.  The necessary and sufficient conditions are established for the characteristics of the economy system  under which
monopoly changes of prices in certain industries of economy  do not lead to the default neither in  certain industries of economy nor in  all the economy system. An algorithm of finding  changes  in the structure of the output vector,\index{output vector} levels of satisfaction of   consumers needs,\index{levels of satisfaction of   consumers needs} and levels of changes of prices for non-monopolists   is constructed;

conditions for a structure matrix of production,\index{structure matrix of production} an  unproductive consumption matrix,\index{unproductive consumption matrix} a matrix of initial supply of goods,\index{matrix of initial supply of goods} an output vector,  levels  of taxation\index{levels  of taxation} and a price vector of monopolists\index{price vector of monopolists} are established under that the vector of levels  of satisfaction    of consumers needs\index{vector of levels  of satisfaction    of consumers needs} is not changed. In this  case  an algorithm of finding    influence on the price vector of non-monopolists is constructed;

The necessary and sufficient conditions about absence of arbitrage opportunities in the case of additional investment in the economy system\index{additional investment in the economy system} are given.
The     general form Theorem giving conditions of absence of arbitrage opportunities in the class of self-financed strategies without short selling\index{class of self-financed strategies without short selling} between periods of economy operation is given.  The Theorem giving sufficient conditions of the existence of non-arbitrage dynamics\index{existence of non-arbitrage dynamics} is  proved.

All results  presented  in the monograph   was first  obtained by the author.

This book is extended English version of the book \cite{92}.

This work  was supported in  part by the State Fund of Fundamental Investigations.





\vspace{\baselineskip}
\begin{flushright}\noindent
Kiev\hfill {\it Nicholas Gonchar}\\
May, 2010 \hfill{\it}\\
\end{flushright}

\tableofcontents

\mainmatter

\chapter{Theory of choice and decisions making in information model of economy}

\abstract*{
The original approach to the description of economy systems is proposed. It is based on the principles: firms supply is primary and consumers choice is secondary, consumers make choice having information about the state of the economy system that is taken into account by them under the choice.  
The notions of the complete  description of consumer choice, random fields of consumer choice
and decision making  by firms are introduced, taking into account information about the state of economy and the fact that supply of firms is primary  and choice of consumers is secondary. Axioms for conditional distribution functions and unconditional distribution functions describing behaviour of consumers and firms  are formulated. The existence Theorem of random fields of consumers choice and decisions making by firms is proved. The Theorem giving a wide class of random fields that can be identified  with random fields of consumers choice and decisions making by firms is established.}

\section{Review of the achievements of the classical mathematical economics}

In this Section, the  original  approach  to the description of  economy systems that differs  cardinally from the classical description of the competitive economy systems  is proposed.
There are two basic principles of the description of the competitive economy system\index{competitive economy system} of the Walras type:\\
1) the maximization  by every firm of its  profit;\\
2) the  maximization by every consumer of its   utility function\index{utility function};\\
To realize  these two principles, it is assumed that the economy system is perfectly competitive\index{perfectly competitive} or, in other words,  no one of  firms can influence  the  prices in the  economy  system, so it takes into consideration them under  decisions making\index{decisions making} relative to  productive processes. Such a model is   ideal, so a question arises about the practical application of it.
It will have the practical application  in the case if the state of economic equilibrium in the model
is unique. Under these conditions, this model can serve for practical computations of the  rational assets allocation\index{rational assets allocation}  that provides the optimal satisfaction of needs of both  consumers and  firms. In the Arrow - Debreu model\index{ Arrow - Debreu model}  \cite{11, 13}, that is the well-known model of the Walras type,\index{model of the Walras type} the problem of the absence of arbitrage is solved. The algorithms construction problem
of computations  of equilibrium price vector\index{equilibrium price vector} that guarantee  the absence of arbitrage\index{absence of arbitrage} is very difficult.
Considerable  intellectual forces were applied to solve it (see, for example, \cite{14, 12, 15}).

The great part of publications is devoted to the  development of the analytical and numerical algorithms
to   find equilibrium price vectors, that is, such  price vectors under which demand does not exceed supply. The complexity of the problem is that it is necessary to solve a set of nonlinear inequalities  and  this is a difficult problem because there are no  algorithms to solve
 the set of nonlinear inequalities generally. In 1962  Uzawa (see \cite{14})  proposed the mathematical construction
on the base of that by the continuous mapping  of  the unit simplex in $R_+^n$ into itself the function of the  surplus demand can be constructed   such that all equilibrium states of such a model of the  economy are fixed points of this mapping.
 It should seem that this assertion with the Scarf algorithm\index{Scarf algorithm} \cite{12}  gives a hope to solve  the problem because for the finding   fixed points of the  nonlinear maps there also exist  other methods. Still on this way difficulties appear   connected with the construction of the function for the  surplus demand  of the real economic  model: consumers, described by certain utility functions, the property that they  have. The inverse procedure of  the  construction of a continuous map
of the  unit simplex in $R_+^n$ into itself by the function of the  surplus demand constructed by the utility functions of consumers and technological maps of firms is not valid.
 Eaves (1971) and Todd (1979) (see \cite{14}) gave a method to construct  the  nonlinear continuous map whose fixed points  are equilibrium states for a certain class of models of the  economy systems. The construction is a general one and the nonlinear continuous map they constructed theoretically  is a solution for a certain nonlinear programming problem with nonlinear
restrictions the  searching  of the solution of that under  the nonlinear  dependence of  profit of prices is practically unsolvable  analytically problem.
The Arrow - Debreu model\index{ Arrow - Debreu model} is the Walras type model  \cite{11, 13} for that  the existence theorem  of  economic equilibrium is proved, that is, the existence of the price vector   is proved under that  demand does not exceed  supply. This theorem is a considerable achievement in justification of the Walras idea on the  economic  equilibrium. Since more than 400 papers
on the problem of  economic equilibrium
were written  that in  various directions generalize ideas  having  appeared in the paper \cite{11}. In the  literature cited  in  \cite{14, 12}, four different approaches  to the problem of the existence of   economic equilibrium, that are close each other,   are considered. The first approach is based on the Kakutani or Brouwer  Theorems (see  \cite{ 13} and the literature cited therein), the second one  is devoted to the development of algorithms of combinatoric nature for the approximate calculation of equilibrium price vectors  \cite{14, 12}, the third approach is based on the theory of degree of map for proving
the existence of  equilibrium state (see  \cite{14}) and the fourth approach to search equilibrium states is developed by Smail  \cite{15}.

The model proposed by  Arrow  and  Debreu describes the economy system whose main assumption  is the supposition of  perfectly competitive economy system. No  firm operating in such an economy system does influence  the price equilibrium vector in the  economy system and so in their activity they make decisions orienting  on the existing equilibrium  price vector. Just under those conditions we can say  of the maximization of the  profit by firms. This is a basic assumption about the  strategies of firms behaviour. The  Arrow-Debreu  Theorem\index{Arrow-Debreu  Theorem}  \cite{11} of the existence of equilibrium regards only such economy systems.

 Certainly, this  mathematical idealization can have the  practical application  for the  optimal assets allocation.\index{optimal assets allocation}
In the  real economy systems  only  separate sectors of economy are   perfectly competitive.\index{perfectly competitive}
The  need  arises  to construct the theory of the description of such economy systems.
 In  such economy systems it is difficult  to say about the maximization by
firms of their profits because they all can influence  the equilibrium price vector so it is not possible  to say   relative to  what  price vector the  optimal strategy of firm behaviour\index{optimal strategy of firm behaviour} must be computed. Therefore  the firm almost never knows whether it realizes  the optimal strategy of firm behaviour under uncertainty. Thus, under   uncertainty conditions the  strategy of firm behaviour is random and  depends on the level of marketing, that is, from the  art to predict actions of competitors and  to evaluate the demand for the  product produced and the other factors, for example,
the possibility of the  firm to have  access to  credit resources.
The  strategy behaviour of the  consumer\index{strategy behaviour of consumer} is also random. Of course, the choice of the consumer\index{choice of the consumer} will depend not only on  his tests  but also on  the supplies of firms. Thus, to describe such economy system it is necessary to go out of the classical
description of the  consumers choice and decisions making by firms.\index{classical
description of the  consumers choice and decisions making by firms}
It is important to have a variant of the theory as the demand of a consumer depends on
the simple economic characteristics such as the profit of the consumer\index{profit of the consumer} and fractions of the profit coming on the consumption of  goods.

The other factor that influences  the price vector\index{price vector} is the demand for goods that varies permanently and determines  an equilibrium price vector.\index{ equilibrium price vector} Under these conditions, the randomness of  behaviour strategies of consumers and firms\index{behaviour strategies of consumers and firms} is more adequate. The urgent need exists to develop the description of economy system that takes into consideration  the stochastic character of the  consumers choice and supply of firms.

Thus, we assume that the consumer choice has stochastic character and we describe it by a random field of consumer choice.\index{random field of consumer choice}  We describe the choice of the  each consumer by the random field of consumer choice defined on the set of possible prices (for example, nonnegative orthant of $n$ dimensional arithmetic space) that takes values in the set of possible goods in the  economy system.  We also describe   decisions making by a firm by a random field of decisions making by the  firm\index{random field of decisions making by the  firm} that takes  values in the set of possible productive processes of the  firm.    We call a realization of  the random field  describing the consumer choice the strategy of consumer behaviour\index{strategy of consumer behaviour} and a realization of the  random field  describing  decisions making by the  firm the strategy of firm behaviour.\index{strategy of firm behaviour}

In the classical theory,  the choice of a consumer is determined by an utility function,   decisions making by a  firm is described  by optimal plans, and the possible influence  of decisions of each firm on the decisions of the other ones is not taken into account. Also the stochastic nature of the consumer choice and the fact  that the choice of consumers are mutually dependent are not taken into consideration. Without such a dependence of consumers choice between themselves, it is not possible to explain that the  distribution of welfare\index{distribution of welfare} in certain societies has the Pareto law\index{Pareto law} and the probability distribution of the  prices of assets has fat  tails \cite{42, 6, 7, 76, 56, 71, 72, 73, 80, 104, 103, 105, 60, 43, 44, 62, 40, 41}.

The stochastic description of a consumer behaviour  more corresponds to his real  economic behaviour. As regards the choice of a productive process by the firm, then it also has  the  stochastic nature
because, for example, the technology line may  fail, the supply of input vector may be disrupted
or unforeseen actions of competitors  bring to make  a  new decision relative to the  productive process in the framework
of the existing set of productive processes. The  proposed description joins macro economy and micro economy, that is,  the governing of strategy of a firm behaviour by credit,  finance and tax policies. By strategy of   firm behaviour we understand the decision that it makes relative to input  and output  vectors or in other words what  productive process one uses under a certain price vector $p=(p_1 ,\dots ,p_n).$ The random fields that describe the economy system  in a certain time interval is determined by the  finite dimensional distribution functions.

\section{Description of  economy with   availability of information about the state of economy}

In this Section on the basis of the  stochastic experiment,\index{stochastic experiment}   notions of the  complete description
of consumers choice,\index{complete description of consumers choice}
random fields of consumers choice\index{random fields of consumers choice}  and decisions making by firms\index{random fields of decisions making by firms}
are introduced. All these notions contain the principle:  firms supply is  primary   and   consumers choice is secondary \cite{ 55, 71, 69}.
 Notions of income  pre-functions of consumers, \index{ income  pre-functions of consumers}  income  functions of consumers\index{ income  functions of consumers} and a productive economic process\index{productive economic process}  are
essential for a further consideration.
The  proved Lemma  \ref{1dl1} permits to construct  productive economic processes. Examples of  the productive economic processes\index{productive economic processes} are given. In the Subsection 1.2, axioms for the introduced conditional distribution functions\index{conditional distribution functions} and unconditional distribution functions\index{unconditional distribution functions}  are formulated. The Theorem \ref{ttl1} guarantees  the existence of continuation of the introduced family   of  finite
dimensional distributions\index{finite
dimensional distributions} to   measures on the introduced measured space  satisfying  conditions of the Kolmogorov theorem.\index{Kolmogorov theorem}  The  existence Theorem \ref{qtl2} of   consumers choice and decisions making by firms random fields  is proved. In the Subsection 1.3 the notion of the  conditionally independent random fields\index{conditionally independent random fields}  playing the fundamental  role in the formulation of that  supply of firms is primary and choice of consumers is secondary
is introduced. In the Lemmas  \ref{pl2} --- \ref{pl4}    random fields and their superposition are investigated  with the help of which in the Theorem  \ref{jl2} conditional independence\index{conditional independence} of the constructed random fields with respect to another family of random fields\index{family of random fields} is proved. The formula for the joint  finite
dimensional distribution  functions for such random fields is presented.
This Theorem,  auxiliary Lemmas \ref{ppl5}, \ref{d2l2},
\ref{d2l3}, \ref{d2l4}, and Theorems \ref{tl2}, \ref{pl5}
 permit to prove Theorems \ref{ptl3}, \ref{ptl4},  \ref{tl4}, \ref{ttl3} with the help of which a wide class of the  economy models that satisfy formulated axioms  are constructed.
Examples of determination of the conditional distribution functions\index{conditional distribution functions} and unconditional distribution functions\index{unconditional distribution functions}  are given.

\subsection{Stochastic model of economy under the availability of information to consumers and firms}

In the  contemporary economy system, the large number of participants of the  economic process act  interacting between themselves.  The chain of economic relations between  participants of the economic process and the interaction between them are not always foreseen. The main aim of the modeling of economic processes is  to construct a mathematical model in that nonessential factors are neglected. A model describes a real state of an economy if it describes observable facts both qualitatively and quantitatively. Goods are all that is produced for  sale. So, if it has a consumption value,\index{consumption value} then the measure of the consumption value\index{measure of the consumption value} is its price. The participants of the economic process are firms and consumers. We call them economic agents.\index{economic agents} By the economic process we understand the activities  of the  economic agents to produce goods and services. Economic agents
interact between themselves to transform one set of goods into another one with the aim to  satisfy needs of   society. A set of goods  transformed is called  input vector\index{input vector} and the set of goods  to that the input vector  is transformed is called  output vector.\index{output vector} A firm is a set of the productive processes and the aim of the firm is to product goods and services for consumption by firms and consumers. A goods, for example, is consumed if it is transformed into the other goods. From this point of view any firm is a consumer.
The main subject of the economic process
is a person. To restore   a labour force and intelligence, the person has to consume  various goods.
Under goods we understand  the material and  intelligent values, technologies, a labour force, services, and so on. Financial operations of banks, intermediary activity, trade, lease are kinds of services. In further consideration,  we assume that the set of possible goods\index{set of possible goods} is ordered. Every set of goods we describe by a vector $x=(x_1 ,\dots ,x_n ),$ where $x_i$ is  a quantity of units of the $i$-th
goods, $e_i $ is a unit of its  measurement,  $x_ie_i$ is the  natural quantity of the goods.
If $p_i $ is the price of the unit of the goods $e_i,$ \ $i=\overline{1,n},$ then
$p=(p_1 ,\dots ,p_n )$ is the price vector\index{ price vector} that corresponds to the vector of goods\index{vector of goods}
 $(e_1 ,\dots ,e_n ).$ The price of vector of goods $x=(x_1 ,\dots ,x_n)$ is given by the formula $\left\langle p,x \right\rangle =\sum \limits_{i=1}^n p_i x_i.$ The set of  possible goods\index{set of  possible goods} in the considered period of the economy system  operation is denoted by $S.$ We assume
that  $S$ is  a convex subset of the set $R_+^n.$  Since for further consideration only
the property of convexity is   important,  we assume,  without loss of generality,
that  $S$ is a certain $n$-dimensional parallelepiped that can coincide with $R_+^n.$
Thus, we assume that in the economy system the set of possible goods $S$   is a convex subset of the non-negative orthant $R_+^n $ of  $n$-dimensional arithmetic space $R^n,$ the set of possible prices\index{set of possible prices} is a certain cone  $K_+^n, $ contained in $R_+^n \setminus \{0\},$ and  that can coincide with $R_+^n \setminus \{0\}.$   Here and further, $R_+^n \setminus \{0\}$ is the cone
formed  from the nonnegative orthant\index{nonnegative orthant} $R_+^n $ by ejection of the null vector  $\{0\}=\{0, \ldots, 0\}.$ Further, the cone  $R_+^n \setminus  \{0\}$ is denoted by $\bar R_+^n .$
The fact that the set of  possible prices is not  $\bar R_+^n$ obligatory  will be understand when we will consider random fields discontinuous on $\bar R_+^n$ whose contraction  on a certain cone
$K_+^n$ is yet continuous random fields of choice. We assume that  the Euclidean metric is introduced in the cone  $K_+^n.$
\begin{definition}
A set $K_+^n \subseteq \bar R_+^n$ is called a nonnegative cone\index{nonnegative cone} if together  with a point $u \in K_+^n $  the point  $t u$ belongs  to the set  $K_+^n $  for every real $t>0.$
\end{definition}
To describe an  economic system,  it is necessary to describe the behaviour of consumers and firms and their interaction.\index{behaviour of consumers and firms and their interaction} We put that in the economy system there are $l$ consumers,  $m$ firms, and $n$ kinds of goods.
\begin{definition} By  a technological map\index{technological map} $F(x), \ x \in X,$ we understand a many-valued map\index{ many-valued map} defined on a  set
$X\subseteq S$  taking  values in the set of all subsets of the set $S.$ Any vector $x\in X$ is called  an input vector\index{input vector} and its image $F(x)$  is called a set of plans of the technological map\index{set of plans of the technological map} $F(x).$ The set $X$  is called the expenditure set\index{expenditure set} and the pair $z=(x,y),$
where  $x\in X,$  $y \in F(x),$ is called  the productive process.\index{productive process} The set $ \Gamma= \{(x,y), \ x \in X, \ y \in F(x) \}$ we call the set of possible productive processes\index{set of possible productive processes for the technological map} for the technological map $F(x).$
\end{definition}
The choice of the   $i$-th consumer we describe by a random field\index{ random field} $\xi_i (p),$ $i=\overline{1,l},$ given on a probability space $\{\Omega, {\cal F}, \bar P\}$  that maps  the set of possible prices\index{set of possible prices} $K_+^n$ into the set of possible goods\index{ set of possible goods} $S\subseteq R_+^n.$ The  economic sense of
the  random field\index{random field} $\xi_i (p)$  is a choice from the set of possible goods $S$ that the $i$-th consumer makes under the price vector $p \in K_+^n$ and the realization $ \omega \in \Omega.$
We assume that $\xi _i (t p)=\xi _i (p)$ for all real $t>0.$ Any $j$-th firm activity we describe by a technological map $F_j(x)$ defined on the expenditure set $X_j$ and taking  values in
$2^S$ and by a random field of decisions making\index{random field of decisions making} $\zeta_j(p)$ mapping the set of possible prices    $K_+^n$ into  the set of  possible productive processes $\Gamma_j=\{(x,y),\  x \in X_j, \ y \in F_j(x)\}$ of the  technological map $F_j(x).$

It is convenient to present the random field $\zeta_j(p)$ by the pair of  random fields $(\zeta_j^{(1)}(p), \zeta_j^{(2)}(p))
=\zeta_j(p)$, where $\zeta_j^{(1)}(p)$ takes  values in the set  $X_j$ and $\zeta_j^{(2)}(p)$
takes  values in the set of plans $F_j (\zeta_j^{(1)}(p)).$
We suppose that $\zeta_j (t p)=\zeta _j(p)$ for all real $t>0.$
To determine a random field, one needs  to give a consistent family of   finite dimensional
distribution functions.\index{consistent family of finite dimensional
distribution functions} The discussion of the question of the definition of  random fields describing the  consumers and firms behaviour will be given later. Now we restrict ourselves
to  consideration  of the case when all random fields are treated  for the one price vector  $p.$
The set of random values $\zeta(p)=\{\zeta_1(p),\dots ,\zeta_m(p)\}$ is described by the probability measure of decisions making\index{probability measure of decisions making} relative to productive processes
\begin{eqnarray*}
\psi_{p}(B)=\bar P(\{\omega,(\zeta_1(p),\dots ,\zeta_m(p))\in B\}),
\end{eqnarray*}
where  $B$ is a certain Borel subset of the direct product of the sets
\begin{eqnarray*}
 \Gamma_i =\{ z=\{x,y\},\ x \in X_i
,\ y\in F_i (x)\}, \quad i=\overline{1,m},
\end{eqnarray*}
that is,
 $B\subseteq \Gamma^m=\prod\limits_{i=1}^m
\Gamma_i.$
The probability measure  $\psi_{p}(B) $ completely characterizes  the interaction of firms for the given price vector  $p\in K_+^n.$   We also describe  every  $i$-th  consumer by the income function\index{ income function}
\begin{eqnarray*}
K_i (p,z),\quad p=\{p_1 ,\dots ,p_n\} \in K_+^n, \quad
z=\{ z^1 ,\dots , z^m \} \in \Gamma^m, \end{eqnarray*}
 given  on the set $K_+^n \times \Gamma^m $ that takes  values in the set $R_+^1 .$
The important characteristic of a consumer is a probability measure of choice\index{probability measure of choice} given on the budget set.\index{budget set} It determines tastes and needs of the consumer.

 Let us come to the  constructive determination of the random fields of  consumers choice that describe consumers behaviour.
In the traditional theory of consumption, any consumer is described by a certain utility function\index{utility function} given on a certain subset of the set of  possible goods \cite{13}. From the theoretical point of view the utility function is a convenient way to think of  the consumer  choice. Both from the theoretical and practical points of view the urgent necessity exists to describe  the real consumer behaviour adequately.

The classical description of an  economy system turned out adequate to the reality partially.
The mismatch is connected with that the choice of a consumer is not rational and the  behaviour of a firm
is such that the profit is not maximal for a long time. Namely, these two aspects:\\
1) the choice of  a consumer is rational;\\
2) firms maximize their profit \\
are fundamental in the classical description of the economy system.

In the stochastic description of the economy system we do not discard  these two aspects they are a special
case among the  other possibilities.

Let us come to the analysis of  the mismatch of the classical description  to the reality. The classical description of the consumer choice means that the choice realized by the consumer is determined
by the solution of the maximization problem for the utility function on a budget set.
In addition, it is assumed that the utility function is known to a consumer on some set of goods and his choice is that to choose the vector of goods maximizing the utility function. It is known that the ordinary consumer does not know his utility function and also does not know how to solve the maximization problem. In this description, it is not taken into consideration that the supply of firms is primary and the choice of consumers is secondary and these choices depend on  what productive processes are realized in the economy system and under what   price vector  choices are made. It is not  also taken into account that the supply of firms can influence  the equilibrium price vector that may be changed under action of supply and demand, the latter one generates the random choice of the consumer.
It means that if  even though there exists  a preference relation  on the set of goods  for the consumer,\index{preference relation  on the set of goods  for the consumer}
 then in view  of the price uncertainty  choice made by the consumer is unknown. It also follows that in the classical approach, information about the state of economy is not taken into account that does not correspond to the reality.  Of course, the choice of consumer takes into consideration  the available information about the state of the economy\index{information about the state of the economy} and those rational expectations   generated by such information. The same is concerned with decisions making by firms relative to productive processes.
 In the classical description, firms maximize their profit on the set of  possible productive processes; moreover, they make this independently  without any  attention to the actions of competitors.
The latter   is possible under the assumption of perfect competition\index{perfect competition} when firms and consumers cannot influence
 the equilibrium price vector.\index{equilibrium price vector} In the opposite case, every firm tries to evaluate actions of  competitors, to take into account the possible fluctuation of the equilibrium price vector, and so  they make decisions relative to  using   productive processes on the basis of the available information of the economy system and the rational waiting of actions of competitors.
Thus, the choice of the productive process by a firm is a random one depending on the equilibrium  price  vector.

On the basis of the analysis made it is revealed that:\\
1) in the reality there are not  perfectly competitive economy systems;\index{ perfectly competitive economy systems}\\
2) on this reason firms can influence  the equilibrium price vector in the economy system and consumers can influence the choice of productive processes by firms that leads to:\\
the consumer choice is a random value depending on the state of economy system that determines decisions made by firms relative to  productive processes at a given price vector, that is,
the consumer choice is a random field defined on the set of possible prices with values in the set of  possible goods. Decisions making  by a firm is also a random value depending on actions of competitors and   evaluated demand   by the firm  for produced goods at a given price vector,  so the decisions making by the firm is a random field defined on the set of the possible prices with values in the set of  possible productive processes.

 The necessity arises:\\
1) to formulate axioms of the stochastic description of the economy system under that the   perfectly competitive economy system\index{perfectly competitive economy system} is a particular case;\\
2) to construct the mathematical theory of the  economic equilibrium\index{ economic equilibrium }  such perfect one as in the classical  model  of the competitive economy;\index{competitive economy}\\
3) to develop  algorithms of searching for the equilibrium price vectors in such economy  models.

\subsection{The complete description of  consumers  choice  and decisions making by firms. Random fields of   consumers choice  and  decisions making by firms}

Let $S$ be a set of possible goods\index{set of possible goods} in an economy system in that the Euclidean metric\index{Euclidean metric} is given and  ${\cal B}(S)$ be the Borel $\sigma$-algebra of subsets of the set $S.$
We assume
\begin{eqnarray*}
S=\{x=\{x_i\}_{i=1}^n, \ x \in R_+^n, \
0 \leq x_i \leq c_i,\  i=\overline{1,n}\}, \quad c_i>0, \quad i=\overline{1,n}.\end{eqnarray*}
We suppose that  there are $m$ firms in the economy system, the structure of production\index{structure of production} of the $i$-th firm is described by a technological map $F_i(x)$ defined on an expenditure set\index{expenditure set} $X_i,\ i=\overline{1,m}.$ It is convenient to assume that the firms are ordered and for the  further consideration,  the set of productive processes
\begin{eqnarray*}
\Gamma_i=\{(x,y), \  x \in X_i \subseteq S, \ y \in F_i(x)\}\end{eqnarray*}
of the $i$-th firm is only important, $i=\overline{1,m}.$
By $\Gamma^m=\prod\limits_{i=1}^m\Gamma_i$  we denote the direct product of the sets of productive processes $\Gamma_i, \
i=\overline{1,m},$ of the $m$ firms and by $[\Gamma^m]^k$ we denote the $k$-multiple  direct product
of the set  $\Gamma^m, \ k=1,2, \ldots  .$  We assume that in $[\Gamma^m]^k, \ k=1,2, \ldots$ the Euclidean metric is given and  ${\cal B}([\Gamma^m]^k), \ k=1,2, \ldots$ are the Borel $\sigma$-algebras of subsets of the set $[\Gamma^m]^k, \ k=1,2, \ldots.$

 The budget sets of the  $i$-th  insatiable\index{budget sets of the  $i$-th  insatiable consumer} and of the  $i$-th non-insatiable consumer\index{budget sets of the  $i$-th  non-insatiable consumer} are correspondingly  given by the formulae
\begin{eqnarray*}
 X_{(p,z)}^i=\{x, \ x \in S, \
 \left\langle p,x \right\rangle=K_i(p,z)\},\quad p \in K_+^n, \quad z \in \Gamma^m, \quad i=\overline{1,l},
 \end{eqnarray*}
\begin{eqnarray*}
  \bar X_{(p,z)}^i=\{x, \ x \in S, \ \left\langle p,x \right\rangle \leq K_i(p,z)\}, \quad p \in K_+^n, \quad z \in \Gamma^m, \quad i=\overline{1,l},
\end{eqnarray*}
where $K_i(p,z)$ is an income function of the $i$-th consumer.

Further, we denote by $\hat X_{(p,z)}^i$ both the budget set of the  $i$-th  insatiable\index{ budget set of insatiable consumer} and the budget set of the  $i$-th  non-insatiable consumer\index{ budget set of non-insatiable consumer} that are determined by  $(p,z)$ and the income function\index{income function} $K_i(p,z).$ Let  ${\cal
B}(\hat X^i_{(p,z)})$ be the Borel $\sigma$-algebra of subsets of the set $\hat X_{(p,z)}^i.$
If for the ordered direct product of  the $k$  budget sets  determined by  $(p_s,z_s), \
 s=\overline{1,k},$ to introduce the notation
 \begin{eqnarray*}
 X_{(p,z)_k}^{k,i}=\prod\limits_{s=1}^k\hat X_{(p_s,z_s)}^i,
 \end{eqnarray*}
  then  ${\cal B}(X_{(p,z)_k}^{k,i})$ is  the Borel  $\sigma$-algebra of subsets of the set $X_{(p,z)_k}^{k,i},$ where we introduced the notation
 \begin{eqnarray*}
 (p,z)_k=\{(p_1,z_1), \ldots, (p_k,z_k)\},
\quad  (p_i, z_i) \in K_+^n\times \Gamma^m, \quad  i=\overline{1,k}.
\end{eqnarray*}

Below we describe a stochastic experiment\index{stochastic experiment} that will help to formulate axioms of the  stochastic model of economy.\index{ axioms of stochastic model of economy} Assume that the $i$-th consumer has the income  $K_i(p,z), \ i=\overline{1,l},$ that, of course, depends on what the  price vector and productive processes
are realized in the economy system. The construction of the income functions $K_i(p,z), \ i=\overline{1,l},$ and the axioms that they  satisfy will be presented in the next section.
We suppose that the consumer receives  this income every month and expends it completely buying  goods. Thus, we assume that the consumer is insatiable. Analogous consideration is valid with respect to the  non-insatiable consumer. The consumer can only  buy the goods that belong to the budget set. Let us divide the budget set on simplexes of  smaller size.
Buying goods  in a shop every month, the  consumer realizes the stochastic experiment.
It is necessary to note the consumer having certain income purchases various goods at different moments of time. That  is natural. To reveal the real behaviour of the consumer,
it is necessary to make  the series of  stochastic experiments. We continue stochastic experiments many  times as long as  the sampling  will be  representative.
Having  made this experiment sufficiently many  times $N,$  we can say about a frequency to choose a certain vector of goods\index{frequency to choose a certain vector of goods} from the   simplex
of the small size by the $i$-th consumer. Having made this experiment infinitely
many  times and reducing the size of the small simplexes to zero,  we assume to come to a certain probability measure on the budget set\index{probability measure on  budget set} of the $i$-th consumer. To reveal the complete behaviour\index{complete behaviour of  consumer} of the $i$-th consumer, we have to change the income of the $i$-th  consumer to determine the  probability measures on  various budget sets of the consumer and their direct products.

We suppose the axiom: the sequence of frequencies to choose a certain vector of goods in the stochastic experiment  made above  tends to a probability measure on the budget set $X_{(p,z)}^i$ when $N$  tends to infinity.

 \begin{definition}\label{chin1}
Let   an income function\index{income function}   $K_i(p,z), \ i=\overline{1,l},$ given on the set $K_+^n\times \Gamma^m$  describe the  income of the  $i$-th consumer,  and let  a technological map\index{technological map} $F_i(x), \ x \in X_i, \ i=\overline{1,m},$ describe the structure of production of the $i$-th firm.\index{structure of production of the $i$-th firm}
The description of   consumers choice  is completely  given if
a probability measure
 \begin{eqnarray*}
 F_{p_1, \ldots, p_k}^s(A^s | z_1, \ldots, z_k),\quad
A^s \in {\cal B}(X_{(p,z)_k}^{k, s}), \quad s=\overline{1,l},
\end{eqnarray*}
on  every direct product
$\{X_{(p,z)_k}^{k,s}, \ {\cal B}(X_{(p,z)_k}^{k,s})\}$ of budget spaces\index{direct product of budget spaces}
\begin{eqnarray*}
\{X_{(p_i,z_i)}^{s}, \ {\cal B}(X_{(p_i,z_i)}^{s})\},  \quad  i=\overline{1,k}, \quad k=\overline{1, \infty},\end{eqnarray*}
is given  for every $s=\overline{1,l}$ and any
\begin{eqnarray*}
\{p_1, \ldots,p_k\} \in [K_+^n]^k, \quad  \{z_1, \ldots,z_k\} \in [\Gamma^m]^k, \quad k=1,2, \ldots.  \end{eqnarray*}

The measure $F_{p_1, \ldots, p_k}^s(A^s | z_1, \ldots, z_k)$ is a probability that the $s$-th
consumer will choose a set of goods from the set $A^s \in {\cal B}(X_{(p,z)_k}^{k, s})$ under condition that  the productive processes
 $\{z_1, \ldots,z_k\} \in [\Gamma^m]^k $
was  realized  in the economy system at the price vectors  $\{p_1, \ldots,p_k\} \in [K_+^n]^k, $ correspondingly.
\end{definition}

\begin{definition}
Let the description of  choice of  $l$  consumers in accordance with the Definition \ref{chin1} be completely given and
\begin{eqnarray*}
F_{p_1, \ldots, p_k}^s\left(A_s\cap X_{(p,z)_k}^{k, s} | z_1, \ldots, z_k\right), \quad s=\overline{1,l}, \end{eqnarray*}
be a measurable mapping\index{ measurable mapping } of the measurable space
$\{[\Gamma^m]^k, {\cal B}([\Gamma^m]^k)\}$ into the measurable space
$\{[0,1], {\cal B}([0,1])\}$ for every fixed   $A_s \in {\cal B}(S^k)$ and $\{p_1, \ldots, p_k\} \in K_+^{nk}.$

If there exist a probability space\index{probability space} $\{\Omega, {\cal F}, \bar P\}$ and  $l$ random fields on it
$\xi_i(p), \ p \in K_+^n, \ i=\overline{1,l},$ being measurable mappings of the measurable space
$\{\Omega, {\cal F}\}$  into the measurable space $\{S, {\cal B}(S)\} $ for every fixed $p \in K_+^n,$
and $m$ random fields  $\zeta_i(p),\ p \in K_+^n, \ i=\overline{1,m},$ where $\zeta_i(p)$ is  a measurable mapping of the measurable space
$\{\Omega, {\cal F}\}$ into the measurable space $\{\Gamma_i, {\cal B}(\Gamma_i)\}, \ i=\overline{1,m}, $ for every fixed $p \in K_+^n,$
such that with probability 1 there hold equalities
\begin{eqnarray*}
 E\{\chi_{A_s}(\xi_s(p_1), \ldots, \xi_s(p_k))|\{\zeta(p_1), \ldots, \zeta(p_k)\}\}
\end{eqnarray*}
\begin{eqnarray}\label{allusja}
 =  F_{p_1, \ldots, p_k}^s\left(A_s \cap X_{(p,z)_k}^{k, s}| z_1, \ldots, z_k\right)|_{z_i=\zeta(p_i),\ i=\overline{1, k}}, \quad   A_s \in {\cal B}(S^k), \quad   s=\overline{1,l},\quad
\end{eqnarray}
\begin{eqnarray*}
 \zeta(p)=\{\zeta_1(p), \ldots, \zeta_m(p)\}, \quad  p \in K_+^n,
\end{eqnarray*}
then we call the random field $\xi_i(p)$ by  the random field of choice of the $i$-th consumer\index{ random field of   consumer choice } described by the ensemble  of  probability measures\index{ensemble  of  probability measures}
\begin{eqnarray*}
F_{p_1, \ldots, p_k}^i(A^i | z_1, \ldots, z_k),
\quad  A^i \in {\cal B}(X_{(p,z)_k}^{k, i}), \quad  i=\overline{1,l},
\end{eqnarray*}
 and the random field $\zeta_s(p),\ s=\overline{1,m},$ we do the random field of the decisions making\index{random field of  decisions making by  firm} by the $s$-th firm relative to productive processes.
\end{definition}

By $\chi_{A_s}(x_1, \ldots, x_k)$ we denote the indicator function of the set  $ A_s \in {\cal B}(S^k).$

Note that in the formula (\ref{allusja})  conditional expectation is taken with respect to the $\sigma$-algebra generated by the family of random values
 $\{\zeta(p_1), \ldots, \zeta(p_k)\}.$
In  other words, the formula  (\ref{allusja}) is an accurate notation of the equality
\begin{eqnarray*}
 \bar P\left(\{\xi_s(p_1), \ldots, \xi_s(p_k)\} \in A_s |\zeta(p_1)=z_1, \ldots, \zeta(p_k)=z_k\right)
\end{eqnarray*}
\begin{eqnarray*}
 =F_{p_1, \ldots, p_k}^s\left(A_s \cap X_{(p,z)_k}^{k, s}| z_1, \ldots, z_k\right), \quad A_s \in {\cal B}(S^k), \quad  s=\overline{1,l},
\end{eqnarray*}
that means that the supply of firms is primary, the choice of consumers is secondary, and
the probability that  the $s$-th consumer will choose a set of goods from the set $A_s \in {\cal B}(S^k)$ under conditions that  the productive processes  $\{z_1, \ldots,z_k\} \in [\Gamma^m]^k $ were realized in the economy system at the price vectors  $\{p_1, \ldots,p_k\} \in [K_+^n]^k $ correspondingly is given by the above formula.

\subsection{Productive economic process\index{productive economic process}}

 Define a set of functions that we  call  income pre-functions  of consumers.\index{income pre-functions  of consumers}
\begin{definition}\label{ddl1}
A set of functions   $K_i^0(p,u), $  $ i=\overline{1,l},$  given on the set   $K_+^n\times \Gamma^m$  with  values in the set  $R^1$ we call  income pre-functions   of consumers if  they satisfy conditions:\\
1. For every $p \in K_+^n$ \ $ K_i^0(p,u), \  i=\overline{1,l},$ is  a measurable mapping\index{measurable mapping} of the  space $\{\Gamma^m, {\cal B}(\Gamma^m)\}$ into the space
$\{R^1, {\cal B}(R^1)\}\ ;$  \\
2. For every  $p \in K_+^n$ the set
$D(p)=\bigcap\limits_{i=1}^l D_i(p)$ is not empty, where
\begin{eqnarray*}
D_i(p)=\{u \in
\Gamma^m,  \ K_i^0(p,u) \geq 0 \}, \quad  i=\overline{1,l}\ ;
\end{eqnarray*}
$3. \ K_i^0(t p,u)=t K_i^0(p,u), \  t>0,
 \ (p,u) \in K_+^n\times \Gamma^m,  \  i=\overline{1,l}\ ;$
\end{definition}

\begin{definition} A technological map\index{technological map}  $F(x), \ x \in X,$ is  the  Kakutani continuous from above\index{Kakutani continuous from above technological map} if for every  sequence
 $x_n \in X,$ $x_n \to x, \  x \in X, $ and for any sequence $y_n \in F(x_n)$ such that
  $y_n \to y,$
it follows that \ $y \in F(x).$
\end{definition}

\begin{definition} A technological map  $F(x), \ x \in X,$ belongs to the CTM
(compact technological maps\index{class of compact technological map   in a wide sense}) class  in a wide sense if the domain of its definition
$X \subseteq S$ is a closed bounded convex set,
$F(x)$ is    Kakutani continuous from above,  takes
 values in the set of   closed bounded convex subsets of the set   $S,$  moreover, there exists a compact set
$Y \subseteq S$ such that $F(x) \subseteq Y, \ x \in X.$
The technological map $F(x), \ x \in X,$ belongs to the CTM class\index{class of compact technological map}  if it belongs to the CTM class  in a wide sense and, in addition,  $0 \in  X, \ 0 \in F(0).$
\end{definition}

\begin{lemma}\label{ly1}
Let a  technological map  $F(x), \ x \in X,$  belong to the  CTM class  in a wide sense,\index{compact technological map in a wide sense}
then the set
\begin{eqnarray*}
\Gamma=\left\{(x,y) \in S^2, \ x \in X, \ y \in F(x) \right\}\end{eqnarray*}
is a closed subset of the set $S^2.$
\end{lemma}
The proof is an  obvious consequence of   belonging $F(x)$ to the  CTM class  in  a wide sense.
Further, we only consider technological maps  from the  CTM class  in a wide sense or
from the  CTM class.

\begin{lemma}\label{Ql1}
Let technological maps
$F_i(x), \ x \in  X_i, \ i=\overline{1,m},$  belong to the CTM class  in a wide sense  and let
nonnegative  property vectors $b_i(p,z)=\{b_{ki}(p,z)\}_{k=1}^n, \ i= \overline{1,l},$ be  measurable maps\index{measurable map} of the   measurable space $\{\Gamma^m,  {\cal B}(\Gamma^m)\}$ into
the   measurable space  $\{S,  {\cal B}(S)\}$ for every $p \in K_+^n,$
then the set
\begin{eqnarray*}
G(p)=\{ z \in \Gamma^m, \  R(p,z) \in S \}
\end{eqnarray*}
belongs to  ${\cal B}(\Gamma^m)$ for every $p \in K_+^n,$
where
\begin{eqnarray*}
R(p,z)= \sum\limits_{i=1}^m[y_i - x_i]+ \sum\limits_{k=1}^lb_k(p, z),
\quad  z^i=(x_i,y_i) \ \in  \ \Gamma_i, \quad z=\{ z^i \}_{i=1}^m.
\end{eqnarray*}
\end{lemma}
\begin{proof}\smartqed
To prove the Lemma, it is sufficient to prove that the map $R(p,z)$
is a measurable one of the space $\{\Gamma^m,  {\cal B}(\Gamma^m)\}$ into the space
$\{R^n,  {\cal B}(R^n)\}$ for every $p \in K_+^n.$ Let us denote by
\begin{eqnarray*}
\varrho_1(x,y)=\{\sum\limits_{s=1}^n(x_s-y_s)^2 \}^{1/2}\end{eqnarray*}
the Euclidean metric in $R^n,$  $x, y \in R^n$ and by
\begin{eqnarray*}
\varrho_2(w_1,w_2)=\{\varrho_1^2(u_1,u_2)+
\varrho_1^2(v_1,v_2)\}^{1/2}\end{eqnarray*}
the Euclidean metric in $R^{2n},$
where $w_1=\{u_1, v_1\},$ \ $w_2=\{u_2, v_2\},$  $u_i, \ v_i  $ $\in R^n, \ i=1,2.$
The map $t_i(z^i)=y_i - x_i,$ $i=\overline{1,m},$ of the space $R^{2n}$  into the space $R^{n}$
 is a continuous one  if for every open set  $D \in {\cal B}(R^n) $ its pre-image $t_i^{-1}(D)$ is an open set in $R^{2n}.$  To prove this, let us to show
that if $(x_i^0, y_i^0) \in
t_i^{-1}(D), $ then $(x_i^0, y_i^0)$ belongs to the set $t_i^{-1}(D)$ with a certain neighborhood.
As the point $y_i^0 - x_i^0 \in D$ it  follows that there exists the ball $C_{\delta}^n(y_i^0 - x_i^0)$ of a  radius $\delta >0$ with the center at the point $y_i^0 - x_i^0$ that belongs to $D.$ But the image of the ball  $C_{\delta/2}^{2n}(\{x_i^0, y_i^0\}) \subset  R^{2n} $
with the center at the point $\{x_i^0, y_i^0\} \in R^{2n}$ of the radius $\delta/2$  under the map $t_i(z^i)$ belongs to  the ball $C_{\delta}^n(y_i^0 - x_i^0) \subseteq D. $
The latter  is  a consequence of the inequality
 \begin{eqnarray*}
  \varrho_1(t_i(w_1), t_i(w_2))=\{\sum\limits_{k=1}^n[(v_1^k -u_1^k) -(v_2^k -u_2^k)]^2\}^{1/2}
\end{eqnarray*}
 \begin{eqnarray*}
 \leq \{2\sum\limits_{k=1}^n(u_1^k- u_2^k)^{2}+ 2\sum\limits_{k=1}^n(v_1^k- v_2^k)^{2}\}^{1/2} \leq 2 \rho_2(w_1, w_2).
\end{eqnarray*}
 Therefore the preimage $t_i^{-1}(C_{\delta}^n(y_i^0 - x_i^0))$ of the ball $C_{\delta}^n(y_i^0 - x_i^0)$ with the center at the point  $y_i^0 - x_i^0 \in D$ contains the ball
$C_{\delta/2}^{2n}(\{x_i^0, y_i^0\}) \subset   t_i^{-1}(D). $
It means openness of $t_i^{-1}(D). $ The contraction of the map\index{contraction of the map} $t_i(z^i)=y_i - x_i$ on the set  $\Gamma_i \subseteq R^{2n}$ that is closed and so belongs to ${\cal B}(R^{2n})$ is a measurable map of the space  $\{ \Gamma_i, {\cal B}(\Gamma_i)\}$ into the space $\{R^n, {\cal B}(R^n)\}.$

Let us consider  the map $t_i^1(z)=t_i(z^i)$ on   $\{ \Gamma^m, {\cal B}(\Gamma^m)\}.$
It is obvious that $t_i^1(z)$ is a measurable  map  of the space $\{ \Gamma^m, {\cal B}(\Gamma^m)\}$ into the space  $\{R^n, {\cal B}(R^n)\}.$
To finish the proof of the Lemma, let us show that the vector sum of the measurable  maps
 $f_1(z)$ and  $f_2(z)$ of the space $\{\Gamma^m,  {\cal B}(\Gamma^m)\}$  into the space
$\{R^n,  {\cal B}(R^n)\}$ is a measurable map of the space $\{\Gamma^m,  {\cal B}(\Gamma^m)\}$ into the space $\{R^n,  {\cal B}(R^n)\}.$
It follows from the equality
\begin{eqnarray*}
 \{z \in \Gamma^m, \ f_1(z) + f_2(z) \in L(b) \}
\end{eqnarray*}
\begin{eqnarray*}
  =\bigcup\limits_{r \in N^n} \{z \in \Gamma^m, \ f_1(z)  \in L(r) \} \cap
\{z \in \Gamma^m, \ f_2(z)  \in L(b - r) \},
\end{eqnarray*}
where $N^n$ is the set of points of the $n$-dimensional arithmetic space  $R^n$ with  rational coordinates and $L(b)=\prod\limits_{i=1}^n(-\infty, b_i), \
b=\{b_i\}_{i=1}^n \in R^n.$  Since the set of open sets $L(b) \subset R^n , \ b=\{b_i\}_{i=1}^n \in  R^n$ generates  ${\cal B}(R^n),$ the  statement is proved.
 \qed \end{proof}

Thus, $ R(p,z)$ is  a measurable  map  of the space $\{\Gamma^m,  {\cal B}(\Gamma^m)\}$ into the space $\{R^n,  {\cal B}(R^n)\}$ for every $p \in K_+^n. $ But $S$ is a Borel subset of $R^n.$ From here it follows the proof of the Lemma because the preimage of it is a measurable set.

Economic sense of the set $G(p)$ is the set of productive processes under   that  firms  produce final products in the economy system at the price vector $p.$

Let us associate with the $i$-th
consumer a certain nonnegative vector
$b_i(p,z)$ $=\{b_{ki}(p,z)\}_{k=1}^n,\ (p,z)  \in K_+^n\times \Gamma^m,$ that we call the supply vector of   goods\index{supply vector of   goods} which the $i$-th consumer has at the beginning of the period of the economy operation and that is a measurable map of the space $\{\Gamma^m, {\cal B}(\Gamma^m)\}$ into the space $\{S, {\cal B}(S)\}$ for every $p \in K_+^n $ and such that  $b_i(t p,z)=b_i(p,z), \ $ $  t > 0, \ i=\overline{1,l}.$
The vector  $b_i(p,z)$ $=\{b_{ki}(p,z)\}_{k=1}^n,\ (p,z)  \in K_+^n\times \Gamma^m,$ we call the   property vector of the $i$-th consumer under the realized price vector $p \in R_+^n $ and the set of productive processes $z \in \Gamma^m.$

\begin{definition}\label{ddl2}
Let  the production of the $i$-th firm in the economy system  be described by technological map
$F_i(x),\ x \in  X_i,\ i=\overline{1,m},$  the   $i$-th consumer have  property vector\index{property vector}
$b_i(p,z), \ i=\overline{1,l},$
and for every $(p,z) \in  K_+^n\times \Gamma^m$  the set of productive processes
\begin{eqnarray*}
(X_i(p,z),Y_i(p, z)),  \quad    X_i(p, z)  \in X_i, \quad
   Y_i(p, z)  \in F_i(X_i(p, z)), \quad  i=\overline{1,m},\end{eqnarray*}
   satisfy
conditions:\\
1. $(X_i(p,z),Y_i(p, z))$ is a measurable  map of the space
 $\{\Gamma^m, {\cal B}(\Gamma^m)\}$ into the space
$\{\Gamma_i, {\cal B}(\Gamma_i)\}, \ i=\overline{1,m},$
for every $p \in K_+^n,$ where the set
\begin{eqnarray*}           \Gamma_i=\{(x,y) \in S^2, \ x \in X_i, \ y \in F_i(x)\}\ ;
\end{eqnarray*}
2.  $(X_i(t p,z),Y_i(t p, z))=(X_i(p,z),Y_i(p, z)), \ t>0, \
(p,z) \in  K_+^n\times \Gamma^m.$\\
A measurable  map   $Q(p,z)$ of the space $\{\Gamma^m, {\cal B}(\Gamma^m)\}$  into itself for every  $p \in K_+^n$  defined by the formula
\begin{eqnarray}\label{qgl1}
Q(p,z)=\{(X_i(p,z),Y_i(p, z))\}_{i=1}^m
\end{eqnarray}
we call  productive economic process if for every $p \in K_+^n$  the set of values  $Q(p,\Gamma^m)$ of the map
$Q(p,z)$ belongs to the set $G(p)$ constructed in the Lemma \ref{Ql1}, where as measurable  map  $b_i(p,z), \ i=\overline{1,l},$
 the set of  supply vectors  of  goods of consumers
$b_i(p,z), \ i=\overline{1,l},$ is chosen at the beginning of the economy operation.
\end{definition}

Since we construct the stochastic model of economy, it is convenient to assume that the  vector of initial supply of goods $b_i(p,z)$ of the  $i$-th consumer, $ i=\overline{1,l},$ depends on  price vector $p \in K_+^n$ and  productive process $z  \in \Gamma^m$  realized  in the economy system. It is justified  by that the  market conjuncture  cannot
 be predicted accurately, therefore the $i$-th consumer can contract to supply a vector of goods
$b_i(p,z)$ from  outside that, of course, depends on  productive process $z \in \Gamma^m$
and  price vector $p \in K_+^n$ that  will be realized in the economy system  in the next period of the economy operation. In a special case we  can assume that the vector $b_i(p,z),$ $ i=\overline{1,l},$  does not depend on $(p,z) \in K_+^n\times \Gamma^m.$

\begin{corollary}
Let technological maps from the CTM class  in  a wide sense $F_i(x),$ $ \ x \in X_i,\ i=\overline{1,m},$  describe the structure of  firms production in the economy system
 and  consumers have  property vectors
$b_i(p,z) \geq 0, \ i=\overline{1,l},$
 satisfying conditions: there exists a strictly positive vector  $b=\{b_i\}_{i=1}^n, \ 0< b_i <\infty, \ i=\overline{1,n}, $ such that  $b_i(p,z) \leq b, \ i=\overline{1,l},$  and the set of possible goods $S$ in the economy system is such that
\begin{eqnarray*}           \sum\limits_{i=1}^{l+m}Y_i \subseteq S,\end{eqnarray*}
where $Y_i=Y, \ i=\overline{1,m}, \ Y_i=b, \ i=\overline{m+1,l+m},$ and the set $Y$  is a common compact for all technological maps $F_i(x),$ $ \ x \in X_i,\ i=\overline{1,m},$ that contains their  set of values.
If for the productive economic process\index{productive economic process} $Q(p,z)$ constructed  the inequality
\begin{eqnarray*}
R(p, Q(p,z))=\sum\limits_{i=1}^m[Y_i(p, z) - X_i(p, z)]+
\sum\limits_{k=1}^lb_k(p, Q(p,z)) \geq 0 
\end{eqnarray*}
is valid, then in such the economy system
\begin{eqnarray*}
R(p, Q(p,z)) \in S.
\end{eqnarray*}
The latter  means that the set of values of the  productive economic process $Q(p,z)$ belongs to the set of productive processes $G(p)$ under that in the economy system the final product can be produced.
\end{corollary}
By the sum of sets  $\sum\limits_{i=1}^sT_i, \ T_i \in S,$ we understand the set of vectors of the kind $\sum\limits_{i=1}^sx_i, \ x_i \in T_i, \ i=\overline{1,s},$ and the inequality $a \leq b $ between two vectors $a=\{a_i\}_{i=1}^n,\ $ b=$ \{b_i\}_{i=1}^n \in R_+^n$ means that $a_i \leq b_i, \  i=\overline{1,n}.$
To verify whether the set of  values of the  productive economic process\index{productive economic process} $Q(p,z)$ belongs to the set $G(p)$ it is sufficient to prove that the set of values of the map $ R(p,Q(p,z))$ belongs to the set  $S.$ Further, we assume that for  the set of possible goods $S$ the conditions of the Consequence  formulated above are valid, so further we only verify the condition
$ R(p,Q(p,z)) \geq 0.$

\begin{definition}\label{dl1}
 A set of functions $K_i(p,z), $ $\ i=\overline{1,l},$ defined on the set
 $K_+^n\times \Gamma^m$  being measurable   maps of the space  $\{\Gamma^m, {\cal B}(\Gamma^m)\}$  into the space
$\{R_+^1, {\cal B}(R_+^1)\}$  for every $p \in K_+^n$ is named  income functions
if there exist a set of income pre-functions
$K_i^0(p,z),\ i=\overline{1,l},$
a productive economic process $Q(p,z)$
defined on the set $ K_+^n\times \Gamma^m$
and such that $Q(p,\Gamma^m)$
belongs to the set  $D(p)$ from the Definition \ref{ddl1}  for every  $p \in K_+^n,$ and the equalities
\begin{eqnarray*}
  K_i(p,z)=K_i^0(p,Q(p,z)), \quad  (p,z) \in K_+^n\times \Gamma^m,
\quad i=\overline{1,l}\  ;
\end{eqnarray*}
\begin{eqnarray}\label{sl0}
 \sum\limits_{i=1}^lK_i(p,z)=\left\langle  p, \sum\limits_{i=1}^m[Y_i(p, z) - X_i(p, z)]
+ \sum\limits_{k=1}^lb_k(p, Q(p,z)) \right\rangle,
\end{eqnarray}
\begin{eqnarray*}
 (p,z) \in K_+^n\times \Gamma^m,
\end{eqnarray*}
are valid.
\end{definition}

\begin{note}
From the Definitions  \ref{ddl2} and   \ref{dl1}  it follows with the necessity that
the set $D(p)\cap G(p) \not = \emptyset $
and $ Q(p,\Gamma^m) \subseteq D(p) \cap G(p).$
\end{note}
\begin{note} To realize in the economy system a certain  productive econo\-mic process\index{productive economic process}
$Q(p,z)=\{(X_i(p,z), Y_i(p,z))\}_{i=1}^m$     supply vectors of goods $b_i(p,z), \  i=\overline{1,l},$  may be needed at the beginning of the economy operation. Therefore, the nonnegative  supply vector of goods $b_i(p,z), \  i=\overline{1,l},$ must be  available  for the $i$-th consumer at the beginning of the economy operation
to realize the  productive economic process $Q(p,z).$
The map $Q(p,z)$ is a characteristic of the way  of production of the final product in the economy system  coming on the final consumption and storage in  the economy operation  period.
The supply vector of goods $b_k(p,z) $ may belong to an external firm that is the  $k$-th consumer  supplying it  into the economy system. The income of the $k$-th consumer from  this supply consists of the part of  firms profits that  use this   supply vector of goods for consumption and production.
\end{note}

Let $F(x)$ be a certain technological map defined on $X \subseteq S.$
The image of the technological map $F(x)$ for every  $x\in X$ is the set of plans of the firm
that it produces under the input vector $x\in X.$ We assume the set of plans $F(x)$ is a convex compact    set for every $x\in X.$ This assumption is a certain  restriction on the  set of values of
the technological map that is not essential in the practical cases. For the very popular models of production such as the Neumann model, the Neumann - Gale model and the  Leontieff model this condition  holds.
Let $X$  be a convex subset of the set $S.$ The requirement of convexity of $X$ has economic basis: if a firm  has costs to buy  certain vectors of goods  $x_1$ and  $x_2,$ then it can buy the vectors of goods  $\alpha x_1+(1-\alpha )x_2,$
$~\alpha\in (0,1).$ By the sum of two sets  $A$ and $B$ we understand the set of vectors of the kind $x_1+x_2,$ where $x_1\in A,$ $x_2\in B.$  The set $\alpha A,$ where $\alpha $ is a real number, is the set of vectors of the kind $\alpha x,~ x\in A.$

\begin{definition} We call a technological map\index{ convex down technological map}
$F(x),\  x \in X,$ convex down if
\begin{eqnarray}\label{1h1l18}
 F(\alpha   x_1+ (1-\alpha ) x_2)
\supseteq \alpha F(x_1)+(1-\alpha )F(x_2),
\end{eqnarray}
\begin{eqnarray*}
          1>\alpha >0,\quad x_1,\ x_2\in X.
\end{eqnarray*}
A technological map $F(x)$ is strictly convex down\index{strictly convex down technological map} if it is  convex down and, moreover,
\begin{eqnarray*}
 \sup\limits _{y\in F(\alpha x_1+(1-\alpha )x_2)}
                                           \left\langle p,y\right\rangle >
   \alpha  \sup\limits _{y\in F(x_1 )} \left\langle p,y\right\rangle +
(1-\alpha )\sup\limits _{y\in F(x_2 )} \left\langle p,y\right\rangle ,
\end{eqnarray*}
\begin{eqnarray*}
  p \not = 0, \quad  1>\alpha >0,\quad x_1\not =x_2 .
\end{eqnarray*}
\end{definition}

It is easy to prove that if a  technological map is convex down, then the inequality
\begin{eqnarray}\label{1h1l19}
\sup\limits _{y\in F(\alpha x_1+(1-\alpha )x_2)} \left\langle
	   p,y\right\rangle \geq \alpha  \sup\limits _{y\in F(x_1 )}
   \left\langle p,y\right\rangle + (1-\alpha )\sup\limits _{y\in F(x_2 )}
\left\langle p,y\right\rangle
\end{eqnarray}
is valid.
Really, from (\ref{1h1l18}) it follows
\begin{eqnarray*}
   \alpha y_1+(1-\alpha )y_2\in F(\alpha
x_1+(1-\alpha )x_2), \quad  y_i \in F(x_i), \quad i=\overline{1,2},
\end{eqnarray*}
therefore
\begin{eqnarray*}
 \sup\limits _{y_1\in
F(x_1), y_2\in F(x_2)} \left\langle p,\alpha y_1+(1-\alpha )y_2\right\rangle
\leq \sup\limits _{y\in F(\alpha x_1+(1-\alpha )x_2)} \left\langle
p,y\right\rangle.
\end{eqnarray*}
 Since
\begin{eqnarray*}               \sup\limits _{y_1\in F(x_1), y_2\in
F(x_2)} \left\langle p,\alpha y_1+(1-\alpha )y_2\right\rangle = \alpha
   \sup\limits _{y_1 \in F(x_1 )} \left\langle p,y_1 \right\rangle +
(1-\alpha )\sup\limits _{y_2 \in F(x_2 )} \left\langle p,y_2 \right\rangle \end{eqnarray*}
the statement is proved.
\begin{note}
If a technological map  $F(x),\  x \in X,$ belongs to the CTM class and is convex down, then for every  $0 \leq \alpha \leq 1$ the pair
$\{\alpha x, \alpha y \} $  is a productive process if $\{ x, y \}$
is also such one, that is, from   $ x  \in X, \ y \in F(x)$
it follows that  $\alpha x \in X, \ \alpha y \in F(\alpha x).$
Indeed, let vectors   $0 $ and $x$ belong to the expenditure set  $X.$ Then from the convexity  down of $F(x)$ we have
\begin{eqnarray*}
\alpha F(x)+ (1-\alpha) F(0) \subseteq F(\alpha x + (1-\alpha) 0).
\end{eqnarray*}
From here it follows that $\alpha F(x) \subseteq F(\alpha x).$ The latter means that the pair  $\{\alpha x, \alpha y \} $ is the productive process if $\{ x, y \}$ is that one.
\end{note}

The next Lemma is important for the  construction of  productive economic processes.

\begin{lemma}\label{1dl1}
Let  technological maps $F_i(x), \ i=\overline{1,m},$ belong to the CTM class and be convex down. If $u_i(p,z), \ i=\overline{1,m}, $ are measurable maps of the space $\{ \Gamma^m, {\cal B}(\Gamma^m)\}$ into the space
$\{ [0,1], {\cal B}([0,1])\}$ for every  $p \in K_+^n,$ then
\begin{eqnarray*}
T(p,z)=\{ u_i(p,z)z^i \}_{i=1}^m
\end{eqnarray*}
is a measurable   map of the space  $\{ \Gamma^m, {\cal B}(\Gamma^m)\}$
into itself for every $p \in  K_+^n.$
\end{lemma}
\begin{proof}\smartqed
It is sufficient to prove the Lemma in the case
\begin{eqnarray*}
u_i(p,z)=\sum\limits_{k=1}^sC_k^i(p)\chi_{B_k^i}(z), \quad
i=\overline{1,m},
\end{eqnarray*}
where
\begin{eqnarray*}
\chi_{B_k^i}(z)=
\left\{\begin{array}{ll}
               1, & \textrm{if} \quad z \in B_k^i,\\
            0, &      \textrm{if}  \quad  z \in \Gamma^m \setminus B_k^i  \textrm{,}
                                                                                                                                                \end{array}
                                                \right.
                                                \end{eqnarray*}
 $0 \leq C_k^i(p) \leq 1, \ p \in K_+^n, \ k=\overline{1,s},$ \ $ B_k^i \in
{\cal B}(\Gamma^m), \ \bigcup\limits_{k=1}^s B_k^i = \Gamma^m.$
In this case the map
\begin{eqnarray*}
 u_i(p,z)z^i=\sum\limits_{k=1}^sC_k^i(p)\chi_{B_k^i}(z)z^i
\end{eqnarray*}
is  a measurable one of the space   $\{ \Gamma^m, {\cal B}(\Gamma^m)\}$
into the space $\{ \Gamma_i, {\cal B}(\Gamma_i)\}$ because
 $u_i(p,z)z^i  \in \Gamma_i$ for any function $0 \leq u_i(p,z) \leq 1$
and as a consequence of the belonging of technological maps to the CTM class and convexity their  down. Really, if $D_i \in {\cal B}(\Gamma_i),$  then
\begin{eqnarray*}  \{z \in \Gamma^m, \ u_i(p,z)z^i  \in D_i\}
=\bigcup\limits_{k=1}^s B_k^i \cap \{z \in \Gamma^m, \ u_i(p,z)z^i  \in D_i\}
\end{eqnarray*}
\begin{eqnarray*}
 =\bigcup\limits_{k=1}^s B_k^i \cap \{z \in \Gamma^m, \ C_k^i(p)z^i  \in D_i\}
=\bigcup\limits_{k=1}^s B_k^i \cap R_k^i  \ \in  \ {\cal B}(\Gamma^m),
\end{eqnarray*}
where $R_k^i=\prod\limits_{r=1}^{i-1}\Gamma_r\times \Delta_i^k \times
\prod\limits_{r=i+1}^{m}\Gamma_r, \
\Delta_i^k =\{z^i \in \Gamma_i, \ C_k^i(p)z^i \in D_i\}.$ The set
$\Delta_i^k $ belongs to  ${\cal B}(\Gamma_i)$ since $C_k^i(p)z^i$ is a measurable map of the measurable space
$\{ \Gamma_i, {\cal B}(\Gamma_i)\}$ into itself. Moreover, it is a continuous map of $ \Gamma_i$ into itself.
Since any measurable map $u_i(p,z)$ of the measurable space $\{ \Gamma^m, {\cal B}(\Gamma^m)\}$ into the measurable space $\{ [0,1], {\cal B}([0,1])\}$ is the pointwise limit of the sequence of the considered above  maps, the pointwise limit $u_i(p,z)z^i$ of the sequence of the measurable maps $u_i^n(p,z)z^i$ is a  measurable map of the measurable space $\{ \Gamma^m, {\cal B}(\Gamma^m)\}$ into the  measurable space $\{ \Gamma_i, {\cal B}(\Gamma_i)\}.$
At last, the direct product of the  measurable maps  $u_i(p,z)z^i$ of the space   $\{ \Gamma^m, {\cal B}(\Gamma^m)\}$ into the space $\{ \Gamma_i, {\cal B}(\Gamma_i)\}, \ i=\overline{1,m}, $
is a  measurable map of the measurable space  $\{ \Gamma^m, {\cal B}(\Gamma^m)\}$  into itself.  \qed \end{proof}           

Let us give an example of income functions adequate to  economic reality.
Here and further in all examples the space of possible prices $K_+^n$ is any subcone of the cone
$\bar R_+^n.$
Denote by
\begin{eqnarray*}            \alpha _i(p,z)=(\alpha _{i1}(p,z),\dots
,\alpha_{im}(p,z)),\quad i=\overline {1,l},\end{eqnarray*}
the vector of  partial shares   of the  $i$-th consumer in  firms profits, then
$\alpha_{is}(p,z)$ is a part of the income of the $i$-th consumer that he obtains from the income of the $s$-th firm.  It may be the part that makes up the  gain of the $i$-th consumer,
share dividend of the  $s$-th firm or, for example, the return of credits with percent to  the $i$-th consumer by  the $s$-th firm and so on.

We describe the economy by a nonnegative continuous  matrix of  partial shares of consumers in   profits of  $m$ firms  $||\alpha _{ij}(p,z)||_{i=1~j=1}^{l\quad m}.$
From  economic point of view it is natural to assume that functions $\alpha _{is} (p,z)$
do not depend on scale of prices, that is, they are homogeneous functions of zero degree.
Therefore, we consider them on the simplex
\begin{eqnarray*}            P=\left\{p \in K_+^n,\ \sum\limits_{i=1}^np_i=1 \right\}\end{eqnarray*}
only and  profits of the $s$-th firm are distributed between all consumers.
Thus, we put that equalities
\begin{eqnarray*}
  \alpha _{is}(tp,z)=\alpha _{is}(p,z), \quad
s=\overline{1,m}, \quad i=\overline{1,l},\quad \forall t>0,
\quad p \in K_+^n,
\end{eqnarray*}
\begin{eqnarray}\label{sl000}
 \sum\limits _{i=1}^l\alpha
_{is}(p,z)=1, \quad s=\overline{1,m},\quad p \in P.
\end{eqnarray}
are valid.
We characterize the economy by two matrices  yet: a nonnegative taxation matrix\index{taxation matrix} $||\pi _{ij}(p,z)||_{i,j=1}^l $ and a nonnegative redistribution matrix\index{redistribution matrix} of profits  $||r_{ij}(p,z)||_{i,j=1}^l.$
 The redistribution matrix of profits may describe investment and  subsidy policies of the state.

Relative to these matrix elements we assume that both $\pi _{ij}(p,z)$ and $r_{ij}(p,z)$  are continuous on $K_+^n\times \Gamma^m$  and homogeneous of zero degree with respect to variable  $p.$
We assume that the equalities
\begin{eqnarray*}
 \sum\limits _{i=1}^l \pi
_{ij}(p,z)=1,\quad \sum\limits _{j=1}^l r_{jk}(p,z)=1,
\quad j,k=\overline{1,l}, \quad p \in P,
\end{eqnarray*}
\begin{eqnarray}\label{sl0000}
 \pi _{ij}(tp,z)=\pi _{ij}(p,z),\quad r_{ij}(tp,z)=r_{ij}(p,z),
\end{eqnarray}
\begin{eqnarray*}
 i,j=\overline{1,l}, \quad
\forall t>0, \quad p \in K_+^n,
\end{eqnarray*}
are valid and  the economic sense of them  are  following:
the profit in the economy system nowhere vanishes, it shares  between all consumers.

We consider a class of the economy model income pre-functions  in that are described by the formula
\begin{eqnarray*}
 K_i^0(p,z)
\end{eqnarray*}
\begin{eqnarray}\label{sl00}
=\sum\limits_{j=1}^l\pi_{ij}(p,z)\sum\limits_{k=1}^l r_{jk}(p,z)\left[
\sum\limits_{s=1}^m\alpha_{ks}(p,z) \left\langle p,y_s-x_s\right\rangle + \left\langle b_k(p,z),p \right\rangle \right],
\end{eqnarray}
\begin{eqnarray*}
i=\overline{1,l}.
\end{eqnarray*}
Here $\pi_{ij}(p,z)$  is the part of the value    remained at the $i$-th consumer after taxation of the value coming from the $j$-th source of the value.

We assume that the economy system consists of $m$ firms and  $l$ consumers. We describe  production of the
$i$-th firm by a technological map $F_i(x), \ x \in X_i, \ i=\overline{1,m},$
that belongs to the CTM class and is convex down. Let the $i$-th consumer have the property vector $b_i(p,z), \ i=\overline{1,l}.$ Assume that $b_i(p,z), \ i=\overline{1,l},$ are continuous functions of variables $(p,z) \in K_+^n\times \Gamma^m$ and such that
\begin{eqnarray*}            b_i(tp,z)=b_i(p,z), \quad t>0,  \quad i=\overline{1,l}.\end{eqnarray*}
Moreover, we assume that there exist nonnegative vectors $b_i \geq 0, \ i=\overline{1,l},$
such that $b_i(p,z) \geq b_i$ for all  $ (p,z) \in K_+^n\times\Gamma^m.$ In all examples considered below, by the set of possible price vectors $K_+^n$ we understand any subcone of the cone $\bar R_+^n.$
 The continuous productive economic process
  $Q(p,z)=\{Q_i(p,z)\}_{i=1}^m$  is given by the formula
\begin{eqnarray*}
Q_i(p,z)=(X_i(p,z),Y_i(p, z)), \quad  (p,z) \in K_+^n\times\Gamma^m,  \quad  i=\overline{1,m},
\end{eqnarray*}
where
\begin{eqnarray*}
X_i(p,z)=\prod\limits_{k=1}^n\varphi_k\left(L_k(p,z) \right)u_i(p,z)x_i,  \quad  i=\overline{1,m},
\end{eqnarray*}
\begin{eqnarray*}
 Y_i(p, z)=\prod\limits_{k=1}^n\varphi_k\left(L_k(p,z)\right)u_i(p,z)y_i,  \quad  i=\overline{1,m},
 \end{eqnarray*}
\begin{eqnarray*}
z^i=(x_i,y_i) \in
\Gamma_i=\{(x_i,y_i) \in S^2, \ x_i \in X_i, \ y_i \in F_i(x_i)\}, \
\end{eqnarray*}
\begin{eqnarray*}
x_i=\{x_{ki}\}_{k=1}^n, \quad   y_i=\{y_{ki}\}_{k=1}^n, \quad  i=\overline{1,m}.
\end{eqnarray*}
By
\begin{eqnarray*}
 L_k(p,z)=\sum\limits_{i=1}^mu_i(p,z)[y_{ki} -x_{ki}]+ \sum\limits_{i=1}^lb_{ki}, \quad k=\overline{1,n},
 \end{eqnarray*}
we denote the $k$-th component of the vector
\begin{eqnarray*}
L(p,z)=\sum\limits_{i=1}^mu_i(p,z)[y_i -x_i]+\sum\limits_{i=1}^lb_i.
\end{eqnarray*}

Functions $\varphi_k(x), \ k=\overline{1,n},$ are continuous nonnegative functions of variable $x$ on the axis
$(-\infty, \infty)$ that equal  unit on the set $[d_k, \infty),$ do zero on the set $[- \infty, 0),$ and  are smaller than  unit on the set $[0, d_k), \ d_k >0.$

We define functions $u_i(p,z),\ i=\overline{1,m},$ by the formula
\begin{eqnarray*}             u_i(p,z)=\chi_{(a_i, \infty)}\left(\left\langle p, y_i- x_i \right\rangle/{\sum\limits_{s=1}^n p_s}\right),
\quad a_i > 0, \quad    i=\overline{1,m},\end{eqnarray*}
where  $\chi_{[a_i, \infty)}(x)$  is a  continuous nonnegative function of the variable $x$
defined on $R^1$ and such that on the interval
 $[a_i, \infty)$ it equals unit,  zero for  $x$ that belongs to $(- \infty, 0],$  and  is smaller than unit on the interval
$(0, a_i), \ a_i >0.$
It is obvious that  $(X_i(p,z),Y_i(p, z))$ is a productive process (that is,
$ Y_i(p, z) \in  F_i(X_i(p, z))$) that is  a continuous function   of the variables
$(p, z) \in K_+^n \times \Gamma^m.$

For any  $p \in K_+^n$ the set of values $Q(p,\Gamma^m)$ of the map
$Q(p,z)$ is a closed  set, so it is the Borel set and belongs to the set $G(p)$ constructed in the Lemma \ref{Ql1}, where as the set of measurable maps  $b_i(p,z), \ i=\overline{1,l},$
it is chosen the set of supply vectors of goods
$b_i(p,z), \ i=\overline{1,l},$  of consumers  at the beginning  of the period of the economy operation that are  continuous functions  of the variables.

Really, for this it is sufficient to prove that  $R(p,Q(p,z)) \in S.$ But

\begin{eqnarray*}            R(p,Q(p,z))=
\prod\limits_{k=1}^n\varphi_k\left(L_k(p,z)\right)\sum\limits_{i=1}^mu_i(p,z)[y_i -x_i] + \sum\limits_{i=1}^lb_i(p,Q(p,z))
 \end{eqnarray*}
\begin{eqnarray*}
   \geq L_1(p,Q(p,z))=
\left\{\begin{array}{ll}
               \sum\limits_{i=1}^lb_i, & \textrm{if} \quad \prod\limits_{k=1}^n\varphi_k\left(L_k(p,z)\right)=0,\\

            L_1(p,Q(p,z))   , &      \textrm{if}  \quad  \prod\limits_{k=1}^n\varphi_k\left(L_k(p,z)\right)>0 \textrm{,}
                                                                                                                                                \end{array}
                                                \right.
                                                \end{eqnarray*}
where
\begin{eqnarray*}
 L_1(p,z)=\sum\limits_{i=1}^m[y_i -x_i]+\sum\limits_{i=1}^lb_i.
\end{eqnarray*}
In both cases above the expressions are nonnegative: in the first case due to nonnegativity of the vector $\sum\limits_{i=1}^lb_i,$  in the second one from the condition
\begin{eqnarray*}            \prod\limits_{k=1}^n\varphi_k\left(L_k(p,z)\right)>0\end{eqnarray*}
it follows the inequality
 $\sum\limits_{i=1}^mu_i(p,z)[y_i -x_i]+ \sum\limits_{i=1}^lb_i\geq 0.$
From the last inequality we have
\begin{eqnarray*}
L_1(p,Q(p,z)) \geq
(1- \prod\limits_{k=1}^n\varphi_k\left(L_k(p,z)\right))\sum\limits_{i=1}^lb_i \geq 0.
\end{eqnarray*}

Let us put $K_i(p,z)=K_i^0(p, Q(p,z)), \ i=\overline{1,l},$ where the pre-functions of income
$K_i^0(p,z)$ are given by the formula
(\ref{sl00}) and $Q(p,z)$  is just now constructed  productive economic process\index{ productive economic process} being continuous function of variables $(p,z) \in K_+^n\times \Gamma^m.$

The  condition 2 of  the Definition  of   income functions  has the form
\begin{eqnarray*}
 \sum\limits_{k=1}^lK_i(p,z)
\end{eqnarray*}
\begin{eqnarray*}
 = \left\langle p, \sum\limits_{i=1}^m[Y_i(p,z) - X_i(p,z)]+
\sum\limits_{k=1}^lb_k(p,Q(p,z)) \right\rangle, \quad (p,z) \in K_+^n\times \Gamma^m.
\end{eqnarray*}

Let us give yet a few examples of  productive economic processes.

Example 1.

Let a  technological map
$F_i(x), \ x \in X_i, \ i=\overline{1,m},$  describe the structure of production of the  $i$-th firm in the  economy system    that belongs to the CTM class and is convex down, and the  $i$-th consumer have a property vector
 $b_i, \ i=\overline{1,l}.$
 We define the continuous productive economic process
 $Q(p,z)=\{Q_i(p,z)\}_{i=1}^m$  by the formula
\begin{eqnarray*}
 Q_i(p,z)=(X_i(p,z),Y_i(p, z)),  \quad  (p,z) \in K_+^n\times\Gamma^m, \quad  i=\overline{1,m},
 \end{eqnarray*}
where
\begin{eqnarray*}
 X_i(p,z)=\prod\limits_{k=1}^n\varphi_k\left(L_k(p,z) \right)u_i(p,z)x_i, \quad  i=\overline{1,m},
\end{eqnarray*}
\begin{eqnarray*}
  Y_i(p, z)=\prod\limits_{k=1}^n\varphi_k\left(L_k(p,z)\right)u_i(p,z)y_i, \quad  i=\overline{1,m},
\end{eqnarray*}
\begin{eqnarray*}
 z^i=(x_i,y_i) \in
\Gamma_i=\{(x_i,y_i) \in S^2, \ x_i \in X_i, \ y_i \in F_i(x_i)\}, \quad
i=\overline{1,m}.
\end{eqnarray*}
By
\begin{eqnarray*}
L_k(p,z)=\sum\limits_{i=1}^mu_i(p,z)[y_{ki} -x_{ki}]+ \sum\limits_{i=1}^lb_{ki}, \quad k=\overline{1,n},
\end{eqnarray*}
we denote the $k$-th component of the vector
\begin{eqnarray*}
L(p,z)=\sum\limits_{i=1}^mu_i(p,z)[y_i -x_i]+\sum\limits_{i=1}^lb_i,
\end{eqnarray*}
\begin{eqnarray*}
 x_i=\{x_{ki}\}_{k=1}^n, \quad   y_i=\{y_{ki}\}_{k=1}^n, \quad  i=\overline{1,m}, \quad b_i=\{b_{ki}\}_{k=1}^n, \quad  i=\overline{1,l}.
\end{eqnarray*}

Functions  $\varphi_k(x), \ k=\overline{1,n},$ are continuous nonnegative functions of the  variable $x$ on the axis  $(-\infty, \infty),$ that equal unit on the set $[d_k, \infty),$   zero on the set  $[- \infty, 0)$ and  are smaller than  unit on the set $[0, d_k), \ d_k >0.$

 We define  functions  $u_i(p,z),\ i=\overline{1,m},$ by the formula
\begin{eqnarray*}
 u_i(p,z)=\chi_{(a_i, \infty)}(t_i(p,z)),\quad a_i > 0,
 \end{eqnarray*}
 \begin{eqnarray*}
 t_i(p,z)=\frac{\left\langle p, y_i- x_i \right\rangle+ \left\langle p,r_i \right\rangle }{\sum\limits_{s=1}^n p_s}, \quad r_i \geq 0,
 \quad    i=\overline{1,m},
 \end{eqnarray*}
where  $\chi_{[a_i, \infty)}(x), \  i=\overline{1,m},$ are  continuous nonnegative functions of  $x$ defined on  $R^1$
 and such that
 they equal unit on the interval  $[a_i, \infty),$  do zero on the interval $(- \infty, 0]$ and  are smaller than  unit on the interval
$(0, a_i), \ a_i >0.$
It is obvious that  $(X_i(p,z),Y_i(p, z))$ is a productive process (that is,
$ Y_i(p, z) \in   F_i(X_i(p, z))$) that is continuous function of variables
$(p, z) \in K_+^n \times \Gamma^m.$
At last,
\begin{eqnarray*}
R(p,Q(p,z))
=\prod\limits_{k=1}^n\varphi_k\left(L_k(p,z)\right)\sum\limits_{i=1}^mu_i(p,z)[y_i -x_i] + \sum\limits_{i=1}^lb_i
\end{eqnarray*}
 \begin{eqnarray*}
=\left\{\begin{array}{ll}
               \sum\limits_{i=1}^lb_i, & \textrm{if} \quad \prod\limits_{k=1}^n\varphi_k\left(L_k(p,z)\right)=0,\\
            R(p,Q(p,z))   , &      \textrm{if}  \quad  \prod\limits_{k=1}^n\varphi_k\left(L_k(p,z)\right)>0 \textrm{.}
                                                                                                                                                \end{array}
                                               \right. \nonumber \end{eqnarray*}

From the condition $\prod\limits_{k=1}^n\varphi_k\left(L_k(p,z)\right)>0$
it follows that
\begin{eqnarray*}
R(p,Q(p,z)) \geq  \left[1 - \prod\limits_{k=1}^n\varphi_k\left(L_k(p,z)\right)\right]\sum\limits_{i=1}^lb_i  \geq 0.\end{eqnarray*}
The last inequality means that   the set of values $R(p,Q(p,z))$ belongs  to the set $S.$
This example is an example of the economy system  in that there may be subsidy under condition that for certain $1 \leq i \leq m,$  $r_i>0.$

Example 2.

Let the economy system consist of  $n$  branches, the Leontieff technological map
\begin{eqnarray*}
 F(x)=\{y \in S, \  Ay \leq x \}, \quad  x \in X=\sum\limits_{i=1}^n X_i,
\end{eqnarray*}
 \begin{eqnarray*}
X_i=\{x \in S, \ x=\{x_k\}_{k=1}^n, \ 0 \leq x_k \leq a_{ki}y_i^0, \ k=\overline{1,n}\},\end{eqnarray*}
describe  the structure of production in the economy system,
where  $y_i^0$  is the  maximally possible output\index{maximally possible output} in the $i$-th branch,  $i=\overline{1,n}.$
Further, by the sum $\sum\limits_{i=1}^n A_i$  of   $n$ sets  $A_i \in R_+^n, \ i=\overline{1,n},$  we understand the set of  vectors  $\sum\limits_{i=1}^n x_i, \ x_i \in A_i,\ i=\overline{1,n}.$

Further, let   $l$ consumers have  property vectors $b_i=\{b_{ki}\}_{k=1}^n, $  $i=\overline{1,l},$  and let the set of possible productive processes\index{set of possible productive processes} be given by the formula
\begin{eqnarray*}
\Gamma=\{z=(x,y), \ x \in X, y \in F(x)\}.
\end{eqnarray*}
The map
\begin{eqnarray*}
 Q(p,z)=\{X(p,z), Y(p,z)\}=\prod\limits_{i=1}^n a_i(p) v_i(p,z)\{Ay,y\},
 \end{eqnarray*}
where
\begin{eqnarray*}
& a_i(p)=\chi_{[0, \infty)}\left(p_i - \sum\limits_{s=1}^na_{si}p_s\right), \quad
v_i(p,z)=\chi_{[0, \infty)}\left(y_i - \sum\limits_{k=1}^na_{ik}y_k + \sum\limits_{k=1}^lb_{ik}\right),
\end{eqnarray*}
\begin{eqnarray*}
\chi_{[0, \infty)}(x)= \left\{\begin{array}{ll}
              1, & \textrm{if} \quad x \geq 0,\\
               0, &      \textrm{if}  \quad   x < 0 \textrm{,}
                             \end{array}\right.
                                \end{eqnarray*}
is a productive economic process\index{productive economic process}  under that the production is stopped if  any branch is unprofitable.
Let us verify the fulfilment of the conditions of Definition \ref{ddl2}. It is obvious that
$\bar Q(p,z)= \{Ay,y\}$ is  a continuous map of  $\Gamma$  into itself.

Further,
$\prod\limits_{i=1}^nv_i(p,z)$  is a measurable map of the measurable space
$\{\Gamma, {\cal B}(\Gamma)\}$ into the space $\{[0,1], {\cal B}([0,1])\}.$
From the Lemma \ref{1dl1}   the measurability  of the map $Q(p,z)$  from
the measurable space $\{\Gamma, {\cal B}(\Gamma)\}$  into itself  follows.

It is obvious that $Q(tp,z)=Q(p,z), \ t > 0.$
At last, let us prove the inequality
\begin{eqnarray*}                Y(p,z)- X(p,z)+\sum\limits_{k=1}^lb_i \geq 0, \quad (p,z) \in   K_+^n\times\Gamma,\end{eqnarray*}
which  means that the set of values $Q(p,\Gamma)$ of the map
$Q(p,z)$ belongs to the set  $G(p)$  constructed in the Lemma \ref{Ql1}.
Let us introduce the notation
\begin{eqnarray*}
a(p,y)=\prod\limits_{i=1}^n a_i(p)v_i(p,z).
\end{eqnarray*}
Then
\begin{eqnarray*}                a(p,y)[y - Ay] + \sum\limits_{i=1}^lb_i=\left\{\begin{array}{ll}
              y - Ay + \sum\limits_{i=1}^lb_i, & \textrm{if} \quad a(p,y)=1,\\
               \sum\limits_{k=1}^lb_i, &      \textrm{if}  \quad   a(p,y)=0 \textrm{.}
                                                                                                                                                \end{array}
                                               \right.\end{eqnarray*}

But for those $y$ for which $a(p,y)=1$ the set of inequalities
\begin{eqnarray*}                 y_i - \sum\limits_{s=1}^na_{ik}y_k + \sum\limits_{k=1}^lb_{ik} \geq 0, \quad i=\overline{1,n},
\end{eqnarray*}
is valid.
This means that the inequality required is valid.

Example 3.

Let the economy system consist of  $n$  branches, the Leontieff technological map\index{Leontieff technological map}
\begin{eqnarray*}
 y=F(x)=\{y \in S, \  Ay \leq x \}, \quad  x \in X=\sum\limits_{i=1}^n X_i,
\end{eqnarray*}
\begin{eqnarray*}
 X_i=\left\{x \in S, \ x=\{x_k\}_{k=1}^n, \ 0 \leq x_k \leq a_{ki}y_i^0, \ k=\overline{1,n}\right\},
\end{eqnarray*}
 describe the structure of production in the economy system,
where $y_i^0$  is maximally possible output in the $i$-th branch,  $i=\overline{1,n}.$

Further, let   $l$ consumers have  property vectors $b_i=\{b_{ki}\}_{k=1}^n, $  $i=\overline{1,l}.$  The set of possible productive processes is
 \begin{eqnarray*}                \Gamma=\{z=(x,y), \ x \in X, \  y \in F(x)\}.\end{eqnarray*}
The map
\begin{eqnarray*}
  Q(p,z)=\{X(p,z), Y(p,z)\}
 \end{eqnarray*}
\begin{eqnarray*}
  =\left\{\{\sum\limits_{i=1}^na_{ki}a_i(p)y_i\}_{k=1}^n, \{a_i(p)y_i\}_{i=1}^n\right\}  \prod\limits_{i=1}^nv_i(p,z), \quad (p,z) \in K_+^n\times \Gamma,
 \end{eqnarray*}
where
\begin{eqnarray*}
a_i(p)=\chi_{[0, \infty)}\left(p_i - \sum\limits_{s=1}^na_{si}p_s\right),
\end{eqnarray*}
\begin{eqnarray*}
  v_i(p,z)=\chi_{[0, \infty)}\left(a_i(p)y_i - \sum\limits_{s=1}^na_{is}a_s(p)y_s + \sum\limits_{k=1}^lb_{ik}\right),
\end{eqnarray*}
\begin{eqnarray*}
               \chi_{[0, \infty)}(x)= \left\{\begin{array}{ll}
              1, & \textrm{if} \quad x \geq 0,\\
               0, &      \textrm{if}  \quad   x < 0 \textrm{,}

                                           \end{array}\right.\end{eqnarray*}
is a productive economic process  under that the production is stopped in those  branches only  that are unprofitable.

Example 4.

We describe the structure of production of the  $i$-th firm in the economy system   by a  technological map  $F_i(x), \ x \in X_i, \ i=\overline{1,m},$    that belongs to the CTM class and is convex down. Let the $i$-th consumer have a property vector   $b_i, $ $ i=\overline{1,l}.$

The map
\begin{eqnarray*}
 Q(p,z)=\{Q_i(p,z)\}_{i=1}^m, \quad (p,z) \in K_+^n\times \Gamma^m,
 \end{eqnarray*}
where \begin{eqnarray*}
Q_i(p,z)=z_ia_i(p,z) \prod\limits_{i=1}^nv_i(p,z),
\quad a_i(p,z)=\chi_{[0, \infty)}(\left\langle p, y_i - x_i \right\rangle),
\end{eqnarray*}
\begin{eqnarray*}
 v_k(p,z)=\chi_{[0, \infty)}(V_k(p,z)), \quad
 z^i=(x_i,y_i) \in \Gamma_i=\{(x,y), \  x \in X_i, \  y \in F_i(x)\},
\end{eqnarray*}
and $ V_k(p,z) $ is the $k$-th  component of the vector
\begin{eqnarray*}
V(p,z)=\sum\limits_{i=1}^m a_i(p,z)[y_i -x_i] + \sum\limits_{i=1}^lb_i,
\end{eqnarray*}
is a productive economic process.

The map $V_k(p,z)=\sum\limits_{i=1}^m a_i(p,z)[y_{ki} -x_{ki}] + \sum\limits_{i=1}^lb_{ki}$
is a measurable map of the space $\{\Gamma^m, {\cal B}(\Gamma^m)\}$ into the space $\{R^1, {\cal B}(R^1)\}$ for every fixed  $p \in K_+^n.$ Really, $[y_{ki} - x_{ki}]$ is a  measurable map of the space $\{\Gamma_i, {\cal B}(\Gamma_i)\}$ into the space $\{R^1,{\cal B}(R^1)\},$
since it is the restriction on the closed set $\Gamma_i$ of the continuous map $[y_{ki} - x_{ki}]$ of
the measurable space $\{R^{2n}, {\cal B}(R^{2n})\}$ into the measurable space $\{R^1,{\cal B}(R^1)\}.$   The map $a_i(p,z)$ is the restriction on the closed set $\Gamma_i$ of the measurable map
$a_i(p,z)$ of the space $\{R^{2n}, {\cal B}(R^{2n})\}$ into the space $\{[0,1],{\cal B}([0,1])\},$
so it is a measurable map of the measurable space $\{\Gamma_i, {\cal B}(\Gamma_i)\}$ into the measurable space $\{[0,1],{\cal B}([0,1])\}.$

Therefore, $a_i(p,z)[y_{ki} - x_{ki}]$ is a measurable map of the space $\{\Gamma_i, {\cal B}(\Gamma_i)\}$ into the space $\{R^1,{\cal B}(R^1)\},$ so $a_i(p,z)[y_{ki} - x_{ki}]$ is a measurable map of the measurable space $\{\Gamma^m, {\cal B}(\Gamma^m)\}$ into the measurable space  $\{R^1,{\cal B}(R^1)\}.$ From here it follows that
\begin{eqnarray*}                V_k(p,z)=\sum\limits_{i=1}^m a_i(p,z)[y_{ki} -x_{ki}] + \sum\limits_{i=1}^lb_{ki}\end{eqnarray*}
is a measurable map of the measurable space  $\{\Gamma^m, {\cal B}(\Gamma^m)\}$ into the measurable space $\{R^1,{\cal B}(R^1)\}.$ As a consequence  $v_k(p,z)=\chi_{[0, \infty)}(V_k(p,z))$ is a measurable map of the measurable space  $\{\Gamma^m, {\cal B}(\Gamma^m)\}$ into the measurable space $\{[0,1],{\cal B}([0,1])\}.$

Therefore, \hfill $a_i(p,z) \prod\limits_{i=1}^nv_i(p,z)$ \hfill
is \hfill a \hfill measurable \hfill map \hfill of \hfill  the \hfill measurable \hfill space \\ $\{\Gamma^m, {\cal B}(\Gamma^m)\}$ into the measurable space $\{[0,1], {\cal B}([0,1])\}, \ i=\overline{1,m}.$
Due to the Lemma  \ref{Ql1}, $Q(p,z)$ is a measurable map of $\{\Gamma^m, {\cal B}(\Gamma^m)\}$
into itself.

It is obvious that $Q(tp,z)=Q(p,z), \ t > 0.$
At last,
\begin{eqnarray*}      \sum\limits_{i=1}^m[Y_i(p,z) - X_i(p,z)] + \sum\limits_{i=1}^lb_i        =\left\{\begin{array}{ll}
             \sum\limits_{i=1}^ma_i(p,z)[y_i - x_i] + \sum\limits_{i=1}^lb_i, &                                                                         \textrm{if} \quad d(p,z)=1,\\
               \sum\limits_{i=1}^lb_i, &      \textrm{if}  \quad   d(p,z)=0 \textrm{,}
                                                                                                                                                \end{array}
                                               \right.
                                                \end{eqnarray*}
where $d(p,z)=\prod\limits_{i=1}^nv_i(p,z).$
But if  $d(p,z)=1,$ then
\begin{eqnarray*}                \sum\limits_{i=1}^ma_i(p,z)[y_i - x_i] + \sum\limits_{i=1}^lb_i \geq 0.\end{eqnarray*}
Let us find the set of values of the map $Q(p,z).$

Consider $\bar Q_i(p,z)=z_ia_i(p,z).$
It is obvious that the set of values of the map $\bar Q(p,z)=\{\bar Q_i(p,z)\}_{i=1}^m$
is \begin{eqnarray*}                \Gamma_1(p)=\prod\limits_{i=1}^m\Gamma_i^1(p),\end{eqnarray*}                   where
 $\Gamma_i^1(p)=
\{z_i \in \Gamma_i, \  \left\langle p, y_i -x_i \right\rangle \ \geq 0\}.$
At last, since
\begin{eqnarray*}                Q(p,z)=\bar Q(p,z) d(p,z),\end{eqnarray*}
we have that the set of values of the map  $Q(p,z)$ is the set
\begin{eqnarray*}                Q(p,\Gamma^m)=\Gamma_1(p) \cap M(p),\end{eqnarray*}
where $M(p)=\{z \in \Gamma^m, \ \sum\limits_{i=1}^m[a_i(p,z)[y_i - x_i] + \sum\limits_{i=1}^lb_i \geq 0\}.$
From this we obtain that the set of values of  $Q(p,z)$ is the Borel set for every $p \in K_+^n.$

Example 5.

In the economy system we describe the structure of   production\index{structure of   production} of the  $i$-th firm   by a technological map
$F_i(x), \ x \in X_i, \ i=\overline{1,m},$ that belongs to the CTM class and is convex down. Let the $i$-th consumer have the property vector
 $b_i,$ $  i=\overline{1,l}.$ The map
 $Q(p,z)=\{Q_i(p,z)\}_{i=1}^m, $ where
\begin{eqnarray*}
Q_i(p,z)=z_ia_i(p,z) \prod\limits_{i=1}^nv_i(p,z),
\end{eqnarray*}
\begin{eqnarray*}
 a_i(p,z)=\chi_{[0, \infty)}(\left\langle p, y_i - x_i \right\rangle + \left\langle p, d_i \right\rangle),
 \end{eqnarray*}
\begin{eqnarray*}
 v_i(p,z)=\chi_{[0, \infty)}(V_i(p,z)), \ V_k(p,z)=\sum\limits_{i=1}^m a_i(p,z)[y_{ki} -x_{ki}] + \sum\limits_{i=1}^lb_{ki},
\end{eqnarray*}
 $z^i=(x_i,y_i) \in \Gamma_i=\{(x,y), \  x \in X_i, \  y \in F_i(x)\},$ and
$ d_i=\{d_{ki}\}_{k=1}^n  $ is the vector with nonnegative components, is a productive economic process.

Example 6.

In the economy system  we describe the structure of the  production of the  $i$-th firm  by a technological map
$F_i(x), \ x \in X_i, \ i=\overline{1,m},$ that belongs to the CTM class and is convex down, and $l$  consumers have  property vectors
$b_i,$ $ i=\overline{1,l}.$ The map
$Q(p,z)=\{Q_i(p,z)\}_{i=1}^m,$
where
\begin{eqnarray*}
 Q_i(p,z)=z_ia_i(p,z) \prod\limits_{i=1}^nv_i(p,z),
\end{eqnarray*}
\begin{eqnarray*}
 a_i(p,z)=\chi_{[0, \infty)}(\left\langle p, y_i - x_i \right\rangle + \left\langle p, d_i \right\rangle),
\end{eqnarray*}
\begin{eqnarray*}
 v_i(p,z)=\chi_{[0, \infty)}(V_i(p,z)), \quad V_k(p,z)=\sum\limits_{i=1}^m a_i(p,z)[y_{ki} -x_{ki}] + \sum\limits_{i=1}^lb_{ki} - b_k^0,
\end{eqnarray*}
 $z^i=(x_i,y_i) \in \Gamma_i=\{(x,y), \  x \in X_i, \  y \in F_i(x)\},$ \
and $ d_i=\{d_{ki}\}_{k=1}^n $ is a vector with nonnegative components, $b^0=\{b_i^0 \}_{i=1}^n, \ b_i^0 > 0,  \ i=\overline{1,n},$
is a productive economic process satisfying the condition
\begin{eqnarray*}
R(p, Q(p,z)) \geq b^0,  \quad R(p,z)=\sum\limits_{i=1}^m[y_i - x_i] + \sum\limits_{i=1}^lb_i,
\end{eqnarray*}
if $\sum\limits_{i=1}^lb_i \geq  b^0.$
Really,
\begin{eqnarray*}
\sum\limits_{i=1}^m[Y_i(p,z) - X_i(p,z)] + \sum\limits_{i=1}^lb_i
 =\left\{\begin{array}{ll}
             \sum\limits_{i=1}^ma_i(p,z)[y_i - x_i] + \sum\limits_{i=1}^lb_i, &                                                                         \textrm{if} \quad d(p,z)=1,\\
               \sum\limits_{i=1}^lb_i, &      \textrm{if}  \quad   d(p,z)=0 \textrm{,}
                                                                                                                                                \end{array}
                                               \right. \nonumber \end{eqnarray*}
where $d(p,z)=\prod\limits_{i=1}^nv_i(p,z).$
But if $d(p,z)=1,$ then
\begin{eqnarray*}                \sum\limits_{i=1}^ma_i(p,z)[y_i - x_i] + \sum\limits_{i=1}^lb_i \geq b^0.\end{eqnarray*}
Together with the condition  $\sum\limits_{i=1}^lb_i \geq  b^0$ we have  $ R(p, Q(p,z)) \geq b^0.$  From the economic point of view  this productive economic process means availability   of the subsidy  in the economy system if $\left\langle p, d_i \right\rangle  > 0.$

Let us find the set of values of the map $Q(p,z).$

Consider $\bar Q_i(p,z)=z_ia_i(p,z).$
It is obvious that the set of values of the map $\bar Q(p,z)=\{\bar Q_i(p,z)\}_{i=1}^m$
is \begin{eqnarray*}                \Gamma_1(p)=\prod\limits_{i=1}^m\Gamma_i^1(p),\end{eqnarray*}                   where
 $\Gamma_i^1(p)=
\{z_i \in \Gamma_i, \ \left\langle p, y_i -x_i +d_i \right\rangle \ \geq 0\}.$
At last, since
\begin{eqnarray*}                Q(p,z)=\bar Q(p,z) d(p,z),\end{eqnarray*}
we have that the set of values of the map  $Q(p,z)$ is the set
\begin{eqnarray*}                Q(p,\Gamma^m)=\Gamma_1(p) \cap M(p),\end{eqnarray*}
where $M(p)=\{z \in \Gamma^m, \ \sum\limits_{i=1}^m[a_i(p,z)[y_i - x_i] + \sum\limits_{i=1}^lb_i \geq b^0\}.$
From this we obtain that the set of values of  $Q(p,z)$ is the Borel set for every $p \in K_+^n.$

\section{Theory of choice and decisions making under the availability of information}

To specify a random field, it is necessary to give a  family of all joint distributions \cite{65, 2, 3, 47, 8}.\index{family of all joint distributions}
As earlier, $S$  is the set of possible goods in the economy system in that Euclidean metric is given and ${\cal B}(S)$ is the Borel $\sigma$-algebra\index{Borel $\sigma$-algebra} of subsets of the set $S.$
Remember that
\begin{eqnarray*}                S=\{x=\{x_i\}_{i=1}^n, \ x \in R_+^n, \
0 \leq x_i \leq c_i,\  i=\overline{1,n}\}, \quad c_i>0, \quad i=\overline{1,n}.\end{eqnarray*}
Assume that in the economy system  $m$ firms operate, we describe  the structure of production of the  $i$-th firm  by  a technological map $F_i(x), \ x \in  X_i,$ $ \
i=\overline{1,m}.$
It is convenient to assume that all firms are ordered and for further consideration only the set of productive processes
\begin{eqnarray*}                \Gamma_i=\{(x,y), \  x \in X_i \subseteq S, \ y \in F_i(x)\}\end{eqnarray*}
of the $i$-th firm, $i=\overline{1,m},$ is important. As earlier, by $\Gamma^m
=\prod\limits_{i=1}^m\Gamma_i$ we denote the ordered direct product of the sets of productive processes $\Gamma_i, \
i=\overline{1,m},$ of $m$ firm and by  $[\Gamma^m]^k$ we denote
$k$-multiple direct product of the sets $\Gamma^m, \ k=1,2,
\ldots  .$  Assume that in $[\Gamma^m]^k, \ k=1,2, \ldots,$ the Euclidean metric is given and  ${\cal B}([\Gamma^m]^k), \ k=1,2, \ldots$ is the Borel $\sigma$-algebra of subsets of $[\Gamma^m]^k, \ k=1,2, \ldots.$ We give the budget set of the $i$-th non-insatiable consumer\index{budget set of  non-insatiable consumer}     by the formula
 \begin{eqnarray*}                \bar
X_{(p,z)}^i=\{x, \ x \in S, \ \left\langle p,x \right\rangle \leq K_i(p,z)\}, \quad p \in K_+^n, \quad
z \in \Gamma^m, \quad i=\overline{1,l}, \end{eqnarray*}
and the budget set of the $i$-th insatiable  consumer\index{budget set of  insatiable  consumer}
 we do by the formula \begin{eqnarray*}                X_{(p,z)}^i=\{x, \ x \in S, \
 \left\langle p,x\right\rangle =K_i(p,z)\},\quad  p \in K_+^n, \quad  z \in \Gamma^m, \quad i=\overline{1,l},  \end{eqnarray*}
where $K_i(p,z)$
is an income function of the $i$-th consumer.

Further, by $\hat X_{(p,z)}^i$ we denote the budget set of the $i$-th  consumer
 both insatiable and non-insatiable, that is determined by $(p,z)$  and the income function  $K_i(p,z).$

Let  ${\cal
B}(\hat X^i_{(p,z)})$ be the Borel $\sigma$-algebra of subsets of the set $\hat X_{(p,z)}^i.$
For the ordered direct product of
$k$ budget sets of the $i$-th consumer that is determined by the ordered set $(p_s,z_s), \
 s=\overline{1,k},$ we introduce the notation
 \begin{eqnarray*}                X_{(p,z)_k}^{k,i}=\prod\limits_{s=1}^k\hat X_{(p_s,z_s)}^i,\end{eqnarray*}                   and by
 ${\cal B}(X_{(p,z)_k}^{k,i})$ we denote the Borel
$\sigma$- algebra of subsets of the set $X_{(p,z)_k}^{k,i},$
where
\begin{eqnarray*}                (p,z)_k=\{(p_1,z_1), \ldots, (p_k,z_k)\}, \quad
(p_i, z_i)  \in K_+^n\times \Gamma^m, \quad i=\overline{1,k}.\end{eqnarray*}

It is obvious that if  $\pi_k$  is a certain permutation in the set of integer numbers
$\{1, \ldots , k\},$ then   $ \Pi_k X_{(p,z)_k}^{k,i}=\prod\limits_{s=1}^k\hat X_{(p_{\pi(s)},z_{\pi(s)})}^i$ is the ordered direct product that corresponds to the permutation
$\pi_k.$

The specific character of the determination of the random  field  of the $i$-th consumer  choice
is that  the choice is realized by him  under  budget constraints and it depends on the realization of  productive processes in the economy system. For  complete description
of consumers choice and decisions making by firms it is sufficient to give the family
of conditional probabilities of each  consumer
\begin{eqnarray*}      \bar P((\xi_i(p_1),
\ldots ,\xi_i(p_k)) \in A^i
\end{eqnarray*}
\begin{eqnarray*}
 |(\zeta_1(p_1), \ldots, \zeta_m(p_1))=z_1,
\ldots, (\zeta_1(p_k), \ldots, \zeta_m(p_k))=z_k)
\end{eqnarray*}
\begin{eqnarray*}
 =F_{p_1, \ldots, p_k}^i( A^i | z_1, \ldots, z_k), \quad  k=1,2,
\ldots , \quad i=\overline{1,l},
\end{eqnarray*}
where $z_j=(z^1_j, \ldots, z^m_j), \ z^s_j \in \Gamma_s,
\ s=\overline{1,m}, \ j=\overline{1,k},$ $A^i \in {\cal B}(
  X_{(p,z)_k}^{k,i}),$
  and to give also the family of unconditional probabilities relative to decisions making  of the choice of  productive processes by firms
 \begin{eqnarray*}             \bar P((\zeta_1(p_1), \ldots, \zeta_m(p_1)), \ldots, (\zeta_1(p_k),
 \ldots, \zeta_m(p_k)) \in B^k)
 \end{eqnarray*}
\begin{eqnarray*}
 =\psi_{p_1, \ldots,
 p_k}(B^k),\quad k=1,2, \ldots ,
\end{eqnarray*}
  where $B^k \in {\cal
 B}([\Gamma^m]^k), \quad  k=1,2, \ldots $
 The economic sense of the conditional probability introduced $F_{p_1, \ldots, p_k}^i( A^i | z_1,
\ldots, z_k)$ is the probability of that the $i$-th consumer will choose a set of goods from the set  $A^i \in {\cal B}(X_{(p,z)_k}^{k,i})$ under condition that in the economy system firms  realized the productive process $z_s \in
 \Gamma^m$ under the price vector $p_s,\
 s=\overline{1,k}, $  and the $i$-th consumer had the income $K_i(p_s,z_s), \ s=\overline{1,k}, \ i=\overline{1,l}.$

\subsection{Axioms of random choice of consumers and decisions making by firms under availability of information to economic agents}

We assume that the description of the  economy system under uncertainty is given  if  a family of  conditional distributions \cite{55, 71, 69, 92, 106}\index{family of  conditional distributions }
\begin{eqnarray*}
 F_{p_1, \ldots, p_k}^s(A^s | z_1, \ldots, z_k), \quad  p_i \in K_+^n, \quad  i=\overline{1,k},  \quad
A^s \in {\cal B}(X_{(p,z)_k}^{k, s}), \quad s=\overline{1,l},
\end{eqnarray*}
\begin{eqnarray*}
   z_i \in \Gamma^m, \quad
i=\overline{1,k}, \quad  k=1,2, \ldots ,
\end{eqnarray*}
 is given that for every fixed $s, \ s=\overline{1,l},$ satisfies the conditions:\\
1)  $F_{p_1, \ldots, p_k}^s( A^s | z_1, \ldots, z_k)$ is a probability measure on the $\sigma$-algebra of the Borel subsets
$A^s \in {\cal B}(X^{k,s}_{(p,z)_{k}})$ for every fixed value of  variables
 \begin{eqnarray*}           \{p_1, \ldots, p_k\} \in K_+^{nk}, \quad  \{z_1, \ldots, z_k\} \in
[\Gamma^m]^k \end{eqnarray*}
and
\begin{eqnarray*}                F_{p_1, \ldots, p_k}^s\left(A_s\cap X_{(p,z)_k}^{k, s} | z_1, \ldots, z_k\right)\end{eqnarray*}
is a measurable map of the measurable space
$\{[\Gamma^m]^k, {\cal B}([\Gamma^m]^k)\}$ into the measurable space
$\{[0,1], {\cal B}([0,1])\}$ for every fixed  $A_s \in {\cal B}(S^k)$ and $\{p_1, \ldots, p_k\} \in K_+^{nk}\ ;$\\
2) for every permutation  $\pi$ of indices $\{1, \ldots, k\}$
the equality \begin{eqnarray*}                F^s_{p_{\pi(1)}, \ldots, p_{\pi(k)}}(\Pi
A^{s} | z_{\pi(1)}, \ldots, z_{\pi(k)})=F^s_{p_1, \ldots, p_k}( A^s
| z_1, \ldots, z_k), \quad k=1,2, \ldots ,
\end{eqnarray*}
holds,
where $\Pi A^{s}$ is the set that is the image of the set  $A^s \in  {\cal B}(X_{(p,z)_k}^{k, s})$ for the transformation
$\Pi $ of the set  $S^k$ into itself:
 \begin{eqnarray*}                \Pi
x=\{x_{\pi(1)}, \ldots, x_{\pi(k)}\},\quad x=\{x_1, \ldots, x_k\}
 \in S^{k}\ ;
 \end{eqnarray*} 
3) for every $1 \leq j < k$ the equality  
\begin{eqnarray*}
  F^s_{p_1, \ldots,
p_k}\left(A^j\times\prod\limits_{i=j+1}^k\hat X^s_{(p_i,z_i)} \left|\right. z_1,
\ldots, z_k\right)
\end{eqnarray*}
 \begin{eqnarray*}
=F^s_{p_1, \ldots, p_j}\left( A^j | z_1, \ldots, z_j\right), \quad   A^j
 \in {\cal B}(X^{j,s}_{(p,z)_j}),
 \end{eqnarray*}
 holds;\\
4) for all strictly positive $t$
 \begin{eqnarray*}
   F^s_{tp_1, \ldots, tp_k}( A^s | z_1,
 \ldots, z_k)=F^s_{p_1, \ldots, p_k}(A^s | z_1, \ldots, z_k), 
 \end{eqnarray*}
\noindent and a family of unconditional distributions\index{family of unconditional distributions}
 \begin{eqnarray*}                \psi_{p_1, \ldots, p_k}(B^k), \ B^k \in
{\cal B}([\Gamma^m]^k), \ p_i \in K_+^n,\quad i=\overline{1,k},
\quad k=1,2, \ldots,\end{eqnarray*}                   that satisfies conditions: \\
1)  ~$\psi_{p_1, \ldots, p_j}(B^j)=\psi_{p_1,
\ldots, p_k}(B^j\times[\Gamma^m]^{k-j}), \ B^j \in {\cal
 B}([\Gamma^m]^j), \quad k=j+1,  j+2, \ldots ,$
 $ \psi_{p_1}(\Gamma^m)=1\ ;$ \\
2) for every permutation $\pi$ of indices
$\{1, \ldots, k\}$
\begin{eqnarray*}                 \psi_{p_{\pi(1)}, \ldots, p_{\pi(k)}}(\Pi
B^{k})=\psi_{p_1, \ldots, p_k}(B^k),\quad k=1,2, \ldots ,\end{eqnarray*}
where $\Pi B^{k}$ is the set that is the image of the set $B^k$ under transformation $\Pi$ of the set   $[\Gamma^m]^k$ into  itself:
$\Pi z=\{z_{\pi(1)},
 \ldots, z_{\pi(k)}\},$  $z=\{z_1, \ldots, z_k\} \in
 [\Gamma^m]^k \ ;$ \\
 3) ~$\psi_{tp_1, \ldots, tp_k}(B^k)=
 \psi_{p_1, \ldots, p_k}(B^k), \ \forall t>0, \ k=1,2, \ldots
   \ ;$ \\
4) If $ b_i(p,z), $  $ i=\overline{1,l},$ is a supply vector of goods\index{supply vector of goods} of the
 $i$-th consumer
at the beginning of the period of the economy operation, that is a measurable map of the space
$\{\Gamma^m, {\cal B}(\Gamma^m)\}$ into the space $\{S, {\cal B}(S)\}$
for every  $p \in K_+^n, $ and
\begin{eqnarray*}                G(p)=\{ z \in \Gamma^m, \  R(p,z) \in S \} \  \in  \ {\cal B}(\Gamma^m),\end{eqnarray*}
where
\begin{eqnarray*}                R(p,z)= \sum\limits_{i=1}^m[y_i - x_i]+ \sum\limits_{k=1}^lb_k(p, z),
\quad  z^i=(x_i,y_i) \ \in  \ \Gamma_i, \end{eqnarray*}
then for all  $ p \in K_+^n$
\begin{eqnarray*}                \int\limits_{\Gamma^m}\chi_{G(p)}(z)\psi_{p}(dz)=1,\end{eqnarray*}
where $\chi_{G(p)}(z)$ is the indicator function of the set $G(p).$

The property  4) for \ the family of conditional distributions and  3) for the family of unconditional distributions  means the  independence of the probability of the choice of consumers and decisions making by firms from the scale of prices.
\begin{definition}
The family of functions of sets
\begin{eqnarray*}
 \Phi_{p_1, \ldots, p_k}({\cal D}\times A_1 \times \ldots \times A_l)
\end{eqnarray*}
\begin{eqnarray*}
 =\int\limits_{{\cal D}}\prod\limits_{i=1}^{l}F_{p_1, \ldots, p_k}^i
\left(A_i\cap X_{(p,z)_{k}}^{k,i}|z_1, \ldots, z_k\right)d\psi_{p_1, \ldots, p_k}(z_1, \ldots, z_k),
\end{eqnarray*}
where
\begin{eqnarray*}           {\cal D} \in  [{\cal B}(\Gamma^m)]^k, \quad
A_i \in  {\cal B}(S^k), \quad i=\overline{1,l},
\end{eqnarray*}
is called  finite-dimensional distributions of  choice of  $l$ consumers and decisions making by $m$ firms,
where \begin{eqnarray*}
A_i=\prod\limits_{s=1}^kA_i^s, \quad {\cal D} =\prod\limits_{i=1}^k
{\cal D}_i, \quad  A_i^s \in  {\cal B}(S),\quad {\cal D}_s \in {\cal B}(\Gamma^m),
 \quad i=\overline{1,l}, \quad
  s=\overline{1,k}.
  \end{eqnarray*}
\end{definition}
$\Phi_{p_1, \ldots, p_k}({\cal D}\times A_1 \times \ldots \times A_l)$ is the simultaneous probability of that  $l$ consumers choose a set of goods  from the set  $A_1 \times \ldots \times A_l,$ $m$ firms make decisions relative to productive processes that belong to the set   ${\cal D},$ where the choice of the $i$-th consumer belongs to the set $A_i^s \in  {\cal B}(S),$ and firms choose  productive processes that belong to the set ${\cal D}_s \in {\cal B}(\Gamma^m) $  under the  price vector \  $p_s \in K_+^n, $ $ \  s=\overline{1,k}, \ i=\overline{1,l}. $

 The  special Theorem on the change of variables in the abstract Lebesgue integral\index{abstract Lebesgue integral} presented below will be used in the proof of the Theorem \ref{ttl1}.
\begin{theorem}\label{ffl1}
Let $\{X, \Sigma, \mu\}$ be a measurable space with a measure.  If a one to one correspondence
$f(x)$ of the space $X$ onto itself is  measurable together with the inverse map $f^{-1}(x),$
then  for every integrable function $ \psi(x) $ the formula
\begin{eqnarray}\label{fffl1}
\int\limits_{f^{-1}(A)}\psi(f(x))d\mu f^{-1}=\int\limits_{A}\psi(x)d\mu, \quad A \in  \Sigma,
\end{eqnarray}
holds, where the notation
$\mu f^{-1}(B)=\mu(f(B))$ is introduced.
\end{theorem}
\begin{proof}\smartqed            By $f^{-1}(B)$ we denote the image of the set $B$ under the map $f^{-1}(x)$ and by $f(B)$ we denote the image of the set $B$ under the map $f(x).$
From the measurability of $f^{-1}(x)$ it follows that  the set $f(B)$  is  measurable  for any $B \in \Sigma$ since the preimage of the set
 $B$  under the map $f^{-1}(x)$ is the set $\{x, \ f^{-1}(x) \in B \}=f(B).$ Therefore,  $\mu(f(B))$  is correctly defined.
  It is sufficient to prove the formula  \ref{fffl1} in the case as $\psi(x)=\chi_{B}(x), \ B \in \Sigma,$ where $ \chi_{B}(x)$ is  an indicator function of the set $B.$ In this case the formula (\ref{fffl1}) takes the form
 \begin{eqnarray*}                 \int\limits_{f^{-1}(A)}\chi_{B}(f(x))d\mu f^{-1}=\int\limits_{A}\chi_{B}(x)d\mu.\end{eqnarray*}
The last equality  holds since  both the left part  and the right one equal
$\mu(A\cap B).$
 \qed \end{proof}           .

The particular case of the equality (\ref{fffl1}) is the equality
\begin{eqnarray}\label{fffl2}
\int\limits_{X}\psi(f(x))d\mu f^{-1}=\int\limits_{X}\psi(x)d\mu,
\end{eqnarray}
as $A=X.$
\begin{theorem}\label{ffl2}
If a family of unconditional distributions
 \begin{eqnarray*}                \psi_{p_1, \ldots, p_k}(B^k), \quad B^k \in
{\cal B}([\Gamma^m]^k), \quad p_i \in K_+^n,\quad i=\overline{1,k},
\quad k=1,2, \ldots,\end{eqnarray*}                   satisfies the axioms formulated and the measurable map
$f(z_1, \ldots,z_r)$ of the measurable  space  $\{[\Gamma^m]^r,\ {\cal B}([\Gamma^m]^r)\}$ into the  measurable space
$\{R^1, {\cal B}(R^1)\}$ is   integrable, then the formula
\begin{eqnarray*} \int\limits_{[\Gamma^m]^k}f(z_1, \ldots,z_r)d\psi_{p_1, \ldots, p_k}(z_1, \ldots,z_k)
\end{eqnarray*}
\begin{eqnarray}\label{fffl3}
 =\int\limits_{[\Gamma^m]^r}f(z_1, \ldots,z_r)d\psi_{p_1, \ldots, p_r}(z_1, \ldots,z_r), \quad k \geq r,
\end{eqnarray}
is valid.
\end{theorem}
\begin{proof}\smartqed
The proof is the consequence of  the validity of (\ref{fffl3}) for the functions of the kind
$f(z_1, \ldots,z_r)=\chi_{B}(z_1, \ldots,z_r), \ B \in {\cal B}([\Gamma^m]^r).$
\qed \end{proof}

\subsection{Probability realization of the random fields  of   consumers   choice  and decisions making by firms  }

The question arises whether a probability space exists  and random fields on it such that
the random choice of the consumer satisfies axioms formulated. The subsequent investigation will clarify this question.

\begin{theorem}\label{ttl1}
Let a family of  conditional distribution functions
\begin{eqnarray*}                    F_{p_1, \ldots, p_k}^i(A^i|z_1, \ldots, z_k), \ A^i \in
{\cal B}(X_{(p,z)_k}^{k,i}), \quad  i=\overline{1,l}, \quad  p_s \in K_+^n,
\quad s=\overline{1,k}, \end{eqnarray*}
and a family of unconditional distribution functions
$\psi_{p_1, \ldots, p_k}(D), \ D \in {\cal B}([\Gamma^m]^k),\
$ $k=\overline{1, \infty},$
satisfy the axioms formulated above.
The function of the sets given by the formula
\begin{eqnarray*}
\Phi_{p_1, \ldots, p_k}({\cal D}\times A_1\times\ldots\times A_l)
\end{eqnarray*}
\begin{eqnarray*}
=\int\limits_{{\cal D}}\prod\limits_{i=1}^{l}F_{p_1, \ldots, p_k}^i
\left(A_i\cap X_{(p,z)_{k}}^{k,i}|z_1, \ldots, z_k\right)
d\psi_{p_1, \ldots, p_k}(z_1, \ldots, z_k) \nonumber  \end{eqnarray*}
on the collection of sets of the kind $ {\cal D}\times A_1\times\ldots\times A_l, $
where
\begin{eqnarray*}                    {\cal D}=\prod\limits_{s=1}^k{\cal D}_s,\quad A_i=\prod\limits_{s=1}^kA_i^s,  \quad
 A_i^s \in {\cal B}(S),\quad  {\cal D}_s \in {\cal B}(\Gamma^m), \quad i=\overline{1,l}, \quad s=\overline{1,k},\end{eqnarray*}
admits   extension to the measurable space
\begin{eqnarray*}                    V_1^k= \{[\Gamma^m\times S^l]^k,\ [{\cal B}(\Gamma^m)\times {\cal B}(S^l)]^k\},\end{eqnarray*}
that is, there exists a family of measures  $\bar \mu_{z_1, \ldots, z_k}^{p_1, \ldots, p_k}( E)$  given on the measurable space  $V_1^k$ that for every fixed
  $\{p_1, \ldots, p_k\} \in K_+^{nk}$    and   $E \in
[{\cal B}(\Gamma^m) \times {\cal B}(S^l)]^{k}$
is a measurable map of the measurable space
$L^k=\{[\Gamma^m]^k,[{\cal B}(\Gamma^m)]^k\}$ into the  measurable space $\{[0,1], {\cal B}([0,1])\}$
such that this extension is given by  the formula
\begin{eqnarray*}                    \bar \Phi_{p_1, \ldots, p_k}(E)= \int\limits_{[\Gamma^m]^k}
\bar \mu_{z_1, \ldots, z_k}^{p_1, \ldots, p_k}( E)
d\psi_{p_1, \ldots, p_k}(z_1, \ldots, z_k), \quad  E \in [{\cal B}(\Gamma^m)\times {\cal B}(S^l)]^k.\end{eqnarray*}
This extension satisfies the conditions
\begin{eqnarray}\label{aal1}
\bar \Phi_{p_{\pi(1)}, \ldots, p_{\pi(k)}}(\Pi_k^1 E)=
\bar \Phi_{p_1, \ldots, p_k}( E), \quad  E \in [{\cal B}(\Gamma^m)\times {\cal B}(S^l)]^k,
\end{eqnarray}
\begin{eqnarray}\label{qal2}
\bar \Phi_{p_1, \ldots, p_k}(A\times (\Gamma^m\times S^l)^{k-r})=
 \bar \Phi_{p_1, \ldots, p_r}(A), \quad
A \in [{\cal B}(\Gamma^m)\times {\cal B}(S^l)]^r,
\end{eqnarray}
where the transformation $\Pi_k^1$ acts on the points $w$ of the measurable space
 $V_1^k$ by the rule
 \begin{eqnarray*}                    \Pi_k^1 w=
\{w_{\pi(1)}, \ldots, w_{\pi(k)} \}, \quad w=\{w_1, \ldots, w_k \}\in [\Gamma^m \times S^l]^k,\end{eqnarray*}
\begin{eqnarray*}                     w_i=
\{z_i, x_1^i, \ldots, x_l^i\},\quad  i=\overline{1,k}, \quad  w_{\pi(i)}=
\{z_{\pi(i)}, x_1^{\pi(i)}, \ldots, x_l^{\pi(i)}\}
 \in \Gamma^m \times S^l,\end{eqnarray*}
and the set $\Pi_k^1E $  is the image of the set $E$ under this transformation, $\{ \pi(1), \ldots, \pi(k)\}$ is a certain permutation in the set of  integer numbers $\{1, 2, \ldots, k\}.$
\end{theorem}
\begin{proof}\smartqed
Introduce the measurable space
\begin{eqnarray*}                    V^k=\{ [\Gamma^m]^k\times [S^k]^l,
[{\cal B}(\Gamma^m)]^k\times [{\cal B}(S^k)]^l\}. \end{eqnarray*}
Let us consider
 the collection  of  functions of  sets
\begin{eqnarray*}
\Phi_{p_1, \ldots, p_k}({\cal D}\times A_1 \times \ldots \times A_l)
\end{eqnarray*}
\begin{eqnarray*}
 =\int\limits_{{\cal D}}\prod\limits_{i=1}^{l}F_{p_1, \ldots, p_k}^i
\left(A_i\cap X_{(p,z)_{k}}^{k,i}|z_1, \ldots, z_k\right)d\psi_{p_1, \ldots, p_k}(z_1, \ldots, z_k)  \end{eqnarray*}
on the sets of the kind  ${\cal D}\times A_1 \times \ldots \times A_l \in
[{\cal B}(\Gamma^m)]^k\times [{\cal B}(S^k)]^l$,
where ${\cal D} \in  [{\cal B}(\Gamma^m)]^k, \
A_i \in  {\cal B}(S^k), \ i=\overline{1,l}.$

Since
$F_{p_1, \ldots, p_k}^i(A^i|z_1, \ldots, z_k),$  $ A^i \in {\cal B}(X_{(p,z)_k}^{k,i}),$ $ i=\overline{1,l},  $
satisfy the axioms formulated,
\begin{eqnarray*}
F_{p_1, \ldots, p_k}^i\left(A_i\cap X_{(p,z)_{k}}^{k,i}|z_1, \ldots, z_k\right),
\quad A_i \in {\cal B}(S^k), \quad  i=\overline{1,l},
\end{eqnarray*}
are  measures on $\{S^k, {\cal B}(S^k)\} $
for every fixed
\begin{eqnarray*}
\{p_1, \ldots, p_k \} \in
[K_+^n]^k, \quad     \{z_1, \ldots, z_k \} \in
[\Gamma^m]^k.
\end{eqnarray*}
There exists a unique measure
$\mu_{z_1, \ldots, z_k}^{p_1, \ldots, p_k}(E), \
E \in   [{\cal B}(\Gamma^m)]^k\times [{\cal B}(S^k)]^l $ on the measurable space
$V^k$ such that its restriction on the sets of the kind
\begin{eqnarray*}
{\cal D}\times A_1 \times \ldots \times A_l \in
[{\cal B}(\Gamma^m)]^k\times [{\cal B}(S^k)]^l
\end{eqnarray*}
coincides with the measure
\begin{eqnarray*}
\chi_{{\cal D}}(z_1, \ldots, z_k)\prod\limits_{i=1}^{l}F_{p_1, \ldots, p_k}^i
\left(A_i\cap X_{(p,z)_{k}}^{k,i}|z_1, \ldots, z_k\right).
\end{eqnarray*}
This measure   $\mu_{z_1, \ldots, z_k}^{p_1, \ldots, p_k}(E)$
is the direct product of the measure $\chi_{{\cal D}}(z_1, \ldots, z_k) $ and the measures
\begin{eqnarray*}                    F_{p_1, \ldots, p_k}^i
\left(A_i\cap X_{(p,z)_{k}}^{k,i}|z_1, \ldots, z_k\right), \quad  i=\overline{1,l},\end{eqnarray*}                       for every fixed
$\{p_1, \ldots, p_k \} \in
[K_+^n]^k$ and   $\{z_1, \ldots, z_k \} \in
[\Gamma^m]^k.$
Let us prove that  $\mu_{z_1, \ldots, z_k}^{p_1, \ldots, p_k}(E)$
is a measurable map of the measurable space  $L^k=\{[\Gamma^m]^k,[{\cal B}(\Gamma^m)]^k\}$ into the measurable space   $\{[0,1], {\cal B}([0,1])\}$
for every fixed  $\{p_1, \ldots, p_k\} \in R_+^{nk}$  and $E \in [{\cal B}(\Gamma^m)]^k\times [{\cal B}(S^k)]^l.$
It is the case on the sets of the kind
 \begin{eqnarray}\label{1l1}
{\cal D}\times A_1\times\ldots\times A_l,\quad  A_i \in {\cal B}(S^k), \quad i=\overline{1,l}, \quad  {\cal D} \in [{\cal B}(\Gamma^m)]^k,
\end{eqnarray}
 in accordance with the axioms for the  conditional distribution functions
\begin{eqnarray*}
F_{p_1, \ldots, p_k}^i
(A^i|z_1, \ldots, z_k), \quad i=\overline{1,l},
\end{eqnarray*}
and due to that $  \chi_{{\cal D}}(z_1, \ldots, z_k)$ is a measurable map of the measurable space $L^k$ into the measurable space $\{[0,1], {\cal B}([0,1])\}.$

The collection of sets of the kind
 $\bigcup\limits_{i\in I}E_i,$  where
$E_i={\cal D}^i\times A_1^i\times \ldots \times A_l^i$ is the sets of the kind  (\ref{1l1}), that is,  $A_s^i \in  {\cal B}(S^k), \ s=\overline{1,l},\ {\cal D}^i \in [{\cal B}(\Gamma^m)]^k,$
and $I$  is an  arbitrary  finite set, $ E_i \cap E_j= \emptyset, \ i \not = j,$
forms the algebra of the sets that we denote by $U_0.$ Because of the countable additivity of   $\mu_{z_1, \ldots, z_k}^{p_1, \ldots, p_k}(E)$ we have
\begin{eqnarray*}                    \mu_{z_1, \ldots, z_k}^{p_1, \ldots, p_k}(\bigcup\limits_{i\in I}E_i)=
\sum\limits_{i \in I}\mu_{z_1, \ldots, z_k}^{p_1, \ldots, p_k}(E_i),\end{eqnarray*}
therefore, $\mu_{z_1, \ldots, z_k}^{p_1, \ldots, p_k}(\bigcup\limits_{i \in I}E_i )$
is a measurable map of  $L^k$ into  $\{[0,1], {\cal B}([0,1])\}.$

Let  $T$ be a class of the sets from  the minimal  $\sigma$-algebra $\Sigma$  generated by
$U_0$  for every subset $E$ of  that
$\mu_{z_1, \ldots, z_k}^{p_1, \ldots, p_k}( E) $ is a measurable map of  $L^k$ into  $\{[0,1], {\cal B}([0,1])\}.$  Let us prove that  $T$ is a monotonic class.\index{monotonic class} Let $E_i, \ i=$ $1,2, \ldots,$ be a monotone sequence that belongs to
 $T$ and, for example, it is monotonically increasing. Then
$\mu_{z_1, \ldots, z_k}^{p_1, \ldots, p_k}( E_i) $ is a monotonically increasing sequence, so it is convergent.
 But
\begin{eqnarray*}                    \mu_{z_1, \ldots, z_k}^{p_1, \ldots, p_k}( E_{i+1} \setminus E_i)=
\mu_{z_1, \ldots, z_k}^{p_1, \ldots, p_k}( E_{i+1} )-
\mu_{z_1, \ldots, z_k}^{p_1, \ldots, p_k}( E_i)\end{eqnarray*}
is a measurable map of $L^k$ into  $\{[0,1], {\cal B}([0,1])\}.$
From this equality it follows that the set  $ E_{i+1} \setminus E_i$ belongs to the class $T.$
Since
\begin{eqnarray*}                    \bigcup\limits_{i=1}^{\infty}E_i = E_1\cup \bigcup\limits_{i=1}^{\infty}[E_{i+1} \setminus E_i],\end{eqnarray*}
we have
\begin{eqnarray*}
 & \lim\limits_{n \to \infty}\mu_{z_1, \ldots, z_k}^{p_1, \ldots, p_k}( E_n)=
\mu_{z_1, \ldots, z_k}^{p_1, \ldots, p_k}(E_1)+\lim\limits_{n \to \infty}\sum\limits_{i=1}^n \mu_{z_1, \ldots, z_k}^{p_1, \ldots, p_k}( E_{i+1} \setminus E_i)
\end{eqnarray*}
\begin{eqnarray*}
& =\mu_{z_1, \ldots, z_k}^{p_1, \ldots, p_k}(E_1)+\sum\limits_{i=1}^{\infty} \mu_{z_1, \ldots, z_k}^{p_1, \ldots, p_k}( E_{i+1} \setminus E_i)
\end{eqnarray*}
\begin{eqnarray*}
& =\mu_{z_1, \ldots, z_k}^{p_1, \ldots, p_k}( E_1\cup \bigcup\limits_{i=1}^{\infty}[E_{i+1} \setminus E_i])= \mu_{z_1, \ldots, z_k}^{p_1, \ldots, p_k}(\bigcup\limits_{i=1}^{\infty}E_i).
\end{eqnarray*}
From this equalities it follows that $\mu_{z_1, \ldots, z_k}^{p_1, \ldots, p_k}(\bigcup\limits_{i=1}^{\infty}E_i)$ is a measurable map of $L^k$ into  $\{[0,1], {\cal B}([0,1])\}.$
Thus, $ E=\bigcup\limits_{i=1}^{\infty} E_i \in T. $

If the sequence $E_i, \ i=1,2, \ldots,$  belongs to $T$ and is monotonically decreasing, then this
case is reduced  to the previous one by the note that the sequence  $\bar E_i= [{\cal B}(\Gamma^m)]^k\times [{\cal B}(S^k)]^l \ \setminus E_i, \ i=1,2, \ldots,$ belongs to $T$  and is  monotonically increasing.
From here it follows that
$ \bar E=\bigcup\limits_{i=1}^{\infty} \bar E_i \in T. $ Therefore,
\begin{eqnarray*}                    \bigcap\limits_{i=1}^{\infty}  E_i=[{\cal B}(\Gamma^m)]^k\times [{\cal B}(S^k)]^l  \setminus
\bigcup\limits_{i=1}^{\infty} \bar E_i \in T.\end{eqnarray*}
Thus, $T$ is a monotone class. But $ U_0 \subset T.$  Hence, $T$ contains
the minimal monotone class generated by the algebra $U_0,$ that is, $m(U_0).$
But $m(U_0)=\Sigma,$  therefore $ \Sigma \subseteq T.$

Thus,
$\mu_{z_1, \ldots, z_k}^{p_1, \ldots, p_k}(A)$ is a measurable map of $L^k$ into  $\{[0,1], {\cal B}([0,1])\}$
for every fixed
 $\{p_1, \ldots, p_k\} \in K_+^{nk}$ and $A \in \Sigma,$  where $\Sigma$ is the minimal $\sigma$-algebra generated by the algebra $U_0.$

Let us introduce in the measurable space $V^k$ the measurable transformation $\Pi_k$ given by the rule: for any point
$u=\{z_1, \ldots, z_k, y_1, \ldots, y_{l}\} \in
[\Gamma^m]^k\times [S^k]^l, \ $
$y_i$ $ =\{x_{i}^1, \ldots, x_{i}^k\},\ $ $i=\overline{1,l},$  we put
\begin{eqnarray*}                    \Pi_k u= \{z_{\pi(1)}, \ldots, z_{\pi(k)}, \pi^1 y_1, \ldots, \pi^l y_{l}\},\end{eqnarray*}
 where $\pi^i y_i=\{ x_{i}^{\pi(1)}, \ldots, x_{i}^{\pi(k)} \},$  and
$\{\pi(1), \ldots, \pi(k)\}$ is a permutation in the set of indices
$\{1, \ldots, k\}.$

The constructed family of  measures $\mu_{z_1, \ldots, z_k}^{p_1, \ldots, p_k}( E)$ satisfies condition
\begin{eqnarray*}                    \mu_{z_{\pi(1)}, \ldots, z_{\pi(k)}}^{p_{\pi(1)}, \ldots, p_{\pi(k)}}(\Pi_k E)
= \mu_{z_1, \ldots, z_k}^{p_1, \ldots, p_k}( E), \quad  E \in [{\cal B}(\Gamma^m)]^k\times [{\cal B}(S^k)]^l,  \end{eqnarray*}
where by  $\Pi_k E$ we denote the image of the set  $E$ under transformation $\Pi_k.$
This property holds for the sets of the kind
$ E={\cal D}\times A_1\times \ldots \times A_l,$
where
\begin{eqnarray*}
 A_i=\prod\limits_{s=1}^k A_i^s, \quad  A_i^s \in {\cal B}(S),  \quad  {\cal D}=\prod\limits_{s=1}^k{\cal D}_s, \quad  {\cal D}_s \in {\cal B}(\Gamma^m), \quad s=\overline{1,k}, \quad i=\overline{1,l},
 \end{eqnarray*}
because  the conditional distribution functions
$F^i_{p_1, \ldots, p_k}(A^k|z_1, \ldots, z_k)$
 satisfy axioms formulated.
The transformation
$\Pi_k$ maps the sets of the kind   $ E={\cal D}\times A_1\times \ldots \times A_l$ into the sets of the same kind
\begin{eqnarray*}                    \Pi_k E=\pi^0{\cal D}\times \pi^1A_1\times \ldots \times\pi^l A_l,\end{eqnarray*}
where
\begin{eqnarray*}                     \pi^0{\cal D}=\prod\limits_{s=1}^k{\cal D}_{\pi(s)}, \quad
\pi^iA_i=\prod\limits_{s=1}^k A_i^{\pi(s)}, \quad i=\overline{1,l}.\end{eqnarray*}
Therefore,
\begin{eqnarray*}
 \mu_{z_{\pi(1)}, \ldots, z_{\pi(k)}}^{p_{\pi(1)}, \ldots, p_{\pi(k)}}(\Pi_k E)
 \end{eqnarray*}
\begin{eqnarray*}
=\chi_{\pi^0{\cal D}}(z_{\pi(1)}, \ldots, z_{\pi(k)})\prod\limits_{i=1}^{l}F_{p_{\pi(1)}, \ldots, p_{\pi(k)}}^i
\left(\pi^iA_i\cap \pi^iX_{(p,z)_{k}}^{k,i}|z_{\pi(1)}, \ldots, z_{\pi(k)}\right),
\end{eqnarray*}
where the notation
$\pi^iX_{(p,z)_{k}}^{k,i}=\prod\limits_{s=1}^k\hat X_{(p_{\pi(s)}, z_{\pi(s)})}^i$
is introduced.
But on the basis of the axioms for the conditional distribution functions we have
\begin{eqnarray*}
   F_{p_{\pi(1)}, \ldots, p_{\pi(k)}}^i
\left(\pi^iA_i\cap \pi^iX_{(p,z)_{k}}^{k,i}|z_{\pi(1)}, \ldots, z_{\pi(k)}\right)
\end{eqnarray*}
\begin{eqnarray*}
 =F_{p_{\pi(1)}, \ldots, p_{\pi(k)}}^i
\left(\pi^i[A_i\cap X_{(p,z)_{k}}^{k,i}]|z_{\pi(1)}, \ldots, z_{\pi(k)}\right)
\end{eqnarray*}
\begin{eqnarray*}
 =F_{p_1, \ldots, p_k}^i
\left(A_i\cap X_{(p,z)_{k}}^{k,i}|z_1, \ldots, z_k\right), \quad i=\overline{1,l},
\end{eqnarray*}
\begin{eqnarray*}
 \chi_{\pi^0{\cal D}}(z_{\pi(1)}, \ldots, z_{\pi(k)})=\chi_{{\cal D}}(z_1, \ldots, z_k).\end{eqnarray*}
Therefore, the asserted equality holds for the considered class of the sets.

If $K$ is a class of the sets for that this property holds, then $K$
contains the algebra $U_0$ and $K$ is a monotone class therefore $\Sigma \subseteq K,$
where $\Sigma$ is the minimal $\sigma$-algebra generated by the algebra $U_0.$
So, $\mu_{z_1, \ldots, z_k}^{p_1, \ldots, p_k}( E)$ satisfies the condition formulated.

Let us give the measurable map $T_k$ of the measurable space $V^k$ into the measurable space
$V_1^k= \{[\Gamma^m\times S^l]^k, [{\cal B}(\Gamma^m)\times {\cal B}(S^l)]^k\}$
by the rule
\begin{eqnarray*}
T_k u=w,\quad
u=\{z_1, \ldots, z_k, y_1,\ldots,  y_l\} \in [\Gamma^m]^k\times [S^k]^l,
\end{eqnarray*}
\begin{eqnarray*}
 w=\{w_i\}_{i=1}^k \in [\Gamma^m\times S^l]^k, \quad w_i=\{z_i, x_1^i, \ldots, x_l^i\} \in \Gamma^m\times S^l,
\quad  i=\overline{1,k},
\end{eqnarray*}
\begin{eqnarray*}
 y_i=\{x_i^1, \ldots, x_i^k\} \in S^k, \quad i=\overline{1,l}.
  \end{eqnarray*}

The map  $T_k$ is a continuous one to one correspondence of the space
 $[\Gamma^m]^k\times [S^k]^l$ on the space  $[\Gamma^m\times S^l]^k.$

Let us define on the measurable space $V_1^k$ the probability measure $\bar \mu_{z_1, \ldots, z_k}^{p_1, \ldots, p_k}(E)$ that is the image of the measure
$\mu_{z_1, \ldots, z_k}^{p_1, \ldots, p_k}(E)$ under map $T_k,$ that is, we put
\begin{eqnarray*}                     \bar \mu_{z_1, \ldots, z_k}^{p_1, \ldots, p_k}(E)= \mu_{z_1, \ldots, z_k}^{p_1, \ldots, p_k}(T_k^{-1}(E)), \quad E \in [{\cal B}(\Gamma^m)\times {\cal B}(S^l)]^k.   \end{eqnarray*}
In accordance with the proved statement
$ \mu_{z_1, \ldots, z_k}^{p_1, \ldots, p_k}(E)$ is a measurable map
of the measurable space $L^k$ into the measurable space   $\{[0,1], {\cal B}([0,1])\}$
for every fixed
 $\{p_1, \ldots, p_k\} \in K_+^{nk}$ and $E \in [{\cal B}(\Gamma^m)]^k\times [{\cal B}(S^k)]^l,$
therefore   $\bar \mu_{z_1, \ldots, z_k}^{p_1, \ldots, p_k}(E)= \mu_{z_1, \ldots, z_k}^{p_1, \ldots, p_k}(T_k^{-1}(E)), \ E \in [{\cal B}(\Gamma^m)\times {\cal B}(S^l)]^k, $ is the same one.
Let us define in the measurable space $V_1^k$ the measurable transformation   $\Pi_k^1,$  that
acts on the points $w=\{w_1, \ldots, w_k \} \in V_1^k$ by the rule
\begin{eqnarray*}                    \Pi_k^1 w= \{w_{\pi(1)}, \ldots, w_{\pi(k)} \},\quad   w_{\pi(i)}=
\{z_{\pi(i)}, x_1^{\pi(i)}, \ldots, x_l^{\pi(i)}\} \in \Gamma^m \times S^l.\end{eqnarray*}
If $E \in [{\cal B}(\Gamma^m)\times {\cal B}(S^l)]^k,$  then it is obvious that
\begin{eqnarray}\label{ksl4}
T_k^{-1}(\Pi_k^1 E)=\Pi_k T_k^{-1}(E),
\end{eqnarray}
where by $\Pi_k^1 E$ we denote the image of the set $E$ under transformation $\Pi_k^1,$ and by $T_k^{-1}(E)$ we do the preimage of the set $E$ under the transformation $T_k$ that in the present case coincides with the image  of the set $E$ under the transformation $T_k^{-1},$ that is, $T_k^{-1}E=T_k^{-1}(E).$

Really, the equality (\ref{ksl4}) holds, because   for any  point $w $ the equality
\begin{eqnarray*}                     T_k^{-1}\Pi_k^1w= \Pi_k T_k^{-1}w, \quad  w \in [\Gamma^m\times S^l]^k,\end{eqnarray*}
 takes place.
From the last pointwise equality we have that for every set $E $
\begin{eqnarray*}                     T_k^{-1}\Pi_k^1E= \Pi_k T_k^{-1}E, \quad E \in [{\cal B}(\Gamma^m)\times {\cal B}(S^l)]^k,\end{eqnarray*}
where by $ T_k^{-1}\Pi_k^1E$ and $\Pi_k T_k^{-1}E$ we understand the image of the set $E$  under the transformations   $ T_k^{-1}\Pi_k^1$  and $\Pi_k T_k^{-1},$ correspondingly.

 Taking into account the equality $T_k^{-1}E=T_k^{-1}(E),$ we obtain the equality (\ref{ksl4}).
 In the last equality from the left side  the image of the set  $E$  under the transformation $T_k^{-1}$ stands  and from the right one the preimage of the set $E$ under the transformation $T_k$ does.  This equality takes place because  the transformation $ T_k $ is a one to one correspondence  of  $[\Gamma^m]^k\times [S^k]^l$ on  $[\Gamma^m\times S^l]^k.$

The constructed family of measures satisfies  the condition
\begin{eqnarray*}
\bar \mu_{z_{\pi(1)}, \ldots, z_{\pi(k)}}^{p_{\pi(1)}, \ldots, p_{\pi(k)}}(\Pi_k^1 E)
=\bar \mu_{z_1, \ldots, z_k}^{p_1, \ldots, p_k}( E),
\end{eqnarray*}
where by $\Pi_k^1 E$ we denote the image of the set $E$ under the transformation $\Pi_k^1.$
Really,
\begin{eqnarray*}
\bar \mu_{z_{\pi(1)}, \ldots, z_{\pi(k)}}
^{p_{\pi(1)}, \ldots, p_{\pi(k)}}(\Pi_k^1 E)
= \mu_{z_{\pi(1)}, \ldots, z_{\pi(k)}}
^{p_{\pi(1)}, \ldots, p_{\pi(k)}}(T_k^{-1}(\Pi_k^1 E))
\end{eqnarray*}
\begin{eqnarray*}
 = \mu_{z_{\pi(1)}, \ldots, z_{\pi(k)}}
^{p_{\pi(1)}, \ldots, p_{\pi(k)}}(\Pi_k(T_k^{-1}(E))=
\mu_{z_1, \ldots, z_k}^{p_1, \ldots, p_k}( T_k^{-1}(E))
=\bar \mu_{z_1, \ldots, z_k}^{p_1, \ldots, p_k}( E).
 \end{eqnarray*}

There holds such an equality
\begin{eqnarray*}
 \bar \mu_{z_1, \ldots, z_k}^{p_1, \ldots, p_k}
(A\times (\Gamma^m\times S^l)^{k-r})=
 \bar \mu_{z_1, \ldots, z_r}^{p_1, \ldots, p_r}(A), \quad
A \in [{\cal B}(\Gamma^m)\times {\cal B}(S^l)]^r, \quad  k > r.
 \end{eqnarray*}

Really, let $B \in  [{\cal B}(\Gamma^m) \times {\cal B}(S^l)]^k$ and have the form
\begin{eqnarray}\label{sl5}
B=A\times (\Gamma^m\times S^l)^{k-r},
\quad  A \in [{\cal B}(\Gamma^m)\times {\cal B}(S^l)]^r, \quad 1 \leq r < k.
\end{eqnarray}
Then the equality
\begin{eqnarray*}
\mu_{z_1, \ldots, z_k}^{p_1, \ldots, p_k}( T_k^{-1}(B)) =
 \mu_{z_1, \ldots, z_r}^{p_1, \ldots, p_r}( T_r^{-1}(A))\end{eqnarray*}
holds.

Really, if   $B$ is the set from the measurable space $V_1^k$
of the form
\begin{eqnarray*}                    B=\prod\limits_{s=1}^k[D_s\times\prod\limits_{i=1}^lA_i^s], \quad  {\cal D}_s \in {\cal B}(\Gamma^m), \quad A_i^s \in {\cal B}(S), \quad i=\overline{1,l}, \quad s=\overline{1,k}, \end{eqnarray*}
then its pre-image
\begin{eqnarray*}                    T_k^{-1}(B)= {\cal D}\times A_1\times\ldots\times A_l,\end{eqnarray*}
where
\begin{eqnarray*}
{\cal D}=\prod\limits_{s=1}^k{\cal D}_s,\quad A_i=\prod\limits_{s=1}^kA_i^s, \quad  A_i^s \in {\cal B}(S),\quad  {\cal D}_s \in {\cal B}(\Gamma^m), \quad i=\overline{1,l}, \quad s=\overline{1,k}.
\end{eqnarray*}
In the case when
\begin{eqnarray*}
B=\prod\limits_{s=1}^r[D_s\times\prod\limits_{i=1}^lA_i^s]
\times [\Gamma^m\times S^l]^{k-r}, \quad 1 \leq r < k,
\end{eqnarray*}
\begin{eqnarray*}
 T_k^{-1}(B)
\end{eqnarray*}
\begin{eqnarray*}
={\cal D}_1\times\ldots \times {\cal D}_r \times [\Gamma^m]^{k-r}
\times A_1^1\times \ldots \times A_1^{r}\times S^{k-r}\times \ldots
\times A_l^1\times \ldots \times A_l^{r}\times S^{k-r}.
\end{eqnarray*}
Therefore, on such a  set
\begin{eqnarray*}
 \mu_{z_1, \ldots, z_k}^{p_1, \ldots, p_k}( T_k^{-1}(B))
\end{eqnarray*}
\begin{eqnarray*}
 =\chi_{{\cal D}_1\times\ldots \times {\cal D}_r \times [\Gamma^m]^{k-r}}
(z_1, \ldots, z_k)
\end{eqnarray*}
\begin{eqnarray*}
\times\prod\limits_{i=1}^l
F_{p_1, \ldots, p_k}^i\left(A_i^1\times \ldots \times A_i^r\times S^{k-r}\cap
X_{(p,z)_k}^{k,i}|z_1, \ldots ,z_k\right)
\end{eqnarray*}
\begin{eqnarray*}
 =\chi_{{\cal D}_1\times\ldots \times {\cal D}_r}(z_1, \ldots, z_r)
\prod\limits_{i=1}^lF_{p_1, \ldots, p_r}^i\left(A_i^1\times \ldots \times A_i^{r}
 \cap X_{(p,z)_k}^{r,i}\left|\right.z_1, \ldots ,z_r\right)
 \end{eqnarray*}
\begin{eqnarray*}
= \mu_{z_1, \ldots, z_r}^{p_1, \ldots, p_r}( T_r^{-1}(A)).
 \end{eqnarray*}
Thus,  such an  equality takes place as  $A$ belongs to the algebra
generated by the  finite  unions of such sets that have  empty intersections.
Denote this algebra by $\Delta_0.$

 Let $G$ be a class of  the sets for that this equality  is valid.
It is obvious that  $G$ is  a monotone class,\index{monotone class} therefore it contains the minimal  monotone class\index{minimal  monotone class}  generated by $\Delta_0,$ so it contains  the   $\sigma$-algebra
 \begin{eqnarray*}
 [{\cal B}(\Gamma^m)\times {\cal B}(S^l)]^r
\times [\Gamma^m\times S^l]^{k-r}, \quad  1 \leq  r < k.
\end{eqnarray*}

By the family of the measures
$\bar \mu_{z_1, \ldots, z_k}^{p_1, \ldots, p_k}( E), $ where
$E  \in [{\cal B}(\Gamma^m)\times {\cal B}(S^l)]^k,$
we introduce the family of measures
\begin{eqnarray*}                    \bar \Phi_{p_1, \ldots, p_k}( E)=\int\limits_{[\Gamma^m]^k}
\bar \mu_{z_1, \ldots, z_k}^{p_1, \ldots, p_k}( E)
d \psi_{p_1, \ldots, p_k}(z_1, \ldots, z_k).\end{eqnarray*}

From the Theorems \ref{ffl1} and  \ref{ffl2}, the properties of the family of the measures constructed
$\bar \mu_{z_1, \ldots, z_k}^{p_1, \ldots, p_k}( E),$
axioms  that $\psi_{p_1, \ldots, p_k}(D),$  $  D \in {\cal B}([\Gamma^m]^k),$ satisfy,
we have the following equalities
\begin{eqnarray*}                    \bar \Phi_{p_{\pi(1)}, \ldots, p_{\pi(k)}}(\Pi_k^1 E)=
  \bar \Phi_{p_1, \ldots, p_k}( E), \quad E \in [{\cal B}(\Gamma^m)\times {\cal B}(S^l)]^k, \end{eqnarray*}
\begin{eqnarray*}                    \bar \Phi_{p_1, \ldots, p_k}
(A\times (\Gamma^m\times S^l)^{k-r})=
 \bar \Phi_{p_1, \ldots, p_r}(A), \
A \in [{\cal B}(\Gamma^m)\times {\cal B}(S^l)]^r, \ 1 \leq r < k.  \end{eqnarray*}
The Theorem \ref{ffl1} is used for the proof of  the first equality for the measurable map  $\Pi_k^1$  that have inverse map being  also measurable,  and the Theorem  \ref{ffl2} is used  for the proof of the second equality.
\qed \end{proof}

From the Theorem   \ref{ttl1} it follows the next Theorem.

\begin{theorem}\label{qtl2}
The family of  countably additive finite dimensional distributions
$\bar \Phi_{p_1, \ldots, p_k}( E),$ $\{p_1, \ldots, p_k\} \in [K_+^n]^k,$  $E \in
[{\cal B}(\Gamma^m) \times {\cal B}(S^l)]^{k},$
 constructed in the Theorem  \ref{ttl1}, satisfies the conditions of the Kolmogorov Theorem \cite{3} with  the full  separable metric space  of  states $X=\Gamma^m\times S^l$
and   the $\sigma$- algebra of subsets
$\Sigma= {\cal B}(\Gamma^m)\times {\cal B}(S^l),$  so it generates  the unique measure
$\bar P$ on the measurable space $\{X^T, \Sigma^T\}$
such that the finite dimensional distributions of the random field
\begin{eqnarray*}                    \nu_p(\omega)=\{\zeta(p), \xi_1(p), \ldots, \xi_l(p)\}=\omega(p), \quad
\omega(p) \in X^T, \quad   p \in K_+^n,\end{eqnarray*}
coincides with the family  $\bar \Phi_{p_1, \ldots, p_k}( E),$ that is,
\begin{eqnarray*}                    \bar P(\omega \in X^T, \{\nu_{p_1}(\omega), \ldots , \nu_{p_k}(\omega)\} \in E)=
\bar \Phi_{p_1, \ldots, p_k}( E), \
E \in [{\cal B}(\Gamma^m) \times {\cal B}(S^l)]^{k}.
\end{eqnarray*}
By  $X^T$ we denote the set of all functions given on the set
 $T=K_+^n$ with  values in the set $X=\Gamma^m\times S^l,$
$\Sigma^T$ is  the minimal   $\sigma$-algebra generated by  cylindric sets of the kind
\begin{eqnarray*}                    \{\omega(p) \in X^T,
\{\nu_{p_1}(\omega), \ldots , \nu_{p_1}(\omega)\} \in E \}, \quad
E \in [{\cal B}(\Gamma^m) \times {\cal B}(S^l)]^{k}.\end{eqnarray*}
\end{theorem}

\section{Conditionally independent random fields}

In this subsection, a new class of random fields is constructed that will be used to construct  random fields of consumers choice in the information model of economy \cite{55, 71, 69, 92, 106}.

\begin{definition}\label{drl1} Let $\{\Omega, {\cal F}, \bar P\}$ be a probability space.
The family of sub $\sigma$-algebras  $ \{B_i, i \in I\}$ of the $\sigma$-algebra ${\cal F}$
is called conditionally independent relative to  a sub $\sigma$-algebra $B \subseteq  {\cal F},$
if almost everywhere
 \begin{eqnarray*}
 E\left \{\prod\limits_{j \in I_{s}}X_j | B\right\}=\prod_{j \in I_{s} }E\{X_j | B\}
 \end{eqnarray*}
for any finite subset $I_s \subseteq I$ and the family of integrable random values
$\{X_j, j \in I_s\},$ where $X_j $ is a $B_j$-measurable nonnegative random value.
\end{definition}
\begin{definition}\label{drl2}
Let  random fields $\xi_1(p), \ldots, \xi_l(p), \ p \in K_+^n,$ of consumers choice   be given on a probability space  $\{\Omega, {\cal F}, \bar P\},$  that take  values in the set of possible goods $S,$  that is, they are  measurable maps  of $\{\Omega, {\cal F}\}$
into the measurable space $\{S, B(S)\}$  for every fixed  $p \in K_+^n,$ and let a random field of decisions making by $m$ firms $\zeta(p)=\{\zeta_1(p), \ldots,  \zeta_m(p)\}$  be a measurable map of $\{\Omega, {\cal F} \}$  into  $\{\Gamma^m, {\cal B}(\Gamma^m)\}$ for every fixed
$p \in K_+^n.$ Random fields of consumers choice are conditionally independent\index{conditionally independent random fields} relative to
the random field of decisions making by firms if
$\sigma$-algebras
  \begin{eqnarray*}
  {\cal F}_i={\cal F}\{\xi_i(p), \ p \in K_+^n\},
\quad i= \overline{1,l},\end{eqnarray*}
are conditionally independent relative to  $\sigma$-algebra
${\cal F}_0={\cal F}\{\zeta(p), \ p \in K_+^n\},$ where the $\sigma$-algebras
${\cal F}_i, \ i=\overline{0,l},$ are the  minimal $\sigma$-algebras generated by the family of the random values
$\{\zeta(p), \ p \in K_+^n\}$ for $i=0$ and by the family of the random values
$\{\xi_i(p), \ p \in K_+^n\}$ for  $ i=\overline{1,l}.$
\end{definition}

\subsection{Construction of the conditionally independent random fields}

Let us consider $(l+1)$ measurable spaces $\{\Omega_i, {\cal F}_i, \bar P_i\}, \
i=\overline{0,l},$  and their direct product $\{\Omega, {\cal F}, \bar P\},$
where $\Omega=\prod\limits_{i=0}^{l}\Omega_i,$ ${\cal F}=\prod\limits_{i=0}^{l}
{\cal F}_i,$ $\bar P=\prod\limits_{i=0}^{l}
\bar P_i.$
\begin{lemma}\label{pl1}
Let  $ \eta_i(p, z, \omega_i)$
be a measurable mapping of the measurable space  $\{\Gamma^m\times\Omega_i,
{\cal B}(\Gamma^m)\times {\cal F}_i\}$  into the measurable space
$\{S, {\cal B}(S)\} $ for every fixed $p \in K_+^n,$  $ i=\overline{1,l},$  and   $\zeta(p, \omega_0)$ be a
 measurable map of the measurable  space  $\{\Omega_0, {\cal F}_0\}$ into the measurable space
$\{\Gamma^m, {\cal B}(\Gamma^m)\}$  for every fixed $ p \in K_+^n.$
Then for every fixed
$p \in K_+^n,$  $ \eta_i(p, \zeta(p, \omega_0), \omega_i)$ is a  measurable map of the measurable space
$\{\Omega, {\cal F}\}$ into the measurable space $\{S, {\cal B}(S)\},\  i=\overline{1,l}.$
\end{lemma}
\begin{proof} \smartqed Consider a measurable map  $f_i(p, \omega) $
of the measurable space  $\{\Omega, {\cal F}\}$
into the measurable space
$\{\Gamma^m\times\Omega_i, {\cal B}(\Gamma^m)\times {\cal F}_i\}$
given by the rule
\begin{eqnarray*}
f_i(p,\omega)=\{\zeta(p,\omega_0), \omega_i\}, \quad
\omega=\{\omega_0, \omega_1, \ldots, \omega_l\}, \quad i=\overline{1,l}.
\end{eqnarray*}
The measurability of $f_i(p,\omega)$ for every fixed  $p \in K_+^n$  follows from the equality
\begin{eqnarray}\label{sl1}
\{\omega, \ f_i(p,\omega) \in {\cal D}\times K_i\}
\end{eqnarray}
\begin{eqnarray*}
 =\zeta^{-1}(p,{\cal D})\times\Omega_1\times\ldots\times\Omega_{i-1}\times
K_i\times\Omega_{i+1}\times\ldots\times\Omega_l,
\end{eqnarray*}
where $\zeta^{-1}(p,{\cal D})=\{\omega_0, \zeta(p,\omega_0) \in {\cal D}\}, \
{\cal D} \in {\cal B}(\Gamma^m), \ K_i \in {\cal F}_i.$
It is evident that the set  (\ref{sl1})  belongs to the $\sigma$-algebra ${\cal F}$
if
${\cal D} \in {\cal B}(\Gamma^m), \ K_i \in {\cal F}_i.$

Let $ T $ be a class of subsets from the set $ {\cal B}(\Gamma^m)\times {\cal F}_i $ such that for every set $Q  \in T$ the set
$\{\omega \in \Omega,  f_i(p,\omega) \in Q\} \in {\cal F}.$ Then the set $T$ contains the algebra of the sets generated by the finite unions of the sets of the kind    ${\cal D}\times K_i,$ where ${\cal D} \in {\cal B}(\Gamma^m), \ K_i \in {\cal F}_i,$ that do not intersect.
It is evident that   the class $T$
forms the
$\sigma$-algebra since
\begin{eqnarray*}
 \{\omega, \ f_i(p,\omega)  \in Q_1 \setminus Q_2 \}=
\{\omega, \ f_i(p,\omega) \in Q_1 \} \setminus
\{\omega, \ f_i(p,\omega) \in Q_2 \},  \end{eqnarray*}
 \begin{eqnarray*}                                   \left\{\omega,\ f_i(p,\omega)  \in \bigcup\limits_{i=1}^{\infty}Q_i\right\}=
\bigcup\limits_{i=1}^{\infty} \{\omega, \ f_i(p,\omega)  \in  Q_i\}.\end{eqnarray*}
From here it follows that $T$ coincides with the $\sigma$-algebra $ {\cal B}(\Gamma^m)\times {\cal F}_i.$  Therefore $f_i(p,\omega)$
is a measurable map of the measurable space $\{\Omega, {\cal F}\}$
into the measurable space
$\{\Gamma^m\times\Omega_i, {\cal B}(\Gamma^m)\times {\cal F}_i\},
\ i=\overline{1,l}.$
So, if
$ \eta_i(p, z, \omega_i)
$ is a measurable  map of the measurable space
$\{\Gamma^m\times\Omega_i, {\cal B}(\Gamma^m)\times {\cal F}_i\}$
into the measurable space  $\{S, {\cal B}(S)\}$ for every fixed  $p \in K_+^n,$ $ i=\overline{1,l},$ then  the superposition of the maps
 \begin{eqnarray*}
 \xi_i(p)=\eta_i\circ f_i=\eta_i(p, \zeta(p,\omega_0), \omega_i),
\quad i=\overline{1,l},\end{eqnarray*}
 is a measurable map of the measurable space $\{\Omega, {\cal F}\}$ into the measurable space  $\{S, {\cal B}(S)\}$ for every fixed $p \in K_+^n.$
\qed
\end{proof}

\begin{note}
In the  conditions  of the Lemma \ref{pl1}, it is  assumed that every map
$ \eta_i(p, z, \omega_i)$
is a measurable map  of the  space
$\{\Gamma^m\times\Omega_i, {\cal B}(\Gamma^m)\times {\cal F}_i\}
$   into the  space  $\{S, {\cal B}(S)\},$ $ i=\overline{1,l}, $
for every fixed   $p \ \in K_+^n.$  If
$ \eta_i(p, z, \omega_i)$ is a measurable map of the space
$\{K_+^n\times\Gamma^m\times\Omega_i,
{\cal B}(K_+^n)\times {\cal B}(\Gamma^m)\times {\cal F}_i\}
$   into the  space $\{S, {\cal B}(S)\},\ i=\overline{1,l},$ then the section, that corresponds to every fixed
$p \in K_+^n,$  is a measurable map  of the measurable space
$\{\Gamma^m\times\Omega_i, {\cal B}(\Gamma^m)\times {\cal F}_i\}
$  into the measurable space $\{S, {\cal B}(S)\}.$
Analogous statement  is concerned  with the map
$\zeta(p, \omega_0)$
of the measurable space $\{\Omega_0, {\cal F}_0\}$  into the measurable space
$\{\Gamma^m, {\cal B}(\Gamma^m)\}$ for every fixed $ p \in K_+^n,$
that is, if
$\zeta(p, \omega_0)$
is a measurable map of the measurable space $\{K_+^n\times\Omega_0,
{\cal B}(\Gamma^m)\times {\cal F}_0\}$  into the measurable space
$\{\Gamma^m, {\cal B}(\Gamma^m)\},$ then the section, that corresponds to every fixed $ p \in K_+^n,$ is a measurable map of the space
$\{\Omega_0,
{\cal F}_0\}$  into the space
$\{\Gamma^m, {\cal B}(\Gamma^m)\}.$
\end{note}
From this Note it follows such a consequence from the Lemma \ref{pl1}.
\begin{lemma}\label{pl2}
Let $ \eta_i(p, z, \omega_i)$  be a measurable mapping of the measurable  space
$\{K_+^n\times\Gamma^m\times\Omega_i,
{\cal B}(K_+^n)\times {\cal B}(\Gamma^m)\times {\cal F}_i\}
$ into the measurable  space $\{S, {\cal B}(S)\},$ $i =\overline{1,l},$ \hfill
and \hfill $\zeta(p, \omega_0)$ \hfill
be \hfill a \hfill measurable \hfill mapping \hfill of \hfill the \hfill measurable \hfill space \\ $\{K_+^n\times\Omega_0,
{\cal B}(\Gamma^m)\times {\cal F}_0\}$ into the measurable space
$\{\Gamma^m, {\cal B}(\Gamma^m)\}.$  For every fixed
$p \in K_+^n,$ $\xi_i(p, \omega)=\eta_i(p, \zeta(p, \omega_0), \omega_i)$
is a measurable map of the measurable space
$\{\Omega, {\cal F}\}$ into the measurable space $\{S, {\cal B}(S)\}$ for every fixed
$p \in K_+^n,\ $ $ i=\overline{1,l}.$
\end{lemma}

\begin{lemma}\label{rtpl3}
Let
$ \eta_i(p, z, \omega_i)$ be a measurable mapping of the measurable space
$\{\Gamma^m\times\Omega_i,  {\cal B}(\Gamma^m)\times {\cal F}_i\}
$  into the measurable space  $\{S, {\cal B}(S)\}$
for every fixed $p \in K_+^n, \ i=\overline{1,l}.$ Then
$ \eta_i(p, z_s, \omega_i), \ i=\overline{1,l},$ for every fixed
$s=\overline{1,k}$ is a measurable map of the measurable space
$\{[\Gamma^m]^k\times\bar \Omega_1,  {\cal B}([\Gamma^m]^k)\times
\bar {\cal F}_1\}$
into the measurable space $\{S, {\cal B}(S)\},$ where
$ \bar \Omega_1=\prod\limits_{i=1}^l\Omega_i, \
\bar {\cal F}_1=\prod\limits_{i=1}^l {\cal F}_i. $
\end{lemma}
\begin{proof}\smartqed            The map $ \eta_i(p, z_s, \omega_i)$ given on
$\{[\Gamma^m]^k\times\bar \Omega_1,  {\cal B}([\Gamma^m]^k)\times
\bar {\cal F}_1\}$  is the superposition of two measurable mappings: the mapping $R_s^i$  of the measurable space   $\{[\Gamma^m]^k\times\bar \Omega_1,  {\cal B}([\Gamma^m]^k)\times
\bar {\cal F}_1\}$ into the space $\{\Gamma^m\times\Omega_i,  {\cal B}(\Gamma^m)\times {\cal F}_i\}$  that is given by the formula
 \begin{eqnarray*}
  R_s^i u=v, \quad u=\{z_1,\ldots,z_k, \omega_1, \ldots, \omega_l\},
\quad v=\{z_s, \omega_i\}, \quad 1 \leq s \leq k,
\end{eqnarray*}
and the map
$ \eta_i(p, z, \omega_i)$
of the measurable space
$\{\Gamma^m\times\Omega_i,  {\cal B}(\Gamma^m)\times {\cal F}_i\}
$  into the measurable space  $\{S, {\cal B}(S)\},\ i=\overline{1,l}.$
The map  $R_s^i$ is a measurable one since the preimage of the  rectangle  $D \times A_i$
is the set
 \begin{eqnarray*}
 [[\Gamma^m]^{s-1}\times D \times  [\Gamma^m]^{k-s}\times \Omega_1 \times \ldots  \times
\Omega_{s-1} \times A_i  \times  \Omega_{s+1}\times \ldots  \times \Omega_{l},
\end{eqnarray*}
that belongs to the $\sigma$-algebra ${\cal B}([\Gamma^m]^k)\times \bar {\cal F}_1.$
As the set of rectangles generates ${\cal B}(\Gamma^m)\times {\cal F}_i$  the measurability of  $R_s^i$ results.  \qed \end{proof}

\begin{lemma}\label{pl3}
Let
$ \eta_i(p, z, \omega_i)$ be a measurable mapping of the measurable space
$\{\Gamma^m\times\Omega_i,  {\cal B}(\Gamma^m)\times {\cal F}_i\}
$ into the measurable space  $\{S, {\cal B}(S)\},\ i=\overline{1,l},$
for every fixed $p \in K_+^n.$  Then
$\chi_{A}(\eta_i(p_1, z_1, \omega_i), \ldots, \eta_i(p_k, z_k, \omega_i))$
 is a measurable map of the measurable space
$\{[\Gamma^m]^k\times\bar \Omega_1,  {\cal B}([\Gamma^m]^k)\times
\bar {\cal F}_1\}$
into the measurable space $\{R^1, {\cal B}(R^1)\}$
for every fixed  $\{p_1, \ldots, p_k\} \in [K_+^n]^k,$
where
\begin{eqnarray*}
A \in {\cal B}(S^k), \quad
 \bar \Omega_1=\prod\limits_{i=1}^l\Omega_i, \quad
\bar {\cal F}_1=\prod\limits_{i=1}^l {\cal F}_i.
\end{eqnarray*}
\end{lemma}
\begin{proof}\smartqed             Let us consider the map
  $ \eta_i(p_s, z_s, \omega_i), \ i=\overline{1,l},$ on the measurable space
$\{[\Gamma^m]^k\times\bar \Omega_1,  {\cal B}([\Gamma^m]^k)\times
\bar {\cal F}_1\}.$
Since $ \eta_i(p, z, \omega_i)$
is a measurable map of the measurable space
$\{\Gamma^m\times\Omega_i,  {\cal B}(\Gamma^m)\times {\cal F}_i\}
$  into the measurable space $\{S, {\cal B}(S)\}, \ i=\overline{1,l}, $
for every fixed $p \in K_+^n,$ the set
 \begin{eqnarray*}
 \Phi_s^i=\{\{z_1, \ldots, z_k, \omega_1, \ldots,\omega_l\}, \eta_i(p_s, z_s, \omega_i)
\in A_s\},\quad  A_s \in {\cal B}(S), \quad  i=\overline{1,l},
\end{eqnarray*}
belongs to the  $\sigma$-algebra ${\cal B}([\Gamma^m]^k)\times \bar {\cal F}_1$ due to the Lemma \ref{rtpl3}.
Further, for
\begin{eqnarray*}
A=\prod\limits_{s=1}^kA_s, \quad A_s \in {\cal B}(S),
\end{eqnarray*}
 \begin{eqnarray*}
 \{\{z_1, \ldots, z_k, \omega_1, \ldots,\omega_l\}, \
\{\eta_i(p_1, z_1, \omega_i), \ldots,  \eta_i(p_k, z_k, \omega_i)\} \in A\}
\end{eqnarray*}
\begin{eqnarray*}
=\bigcap\limits_{s=1}^k \Phi_s^i \in
{\cal B}([\Gamma^m]^k)\times \bar {\cal F}_1, \quad  i=\overline{1,l}.
\end{eqnarray*}
Therefore $\zeta^i_{p_1,\ldots, p_k}(z_1, \ldots, z_k)=
\{\eta_i(p_1, z_1, \omega_i), \ldots,  \eta_i(p_k, z_k, \omega_i)\}$
is a measu\-rable map of the measurable space
$\{[\Gamma^m]^k\times\bar \Omega_1,  {\cal B}([\Gamma^m]^k)\times
\bar {\cal F}_1\}$
 into the measurable space $\{S^k, {\cal B}(S^k)\},$
since the class  $T$ of those sets $B \in {\cal B}(S^k)$ for that
\begin{eqnarray*}
\{\{z_1, \ldots, z_k, \omega_1, \ldots,\omega_l\},\
\zeta^i_{p_1, \ldots, p_k}(z_1,\ldots, z_k,) \in B\} \in
{\cal B}([\Gamma^m]^k)\times\bar {\cal F}_1
\end{eqnarray*}
forms the  $\sigma$-algebra as  it contains
the set $B_1\setminus B_2$ together with the sets $B_1$ and $B_2$ and also the set $\bigcup\limits_{i=1}^{\infty}B_i,$ if
 $B_i \in T, \  i=\overline{1,\infty}.$ However, the class $T$ contains the  algebra of the sets that is finite unions of the sets of the form
$\prod\limits_{i=1}^{k}A_i, \ A_i \in {\cal B}(S),$  that do not intersect. Therefore it coincides with
the $\sigma$-algebra ${\cal B}(S^k).$
The Lemma is proved, since
$\chi_{A}(\eta_i(p_1, z_1, \omega_i), \ldots, \eta_i(p_k, z_k, \omega_i))$
is the superposition of the measurable maps
$\zeta^i_{p_1, \ldots p_k}(z_1,\ldots, z_k,) $
and
$\chi_{A}(u_1, \ldots, u_k)$ that are   correspondingly the  measurable maps
of the   space $\{[\Gamma^m]^k\times\bar \Omega_1,  {\cal B}([\Gamma^m]^k)\times
\bar {\cal F}_1\}$
into the  space  $\{S^k, {\cal B}(S^k)\}$ and  the  space
 $\{S^k, {\cal B}(S^k)\}$
into  the  space $\{R^1, {\cal B}(R^1)\}.$
\qed \end{proof}
\begin{corollary}\label{cl1} The random function
\begin{eqnarray*}
 \prod_{i=1}^lf_i(\eta_i(p_1, z_1, \omega_i), \ldots,
\eta_i(p_k, z_k, \omega_i))
\end{eqnarray*}
is a   measurable map of the space
$\{[\Gamma^m]^k\times\bar \Omega_1,  {\cal B}([\Gamma^m]^k)\times
\bar {\cal F}_1\}$
into the space  $\{R^1, {\cal B}(R^1)\}$
for every fixed $\{p_1, \ldots, p_k\} \in [K_+^n]^k,$
if  $f_i(u_1, \ldots, u_k), \  i=\overline{1,l},$  is a   measurable map of
the space $\{S^k, {\cal B}(S^k)\}$
into the space $\{R^1, {\cal B}(R^1)\}.$
\end{corollary}
\begin{lemma}\label{pl4}
Let a function  ${\cal U}={\cal U}(z_1,\ldots, z_k,\omega_1, \ldots, \omega_l)$
be  a   measurable map of the  measurable space
$\{[\Gamma^m]^k\times\bar \Omega_1,  {\cal B}([\Gamma^m]^k)\times
\bar {\cal F}_1\}$
into the  measurable space  $\{R^1, {\cal B}(R^1)\},$  $\zeta(p, \omega_0) $
be  a   measurable map of the  measurable  space $\{\Omega_0, {\cal F}_0\}$
into  the  measurable  space $\{\Gamma^m, {\cal B}(\Gamma^m)\}$ for every fixed $ p \in K_+^n.$
Then
\begin{eqnarray*}
 {\cal U}(\zeta(p_1, \omega_0),\ldots, \zeta(p_k,
\omega_0),\omega_1, \ldots, \omega_l)
\end{eqnarray*}
is a measurable map of the measurable  space $\{\Omega, {\cal F}\}$ into the measurable space $\{R^1, {\cal B}(R^1)\}.$
If \begin{eqnarray*}
E|{\cal U}(\zeta(p_1, \omega_0),\ldots, \zeta(p_k, \omega_0),\omega_1,
\ldots, \omega_l)| < \infty,
\end{eqnarray*}
then
\begin{eqnarray}\label{uml1}
& E\{{\cal U}(\zeta(p_1, \omega_0),\ldots, \zeta(p_k,
\omega_0),\omega_1, \ldots, \omega_l)|{\cal F}^0\}
\end{eqnarray}
\begin{eqnarray*}
 =E{\cal U}(z_1,\ldots, z_k,\omega_1, \ldots, \omega_l)|
_{z_i=\zeta(p_i,\omega_0), \ i=\overline{1,k}},
\end{eqnarray*}
where ${\cal F}^0={\cal F}\{\zeta(p, \omega_0), p  \in K_+^n\}.$
\end{lemma}
\begin{proof}\smartqed
Prove that $ {\cal U}(\zeta(p_1, \omega_0),\ldots, \zeta(p_k,
\omega_0),\omega_1, \ldots, \omega_l)$ is a measurable map of the measurable  space $\{\Omega, {\cal F}\}$ into the measurable space $\{R^1, {\cal B}(R^1)\}.$ It follows from the fact that
 $ {\cal U}(\zeta(p_1, \omega_0),\ldots, \zeta(p_k,
\omega_0),\omega_1, \ldots, \omega_l)$ is a superposition  of  the measurable map
\begin{eqnarray*}                     \varphi_{p_1, \ldots, p_k}(\omega_0, \omega_1,\ldots, \omega_l)=\{\zeta(p_1, \omega_0), \ldots, \zeta(p_k, \omega_0),  \omega_1,\ldots, \omega_l\} \end{eqnarray*}
from the   space  $\{\Omega, {\cal F}\}$ into the  space
$\{[\Gamma^m]^k\times\bar \Omega_1,  {\cal B}([\Gamma^m]^k)\times
\bar {\cal F}_1\}$ and the map ${\cal U}(z_1,\ldots, z_k,\omega_1, \ldots, \omega_l)$ from the  space $\{[\Gamma^m]^k\times\bar \Omega_1,  {\cal B}([\Gamma^m]^k)\times
\bar {\cal F}_1\}$ into the space $\{R^1, {\cal B}(R^1)\}.$
The measurability of the map
\begin{eqnarray*}                     \varphi_{p_1, \ldots, p_k}(\omega_0, \omega_1,\ldots, \omega_l)=\{\zeta(p_1, \omega_0), \ldots, \zeta(p_k, \omega_0),  \omega_1,\ldots, \omega_l\} \end{eqnarray*}
from the  space  $\{\Omega, {\cal F}\}$ into the  space
$\{[\Gamma^m]^k\times\bar \Omega_1,  {\cal B}([\Gamma^m]^k)\times
\bar {\cal F}_1\}$ follows from the equality
\begin{eqnarray*}
 \{\omega \in \Omega, \ \varphi_{p_1, \ldots, p_k}(\omega_0, \omega_1,\ldots, \omega_l) \in D_1\times\ldots \times D_k \times A_1\times \ldots \times A_l \}
\end{eqnarray*}
\begin{eqnarray*}
 =\bigcap\limits_{i=1}^k\{\omega_0 \in \Omega_0, \  \zeta(p_i, \omega_0) \in D_i\}\times \prod\limits_{i=1}^l A_i  \in   {\cal F},
\end{eqnarray*}
\begin{eqnarray*}
 A_i \in {\cal F}_i, \quad i=\overline{1,l}, \quad  D_i \in  {\cal B}(\Gamma^m), \quad i=\overline{1,k},
\end{eqnarray*}
and the fact that rectangles of the kind $D_1\times\ldots \times D_k \times A_1\times \ldots \times A_l $ generate the $\sigma$-algebra  $ {\cal B}([\Gamma^m]^k)\times
\bar {\cal F}_1.$

Formula (\ref{uml1}) holds for  functions of the kind
\begin{eqnarray*}
{\cal U}(z_1,\ldots, z_k,
\omega_1, \ldots, \omega_l)=\chi_{A}(z_1, \ldots , z_k)\chi_{B}(\omega_1, \ldots , \omega_l),\end{eqnarray*}
where  $A \in [{\cal B}(\Gamma^m)]^k, \ B \in \bar {\cal F}^1.$
Let  $T$ be a class of the sets of the direct product of $\sigma$-algebras $[{\cal B}(\Gamma^m)]^k\times\bar {\cal F}^1$ for every set $G$  of that and functions of the kind
\begin{eqnarray*}                     {\cal U}(z_1,\ldots, z_k,
\omega_1, \ldots, \omega_l)=\chi_{G}(z_1, \ldots , z_k, \omega_1, \ldots , \omega_l),\end{eqnarray*}
where $\chi_{G}(z_1, \ldots , z_k, \omega_1, \ldots , \omega_l)$ is the  indicator function of the set $G,$ the formula  (\ref{uml1}) holds. It is evident that  $T$  is a monotonous class
that contains the algebra of the sets $V_0,$  formed by the finite unions  of rectangles
$A \times B,$ that do not intersect, where $A \in [{\cal B}(\Gamma^m)]^k, \ B \in \bar {\cal F}^1.$
From the last it follows that the minimal monotonous class coincides  with the  $\sigma$-algebra
$[{\cal B}(\Gamma^m)]^k\times\bar {\cal F}^1.$  So,  for any set
\begin{eqnarray*}
G \in [{\cal B}(\Gamma^m)]^k\times\bar {\cal F}^1,
\end{eqnarray*}
\begin{eqnarray*}
 E\left\{\chi_{G}(\zeta(p_1, \omega_0),\ldots, \zeta(p_k,
\omega_0),\omega_1, \ldots, \omega_l)|{\cal F}^0\right\}
\end{eqnarray*}
\begin{eqnarray}\label{uml2}
  =E\chi_{G}(z_1,\ldots, z_k,\omega_1, \ldots, \omega_l)|
_{z_i=\zeta(p_i,\omega_0), \ i=\overline{1,k}} \ .
\end{eqnarray}
Let  ${\cal U}(z_1,\ldots, z_k,\omega_1, \ldots, \omega_l)$ be a nonnegative function that satisfies the conditions of the Lemma. Then the sequence of functions
\begin{eqnarray*}                     {\cal U}_n(z_1,\ldots, z_k,\omega_1, \ldots, \omega_l)=\sum\limits_{s=0}^{\infty}
\frac{s}{2^n}\chi_{G_{n,s}}(z_1,\ldots, z_k,\omega_1, \ldots, \omega_l),\end{eqnarray*}
where
\begin{eqnarray*}                     G_{n,s}=\left\{\{z_1,\ldots, z_k,\omega_1, \ldots, \omega_l\}, \  \frac{s}{2^n} \leq {\cal U}(z_1,\ldots, z_k,\omega_1, \ldots, \omega_l) < \frac{s+1}{2^n} \right\},\end{eqnarray*}
is monotonically increasing and converges to
${\cal U}(z_1,\ldots, z_k,\omega_1, \ldots, \omega_l)$
every\-where. For ${\cal U}_n(z_1,\ldots, z_k,\omega_1, \ldots, \omega_l)$
the formula  (\ref{uml1}) holds on the basis of the proved one and the possibility to come to the limit  under the sign of the conditional expectation   due to the  monotone convergence of the integrable sequence.
Owing to the monotone convergence  of the sequence  ${\cal U}_n(z_1,\ldots, z_k,\omega_1, \ldots, \omega_l)$  to the function ${\cal U}(z_1,\ldots, z_k,\omega_1, \ldots, \omega_l)$
everywhere, the formula  (\ref{uml1}) holds for  the function ${\cal U}(z_1,\ldots, z_k,\omega_1, \ldots, \omega_l)$   too.
If now ${\cal U}(z_1,\ldots, z_k,\omega_1, \ldots, \omega_l)$  has  any sign, then it can be presented as the algebraic  sum of the considered ones.
 \qed \end{proof}
\begin{theorem}\label{tl1}
Let
$ \eta_i(p, z, \omega_i),\ i=\overline{1,l},$ and
$\zeta(p, \omega_0)$  satisfy the conditions of the Lemma \ref{pl1}.
Then the random fields
\begin{eqnarray}\label{jl2}
\xi_i(p, \omega)= \eta_i(p, \zeta(p, \omega_0), \omega_i),
\quad  i=\overline{1,l},
\end{eqnarray}
on the probability space $\{\Omega, {\cal F}, \bar P\}$
are conditionally independent relative to the random field   $\zeta(p, \omega_0).$ If, moreover, the random function    $ \eta_i(p, z, \omega_i)$
 takes values in the budget set $  X^{i}_{(p,z)}$ for every fixed $(p, z) \in K_+^n\times\Gamma^m,$  $ i=\overline{1,l},$  then
the formula
\begin{eqnarray*}
\bar P(\{\xi_1(p_1,\omega), \ldots, \xi_1(p_k,\omega)\} \in A_1, \ldots,
\{\xi_l(p_1,\omega), \ldots, \xi_l(p_k,\omega)\} \in A_l,
\end{eqnarray*}
\begin{eqnarray*}
 \{\zeta(p_1,\omega_0),
\ldots, \zeta(p_k,\omega_0) \} \in {\cal D})
\end{eqnarray*}
\begin{eqnarray}\label{sl2}
 =\int\limits_{{\cal D}}\prod\limits_{i=1}^{l}F_{p_1, \ldots, p_k}^i
\left(A_i\cap X_{(p,z)_{k}}^{k,i}|z_1, \ldots, z_k\right)
d\psi_{p_1, \ldots, p_k}(z_1, \ldots, z_k),
\end{eqnarray}
is valid,
where  $A_i \in {\cal B}(S^k), \ i=\overline{1,l},$
\begin{eqnarray*}
F_{p_1, \ldots, p_k}^i(B^i|z_1, \ldots, z_k)
\end{eqnarray*}
\begin{eqnarray*}
 =E\chi_{B^i} (\eta_i(p_1,z_1, \omega_i), \ldots
\eta_i(p_k,z_k, \omega_i) ), \quad
B^i \in {\cal B}(X_{(p,z)_{k}}^{k,i}),\quad i=\overline{1,l},
\end{eqnarray*}
\begin{eqnarray*}
 \psi_{p_1, \ldots, p_k}({\cal D})=\bar P(\{\zeta(p_1, \omega_0), \ldots,
 \zeta(p_k, \omega_0)\} \in {\cal D}),  \quad {\cal D} \in {\cal B}([\Gamma^m]^k).
 \end{eqnarray*}
\end{theorem}
\begin{proof}\smartqed  Prove that $\sigma$-algebras
${\cal F}^i= {\cal F}\{\xi_i(p, \omega),\ p \in K_+^n\}, \ i=\overline{1,l},$
are conditionally independent relative to   $\sigma$-algebra
 ${\cal F}^0= {\cal F}\{\zeta(p, \omega_0),\ p \in K_+^n\}.$
Consider the cylindrical functional
$f_i^k(\xi_i(p_1, \omega), \ldots, \xi_i(p_k, \omega)), \ i=\overline{1,l},$ where
$f_i^k(u_1, \ldots, u_k)), \ i=\overline{1,l},$ is a measurable map of the measurable space $\{S^k, {\cal B}(S^k)\}$
into the measurable space $\{R^1, {\cal B}(R^1)\}.$
Then the function
\begin{eqnarray*}
 f_i^k(\xi_i(p_1, \omega), \ldots, \xi_i(p_k, \omega))
\end{eqnarray*}
\begin{eqnarray*}
=f_i^k(\eta_i(p_1, \zeta(p_1,\omega_0), \omega_i), \ldots,
\eta_i(p_k, \zeta(p_k,\omega_0), \omega_i)),
\quad i=\overline{1,l},
\end{eqnarray*}
is a measurable map of
$\{ \Omega,   {\cal F}\}$
into $\{R^1, {\cal B}(R^1)\}.$
Since
\begin{eqnarray*}
 f_i^k(\eta_i(p_1, z_1, \omega_i), \ldots, \eta_i(p_k, z_k, \omega_i)),
 \quad i=\overline{1,l},
 \end{eqnarray*}
is a measurable map of
$\{ [\Gamma^m]^k\times \bar \Omega_1,
{\cal B}([\Gamma^m]^k)\times \bar {\cal F}_1 \}$
into $\{R^1, {\cal B}(R^1)\},$
then on the basis of the Lemma \ref{pl1}, the  consequence \ref{cl1}, and the Lemma \ref{pl4}
\begin{eqnarray}\label{sl3}
  E\left\{\prod\limits_{i=1}^l f_i^k(\xi_i(p_1, \omega), \ldots, \xi_i(p_k, \omega))|{\cal F}^0\right\}
\end{eqnarray}
\begin{eqnarray*}
 =\prod\limits_{i=1}^l E\{f_i^k(\xi_i(p_1, \omega), \ldots, \xi_i(p_k, \omega))|{\cal F}^0\}.
\end{eqnarray*}
Among the  measurable maps that satisfy equalities (\ref{sl3}) there is a function \\ $\chi_{A_i}(\xi_i(p_1, \omega), \ldots, \xi_i(p_k, \omega)),$
where $A_i \in {\cal B}(S^k),  \ i=\overline{1, l}.$
The set
\begin{eqnarray*}
 C_{p_1, \ldots, p_k}(A_i)=\{\omega \in \Omega, \ \chi_{A_i}(\xi_i(p_1, \omega), \ldots, \xi_i(p_k, \omega)) > 0\}
\end{eqnarray*}
\begin{eqnarray*}
 =\{\omega \in \Omega, \{\xi_i(p_1, \omega), \ldots, \xi_i(p_k, \omega)\} \in A_i\}
\end{eqnarray*}
is a cylindrical set.
Let us show that for the collection of the sets $B_1, \ldots , B_l,$
where $B_i \in {\cal F}^i,$ the equality
\begin{eqnarray}\label{csl3}
E\left\{\prod\limits_{i=1}^l\chi_{B_i}(\omega)|{\cal F}^0\right\}
 =\prod\limits_{i=1}^l E\{\chi_{B_i}(\omega)|{\cal F}^0\}
\end{eqnarray}
holds.
The collection of the  cylindrical sets $C_{p_1, \ldots, p_k}(A_i)$ for every $i=\overline{1,l},$
as $\{p_1, \ldots , p_k\} \in [K_+^n]^k,\  k=\overline{1,\infty},$ forms the algebra of  sets $V_i$ and
the minimal  $\sigma$-algebra generated by the algebra $V_i$ is the $\sigma$-algebra ${\cal F}^i.$
The equality (\ref{csl3}) is valid for cylindrical sets   $C_{p_1, \ldots, p_k}(A_i),$ that is,
\begin{eqnarray*}
 E\left\{\prod\limits_{i=1}^l \chi_{A_i}(\xi_i(p_1, \omega), \ldots, \xi_i(p_k, \omega)) |{\cal F}^0\right\}
\end{eqnarray*}
\begin{eqnarray}\label{ccsl3}
& =\prod\limits_{i=1}^l E\{\chi_{A_i}(\xi_i(p_1, \omega), \ldots, \xi_i(p_k, \omega))|{\cal F}^0\}.
\end{eqnarray}
For the proof we use the induction. Let $T_1$ be a class of the sets that belongs to  ${\cal F}^1$  and for that the equality
\begin{eqnarray*}
 E\left\{\chi_{B_1}(\omega)\prod\limits_{i=2}^l \chi_{A_i}(\xi_i(p_1, \omega), \ldots, \xi_i(p_k, \omega)) |{\cal F}^0\right\}
 \end{eqnarray*}
\begin{eqnarray}\label{sl31}
 = E\left\{\chi_{B_1}(\omega)|{\cal F}^0\right\}
  \prod\limits_{i=2}^l E\left\{\chi_{A_i}(\xi_i(p_1, \omega), \ldots, \xi_i(p_k, \omega))|{\cal F}^0\right\},\quad  B_1 \in T_1,
\end{eqnarray}
is valid.
It is evident that $T_1$ is a monotone class since if the sequence of the sets $B_1^n $
is monotonically increasing, that is, $B_1^n \subseteq B_1^{n+1},$ then   we can go to the limit under the sign of the conditional expectation in the equality (\ref{sl31}) if   we substitute the set $B_1^n$ instead of the set $B_1.$
The analogous statement is valid for monotonically decreasing  sequences. So, $T_1$ is a monotone class that contains the algebra  $V_1,$ but the minimal monotone class that contains  $V_1$ coincides with the $\sigma$-algebra ${\cal F}^1.$ Thus, the equality (\ref{sl31}) is proved for any set $B_1 \in {\cal F}^1.$ Having made the analogous proof with respect to the rest indices,
we obtain  the equality
 (\ref{csl3}) we need.  Let now  $ X_1, \ldots , X_l $ be arbitrary nonnegative random values  measurable  correspondingly with respect to the $\sigma$-algebras ${\cal F}^1, \ldots , , {\cal F}^l$ and such that  $EX_i < \infty.$ Then the sequence of random values
\begin{eqnarray*}                     X_i^n=\sum\limits_{s=0}^\infty\frac{s}{2^n}\chi_{A_{s,n}^i}(\omega)
\end{eqnarray*}
monotonously and pointwise  converges  to the random value $X_i,$ where
\begin{eqnarray*}
 A_{s,n}^i=\left\{\omega \in \Omega,\quad \frac{s}{2^n} \leq X_i < \frac{s+1}{2^n}\right\}
 \end{eqnarray*}
  are the sets that belong to  ${\cal F}^i.$
Since   the relation  (\ref{csl3}) is multilinear and there is the possibility to come to the limit  under the sign of the  conditional expectation owing to the monotone convergence,  it is easy to prove  that the equality
\begin{eqnarray}\label{sl32}
 E\left \{\prod\limits_{i=1}^l X_i^n|{\cal F}^0\right\}=\prod\limits_{i=1}^l E\{X_i^n|{\cal F}^0\}
\end{eqnarray}
takes place.
Using the monotone convergence of the sequences  $X_i^n$  and coming to the limit in the equality (\ref{sl32}), we obtain the needed proof.

Let us prove the second part of the Theorem.
Let $A_i \in {\cal B}(S^k),\ i=\overline{1,l},$ and the set  $ D \in [{\cal B}(\Gamma^m)]^k.$
Then
\begin{eqnarray*}
 \bar P(\{\xi_1(p_1,\omega), \ldots, \xi_1(p_k,\omega)\} \in A_1, \ldots,
\{\xi_l(p_1,\omega), \ldots, \xi_l(p_k,\omega)\} \in A_l,
 \end{eqnarray*}
\begin{eqnarray*}
 \{\zeta(p_1,\omega_0), \ldots, \zeta(p_k,\omega_0) \} \in {\cal D})
 \end{eqnarray*}
\begin{eqnarray*}
 =E\prod\limits_{i=1}^l\chi_{A_i}(\xi_i(p_1, \omega), \ldots, \xi_i(p_k, \omega))
\chi_{D}(\zeta(p_1, \omega_0), \ldots, \zeta(p_k, \omega_0)) \\
 =EE\left\{\prod\limits_{i=1}^l\chi_{A_i}(\xi_i(p_1, \omega), \ldots, \xi_i(p_k, \omega))
\chi_{D}(\zeta(p_1, \omega_0), \ldots, \zeta(p_k, \omega_0))|{\cal F}^0\right\}
 \end{eqnarray*}
\begin{eqnarray*}
 =E\prod\limits_{i=1}^lE\{\chi_{A_i}(\xi_i(p_1, \omega), \ldots, \xi_i(p_k, \omega))
|{\cal F}^0\}
\chi_{D}(\zeta(p_1, \omega_0), \ldots, \zeta(p_k, \omega_0)).
 \end{eqnarray*}
 \begin{eqnarray*}
 E\{\chi_{A_i}(\xi_i(p_1, \omega), \ldots, \xi_i(p_k, \omega))
|{\cal F}^0\}
 \end{eqnarray*}
\begin{eqnarray*}
 =E\chi_{A_i}(\eta_i(p_1, z_1, \omega_i), \ldots, \eta_i(p_k, z_k, \omega_i))
|_{z_i=\zeta(p_i, \omega_0), \ i=\overline{1,k}} \ .
\end{eqnarray*}
Introduce into consideration the set of functions of the sets
\begin{eqnarray*}
 F_{p_1, \ldots, p_k}^i\left(A_i|z_1, \ldots, z_k\right)=
E\chi_{A_i}(\eta_i(p_1,z_1, \omega_i), \ldots
\eta_i(p_k,z_k, \omega_i) ), \quad i=\overline{1,l}.
\end{eqnarray*}
If we  take into account that
$\eta_i(p_s, z_s, \omega_i)$ takes  values in the set
$X_{(p_s, z_s)}^i, $ then $\zeta^i_{p_1,\ldots, p_k}(z_1, \ldots, z_k)=
\{\eta_i(p_1, z_1, \omega_i), \ldots,  \eta_i(p_k, z_k, \omega_i)\}$
takes  values in the set $X_{(p, z)_k}^{k,i}. $
Therefore,
\begin{eqnarray*}
 \chi_{A_i}(\eta_i(p_1,z_1, \omega_i), \ldots
\eta_i(p_k,z_k, \omega_i) )
 \end{eqnarray*}
\begin{eqnarray*}
 =\chi_{A_i\cap X_{(p, z)_k}^{k,i}}(\eta_i(p_1,z_1, \omega_i), \ldots
\eta_i(p_k,z_k, \omega_i) ),
\end{eqnarray*}
thus,
\begin{eqnarray*}
F_{p_1, \ldots, p_k}^i\left(A_i|z_1, \ldots, z_k\right)=
F_{p_1, \ldots, p_k}^i
\left(A_i\cap X_{(p,z)_{k}}^{k,i}|z_1, \ldots, z_k\right).
\end{eqnarray*}
From here it follows that  we can consider   $F_{p_1, \ldots, p_k}^i(B^i|z_1, \ldots, z_k) $ only
on the sets $B^i$ that belong to the $\sigma$-algebra  ${\cal B}(X_{(p,z)_{k}}^{k,i}).$
Then
\begin{eqnarray*}
 E\{\chi_{A_i}(\xi_i(p_1, \omega), \ldots, \xi_i(p_k, \omega))
|{\cal F}^0\}=
F_{p_1, \ldots, p_k}^i(A_i|\zeta(p_1,\omega_0), \ldots, \zeta(p_k, \omega_0)).
\end{eqnarray*}
So, for $A_i \in {\cal B}(S^k), \ i=\overline{1,l},  \  D \in  [{\cal B}(\Gamma^m)]^k,$
\begin{eqnarray*}
 \bar P(\{\xi_1(p_1,\omega), \ldots, \xi_1(p_k,\omega)\} \in A_1, \ldots,
\{\xi_l(p_1,\omega), \ldots, \xi_l(p_k,\omega)\} \in A_k,
 \end{eqnarray*}
\begin{eqnarray*}
 \{\zeta(p_1,\omega_0),
\ldots, \zeta(p_k,\omega_0) \} \in {\cal D})
 \end{eqnarray*}
\begin{eqnarray*}
 =\int\limits_{{\cal D}}\prod\limits_{i=1}^{l}F_{p_1, \ldots, p_k}^i
\left(A_i\cap X_{(p,z)_{k}}^{k,i} |z_1, \ldots, z_k\right)d\psi_{p_1, \ldots, p_k}(z_1, \ldots, z_k).
\end{eqnarray*}
\qed \end{proof}

Let us consider  examples of the determination of  measurable maps.
Denote by
 \begin{eqnarray*}
 N=\{ \{p_i, z_i\} \in K_+^n\times\Gamma^m,\
i=\overline{1, \infty}\}\end{eqnarray*}
a countable and  everywhere  dense set of points in the metric space
$K_+^n\times\Gamma^m.$
For given  $\varepsilon > 0 $ by $B_n^{\varepsilon}(x_n)$
we denote the ball
\begin{eqnarray*}
 B_n^{\varepsilon}(x_n)=\{y=(p,z) \in K_+^n\times\Gamma^m, \ \rho(y, x_n)
< \varepsilon\},\quad x_n=(p_n,z_n) \in N,
\end{eqnarray*}
where $\rho(x, y)$ is a metric in the space $K_+^n\times\Gamma^m.$ It is evident that the equality $\bigcup\limits_{n=1}^{\infty} B_n^{\varepsilon}(x_n)
=K_+^n\times\Gamma^m$ holds.
Introduce into consideration the collection of the Borel sets $C_n^{\varepsilon}=
B_n^{\varepsilon}(x_n)\setminus \bigcup\limits_{s < n}B_s^{\varepsilon}(x_s).$
This collection of  sets is such that
$\bigcup\limits_{n=1}^{\infty} C_n^{\varepsilon}=K_+^n\times\Gamma^m$ and
$C_n^{\varepsilon}\cap C_j^{\varepsilon}=\emptyset, \ n \neq j, \
C_n^{\varepsilon} \in {\cal B}(K_+^n)\times {\cal B}(\Gamma^m).$ Having  the sequence of  random values $\xi_n, \ n=\overline{1, \infty},$
on a probability space  $\{\Omega, {\cal F}, \bar P\},$ where
$\xi_n$ is a measurable map of the space $\{\Omega, {\cal F}\}$  into the space
$\{S, {\cal B}(S)\}, \  n=\overline{1, \infty}, $
let us determine a random function on  $K_+^n\times\Gamma^m$ with values in  $S$ by the rule
\begin{eqnarray}\label{sl4}
 \xi(p,z)=\xi_n, \quad  (p,z) \in  C_n^{\varepsilon},\quad
n=\overline{1, \infty}.
\end{eqnarray}

\begin{lemma}\label{ppl5} The random field $\xi(p,z),$ $(p,z) \in K_+^n\times\Gamma^m,$ on a probability space
$\{\Omega, {\cal F}, \bar P\}$ given
by the formula (\ref{sl4})  and taking  values in the set $S$ is a measurable map of the space $\{K_+^n\times \Gamma^m\times \Omega,
{\cal B}(K_+^n) \times {\cal B}(\Gamma^m)\times {\cal F}\}$ into the space
$\{S, {\cal B}(S)\}.$
\end{lemma}
\begin{proof}\smartqed
 The proof follows  from the equality
\begin{eqnarray*}                     \{ \{p, z, \omega\}, \xi(p,z) \in A\}= \bigcup\limits_{n=1}^{\infty}
C_n^{\varepsilon} \times K_n,\end{eqnarray*}
where $K_n=\{\omega, \ \xi_n \in A\}, \ A \in
{\cal B}(S).$ But $C_n^{\varepsilon} \times K_n \in
{\cal B}(K_+^n) \times {\cal B}(\Gamma^m)\times {\cal F},$  therefore
$\bigcup\limits_{n=1}^{\infty} C_n^{\varepsilon} \times K_n \in
{\cal B}(K_+^n) \times {\cal B}(\Gamma^m)\times {\cal F}.$
\qed \end{proof}
\begin{theorem}\label{tl2}
Let a random field $\eta_i(p,z, \omega_i), (p,z) \in  K_+^n\times \Gamma^m,$
on a probability space $\{\Omega_i, {\cal F}_i, \bar P_i\}$ be such that for every $\omega_i \in \Omega_i,$  $\eta_i(p,z, \omega_i)$ is a continuous function of $(p,z) \in  K_+^n\times \Gamma^m$ with  values in  the set $S, \ i=\overline{1,l}, $ and let a random field
$\zeta(p, \omega_0), \  p \in K_+^n, $  on a probability space $\{\Omega_0, {\cal F}_0, \bar P_0\}$
take  values in the set  $\Gamma^m$ and  every its realization be a continuous function of
$  p \in K_+^n.$

Under these conditions, the random field $\eta_i(p,z, \omega_i)$ is a measurable
map of the measurable space
$\{K_+^n \times \Gamma^m\times \Omega_i,
{\cal B}(K_+^n) \times {\cal B}(\Gamma^m)\times {\cal F}_i\}$
into the measurable space $\{S, {\cal B}(S)\},  \ i=\overline{1,l}, $ and $\zeta(p, \omega_0)$
is a measurable map of the measurable space   $\{ K_+^n \times \Omega_0,
{\cal B}(\Gamma^m)\times {\cal F}_0\}$ into the measurable space $\{\Gamma^m, {\cal B}(\Gamma^m)\}.$
The random fields $\xi_i(p, \omega)=\eta_i(p,\zeta(p, \omega_0), \omega_i),  \ i=\overline{1,l},$
are conditionally independent relative to the random field $\zeta(p, \omega_0)$ on a probability space $\{\Omega, {\cal F}, \bar P\},$ where
 $\Omega=\prod\limits_{i=0}^{l}\Omega_i,$ ${\cal F}=\prod\limits_{i=0}^{l}
{\cal F}_i,$ $\bar P=\prod\limits_{i=0}^{l}
\bar P_i.$
\end{theorem}
\begin{proof}\smartqed
  Consider the sequence of  measurable maps
\begin{eqnarray*}                     \eta_i^{\varepsilon^k}(p,z, \omega_i)=\eta_i(
p_s^{\varepsilon^k}, z_s^{\varepsilon^k} , \omega_i),
\quad (p,z) \in  C_s^{\varepsilon^k}, \quad k=\overline{1, \infty}, \end{eqnarray*}
where $(p_s^{\varepsilon^k}, z_s^{\varepsilon^k}) \in N \cap C_s^{\varepsilon^k},
\ s=\overline{1,\infty},$ and
the sequence  $\varepsilon^k \to 0,$ as $ k \to \infty.$
On the basis of the previous Lemma $\eta_i^{\varepsilon^k}(p,z, \omega_i)$
is a measurable map of the space
$\{K_+^n \times \Gamma^m\times \Omega_i,
{\cal B}(K_+^n) \times {\cal B}(\Gamma^m)\times {\cal F}_i\}$
into the space   $\{S,  {\cal B}(S)\}.$
The sequence  $\eta_i^{\varepsilon^k}(p,z, \omega_i)$ converges pointwise  to  $\eta_i(p, z, \omega_i)$  as $ k \to \infty$  that means the stated measurability.
Really,
\begin{eqnarray*}
 \{\{p,z, \omega_i\}, \ \eta_i(p, z, \omega_i)  \in L(b)\cap S \}
 \end{eqnarray*}
\begin{eqnarray*}  =\bigcup\limits_{k=1}^{\infty}\bigcup\limits_{s=1}^{\infty}\bigcup\limits_{j=s}^{\infty}
\{\{p,z, \omega_i\}, \ \eta_i^{\varepsilon^j}(p,z, \omega_i) \in L(b -   e/k) \cap S \} \in
   {\cal B}(K_+^n) \times {\cal B}(\Gamma^m)\times {\cal F}_i  \nonumber
\end{eqnarray*}
for every  set $L(b)=\{x=\{x_1, \ldots, x_n\} \in R^n, \ x_i < b_i, \ i=\overline{1,n}\}, \ $ where $b=\{b_i\}_{i=1}^n \in R^n,$  $b - e/k =\{b_i - 1/k \}_{i=1}^n, \ e=\{1, \ldots, 1 \}.$
It is evident that   the class of sets $T$  for which
\begin{eqnarray*}                     \{\{p,z, \omega_i\}, \ \eta_i(p, z, \omega_i)  \in Q \} \in  {\cal B}(K_+^n) \times {\cal B}(\Gamma^m)\times {\cal F}_i, \quad Q \in T,  \end{eqnarray*}
forms the
$\sigma$-algebra.
From the latter   and that the sets $S \cap L(b), \ b \in R^n,$ generates the $\sigma$-algebra ${\cal B}(S),$ we have that
 $\eta_i(p, z, \omega_i)$ is a measurable map of the space
$\{K_+^n \times \Gamma^m\times \Omega_i,
{\cal B}(K_+^n) \times {\cal B}(\Gamma^m)\times {\cal F}_i\}$
into the space   $\{S,  {\cal B}(S)\}$ \cite{4}.
It can be proved  analogously that the random field
$\zeta(p, \omega_0)$ is a measurable map of the space   $\{ K_+^n \times \Omega_0,
{\cal B}(\Gamma^m)\times {\cal F}_0\}$ into the space
$\{\Gamma^m, {\cal B}(\Gamma^m)\}.$
The rest statements  of the Theorem are consequences of the previous consideration.
\qed \end{proof}
\begin{theorem}\label{pl5}
Let a random field $\eta_i(p,z, \omega_i), \ (p,z) \in  K_+^n\times \Gamma^m,  \ i=\overline{1,l},$ on a probability space $\{\Omega_i, {\cal F}_i, \bar P_i\}$ with values in the set  $S$  be  continuous with probability one and let a random field
$\zeta(p, \omega_0), \  p \in K_+^n,$  on a probability space
$\{\Omega_0, {\cal F}_0, \bar P_0\}$ take values in the set $\Gamma^m$ and be  continuous with probability one. Then there exists  a modification $\tilde \eta_i(p,z, \omega_i)$ of the random field  $\eta_i(p,z, \omega_i), \ i=\overline{1,l}, $ and a  modification  $ \tilde \zeta(p, \omega_0)$ of the random field $ \zeta(p, \omega_0)$ that are correspondingly  measurable  maps of the measurable space
$\{K_+^n \times \Gamma^m\times \Omega_i,
{\cal B}(K_+^n) \times {\cal B}(\Gamma^m)\times {\cal F}_i\}$
into the measurable  space   $\{S, {\cal B}(S)\},  \ i=\overline{1,l}, $  and the measurable  space  $\{ K_+^n \times \Omega_0,
{\cal B}(\Gamma^m)\times {\cal F}_0\}$ into the   measurable space $\{\Gamma^m, {\cal B}(\Gamma^m)\}.$
\end{theorem}
\begin{proof}\smartqed            If   $\eta_i(p,z, \omega_i)$ is a continuous random field  with probability one on the probability space
$\{\Omega_i, {\cal F}_i, \bar P_i\},$  then  we put
\begin{eqnarray*}
\tilde \eta_i(p,z, \omega_i) =
\left\{\begin{array}{ll}
 \eta_i(p,z, \omega_i), &  \mbox{if ~ $ \omega_i \in \Omega_i^{'}$ } \\
                                   0,  & \mbox{if ~ $\omega_i \in \Omega_i\setminus \Omega_i^{'}$}.
                                      \end{array}
                                       \right.
                                       \end{eqnarray*}
By $\Omega_i^{'}$ we denoted the set of elementary events of probability one   for which the realizations  of the random field
$\eta_i(p,z, \omega_i)$  are continuous functions of variables $(p,z) \in
K_+^n \times \Gamma^m.$
Then   $ \tilde \eta_i(p,z, \omega_i)$  is a random field on
$\{\Omega_i, {\cal F}_i, \bar P_i\}.$ Really,
\begin{eqnarray*}                     \{\omega_i, \tilde \eta_i(p,z, \omega_i) \in B\}=
\{\omega_i, \eta_i(p,z, \omega_i) \in B\}\cap \Omega_i^{'} \cup U \in {\cal F},\end{eqnarray*}
where
\begin{eqnarray*}
 U=\left\{\begin{array}{ll}
\Omega_i\setminus \Omega_i^{'}, &\mbox{if ~$ 0 \in B, $ } \\
                                  \emptyset,  &\mbox{if ~$ 0 \in S \setminus B. $ }
                                      \end{array}
                                       \right.  \nonumber
                                       \end{eqnarray*}
So, $\tilde \eta_i(p,z, \omega_i)$ is a measurable map of the measurable space $\{\Omega_i, {\cal F}_i\}$ into the  measurable space $\{S, {\cal B}(S)\}$
for every fixed $(p,z) \in K_+^n\times \Gamma^m.$ Every realization of the random field $\tilde \eta_i(p,z, \omega_i)$ is  continuous. A modification $\tilde \zeta(p, \omega_0)$  of the random field $ \zeta(p, \omega_0)$  is constructed analogously.
\qed \end{proof}
As a consequence of the Theorem \ref{pl5}, we assume, without loss of generality, that every realization of the considered random fields is continuous.

\begin{lemma}\label{d2l2}
Let \hfill $f(z, \omega_i)$ \hfill be \hfill a \hfill measurable \hfill mapping \hfill  of \hfill the \hfill measurable \hfill space \\
$\{\Gamma^m\times \Omega_i,{\cal B}(\Gamma^m)\times {\cal F}_i\}$ into the  measurable space $\{S, {\cal B}(S)\}$  and  $\varphi(z, \omega_i)$ be a measurable map of the measurable space
$\{\Gamma^m\times \Omega_i,{\cal B}(\Gamma^m)\times {\cal F}_i\}$
into the measurable space $\{R^1, {\cal B}(R^1)\}.$  Then  $\varphi(z, \omega_i)
f(z, \omega_i) $ is a measurable map of the measurable space
$\{\Gamma^m\times \Omega_i,{\cal B}(\Gamma^m)\times {\cal F}_i\}$
into the measurable space $\{S, {\cal B}(S)\}.$
\end{lemma}
\begin{proof}\smartqed            If  $f(z, \omega_i)$ is a measurable map of the space
$\{\Gamma^m\times \Omega_i,{\cal B}(\Gamma^m)\times {\cal F}_i\}$  into the space $\{S, {\cal B}(S)\},$  then  $c f(z, \omega_i)$  is also a measurable map for all positive number $c >0,$  since for an  open  set $A \in {\cal B}(S)$
\begin{eqnarray*}
 \{(z, \omega_i) \in   \Gamma^m\times \Omega_i, \  c f(z, \omega_i)  \in A\}
\end{eqnarray*}
\begin{eqnarray*}
 =\left\{(z, \omega_i) \in   \Gamma^m\times \Omega_i, \   f(z, \omega_i)
\in \frac{1}{c} A\right\} \in {\cal B}(\Gamma^m)\times {\cal F}_i,
\end{eqnarray*}
because   $\frac{1}{c} A =\left\{ \frac{1}{c} x \in S,\ x \in A\right\}$ is an open set. The open sets generate   ${\cal B}(S),$ therefore the needed statement is proved.

Let $\varphi(z, \omega_i)$ be a map of the form
\begin{eqnarray*}                     \varphi_N(z, \omega_i)=\sum\limits_{s=1}^Nc_s^N\chi_{D_s}(p,\omega_i), \quad c_s^N >0, \quad s=\overline{1,N},\end{eqnarray*}
\begin{eqnarray*}
D_s \in {\cal B}(\Gamma^m)\times {\cal F}_i, \quad  c_s^N \in R_+^1, \quad
\bigcup\limits_{s=1}^ND_s= \Gamma^m\times \Omega_i.\end{eqnarray*}
Then
\begin{eqnarray*}
\varphi_N(z, \omega_i) f(z, \omega_i) = c_s^Nf(z,\omega_i), \quad  (z, \omega_i) \in D_s, \end{eqnarray*}
and
\begin{eqnarray*}
 \{(z, \omega_i) \in \Gamma^m\times \Omega_i, \
\varphi_N(z, \omega_i) f(z,\omega_i) \in A\}
\end{eqnarray*}
\begin{eqnarray*}
 =\bigcup\limits_{s=1}^N
\{(z, \omega_i) \in \Gamma^m\times \Omega_i, \
c_s^N f(z,\omega_i) \in A\}\cap D_s \in  {\cal B}(\Gamma^m)\times {\cal F}_i.
\end{eqnarray*}
The Lemma is proved since  $\varphi(z, \omega_i) f(z,\omega_i)$ is a pointwise limit of the sequence of the kind $ \varphi_N(z, \omega_i) f(z, \omega_i).$
\qed \end{proof}
\begin{lemma}\label{d2l3}
Let \hfill $f(z,\omega_i)$  \hfill  be  \hfill  a  \hfill  measurable  \hfill  map  \hfill  of  \hfill  the  \hfill  measurable   \hfill  space\\
$\{\Gamma^m\times \Omega_i,{\cal B}(\Gamma^m)\times {\cal F}_i\}$ into the measurable space $\{S, {\cal B}(S)\}$  and $Q(z)$  be a measurable map of the  measurable space
$\{\Gamma^m,{\cal B}(\Gamma^m)\}$ into itself, then $f(Q(z),\omega_i)$ is  a measurable map of the space $\{\Gamma^m\times \Omega_i,{\cal B}(\Gamma^m)\times {\cal F}_i\}$ into the  space $\{S, {\cal B}(S)\}.$
\end{lemma}
\begin{proof}\smartqed             It follows from that the map $f(Q(z),\omega_i)$ is a superposition of the measurable map $\varphi_0(z, \omega_i)=\{Q(z), \omega_i\}$
of the measurable space $\{\Gamma^m\times \Omega_i,{\cal B}(\Gamma^m)\times {\cal F}_i\}$
into \hfill itself \hfill and \hfill of \hfill  the \hfill measurable \hfill mapping \hfill $f(z,\omega_i)$ \hfill of \hfill  the \hfill measurable \hfill space\\
$\{\Gamma^m\times \Omega_i,{\cal B}(\Gamma^m)\times {\cal F}_i\}$ into the measurable space $\{S, {\cal B}(S)\}.$
\qed \end{proof}
\begin{lemma}\label{d2l4}
Let $\zeta_0(p, \omega_0)$ be a measurable map of the measurable space $\{\Omega_0,
{\cal F}_0\}$ into the measurable space  $\{\Gamma^m,{\cal B}(\Gamma^m)\}$  for every $p \in K_+^n,$ $Q(p,z)$  be a productive economic process.
Then  $\zeta(p, \omega_0)=Q(p, \zeta_0(p,\omega_0))$ is a measurable map of the measurable space   $\{\Omega_0,{\cal F}_0\}$ into the measurable space
$\{\Gamma^m,{\cal B}(\Gamma^m)\}$ for every $p \in K_+^n.$
\end{lemma}
\begin{proof}\smartqed   Proof follows from that $\zeta(p, \omega_0)$ is a superposition of the measurable map
  $\zeta_0(p, \omega_0)$ of the space $\{\Omega_0,
{\cal F}_0\}$ into the space  $\{\Gamma^m,{\cal B}(\Gamma^m)\}$ and of
the measurable map  $Q(p,z)$ of the space $\{\Gamma^m,{\cal B}(\Gamma^m)\}$
into itself for every $p \in K_+^n.$  \qed \end{proof}
\begin{definition}\label{dl2}
Let  $Q(p,z)$ be a  productive economic process given on
$K_+^n \times \Gamma^m,$  and $\zeta_0(p, \omega_0)$ be  a measurable map of the measurable space $\{\Omega_0,
{\cal F}_0\}$ into the measurable space   $\{\Gamma^m,{\cal B}(\Gamma^m)\}$ for every $p \in K_+^n.$
The  measurable map   $\zeta(p, \omega_0)= Q(p, \zeta_0(p, \omega_0))$
of the space $\{\Omega_0,
{\cal F}_0\}$ into the space  $\{\Gamma^m,{\cal B}(\Gamma^m)\}$  for every
$ p \in K_+^n,$ where
\begin{eqnarray}\label{pl6}
\zeta(p, \omega_0)=\{\zeta_i(p, \omega_0) \}_{i=1}^m, \quad
 \zeta_i(p, \omega_0)=
\{\zeta_i^{1}(p, \omega_0), \zeta_i^{2}(p, \omega_0)\},
\end{eqnarray}
\begin{eqnarray*}
 \zeta_i^{1}(p, \omega_0)= X_i(p,\zeta_0(p, \omega_0)), \quad
\zeta_i^{2}(p, \omega_0)=Y_i(p,\zeta_0(p, \omega_0))\},
\quad i=\overline{1,m},\end{eqnarray*}
is called the random field  of decisions making by firms relative to productive processes.
\end{definition}
The next Theorem gives an algorithm of the construction of random fields of consumers choice and decisions making by firms \cite{55, 71, 69, 92, 106}.
\begin{theorem}\label{ptl3}
Let a random field $\eta_i^0(p,z, \omega_i),\  (p,z) \in  K_+^n\times \Gamma^m,$ on a  probability \hfill  space \hfill  $\{\Omega_i, {\cal F}_i, \bar P_i\},$ \hfill   be \hfill  a \hfill  measurable \hfill  mapping \hfill   of \hfill  the \hfill  measurable \hfill  space\\
$\{\Gamma^m\times \Omega_i,{\cal B}(\Gamma^m)\times {\cal F}_i\}$ into the measurable  space
$\{S, {\cal B}(S)\} $ for every  $p \in K_+^n,$ $ \overline{1,l},$ and a random field
$\zeta_0(p, \omega_0),$  on a  probability   space   $\{\Omega_0, {\cal F}_0, \bar P_0\},$ be  a measurable map of  the measurable space
$\{\Omega_0, {\cal F}_0\}$ into the measurable  space $\{\Gamma^m,{\cal B}(\Gamma^m)\}$
for every  $p \in K_+^n.$ Let, moreover,
 $\eta_i^0(tp,z, \omega_i)=\eta_i^0(p,z, \omega_i), \ i=\overline{1,l}, $
$\zeta_0(tp, \omega_0)=\zeta_0(p, \omega_0), \  p \in K_+^n,\ t > 0,$
and $K_i(p,z), \ i=\overline{1,l}, $ be  income functions of   consumers that satisfy the conditions of the Definition \ref{dl1}.
If
\begin{eqnarray*}
 \left\langle\eta_i^0(p,z, \omega_i), p \right\rangle  > 0, 	\quad (p,z, \omega_i)  \in
K_+^n\times \Gamma^m\times \Omega_i, \quad  i=\overline{1,l},
\end{eqnarray*}
\begin{eqnarray*}
 \bar P_i\left(\left\langle\eta_i^0(p,z, \omega_i), p \right\rangle < \infty\right)=1,		 \quad (p,z)  \in
K_+^n\times \Gamma^m, \quad  i=\overline{1,l},
\end{eqnarray*}
then  random fields
\begin{eqnarray}\label{dppl7}
\xi_i(p, \omega)
=\frac{K_i(p,\zeta_0(p, \omega_0)) \eta_i(p,\zeta_0(p, \omega_0),\omega_i)}
{\left\langle\eta_i(p,\zeta_0(p, \omega_0),\omega_i),p \right\rangle }, \quad  i=\overline{1,l},
\end{eqnarray}
where
\begin{eqnarray*}
\eta_i(p,z, \omega_i)=\eta_i^0(p,Q(p,z),\omega_i), \quad  i=\overline{1,l},
\end{eqnarray*}
 are measurable maps of the measurable space $\{\Omega, {\cal F}\}$ into the measurable space $\{S, {\cal B}(S)\}$ for every
$p \in K_+^n$   that can be identified with  random fields of insatiable consumers choice
 on the probability space $\{\Omega, {\cal F}, \bar P\},$ where
 \begin{eqnarray*}                     \Omega=\prod\limits_{i=0}^{l}\Omega_i,\quad {\cal F}=\prod\limits_{i=0}^{l}
{\cal F}_i, \quad \bar P=\prod\limits_{i=0}^{l}
\bar P_i,\end{eqnarray*}
under condition that
$ \zeta(p, \omega_0)=Q(p,\zeta_0(p, \omega_0))$
is identified with the random field of decisions making by firms relative to productive processes and the image  of the set  $\Gamma^m$ under the map $Q(p,z)$ is a Borel set for every  $p \in K_+^n,$ that is,  $Q(p, \Gamma^m) \in {\cal B}(\Gamma^m), \ p \in K_+^n.$
\end{theorem}
\begin{proof}\smartqed
To prove the Theorem, first we show that there exists a family of conditional distribution functions
 \begin{eqnarray*}
 F_{p_1, \ldots, p_k}^{i}
(B^i|u_1, \ldots, u_k), \quad
\{p_1, \ldots, p_k\} \in [K_+^n]^k, \quad \{u_1, \ldots, u_k\} \in [\Gamma^m]^k,\end{eqnarray*}
given on the measurable spaces
$\{ \tilde X_{(p,u)_{k}}^{k,i}, {\cal B}(\tilde X_{(p,z)_k}^{k,i})\},$ $\  i=\overline{1,l}, \
k=\overline{1, \infty},$
\begin{eqnarray*}
 B^i \in {\cal B}(\tilde X_{(p,u)_{k}}^{k,i}),\quad
\tilde X_{(p,u)_{k}}^{k,i}=\prod\limits_{s=1}^k\tilde X_{(p_s,u_s)}^i, \end{eqnarray*}
and a family of unconditional distribution functions
\begin{eqnarray*}
\psi_{p_1, \ldots, p_k}({\cal D}),\quad
\{p_1, \ldots, p_k\} \in  [K_+^n]^k, \quad
k=\overline{1, \infty}, \quad  {\cal D} \in {\cal B}( [\Gamma^m]^k), \end{eqnarray*}
given on the measurable spaces $\{[\Gamma^m]^k, {\cal B}( [\Gamma^m]^k)\}, \ k=\overline{1, \infty},$ that satisfy the axioms,
the formula
\begin{eqnarray*}
\bar P(\{\xi_1(p_1,\omega), \ldots, \xi_1(p_k,\omega)\} \in A_1, \ldots,
\{\xi_l(p_1,\omega), \ldots, \xi_l(p_k,\omega)\} \in A_l,
 \end{eqnarray*}
\begin{eqnarray*}
\{\zeta(p_1,\omega_0),
\ldots, \zeta(p_k,\omega_0) \} \in {\cal D})
\end{eqnarray*}
\begin{eqnarray}\label{qmsl4}
& =\int\limits_{{\cal D}}\prod\limits_{i=1}^{l}F_{p_1, \ldots, p_k}^{i}
\left(A_i\cap \tilde X_{(p,u)_{k}}^{k,i}|u_1, \ldots, u_k\right)
d\psi_{p_1, \ldots, p_k}(u_1, \ldots, u_k)
\end{eqnarray}
is valid for $A_i \in {\cal B}(S^k), \ {\cal D} \in {\cal B}([\Gamma^m]^k),$ where the family of  budget sets
$\tilde X_{(p,u)}^i$  is generated by a certain so-called income function
$\bar K_i(p,u), \ (p,u) \in K_+^n\times \Gamma^m,$ built later  and this family of  budget sets contains  the family of  budget sets   $X_{(p,z)}^{i},$  generated by the income function
$K_i(p,z), \ i=\overline{1,l}, $
  and
\begin{eqnarray*}                     \psi_{p_1, \ldots, p_k}({\cal D})=\bar P(\{\zeta(p_1, \omega_0), \ldots,
 \zeta(p_k, \omega_0)\} \in {\cal D}), \quad
{\cal D} \in {\cal B}([\Gamma^m]^k).\end{eqnarray*}
Let  $K_i^0(p,z), \ i=\overline{1,l},$ be a collection of income pre-functions  of consumers  generating  the family of income functions of consumers $K_i(p,z), \ i=\overline{1,l},$  that figures in the Theorem.

Let $\chi_{Q(p, \Gamma^m)}(u)$ be an indicator function of the set $Q(p, \Gamma^m)$
that is the image of the set  $\Gamma^m$ under the map $Q(p,z)$  and that is a Borel set under the assumption of the Theorem.
On a probability space $\{\Omega_i, {\cal F}_i, \bar P_i\}$ the random field
\begin{eqnarray*}                     \eta_i^2(p,u,\omega_i)
=\frac{ K_i^0(p,u)\chi_{Q(p, \Gamma^m)}(u) \eta_i^0(p,u,\omega_i)}
{\left\langle\eta_i^0(p,u,\omega_i) ,p\right\rangle }, \quad (p,u)\in K_+^n\times \Gamma^m, \quad i=\overline{1,l}, \end{eqnarray*}
 is a measurable map of the space
$\{\Gamma^m\times \Omega_i,{\cal B}(\Gamma^m)\times {\cal F}_i\}$
into the space $\{S, {\cal B}(S)\}, $ $ i=\overline{1,l},$ since
$\left\langle\eta_i^0(p,u,\omega_i), p \right\rangle $ is a measurable mapping  of the  measurable space  $\{\Gamma^m\times \Omega_i,{\cal B}(\Gamma^m)\times {\cal F}_i\}$
into the  measurable space  $\{R^1, {\cal B}(R^1)\}, \ p \in K_+^n,$ as superposition of the continuous map $\left\langle x,p \right\rangle $ of the  measurable space  $\{S, {\cal B}(S)\}$
into the  measurable space  $\{R^1, {\cal B}(R^1)\}, \ p \in K_+^n,$ and
of the measurable map  $\eta_i^0(p,u,\omega_i)$  of the  measurable space
$\{\Gamma^m\times \Omega_i,{\cal B}(\Gamma^m)\times {\cal F}_i\}$
into the  measurable space  $\{S, {\cal B}(S)\}, \ i=\overline{1,l}.$ On the basis of  Lemmas \ref{ppl5} --- \ref{d2l3} we obtain the needed statement.

In accordance with the Lemma
\ref{d2l4} the random field   $\zeta(p,\omega_0)=Q(p,\zeta_0(p,\omega_0))$
is a measurable map of the  measurable space
$\{\Omega_0,
{\cal F}_0\}$ into the  measurable  space  $\{\Gamma^m,{\cal B}(\Gamma^m)\}$
for every $p \in K_+^n.$

Let us connect with random field $\eta_i^2(p,u,\omega_i)$ an auxiliary economy system.
Introduce on the set $K_+^n\times \Gamma^m$  the collection of auxiliary   income pre-functions of consumers $\bar K_i^0(p,u)=\chi_{Q(p, \Gamma^m)}(u)K_i^0(p,u), \ ( p,u)  \in
K_+^n\times \Gamma^m, \ i=\overline{1,l}, $
where  $K_i^0(p,u),  \ i=\overline{1,l},$ is a collection of income pre-functions that generates the  family of  income functions  $K_i(p,z), \ i=\overline{1,l},$
a productive economic process
\begin{eqnarray*} Q_1(p,u)= \chi_{Q(p, \Gamma^m)}(u) u,  \quad  ( p,u)  \in
K_+^n\times \Gamma^m,
\end{eqnarray*}
 a vector of initial supply of goods of the  $k$-th consumer
\begin{eqnarray*}   \bar b_k(p,u)= b_k(p,u)\chi_{Q(p, \Gamma^m)}(u), \quad  (p,u)  \in
K_+^n\times \Gamma^m,
\end{eqnarray*}  
where $b_k(p,u)$ is a vector of initial supply of goods of the  $k$-th consumer in considered economy system. Under such conditions a set of  the  family of  income functions of consumers
\begin{eqnarray*}  \bar K_i(p,u)=\bar K_i^0(p,Q_1(p,u))=
\chi_{Q(p, \Gamma^m)}(u)K_i^0(p,u)= \bar K_i^0(p,u)
\end{eqnarray*}
satisfy condition
\begin{eqnarray*} \sum\limits_{i=1}^l\bar K_i(p,u)=\left\langle\sum\limits_{i=1}^m
[\bar Y_i(p,u) -\bar X_i(p,u) ] + \sum\limits_{i=1}^l\bar b_k(p,Q_1(p,u)), p \right\rangle, \end{eqnarray*}
where $ \{(\bar X_i(p,u), \bar Y_i(p,u))\}_{i=1}^m= \chi_{Q(p, \Gamma^m)}(u) u. $
It is evident that
\begin{eqnarray*} \sum\limits_{i=1}^m
[\bar Y_i(p,u) -\bar X_i(p,u) ] + \sum\limits_{i=1}^l\bar b_k(p,Q_1(p,u)) \geq 0.
\end{eqnarray*}
For every auxiliary income function $\bar K_i^0(p,u), \ i=\overline{1,l}, $ let us build the family of budget sets $\tilde X_{(p,u)}^{i}$
for the $i$-th insatiable consumer
by the formula
\begin{eqnarray*}   \tilde X_{(p,u)}^{i}=\left\{x, \ x \in S, \
 \left\langle p,x\right\rangle =\bar K_i(p,u)\right\},\quad (p, u) \in K_+^n\times\Gamma^m, \quad i=\overline{1,l}.  \end{eqnarray*}
For $(p,u) \in K_+^n \times [\Gamma^m \setminus Q(p, \Gamma^m)] $ \
$\eta_i^2(p,u,\omega_i)=0 \in S.$ Put that for such  $(p,u)$ the budget set
 $\tilde X_{(p,u)}^i$ consists of zero vector $0 \in S$ only.

The relations
$\tilde X_{(p,\ Q(p,z))}^i=  X_{(p,z)}^i, \ i=\overline{1,l},$
take place.
From this relations it follows that the collection of  budget sets
$\tilde X_{(p,u)}^i$ generated by the family of  income functions   $\bar K_i(p,u)$
contains the collection of  budget sets $\hat X_{(p,z)}^i$ generated by the family of  income functions   $ K_i(p,z).$

Further, $\eta_i^2(t p, u, \omega_i)= \eta_i^2(p, u, \omega_i),$
$ \zeta(t p, \omega_0)= \zeta(p, \omega_0).$
On the basis of the Lemma \ref{pl1} and the Theorem  \ref{tl1},
the random fields   $\xi_i(p, \omega),\ i=\overline{1,l},$ for that the representation
\begin{eqnarray*}                     \xi_i(p, \omega)=\eta_i^2(p, \zeta(p, \omega_0),\omega_i),
\quad i=\overline{1,l},\end{eqnarray*}
is valid, are conditionally independent relative to the random field
 $\zeta(p, \omega_0)=Q(p,\zeta_0(p, \omega_0)) $
on the probability space
$\{\Omega, {\cal F}, \bar P\}.$  Moreover, the formula
\begin{eqnarray*}
 \bar P(\{\xi_1(p_1,\omega), \ldots, \xi_1(p_k,\omega)\} \in A_1, \ldots,
\{\xi_l(p_1,\omega), \ldots, \xi_l(p_k,\omega)\} \in A_l,
\end{eqnarray*}
\begin{eqnarray*}
 \{\zeta(p_1,\omega_0),
\ldots, \zeta(p_k,\omega_0) \} \in {\cal D})
\end{eqnarray*}
\begin{eqnarray}\label{qqsl4}
 =\int\limits_{{\cal D}}\prod\limits_{i=1}^{l}F_{p_1, \ldots, p_k}^{i}
\left(A_i\cap \tilde X_{(p,u)_{k}}^{k,i}|u_1, \ldots, u_k\right)
d\psi_{p_1, \ldots, p_k}(u_1, \ldots, u_k),
\end{eqnarray}
is valid,
where
\begin{eqnarray*}
 F_{p_1, \ldots, p_k}^{i}
(A^i|u_1, \ldots, u_k)=
E\chi_{A^i}(\eta_i^2(p_1,u_1, \omega_i), \ldots
\eta_i^2(p_k,u_k, \omega_i) ),
\end{eqnarray*}
\begin{eqnarray*}
  A^i \in {\cal B}(\tilde X_{(p,u)_{k}}^{k,i}),\quad
\tilde X_{(p,u)_{k}}^{k,i}=\prod\limits_{s=1}^k\tilde X_{(p_s,u_s)}^i,
\end{eqnarray*}
\begin{eqnarray*}
 \psi_{p_1, \ldots, p_k}({\cal D})=\bar P(\{\zeta(p_1, \omega_0), \ldots,
 \zeta(p_k, \omega_0)\} \in {\cal D}), \quad
{\cal D} \in {\cal B}([\Gamma^m]^k).
\end{eqnarray*}
Let us show that  functions
$F_{p_1, \ldots, p_k}^{i}(A^i|u_1, \ldots, u_k),
\ i=\overline{1,l},$
satisfies the axioms for   conditional  distribution functions
and the set of functions  $\psi_{p_1, \ldots, p_k}({\cal D}), \ {\cal D} \in {\cal B}([\Gamma^m]^k$
satisfies the axioms for   unconditional  distribution functions.
Let us prove that the  functions
\begin{eqnarray*}                     F_{p_1, \ldots, p_k}^{i}
\left(A_i\cap \tilde X_{(p,u)_{k}}^{k,i}|u_1, \ldots, u_k\right), \quad  A_i \in {\cal B}(S^k),\end{eqnarray*}
are  measurable maps of the measurable space  $\{[\Gamma^m]^k, {\cal B}([\Gamma^m]^k)\}$
into the measurable space $\{[0,1], {\cal B}([0,1])\}.$
Really,
\begin{eqnarray*}
 F_{p_1, \ldots, p_k}^{i}
\left(A_i\cap \tilde X_{(p,u)_{k}}^{k,i}|u_1, \ldots, u_k\right)
\end{eqnarray*}
\begin{eqnarray*}
 =E\chi_{A_i\cap \tilde X_{(p,u)_{k}}^{k,i}}(\eta_i^2(p_1,u_1, \omega_i), \ldots
\eta_i^2(p_k,u_k, \omega_i) )
\end{eqnarray*}
\begin{eqnarray*}
 =E\chi_{A_i}(\eta_i^2(p_1,u_1, \omega_i), \ldots
\eta_i^2(p_k,u_k, \omega_i)), \quad A_i \in {\cal B}(S^k).
\end{eqnarray*}
The last equality holds since
\begin{eqnarray*}
\chi_{\tilde X_{(p,u)_{k}}^{k,i}}(\eta_i^2(p_1,u_1, \omega_i), \ldots
\eta_i^2(p_k,u_k, \omega_i) )=1.
\end{eqnarray*}
 In accordance with the Lemma \ref{pl3},
$\chi_{A_i}(\eta_i^2(p_1,u_1, \omega_i), \ldots
\eta_i^2(p_k,u_k, \omega_i))$ is a measurable map of the measurable space
$\{[\Gamma^m]^k\times\bar \Omega_1,  {\cal B}([\Gamma^m]^k)\times
\bar {\cal F}_1\}$
into the measurable space $\{[0,1],  {\cal B}([0,1])\}.$
Therefore,
\begin{eqnarray*}
 E\chi_{A_i}(\eta_i^2(p_1,u_1, \omega_i), \ldots
\eta_i^2(p_k,u_k, \omega_i))\end{eqnarray*}
is a measurable map of the measurable space $\{[\Gamma^m]^k, {\cal B}([\Gamma^m]^k)\}$
into the measurable space $\{[0,1], {\cal B}([0,1])\}.$
The verification of the rest axioms presents no  difficulties.
To complete the proof of the Theorem, we note  that, in fact, the
formula
\begin{eqnarray*}
 \bar P(\{\xi_1(p_1,\omega), \ldots, \xi_1(p_k,\omega)\} \in A_1, \ldots,
\{\xi_l(p_1,\omega), \ldots, \xi_l(p_k,\omega)\} \in A_l,
\end{eqnarray*}
\begin{eqnarray*}
  \{\zeta(p_1,\omega_0),
\ldots, \zeta(p_k,\omega_0) \} \in {\cal D})
\end{eqnarray*}
\begin{eqnarray}\label{potqqsl4}
 =\int\limits_{{\cal D}}\prod\limits_{i=1}^{l}F_{p_1, \ldots, p_k}^{i}
\left(A_i\cap X_{(p,u)_{k}}^{k,i}|u_1, \ldots, u_k\right)
d\psi_{p_1, \ldots, p_k}(u_1, \ldots, u_k)
\end{eqnarray}
is valid,
where
\begin{eqnarray*}
 X_{(p,u)_{k}}^{k,i}=\prod\limits_{s=1}^k  X_{(p_s,u_s)}^i,
\end{eqnarray*}
\begin{eqnarray*}
 X_{(p_s,u_s)}^i=\left\{x \in S, \  \left\langle p_s, x\right\rangle= K_i^0(p_s, u_s)\right\}, \quad  u_s \in Q(p_s,\Gamma^m),
 \end{eqnarray*}
\begin{eqnarray*}
 p_s \in K_+^n \quad  s=\overline{1,k},  \quad i=\overline{1,l}.
\end{eqnarray*}
The last equality follows from   that the  measure $\psi_{p_1, \ldots, p_k}({\cal D}), \ {\cal D} \in  {\cal B}([\Gamma^m]^k),$ is concentrated on the set $ \prod\limits_{s=1}^kQ(p_s,\Gamma^m)  \in  {\cal B}([\Gamma^m]^k),$
the collection of  sets
\begin{eqnarray*}                      X_{(p,u)_{k}}^{k,i}=\prod\limits_{s=1}^k X_{(p_s,u_s)}^i, \quad  (u_1, \ldots, u_k) \in  \prod\limits_{s=1}^kQ(p_s,\Gamma^m), \quad i=\overline{1,l}, \end{eqnarray*}
coincides with the  collection of sets
\begin{eqnarray*}                      X_{(p,z)_{k}}^{k,i}=\prod\limits_{s=1}^k X_{(p_s,z_s)}^i, \quad (z_1, \ldots, z_k) \in  [\Gamma^m]^k,  \quad i=\overline{1,l}, \end{eqnarray*}
where $ X_{(p_s,z_s)}^i=\{x \in S, \  \left\langle p_s, x\right\rangle= K_i(p_s, z_s)\}, $
there hold the equalities
\begin{eqnarray*}
\chi_{A_i \cap \tilde X_{(p,u)_{k}}^{k,i}}(\eta_i^2(p_1,u_1, \omega_i), \ldots
\eta_i^2(p_k,u_k, \omega_i) )
\end{eqnarray*}
\begin{eqnarray*}
 =\chi_{A_i \cap  X_{(p,u)_{k}}^{k,i}}(\eta_i^2(p_1,u_1, \omega_i), \ldots
\eta_i^2(p_k,u_k, \omega_i)),
\end{eqnarray*}
\begin{eqnarray*}
 A_i \in {\cal B}(S^k),\quad (u_1, \ldots, u_k) \in  \prod\limits_{s=1}^kQ(p_s,\Gamma^m), \quad i=\overline{1,l}.
 \end{eqnarray*}
 From here and from the definition of $ F_{p_1, \ldots, p_k}^{i}
(A^i|u_1, \ldots, u_k),$ there hold the equalities
\begin{eqnarray*}
F_{p_1, \ldots, p_k}^{i}
\left(A_i\cap \tilde X_{(p,u)_{k}}^{k,i}|u_1, \ldots, u_k\right)=
F_{p_1, \ldots, p_k}^{i}
\left(A_i \cap X_{(p,u)_{k}}^{k,i}|u_1, \ldots, u_k\right),
\end{eqnarray*}
\begin{eqnarray*}
 A_i \in {\cal B}(S^k), \quad i=\overline{1,l}, \quad (u_1, \ldots, u_k) \in  \prod\limits_{s=1}^kQ(p_s,\Gamma^m).
\end{eqnarray*}
 \qed \end{proof}
Under the presence of non-insatiable consumers the analog of the Theorem  \ref{ptl3} is formulated as follows:
\begin{theorem}\label{ptl4}
Let  random fields $\eta_i^1(p,z, \omega_i), \ \eta_i^0(p,z, \omega_i),$ on a probability space  $\{\Omega_i, {\cal F}_i, \bar P_i\}, \
(p,z) \in  K_+^n\times \Gamma^m,$ be  measurable  mappings of  the measurable  space
$\{\Gamma^m\times \Omega_i,{\cal B}(\Gamma^m)\times {\cal F}_i\}$ into the measurable  space $\{S, {\cal B}(S)\}$ for every $p \in K_+^n,\ \overline{1,l},$
and a random field
$\zeta_0(p, \omega_0), \  p \in K_+^n,$ on a probability space  $\{\Omega_0, {\cal F}_0, \bar P_0\},$ be a measurable map of the measurable space
$\{\Omega_0, {\cal F}_0\}$ into the measurable  space $\{\Gamma^m,{\cal B}(\Gamma^m)\}$
for every $p \in K_+^n.$
Let, moreover,
\begin{eqnarray*}
 \eta_i^1(tp,z, \omega_i)=\eta_i^1(p,z, \omega_i), \quad
\eta_i^0(tp,z, \omega_i)=\eta_i^0(p,z, \omega_i),  \quad t> 0, \quad
i=\overline{1,l},
\end{eqnarray*}
\begin{eqnarray*}
 \zeta_0(tp, \omega_0)=\zeta_0(p, \omega_0), \quad  p \in K_+^n,\quad t > 0,
\end{eqnarray*}
and $K_i(p,z), \ i=\overline{1,l}, $  be  income functions of  consumers that satisfy all conditions of the Definition \ref{dl1}. If
\begin{eqnarray*}
 \left\langle\eta_i^1(p,z, \omega_i), p \right\rangle   > 0, \quad  \eta_i^0(p,z, \omega_i)
\leq \eta_i^1(p,z, \omega_i),
\end{eqnarray*}
\begin{eqnarray*}
(p,z, \omega_i) \in
K_+^n\times \Gamma^m\times \Omega_i, \quad  i=\overline{1,l},
\end{eqnarray*}
\begin{eqnarray*}
 \bar P_i(\left\langle\eta_i^1(p,z, \omega_i), p \right\rangle < \infty)=1,	
  \quad  (p,z) \in
K_+^n\times \Gamma^m, \quad i=\overline{1,l},
\end{eqnarray*}
then the random fields
\begin{eqnarray}\label{pdl7}
\xi_i(p, \omega)
=\frac{K_i(p,\zeta_0(p, \omega_0)) \eta_i^0(p,\zeta(p, \omega_0),\omega_i)}
{\left\langle\eta_i^1(p,\zeta(p, \omega_0),\omega_i),p \right\rangle}, \quad  i=\overline{1,l},
\end{eqnarray}
are  measurable maps of a measurable space $\{\Omega, {\cal F}\}$ into the measurable space $\{S, {\cal B}(S)\}$ for every
$p \in K_+^n$
   that can be identified with the random fields  of  non-insatiable consumers choice on the probability space $\{\Omega, {\cal F},\bar P\},$ where
\begin{eqnarray*}                     \Omega=\prod\limits_{i=0}^{l}\Omega_i, \quad {\cal F}=\prod\limits_{i=0}^{l}
{\cal F}_i, \quad \bar P=\prod\limits_{i=0}^{l}\bar P_i,\end{eqnarray*}
under the condition that
$ \zeta(p, \omega_0)=Q(p,\zeta_0(p, \omega_0))$
is identified with a random field  of decisions making by firms relative to productive processes and the image of the set  $\Gamma^m$ under the map $Q(p,z)$ is a Borel set for every $p \in K_+^n,$ that is,  $Q(p, \Gamma^m) \in {\cal B}(\Gamma^m).$
\end{theorem}

We call the random field
$ \eta_i^0(p,z, \omega_i)$  the random field of evaluation of information\index{random field of evaluation of information by  consumer} by the $i$-th  consumer and   we do the random field    $ \eta_i^1(p,z, \omega_i)$ the random field of   overestimation of information\index{random field of overestimation of information by non-insatiable consumer} by the $i$-th non-insatiable  consumer.

In the  next two Theorems we additionally assume that the set of income pre-functions of consumers $K_i^0(p,z), \ i=\overline{1,l},$
a productive economic process $Q(p,z)$ and  vectors of initial  supply of goods
$b_k(p,z), \  k=\overline{1,l},$
are continuous functions of variables   $(p,z) \in K_+^n\times \Gamma^m$ with values in the sets  $R^1, \ \Gamma^m, \ S$ correspondingly.  Namely, these two Theorems are important for the construction of the theory of economic equilibrium.
\begin{theorem}\label{tl4}
Let a random field $\eta_i^0(p,z, \omega_i), (p,z) \in  K_+^n\times \Gamma^m,$
 on a probability space $\{\Omega_i, {\cal F}_i, \bar P_i\},$ be such that for every
$\omega_i \in \Omega_i,$  $\eta_i^0(p,z, \omega_i)$ is a continuous function of variables
$(p,z) \in  K_+^n\times \Gamma^m, \ i=\overline{1,l},  $ with  values in the set $S$ and let  a random field $\zeta_0(p, \omega_0), \  p \in K_+^n,$  on a probability space
$\{\Omega_0, {\cal F}_0, \bar P_0\},$ take  values in the set $\Gamma^m$ and every its realization
be a continuous  function of  $  p \in K_+^n.$
Let, moreover,
 $\eta_i^0(tp,z, \omega_i)=\eta_i^0(p,z, \omega_i),\ i=\overline{1,l}, \ t> 0,$
$\zeta_0(tp, \omega_0)=\zeta_0(p, \omega_0), \  p \in K_+^n,\ t > 0,$
and $K_i(p,z), \ i=\overline{1,l}, $ be  income functions of consumers  that satisfy all conditions of the Definition \ref{dl1} and they additionally are  continuous functions  of variables $(p,z) \in  K_+^n\times \Gamma^m$.
If
\begin{eqnarray*}
\left\langle\eta_i^0(p,z, \omega_i), p \right\rangle \ > 0, \quad
  (p,z, \omega_i)  \in
K_+^n\times \Gamma^m\times \Omega_i, \quad  i=\overline{1,l},
\end{eqnarray*}
\begin{eqnarray*}
 \bar P_i\left(\left\langle\eta_i^0(p,z, \omega_i), p \right\rangle < \infty\right)=1,		 \quad (p,z)  \in
K_+^n\times \Gamma^m, \quad  i=\overline{1,l},
\end{eqnarray*}
then  the random fields
\begin{eqnarray}\label{pl7}
\xi_i(p, \omega)
=\frac{K_i(p,\zeta_0(p, \omega_0)) \eta_i(p,\zeta_0(p, \omega_0),\omega_i)}
{\left\langle\eta_i(p,\zeta_0(p, \omega_0),\omega_i),p \right\rangle}, \quad  i=\overline{1,l},
\end{eqnarray}
where
\begin{eqnarray*}
\eta_i(p,z, \omega_i)=\eta_i^0(p,Q(p,z), \omega_i),	 \quad  i=\overline{1,l},
\end{eqnarray*}
are continuous random fields for every realization on a probability space $\{\Omega, {\cal F}, \bar P\},$  where
 $\Omega=\prod\limits_{i=0}^{l}\Omega_i,$ ${\cal F}=\prod\limits_{i=0}^{l}
{\cal F}_i,$ $\bar P=\prod\limits_{i=0}^{l} \bar P_i$ that can be identified with random fields of insatiable  consumers choice  on the same probability space under condition that
$ \zeta(p, \omega_0)=Q(p,\zeta_0(p, \omega_0))$
is identified with a random field of decisions making by firms relative to productive processes.
\end{theorem}
\begin{theorem}\label{ttl3}
Let  random fields
$\eta_i^1(p,z, \omega_i), \eta_i^0(p,z, \omega_i),$
 on a probability space $\{\Omega_i, {\cal F}_i, \bar P_i\},$ be such that for every event
 $\omega_i \in \Omega_i,$  $\eta_i^1(p,z, \omega_i)$  and $\eta_i^0(p,z, \omega_i)$
are continuous functions of variable $(p,z) \in  K_+^n\times \Gamma^m, \ i=\overline{1,l},  $
with  values in the set $S,$ and let a random field $\zeta_0(p, \omega_0), \  p \in K_+^n,$ on a probability space
$\{\Omega_0, {\cal F}_0, \bar P_0\},$
take  values in the set $\Gamma^m$
and every its realization  be a continuous function of
$  p \in K_+^n.$
Let, moreover,
 $\eta_i^1(tp,z, \omega_i)=\eta_i^1(p,z, \omega_i),\
\eta_i^0(tp,z, \omega_i)=\eta_i^0(p,z, \omega_i), \
   i=\overline{1,l},$   $  t> 0,$
$\zeta_0(tp, \omega_0)=\zeta_0(p, \omega_0), \  p \in K_+^n,\ t > 0,$
and $K_i(p,z), \ i=\overline{1,l}, $  be  income functions of consumers that satisfy all conditions of the Definition  \ref{dl1}
and  they additionally are continuous functions of variables $(p,z) \in  K_+^n\times \Gamma^m.$
If
\begin{eqnarray*}
 \left\langle\eta_i^1(p,z, \omega_i), p \right\rangle  \ > 0, \quad  \eta_i^0(p,z, \omega_i)
\leq \eta_i^1(p,z, \omega_i),
\end{eqnarray*}
\begin{eqnarray*}
 (p,z, \omega_i) \in
K_+^n\times \Gamma^m\times \Omega_i, \quad  i=\overline{1,l},
\end{eqnarray*}
\begin{eqnarray*}
  \bar P_i\left(\left\langle\eta_i^1(p,z, \omega_i), p \right\rangle < \infty\right)=1,	    \quad
 (p,z) \in
K_+^n\times \Gamma^m, \quad i=\overline{1,l},
\end{eqnarray*}
then the  random fields
\begin{eqnarray}\label{ppl7}
\xi_i(p, \omega)
=\frac{K_i(p,\zeta_0(p, \omega_0)) \eta_i^0(p,\zeta(p, \omega_0),\omega_i)}
{\left\langle \eta_i^1(p,\zeta(p, \omega_0),\omega_i),p \right\rangle }, \quad  i=\overline{1,l},
\end{eqnarray}
are continuous random fields for every realization on a probability space $\{\Omega, {\cal F}, \bar P\},$  where
 $\Omega=\prod\limits_{i=0}^{l}\Omega_i,$ ${\cal F}=\prod\limits_{i=0}^{l}
{\cal F}_i,$ $\bar P=\prod\limits_{i=0}^{l} \bar P_i$ that can be identified with  random fields of non-insatiable consumers choice on the same probability space under the condition that  $ \zeta(p, \omega_0)= Q(p,\zeta_0(p, \omega_0))$
is identified with a random field  of decisions making by firms relative to productive processes.
\end{theorem}

We can define a family of  conditional distributions  for the $s$-th consumer, for example, by conditional densities  of distributions.\index{conditional densities  of distributions}
Let a certain family of income functions $K_s(p,z), \ s=\overline{1,l},$ be given  that satisfies  conditions of Definition \ref{dl1} and such that every income  function is  a continuous function of  variable $z \in \Gamma^m$ for every fixed  $p \in K_+^n.$
Consider on the sets  $S^k \times [\Gamma^m]^k,$ \
$k=\overline{1,\infty},$ a family of maps
\begin{eqnarray*} f_k^s(x_1, \ldots,
x_k|(p_1,z_1), \ldots, (p_k,z_k)), \quad k=1,2, \ldots, \quad s=\overline{1,l},  \end{eqnarray*}                        
that take  values  in  $R_+^1,$ where $x_i \in S,  \ (p_i, z_i) \in K_+^n\times
 \Gamma^m,$ $\ i=\overline{1,k},$ and satisfy conditions:\\
1) $f_k^s(x_1, \ldots, x_k|(p_1,z_1), \ldots, (p_k,z_k))$ is a continuous mapping of the set
$S^k\times [\Gamma^m]^k$  into the set $R_+^1$ for every fixed  $(p_1, \ldots , p_k) \in [K_+^n]^k\  ;$ \\
2) for an arbitrary permutation $\pi$ of indices $\{1, \ldots, k\}$ the  equalities
\begin{eqnarray*} f_k^s(x_{\pi(1)}, \ldots,
 x_{\pi(k)}|(p_{\pi(1)},z_{\pi(1)}), \ldots ,
 (p_{\pi(k)},z_{\pi(k)}))\end{eqnarray*} 
\begin{eqnarray*}
 =f_k^s(x_1, \ldots, x_k|(p_1,z_1),
 \ldots, (p_k,z_k)), \quad  k=1,2, \ldots,\quad s=\overline{1,l}, 
 \end{eqnarray*}  
 are valid;\\
3) for $\ k=1,2,  \ldots, \ j <k,$ the  conditions of consistency
\begin{eqnarray*} f_{j}^s(x_1, \ldots,
 x_{j}|(p_1,z_1), \ldots , (p_{j},z_{j}))\end{eqnarray*}
\begin{eqnarray*} 
=\int\limits_{
X_{(p_{j+1},z_{j+1})}^{1,s}}\ldots \int\limits_{ X_{(p_k,z_k)}^{1,s}}
f_k^s(x_1, \ldots, x_{k}|(p_1,z_1), \ldots ,
 (p_k,z_k))dX_{(j+1),k},
 \end{eqnarray*}
are valid, where $dx_i$ is the Lebesgue measure on a budget set $ X_{(p_i,z_i)}^{1,s}$ of the $s$-th consumer,  \
 $dX_{(j+1),k}=dx_{(j+1)}\ldots dx_k$ is the  direct product of the Lebesgue measures on the direct product of  budget sets
 $\prod\limits_{i=j+1}^k  X_{(p_i,z^i)}^{1,s}$ of the $s$-th consumer;\\
4)  For all strictly positive $ t$
\begin{eqnarray*} 
 f_k^s(x_1,
 \ldots, x_k|(tp_1,z_1), \ldots, (tp_k,z_k))\end{eqnarray*}
\begin{eqnarray*}
=f_k^s(x_1, \ldots,
 x_k|(p_1,z_1), \ldots, (p_k,z_k)),  \quad k=1,2,
 \ldots,\quad s=\overline{1,l}.\end{eqnarray*}
 Let us put
 \begin{eqnarray}\label{nas1}
 F_{p_1, \ldots, p_k}^s(A^k | z_1, \ldots,
 z_k)
 \end{eqnarray}
\begin{eqnarray*}
=\int\limits_{A^k} f_k^s(x_1, \ldots, x_{k}|(p_1,z_1),
 \ldots, (p_k,z_k))dx_1\ldots dx_k,
 \quad  k=1,2,
 \ldots, \quad s=\overline{1,l}, 
 \end{eqnarray*}
  where $A^k \in {\cal
 B}(X_{(p,z)_k}^{k,s}),$  $dx_1\ldots dx_k $  is the direct product of the Lebesgue measures on  the direct product of  budget sets
$X_{(p,z)_k}^{k,s}.$
\begin{theorem}\label{fl1}
The conditional distribution functions given by the formula (\ref{nas1}) satisfy the  axioms formulated above.
\end{theorem}
\begin{proof}\smartqed
    Let us verily the fulfillment of axioms as all consumers are insatiable and the set of possible goods  is the set
\begin{eqnarray*}
S=\{x \in R_+^n, \ 0 \leq x_k \leq c_k, \ 0 \leq c_k < \infty, \ k=\overline{1,n}\}. \end{eqnarray*}
The expression  (\ref{nas1}) can be  rewritten  in the form
\begin{eqnarray*}
 F_{p_1, \ldots, p_k}^s\left(A \cap X_{(p,z)_k}^{k,s} | z_1, \ldots,
 z_k\right)
  \end{eqnarray*}
\begin{eqnarray*}
  =\int\limits_{X_{(p,z)_k}^{k,s}} \chi_{A}(x_1, \ldots, x_k)f_k^s(x_1, \ldots, x_{k}|(p_1,z_1),
 \ldots, (p_k,z_k))dx_1\ldots dx_k,
  \end{eqnarray*}
\begin{eqnarray*}
  k=1,2, \ldots, \quad  s=\overline{1,l}, \quad A \in {\cal
 B}(S^k).
  \end{eqnarray*}
For a set  $A$ of the form $A=\prod\limits_{i=1}^kA_i, $ where $A_i \in {\cal B}(S), \ i=\overline{1,k},$ the equality
 \begin{eqnarray*}
 F_{p_1, \ldots, p_k}^s\left(A \cap X_{(p,z)_k}^{k,s}| z_1, \ldots, z_k\right)
 \end{eqnarray*}
\begin{eqnarray*}
 =\prod\limits_{i=1}^kK_s^{n-1}(p_i,z_i)\int\limits_{\tilde X_{p_1} }\ldots \int\limits_{\tilde X_{p_k}}\chi_{A}(K_s(p_1,z_1)x_1, \ldots, K_s(p_k,z_k)x_k)
 \end{eqnarray*}
\begin{eqnarray*}
 \times \chi_{S^k}(K_s(p_1,z_1)x_1, \ldots,K_s(p_k,z_k) x_k)
  \end{eqnarray*}
\begin{eqnarray*}
  \times  f_k^s(K_s(p_1,z_1)x_1, \ldots, K_s(p_k,z_k)x_{k}|(p_1,z_1)
 \ldots, (p_k,z_k))dx_1\ldots dx_k,
  \end{eqnarray*}
\begin{eqnarray*}
 k=1,2, \ldots, \quad s=\overline{1,l},
 \end{eqnarray*}
 takes place, where
$\tilde X_{p_i}=\{x_i \in R_+^n, \ \left\langle x_i, p_i\right\rangle =1\}.$

Now let $\chi_{A}^r(x_1, \ldots, x_k)$ and $\chi_{S^k}^r(x_1, \ldots, x_k), \ r=\overline{1,\infty},$ be sequences of continuous functions on the set $[R_+^n]^k$ that converge pointwise  to   indicator functions  $\chi_{A}(x_1, \ldots, x_k)$ and  $\chi_{S^k}(x_1, \ldots, x_k),$ correspondingly, as $r \to \infty.$
Consider the sequence of  conditional distribution functions of  $l$ consumers
\begin{eqnarray*}
 F_{p_1, \ldots, p_k}^{s,r}\left(A \cap X_{(p,z)_k}^{k,s}| z_1, \ldots, z_k\right)
 \end{eqnarray*}
\begin{eqnarray*}
 =\prod\limits_{i=1}^kK_s^{n-1}(p_i,z_i)\int\limits_{\tilde X_{p_1} }\ldots \int\limits_{\tilde X_{p_k}}\chi_{A}^r(K_s(p_1,z_1)x_1, \ldots, K_s(p_k,z_k)x_k)
 \end{eqnarray*}
\begin{eqnarray*}
 \times \chi_{S^k}^r(K_s(p_1,z_1)x_1, \ldots,K_s(p_k,z_k) x_k)
 \end{eqnarray*}
\begin{eqnarray*}
 \times  f_k^s(K_s(p_1,z_1)x_1, \ldots, K_s(p_k,z_k)x_{k}|(p_1,z_1)
 \ldots, (p_k,z_k))dx_1\ldots dx_k,
  \end{eqnarray*}
\begin{eqnarray*}
  k=1,2, \ldots, \quad s=\overline{1,l}, \quad r=\overline{1,\infty}.\nonumber
\end{eqnarray*}
Since the set of functions
\begin{eqnarray*}
f_k^s(K_s(p_1,z_1)x_1, \ldots, K_s(p_k,z_k)x_{k}|(p_1,z_1), \ldots, (p_k,z_k))
\end{eqnarray*}
 is continuous and bounded  on the set of integration,  every member of the sequence $F_{p_1, \ldots, p_k}^{s,r}\left(A \cap X_{(p,z)_k}^{k,s}| z_1, \ldots, z_k\right)$ is a continuous map of the set $[\Gamma^m]^k$ into the set $[0,1]$
and the sequence pointwise converges to a function $F_{p_1, \ldots, p_k}^s\left(A \cap X_{(p,z)_k}^{k,s}| z_1, \ldots, z_k\right),$  since the conditions of the Lebesgue dominated   convergence  theorem   are satisfied. Therefore, the limit function is a measurable map of the measurable space  $\{[\Gamma^m]^k, \ [{\cal B}(\Gamma^m)]^k\}$
into the measurable space $\{[0,1],{\cal B}([0,1])\}.$
For such a kind of  sets  $A$ the Theorem is proved.
Consider  finite unions of the sets of the kind $A=\prod\limits_{i=1}^kA_i$ that do not intersect, where $A_i \in S, \ i=\overline{1,k},$
and denote  finite unions of the considered kind of sets by   $V.$  It is evident that  $V$ forms the algebra of  sets. Let  $T$ be a class of sets for that the map
$F_{p_1, \ldots, p_k}^s\left(A \cap X_{(p,z)_k}^{k,s} | z_1, \ldots, z_k\right)$
is a measurable map of the measurable space $\{[\Gamma^m]^k, \ [{\cal B}(\Gamma^m)]^k\}$
into the measurable space $\{[0,1],{\cal B}([0,1]\}.$ It is easy to show that  $T$ is a monotone class.
The last  means that  $T$ contains the minimal  $\sigma$-algebra generated by the algebra of  sets $V$ that coincides with the $\sigma$-algebra  ${\cal B}(S^k).$ The verification of the rest axioms is obvious.  \qed \end{proof}

{\bf Examples.}\\ Below we give examples
of the definition of random fields of consumers choice.

1). In this example we give only a family of  conditional distribution functions, since a family of  unconditional distribution functions is not difficult  to be given.
Let the cone $K_+^n$ be given by the formula \begin{eqnarray*}                     K_+^n=\{p \in R_+^n, \ p_i>0, \ i=\overline{1,n} \},\end{eqnarray*}                        and let  a set of income  functions  $K_s(p,z), \ (p,z) \in K_+^n\times\Gamma^m, \ s=\overline{1,l},$
be continuous maps of their  variables and such that
\begin{eqnarray*}                     K_s(p,z)> d >0, \quad (p,z) \in K_+^n\times\Gamma^m, \quad s=\overline{1,l}.\end{eqnarray*}

Assume that a family of  strictly positive continuous maps $f^s(p,x,z), $ $ s=\overline{1,l},$
of the measurable  space $\{K_+^n\times S \times \Gamma^m, {\cal B}(K_+^n)\times
{\cal B}(S)\times {\cal B}(\Gamma^m) \} $
into the  measurable  space  $\{R_+^1, {\cal B}(R_+^1)\} $ is given that satisfy the condition
 $f^s(tp,x,z)=f^s(p,x,z),\ t>0.$  We assume that the set  $S=R_+^n,$ and the inequalities $ f^s(p,x,z)>c>0,\ (p,x,z) \in K_+^n\times R_+^n \times \Gamma^m $ hold.
 Let us put
 \begin{eqnarray*}                     F_{p_1, \ldots, p_k}^s(A^k | z_1, \ldots, z_k)=\int\limits_{A^k}
 \frac{\prod\limits_{i=1}^kf^s(p_i,x_i,z_i)dx_i}
 {\prod\limits_{i=1}^k
 \int\limits_{ X_{(p_i,z_i)}^{1,s}}f^s(p_i,x_i,z_i)dx_i},\quad k=1,2,
 \ldots , \end{eqnarray*}
for $A^k \in {\cal B}(X_{(p,z)_k}^{k, s}), \ k=1,2, \ldots,  $  $X_{(p_i,z_i)}^{1,s}$  is a budget set of the $s$-th consumer.

The set of conditional distribution functions given above satisfies the axioms  formulated above.
For this it is sufficient to show the fulfilment of the conditions of the Theorem \ref{fl1}.
Let us show the continuity of the set of functions

\begin{eqnarray*}                    
 f_k^s(x_{1}, \ldots,
 x_{k}|(p_{1},z_{1}), \ldots ,
 (p_{k},z_{k}))
 \end{eqnarray*}
 \begin{eqnarray*}
 = \frac{\prod\limits_{i=1}^kf^s(p_i,x_i,z_i)dx_i}
 {\prod\limits_{i=1}^k
 \int\limits_{ X_{(p_i,z_i)}^{1,s}}f^s(p_i,x_i,z_i)dx_i},\quad  k=1,2,
 \ldots , \quad   s=\overline{1,l},
 \end{eqnarray*}
of variables  $(x_1, \ldots , x_k,z_1, \ldots , z_k)$
for every fixed  $(p_1, \ldots , p_k) \in [K_+^n]^k.$
It is sufficient to show the continuity of $\int\limits_{ X_{(p_i,z_i)}^{1,s}}f^s(p_i,x_i,z_i)dx_i$
in variable $z_i$  and boundedness from below by a positive number.
The inequality
\begin{eqnarray*}
\int\limits_{ X_{(p_i,z_i)}^{1,s}}f^s(p_i,x_i,z_i)dx_i=
[K_s(p_i,z_i)]^{n-1}\int\limits_{\tilde X_{p_i}}f^s(p_i, x_iK_s(p_i,z_i),z_i)dx_i
 \end{eqnarray*}
 \begin{eqnarray*}
 > c[K_s(p_i,z_i)]^{n-1} \int\limits_{\tilde X_{p_i}}dx_i>  cd^{n-1} \int\limits_{\tilde X_{p_i}}dx_i
\end{eqnarray*}
is valid.
Since
\begin{eqnarray*}
\int\limits_{\tilde X_{p_i}}dx_i=\frac{|p_i|}{(n-1)! \prod\limits_{s=1}^np_s^i}>0
\end{eqnarray*}
 and does not depend on  $z_i,$ the needed estimation is proved. By $|p_i|$ we denoted $\{\sum\limits_{s=1}^n[p_s^i]^2\}^{1/2}.$
From the last inequality the needed  continuity with respect to the variable $z_i$ follows  for  a fixed price vector  $p_i \in K_+^n$ and also does  the boundedness from below by a positive constant.
We can remove the restriction with respect to  strict boundedness  from below of $K_i(p,z)$  using the formula
\begin{eqnarray*}
 F_{p_1, \ldots, p_k}^s(A^k | z_1, \ldots, z_k)=\int\limits_{A^k}
 \frac{\prod\limits_{i=1}^kf^s(p_i,x_i,z_i)dx_i}
 {\prod\limits_{i=1}^k
 \int\limits_{ X_{(p_i,z_i)}^{1,s}}f^s(p_i,x_i,z_i)dx_i}  \end{eqnarray*}
\begin{eqnarray*}
 =\int\limits_{\tilde X_{p_1}}\ldots\int\limits_{\tilde X_{p_k}}
 \chi_{A^k}(x_1K_s(p_1,z_1), \ldots , x_kK_s(p_k,z_k)) \end{eqnarray*}
\begin{eqnarray*}
 \times \frac{\prod\limits_{i=1}^kf^s(p_i,x_iK_s(p_i,z_i),z_i)dx_i}
 {\prod\limits_{i=1}^k
 \int\limits_{\tilde X_{p_i}}f^s(p_i, x_iK_s(p_i,z_i), z_i)dx_i},\quad k=1,2,
 \ldots ,\quad  s=\overline{1,l}.
\end{eqnarray*}
The measurability of the numerator from the space $\{[\Gamma^m]^k, \ [{\cal B}(\Gamma^m)]^k\}$ into the  space
$\{R^1, {\cal B}(R^1)\}$ in the last formula can be proved as it was  made in the Theorem \ref{fl1}. The denominator is a continuous map  of the space
 $[\Gamma^m]^k$ into  the space  $R^1$ and is strictly bounded from below.
The last removes the restriction of  strict  boundedness of income functions from below.

A random field of consumer choice defined by such a set of conditional distribution functions
is   not sufficiently regular.

 2). Next example is important and it is based on the Theorem \ref{tl4}.
This is an example of  random fields of consumers choice  concentrated  on a certain set of realizations.
Let a certain  family of  income pre-functions  $K_s^0(p,z), \ s=\overline{1,l},$ be given that satisfy conditions of Definition  \ref{ddl1}, and every of them is a continuous function of variables $(p,z) \in  K_+^n \times \Gamma^m.$ We assume that a productive economic process
$Q(p,z)$ is a  continuous  function of variables  $(p,z) \in K_+^n \times \Gamma^m.$
Therefore, for every fixed  $p \in K_+^n$ the set  $Q(p,\Gamma^m)$ is a closed one, so it is a Borel set.

Let $y_i(p,z,j), \ j=\overline{1, \infty},
\ i=\overline{1,l},$
 be a family of maps  given on the set $K_+^n\times \Gamma^m$  with  values in the  set $S.$  We assume  that  maps $y_i(p,z, j),$  $ j=\overline{1, \infty},
\ i=\overline{1,l},$  are continuous ones of   $K_+^n \times \Gamma^m$ into the set
 $S.$ As to the set  of  maps   $y_i(p,z,j), \ j=\overline{1,
 \infty},$  we  assume  that  $y_i(p,z,j) \neq y_i(p,z,k),$  $\ j \neq k,$
 and, moreover, \begin{eqnarray*}
 \sum\limits_{k=1}^ny_{ki}(p,z,j)p_k >0, \quad p \in
 K_+^n, \quad  p \neq 0, \quad  z  \in \Gamma^m, \quad j=1,2, \ldots ,
 \end{eqnarray*}
 \begin{eqnarray*}
 y_i(tp,z,k)=y_i(p,z,k), \quad \forall t>0, \quad
\ i=\overline{1,l}, \quad k=1,2, \ldots.
 \end{eqnarray*}
 Let, further, $K_i(p,z)= K_i^0(p, Q(p,z))$  be an income function of the  $i$-th consumer, $\ i=\overline{1,l}.$
By this family of maps, we construct the family of maps
\begin{eqnarray}\label{h1l17}
 x_i(p,z,j) \end{eqnarray}
 \begin{eqnarray*}
  =K_i(p,z)\left\{\frac{y_{ki}(p,Q(p,z),j)}{\sum\limits_{k=1}^n
y_{ki}(p,Q(p,z),j)p_k} \right\}_{k=1}^n,\quad \ i=\overline{1,l}, \quad  j=1,2, \ldots
\end{eqnarray*}
and the family of maps
\begin{eqnarray*}
 \bar x_i(p,u,j)
\end{eqnarray*}
\begin{eqnarray}\label{nh1l17}
=K_i^0(p,u)\chi_{Q(p,\Gamma^m)}(u)\left\{\frac{y_{ki}(p,u,j)}
{\sum\limits_{k=1}^ny_{ki}(p,u,j)p_k} \right\}_{k=1}^n,\quad  i=\overline{1,l},
\quad  j=1,2, \ldots.
\end{eqnarray}
Define on  $ A^k \in {\cal B}( X_{(p,z)_k}^{k,i})$ the family of  conditional distribution functions
 \begin{eqnarray*}
 F_{p_1, \ldots, p_k}^{k,i}( A^k | u_1, \ldots ,
 u_k)
  \end{eqnarray*}
\begin{eqnarray*}
 = \sum\limits_{j=1}^{\infty}\pi_j^i
\chi_{A^k}(\bar x_j^i(p_1,u_1),\ldots,   \bar x_j^i(p_k,u_k)), \quad
 i=\overline{1,l}, \quad  k=1,2, \ldots ,
\end{eqnarray*}
where $\chi_{A^k}(x_1,\ldots,  x_k)$ is the indicator function of the set $A^k \in  {\cal B}(
X_{(p,z)_k}^{k,i}),$ and $\sum\limits_{j=1}^{\infty}\pi_j^i=1, \ \pi_j^i
\geq 0, \  i=\overline{1,l}.$

The family of  conditional distribution functions defined above satisfies all axioms formulated above.
Define now a family of  unconditional distribution functions  $\psi_{p_1, \ldots, p_k}(B^k).$
Let $z_j(p), \ j=\overline{1,
 \infty},$ be a family of continuous maps of  $K_+^n$ into
 $\Gamma^m,$ \ $z_j(t p)=z_j(p).$  Suppose that  $\gamma_l \geq 0, \
 \sum\limits_{l=1}^{\infty}\gamma_l=1.$
 Let us put
\begin{eqnarray*}
 \bar z_j(p)=Q(p, z_j(p)),
  \end{eqnarray*}
\begin{eqnarray*}   \psi_{p_1, \ldots,
 p_k}(B^k)
=\sum\limits_{j=1}^{\infty}\gamma_j
\chi_{B^k}(\bar z_j(p_1),\ldots,   \bar z_j(p_k)), \ k=1,2, \ldots ,
\end{eqnarray*}
where $\chi_{B^k}(z_1,\ldots,   z_k)$ is an indicator function of the set $B^k \in {\cal B}([\Gamma^m]^k).$

We can interpret the last example  as random fields of consumers choice on the discrete probability space of the following kind:
$\Omega_0=\{1,2, \ldots, k, \ldots\}\  ;$   $\Omega_i=\{1,2, \ldots,
k, \ldots\}\  ;$ $\bar P_i$  is an atomic measure on
$\Omega_i\  ;$  $\bar P_i(k)=\pi_k^i, \ i=\overline{1,l}, \  k=1,2,3,\ldots \  ;$
$\bar P_0(i)=\gamma_i, \ i=1,2,3, \ldots\  ;$
$\sigma$-algebra ${\cal F}_i $ coincides with $\sigma$-algebra of all subsets  $\Omega_i, \  i=\overline{0,l}.$
In this case
  $\eta_i^0(p,z,\omega_i)=\eta^0_i(p,z,j)=y_i(p,z, j),$
$\zeta_0(p, \omega_0)=\zeta_0(p,k)=z_k(p),$  and  $\xi_i(p,\omega_0,
\omega_j)=x_i(p,z_k(p), j),  \ \omega_j=j,$ \ $ \omega_0=k.$

In the following example  random fields  $\eta_i(p,z, \omega_i), \ i= \overline{1,l},$
are correspondingly generated by  a certain set of  random values  $\xi_i, \ \overline{1,l},$
on a  probability space $\{\Omega_i, {\cal F}_i, \bar P_i\}, \ \overline{1,l},$ that takes  values in the set of the possible goods $S$ and $\xi_i $ is a measurable map of the measurable space  $\{\Omega_i, {\cal F}_i\}$ into the measurable space  $\{S, {\cal B}(S)\}, \ \overline{1,l}.$

Consider a certain family of income pre-functions  $K_s^0(p,z), \ s=\overline{1,l},$
that satisfy conditions of the  Definition  \ref{ddl1} and they are continuous functions of variables
$(p,z) \in K_+^n \times \Gamma^m.$

We assume that a productive economic process
$Q(p,z)$ is a continuous function of  variables $(p,z) \in K_+^n\times\Gamma^m,$
therefore, for every fixed  $p \in K_+^n$ the set  $Q(p,\Gamma^m)$ is a closed one,
so it is a  Borel set. Let  $\mu_i(A), \ i= \overline{1,l},$  be probability measures on
$\{S, {\cal B}(S)\}$ generated by  random values $\xi_i=\{\xi_{ki}\}_{k=1}^n, \ \overline{1,l}.$
Let, further, $h_i(p, x)=\{h_{ki}(p, x)\}_{k=1}^n$ be a set of continuous maps of
$K_+^n\times S$ into  $S$ that satisfy conditions: for all $\omega_i \in \Omega_i,$~  $\left\langle r_i(p), p \right\rangle \ > 0, \  p \in K_+^n, \ i=\overline{1,l}, $  where $ r_i(p)=\{h_{ki}(p, \xi_i)\}_{k=1}^n$
be a random field generated by the random value $\xi_i.$
Let  $\zeta_0(p,\omega_0)$ be a certain random field that is  continuous with probability 1 on a probability space $\{\Omega_0, {\cal F}_0, \bar P_0\}$ taking  values in the set $\{\Gamma^m, {\cal B}(\Gamma^m)\}.$
Let us consider  the random fields
\begin{eqnarray*}
 \xi_i(p)=\frac{K_i(p,\zeta_0(p, \omega_0))r_i(p)}{\left\langle r_i(p), p \right\rangle}, \quad  \overline{1,l},\end{eqnarray*}
on a probability space   $\{\Omega, {\cal F}, \bar P\},$  where
\begin{eqnarray*}
\Omega=\prod\limits_{i=0}^l\Omega_i,\quad  {\cal F}= \prod\limits_{i=0}^l{\cal F}_i,\quad \bar P=\prod\limits_{i=0}^l\bar P_i,
\end{eqnarray*}

On the basis of the Theorem  \ref{tl4} they can be identified with  random fields of consumers choice if
$\zeta(p)=Q(p,\zeta_0(p, \omega_0))$ is identified  with a random field of decisions making by firms.

This example is important because  it gives the possibility for every utility function  of consumer   to correspond  to a certain random field of consumer choice.

For this purpose, we interpret every utility function  on the set of possible goods  $S$ as  the probability  density to choose a certain set of goods  from the set on that the utility function is given. Define by such probability measure the random value $\xi$ with values in the set of the possible goods and construct for every consumer such random value. Having used  the method of the construction of random fields  by  random values  described above,  we refer  the random field of consumer choice to every consumer.

\chapter{The structure of production technology}

\abstract*{A wide class of compact technological maps (CTM) are introduced. Sufficient conditions of the continuity of the optimal   strategy  of firm behaviour  are given. The Theorems giving an algorithm  of construction of  optimal strategy of firm behaviour for Leontieff and Neumann technological maps are proved.   For  technological map from CTM class in wide sense  and convex down, the Theorem guaranteeing  existence  of continuous strategy of firm behaviour arbitrary close, in income,  to optimal one is established. The structure of Kakutani  continuous  technological maps are obtained. The Theorem of the existence  of continuation of two  Kakutani  continuous  technological maps on the set that are minimal  linear convex span of sets on which extended maps are defined. As a result, the Theorem about  the possibility  of continuation of every  technological map from CTM class in a wide sense  to  technological map from CTM class is obtained. }

This section is devoted to the investigation of   technological maps that describe  production of firms. The most important among  them are  technological maps from the CTM class\index{ technological maps from the CTM class} and from the CTM class in a wide sense.\index{ technological maps from the CTM class in a wide sense} The notion of  optimal strategy of firm behavior is introduced and the proposition    \ref{strat1} of the existence of  optimal  strategy of firm behavior for   technological maps from the CTM class in a wide sense is proved.
In the Lemma \ref{ol1} sufficient conditions are presented that provide the continuity of optimal strategy of firm behavior. The Theorem \ref{krt1} gives the full  description
of the set of  extreme points for the Leontieff  technological map  and in the Theorems
 \ref{algs1},   \ref{algs2} an algorithm of the construction of  optimal  strategy of firm behaviour\index{optimal  strategy of firm behavior} for  the Leontieff  technological map\index{optimal  strategy of firm behavior for  the Leontieff  technological map }  is presented. In the Theorem  \ref{naj1} an algorithm of the construction of optimal  strategy of firm behavior for the case of the Neumann   technological map\index{ optimal  strategy of firm behavior for Neumann   technological map} is presented.
A technological map given by the formula (\ref{sl6}) is introduced, the Lemma \ref{start2} on the belonging of this technological map  to the CTM class in a wide sense and its convexity  down  is proved. In the Lemma \ref{s1} an algorithm of the construction of  optimal  strategy of firm behaviour\index{optimal  strategy of firm behavior} for the  technological map given by the formula (\ref{sl6}) is presented.
The Lemma  \ref{s2} is fundamental for further investigation and proves   for the technological map, given by the formula (\ref{sl6}), the existence of   continuous strategy of firm behavior that is  arbitrary close  in income to the optimal one.
By a convex down technological map  from the CTM class in  a wide sense, a technological map given by the formula (\ref{ssl6}) is constructed and the Lemma  \ref{strat2} about the belonging of the constructed technological map to the CTM class and its convexity  down is proved.
In the Lemma \ref{s100} the structure of  optimal strategy of firm behavior for a technological map given by the formula  (\ref{ssl6}) is found. The Lemma \ref{s101}
guarantee the convergence of the sequence of  optimal incomes for  technological maps of the type (\ref{ssl6}) to optimal income of the technological map  by that the sequence of  technological maps is constructed.
The basic result of the subsection 2.1 is the Theorem  \ref{nnl1}, in which under   wide assumptions the existence of    continuous strategy of firm behavior that is arbitrarily close in income  to  optimal strategy of firm behavior  is proved.
In the  subsection 2.2 the structure of  technological maps\index{structure of Kakutani  continuous from above technological maps} that are the Kakutani  continuous from above is investigated  and  conditions under that they can be extended to a  wider set of the definition are found.
In the Theorem \ref{prodov1} the conditions
for  collection of   probability measures are obtained  under that the representation for the Kakutani  continuous from above    technological maps\index{representation for the Kakutani  continuous from above    technological maps} relative to this collection of   probability measures
are valid.
In the Theorem  \ref{prodov2} a  specific collection of   probability measures are constructed  and maps with the help of which  every  Kakutani  continuous  technological map can be   represented by means of this  collection of   probability measures and constructed maps.
The Theorem \ref{prodov3} guarantees the existence of continuation of  two   Kakutani  continuous  technological\index{continuation of  two   Kakutani  continuous  technological maps} maps on the set that are minimal linear convex span  of   sets on which  extended maps are defined. As a result of the proved Theorem,  the statement that every  technological map  from the CTM class in a wide sense can be extended  to a technological map from the CTM class is proved. The rest  statements of this subsection
has constructive character and
with the help of them one can construct technological maps with  given properties.

\section{Convex  down technological maps}

Let $F(x), \ x \in X,$ be a technological map  describing the production structure of the firm.
 Strategy of firm behaviour\index{ strategy of firm behavior}  is a map of the set possible prices  $K_+^n$ into the set of possible productive processes  of the firm $\Gamma=\{(x,y) \in S^2, \ x \in X, \ y \in F(x) \}.$  We denote strategy of firm behavior by $(x(p),y(p)), \  p \in K_+^n.$
\begin{definition} A strategy of firm behavior  $(x(p),y(p)), \  p \in K_+^n,$ is  optimal one if
\begin{eqnarray*}          \sup\limits _{x\in X} \sup\limits
_{y\in F(x)}\left\langle y-x,p\right\rangle = \left\langle y(p)-x(p),p\right\rangle, \quad
p \in K_+^n. \end{eqnarray*}
\end{definition}
\begin{proposition}\label{strat1}
For  technological map  $F(x),\  x \in X,$  belonging to the CTM class in a wide sense an optimal strategy of firm  behaviour\index{optimal strategy of firm  behavior} exists.
\end{proposition}

\begin{proof}\smartqed
Prove the existence of optimal productive process
$\{x^0(p), y^0(p)\}$ such that
\begin{eqnarray*}           \sup_{x \in X}\sup_{y \in F(x)} \left\langle y -x, p \right\rangle = \left\langle y^0(p)- x^0(p),p \right\rangle \end{eqnarray*}
for every vector  $p \in K_+^n.$
If
\begin{eqnarray*}          \Gamma=\{(x,y) \in S^2, \ x \in X, \ y \in F(x) \},\end{eqnarray*}           then in accordance with the Lemma
\ref{ly1} the set $\Gamma$  is  closed and bounded one, moreover, the equality
\begin{eqnarray*}          \bigcup\limits_{x \in X}\{x, F(x)\}=\Gamma\end{eqnarray*}
is valid,
where $\{x, F(x)\}=\{(x,y), y \in F(x)\}$
is a closed and bounded subset of the set  $\Gamma.$
Because of continuity of the function $ \left\langle y -x, p \right\rangle  $ of variable  $(x,y)$ on the set $\Gamma$ for every fixed  $p \in K_+^n,$   in accordance with the Weierstrass theorem,\index{Weierstrass theorem} there exists a pair  $\{x^0(p), y^0(p)\} \in \Gamma$ such that
\begin{eqnarray*}           \sup_{(x,y) \in \Gamma} \left\langle y -x, p \right\rangle= \left\langle y^0(p)- x^0(p),p \right\rangle.\end{eqnarray*}
Let us show that
\begin{eqnarray*}           \sup_{x \in X}\sup_{y \in F(x)} \left\langle y -x, p \right\rangle=  \sup_{(x,y) \in \Gamma}\left\langle y -x, p \right\rangle.\end{eqnarray*}
It is evident that
\begin{eqnarray*}
\sup_{y \in F(x)}\left\langle y -x, p \right\rangle=\sup_{(x,y) \in \{x, F(x)\}}\left\langle y -x, p \right\rangle \leq
\sup_{(x,y) \in \Gamma}\left\langle y -x, p \right\rangle,
\end{eqnarray*}
therefore
\begin{eqnarray*}
 \sup_{x \in X}\sup_{y \in F(x)} \left\langle y -x, p \right\rangle \leq
\sup_{(x,y) \in \Gamma}\left\langle y -x, p \right\rangle.
\end{eqnarray*}
Prove the inverse inequality. If   $\{x^0(p), y^0(p)\} \in \Gamma$ is  a pair that gives maximum, in accordance with the Weierstrass theorem, then
\begin{eqnarray*}
   \sup\limits_{x \in X}\sup\limits_{y \in F(x)} \left\langle y -x, p \right\rangle \geq
\sup\limits_{y \in F(x^0(p))} \left\langle y -x^0(p), p \right\rangle
\end{eqnarray*}
\begin{eqnarray*}
 \geq \left\langle y^0(p) -x^0(p), p \right\rangle= \sup\limits_{(x,y) \in \Gamma}\left\langle y -x, p \right\rangle.
\end{eqnarray*}
\qed
\end{proof}
For  further    investigation, the continuity of  strategy of firm behavior and its closeness to  optimal strategy of behavior is very important.

The most essential in the description of a firm is its technological map  and strategy of its behavior.

In the next Lemma, sufficient conditions are given under fulfilment  of which the  optimal strategy of firm behavior is continuous one.

\begin{lemma}\label{ol1}
Let a technological map  $F(x), \ x \in X,$ belong to the CTM class in a wide sense and be strictly convex down.
The solution $y(p,x)$ of the problem
\begin{eqnarray}\label{h1l20}
\max\limits _{y\in F(x)} \left\langle y-x,p\right\rangle =\left\langle
y(p,x)-x,p\right\rangle
\end{eqnarray}
is a continuous function of variables   $(p,x).$    Then  optimal input vector,\index{optimal input vector} $x(p)$
that is a solution of the problem
\begin{eqnarray}          \label{h1l21} \max\limits _{x\in X}\max\limits
_{y\in F(x)}\left\langle y-x,p\right\rangle = \left\langle
y(p,x(p))-x(p),p\right\rangle,
\end{eqnarray}
 and  optimal strategy of firm behavior $(x(p),y(p,x(p)))$ are continuous on $\bar R_+^n.$
\end{lemma}
\begin{proof}\smartqed          From the  conditions of the Lemma \ref{ol1}, it follows that
the problem  (\ref{h1l20}) always has the solution $y(p,x),$ but  the requirement of its continuity  is  restrictive. Let us give an example as this restriction is  satisfied. Let $f_i(x) = f_i (x_1,\dots
,x_n),$\ $i=\overline{1,n},$ \ $  x=\{x_i\}_{i=1}^n,$ be functions that are strictly convex  up and continuous on $X.$ By these functions,  let us construct the technological map $F(x)$ by the law
\begin{eqnarray*}          F(x)=\{y'=(y'_1,\dots ,y'_n),\quad 0\leq y'_i\leq f_i
(x_1,\dots ,x_n), \quad i=\overline{1,n}\} .\end{eqnarray*}           Then the solution of the problem (\ref{h1l20}) is the vector
\begin{eqnarray*}          y(p,x)=\{ f_1 (x_1,\ldots,x_n),\ldots, f_n (x_1,\ldots ,x_n)\}, \end{eqnarray*}
that does not depend on  $p.$
From the continuity of $f_i(x_1,\dots ,x_n)$ on  $X$   the continuity of $y(p,x)$ follows.  This technological map is the Kakutani continuous from above.  Let us show that it is convex down.  From the convexity   up of  $f_i(x)$  we have
\begin{eqnarray*}          f_i (\alpha x_1+(1-\alpha
)x_2)\geq\alpha f_i(x_1)+(1-\alpha )f_i(x_2),\end{eqnarray*}
\begin{eqnarray*}          x_1=\{x_i^1\}_{i=1}^n, \quad x_2=\{x_i^2\}_{i=1}^n.\end{eqnarray*}
From the last inequality, it follows that the set
  $F(\alpha x_1
+(1-\alpha ) x_2 ) $ contains the set $\alpha F(x_1 )+(1-\alpha)
F(x_2 ),$ therefore  \begin{eqnarray*}          F(\alpha x_1+(1-\alpha )x_2)\supseteq\alpha
F(x_1)+(1-\alpha )F(x_2).\end{eqnarray*}            From the strict convexity  of
$f_i(x),~i=\overline {1,n},$
the strict convexity  down of   $F(x)$ follows.   Really,
\begin{eqnarray*}          \sum\limits
_{i=1}^n p_i f_i (\alpha x_1+(1-\alpha )x_2) > \alpha\sum\limits
_{i=1}^n f_i (x_1)p_i+(1-\alpha) \sum\limits _{i=1}^n     f_i
      (x_2)p_i,\quad 0<\alpha <1,\end{eqnarray*}
if  at least for one  $p_i \not=0,~x_1 \not= x_2 .$

Let us prove the Lemma  \ref{ol1}. The solution of the problem  (\ref{h1l21})
exists due to the continuity of $\left\langle
y(p,x)-x,p\right\rangle $  and compactness of $X.$   Denoting it by
$x_1(p),$ let us show that it is unique.
Assume that there exists one more solution  $x_2(p)$
of the problem  (\ref{h1l21}) that is different from $x_1(p)$.
From the convexity  down of $F(x)$ it follows that
the vector  $\alpha x_1(p)+(1-\alpha )x_2 (p)$ is a solution of the problem
(\ref{h1l21}) for any  $0<\alpha <1.$ Really, from
(\ref{1h1l19}) \begin{eqnarray*}          \left\langle p,y(p,\alpha x_1+(1-\alpha
      )x_2)\right\rangle\geq \alpha\left\langle p,y(p,x_1)
  \right\rangle +(1-\alpha ) \left\langle p,y(p,x_2)
        \right\rangle. \end{eqnarray*}            Taking  instead  of  $x_1$ and $x_2$ the solutions $x_1(p)$ and
$x_2(p)$ we obtain the inequality
\begin{eqnarray*}          \left\langle p,y(p,\alpha x_1(p)+(1-\alpha
)x_2(p)) - \alpha x_1(p)- (1- \alpha) x_2(p)\right\rangle
\end{eqnarray*}
\begin{eqnarray*}
 \geq
\left\langle p,y(p,x_1(p))- x_1(p)\right\rangle ,
\end{eqnarray*}           since
\begin{eqnarray*}          \left\langle p,y(p_,x_1(p))- x_1(p)\right\rangle =\left\langle p,y(p_,x_2(p))- x_2(p)\right\rangle =
  \max\limits _{x\in X}\left\langle p,y(p,x)- x \right\rangle .\end{eqnarray*}
On the other hand,
\begin{eqnarray*}          \left\langle p,y(p,\alpha x_1(p)+(1-\alpha )x_2(p))
- \alpha x_1(p)- (1- \alpha) x_2(p) \right\rangle\leq
\end{eqnarray*}
\begin{eqnarray*}           \leq
  \max\limits _{x\in X}\left\langle p,y(p,x) - x \right\rangle =
  \left\langle p,y(p,x_1(p))- x_1(p) \right\rangle,
\end{eqnarray*}
therefore, for all $\alpha\in (0,1)$
\begin{eqnarray*}
\left\langle p,y(p,\alpha x_1(p)+(1-\alpha )x_2(p))\right\rangle =
  \alpha \left\langle p,y(p,x_1(p))\right\rangle +
(1-\alpha) \left\langle p,y(p,x_2(p))\right\rangle .
\end{eqnarray*}
 The consequence of a strict convexity down of  technological map  $F(x)$
is the inequality
\begin{eqnarray*}          \left\langle p,y(p,\alpha
x_1(p)+(1-\alpha )x_2(p))\right\rangle > \alpha\left\langle
p,y(p,x_1(p))\right\rangle +(1-\alpha )\left\langle p,y(p,x_2(p))\right\rangle. \end{eqnarray*}
The last inequality contradicts the  previous equality. Hence, it follows that $x_1(p)=x_2(p).$
Let us show the continuity of  $x_1(p).$  Let  $p_n$ be a convergent sequence  the limit of that is  $p_0.$
Consider the sequence  $x_1(p_n).$  This sequence is  compact one, therefore, there exists a subsequence
$x_1 (p_{n_m})$ that converges to
$\overline x_1.$ For every  $y \in F(x),\  x\in X, $
\begin{eqnarray}          \label{h1l22}
\left\langle y-x,p_{n_m}\right\rangle\leq\left\langle y(p_{n_m}, x_1(p_{n_m}))-
 x_1(p_{n_m}),p_{n_m}\right\rangle.
\end{eqnarray}
This inequality follows from the definition  of the maximum.
Taking the limit in (\ref{h1l22}),  we obtain
\begin{eqnarray*}          \left\langle
y-x,p_0\right\rangle\leq\left\langle y(p_0,\bar x_1 )-\bar x_1 ,p_0\right\rangle.\end{eqnarray*}
From the last inequality, we obtain inequality
\begin{eqnarray}          \label{h1l23}
\max\limits _{x\in X}\max\limits _{y\in F(x)}\left\langle y-x,p_0\right\rangle
\leq\left\langle y(p_0,\bar x_1)-x_1,p_0\right\rangle.
\end{eqnarray}
However,
\begin{eqnarray*}          \max\limits _{x\in X}\max\limits _{y\in F(x)}\left\langle y-x,p_0\right\rangle =
  \left\langle y(p_0,x_1(p_0))-x_1(p_0),p_0\right\rangle, \end{eqnarray*}
therefore,
\begin{eqnarray}          \label{h1l24}
\left\langle y(p_0,x_1(p_0))-x_1(p_0),p_0\right\rangle\leq
\left\langle y(p_0,\bar x_1)-\bar x_1,p_0\right\rangle .
\end{eqnarray}
From the inequality  (\ref{h1l24}), it follows that $\bar x_1=x_1(p_0),$
since the maximum is realized only on one vector and for the rest vectors the inequality
\begin{eqnarray*}          \left\langle y(p_0,     x_1(p_0))-
  x_1(p_0),p_0\right\rangle\geq \left\langle y(p_0,\bar x_1 )-\bar
    x_1,p_0\right\rangle \end{eqnarray*}
is valid because of  $y(p_0,\bar x_1 )\in F(\bar x_1).$
The last one is a consequence of the Kakutani continuity from above of  $F(x).$
Really, since
\begin{eqnarray*}          y(p_{n_m},x_1(p_{n_m}))\in F(x_1(p_{n_m}))\end{eqnarray*}
and $F(x)$ is  Kakutani continuous from above, we obtain
\begin{eqnarray*}          y(p_0,\bar x_1)\in F(\bar x_1). \end{eqnarray*}
 From  (\ref{h1l23})
and (\ref{h1l24}), we have
\begin{eqnarray*}          \left\langle y(p_0,     x_1(p_0))-
x_1(p_0),p_0\right\rangle = \left\langle y(p_0,\bar x_1)-\bar x_1
  ,p_0\right\rangle, \end{eqnarray*}
therefore, $\bar x_1=x_1(p_0).$
 From the arbitrariness  of the subsequence  $x_1 (p_{n_m}),$  it follows that
$x_1(p_n)$ tends to  $x_1 (p_0).$
\qed
\end{proof}

As $f_i(x)$ we can choose, for example,  the Cobb-Duglas functions\index{Cobb-Duglas functions}
\begin{eqnarray*}          f_i(x)=A_i\prod\limits _{k=1}^n {x_k}^{\lambda
_k},\quad \lambda _k\geq 0,\quad \sum\limits _{k=1}^n\lambda
_k=1.\end{eqnarray*}
If all  $\lambda _i >0,$ then the Cobb-Duglas function is strictly convex  up, for example, on the set $X=\{x=\{x_i\}_{i=1}^n \in R_+^n, \  \sum\limits_{i=1}^nx_i=1 \}.$ If for certain   $\lambda _i =0,$ then the Cobb-Duglas function is  convex up.  Really,
\begin{eqnarray*}          f_i(x)=\min\limits_{y \in Y}A_i \sum\limits_{k=1}^n\lambda_k
x_ky_k,\end{eqnarray*}
where
\begin{eqnarray*}          Y=\{y \in R_+^n,\ \prod\limits_{i=1}^ny_i^{\lambda_i}=1\}.\end{eqnarray*}
First of all,
\begin{eqnarray*}          f_i(x)=A_i\prod\limits _{k=1}^n {x_k}^{\lambda
_k}=A_i\prod\limits _{k=1}^n (x_ky_k)^{\lambda_k} \leq
A_i \sum\limits_{k=1}^n\lambda_k x_k y_k.\end{eqnarray*}
The last inequality follows from the convexity of  logarithmic function
\begin{eqnarray*}          -\ln(\sum\limits_{i=1}^n\lambda_i x_i) \leq -
\sum\limits_{i=1}^n\lambda_i \ln x_i=-\ln
(\prod\limits _{i=1}^n {x_i}^{\lambda _i}) \end{eqnarray*}
or
\begin{eqnarray*}          \prod\limits _{i=1}^n {x_i}^{\lambda _i}\leq
\sum\limits_{i=1}^n\lambda_i x_i.\end{eqnarray*}
Let us put
\begin{eqnarray*}          \bar y=\{\bar y_i\}_{i=1}^n=\prod\limits _{i=1}^n {x_i}^{\lambda
_i}\{\frac{1}{x_1}, \ldots,\frac{1}{x_n}\},\end{eqnarray*}
then $\bar y \in Y$ and
\begin{eqnarray*}          f_i(x)=A_i\sum\limits_{k=1}^n\lambda_k x_k\bar y_k.\end{eqnarray*}
The function
\begin{eqnarray*}          \psi_i(x,y)=A_i\sum\limits_{k=1}^n\lambda_k x_k y_k\end{eqnarray*}
is a convex function of the variable  $x$ for every fixed   $y \in Y.$  Therefore,
\begin{eqnarray*}
   f_i(\alpha x_1+(1-\alpha) x_2)=\inf\limits_{y \in Y}\psi_i(
\alpha x_1+(1-\alpha) x_2, y)
\end{eqnarray*}
\begin{eqnarray*}
   =\inf\limits_{y \in Y}[\alpha
\psi_i(x_1,y)+(1-\alpha)\psi_i(x_2,y)]
\end{eqnarray*}
\begin{eqnarray*}
          \geq \alpha
\inf\limits_{y \in Y}\psi_i(x_1,y)+(1-\alpha)
\inf\limits_{y \in Y}\psi_i(x_2,y)
\end{eqnarray*}
\begin{eqnarray*}
  = \alpha f_i(x_1)+(1-\alpha)
f_i(x_2). \end{eqnarray*}
Introduce a partial  ordering  $ \leq $ between two vectors $x_1, \ x_2$ from $n$-dimensional space $R^n.$ The inequality $x_1 \leq x_2$  means corresponding inequalities between all components of these vectors.

Consider now the case as the technological map is given by the Leontieff matrix\index{Leontieff matrix} on an expenditure set  $X.$ Assume that the expenditure set $X$ is of the kind
\begin{eqnarray*}          X=\{x=\{x_i\}_{i=1}^n  \in R_+^n, \ x_i \leq x_i^0, \ i=\overline{1,n} \},\end{eqnarray*}
where the vector  $x^0 \in R_+^n$ and has strictly positive components. The technological map given by the Leontieff matrix $A$ has the form
\begin{eqnarray}          \label{h1l25}
F(x)=\{ y\in
R_+^n ,\ Ay\leq x\}, \quad x \in X \subseteq R_+^n.
 \end{eqnarray}
\begin{lemma} Let the matrix $A$ do not contain  zero columns.
If the set of  extreme points of the set $F(x^0)$
 consists of  $m$ vectors $y_1,\dots ,y_m\in R_+^n,$
then
\begin{eqnarray*}          \max\limits _{y\in F(x^0)}\left\langle y- Ay,p\right\rangle =
\max\{\left\langle p,y_1-Ay_1\right\rangle ,\dots ,\left\langle p,y_m-Ay_m\right\rangle\} .\end{eqnarray*}
\end{lemma}
\begin{proof}\smartqed
The economic sense of the  matrix element $a_{ij}$ is the quantity of units of the $i$-th goods needed
for production of one unit of the  $j$-th goods.
The  assumption made relative to  matrix elements of the matrix $A$ means that  $F(x)$ is a bounded and closed polyhedron in  $R_+^n$ for every  $ x \in X $ that is convex.

Since   $F(x^0)$  is a  convex bounded and closed polyhedron in  $R_+^n,$ then there exists a finite number of extreme points\index{extreme point} $(y_1,\dots ,y_m)$ of the set $F(x^0)$ such that for any $y\in F(x^0)$ there exist $m$ numbers $\tau_i\geq 0, \ i=\overline{1,m},$ such that $ \sum\limits
_{i=1}^m\tau _i =1$ and the representation
\begin{eqnarray*}          y=\sum\limits
_{i=1}^m\tau _iy_i\end{eqnarray*}
is valid.
Substituting the expression for $y$ in
$\left\langle
y-Ay,p\right\rangle, $
 we obtain
\begin{eqnarray*}          \left\langle
y - Ay,p\right\rangle =\sum\limits _{i=1}^m \left\langle y_i - Ay_i,p\right\rangle ~\tau
_i=\psi (\tau _1 ,\dots ,\tau _m), \quad \tau =\{\tau_1, \ldots, \tau_m\} \in P_m, \end{eqnarray*}
where
\begin{eqnarray*}           P_m = \left\{\tau=\{\tau_i\}_{i=1}^m, \ \tau _i \geq
0, \ i=\overline{1,m}, \ \sum\limits _{i=1}^m\tau _i =1\right\}.\end{eqnarray*}
From the representation for  $ \left\langle
y - Ay,p\right\rangle,$
the inequality
\begin{eqnarray*} \max\limits _{y\in F(x^0)}\left\langle y- Ay,p\right\rangle \leq \max\limits
 _i\left\langle p,y_i-Ay_i\right\rangle\end{eqnarray*}
 is valid.
 Since    $\max\limits
 _i\left\langle p,y_i-Ay_i\right\rangle=\left\langle
y_{j_0}- Ay_{j_0},p\right\rangle $ for a certain  $j_0,$ then we have
\begin{eqnarray*}
\max\limits _{y\in F(x^0)}\left\langle y- Ay,p\right\rangle \geq
\left\langle y_{j_0}- Ay_{j_0},p\right\rangle=\max\limits
 _i\left\langle p,y_i-Ay_i\right\rangle.
\end{eqnarray*}
\qed
\end{proof}

\begin{theorem}\label{krt1}
The necessary and sufficient conditions for  $y_0$ to be  a nonzero extreme point of the set\begin{eqnarray*}          V_{x_0}=\{y \in R_+^n, \ Ay \leq x_0,\}, \quad x_0 \in R^m,\quad  x_0 > 0, \quad x_0  < \infty,\end{eqnarray*}
where  $A$ is a rectangular nonnegative matrix of dimensionality $n \times m$ that does not contain  zero columns is the existence of nondegenerate square minor
$A_{I,J}=|a_{ij}|_{i \in I, j \in J}, |I|=|J|, $
such that  strictly positive components of the vector $y_0,$  indices of that belong to the set  $J,$
satisfy the set of equations
\begin{eqnarray*}           \sum\limits_{j \in J}a_{ij}y_j^0=x_i^0, \quad i \in I, \end{eqnarray*}
and the rest of components of the vector  $y_0,$ the indices of that do not belong to the set $J,$ equal zero.
\end{theorem}
\begin{proof}\smartqed
Necessity. Since $y_0 \neq 0$ is an extreme point of  $V_{x_0},$   the inequalities
\begin{eqnarray*}           \sum\limits_{j=1}^ma_{ij}y_j^0<x_i^0, \quad i=\overline{1,n},\end{eqnarray*}
are not possible because  there exists a nonzero vector  $\varepsilon=\{\varepsilon_i\}_{i=1}^m$ with $\varepsilon_i \geq 0,$ $ i=\overline{1,m},$ such that
\begin{eqnarray*}           y_0 - \varepsilon \geq 0, \quad A(y_0 + \varepsilon) \leq x_0.\end{eqnarray*}
Then  $y_0=\frac{1}{2}[(y_0 + \varepsilon)+(y_0 - \varepsilon)]$  that contradicts that
$y_0$ is  an  extreme point of the set $V_{x_0}.$ So, there exists a subset $I$ of the set $N=\{1,2, \ldots, n\}$ and a subset $J$ of the set $M=\{1,2, \ldots, m\}$
that equalities
\begin{eqnarray*}           \sum\limits_{j \in J}a_{ij}y_j^0=x_i^0, \quad i \in I, \end{eqnarray*}
are valid, moreover,
 $y_j^0 >0,\  j \in J,$ \ $x_i^0 >0,\  i \in I,$
and $y_j^0 =0,\  j \in M\setminus J.$
For the rest of indices the strict inequalities
\begin{eqnarray*}           \sum\limits_{j \in J}a_{ij}y_j^0< x_i^0, \quad i \in N \setminus I, \end{eqnarray*}
are valid.
Let  the capacity of the sets   $I$ and $J$  be correspondingly   $|I|$ and $|J|.$
Assume that  $|I| < |J|.$  From this it follows that there exists a nonzero solution   $u_0=\{u_j^0\}_{j \in J}$ of the set of equations
\begin{eqnarray*}          \sum\limits_{j \in J}a_{ij}u_j^0= 0, \quad i \in I. \end{eqnarray*}
Construct two vectors   $y_1$ and $y_2$ that belong to  $V_{x_0}$ and show that the vector $y_0$ is a linear combination of that. Choose $t>0$ such that the inequalities
\begin{eqnarray*}            y_1=y_0 - tu \geq 0, \quad  y_2=y_0 + tu \geq 0, \quad Ay_1 \leq x_0, \quad Ay_2 \leq x_0,\end{eqnarray*}
are satisfied,
where $u=\{u_i\}_{i=1}^m, \ u_i=u_i^0, \ i \in J, \ u_i=0, \ i \in N \setminus J.$

It always can be done since
\begin{eqnarray*}
 \sum\limits_{j \in N}a_{ij}y_j^1=x_i^0, \quad i \in I, \quad  \sum\limits_{j \in N}a_{ij}y_j^2=x_i^0, \quad i \in I,
 \end{eqnarray*}
\begin{eqnarray*}
 \sum\limits_{j \in N}a_{ij}y_j^0<x_i^0, \quad i \in M\setminus I,
 \end{eqnarray*}
$y_1=\{y_i^1\}_{i=1}^n, \ y_2=\{y_i^2\}_{i=1}^n.$
Hence, it follows that
\begin{eqnarray*}          y_0=\frac{y_1+y_2}{2},\end{eqnarray*}
but it contradicts   that $y_0$ is an  extreme point.
Consider the case  $|I| \geq |J|.$ Let us prove that in the set $I$ there exists a subset $I_1$ that contains  precisely $|J|$ rows such that the determinant of the minor
 $A_{I_1,J}$  does not equal zero. The supposition that all minors of the order
$|J|$ have zero determinant leads to the existence of a nonzero solution of the set of  equations
\begin{eqnarray*}          \sum\limits_{j \in J}a_{ij}u_j^0= 0, \quad i \in I, \quad |I| \geq |J|. \end{eqnarray*}
Therefore, we can use the previous construction  and prove that  $y_0$ is not  extreme point.
So, there exists  a set  $I_1 \subseteq I, \ |I_1| = |J|,$ that the determinant  of $A_{I_1,J}$
does not equal zero. The necessity is proven.

Let us prove the sufficiency. Let there exist a minor  $A_{I,J}, |I|=|J|  $
such that the determinant of it does not equal zero  and the vector $y_0$
satisfies conditions
\begin{eqnarray*}          Ay_0 \leq x_0, \quad \sum\limits_{j \in J}a_{ij}y_j^0=x_i^0, \quad i \in I, \quad y_j=0, \ j \in N\setminus J. \end{eqnarray*}
Assume that $y_0$ is not an extreme point. Therefore, there exist two vectors  $y_1$ and  $y_2$ that belong to  $V_{x_0}$ and such that   $y_0=\alpha y_1 + (1- \alpha) y_2$ for a certain   $0 < \alpha <1.$  It is obvious that
$Ay_1 \leq x_0, \ Ay_2 \leq x_0,$ moreover, $y_j^1=y_j^2=0,\  j \in N \setminus J.$
It means that
\begin{eqnarray*}
   \sum\limits_{j \in J}a_{ij}y_j^1\leq x_i^0, \quad i \in I,
\end{eqnarray*}
\begin{eqnarray*}
 \sum\limits_{j \in J}a_{ij}y_j^2\leq x_i^0, \quad i \in I,
 \end{eqnarray*}
$y_1=\{y_i^1\}_{i=1}^n, \ y_2=\{y_i^2\}_{i=1}^n.$
In these inequalities  a strict inequality for a certain $i \in I$  is not possible  since if it were the case  for a certain  $ i \in I,$ then for the vector
$y_0=\alpha y_1 + (1- \alpha) y_2$
the strict inequality
\begin{eqnarray*}           \sum\limits_{j \in J}a_{ij}y_j^0 <  x_i^0, \quad i \in I, \end{eqnarray*}
should take place  for the same $i,$ that is  not possible. So, only the equalities
\begin{eqnarray*}          y_j^1=y_j^2=y_j^0, \quad j \in J,\end{eqnarray*}
 are  valid
because of the uniqueness of the solution. The sufficiency is proven.
\qed
\end{proof}
\begin{theorem}\label{algs1}
Let    a nonnegative matrix  $A$ of dimensionality  $n \times n$  do  not contain   zero  columns and give  the Leontieff technological map\index{Leontieff technological map}
\begin{eqnarray*}          F(x)=\{y \in R_+^n, \  Ay \leq x\}\end{eqnarray*}
 on a closed   bounded set
 $X=\bigcup\limits_{\alpha \in I}X_{\alpha}, $
 where  $I$ is a  nonempty set,
and
\begin{eqnarray*}           X_{\alpha}=\{ x \in R_+^n, \ x \leq x_{\alpha}\},\end{eqnarray*}
for a certain  $x_{\alpha} \in R_+^n, \ x_{\alpha} > 0.$
Then the following equalities
\begin{eqnarray*}          \sup\limits_{x \in X}\sup\limits_{y \in F(x)}\left\langle y - x, p \right\rangle= \sup\limits_{\alpha \in I}\sup\limits_{x \in X_{\alpha}}\sup\limits_{y \in F(x)}\left\langle y - x, p \right\rangle
\end{eqnarray*}
\begin{eqnarray*}
 =\sup\limits_{\alpha \in I}\sup\limits_{y \in F(x_{\alpha})}\left\langle y - Ay, p \right\rangle
=\sup\limits_{\alpha \in I}\max\limits_{1\leq j \leq m(\alpha)}\left\langle y_j^{\alpha} - Ay_j^{\alpha}, p \right\rangle \end{eqnarray*}
are valid,
where $\{y_j^{\alpha}\}_{j=1}^{m(\alpha)}$ is the set of extreme points of the set $F(x_{\alpha}).$
\end{theorem}
\begin{proof}\smartqed
Let us prove the equality
\begin{eqnarray*}          \sup\limits_{x \in X}\sup\limits_{y \in F(x)}\left\langle y - x, p \right\rangle=
\sup\limits_{\alpha \in I}\sup\limits_{x \in X_{\alpha}}\sup\limits_{y \in F(x)}\left\langle y - x, p \right\rangle.\end{eqnarray*}
Because of the convexity  down of the Leontieff technological map,
the solution $y(p,x)$ of the problem
\begin{eqnarray*}          \sup\limits_{y \in F(x)}\left\langle y - x, p \right\rangle= \left\langle y(p,x) - x, p \right\rangle \end{eqnarray*}
is such that the function  $ \left\langle y(p,x) - x, p \right\rangle  $ is a convex  up function of the argument  $x $ on every closed convex set  $X_1$ containing the set  $X$ for every fixed  $p \in \bar R_+^n.$
From the convexity  up we have  that the  function  $ \left\langle y(p,x) - x, p \right\rangle  $ is  continuous one of the argument
$x \in   X_1.$ Therefore, the contraction of the function  $ \left\langle y(p,x) - x, p \right\rangle  $ on the closed bounded set  $X$ is a continuous function of the argument  $x \in X.$  By the Weierstrass theorem, the supremum of this function is realized at  certain point
$x_0 \in X$ that gets into a certain set $X_{\alpha_0}.$ The last  proves the  needed equality since if
$\varphi(x)=\left\langle y(p,x) - x, p \right\rangle,$  then
\begin{eqnarray*}          \sup\limits_{x \in X}\varphi(x)=\varphi(x_0)=\sup\limits_{x \in X_{\alpha_0}}\varphi(x) \leq
\sup\limits_{\alpha \in I}\sup\limits_{x \in X_{\alpha}}\varphi(x).\end{eqnarray*}
However, for any  $\alpha \in I$
\begin{eqnarray*}          \sup\limits_{x \in X_{\alpha}}\varphi(x) \leq \sup\limits_{x \in X}\varphi(x).\end{eqnarray*}
Therefore,
\begin{eqnarray*}          \sup\limits_{\alpha \in I}\sup\limits_{x \in X_{\alpha}}\varphi(x) \leq \sup\limits_{x \in X}\varphi(x).\end{eqnarray*}
So,
\begin{eqnarray*}          \sup\limits_{x \in X}\varphi(x)=\sup\limits_{\alpha \in I}\sup\limits_{x \in X_{\alpha}}\varphi(x).\end{eqnarray*}
Prove the rest equalities.
Let us obtain the inequality from above. We have
\begin{eqnarray*}
  \sup\limits_{x \in X}\sup\limits_{y \in F(x)}\left\langle y - x, p \right\rangle =
\sup\limits_{\alpha \in I}\sup\limits_{x \in X_{\alpha}}\sup\limits_{y \in F(x)}\left\langle y - x, p \right\rangle
\end{eqnarray*}
\begin{eqnarray*}
  \leq \sup\limits_{\alpha \in I}\sup\limits_{x \in X_{\alpha}}\sup\limits_{y \in F(x)}\left\langle y - Ay, p \right\rangle \leq \sup\limits_{\alpha \in I}\sup\limits_{x \in X_{\alpha}}\sup\limits_{y \in F(x_{\alpha})}\left\langle y - Ay, p \right\rangle
\end{eqnarray*}
\begin{eqnarray*}
 =\sup\limits_{\alpha \in I}\sup\limits_{y \in F(x_{\alpha})}\left\langle y - Ay, p \right\rangle=
\sup\limits_{\alpha \in I}\max\limits_{1\leq j \leq m(\alpha)}\left\langle y_j^{\alpha} - Ay_j^{\alpha}, p \right\rangle.
\end{eqnarray*}
Here we used that if a vector $x \in X, \ y \in F(x),$ then the vector  $Ay \in X$ is  an expenditure vector.
The second inequality is valid since $ F(x) \subseteq  F(x _{\alpha}).$
Further, if  $y_j^{\alpha}$ is an extreme point of the set $F(x_{\alpha}),$
then since   $Ay_j^{\alpha} \leq x_{\alpha}$ such  set of inequalities
\begin{eqnarray*}          \sup\limits_{x \in X}\sup\limits_{y \in F(x)}\left\langle y - x, p \right\rangle \geq \sup\limits_{y \in F(Ay_j^{\alpha})}\left\langle y - Ay_j^{\alpha}, p \right\rangle\geq \left\langle y_j^{\alpha} - Ay_j^{\alpha}, p \right\rangle\end{eqnarray*}
 are valid.
The last inequality takes place since $y_j^{\alpha} \in F(Ay_j^{\alpha}),$
$Ay_j^{\alpha} \in X.$
Taking firstly maximum in $j$ and after that  supremum in  $\alpha,$ we obtain the inequality from below
\begin{eqnarray*}          \sup\limits_{x \in X}\sup\limits_{y \in F(x)}\left\langle y - x, p \right\rangle\geq
\sup\limits_{\alpha \in I}\max\limits_{1\leq j \leq m(\alpha)}\left\langle y_j^{\alpha} - Ay_j^{\alpha}, p \right\rangle.\end{eqnarray*}
\qed
\end{proof}

The consequence of the  previous Theorem \ref{algs1} is the next Theorem.
\begin{theorem}\label{algs2}
Let  a nonnegative matrix $A$  of the dimensionality  $n \times n$  do not contain  zero columns and give the Leontieff  technological map
\begin{eqnarray*}          F(x)=\{y \in R_+^n, \  Ay \leq x\}\end{eqnarray*}            on a bounded and closed set $X=\bigcup\limits_{i=1}^kX_i, \ k < \infty, $
where
\begin{eqnarray*}          X_i=\{x \in R_+^n, \ x \leq x_i\}\end{eqnarray*}
for a certain  $x_i \in R_+^n, \ x_i > 0.$
Then
\begin{eqnarray*}          \sup\limits_{x \in X}\sup\limits_{y \in F(x)}\left\langle y - x, p \right\rangle=
\max\limits_{1 \leq i \leq k}\max\limits_{1\leq j \leq m(i)}\left\langle y_j^i - Ay_j^i, p \right\rangle,\end{eqnarray*}
where $\{y_j^i\}_{j=1}^{m(i)}$ is  the set of extreme points of the set $F(x_i).$
\end{theorem}

\begin{theorem}\label{naj1}
Let on a bounded  set
\begin{eqnarray*}          X=\{x \in R_+^n, \  x \leq x_0, \  x_0 =\{x_i^0\}_{i=1}^n, \ x_i^0 >0, \ i=\overline{1,n}  \}\end{eqnarray*}
 the Neumann technological map
\begin{eqnarray*}          F(x)=\{y \in R_+^n,\ \mbox{exists} \  \xi  \in R_+^n, \  y=B\xi, \  A\xi \leq x \},\end{eqnarray*}
be given, where  $A, B$ are nonnegative matrices of the dimensionality  $n \times n,$
and let  the matrix  $A$ do not contain zero columns.
Then
\begin{eqnarray*}          \sup\limits_{x \in X}\sup\limits_{y \in F(x)}\left\langle y - x, p \right\rangle=
\max\limits_{1 \leq j \leq k} \left\langle[B - A]\xi_j, p \right\rangle,\end{eqnarray*}
where $\{\xi_j\}_{j=1}^{k}$ is the set of extreme points of the set
\begin{eqnarray*}          F_1(x_0)=\{\xi \in R_+^n, \ A\xi \leq x_0\} .\end{eqnarray*}
\end{theorem}

\begin{proof}\smartqed

Let us prove first equality
\begin{eqnarray*}          \sup\limits_{x \in X}\sup\limits_{y \in F(x)}\left\langle y - x, p \right\rangle=\sup\limits_{x \in X}\sup\limits_{\xi \in F_1(x)} \left\langle[B - A]\xi, p \right\rangle, \end{eqnarray*}
where
\begin{eqnarray*}          F_1(x)=\{\xi \in R_+^n, \ A\xi \leq x\} .\end{eqnarray*}
For this let us prove the inequality from above
\begin{eqnarray}          \label{naj2}
\sup\limits_{x \in X}\sup\limits_{y \in F(x)}\left\langle y - x, p \right\rangle \leq \sup\limits_{x \in X} \sup\limits_{\xi \in F_1(x)}\left\langle [B -A]\xi, p \right\rangle.
\end{eqnarray}
Let  $(x, y) \in \Gamma, $  where $ \Gamma=\{(x,y), \  x \in X, \ y \in  F(x)\}.$
Then there exists  $\xi \in R_+^n$ such that the inequality and equality  $A\xi \leq x, \ y=B\xi$ are valid. Hence,
\begin{eqnarray*}
 \left\langle y - x, p \right\rangle \leq \left\langle [B -A]\xi, p \right\rangle  \leq \sup\limits_{\xi \in F_1(x)} \left\langle[B - A]\xi, p \right\rangle
\end{eqnarray*}
\begin{eqnarray*}
 \leq  \sup\limits_{x \in X}\sup\limits_{\xi \in F_1(x)} \left\langle[B - A]\xi, p \right\rangle. \end{eqnarray*}
The right part of the last inequality does not depend on  $(x,y) \in \Gamma,$ therefore, taking supremum in
$y \in F(x)$ and after supremum in
$x \in X,$ we will obtain the inequality (\ref{naj2}).
Let us prove the contrary inequality
\begin{eqnarray}          \label{naj3}
\sup\limits_{x \in X}\sup\limits_{y \in F(x)}\left\langle y - x, p \right\rangle \geq \sup\limits_{x \in X}\sup\limits_{\xi \in F_1(x)}\left\langle [B -A]\xi, p \right\rangle.
\end{eqnarray}
Since for every  $x \in X, \  \xi \in F_1(x)$ the inequality
$A\xi \leq x$ and inclusion $A\xi \in X$
take place, then
\begin{eqnarray*}          \sup\limits_{x \in X}\sup\limits_{y \in F(x)}\left\langle y - x, p \right\rangle \geq \sup\limits_{y \in F(A\xi)}\left\langle y - A\xi, p \right\rangle \geq \left\langle [B -A]\xi, p \right\rangle.\end{eqnarray*}
Taking supremum in  $\xi \in F_1(x)$ and after in  $x \in X$  from the left and the right part  in the last inequality,  we obtain the  needed inequality from below.
Let us prove the equality
\begin{eqnarray*}          \sup\limits_{x \in X}\sup\limits_{\xi \in F_1(x)}\left\langle [B -A]\xi, p \right\rangle= \sup\limits_{\xi \in F_1(x_0)}\left\langle [B -A]\xi, p \right\rangle. \end{eqnarray*}
Really, on the set
\begin{eqnarray*}
X=\{x \in R_+^n, \  x \leq x_0, \  x_0 =\{x_i^0\}_{i=1}^n, \ x_i^0 >0, \ i=\overline{1,n}  \}
\end{eqnarray*}
\begin{eqnarray*}
  \sup\limits_{\xi \in F_1(x)}\left\langle [B -A]\xi, p \right\rangle \leq \sup\limits_{\xi \in F_1(x_0)}\left\langle [B -A]\xi, p \right\rangle,
  \end{eqnarray*}
since if     $\xi \in F_1(x),$ then  $\xi \in F_1(x_0).$
Therefore,
\begin{eqnarray*}          \sup\limits_{x \in X}\sup\limits_{\xi \in F_1(x)}\left\langle [B -A]\xi, p \right\rangle \leq \sup\limits_{\xi \in F_1(x_0)}\left\langle [B -A]\xi, p \right\rangle.
\end{eqnarray*}
On the other hand,
\begin{eqnarray*}
\sup\limits_{x \in X}\sup\limits_{\xi \in F_1(x)}\left\langle [B -A]\xi, p \right\rangle \geq \sup\limits_{\xi \in F_1(x_0)}\left\langle [B -A]\xi, p \right\rangle
\end{eqnarray*}
that proves the needed formula.
Let now   $\{\xi_j\}_{j=1}^{k}$ be the set of extreme points of the set
\begin{eqnarray*}
F_1(x_0)=\{\xi \in R_+^n, \ A\xi \leq x_0\} .
\end{eqnarray*}
Then
\begin{eqnarray*}
\sup\limits_{\xi \in F_1(x_0)}\left\langle [B -A]\xi, p \right\rangle= \max\limits_{1 \leq j \leq k}\left\langle [B -A]\xi_j, p \right\rangle.
\end{eqnarray*}
\qed
\end{proof}

The obvious generalization  of the previous Theorem is the next Theorem.
\begin{theorem}\label{naj4}
Let the Neumann  technological map\index{Neumann  technological map}
\begin{eqnarray*}          F(x)=\{y \in R_+^n,\ \mbox{there exists} \  \xi  \in R_+^n, \  y=B\xi, \  A\xi \leq x \} \end{eqnarray*}
be given on a  bounded closed set  $X=\bigcup\limits_{\alpha \in I}X_{\alpha},$ where
\begin{eqnarray*}          X_{\alpha}=\{x \in R_+^n, \  x \leq x_{\alpha}, \ x_{\alpha} =\{x_i^{\alpha}\}_{i=1}^n, \ x_i^{\alpha} >0, \ i=\overline{1,n}  \},\end{eqnarray*}
and $A, B$ are nonnegative matrices of the dimensionality   $n \times n,$
 the matrix $A$ do not contain  zero   columns.
Then
\begin{eqnarray*}          \sup\limits_{x \in X}\sup\limits_{y \in F(x)}\left\langle y - x, p \right\rangle=
\sup\limits_{\alpha \in I}\max\limits_{1 \leq j \leq k(\alpha)} \left\langle[B - A]\xi_j^{\alpha}, p \right\rangle,\end{eqnarray*}
where $\{\xi_j^{\alpha}\}_{j=1}^{k(\alpha)}$ is the set of  extreme points of the set
\begin{eqnarray*}          F_1(x_{\alpha})=\{\xi \in R_+^n, \ A\xi \leq x_{\alpha}\} .\end{eqnarray*}
\end{theorem}

\begin{proof}\smartqed

Proof follows from the equality
\begin{eqnarray*}          \sup\limits_{x \in X}\sup\limits_{y \in F(x)}\left\langle y - x, p \right\rangle=
\sup\limits_{\alpha \in I}\sup\limits_{x \in X_{\alpha}}\sup\limits_{y \in F(x)}\left\langle y - x, p \right\rangle \end{eqnarray*}
that is also  valid  for the Neumann technological map since it is convex down  and from  the  Theorem  \ref{naj1} proved above.
\qed
\end{proof}

\begin{definition}   A set of points    $(x_1, \dots ,x_k),$  where
$x_i \in R_+^n,\ i=\overline{1,k},$ generates a set
$X,$  if the set $X$ coincides with linear convex span constructed by the set of points
$(x_1,\dots ,x_k),$ that is,
\begin{eqnarray*}          X=\left\{x \in R_+^n, \  x=\sum\limits_{i=1}^k\alpha_ix_i, \ \alpha \in P_1\right\},\end{eqnarray*}
where
\begin{eqnarray*}          P_1=\left\{\alpha=\{\alpha_i\}_{i=1}^k \in  R_+^k, \  \sum\limits_{i=1}^k\alpha_i=1\right\}. \end{eqnarray*}
We call  the set $X$ the linear convex span\index{linear convex span} generated by the set of points $(x_1,\dots ,x_k).$
\end{definition}
\begin{note} The same set $X$ can be generated by various  sets of points. If to the set of points that generates  $X$ to add any set of points that belongs to  $X,$ then they  will also generate the set  $X.$
But there always exists the  minimal number of points from the set of points
$(x_1,\dots ,x_k)$ that generates  $X.$ These points are called extreme points.\index{extreme points}
\end{note}
Let  $X \subset R_+^n$ be a convex closed bounded  polyhedron\index{convex closed bounded  polyhedron} and the set of points   $(x_1,\dots ,x_k)$ generates it.
Further, let $Y_i, \ i=\overline{1,k},$ be a convex closed bounded polyhedrons from $R_+^n$ and $\{y_j^{(i)}\}_{j=1}^{m(i)}$ be a set of points that generates the polyhedron $Y_i, \ i=\overline{1,k}.$ Let us give a  technological map
$F(x)$ on  $X$ by the formula
\begin{eqnarray}          \label{sl6}
F(x)=\bigcup\limits_{\alpha \in \Delta(x)}\sum\limits_{i=1}^k\alpha_i F_1(x_i),
\quad  x \in X,
\end{eqnarray}
where  $F_1(x_i)=Y_i,$ and by $\Delta(x)$ we denote the set
\begin{eqnarray*}           \Delta(x)=\left\{\alpha=\{\alpha_i\}_{i=1}^k \in P_1, \
x=\sum\limits_{i=1}^k\alpha_ix_i \right\}. \end{eqnarray*}
The set $\sum\limits_{i=1}^k\alpha_i F_1(x_i) $ is the set of all points of the kind
$\sum\limits_{i=1}^k\alpha_i y_i,$  where the point $ y_i $ runs over the set
$ F_1(x_i), \ i=\overline{1,k}.$
\begin{lemma}\label{start2} A technological map defined by the formula  (\ref{sl6})
takes  values in the set of closed bounded  and convex sets, is convex down,  and Kakutani continuous from above, that is, it belongs to  the CTM class in a wide sense and is convex down.
\end{lemma}
\begin{proof}\smartqed          Let us prove that $F(x)$ is a convex and closed set for every $x \in X.$
First prove the convexity of $ F(x).$
Let  $y_1$ and  $y_2$ belong  to $F(x).$ It means that there  exist
$\alpha^{'}=\{\alpha_i^{'}\}_{i=1}^k, \
\alpha^{''}=\{\alpha_i^{''}\}_{i=1}^k \in \Delta(x)$  such that
$x=\sum\limits_{i=1}^k\alpha_i^{'}x_i,$
$x=\sum\limits_{i=1}^k\alpha_i^{''}x_i,$ moreover, for  $y_1$ and  $y_2$
the representations
\begin{eqnarray*}
 y_1=\sum\limits_{i=1}^k\alpha_i^{'}
\sum\limits_{j=1}^{m(i)} \gamma_j^{i,1}y_j^{(i)}, \quad
\gamma_j^{i,1} \geq 0,
\quad \sum\limits_{j=1}^{m(i)} \gamma_j^{i,1} =1,
\end{eqnarray*}
\begin{eqnarray*}
 y_2=\sum\limits_{i=1}^k\alpha_i^{''}
\sum\limits_{j=1}^{m(i)} \gamma_j^{i,2}y_j^{(i)}, \quad
 \gamma_j^{i,2} \geq 0,   \quad
\sum\limits_{j=1}^{m(i)} \gamma_j^{i,2} =1,
\end{eqnarray*}
are valid.
Hence,
\begin{eqnarray*}          \alpha y_1 +(1 - \alpha)y_2=
\sum\limits_{i=1}^k\alpha_i^{'''}
\sum\limits_{j=1}^{m(i)} \gamma_j^{i,3}y_j^{(i)}, \quad 0 < \alpha < 1,\end{eqnarray*}
where
 \begin{eqnarray*}           \gamma_j^{i,3}=\frac{\alpha\alpha_i^{'} \gamma_j^{i,1}+
(1- \alpha)\alpha_i^{''} \gamma_j^{i,2}}{ \alpha\alpha_i^{'}+
(1- \alpha)\alpha_i^{''}}, \quad   \alpha_i^{'''}=\alpha\alpha_i^{'}+
(1- \alpha)\alpha_i^{''}, \end{eqnarray*}
if only $\alpha_i^{'}$ or $\alpha_i^{''}$ do not equal zero and
$\alpha_i^{'''}=0,$ if $\alpha_i^{'}=0$ and $\alpha_i^{''}=0.$
Owing to   \begin{eqnarray*}          x=\sum\limits_{i=1}^k\alpha_i^{'''}x_i, \quad
\sum\limits_{i=1}^k\alpha_i^{'''}=1, \quad
\sum\limits_{j=1}^{m(i)} \gamma_j^{i,3}=1, \quad  \alpha_i^{'''} \geq 0,
\quad  \gamma_j^{i,3} \geq 0,\end{eqnarray*}
we have   $\alpha y_1 +(1 - \alpha)y_2  \in F(x)$
for  any $\alpha \in (0,1).$

Prove the  closure of  $F(x).$ Let  a sequence  $y_n \in F(x)$ and,  furthermore,
$y_n \to y_0.$ Prove that $y_0 \in F(x).$
From the  fact that  $y_n \in F(x)$ it follows that there exist sequences
$\alpha^n=\{\alpha_i^n\}_{i=1}^k, \in P_1, \
\gamma^{i,n}=\{\gamma_j^{i,n}\}_{j=1}^{m(i)},$  that satisfy conditions
\begin{eqnarray*}          \gamma_j^{i,n} \geq 0, \quad j=\overline{1,m(i)}, \quad
\sum\limits_{j=1}^{m(i)}\gamma_j^{i,n} =1, \quad  i=\overline{1,k}, \quad
n=1, 2, 3, \ldots,\end{eqnarray*}
and for  $y_n \in F(x)$  and $x \in X$ the representations
\begin{eqnarray*}           x=\sum\limits_{i=1}^k\alpha_i^nx_i, \quad
 y_n=\sum\limits_{i=1}^k\alpha_i^n
\sum\limits_{j=1}^{m(i)}\gamma_j^{i,n}y_j^{(i)}\end{eqnarray*}
are valid.
From the compactness of  the  considered sequences  the  existence of subsequence
  $n_m$  follows such that
$\alpha_i^{n_m} \to \alpha_i^0,  \
\gamma_j^{i,n_m} \to \gamma_j^{i,0},\ j=\overline{1,m(i)},$ $ \ i=\overline{1,k}, $
as  $ m \to \infty,$
besides for  the limit point  $y_0$ the representation
\begin{eqnarray*}          y_0=\sum\limits_{i=1}^k\alpha_i^0
\sum\limits_{j=1}^{m(i)}\gamma_j^{i,0}y_j^{(i)}, \quad
\gamma_j^{i,0} \geq 0, \quad j=\overline{1,m(i)},  \quad
\sum\limits_{j=1}^{m(i)}\gamma_j^{i,0} =1, \quad  i=\overline{1,k}, \end{eqnarray*}
is valid, where
\begin{eqnarray*}            x=
\sum\limits_{i=1}^k\alpha_i^0x_i, \quad \alpha^0=\{\alpha_i^0\}_{i=1}^k \in P_1.\end{eqnarray*}
The latter  means that  $y_0 \in F(x).$
The closure of   $F(x)$ follows from that
$y_0$ is an arbitrary limit point of the sequence $y_n.$ The boundedness of $F(x)$
is obvious.

Prove the Kakutani continuity from above\index{Kakutani continuity from above} of   $F(x).$
Let a sequence  $y_n \in F(x_n),$  and  $x_n \to x_0,$
$y_n \to y_0.$ Prove that $y_0 \in F(x_0).$
From the fact that  $y_n \in F(x_n),$  the existence of sequences $\alpha^n=\{\alpha_i^n\}_{i=1}^k \in P_1$ and
$\gamma^{i,n}=\{\gamma_j^{i,n}\}_{j=1}^{m(i)}, \ i=\overline{1,k}, $  follows that satisfy conditions
\begin{eqnarray*}          \gamma_j^{i,n} \geq 0, \quad j=\overline{1,m(i)}, \quad
\sum\limits_{j=1}^{m(i)}\gamma_j^{i,n} =1, \quad  i=\overline{1,k},
\end{eqnarray*}
and for $y_n \in F(x_n)$  and $x_n \in X$ the representations
\begin{eqnarray*}           x_n=\sum\limits_{i=1}^k\alpha_i^nx_i, \quad
 y_n=\sum\limits_{i=1}^k\alpha_i^n
\sum\limits_{j=1}^{m(i)}\gamma_j^{i,n}y_j^{(i)}\end{eqnarray*}
are valid.
From the compactness of the  considered sequences  the existence of subsequence
  $n_m$  follows such that
$\alpha_i^{n_m} \to \alpha_i^0,  \
\gamma_j^{i,n_m} \to \gamma_j^{i,0},\ j=\overline{1,m(i)}, $ $\ i=\overline{1,k}, $
as $ m \to \infty,$
and for limit points $x_0$ and $y_0$ the representations
\begin{eqnarray*}          y_0=\sum\limits_{i=1}^k\alpha_i^0
\sum\limits_{j=1}^{m(i)}\gamma_j^{i,0}y_j^{(i)}, \quad
\gamma_j^{i,0} \geq 0, \quad j=\overline{1,m(i)},  \quad
\sum\limits_{j=1}^{m(i)}\gamma_j^{i,0} =1, \quad  i=\overline{1,k},
\end{eqnarray*}
\begin{eqnarray*}
x_0=
\sum\limits_{i=1}^k\alpha_i^0x_i, \quad \alpha^0=\{\alpha_i^0\}_{i=1}^k \in P_1,
\end{eqnarray*}
are valid.
The last  means that  $y_0 \in F(x_0).$

Prove that $F(x)$ is a convex down technological map.
Let
\begin{eqnarray*}
x_1=\sum\limits_{i=1}^k\alpha_i^{'}x_i,\quad
x_2=\sum\limits_{i=1}^k\alpha_i^{''}x_i,
\quad \alpha^{'}=\{\alpha_i^{'}\}_{i=1}^k \in P_1,
\quad \alpha^{''}=\{\alpha_i^{''}\}_{i=1}^k \in P_1,
\end{eqnarray*}
 \begin{eqnarray*}
 y_1=\sum\limits_{i=1}^k\alpha_i^{'}
\sum\limits_{j=1}^{m(i)} \gamma_j^{i,1}y_j^{(i)}, \quad
\gamma_j^{i,1} \geq 0,
\quad \sum\limits_{j=1}^{m(i)} \gamma_j^{i,1} =1,
\end{eqnarray*}
\begin{eqnarray*}
  y_2=\sum\limits_{i=1}^k\alpha_i^{''}
\sum\limits_{j=1}^{m(i)} \gamma_j^{i,2}y_j^{(i)}, \quad
\gamma_j^{i,2} \geq 0, \quad
\sum\limits_{j=1}^{m(i)} \gamma_j^{i,2} =1.
 \end{eqnarray*}
Then $y_1 \in F(x_1), \  y_2 \in F(x_2) $ and
\begin{eqnarray*}          \alpha y_1 +(1 - \alpha)y_2=
\sum\limits_{i=1}^k\alpha_i^{'''}
\sum\limits_{j=1}^{m(i)} \gamma_j^{i,3}y_j^{(i)}, \quad 0 < \alpha < 1,\end{eqnarray*}
where
 \begin{eqnarray*}          \gamma_j^{i,3}=\frac{\alpha\alpha_i^{'} \gamma_j^{i,1}+
(1- \alpha)\alpha_i^{''} \gamma_j^{i,2}}{ \alpha\alpha_i^{'}+
(1- \alpha)\alpha_i^{''}}, \quad  \alpha_i^{'''}=\alpha\alpha_i^{'}+
(1- \alpha)\alpha_i^{''},  \end{eqnarray*}
if only $\alpha_i^{'}$  or $\alpha_i^{''}$ do not equal zero and
$\alpha_i^{'''}=0,$ if  $\alpha_i^{'}=\alpha_i^{''}=0.$

It is obvious that
\begin{eqnarray*}          \sum\limits_{i=1}^k\alpha_i^{'''}=1, \quad
\sum\limits_{j=1}^{m(i)} \gamma_j^{i,3} =1, \quad
\alpha x_1+(1- \alpha)x_2=\sum\limits_{i=1}^k\alpha_i^{'''}x_i.\end{eqnarray*}
The last means that  $\alpha y_1+(1 - \alpha)y_2 \in
F(\alpha x_1 + (1 - \alpha)x_2),$ or the same
\begin{eqnarray*} \alpha F(x_1) +(1 - \alpha) F(x_2) \subseteq  F(\alpha x_1 + (1 - \alpha)x_2). \end{eqnarray*}
\qed
\end{proof}
\begin{lemma}\label{s1}
Let   $X \subset R_+^n$ be a convex bounded closed polyhedron and the set of points
  $(x_1,\dots ,x_k)$ generate  it.
Further, let  $Y_i, \ i=\overline{1,k},$ be  convex bounded closed polyhedrons from
$R_+^n,$ and a technological map  $F(x)$ given on  $X$
be defined by the formula  (\ref{sl6}).
If the set of points  $\left\{ y_1^{(i)},\dots ,y_{m(i)}^{(i)}\right\} $  generates the set  $Y_i, \ i=\overline{1,k},$
then
\begin{eqnarray*}          \max\limits _{x\in X}\max\limits _{y\in
F(x)}\left\langle y-x,p\right\rangle = \max\limits _{1\leq i\leq k}
\max\limits _{1\leq s\leq m(i)}\left\langle
p,y_s^{(i)}-x_i\right\rangle.\end{eqnarray*}
\end{lemma}
\begin{proof}\smartqed          Any point  $x\in X$ can be represented in the form
\begin{eqnarray*}          x=\sum\limits_{i=1}^k\alpha_ix_i,\end{eqnarray*}
and any point of the  technological map
$F(x)$  can be represented in the form
\begin{eqnarray}          \label{h1l26}
y=\sum\limits
 _{i=1}^k\alpha _i~\sum\limits _{j=1}^{m(i)}\gamma _j^i ~
y_j^{(i)},
\end{eqnarray}
\begin{eqnarray*}          \sum\limits _{i=1}^k\alpha _i
=1,\quad \sum\limits _{j=1}^{m(i)}\gamma _j^i=1,\quad \alpha _i
\geq 0,\quad \gamma _j^i\geq 0, \quad j=\overline{1,m(i)},\quad i=\overline{1,k}.\end{eqnarray*}
Substituting the representations for  $x$ and $y$ in the expression  $\psi(y,x)=\left\langle y-x,p\right\rangle,$
 we obtain
\begin{eqnarray*}          \left\langle y-x,p\right\rangle=\sum\limits _{i=1}^k\alpha _i
\sum\limits _{j=1}^{m(i)}\gamma _j^i\left\langle
y_j^{(i)}-x_i,p\right\rangle=\phi(\alpha,\gamma).\end{eqnarray*}
The estimation
\begin{eqnarray*}          \left\langle y-x,p\right\rangle \leq \beta\end{eqnarray*}
is valid, where
\begin{eqnarray*}          \beta=\max\limits _{1\leq i\leq k}
\max\limits _{1\leq s\leq m(i)}\left\langle
p,y_s^{(i)}-x_i\right\rangle.\end{eqnarray*}
From here,
\begin{eqnarray*}          \max\limits _{x\in X}\max\limits _{y\in
F(x)}\left\langle y-x,p\right\rangle \leq \max\limits _{1\leq i\leq k}
\max\limits _{1\leq s\leq m(i)}\left\langle
p,y_s^{(i)}-x_i\right\rangle.\end{eqnarray*}
Let us prove the contrary inequality.

 There exist  $i_0$ and $s_0$ such that  $\beta=\left\langle y_{s_0}^{(i_0)}-x_{i_0},p\right\rangle.$
Choose  \begin{eqnarray*}          \alpha_{i_0}=1, \quad \alpha_i=0, \quad  i\neq i_0, \quad \gamma_{s_0}^{i_0}=1,\quad \gamma_{j}^{i}=0, \quad j\neq s_0, \quad i\neq i_0.\end{eqnarray*}   It is evident that $  y_{s_0}^{(i_0)} \in Y_{i_{0}}, $ therefore  $  y_{s_0}^{i_0} \in F(x_{i_0}). $  So, we have
\begin{eqnarray*}          \max\limits _{x\in X}\max\limits _{y\in
F(x)}\left\langle y-x,p\right\rangle \geq   \max\limits _{y\in
F(x_{i_0})}\left\langle y-x_{i_0},p\right\rangle \geq   \left\langle
y_{s_0}^{(i_0)}-x_{i_0}, p\right\rangle
\end{eqnarray*}
\begin{eqnarray*}
=  \max\limits _{1\leq i\leq k}
\max\limits _{1\leq s\leq m(i)}\left\langle
p,y_s^{(i)}-x_i\right\rangle.
\end{eqnarray*}
The last proves the Lemma \ref{s1}.
\qed
\end{proof}
\begin{lemma}\label{ksl1}
Let $X \subset R_+^n$  be a compact set, a  technological map  $F(x)$ maps $X$ into bounded closed subsets of  $R_+^n.$  There exists a compact  $Y \subset R_+^n$ such that
\begin{eqnarray*}           F(x)\subseteq Y, \quad x\in X.\end{eqnarray*}
Then
\begin{eqnarray*}          \varphi (p)=
\sup\limits _{x\in X}\sup\limits _{y\in
F(x)} \left\langle y-x,p\right\rangle,\quad p\in R_+^n ,\end{eqnarray*}
is positively homogeneous subadditive continuous function on
$R^n_+.$
\end{lemma}
\begin{proof}\smartqed          From the definition of
$\varphi(p),$ the  positive homogeneity of
\begin{eqnarray*}          \varphi(\lambda p)=\lambda\varphi(p),\quad \lambda \geq 0,\end{eqnarray*}
follows.
Subadditive property is a consequence of inequalities
\begin{eqnarray*}           \varphi(p_1+p_2)=\sup\limits_{x \in X}\sup\limits_{y \in F(x)}(
\left\langle y-x,p_1\right\rangle+\left\langle y-x,p_2\right\rangle)
\end{eqnarray*}
\begin{eqnarray*}
 \leq \sup\limits_{x \in X}\sup\limits_{y \in F(x)}
\left\langle y-x,p_1\right\rangle+\sup\limits_{x \in X}\sup\limits_{y \in F(x)}
\left\langle y-x,p_2\right\rangle =\varphi(p_1)+\varphi(p_2).
\end{eqnarray*}
 From subadditive property and from positive homogeneity  the  convexity  down of $\varphi(p)$  follows:
\begin{eqnarray*}          \varphi(\alpha p_1+(1-\alpha)p_2)\leq \alpha\varphi(p_1)+
(1-\alpha)\varphi(p_2).\end{eqnarray*}
Prove the continuity of
$\varphi(p).$  First of all, $\varphi(0)=0.$ Further,
\begin{eqnarray*}          \varphi(p_1)\leq \varphi_1(\pi(p_1,p_{0}))+
\varphi(p_{0}), \end{eqnarray*}
or
\begin{eqnarray*}          \varphi(p_1)- \varphi(p_{0}) \leq
\varphi_1(\pi(p_1,p_{0})),\end{eqnarray*}           where
\begin{eqnarray*}          \varphi_1(p)=\sup\limits_{x \in
X}\sup\limits_{y \in F(x)} \left\langle y+x,p\right\rangle,\end{eqnarray*}
and the vector
$\pi(p_1,p_{0})$ has the form
$\pi(p_1,p_{0})=\{|p_i^1-p_i^{0}|\}_{i=1}^n,$ where
$p_1=\{p_i^1\}_{i=1}^n,$ \
$p_{0}=\{p_i^{0}\}_{i=1}^n.$
Analogously,  \begin{eqnarray*}          \varphi(p_{0})- \varphi(p_1) \leq \varphi_1(\pi(
p_1,p_{0})).\end{eqnarray*}
Therefore,
\begin{eqnarray*}          |\varphi(p_{0})- \varphi(p_1)| \leq
\varphi_1(\pi(p_1,p_{0})).\end{eqnarray*}           However,
\begin{eqnarray*}          \varphi_1(\pi(p_1,p_{0})) \leq |p_1-p_{0}|
\sup\limits_{x \in X}\sup\limits_{y \in F(x)}
\left\langle y+x,e\right\rangle,\end{eqnarray*}
where $e=(1,\ldots,1),$ \
$|p_1-p_{0}|=\max\limits_{i}|p_i^1-p_i^{0}|.$
As a consequence of finiteness of $\sup\limits_{x \in X}\sup\limits_{y \in F(x)}
\left\langle y+x,e\right\rangle$ we obtain the proof of the statement.
\qed
\end{proof}
Hereinafter, we are only interested in the restriction of  $\varphi(p)$
on the simplex
\begin{eqnarray*}          P=\left\{p=(p_1, \ldots,p_n) \in R_+^n, \
\sum\limits_{i=1}^np_i=1\right\},\end{eqnarray*}           due to  positive homogeneity of
$\varphi(p).$
\begin{definition} We call a productive process   $(x(p),y(p))$ of  technological map  $F(x)$ given on  $X$
 optimal one for the price vector  $p$  if
$x(p) \in X, \ y(p) \in F(x(p))$ and the equality \begin{eqnarray*}          \left\langle
y(p)-x(p),p \right\rangle=\varphi(p)\end{eqnarray*}
is valid, where \begin{eqnarray*}          \varphi (p)= \sup\limits
_{x\in X}\sup\limits _{y\in F(x)} \left\langle y-x,p\right\rangle,\quad p\in
\bar R_+^n.  \end{eqnarray*}
\end{definition}
In the next Lemma, we give the sufficient conditions for a technological map of the firm  under the realization of that there exists a continuous strategy of  firm behavior which is arbitrary close to  optimal one. Further, this Lemma will be generalized onto a wide class of  technological maps. The need in such assertion exists at least  because  the optimal behavior strategies  for a quite wide class of technological maps   are not continuous. In further construction, this result will play the very important role.

\begin{lemma}\label{s2} \cite{55, 71, 92, 106} Let points  $x_i,$  $i=\overline{1,k},$ generate convex bounded closed polyhedron  $X$  and
    $Y_i, \ i=\overline{1,k},$ be convex bounded closed polyhedrons from
$R_+^n$  generated, correspondingly, by points   $~\{ y_1^{(i)},\dots ,y_{m(i)}^{(i)}\}, \  i=\overline{1,k}.$  If  technological map  $F(x)$  given on  $X$ by the formula (\ref{sl6}),
then for every sufficiently small $\delta>0$  there exists a firm behavior strategy    $(x^0 (p),y^0 (p)),$  $y^0
(p)\in F(x^0 (p)),$
where  $x^0 (p)$ is an input vector and $y^0 (p)$ is an output vector such that
$(x^0 (p),y^0 (p))$ is a continuous map  of
\begin{eqnarray*}          P=\left\{ p=(p_1\dots p_n)\in
R_+^n,\ \sum\limits _{i=1}^n p_i=1\right\} \end{eqnarray*}
into $R_+^{2n}$
and, furthermore,
\begin{eqnarray*}          \sup\limits _{p\in P}\left|\varphi (p)-\left\langle
y^0 (p)- x^0 (p),p\right\rangle \right|<\delta .\end{eqnarray*}
\end{lemma}
\begin{proof}\smartqed
The simplex  $P$ is a complete metrical space with the Euclidean metric $\rho(p_1,p_2)=|p_1 - p_2|,\ p_1,\  p_2 \in P. $
In accordance with the Lemma  \ref{s1}
\begin{eqnarray}          \label{gst1}
\varphi (p)= \max\limits
_{1 \leq i \leq k} \max\limits _{1\leq s\leq m(i)}\left\langle
p,y_s^{(i)}-x_i\right\rangle .
\end{eqnarray}
Further, without loss of generality, we assume that in the formula  (\ref{gst1}) all vectors  $ y_s^{(i)}-x_i, \ 1 \leq i \leq k, \ 1 \leq s\leq m(i), $ are different because if it is not the case,  then it is necessary to leave only different vectors from this set of vectors  throwing out those vectors that are more than two and leaving only one vector from those  that coincide.
By such  set of vectors,  the function $\varphi (p) $ is determined unambiguously.
If there exist  $s$ and  $i$ such that
for all $p\in P$
\begin{eqnarray}          \label{h1l27}
\left\langle
p,y_s^{(i)}-x_i \right\rangle > \left\langle p,y_n^{(j)}-x_j \right\rangle , \quad
\forall n\neq s,\quad \forall j \neq i,
\end{eqnarray}
then the  optimal output vector  for every price vector  $p\in P$
is the vector $y_s^{(i)}$  and corresponding to it  the input vector  is
$x_i,$ where  $y_s^{(i)}$ is one of    generating points of the set  $F(x_i).$  In this case, the Lemma is proved since the optimal productive process $(x_i,y_s^{(i)})$ is continuous on  $P.$ Now, suppose that the inequalities  (\ref{h1l27})
are valid for certain  $p\in P,$ then for this price vector the optimal output vector is  $y_s^{(i)},$ and  corresponding to it the input vector is
$x_i.$

Let us  make a note that  we will use after. Suppose that a  price vector
$p$ satisfies the equalities
\begin{eqnarray*}          \left\langle p,y_{k_1}^{(j_1)}-x_{j_1}\right\rangle =
  \left\langle p,y_{k_2}^{(j_2)}-x_{j_2}\right\rangle =\dots =
  \left\langle p,y_{k_t}^{(j_t)}-x_{j_t}\right\rangle, \end{eqnarray*}
and for the rest  $s$ and  $i$ it satisfies inequalities
\begin{eqnarray*}          \left\langle p,y_{k_1}^{(j_1)} -x_{j_1}\right\rangle > \left\langle
  p,y_s^{(i)}-x_i\right\rangle, \quad s \neq k_m, \quad i \neq j_m, \quad m=\overline{1,t}.\end{eqnarray*}
 From this, it follows that the productive processes
$\left(x_{j_s},y_{k_s}^{(j_s)}\right),\ s=\overline
{1,t},$
are optimal ones for the  price vector  $p.$
Let  $\chi _i(p),$ $~i=\overline {1,t},$ be
the set of positive numbers such that
\begin{eqnarray*}          \sum\limits _{i=1}^t \chi _i(p)=1,\end{eqnarray*}
then the  productive process
\begin{eqnarray*}          \left( \sum\limits _{i=1}^t \chi _i (p)~x_{j_i},
\sum\limits _{i=1}^t \chi _i (p)~y_{k_i}^{(j_i)}\right) \end{eqnarray*}
is also optimal.

Continue the proof of the Lemma.
Let a vector  $\bar p=\{\bar p_1, \ldots, \bar p_n\} \in P $
 be such that  $\bar p_i> 0, \ i=\overline{1,n}.$
In the formula  (\ref{gst1}), the set of vectors
\begin{eqnarray*}          y_s^{(i)} -x_i, \quad   1 \leq i \leq  k, \quad  1 \leq s \leq m(i),\end{eqnarray*}
that determine the function $\varphi(p)$ is such that all these vectors are different.

Two cases are possible:\\
1) there exists a pair $\left(x_{i_s},y_{k_s}^{(i_s)}\right)$ such that  inequalities
 \begin{eqnarray*}          \left\langle y_{k_s}^{(i_s)} - x_{i_s},\bar p \right\rangle  > \left\langle y_{j}^{(i)} - x_i,\bar p \right\rangle, \quad  \forall j \neq k_s, \quad  \forall  i\neq i_s, \end{eqnarray*}
 are satisfied;\\
2) there exists the set of pairs  $(x_{i_1}, y_{k_1}^{(i_1)}), \ldots, (x_{i_t}, y_{k_t}^{(i_t)})$ such that  equalities
\begin{eqnarray*}          \left\langle \bar p,y_{k_1}^{(i_1)}-x_{i_1}\right\rangle =
  \left\langle \bar p,y_{k_2}^{(i_2)}-x_{i_2}\right\rangle =\dots =
  \left\langle \bar p,y_{k_t}^{(i_t)}-x_{i_t}\right\rangle \end{eqnarray*}
  are satisfied
and for the rest  $s$ and $i$ the inequalities
\begin{eqnarray*}          \left\langle \bar p,y_{k_1}^{(i_1)} -x_{i_1}\right\rangle > \left\langle
 \bar p,y_s^{(i)}-x_i\right\rangle, \quad s \neq k_v, \quad  i \neq  i_v, \quad   v=\overline{1,t}, \end{eqnarray*}
are valid.
In the first case,  productive process  $(x_{i_s}, y_{k_s}^{(i_s)})$ is optimal for the  price  vector $\bar p.$ Define an open set
\begin{eqnarray*}          A_{i_s,k_s}=\left\{ p \in P,  \ \left\langle
p,y_{k_s}^{(i_s)}-x_{i_s}\right\rangle > \left\langle
p,y_{j}^{(i)}-x_{i}\right\rangle, \ \forall i \neq i_s, \ \forall j \neq k_s \right\}.\end{eqnarray*}
This set is not empty since the vector  $\bar p$ belongs to it.

In the  second case,  productive processes
$\left(x_{i_s},y_{k_s}^{(i_s)}\right),\ s=\overline
{1,t},$
are optimal for the price vector  $\bar p.$  Show that there exists a productive process $\left(x_{i_v}, y_{k_v}^{(i_v)}\right),  \ v \in [1, \ldots, t],$ and  a vector
$p$ with strictly positive components that is arbitrarily close  to the vector
$\bar p$  and such that
\begin{eqnarray*}          p \in A_{i_v,k_v}=\left\{ p \in P,  \ \left\langle
p,y_{k_v}^{(i_v)}-x_{i_v}\right\rangle > \left\langle
p,y_{j}^{(i)}-x_{i}\right\rangle, \ \forall i \neq i_v, \ \forall j \neq k_v \right\}.\end{eqnarray*}

The last means that the vector $\bar p \in \bar A_{i_v,k_v}, $
where  $\bar A_{i_v,k_v}$ is the closure of the open set $A_{i_v,k_v}.$
Let us put  $p=\bar p +w\left(y_{k_v}^{(i_v)}-x_{i_v}\right)$
and determine the pair  $\left(x_{i_v}, y_{k_v}^{(i_v)}\right)$ from the condition
\begin{eqnarray*}           \max_{u \in [1,\ldots t]}\left\langle y_{k_u}^{(i_u)}-x_{i_u}, y_{k_u}^{(i_u)}-x_{i_u} \right\rangle
=\left\langle y_{k_v}^{(i_v)}-x_{i_v}, y_{k_v}^{(i_v)}-x_{i_v} \right\rangle.\end{eqnarray*}
The set of inequalities
\begin{eqnarray*}          \left\langle p,y_{k_v}^{(i_v)}-x_{i_v}\right\rangle > \left\langle
p,y_{j}^{(i)}-x_{i}\right\rangle, \quad  i=i_m, \quad  j=k_m, \quad m=\overline{1,t}, \quad  m \neq v, \end{eqnarray*}
is satisfied  for all $ w >0$ if the set of inequalities
\begin{eqnarray*}          \left\langle y_{k_v}^{(i_v)}-x_{i_v}, y_{k_v}^{(i_v)}-x_{i_v} \right\rangle
 > \left\langle y_{k_v}^{(i_v)}-x_{i_v}, y_{k_r}^{(i_r)}-x_{i_r} \right\rangle, \quad r \neq v, \quad
 r \in [1, \ldots, t],\end{eqnarray*}
 is valid.
The last set of inequalities  is always satisfied if all the  vectors
$y_{k_u}^{(i_u)}-x_{i_u},\ u \in [1, \ldots, t],$ are different that takes place.
Now choose a positive number $w $ such small that the components of the vector $p$ are strictly positive and the set of inequalities
\begin{eqnarray*}          \left\langle
p,y_{k_v}^{(i_v)}-x_{i_v}\right\rangle > \left\langle
p,y_{j}^{(i)}-x_{i}\right\rangle, \quad \forall i \neq i_v, \quad  \forall j \neq k_v. \end{eqnarray*}
is satisfied.
We can reach it because of  continuity of functions
$\left\langle p,y_{j}^{(i)}-x_{i}\right\rangle  $ of the  vector  $p.$

From the last, it follows that the vector $\bar p$ is the limit point of the set $A_{i_v,k_v}.$
Thereby, it is proved that there exists the finite set of the  closed sets  $\bar A_{i_s,k_s}$
 constructed as the closure of the open sets $A_{i_s,k_s}$ that cover the set
\begin{eqnarray*}          \tilde P=\left\{ p= \{p_1, \ldots,  p_n\} \in P, \    p_i >0, \  i=\overline{1,n}\right\}. \end{eqnarray*}
But the union of the finite number of the sets  $\bar A_{i_u,k_u}$ is a closed set, therefore
$\tilde P \subseteq \bigcup\limits_{u}\bar A_{i_u,k_u}.$  Hence, it immediately  follows  that
 $ P = \bigcup\limits_{u}\bar A_{i_u,k_u}.$

Construct  the minimal set of productive processes that are optimal ones in certain domains of the simplex $P$ and such that the union of the closures of these domains  cover simplex  $P$ exactly. More accurately, let
$\left\{ \left(x_{i_s},y_{k_s}^{(i_s)}\right) \right\},$ $ s=\overline{1,m_0}, $ be
a minimal subset of the set of productive processes $\left\{ \left(x_i,y_s^{(i)}\right)\right\}
_{i=\overline {1,k}, \ 1\leq s\leq m(i)}$ that have the property: for every fixed  $s,\ s=\overline{1,m_0}, $ the productive process  $\left(x_{i_s},y_{k_s}^{(i_s)}\right)$ is optimal one in  nonempty domain  $A_{i_s,k_s} \subseteq P$ determined by the formula
\begin{eqnarray*}          A_{i_s,k_s}=\left\{ p \in P, \  \left\langle
p,y_{k_s}^{(i_s)}-x_{i_s}\right\rangle > \left\langle
p,y_{j}^{(i)}-x_{i}\right\rangle, \ \forall i \neq i_s, \ \forall j \neq k_s \right\}.\end{eqnarray*}
We determine  the minimal number  $m_0$ of  optimal productive processes by the condition
\begin{eqnarray*}          \bigcup\limits _{s=1}^{m_0}\overline A_{i_s,k_s}=P, \quad
\bigcup\limits _{s=1}^{{m_0}-1}\overline A_{i_s,k_s}\subset P,\end{eqnarray*}
where the set  $\overline A_{i_s,k_s}$ is the closure of the set $A_{i_s,k_s}$ in the Euclidean metric.
For any set $A \subseteq P,$ we define a function
\begin{eqnarray*}           d(p,A)=
\inf\limits _{\tilde p\in A}\rho(p,\tilde p), \quad p \in P,\end{eqnarray*}
where  $\rho (p,\tilde p)$ is the Euclidean distance between vectors
$p$ and $\tilde p.$
It is evident that $~d(p,A)$ is a continuous function on $P.$
For sufficiently small
$0 < \varepsilon < 1/2,$ we construct  sets
$A_{i_s,k_s}^{\varepsilon +}$ and $ A_{i_s,k_s}^{\varepsilon -}.$
Let us put
\begin{eqnarray*}          A_{i_s,k_s}^{\varepsilon +}= \{
p \in P,\ d(p,\bar A_{i_s,k_s})\leq\varepsilon\} .\end{eqnarray*}
Denote by  $\partial\bar A_{i_s,k_s}$ the  boundary of the set
$\overline A_{i_s,k_s}$ and choose
$\varepsilon >0$ such small that for every
$s=\overline {1,m_0}$ the set $A_{i_s,k_s}^{\varepsilon -}$ is
nonempty closed subset of the set
$\overline A_{i_s,k_s}$  that contains  at least    one internal point, where
\begin{eqnarray*}          A_{i_s,k_s}^{\varepsilon -}=\{ p\in\overline
A_{i_s,k_s},\  d(p,\partial\overline
A_{i_s,k_s})\geq\varepsilon\}, \quad s=\overline{1, m_0}.\end{eqnarray*}
For every set
$A_{i_s,k_s}^{\varepsilon +},$
we construct a continuous function
$\chi _{i_s,k_s}^{\varepsilon}(p) .$
 For chosen small  $0< \varepsilon<1/2,$ we put
 \begin{eqnarray*}          \chi _{i_s,k_s}^{\varepsilon}(p)
  =\left\{
\begin{array}{lr}{[1-d(p,A_{i_s,k_s}^{\varepsilon
-})]}-\varepsilon^{-1}[1-d(p, A_{i_s,k_s}^{\varepsilon-})]d(p,\overline A_{i_s,k_s}), &
p\in A_{i_s,k_s}^{\varepsilon+}\  ; \\
0,& p\in P\setminus A_{i_s,k_s}^{\varepsilon+}. \end{array}
\right.\end{eqnarray*}
It is obvious that $\chi _{i_s,k_s}^{\varepsilon} (p)$ is a continuous function on $P$ that equals  1 for  $p\in
A_{i_s,k_s}^{\varepsilon -},$
does not exceed 1 and is nonnegative on the set
$A_{i_s,k_s}^{\varepsilon +}\setminus
 A_{i_s,k_s}^{\varepsilon -},$
and equals zero for  $p\in P\setminus
A_{i_s,k_s}^{\varepsilon+}.$
Furthermore, the inequality
\begin{eqnarray*}           \chi _{i_s,k_s}^{\varepsilon}(p) =1-d(p,A_{i_s,k_s}^{\varepsilon
-}) \geq  1-  \varepsilon, \quad p \in \overline A_{i_s,k_s},\end{eqnarray*}
is valid.

Let $\varphi _{i_s,k_s}(p), \ s=\overline{1, m_0},$ be  continuous strictly positive functions on
$P.$ The set of functions
\begin{eqnarray*}          \tau _{i_s,k_s}^\varepsilon (p)=\frac { \chi
_{i_s,k_s}^\varepsilon (p)~\varphi _{i_s,k_s} (p)}{
\sum\limits _{s=1}^{m_0} \chi _{i_s,k_s}^\varepsilon (p)~\varphi
_{i_s,k_s} (p)}, \quad s=\overline{1, m_0}, \end{eqnarray*}
is a decomposition of unit, that is,
\begin{eqnarray*}          \sum\limits _{s=1}^{m_0}\tau _{i_s,k_s}^\varepsilon (p)=1 .\end{eqnarray*}
The introduced set of functions   $\tau _{i_s,k_s}^\varepsilon (p)$ is continuous on the simplex $P.$
Really, since for every  $ p \in P$ there exists a set $\overline A_{i_s,k_s}$ such that $p \in \overline A_{i_s,k_s},$ the bound
\begin{eqnarray*}          \sum\limits _{s=1}^{m_0} \chi _{i_s,k_s}^\varepsilon (p)~\varphi
_{i_s,k_s} (p) \geq (1- \varepsilon)\varphi
_{i_s,k_s} (p) \geq (1- \varepsilon) \min\limits_{1 \leq s \leq m_0}\inf\limits_{p \in P}\varphi
_{i_s,k_s} (p) > 0\end{eqnarray*}
is valid.
From the last,   the  continuity of $\tau _{i_s,k_s}^\varepsilon (p)$ follows.
Consider the strategy of firm behavior
\begin{eqnarray*}          \left( \sum\limits _{s=1}^{m_0}\tau _{i_s,k_s}^\varepsilon (p)x_{i_s},
         \sum\limits _{s=1}^{m_0}\tau _{i_s,k_s}^{\varepsilon}(p)y_{k_s}^{(i_s)}
  \right)=
\left(\bar x^\varepsilon (p),\bar y^\varepsilon (p)\right),\quad
 \bar y^\varepsilon (p)\in F(\overline x^\varepsilon (p)).\end{eqnarray*}
The strategy
$(\bar x^\varepsilon (p),\bar y^\varepsilon (p))$
is continuous on $P$ for $\varepsilon >0.$
Let us find the limit of the sequence of productive processes
$(\bar x^\varepsilon (p),\bar y^\varepsilon (p)),$
as $\varepsilon\rightarrow 0.$
Note that


\begin{eqnarray*}           \tau _{i_s,k_s}(p)=\lim\limits _{\varepsilon\rightarrow 0}
\tau _{i_s,k_s}^\varepsilon (p)=
\left\{\begin{array}{lr}
1,& p\in A_{i_s,k_s},\\
\frac{\varphi _{i_s,k_s}(p)}{\sum\limits _{r=1}^{m_0}
 \varphi _{i_r,k_r}(p)~\chi _{\overline A_{i_r,k_r}}(p)},&
    p\in \partial\overline A_{i_s,k_s},\\
0,& p\in P\setminus\overline A_{i_s,k_s},\\
\end{array} \right.\end{eqnarray*}
\begin{eqnarray*}
\chi _{\overline A_{i_s,k_s}}(p)=
\left\{\begin{array}{lr}
1,& p\in            \overline A_{i_s,k_s}, \\
0,& p\in P\setminus\overline A_{i_s,k_s}.
\end{array}
 \right.
\end{eqnarray*}
The limit productive process  $(x(p),y(p))$ is determined by the formulas
\begin{eqnarray*}
 x(p)=\lim\limits
_{\varepsilon\rightarrow 0}\overline x^ \varepsilon
(p)=\sum\limits _{s=1}^{m_0}\tau _{i_s,k_s} (p)x_{i_s},\quad
y(p)=\lim\limits _{\varepsilon\rightarrow 0}\overline y^
  \varepsilon (p)=\sum\limits _{s=1}^{m_0}\tau
_{i_s,k_s}(p)y_{k_s}^{(i_s)},
\end{eqnarray*}
\begin{eqnarray*}
& y(p) \in F(x(p)).
\end{eqnarray*}
As a consequence of the explicit form of the productive process
$(x(p),y(p)),$   its optimality  follows for all  $p\in P.$
Really,
for  $p\in A_{i_s,k_s}$ the optimality is evident. If
$p\in\partial \bar {A}_{i_s,k_s},$ then from that the functions
$\tau _{i_s,k_s}(p),$~ ${s=\overline{1,m_0}},$ form  the  decomposition of  the  unit and the note before the proof of this Lemma   the optimality of $(x(p),y(p))$ follows.
Thus,
\begin{eqnarray*}          \varphi (p)=\left\langle y(p)-x(p),p\right\rangle .\end{eqnarray*}
Let us estimate the absolute value  of the  difference
\begin{eqnarray*}          \left|\left\langle \overline y^\varepsilon (p)-\overline x^
   \varepsilon (p),p\right\rangle - \varphi (p)\right|=
\left|\sum\limits _{s=1}^{m_0} \left[\tau _{i_s,k_s}^\varepsilon (p)
- \tau _{i_s,k_s}(p)\right]\left\langle p,y_{k_s}^{(i_s )}-x_{i_s}\right\rangle \right|.\end{eqnarray*}
For every point  $\tilde p\in\partial \bar A_{i_s,k_s}$ the equalities
\begin{eqnarray*}          \left\langle\tilde
  p,y_{k_s}^{(i_s)}-x_{i_s}\right\rangle = \left\langle\tilde
  p,y_{k_1}^{(i_1)}-x_{i_1}\right\rangle =\dots = \left\langle\tilde
  p,y_{k_t}^{(i_t)}-x_{i_t}\right\rangle \end{eqnarray*}
  are valid,
where $t$  is the maximal number of equalities written above.
In the last equalities,  we  choose the first $t$ pairs of  productive  processes that correspond to the numeration chosen by us. The last does not restrict  the generality of consideration because, if it is not so,  we can reach
 this by the renumbering of  productive processes for every point
 $\tilde p \in \partial \bar {A}_{i_s,k_s}.$
It is obvious that
 \begin{eqnarray*}        \tau _{i_s,k_s}(\tilde p) \neq 0,\quad
\tau _{i_1,k_1}(\tilde p) \neq 0, \quad \ldots,\quad
\tau _{i_{t},k_{t}}(\tilde p) \neq 0,
\end{eqnarray*}
\begin{eqnarray*}
 \tau _{i_s,k_s}^{\varepsilon}(\tilde p) \neq 0,\quad
\tau _{i_1,k_1}^{\varepsilon}(\tilde p) \neq 0,\quad \ldots,\quad
\tau _{i_{t},k_{t}}^{\varepsilon}(\tilde p) \neq 0.
\end{eqnarray*}
For the rest  $v$ that belong to the set
\begin{eqnarray*}
 T=\{1, \ldots, m_0\} \setminus \{s, 1, 2, \ldots, t\},
 \end{eqnarray*}
 $\tau _{i_v,k_v}(\tilde p)=0.$
The point  $\tilde p \in \partial \bar A_{i_s,k_s}$ is situated at a positive distance from the closed sets $\bar A_{i_v,k_v}, \ v \in T.$
Let this distance equals  $r(\tilde p).$
Denote by $c(\delta(\tilde p), \tilde p)$  the ball of the radius  $\delta(\tilde p)$
with the center at the point  $\tilde p,$
\begin{eqnarray*}          c(\delta(\tilde p), \tilde  p) =\{ p \in P,~|p-\tilde p|<\delta(\tilde p)\}, \quad 0 < \delta(\tilde p)\leq \frac{r(\tilde p)}{4}.\end{eqnarray*}
Taking into account the above,   the following equalities
\begin{eqnarray*}          \tau_{i_v,k_v}(p)=0, \quad  p \in c(\tilde p, \delta(\tilde p)), \quad  v \in T,\end{eqnarray*}
are valid.
Cover the set
$\partial \overline A = \bigcup\limits _{s=1}^{m_0} \partial\overline A_{i_s,k_s}$
by the balls   $c(\delta(\tilde p), \tilde p)$ such that the radius
$\delta(\tilde p) \leq \delta_0, \ \tilde p \in \partial\overline A, $
and the number $\delta_0 > 0$ we will choose later.
Thus,
\begin{eqnarray*}           \partial\overline A
  \subseteq\bigcup\limits _{\tilde p\in
           \bigcup\limits _{s=1}^{m_0} \partial\overline A_{i_s,k_s}}
c(\delta(\tilde p) ,\tilde p).\end{eqnarray*}
Since  $\partial\overline A$
is a compact set in $P,$  there exists a finite subcovering
\begin{eqnarray*}          c(\delta (\tilde p_1),\tilde p_1),\dots,
c\left(\delta (\tilde p_r), \tilde p_r\right)\end{eqnarray*}           such that
\begin{eqnarray*}          \partial\overline A \subseteq \bigcup\limits
  _{i=1}^r c(\delta (\tilde p_i),\tilde p_i)= C,\end{eqnarray*}
and $\delta (\tilde p_i)>0, \ i=\overline{1,r}.$
Denote the boundary of the set $\overline C,$ that is the closure of the set  $C,$ by
$\partial \overline C= \overline C \setminus  C.$ Then the set $\partial \overline A$ and the set  $\partial \overline C$
are at a positive distance as closed sets that do not intersect. Let this distance equals  $\rho >0.$ Then
\begin{eqnarray*}          \rho=\inf\limits_{p \in \partial \overline A, \ \tilde p \in \partial \overline C}\rho(p, \tilde p).\end{eqnarray*}
Choose  $\varepsilon < \rho/4.$ Then the enclosure
\begin{eqnarray*}          W_{\varepsilon}=\bigcup\limits _{s=1}^{m_0} \left [ A_{i_s,k_s}^{\varepsilon+} \setminus
A_{i_s,k_s}^{\varepsilon-}\right] \subset C\end{eqnarray*}
is valid.
To prove this, let us show that the set of balls  $c(\rho/4,\tilde p)$ with centers at points
$\partial\overline A$ of the radius  $\rho/4$
cover the set $W_{\varepsilon}. $
Really, any point $x \in W_{\varepsilon}$ is situated at the distance from the set $\partial\overline A$ that does not exceed  $\varepsilon, $ that is,
there exists a point  $x_0$ from the set $ \partial\overline A$
such that  $\rho(x, x_0) \leq \varepsilon.$
 Let us consider the open ball of the radius  $ 2 \varepsilon$ with the center at the point  $x_0.$ As a result of arbitrariness of $x,$ we prove that the set of balls of the radius  $\rho/2$ with the centers at the points of the set $\partial\overline A$ cover the set
$W_{\varepsilon}.$
But all these balls belong to the set $C.$

Let a point  $\tilde p_j \in \partial\overline A_{i_s,k_s} $ be one of those that figure in the finite subcovering  of the set $\partial\overline A.$
Suppose that at this point the equalities
\begin{eqnarray*}          \left\langle\tilde
  p_j,y_{k_s}^{(i_s)}-x_{i_s}\right\rangle = \left\langle\tilde
  p_j,y_{k_1}^{(i_1)}-x_{i_1}\right\rangle =\dots = \left\langle\tilde
  p_j,y_{k_{t_{j}}}^{(i_{t_{j}})}-x_{i_{t_{j}}}\right\rangle \end{eqnarray*}
are satisfied, where
 $t_{j}$ is the maximal number of the above equalities.
In the last equalities,  we choose the first  $t_{j}$ pairs of productive processes that correspond to chosen by us numeration. This does not restrict the  generality of consideration since we can  always  achieve  this  by the renumbering  of  productive processes for every point  $\tilde p_j \in \partial \bar {A}_{i_s,k_s}.$
It is evident that
 \begin{eqnarray*}
    \tau _{i_s,k_s}(\tilde p_j) \neq 0,\quad \tau _{i_1,k_1}(\tilde p_j) \neq 0,\quad \ldots,\quad
\tau _{i_{{t_j}},k_{{t_j}}}(\tilde p_j) \neq 0,
\end{eqnarray*}
 \begin{eqnarray*}
  \tau _{i_s,k_s}^{\varepsilon}(\tilde p_j) \neq 0,\quad
\tau _{i_1,k_1}^{\varepsilon}(\tilde p_j) \neq 0,\quad \ldots,\quad
\tau _{i_{{t_j}},k_{{t_j}}}^{\varepsilon}(\tilde p_j) \neq 0.
\end{eqnarray*}
For the rest  $v$ that belong to the set
$T_j=\{1, \ldots, m_0\} \setminus \{s, 1, 2, \ldots, t_j\},$ 
 $\tau _{i_v,k_v}(p)=0, \ p \in c(\delta(\tilde p_j), \tilde p_j).$
Let us show that under chosen by us  $ \varepsilon$
\begin{eqnarray*}          \tau _{i_v,k_v}^{\varepsilon}(p)=0,
\quad  p \in c(\delta(\tilde p_j), \tilde p_j).\end{eqnarray*}
Really, the distance from the ball $c(\delta(\tilde p_j), \tilde p_j)$ to the closed set
$\bar {A}_{i_v,k_v} $ is greater than or equals
\begin{eqnarray*}          \frac{3 r(\tilde p_j)}{4}\geq 3 \delta(\tilde p_j).\end{eqnarray*}
But \hfill $\rho \leq \min\limits_{1 \leq j \leq r}\delta(\tilde p_j).$ \hfill  From \hfill here, \hfill the \hfill distance \hfill from \hfill the \hfill set \hfill
$\bar {A}_{i_v,k_v} $ \hfill to \hfill the \hfill ball \\ $c(\delta(\tilde p_j), \tilde p_j)$ is greater than or equals
 $ 3 \rho  \geq 12 \varepsilon.$ So,
\begin{eqnarray*}           \tau _{i_v,k_v}^{\varepsilon}(p)=0,
\quad  p \in c(\delta(\tilde p_j), \tilde p_j), \quad v \in T_j,\end{eqnarray*}           since   $\chi _{i_v,k_v}^{\varepsilon}(p)=0,$ as only the point $p$ is situated  at the distance  greater than  $ \varepsilon$ from the set  $\bar {A}_{i_v,k_v}.$
Therefore,
\begin{eqnarray*}
   1=\sum\limits
_{s=1}^{m_0}\tau _{i_s,k_s}^\varepsilon (\tilde p_j )= \tau
_{i_s,k_s}^\varepsilon (\tilde p_j)+\tau _{i_1,k_1}^\varepsilon
(\tilde p_j)+\dots +\tau _{{i_{t_j}},{k_{t_j}}}^\varepsilon
(\tilde p_j)
\end{eqnarray*}
\begin{eqnarray*}
=\tau _{i_s,k_s}^\varepsilon (p)+\dots +\tau
_{{i_{t_j}},{k_{t_j}}}^ \varepsilon (p),\quad p \in c(\delta(\tilde p_j), \tilde p_j).\end{eqnarray*}
Furthermore,
\begin{eqnarray*}
   1=\sum\limits _{s=1}^{m_0}\tau
_{i_s,k_s} (\tilde p_j)= \tau _{i_s,k_s}(\tilde p_j)+\dots +\tau
    _{i_{t_j},k_{t_j}}(\tilde p_j)= \tau _{i_s,k_s}(p)
    +\dots +\tau_{i_{t_j},k_{t_j}}(p)
\end{eqnarray*}
for
\begin{eqnarray*}
     p \in c(\delta(\tilde p_j), \tilde p_j).
     \end{eqnarray*}
For those  $p \in P$  that belong to the set $c(\delta(\tilde p_j), \tilde p_j)$
the difference under consideration is written as follows
\begin{eqnarray*}
 \left|\sum\limits _{s=1}^{m_0} \left[\tau
_{i_s,k_s}^ \varepsilon (p)-\tau _{i_s,k_s}(p)\right]\left\langle
p,y_{k_s}^{(i_s)}-x_{i_s}\right\rangle \right|
 \end{eqnarray*}
\begin{eqnarray*}
 =\left|\tau_{i_s,k_s}^\varepsilon (p) \left\langle p-\tilde
p_j,y_{k_s}^{(i_s)}-x_{i_s}\right\rangle +\dots +\tau ^\varepsilon
_{i_{t_j},k_{t_j}}(p)\left\langle p-\tilde p_j,
y_{k_{t_j}}^{(i_{t_j})}-x_{i_{t_j}}\right\rangle \right.
\end{eqnarray*}
\begin{eqnarray*}
 +\left[\tau _{i_s,k_s}^\varepsilon ( p)\left\langle\tilde p_j,
y_{k_s}^{(i_s)}-x_{i_s}\right\rangle + \dots +
\tau _{i_{t_j},k_{t_j}}^\varepsilon (p) \left\langle
\tilde p_j,y_{k_{t_j}}^{(i_{t_j})}-x_{i_{t_j}}\right\rangle \right]
\end{eqnarray*}
\begin{eqnarray*} - \left[\tau _{i_s,k_s}( p)\left\langle\tilde p_j,y_{k_s}^{(i_s)}-x_{i_s}
\right\rangle +\dots +\tau _{i_{t_j},k_{t_j}}(p)
\left\langle\tilde p_j,y_{k_{t_j}}^{(i_{t_j})}-x_{i_{t_j}}\right\rangle \right]
\end{eqnarray*}
\begin{eqnarray*}
+\left. \left [\tau _{i_s,k_s}(p)\left\langle\tilde p_j-p,y_{k_s}^{(i_s)}-x_{i_s}
\right\rangle +\dots +\tau _{i_{t_j},k_{t_j}}(p)\left\langle
\tilde p_j-p,y_{k_{t_j}}^{(i_{t_j})}-x_{i_{t_j}}\right\rangle \right]\right|
 \end{eqnarray*}
\begin{eqnarray*}
 \leq \sum\limits_{s=1}^{m_0} \tau _{i_s,k_s}^\varepsilon ( p)
\left|\left\langle p - \tilde p_j,y_{k_s}^{(i_s)}-x_{i_s}\right\rangle \right|+
\sum\limits_{s=1}^{m_0} \tau _{i_s,k_s}( p)
\left|\left\langle p - \tilde p_j,y_{k_s}^{(i_s)}-x_{i_s}\right\rangle\right|
 \end{eqnarray*}
\begin{eqnarray*}
 \leq 2 \delta_0 \max\limits _s |y_{k_s}^{i_s}-x_{i_s}| = \delta_0 A,
\quad  A=2\max\limits _s |y_{k_s}^{i_s}-x_{i_s}|.
\end{eqnarray*}
Since the choice of  $\delta_0$ is arbitrary, then it  can be  chosen arbitrarily small. Choose it so that $\delta_0 A  < \delta.$
Outside of the covering of $C,$ equalities
$\tau_{i_sk_s}(p)- \tau_{i_sk_s}^{\varepsilon}(p)=0, \  s=\overline{1,m_0}$
hold.
Really,
$\tau_{i_sk_s}(p)- \tau_{i_sk_s}^{\varepsilon}(p)=0, $ $  s=\overline{1,m_0},$
outside of the covering of $C$
since
$\chi_{i_sk_s}(p)= \chi_{i_sk_s}^{\varepsilon}(p)=1$
on the set $A^{\varepsilon-}_{i_sk_s}, \ s=\overline{1,m_0},$
and $\chi_{i_sk_s}(p)= \chi_{i_sk_s}^{\varepsilon}(p)=0$ on the set
$A^{\varepsilon-}_{i_jk_j},$  if $ j \neq s.$
As far as,
\begin{eqnarray*}          W_{\varepsilon} \subset
\bigcup\limits
  _{i=1}^r c(\delta (\tilde p_i),\tilde p_i)\end{eqnarray*}
and
\begin{eqnarray*}          \bigcup\limits _{s=1}^{m_0}  A_{i_s,k_s}^{\varepsilon-} \cup
W_{\varepsilon} =
\bigcup\limits _{s=1}^{m_0}  A_{i_s,k_s}^{\varepsilon+} \supseteq  P
\end{eqnarray*}
 on the set
 $\bigcup\limits _{s=1}^{m_0}  A_{i_s,k_s}^{\varepsilon-},  $
$\tau_{i_sk_s}(p)= \tau_{i_sk_s}^{\varepsilon}(p)$ for considered   $\varepsilon < \rho/4.$
Choose, as strategy of firm  behavior  $(x^0(p), y^0(p)),$
the strategy
\begin{eqnarray*}          (\bar x^\varepsilon (p),\bar y^\varepsilon (p))=
\left( \sum\limits _{s=1}^{m_0}\tau _{i_s,k_s}^\varepsilon (p)~x_{i_s},
         \sum\limits _{s=1}^{m_0}\tau _{i_s,k_s}^{\varepsilon}(p)~y_{k_s}^{(i_s)}
  \right), \quad
 \bar y^\varepsilon (p)\in F(\overline x^\varepsilon (p)),\end{eqnarray*}
with such  $\varepsilon >0 $  that satisfies the  condition $\varepsilon < \rho/4.$
From bounds made,  it follows that such strategy satisfies the Lemma conditions.

\qed
\end{proof}

Let $X$ be a convex linear span of the set of  points
$(x_1,\dots ,x_k),$  the set of that is not obligatory minimal, and let $F(x)$ be a convex down technological map given on  $X_1$ belonging to the CTM class in a wide sense. We assume that every point of $X$  is  an internal point of the set $X_1.$
Let us give on $X$ a technological map
\begin{eqnarray}          \label{ssl6}
F_1(x)=\bigcup\limits_{\alpha \in \Delta(x)}\sum\limits_{i=1}^k\alpha_i F(x_i),
\quad  x \in X,
\end{eqnarray}
where by  $\Delta(x)$ we denote the set
\begin{eqnarray*}           \Delta(x)=\left\{\alpha=\{\alpha_i\}_{i=1}^k \in P_1, \
x=\sum\limits_{i=1}^k\alpha_ix_i \right\}. \end{eqnarray*}
The set $\sum\limits_{i=1}^k\alpha_i F(x_i) $ is the set of all points of the kind
$\sum\limits_{i=1}^k\alpha_i y_i,$  where the point $ y_i \in  F(x_i),$ and
\begin{eqnarray*}          P_1=\left\{\alpha=\{\alpha_i\}_{i=1}^k \in  R_+^k, \
\sum\limits_{i=1}^k\alpha_i=1\right\}. \end{eqnarray*}

\begin{lemma}\label{strat2} The technological map $F_1(x)$ defined on $X$
by the formula (\ref{ssl6}) is convex down and belongs to the CTM class in a wide sense if $F(x)$ belongs to the CTM class in a wide sense and  is convex down, furthermore, $F_1(x) \subseteq F(x), \ x \in X.$
\end{lemma}
\begin{proof}\smartqed          Prove the convexity of  $F_1(x)$ for every $x \in X.$
Let $y_1$ and $y_2$ belong to $F_1(x).$ It means that there exist
\begin{eqnarray*}          \alpha^{'}=\{\alpha_i^{'}\}_{i=1}^k, \quad
\alpha^{''}=\{\alpha_i^{''}\}_{i=1}^k \in \Delta(x)\end{eqnarray*}
such that
$x=\sum\limits_{i=1}^k\alpha_i^{'}x_i, \
x=\sum\limits_{i=1}^k\alpha_i^{''}x_i,$ and for $y_1$ and $y_2$
the representations
\begin{eqnarray*}          y_1=\sum\limits_{i=1}^k\alpha_i^{'}
y_j^1, 
 \quad y_2=\sum\limits_{i=1}^k\alpha_i^{''}y_i^2, \quad  y_i^1, \  y_i^2 \in F(x_i), \quad i=\overline{1,k}, \end{eqnarray*}
are valid.
For arbitrary  $ 0 < \alpha <1 $ and for those  $i$ for which
$\alpha_i^{'}, \alpha_i^{''}$
do not equal zero simultaneously
\begin{eqnarray*}          \alpha y_1 +(1 - \alpha)y_2=
\sum\limits_{i=1}^k\alpha_i^{'''} \left [
\frac{\alpha\alpha_i^{'}}{ \alpha\alpha_i^{'}+
(1- \alpha)\alpha_i^{''}}
y_i^1+
\frac{(1- \alpha)\alpha_i^{''}}{ \alpha\alpha_i^{'}+
(1- \alpha)\alpha_i^{''}}
 y_i^2 \right],\end{eqnarray*}
 where $\alpha_i^{'''}= \alpha\alpha_i^{'}+
(1- \alpha)\alpha_i^{''}.$
 Due to the convexity of the set $F(x_i),$  the point
\begin{eqnarray*}           \frac{\alpha\alpha_i^{'}}{\alpha\alpha_i^{'} +(1- \alpha)\alpha_i^{''}}y_i^1
+
  \frac{(1- \alpha)\alpha_i^{''}}{ \alpha\alpha_i^{'}+
(1- \alpha)\alpha_i^{''}} y_i^2 \end{eqnarray*}
belongs to  this set too.
Because   \begin{eqnarray*}          x=\sum\limits_{i=1}^k\alpha_i^{'''}x_i, \quad
\sum\limits_{i=1}^k\alpha_i^{'''}=1,\end{eqnarray*}
we have  $\alpha y_1 +(1 - \alpha)y_2  \in F_1(x)$
for any $\alpha \in (0,1).$

Prove the completeness of $F_1(x).$

Let the sequence  $y_n \in F_1(x)$  and
$y_n \to y_0.$ Prove  that $y_0 \in F_1(x).$
From that  $y_n \in F_1(x)$  the existence of sequences follows
\begin{eqnarray*}          \alpha^n=\{\alpha_i^n\}_{i=1}^k  \in P_1, \quad \bar y_n=\{y_i^n\}_{i=1}^k, \quad y_i^n \in F(x_i), \quad i=\overline{1,k}, \end{eqnarray*}
that satisfy conditions:
 for  $y_n \in F_1(x)$  and  $x \in X$ the representations
\begin{eqnarray*}           x=\sum\limits_{i=1}^k\alpha_i^nx_i, \quad
 y_n=\sum\limits_{i=1}^k\alpha_i^ny_i^n\end{eqnarray*}
hold.
From the compactness of the  considered  sequences,   the existence of a subsequence
 $n_m$ follows such that
$\alpha^{n_m}= \{\alpha_i^{n_m}\}_{i=1}^k,\
\bar y_{n_m}=\{ y_i^{n_m} \}_{i=1}^k$
are convergent correspondingly  to
$\alpha^0 =\{\alpha_i^{0}\}_{i=1}^k,\  \bar y_0=\{ y_i^0 \}_{i=1}^k, $
as $ m \to \infty,$
and for the  limit point $y_0$ of the subsequence $ y_{n_m}$ the representation
\begin{eqnarray*}          y_0=\sum\limits_{i=1}^k\alpha_i^0
y_i^0,\quad  y_i^0 \in F(x_i),  \quad  i=\overline{1,k}, \end{eqnarray*}
holds, where
\begin{eqnarray*}            x=
\sum\limits_{i=1}^k\alpha_i^0x_i, \quad \alpha^0=\{\alpha_i^0\}_{i=1}^k \in P_1.\end{eqnarray*}
The latter  means that $y_0 \in F_1(x).$
The closure of  $F_1(x)$  follows  from that
$y_0$  is  an arbitrary limit point of the  sequence  $y_n.$ The boundedness of $F_1(x)$
is obvious.

Prove the Kakutani  continuity from above of $F_1(x).$
Let the sequence  $y_n \in F_1(x_n)$  and $x_n \to x_0,$
$y_n \to y_0.$ Show that $y_0 \in F_1(x_0).$
From that  $y_n \in F_1(x_n),$   the existence of such sequences
$\alpha^n=\{\alpha_i^n\}_{i=1}^k \in P_1$ and $\bar y_n=\{y_i^n\}_{i=1}^k$
follows  that for  $y_n \in F_1(x_n)$  and  $x_n \in X$ the representations
\begin{eqnarray*}           x_n=\sum\limits_{i=1}^k\alpha_i^nx_i, \quad
 y_n=\sum\limits_{i=1}^k\alpha_i^n
y_i^n\end{eqnarray*}
hold.
From the compactness of the considered  sequences,  the existence of  such subsequence
  $n_m$  follows  that
$\alpha^{n_m}=\{\alpha_i^{n_m}\}_{i=1}^k,$ and $
\bar y_{n_m}=\{ y_i^{n_m} \}_{i=1}^k$
are correspondingly convergent to
$\alpha^0=\{\alpha_i^{0}\}_{i=1}^k$ and $
\bar y_0=\{ y_i^0 \}_{i=1}^k, $
as $ m \to \infty,$
and for the limit points  $y_0$ and  $x_0$ of subsequences $y_{n_m}$ and  $x_{n_m},$ correspondingly,   the representations
\begin{eqnarray*}          y_0=\sum\limits_{i=1}^k\alpha_i^0
y_i^0, \quad y_i^0 \in F(x_i),  \quad  i=\overline{1,k},
 \end{eqnarray*}
\begin{eqnarray*}            x_0=
\sum\limits_{i=1}^k\alpha_i^0x_i, \quad \alpha^0=\{\alpha_i^0\}_{i=1}^k \in P_1, \end{eqnarray*}
are valid.
The latter means that  $y_0 \in F_1(x_0).$

Prove that  $F_1(x)$ is a convex down technological map.
Let
\begin{eqnarray*}
  x_1=\sum\limits_{i=1}^k\alpha_i^{'}x_i,\quad
x_2=\sum\limits_{i=1}^k\alpha_i^{''}x_i \in X
\end{eqnarray*}
\begin{eqnarray*} \alpha^{'}=\{\alpha_i^{'}\}_{i=1}^k \in P_1, \quad
 \alpha^{''}=\{\alpha_i^{''}\}_{i=1}^k \in P_1,
 \end{eqnarray*}
\begin{eqnarray*}  y_1=\sum\limits_{i=1}^k\alpha_i^{'}
y_i^1  \in F_1(x_1),
\quad y_2=\sum\limits_{i=1}^k\alpha_i^{''}
y_i^2  \in F_1(x_2).
\end{eqnarray*}
Then   for arbitrary
$ 0 < \alpha < 1,$
\begin{eqnarray*}          \alpha y_1 +(1 - \alpha)y_2=
\sum\limits_{i=1}^k\alpha_i^{'''} \left [
\frac{\alpha\alpha_i^{'}}{ \alpha\alpha_i^{'}+
(1- \alpha)\alpha_i^{''}}
 y_i^1+
\frac{(1- \alpha)\alpha_i^{''}}{ \alpha\alpha_i^{'}+
(1- \alpha)\alpha_i^{''}}
 y_i^1 \right].\end{eqnarray*}
  Due to the convexity of the set  $F(x_i),$   the point
\begin{eqnarray*}          y_i^3= \frac{\alpha\alpha_i^{'}}{\alpha\alpha_i^{'} +(1- \alpha)\alpha_i^{''}}
y_i^1+
  \frac{(1- \alpha)\alpha_i^{''}}{ \alpha\alpha_i^{'}+
(1- \alpha)\alpha_i^{''}} y_i^2 \end{eqnarray*}
belongs to this set too, where $\alpha_i^{'''}= \alpha\alpha_i^{'}+
(1- \alpha)\alpha_i^{''}$ for those $i$ for which  $\alpha_i^{'} $ or $\alpha_i^{''} $ do not equal zero.
So,
\begin{eqnarray*}          \alpha y_1 +(1 - \alpha)y_2=
\sum\limits_{i=1}^k\alpha_i^{'''}
y_i^3, \quad 0 < \alpha < 1.\end{eqnarray*}
Since   \begin{eqnarray*}          \alpha x_1+(1- \alpha)x_2=\sum\limits_{i=1}^k\alpha_i^{'''}x_i, \quad
\sum\limits_{i=1}^k\alpha_i^{'''}=1, \end{eqnarray*}
we have   $\alpha y_1 +(1 - \alpha)y_2  \in F_1(\alpha x_1+(1- \alpha)x_2)$
for any  $\alpha \in (0,1).$

The latter means that  $\alpha y_1+(1 - \alpha)y_2 \in
F_1(\alpha x_1 + (1 - \alpha)x_2),$ or the same that
$\alpha F_1(x_1) +(1 - \alpha) F_1(x_2) \subseteq
F_1(\alpha x_1 + (1 - \alpha)x_2). $
At last,
\begin{eqnarray*}          F_1(x)=\bigcup\limits_{\alpha \in \Delta(x)}\sum\limits_{i=1}^k\alpha_i F(x_i)
\subseteq
\bigcup\limits_{\alpha \in \Delta(x)}F(\sum\limits_{i=1}^k\alpha_ix_i)=F(x),
\end{eqnarray*}
\begin{eqnarray*}
 x =\sum\limits_{i=1}^k\alpha_ix_i, \quad  x \in X.
\end{eqnarray*}
\qed
\end{proof}

\begin{lemma}\label{s100}
Let points   $(x_1,\dots ,x_k)$ generate  a convex bounded closed polyhedron $X \subset R_+^n$
and a technological map  $F_1(x),$  given on  $X,$
be defined by the formula  (\ref{ssl6}).
Then
\begin{eqnarray*}          \sup\limits_{x \in X} \sup\limits_{y \in F_1(x)}\left\langle p,y - x\right\rangle =
\max\limits_{1 \leq i \leq k}
\sup\limits_{y \in F(x_i)}\left\langle p,y - x_i\right\rangle.\end{eqnarray*}

\end{lemma}
\begin{proof}\smartqed          Any points $x \in X$ and  $y \in F_1(x)$  can be represented in the form
\begin{eqnarray}\label{h1l126}
 x=\sum\limits
_{i=1}^k\alpha_ix_i, \quad y=\sum\limits
 _{i=1}^k\alpha_i y_i,
 \end{eqnarray}
 \begin{eqnarray*}         \sum\limits _{i=1}^k\alpha _i
=1,\quad \alpha _i
\geq 0,\quad i=\overline{1,k}.
\end{eqnarray*}
Substituting the representations for  $x$ and $y$ into the expression
\begin{eqnarray*}          \psi(y,x)=\left\langle y-x,p\right\rangle,\end{eqnarray*}
 we obtain
\begin{eqnarray*}          \left\langle y-x,p\right\rangle=\sum\limits _{i=1}^k\alpha _i
\left\langle
y_i-x_i,p\right\rangle.\end{eqnarray*}
There hold the inequalities
\begin{eqnarray*}          \left\langle y-x_i,p\right\rangle \leq  \sup\limits _{y\in F(x_i)}\left\langle
p,y - x_i\right\rangle \leq \max\limits _{1\leq i\leq k}\sup\limits _{y\in F(x_i)}\left\langle
p,y - x_i\right\rangle.\end{eqnarray*}
Therefore, for any $x \in X$ and $y \in F_1(x)$
\begin{eqnarray*}          \left\langle y-x,p\right\rangle \leq \max\limits _{1\leq i\leq k}\sup\limits _{y\in F(x_i)}\left\langle
p,y - x_i\right\rangle,\end{eqnarray*}
or
\begin{eqnarray*}          \sup\limits_{x \in X} \sup\limits_{y \in F_1(x)}\left\langle p,y - x\right\rangle \leq \max\limits _{1\leq i\leq k}\sup\limits _{y\in F(x_i)}\left\langle
p,y - x_i\right\rangle.\end{eqnarray*}
Prove the inverse inequality.
It is obvious that
$F_1(x_i)=F(x_i), \ i=\overline{1,k}.$ Therefore,
\begin{eqnarray*}          \sup\limits_{x \in X} \sup\limits_{y \in F_1(x)}\left\langle p,y - x\right\rangle  \geq  \sup\limits_{y \in F_1(x_i)}\left\langle p,y - x_i\right\rangle= \sup\limits_{y \in F(x_i)}\left\langle p,y - x_i\right\rangle, \quad i=\overline{1,k}.\end{eqnarray*}
 Taking maximum over all $1 \leq i \leq k$ from the left and right side of the last inequality, we obtain the needed inequality.

\qed
\end{proof}
\begin{definition} A set of points $\{x_i, \ i=\overline{1, \infty} \}, \  x_i \in R_+^n, $ generates a set $X \subseteq R_+^n$ if the set $X$ is the closure of the set of points  of the form
\begin{eqnarray*}          V= \left\{ x=\sum\limits_{i=1}^k\alpha_ix_i \in R_+^n, \ \{\alpha=\{\alpha_i\}_{i=1}^k \in P_1^k, \ k=
\overline{1, \infty}\right\},\end{eqnarray*}
where
\begin{eqnarray*}          P_1^{k}=\left\{\alpha=\{\alpha_i\}_{i=1}^{k} \in  R_+^{k}, \
\sum\limits_{i=1}^{k}\alpha_i=1\right\}. \end{eqnarray*}
\end{definition}
 \begin{lemma}\label{s101}
Let   $X \subset R_+^n$ be a convex bounded closed set whose every point  is  internal for a set $X_1$ and
$\{x_i, \ i=\overline{1, \infty} \} $ be  dense  in $X$  set of points that generate it.
If the technological map $F_1^k(x)$
is given by the formula   (\ref{ssl6}) on the set $X_k$ generated by the first $k$  points
$\{x_i, \ i=\overline{1,k} \} $
from the set of points
$\{x_i, \ i=\overline{1, \infty} \}, $ that generate $X,$
then
\begin{eqnarray*}          \lim\limits_{k \to \infty}\sup\limits_{x \in X_k}
\sup\limits_{y \in F_1^k(x)}\left\langle p,y - x \right\rangle=
\sup\limits_{x \in X}
\sup\limits_{y \in F(x)}\left\langle p,y - x \right\rangle.\end{eqnarray*}
\end{lemma}
\begin{proof}\smartqed
We assume that the set of points
$\{x_i, \ i=\overline{1, \infty} \}$  are ordered and the set $X_{k+1}$ is generated  by the set of points $\{x_i, \ i=\overline{1,k+1} \}. $
There holds the inclusion
\begin{eqnarray*}          F_1^k(x)=
\bigcup\limits_{\alpha \in \Delta(x)}\sum\limits_{i=1}^k\alpha_i F(x_i)
\subseteq
\bigcup\limits_{\alpha \in \Delta_1(x)}\sum\limits_{i=1}^{k+1}\alpha_i F(x_i)=
F_1^{k+1}(x),\end{eqnarray*}
where
\begin{eqnarray*}           \Delta(x)=\left\{\alpha=\{\alpha_i\}_{i=1}^k \in P_1^k, \
x=\sum\limits_{i=1}^k\alpha_ix_i \right\}, \end{eqnarray*}
\begin{eqnarray*}           \Delta_1(x)=\left\{\alpha=\{\alpha_i\}_{i=1}^{k+1} \in P_1^{k+1}, \
x=\sum\limits_{i=1}^{k+1}\alpha_ix_i\right \}, \end{eqnarray*}
\begin{eqnarray*}          P_1^k=\left\{\alpha=\{\alpha_i\}_{i=1}^k \in  R_+^k, \
\sum\limits_{i=1}^k\alpha_i=1\right\}, \end{eqnarray*}
\begin{eqnarray*}          P_1^{k+1}=\left\{\alpha=\{\alpha_i\}_{i=1}^{k+1} \in  R_+^{k+1}, \
\sum\limits_{i=1}^{k+1}\alpha_i=1\right\}. \end{eqnarray*}
So, the sequence of functions
$\varphi_k(p)=\sup\limits _{x\in X_k}
\sup\limits _{y \in F_1^k(x)}\left\langle y-x,p\right\rangle$ is monotonously  non decreasing,  that is, $\varphi_k(p) \leq \varphi_{k+1}(p).$
Consider $\sup\limits _{y\in F(x)}\left\langle y-x,p\right\rangle.$
Due to the continuity of the function  $\left\langle y-x,p\right\rangle$ in argument  $y$
and the compactness of the set $F(x),$  there exists a point $y(x,p) \in F(x)$
such that
\begin{eqnarray*}          \sup\limits _{y\in F(x)}\left\langle y-x,p\right\rangle=\left\langle y(x,p)-x,p\right\rangle.\end{eqnarray*}
From the convexity  down of $F(x),$ the function  $\left\langle y(x,p)-x,p\right\rangle $ is a convex  up function of the argument $x$ and so it is  continuous one by argument  $x$ on the set $X.$ On the basis of  the Weierstrass theorem,
\begin{eqnarray*}          \sup\limits _{x\in X}
\sup\limits _{y \in F(x)}\left\langle y-x,p\right\rangle=
\sup\limits _{x\in X} \left\langle y(x,p)-x,p\right\rangle=
\left\langle y(x_0,p)-x_0,p\right\rangle, \quad x_0 \in X.\end{eqnarray*}
Consider the sequence
$\varphi_k(p)=\sup\limits _{x\in X_k}
\sup\limits _{y \in F_1^k(x)}\left\langle y-x,p\right\rangle$
and show that for every  $p \in P$
\begin{eqnarray*}          \lim\limits_{k \to \infty}\varphi_k(p)=\sup\limits_{x\in X}
\sup\limits_{y \in F(x)}\left\langle y-x,p\right\rangle,\end{eqnarray*}
where
\begin{eqnarray*}          P=\left\{p=\{p_i\}_{i=1}^n \in R_+^n, \ \sum\limits_{i=1}^np_i=1\right\}.\end{eqnarray*}
Because  the sequence of points
$\{x_i, \ i=\overline{1, \infty}\}$
of the set  $X$ is dense in  $X,$  there exists subsequence  $x_{n_m}$ of this sequence that converges to a point  $x_0$ in which the supremum is realized
$\sup\limits _{x \in X}\left\langle y(x,p)-x,p\right\rangle=
\left\langle y(x_0,p)-x_0,p\right\rangle. $ It is evident that for $k > n_m$
there holds the inequality
\begin{eqnarray*}          \varphi_k(p)=\left\langle y(x_{i_0^k},p)-x_{i_0^k},p\right\rangle \geq
\left\langle y(x_{n_m},p)-x_{n_m},p\right\rangle, \quad x_{n_m}, \ x_{i_0^k} \in X_k.\end{eqnarray*}
Tending  from the beginning  $k$  and then  $m$  to the infinity, we obtain
 \begin{eqnarray*}          \lim\limits_{k \to \infty}\varphi_k(p) \geq
\left\langle y(x_0,p)-x_0,p\right\rangle=
\sup\limits _{x \in X}\left\langle y(x,p)-x,p\right\rangle.\end{eqnarray*}
On the other hand,
\begin{eqnarray*}          \varphi_k(p)=\sup\limits _{x \in X_k}\sup\limits _{y \in F_1^k(x)}\left\langle y-x,p\right\rangle
\leq \sup\limits _{x \in X}\sup\limits _{y \in F(x)}\left\langle y-x,p\right\rangle=
\varphi(p).\end{eqnarray*}
Therefore, $\lim\limits_{k \to \infty}\varphi_k(p)=\varphi(p).$
\qed
\end{proof}

Let  $F(x)$ be a convex down technological map from the CTM class in a wide sense defined on  $X_1.$
Cover the set $ P$ by balls
\begin{eqnarray*}          C(\delta, \bar p)=\{p \in P,\ |p-\bar p< \delta\}\end{eqnarray*}
of  a radius $\delta>0.$   Thanks to the compactness
of  $P,$ there exists a finite subcovering with the  center at the points
$\{\tilde p_1, \ldots, \tilde p_{m(\delta)}\},$ that is,
\begin{eqnarray*}           \bigcup\limits_{i=1}^{m(\delta)}C(\delta, \tilde  p_i)=P.\end{eqnarray*}            Denote by
 $y(x_s, \tilde p_i)$ the point of the set  $F(x_s),$ in which the maximum of the problem
 \begin{eqnarray*}          \sup\limits _{y \in F(x_s)}\left\langle y-x_s,\tilde p_i\right\rangle=
\left\langle y(x_s, \tilde p_i)-x_s,\tilde p_i\right\rangle \end{eqnarray*}
is reached, where the set of points $\{x_i, \ i=\overline{1,k}\}$  generates the set
$X  \subset X_1.$

Let  $\{y(x_s, \tilde p_i), \  i=\overline{1, m(\delta)}\} $
be a set of points that generates the set
$Y_s^{\delta}, \ s=\overline{1, k}, $ and  $F_1(x)$ be a technological map given on the set  $X$ by the formula  (\ref{ssl6}). Define on  $X$
 a technological map
\begin{eqnarray}          \label{ssl7}
F_1^{\delta}(x)=\bigcup\limits_{\alpha \in \Delta(x)}
\sum\limits_{i=1}^k\alpha_i Y_i^{\delta}, \quad
 x \in X,
\end{eqnarray}
where by  $\Delta(x)$  we denote the set
\begin{eqnarray*}           \Delta(x)=\left\{\alpha=\{\alpha_i\}_{i=1}^k \in P_1, \
x=\sum\limits_{i=1}^k\alpha_ix_i \right\}. \end{eqnarray*}
The set  $\sum\limits_{i=1}^k\alpha_i Y_i^{\delta}  $ is the set of points of the form
$\sum\limits_{i=1}^k\alpha_i y_i,$  where the point  $ y_i \in Y_i^{\delta}$ and
the vector $\alpha=\{\alpha_i\}_{i=1}^k $ runs over the set
\begin{eqnarray*}          P_1=\left\{\alpha=\{\alpha_i\}_{i=1}^k \in  R_+^k, \
\sum\limits_{i=1}^k\alpha_i=1\right\}. \end{eqnarray*}
Owing to the points
 $\{y(x_s, \tilde p_i), \  i=\overline{1, m(\delta)}\} $
 generate the set
$Y_s^{\delta}, \ s=\overline{1, k}, $ and $Y_s^{\delta} \subseteq F(x_s),$
then
\begin{eqnarray*}          \sum\limits_{i=1}^k\alpha_i Y_i^{\delta} \subseteq \sum\limits_{i=1}^k\alpha_i F(x_i) \subseteq F(x).\end{eqnarray*}
Thus, $F_1^{\delta}(x)  \subseteq F(x).$

\begin{lemma}\label{ss101} Let technological maps $F_1(x)$  and $F_1^{\delta}(x)$ be  given, respectively, by the formulas
(\ref{ssl6}) and   (\ref{ssl7}) on the set $X$
generated by the set of points
$(x_1,\dots ,x_k)$ that non obligatory is minimal and every point of the set  $X$
be an internal for the set  $X_1,$ on which  a convex down technological map $F(x)$ from the CTM class in a wide sense is given. Then, for every
$\varepsilon>0,$ there exists  $\delta>0 $ such  that
 \begin{eqnarray*}          \sup\limits_{p \in P}|
\sup\limits_{x \in X}\sup\limits _{y \in F_1(x)}\left\langle y-x,p\right\rangle -
\sup\limits_{x \in X}
\sup\limits _{y \in F_1^{\delta}(x)}\left\langle y-x,p\right\rangle| <  \varepsilon.
\end{eqnarray*}
\end{lemma}
\begin{proof}\smartqed          Let us show the equality
\begin{eqnarray*}          \sup\limits _{x\in X}\sup\limits _{y\in
F_1(x)}\left\langle y-x,\tilde p_i\right\rangle=
\sup\limits _{x\in X}\sup\limits _{y\in
F_1^{\delta}(x)}\left\langle y-x,\tilde p_i\right\rangle, \quad i=\overline{1, m(\delta)}.
\end{eqnarray*}
At first, show that
\begin{eqnarray*}          \sup\limits _{x\in X}\sup\limits _{y\in
F_1(x)}\left\langle y-x,\tilde p_i\right\rangle=\max\limits_{1 \leq s \leq k}
\left\langle y(x_s, \tilde p_i)-x_s,\tilde p_i\right\rangle,
 \quad  i=\overline{1, m(\delta)}.\end{eqnarray*}
This equality follows from the Lemma  \ref{s100} and the fact that
\begin{eqnarray*}          \sup\limits _{y\in
F(x_s)}\left\langle y-x_s,\tilde p_i\right\rangle=
\left\langle y(x_s, \tilde p_i)-x_s,\tilde p_i\right\rangle.\end{eqnarray*}
At last, on the basis of the Lemma  \ref{s1}
\begin{eqnarray*}          \sup\limits _{x\in X}\sup\limits _{y\in
F_1^{\delta}(x)}\left\langle y-x,p\right\rangle=
\max\limits _{1\leq s \leq k}
\max\limits _{1\leq j \leq m(\delta)}\left\langle
p,y(x_s, \tilde p_j) - x_s \right\rangle.\end{eqnarray*}
Since
\begin{eqnarray*}           \left\langle y(x_s, \tilde p_j) -x_s, \tilde p_i \right\rangle \leq
\sup\limits _{y\in
F(x_s)}\left\langle y-x_s,\tilde p_i\right\rangle=
\left\langle y(x_s, \tilde p_i) -x_s, \tilde p_i \right\rangle, \quad
\tilde p_j \neq \tilde p_i,\end{eqnarray*}
we have
\begin{eqnarray*}          \max\limits _{1\leq j\leq m(\delta)}\left\langle
y(x_s, \tilde p_j)-x_s, \tilde p_i \right\rangle=
\left\langle y(x_s, \tilde p_i)-x_s, \tilde p_i \right\rangle,
\quad  i=\overline{1, m(\delta)}.   \end{eqnarray*}
So,
\begin{eqnarray*}          \sup\limits _{x\in X}\sup\limits _{y\in
F_1^{\delta}(x)}\left\langle y-x,\tilde p_i \right\rangle=
\max\limits _{1\leq s\leq k}
\left\langle y(x_s, \tilde p_i)-x_s, \tilde p_i \right\rangle,
\quad  i=\overline{1, m(\delta)}.  \end{eqnarray*}
The needed equality is proved.
Let us estimate the difference
 \begin{eqnarray*}         T=\sup\limits_{p \in P}|
\sup\limits_{x \in X}\sup\limits _{y \in F_1(x)}\left\langle y-x,p\right\rangle -
\sup\limits_{x \in X}\sup\limits _{y \in F_1^{\delta}(x)}\left\langle y-x,p\right\rangle|
\end{eqnarray*}
 \begin{eqnarray*}
        \leq \sup\limits_{p \in P}
|\sup\limits_{x \in X}\sup\limits _{y \in F_1(x)}\left\langle y-x,p\right\rangle -
 \sup\limits_{x \in X}\sup\limits _{y \in F_1(x)}\left\langle y-x,\tilde p_i\right\rangle|
  \end{eqnarray*}
 \begin{eqnarray*}
  + \sup\limits_{p \in P}
|\sup\limits_{x \in X}\sup\limits _{y \in F_1^{\delta}(x)}\left\langle y-x,p\right\rangle -
 \sup\limits_{x \in X}\sup\limits _{y \in F_1^{\delta}(x)}\left\langle y-x,\tilde p_i\right\rangle|.
 \end{eqnarray*}
Since \hfill $F_1(x)$ \hfill and \hfill  $F_1^{\delta}(x)$ \hfill satisfy \hfill conditions \hfill of \hfill the \hfill Lemma \hfill
\ref{ksl1}, \hfill then \hfill for \hfill $p \in C(\delta, \tilde p_i),$~    \\ $T \leq \delta A,$~
$A=4 \sup\limits_{x \in X}\sup\limits _{y \in F(x)}\left\langle y+x,e\right\rangle.$
Choosing  $\delta>0 $  such that for  given  $\varepsilon>0$ the inequality  $\delta A < \varepsilon$  would be valid, we obtain the proof of the Lemma.
\qed
\end{proof}
Now, prove the main statement of this Chapter.
\begin{theorem}\label{nnl1}  \cite{55, 71, 92, 106}
Let  $X$ be a bounded closed convex set\index{bounded closed convex set} whose every point  is  internal for  $X_1$ and $F(x)$ be a convex down technological map  from the CTM class in a wide sense\index{convex down technological map  from the CTM class in a wide sense} given on a convex  compact set\index{convex  compact set }  $X_1.$ Then,  for every sufficiently small
$\varepsilon >0,$ there exists a continuous firm   behavior strategy\index{continuous firm   behavior strategy}
$(x^0(p),y^0(p)),$~  $y^0(p)  \in   F(x^0(p))$
such that
\begin{eqnarray*}          \sup\limits_{p\in P}\left|\varphi (p)-\left\langle y^0(p)- x^0(p),p\right\rangle \right|
<\varepsilon ,\end{eqnarray*}           where \begin{eqnarray*}          \varphi (p)=\sup\limits
_{x\in X}\sup\limits _{y\in F(x)} \left\langle y-x,p\right\rangle .\end{eqnarray*}
\end{theorem}
\begin{proof}\smartqed          Let $\{x_i, \ i=\overline{1, \infty}\}$ be a dense countable set of points that generates $X$ and such that it contains a dense set of extreme points of $X.$ Further, let $X_k$ be a polyhedron generated  by the first  $k$ points  $(x_1,\dots ,x_k).$ There hold such enclosures $X_1 \subseteq X_2 \ldots \subseteq X_k
\ldots \subseteq  X.$ On the basis of the Lemma  \ref{s101},  the sequence of continuous functions on  $P$
$\varphi_k(p) = \sup\limits _{x\in X_k}\sup\limits _{y\in F_1^k(x)}\left\langle y-x,p\right\rangle $
is monotonically non decreasing  and converges to a continuous function
$\varphi(p)=\sup\limits _{x \in X}\sup\limits _{y \in F(x)}\left\langle y-x,p\right\rangle.$
According to the  Dini theorem,\index{ Dini theorem} the sequence $\varphi_k(p)$ converges to
$\varphi(p)$   uniformly.  Therefore, there exists a number $k_0$  such that for every $k \geq k_0$
\begin{eqnarray*}          \sup\limits_{p\in P}|\varphi_k(p)- \varphi(p)|
<\varepsilon/3. \end{eqnarray*}
From the Lemma  \ref{ss101}
 \begin{eqnarray*}          \sup\limits_{p \in P}\left|\varphi_k(p)- \varphi_k^{\delta}(p)\right| < \varepsilon/3
\end{eqnarray*}
for sufficiently small $\delta,$  where
\begin{eqnarray*}          \varphi_k^{\delta}(p)=\sup\limits
_{x\in X_k}\sup\limits _{y\in F_1^{\delta}(x)} \left\langle y-x,p\right\rangle .\end{eqnarray*}
According to the Lemma \ref{s2}, there exists a continuous firm behavior strategy  $(x_0(p), y_0(p)),$ \
$x_0(p) \in X_k \subseteq X, \   y_0(p)  \in F_1^{\delta}(x_0(p))
\subseteq F(x_0(p))$ such that
\begin{eqnarray*}          \sup\limits_{p \in P}
|\varphi_k^{\delta}(p) - \left\langle y_0(p) - x_0(p),p \right\rangle|
< \varepsilon/3.\end{eqnarray*}
The latter  proves the Theorem.
\qed
\end{proof}

Let us give some examples of technological maps that satisfy conditions of the Theorem  \ref{nnl1}. Let  $X$ be a compact convex set from $R_+^n,$  $A=|a_{ij}|_{i=1,j=1}^{n,m}$ be a matrix of the Leontieff technological coefficients  of the dimensionality  $n \times m$
  that have such economic sense: $a_{ij}$ is the quantity of units of the $i$-th goods needed for production of one unit of the $j$-th goods. Assume that the matrix $A$ does not contain zero columns.
Define for every  $x \in X$ a technological map
\begin{eqnarray*}          F(x)=\{y \in R_+^m, \ Ay \leq x\}.\end{eqnarray*}
Then  $F(x)$ is a convex down technological map  whose images  are bounded convex closed sets.  Moreover, it is easy to construct a compact set in that all these images are contained. So, all conditions of the Theorem  \ref{nnl1} for this technological map are satisfied.

Over again, let  $X$ be a compact convex set from $R_+^n,$  a matrix $A$ satisfies the same conditions  as earlier,  that is, it is the matrix with non negative coefficients  without zero columns. Further, let  $B$ be a nonnegative matrix $|b_{ij}|_{i=1,j=1}^{s,m}$ of the dimensionality $s
\times m.$
Consider the  Neumann technological map defined for  $x \in X$ by the rule
 \begin{eqnarray*}          F(x)=\{y \in R_+^s, \mbox{there exists} \  \xi \in R_+^m, \  y=B\xi, \ A\xi \leq
x\}.\end{eqnarray*}           Then the  Neumann technological  map  satisfies all conditions of the Theorem \ref{nnl1}.

Construct a nonlinear technological map that satisfies conditions of the Theorem \ref{nnl1}. Consider a vector function
\begin{eqnarray*}          f(y)=(f_1(y), \ldots,f_n(y)),\end{eqnarray*}
given on $R_+^n,$ whose every component  is a convex   down function, that is, there holds inequality
\begin{eqnarray*}          f_i(\alpha_1y_1+\ldots+ \alpha_sy_s) \leq
\alpha_1f_i(y_1)+\ldots+ \alpha_sf_i(y_s)\end{eqnarray*}
for all
\begin{eqnarray*}          0 \leq \alpha_j \leq 1,\quad  \sum\limits_{i=1}^s\alpha_i=1, \quad
y_i \in R_+^n.\end{eqnarray*}
Then
\begin{eqnarray*}          f(\alpha_1y_1+\ldots+ \alpha_sy_s) \leq
\alpha_1f(y_1)+\ldots+ \alpha_sf(y_s),\end{eqnarray*}
where the vector inequality  is  understood  for every component.
Let a vector function   $g(y)=(g_1(y), \ldots,g_n(y))$  be convex  up, that is,
\begin{eqnarray*}          g_i(\alpha_1y_1+\ldots+ \alpha_sy_s) \geq
\alpha_1g_i(y_1)+\ldots+ \alpha_sg_i(y_s)\end{eqnarray*}
for all
\begin{eqnarray*}          0 \leq \alpha_j \leq 1,\quad \sum\limits_{i=1}^s\alpha_i=1, \quad
y_i \in R_+^n.\end{eqnarray*}
Then
\begin{eqnarray*}          g(\alpha_1y_1+\ldots+ \alpha_sy_s) \geq
\alpha_1g(y_1)+\ldots+ \alpha_sg(y_s),\end{eqnarray*}
\begin{eqnarray*}          0 \leq \alpha_j \leq 1,\quad \sum\limits_{i=1}^s\alpha_i=1, \quad
y_i \in R_+^n.\end{eqnarray*}
Let us define a technological map given on a compact convex set  $X$ by the formula
\begin{eqnarray*}          F(x)=\{y \in R_+^n,  \ f(y) \leq g(x)\}.\end{eqnarray*}
Assume that $f(0)=0$ and $f_i(y),\ i=\overline{1,n},$
are continuous and  strictly increasing    to the infinity
for every component $y_i,\
i=\overline{1,n},$ that increase to the infinity. Then
\begin{eqnarray*}        \alpha_1F(x_1)+\ldots+ \alpha_sF(x_s)
\end{eqnarray*}
\begin{eqnarray*} =\{\alpha_1y_1+\ldots+
\alpha_sy_s, f(y_1) \leq g(x_1), \ldots, f(y_s) \leq g(x_s)\}
\end{eqnarray*}
\begin{eqnarray*}         \subseteq \{ \alpha_1y_1+\ldots+ \alpha_sy_s,\
\alpha_1f(y_1)+\ldots+ \alpha_sf(y_s)\leq
\alpha_1g(x_1)+\ldots+ \alpha_sg(x_s)\}
\end{eqnarray*}
\begin{eqnarray*} \subseteq \{\alpha_1y_1+\ldots+ \alpha_sy_s, \
f(\alpha_1y_1+\ldots+ \alpha_sy_s) \leq g(\alpha_1x_1+\ldots+
\alpha_sx_s)\}
\end{eqnarray*}
\begin{eqnarray*}
\subseteq \{y=\alpha_1y_1+\ldots+ \alpha_sy_s, \
f(y) \leq g(\alpha_1x_1+\ldots+
\alpha_sx_s)\}
\end{eqnarray*}
\begin{eqnarray*}
\subseteq F(\alpha_1x_1+\ldots+ \alpha_sx_s).
\end{eqnarray*}
From the assumption  for  $f(y),$  the existence of a compact $Y$ that contains convex compact images  $F(x)$  follows.
So, the constructed technological map satisfies the  conditions of the Theorem \ref{nnl1}.

\section{Structure of the Kakutani continuous from above  technological maps\index{structure of the Kakutani continuous from above  technological maps} }

In this section, the general formula for  the Kakutani continuous from above  technological maps  and the theorem of continuation of this class of technological maps  will be proved  \cite{ 92, 106}.
As earlier, we assume that the set of  possible goods $S$ is a convex set.
\begin{definition}
Let  $T$ be a compact set\index{compact set} of a separable metric space\index{separable metric space} $K,$  $X$ be a closed bounded set from $S,$  $\Gamma$ be a certain closed  subset of the set $X\times T,$ and let $M$ be  a closed in coarse  topology  set of  probability measures on  $\Gamma$
 such that for  $X$ there holds the representation
\begin{eqnarray}          \label{prod1}
X=\left\{x \in S, \ x=\int\limits_{\Gamma}e(u, t) d\mu, \ \mu \in M\right\},
\end{eqnarray}
where $e(u, t)=u$ is a continuous map of the set  $\Gamma$ into the set $X.$
We say that the set  $M$ is local convex concerning the set  $X$ if the set
\begin{eqnarray}\label{prod2}
 \Delta(x)=\left\{\mu \in M, \ x=\int\limits_{\Gamma}u d\mu \right\}
\end{eqnarray}
is  convex one for all $x \in X.$
\end{definition}
The set $ \Delta(x) $ is the set of all probability measures  from the set  $M$  such  that for a point   $x \in X$ there holds the representation
\begin{eqnarray*}          x=\int\limits_{\Gamma}u d\mu.\end{eqnarray*}
\begin{theorem}\label{prodov1}
Let ~$T$ be a compact set of a separable metric space $K,$  $X$ be a closed bounded set from $S,$  $\Gamma$ be a certain closed subset of the set $X\times T,$
and let $M$ be a closed in coarse  topology\index{coarse  topology}  set of  probability measures on $\Gamma$
 such that for  $X$ there holds the representation (\ref{prod1}). If
 $g(u,t)$ is a continuous map of  the set  $\Gamma$ into the set $S$ the values of which lies in a certain compact  $Y \subseteq S,$
then the technological map given on the set  $X$ by the formula
\begin{eqnarray*}           F(x)=\bigcup\limits_{\mu \in \Delta(x)}\left\{y \in S, \  y=\int\limits_{\Gamma}g(u,t)d\mu\right\}\end{eqnarray*}
is the Kakutani continuous from above  with values in the set of closed bounded subsets of the set $S,$ where
\begin{eqnarray*}            \Delta(x)=\left\{\mu \in M, \ x=\int\limits_{\Gamma}u d\mu \right\}.\end{eqnarray*}
If the set  $X$ is  convex one and the set  $M$ is local convex concerning  the set $X$ and such that for any
\begin{eqnarray*}          y_1 \in F(x_1), \quad  y_2 \in F(x_2), \quad 0 < \alpha < 1, \quad x_1, \ x_2 \in X, \end{eqnarray*}
there exists  a measure
$\mu \in \Delta(\alpha x_1+(1 -\alpha)x_2)$ such that the representation
 \begin{eqnarray*}           \alpha y_1+(1 -\alpha)y_2=
\int\limits_{\Gamma}g(u,t)d\mu, \end{eqnarray*}
is valid, then
$F(x), \ x \in X,$ is a convex down technological map given on the  closed bounded set $X$ and takes values in the set of closed bounded convex subsets, there exists a compact  $Y_1$ such that  $ F(x) \subseteq Y_1 \subset S.$
\end{theorem}
\begin{proof}\smartqed
Prove the completeness of $F(x).$
Let a sequence  $y_n, \ n=  1,2, \ldots$ belong to $F(x)$ and converge to  $y_0.$ Show that
$y_0 \in F(x).$
If  $y_n \in F(x),$ then there exists a sequence of probability measures  $\mu_n$ such that
\begin{eqnarray*}          x=\int\limits_{\Gamma}u d\mu_n, \quad y_n=
\int\limits_{\Gamma}g(u,t)d\mu_n.\end{eqnarray*}
Since  $\Gamma$ is a compact set, the sequence of probability measures is weakly compact. It means that there exist a subsequence  $\mu_{n_k}$ of probability measures, a certain probability measure $\mu_0$ on  $\Gamma$  such that for any continuous bounded function  $f(u,t)$ on $\Gamma$ with  values in  $S$ there holds equality
\begin{eqnarray*}           \lim\limits_{k \to \infty}\int\limits_{\Gamma}f(u,t) d\mu_{n_k}= \int\limits_{\Gamma}f(u,t) d\mu_0. \end{eqnarray*}
Due to completeness of the set $M, $ $\mu_0 \in M.$
The last guarantees that
\begin{eqnarray*}
x=\lim\limits_{k \to \infty}\int\limits_{\Gamma}u d\mu_{n_k}=\int\limits_{\Gamma}u d\mu_0 ,\end{eqnarray*}
\begin{eqnarray*}   y_0 = \lim\limits_{k \to \infty}y_{n_k}=
\lim\limits_{k \to \infty}\int\limits_{\Gamma}g(u,t)d\mu_{n_k}=\int\limits_{\Gamma}g(u,t)d\mu_0.
\end{eqnarray*}
So, $y_0 \in F(x).$
Show the Kakutani continuity from above of  $F(x).$ Let  $x_n \in X, \ $ $ \lim\limits_{n \to \infty}x_n=x_0, $ and  $y_n \in F(x_n), \ \lim\limits_{n \to \infty}y_n = y_0.$ Show that
$y_0 \in F(x_0).$ As before, we have that there exists a sequence  of probability measures  $\mu_n$ such that
\begin{eqnarray*}          x_n=\int\limits_{\Gamma}u d\mu_n, \quad y_n=
\int\limits_{\Gamma}g(u,t)d\mu_n.\end{eqnarray*}
As a result of  the weak compactness of measures, we obtain the existence of a subsequence of probability measures
 $\mu_{n_k}$ and a probability measure  $\mu_0$ on $\Gamma$ such that
\begin{eqnarray*}
&    x_0=\lim\limits_{k \to \infty}x_{n_k}= \lim\limits_{k \to \infty}\int\limits_{\Gamma}u d\mu_{n_k}=\int\limits_{\Gamma}u d\mu_0 ,  \end{eqnarray*}
 \begin{eqnarray*}       y_0 = \lim\limits_{k \to \infty}y_{n_k}=
\lim\limits_{k \to \infty}\int\limits_{\Gamma}g(u,t)d\mu_{n_k}=\int\limits_{\Gamma}g(u,t)d\mu_0.
\end{eqnarray*}
Thus, $y_0 \in F(x_0).$
The last   means the Kakutani continuity from above.

Show the convexity of $F(x).$  If $y_1$ and  $y_2$ belong $F(x),$ then there exist measures
$\mu_1$ and  $\mu_2$ such that
\begin{eqnarray*}          x=\int\limits_{\Gamma}u d\mu_1, \quad x=\int\limits_{\Gamma}u d\mu_2,\end{eqnarray*}
\begin{eqnarray*}          y_1=\int\limits_{\Gamma}g(u,t)d\mu_1, \quad   y_2=\int\limits_{\Gamma}g(u,t)d\mu_2.\end{eqnarray*}
Hence, we have the representations
\begin{eqnarray*}          x=\int\limits_{\Gamma}u d(\alpha\mu_1+(1 -\alpha)\mu_2), \quad  (\alpha\mu_1+(1 -\alpha)\mu_2) \in \Delta(x),\end{eqnarray*}
\begin{eqnarray*}           \alpha y_1+(1 -\alpha)y_2=
\int\limits_{\Gamma}g(u,t)d(\alpha\mu_1+(1 -\alpha)\mu_2), \quad 0 \leq \alpha \leq 1.\end{eqnarray*}
From these representations, it follows that  $\alpha y_1+(1 -\alpha)y_2 \in F(x).$

Show the convexity  down of  $F(x).$
If  $y_1 \in F(x_1), \ y_2 \in F(x_2),$  then there exist measures
$\mu_1 \in \Delta(x_1)$ and $\mu_2 \in \Delta(x_2)$  such that
\begin{eqnarray*}
 x_1=\int\limits_{\Gamma}u d\mu_1, \quad x_2=\int\limits_{\Gamma}u d\mu_2, \end{eqnarray*}      \begin{eqnarray*}  y_1=\int\limits_{\Gamma}g(u,t)d\mu_1, \quad   y_2=\int\limits_{\Gamma}g(u,t)d\mu_2.
 \end{eqnarray*}
From here, we have the representations
\begin{eqnarray*}         \alpha x_1+(1 -\alpha)x_2=\int\limits_{\Gamma}u d(\alpha\mu_1+(1 -\alpha)\mu_2),
\end{eqnarray*}
\begin{eqnarray*}
 \alpha y_1+(1 -\alpha)y_2=
\int\limits_{\Gamma}g(u,t)d(\alpha\mu_1+(1 -\alpha)\mu_2)= \int\limits_{\Gamma}g(u,t)d\mu, \quad 0 \leq \alpha \leq 1.\end{eqnarray*}
Because   $\mu \in \Delta(\alpha x_1+(1 -\alpha) x_2)$ we have
\begin{eqnarray*}          \alpha y_1+(1 -\alpha)y_2 \in F(\alpha x_1+(1 -\alpha)x_2).\end{eqnarray*}
The compact $Y_1,$     the theorem said about, is the smallest convex  linear span generated by the points of the compact  $Y.$

\qed
\end{proof}

\begin{theorem}\label{prodov2}
Let  $F(x),\  x \in X,$ be a  technological map given on a closed bounded set $X$ that takes values in the set of closed bounded subsets of the set  $S,$
there exist a compact $Y$ such that $F(x) \subseteq Y, \  x \in X.$
If $F(x)$  is  the Kakutani continuous from above, then there exists  a closed set of probability measures\index{closed set of probability measures}  $M_0,$  a compact $T$ in  a separable metric space,  a closed subset $\Gamma \subseteq X\times T,$  and a continuous map  $g(u,t)$ on  $\Gamma$ that satisfy conditions of the Theorem \ref{prodov1} and such that
\begin{eqnarray*}           F(x)=\bigcup\limits_{\mu \in \Delta(x)}\left\{y \in S, \  y=\int\limits_{\Gamma}g(u,t)d\mu\right\}, \quad  \Delta(x)=\left\{\mu \in M_0, \ x=\int\limits_{\Gamma}u d\mu \right\}.\end{eqnarray*}
If  $X$ is  a convex closed bounded set and  $F(x),\  x \in X,$ is additionally convex down,  then  for any
\begin{eqnarray*}          y_1 \in F(x_1), \quad  y_2 \in F(x_2), \quad 0 < \alpha < 1, \quad x_1, \ x_2 \in X, \end{eqnarray*}
there exists a probability measure
$\mu \in \Delta(\alpha x_1+(1 -\alpha)x_2)$ such that there holds the representation
 \begin{eqnarray*}           \alpha y_1+(1 -\alpha)y_2=
\int\limits_{\Gamma}g(u,t)d\mu, \end{eqnarray*}
if, moreover,  $ F(x)$  is a convex closed bounded set for all $ x \in X,$ then $M_0$ is local convex concerning the set $X.$
 \end{theorem}
\begin{proof}\smartqed          Point out a family of measures  $M_0,$ a compact  $T$ in a separable metric space, a closed set $\Gamma,$ and a map  $g(u,t)$  that satisfy the conditions of the Theorem \ref{prodov1} and under that the statement of the Theorem is valid. Choose as a compact  $T,$ the set $Y$ that figures in the condition of the Theorem and the map  $g(u,t)=t, \ (u, t) \in \Gamma=\{(u,t), \ u \in X, \ t \in F(u)\} .$  We choose the set of probability measures\index{set of probability measures} in the form
$M_0=\bigcup\limits_{x \in X}{M_0^x},$ where  $M_0^x$ is a set of probability measures concentrated at the points of the form
$(x, y), \ x \in X, \ y \in F(x),$

\begin{eqnarray*}            \mu_{(x,y)}(A)=\left\{\begin{array}{ll} 1, &  \textrm{if} \quad (x, y) \in  A,  \\
                          0, &\textrm{if}  \quad (x, y) \in [X\times Y] \setminus A   \textrm{,}
                                  \end{array}
                                  \right. \quad  A \in {\cal B}(X\times Y).  \end{eqnarray*}
The introduced family of measures  $M_0$ is a closed set concerning the weak convergence of measures  because of the Kakutani continuity  from above of  $F(x).$ The set $\Delta(x) = M_0^x$ is a convex set for every $ x \in X,$ and so $M_0$ is local convex concerning the set $X.$
Under these conditions, the validity  of the stated formula is obvious.
The declared measure  is expressed as
\begin{eqnarray*}          \mu(A)=\mu_{(\alpha x_1+(1 -\alpha)x_2,\ \alpha y_1+(1 -\alpha)y_2 )}(A), \quad A \in {\cal B}(X\times Y).\end{eqnarray*}
\qed
\end{proof}

\begin{definition}
Let $F_1(x), \ x \in X_1,$ and $F_2(x), \ x \in X_2,$ be two technological maps given on closed bounded sets $X_1$  and $X_2, \ X_1 \subset X_2,$ that take  values in the set of closed bounded subsets. The technological map  $F_2(x), \ x \in X_2$ is called the continuation of the technological map\index{continuation of the technological map} $F_1(x), \ x \in X_1,$ if  $F_1(x) \subseteq F_2(x), \ x \in X_1.$
\end{definition}
We   call the next Theorem  the Theorem of continuation.\index{Theorem of continuation}
\begin{theorem}\label{prodov3}
Let $F_1(x), \ x \in X_1,$ and $F_2(x), \ x \in X_2,$ be two  Kakutani continuous from above technological maps given on closed bounded sets  $X_1$  and $X_2$  and take values in the set of closed bounded subsets of a certain compact set  $Y \subseteq S.$ Then there exists the Kakutani continuous from above technological map $F(x)$ given on the minimal convex linear span $X$ of the sets $X_1$  and $X_2$ such that it is convex down on  $X,$ takes values in the set of closed bounded subsets that belong to a certain compact  $Y_1,$ and inclusions $ F(x) \supseteq F_1(x), \  x \in X_1, $ $ F(x) \supseteq F_2(x), \  x \in X_2,$
take place.
\end{theorem}
\begin{proof}\smartqed          Denote by  $M^{\Gamma_1}$ and  $M^{\Gamma_2}$ the set of atomic probability measures\index{set of atomic probability measures} concentrated at points  $(x,y) \in \Gamma_1$ and at points
 $(x,y) \in \Gamma_2,$ correspondingly, where
\begin{eqnarray*}          \Gamma_1=\{(x,y), \ x \in X_1,\  y \in F_1(x)\}, \quad \Gamma_2=\{(x,y), \ x \in X_2,\ y \in F_2(x)\}.\end{eqnarray*}
We assume that  $M^0$ is a completion of linear convex span of probability measures concerning a weak convergence of measures generated by the set of atomic probability measures  $M^{\Gamma_1}\cup M^{\Gamma_2}$ defined on the  measurable space
 $\{\Gamma_1\cup \Gamma_2, {\cal B}(\Gamma_1\cup \Gamma_2)\}.$
Let  $D$ be a compact of a certain separable metric space.
By linear convex span of a certain set of measures\index{linear convex span of a certain set of measures}  $M_1$ on a certain measurable space  $\{D,\ \Sigma \},$ we understand the set of measures of the form
\begin{eqnarray*}          M_1^{lcs}=\left\{\mu=\sum\limits_{i=1}^n\gamma_i\mu_i, \ \mu_i \in M_1, \  \{\gamma_i\}_{i=1}^n \in P_n, \ n=1, 2, 3, \ldots\right\}.\end{eqnarray*}           As before,
 $P_n=\left\{\{\gamma_i\}_{i=1}^n  \in R_+^n, \ \sum\limits_{i=1}^n\gamma_i=1\right\}.$
It is evident that such completion exists  and it is an intersection of all closed relative to weak convergence of measures\index{weak convergence of measures}  convex sets of probability measures that contains the set of probability measures  $M_1.$ Really, the set of all probability measures on  $\{D,\ \Sigma \}$ is a convex set that is closed relative to weak convergence of measures  and contains the set $M_1.$ Therefore, there exists  the closure of the minimal linear convex span relative to a weak convergence of measures  that contains  $M_1$ and which is an intersection of all convex  sets of probability measures closed relative to weak convergence of measures.

Consider a technological map  $F(x), \ x \in X,$ given by the formula
\begin{eqnarray*}           F(x)=\bigcup\limits_{\mu \in \Delta(x)}\left\{y, \ y=\int\limits_{\Gamma_1\cup \Gamma_2}t d \mu\right\}, \quad \Delta(x)=\left\{\mu \in M^0, \ x=\int\limits_{\Gamma_1\cup \Gamma_2}u d \mu\right\},\end{eqnarray*}
where   $X$ is a set of the form
\begin{eqnarray*}          X=\left\{x \in S, \ x=\int\limits_{\Gamma_1\cup \Gamma_2}u d \mu, \  \mu \in M^0\right\}.\end{eqnarray*}
The constructed technological map satisfies all assertions of the Theorem, thanks to the validity of  conditions of the Theorem   \ref{prodov1}.
\qed
\end{proof}

\begin{corollary}\label{prod3}
Any  technological map   $F(x), \ x \in X,$  from the CTM class in a wide sense admits a continuation  to  a technological map  from the CTM class.
\end{corollary}
\begin{proof}\smartqed          Consider two technological maps   $F_1(x), \ x \in X_1,$ where
 $F_1(x)=F(x),\ X_1=X,$ and  $F_2(x), \ x \in X_2,$ where the set  $X_2$ contains only one point $x=0,$  and $F_2(0)=0.$  These two technological maps satisfy  the conditions of the Theorem  \ref{prodov3}.
Then there exists  the Kakutani continuous from above  technological map  $F_3(x)$ given on the minimal linear convex span  $X_3$ of the sets  $X_1$  and $X_2$ such that it is a convex down on  $X_3,$ takes  values in the set of closed bounded convex subset that belong to a certain compact  $Y_1$ and inclusions   $ F_3(x) \supseteq F_1(x), \  x \in X_1, $ $ F_3(x) \supseteq F_2(x), \  x \in X_2$ take place.

\qed
\end{proof}

In the next Theorem, we construct  convex down technological maps  on a  wider set of closed convex sets of measures. We assume that a certain closed bounded convex  set  $X$  is given and the set of all finite  measures  $N$ on the measurable space   $\{T, {\cal B}(T)\}$ such that $m(T) \leq B_0 < \infty, \ m(t) \leq 1,$ $ \ t \in T, \  m  \in N,$ where  $T$ is a compact  set of a separable metric space  $K.$  It is obvious that the set of measures  $N$ is a closed relative to the weak topology  and is convex.
\begin{theorem}\label{prodov4}
Let ~$T$ be a compact set of a separable metric space  $K,$  $X$ be  a closed bounded convex set  from  $S,$  $\Gamma= X\times T,$ \ $\varphi (u, t)$ be a continuous map of the set $\Gamma$ into the set  $R_+^1$ that is strictly positive.
The technological map given on the set $X$ by the formula
\begin{eqnarray*}           F(x)=\bigcup\limits_{m \in N}\left\{y \in S, \  y=\int\limits_{T}\varphi(x,t)g(t)dm\right\}, \end{eqnarray*}
where $g(t)$ is a continuous map on $T$ with  values in the set $S,$
is the Kakutani  continuous from above  with  values in the set of closed bounded convex  subsets  of the set $S$ that is convex down if the  next condition is satisfied:
for every  $t \in T$ the map  $\varphi(x,t)$ is convex  up on the set $X.$
There exists a compact  $Y_1$ such that   $ F(x)  \subseteq Y_1 \subset S.$
\end{theorem}
\begin{proof}\smartqed          We prove the closure and convexity of  $F(x),$ as in the Theorem  \ref{prodov1}.
Show the Kakutani continuity from above of  $F(x).$ Let  $x_n \in X, \ \lim\limits_{n \to \infty}x_n=x_0, $ and  $y_n \in F(x_n), \ \lim\limits_{n \to \infty}y_n = y_0.$ Show that
$y_0 \in F(x_0).$ As before, we have that there exists a sequence of probability measures  $m_n$ such that
\begin{eqnarray*}           y_n=
\int\limits_{T}\varphi(x_n,t)g(t)dm_n.\end{eqnarray*}
Due to the  weak compactness of the measures, we obtain the existence of  a subsequence of probability measures  $m_{n_k}$ and a measure    $m_0$ on  $T$ such that
\begin{eqnarray*}
\lim\limits_{k \to \infty}\int\limits_{T}\varphi(x_0,t)g(t)dm_{n_k}=\int\limits_{T}\varphi(x_0,t)g(t)dm_0.\end{eqnarray*}
Show that  $y_0=\int\limits_{T}\varphi(x_0,t)g(t)dm_0.$
Really, if
$g(t)=\{g_i(t)\}_{i=1}^n \in S,$ the norm of a vector  $y \in S$ is given by the formula $||y||=\max\limits_{i}|y_i|,$ then if we choose  $k_0$ such that for  $k > k_0$ the inequalities  \begin{eqnarray*}
     \left|\left|\int\limits_{T}\varphi(x_0,t)g(t)dm_0 - \int\limits_{T}\varphi(x_0,t)g(t)dm_{n_k}\right|\right| \leq \varepsilon, \quad ||x_{n_k} - x_0|| <\delta,
\end{eqnarray*}
 \begin{eqnarray*}
 |\varphi(x_{n_k},t) - \varphi(x_0,t)| < \frac{\varepsilon}{R_0}, \quad R_0=\sup\limits_{t \in T}||g(t)||,
 \end{eqnarray*}
would be valid, then we have for
  $k > k_0$
\begin{eqnarray*}           \left|\left|y_{n_k} - y_0\right|\right| =\left|\left|\int\limits_{T}\varphi(x_{n_k},t)g(t)dm_{n_k} - \int\limits_{T}\varphi(x_0,t)g(t)dm_0\right|\right|\end{eqnarray*}
\begin{eqnarray*}
 =\left|\left|\int\limits_{T}[\varphi(x_{n_k},t) - \varphi(x_0,t)]g(t)dm_{n_k}\right.\right.
 \end{eqnarray*}
 \begin{eqnarray*} \left.\left.   - \int\limits_{T}\varphi(x_{0},t)g(t)dm_0 + \int\limits_{T}\varphi(x_0,t)g(t)dm_{n_k}\right|\right|
 \end{eqnarray*}
\begin{eqnarray*} \leq \int\limits_{T}|\varphi(x_{n_k},t) - \varphi(x_0,t)| ||g(t)||dm_{n_k} \end{eqnarray*}
\begin{eqnarray*} +
\left|\left|\int\limits_{T}\varphi(x_0,t)g(t)dm_0 - \int\limits_{T}\varphi(x_0,t)g(t)dm_{n_k}\right|\right| \leq 2 \varepsilon. \end{eqnarray*}
Thus,  due to the arbitrariness of  $ \varepsilon, $ \ $y_0 \in F(x_0).$

Note that
\begin{eqnarray*}          |\varphi(x_{n_k},t) - \varphi(x_0,t)| \leq  \frac{\varepsilon}{R_0}\end{eqnarray*}
uniformly on the compact
$X \times T,$ on the basis of  the Cantor Theorem\index{Cantor Theorem} about uniform continuity,   as soon as for sufficiently small  $\delta> 0,$
$||x_{n_k} - x_0|| <\delta.$
Show the convexity  down of  $F(x).$
If  $y_1 \in F(x_1), \ y_2 \in F(x_2),$ then there exist measures
$m_1 \in N$ and $m_2 \in N$ such that
\begin{eqnarray*}          y_1=\int\limits_{T}\varphi(x_1,t)g(t)dm_1, \quad   y_2=\int\limits_{T}\varphi(x_2,t)g(t)d m_2.\end{eqnarray*}
Hence, we have such representation
\begin{eqnarray*}           \alpha y_1+(1 -\alpha)y_2=
\int\limits_{T}\varphi(\alpha x_1+ (1 -\alpha)x_2,t)g(t)dm_3, \quad 0 \leq \alpha \leq 1,\end{eqnarray*}
where
\begin{eqnarray*}           m_3(A)=\int\limits_{A}\frac{\alpha \varphi(x_1,t)}{ \varphi(\alpha x_1+ (1 -\alpha)x_2,t)}d m_1 +  \int\limits_{A}\frac{(1 -\alpha) \varphi(x_2,t)}{ \varphi(\alpha x_1+ (1 -\alpha)x_2,t)}d m_2.\end{eqnarray*}
Show that  $m_3$ belongs to $N.$ For this, it is sufficient to prove that
$m_3(t) \leq 1, \ t \in T,$ $m_3(T)\leq B_0.$
Really,
\begin{eqnarray*}          m_3(t)=\frac{\alpha \varphi(x_1,t)}{ \varphi(\alpha x_1+ (1 -\alpha)x_2,t)} m_1(t) +
\frac{(1 -\alpha) \varphi(x_2,t)}{ \varphi(\alpha x_1+ (1 -\alpha)x_2,t)} m_2(t) \end{eqnarray*}
\begin{eqnarray*}
 \leq \frac{\alpha \varphi(x_1,t)}{ \varphi(\alpha x_1+ (1 -\alpha)x_2,t)} +
\frac{(1 -\alpha) \varphi(x_2,t)}{ \varphi(\alpha x_1+ (1 -\alpha)x_2,t)}  \leq 1.
\end{eqnarray*}
There holds the bound
\begin{eqnarray*}           m_3(T) \leq B_0 \left[\sup\limits_{t \in T}\frac{\alpha \varphi(x_1,t)}{ \varphi(\alpha x_1+ (1 -\alpha)x_2,t)}+\sup\limits_{t \in T}\frac{(1 -\alpha) \varphi(x_2,t)}{ \varphi(\alpha x_1+ (1 -\alpha)x_2,t)}\right]
 \end{eqnarray*}
 \begin{eqnarray*}
 =  B_0 \left[\frac{\alpha \varphi(x_1,t_1)}{ \varphi(\alpha x_1+ (1 -\alpha)x_2,t_1)}+\frac{(1 -\alpha) \varphi(x_2,t_2)}{ \varphi(\alpha x_1+ (1 -\alpha)x_2,t_2)}\right].\end{eqnarray*}
Here, $t_i$ is a point in which the function
\begin{eqnarray*}           \frac{ \varphi(x_i,t)}{ \varphi(\alpha x_1+ (1 -\alpha)x_2,t)}, \quad i=1,2.\end{eqnarray*}
reaches the  maximal value.
Such maximum exists since this function is  continuous on the compact  in accordance with the assumptions of the Theorem.
However,
\begin{eqnarray*}          \frac{\alpha \varphi(x_1,t_1)}{ \varphi(\alpha x_1+ (1 -\alpha)x_2,t_1)}+\frac{(1 -\alpha) \varphi(x_2,t_2)}{ \varphi(\alpha x_1+ (1 -\alpha)x_2,t_2)} \leq  1\end{eqnarray*}
for any  $t_1, t_2 \in T.$
Really, at first  let the inequality
\begin{eqnarray*}          \frac{\varphi(x_1,t_1)}{ \varphi(\alpha x_1+ (1 -\alpha)x_2,t_1)} \leq \frac{ \varphi(x_1,t_2)}{ \varphi(\alpha x_1+ (1 -\alpha)x_2,t_2)}\end{eqnarray*}
take place.
Then
\begin{eqnarray*}          \frac{\alpha \varphi(x_1,t_1)}{ \varphi(\alpha x_1+ (1 -\alpha)x_2,t_1)}+\frac{(1 -\alpha) \varphi(x_2,t_2)}{ \varphi(\alpha x_1+ (1 -\alpha)x_2,t_2)} \end{eqnarray*}
 \begin{eqnarray*}        \leq \frac{\alpha \varphi(x_1,t_2)}{ \varphi(\alpha x_1+ (1 -\alpha)x_2,t_2)}+
 \frac{(1 -\alpha) \varphi(x_2,t_2)}{ \varphi(\alpha x_1+ (1 -\alpha)x_2,t_2)}.
 \end{eqnarray*}
In consequence of the convexity  up of $\varphi(x,t_2),$ we have the  inequality
\begin{eqnarray*}          \frac{\alpha \varphi(x_1,t_2)}{ \varphi(\alpha x_1+ (1 -\alpha) x_2 ,t_2)}+
 \frac{(1 -\alpha) \varphi(x_2,t_2)}{ \varphi(\alpha x_1+ (1 -\alpha)x_2,t_2)} \leq  1 .\end{eqnarray*}
If there holds a contrary inequality, that is,
\begin{eqnarray*}          \frac{ \varphi(x_1,t_2)}{ \varphi(\alpha x_1+ (1 -\alpha)x_2,t_2)} \leq \frac{\varphi(x_1,t_1)}{ \varphi(\alpha x_1+ (1 -\alpha)x_2,t_1)},\end{eqnarray*}
then we act similarly. Therefore, $m_3(A) \in N.$
So,
\begin{eqnarray*}          \alpha y_1+(1 -\alpha)y_2 \in F(\alpha x_1+(1 -\alpha)x_2).\end{eqnarray*}
The compact  $Y_1,$    the Theorem said about, is the smallest linear convex span generated by the  points of the compact  $Y.$

\qed
\end{proof}

There holds a more general statement than the Theorem  \ref{prodov4}.
\begin{theorem}\label{prodov5}
Let ~$T$ be a compact set of a separable metric space  $K,$  $X$ be a closed bounded convex  set from  $S,$  $\Gamma= X\times T,$ \ $ g(u, t)$ be a continuous map of the set $\Gamma$ into the set $S.$
The technological map given on the set $X$ by the formula
\begin{eqnarray*}           F(x)=\bigcup\limits_{m \in N}\left\{y \in S, \  y=\int\limits_{T}g(x,t)dm\right\} \end{eqnarray*}
is the Kakutani continuous  with  values  in the set  of closed bounded  convex subsets of the set $S,$ that is convex down, there exists a compact  $Y_1$ such that   $ F(x) \subseteq Y_1 \subset S$ if the conditions are satisfied:
the set of uniformly bounded measures $N$ is a convex set and a continuous map  $g(x,t)$  for every $x_1, x_2 \in X, \ 0 \leq \alpha \leq 1$ and arbitrary  $m_1, m_2 \in N$ is such that there exists a measure
$m_3 \in N$ for which there holds equality
\begin{eqnarray*}            \alpha \int\limits_{T}g(x_1,t)dm_1 + (1 -\alpha)\int\limits_{T}g(x_2,t)dm_2= \int\limits_{T}g(\alpha x_1+(1 -\alpha)x_2,t)dm_3.\end{eqnarray*}
\end{theorem}
The proof of this statement follows argumentation of the previous statements.

As a consequence of the Theorem  \ref{prodov5}, there holds the following statement. Let a compact  $T=N_0,$ where
$N_0=[1,2, \ldots, N].$ Let us give on  $T$ a family of measures  $M_0$ that satisfy condition
\begin{eqnarray*}           \sum\limits_{i=1}^Nm(i)\leq 1, \quad m(i) \geq 0, \quad i=\overline{1,N}.\end{eqnarray*}
Let us put
\begin{eqnarray}          \label{prodov7}
 F(x)=\bigcup\limits_{m \in M_0}\left\{y \in S,  \ y=\sum\limits_{j=1}^Ng(x,j)m(j)\right\},
\end{eqnarray}
where the map  $g(x,j)$ is a continuous function of the variable $x \in X$  into the set $S$ for every fixed  $j \in T$ and, besides, the map  $g(x,j)$ for every  $x_1, \  x_2 \in X, \ 0 \leq \alpha \leq 1$ and arbitrary $m_1, m_2 \in M_0$ is such that there exists a measure $m_3 \in M_0$ that the equality holds
\begin{eqnarray*}         &   \alpha \sum\limits_{j=1}^Ng(x_1,j)m_1(j) + (1 -\alpha)\sum\limits_{j=1}^Ng(x_2,j)m_2(j) \\
& = \sum\limits_{j=1}^Ng(\alpha x_1+(1 -\alpha)x_2,j)m_3(j).\end{eqnarray*}
Assume that  $X$ is a closed convex bounded subset of the set $S.$ Under these conditions all conditions of the previous Theorem are satisfied, therefore  $F(x), \ x \in X,$ is a convex down technological map that is  the Kakutani continuous  from above.

Show that among such maps there are the Leontieff and Neumann technological maps.
Really, if  $F(x), \ x \in X,$  is the Leontieff  technological map given by a certain expenditure matrix $A=||a_{ij}||_{i,j=1}^n,$ then $F(x)=\{y \in S, \ Ay \leq x \}, \  x \in X.$  Let $N$ be  the number of non degenerate square minors of the matrix  $A.$ Let us number them in a certain order. Consider one of  such  minors  $a_{I,J}=||a_{ij} ||_{i \in I, j \in J}$ that corresponds to index   $s.$  In accordance with the Theorem \ref{krt1},
the strictly positive components of the extreme point  $y_s(x)$ of the technological map $F(x),$ whose indices  belong to the set $J,$
satisfy the set of equations
\begin{eqnarray}\label{prodov8}
 \sum\limits_{j \in J}a_{ij}y_j^s(x)=x_i, \quad i \in I,
\end{eqnarray}
the rest components of the vector $y_s(x)$ that belong to the set $J$ equal  zero.
Two cases are possible. For the given vector  $x \in X$  a solution to the set of equations  (\ref{prodov8})
exists and satisfies inequalities
\begin{eqnarray}\label{prodov9}
 \sum\limits_{j \in J}a_{ij}y_j^s(x) < x_i, \quad i \in N_0 \setminus I,
\end{eqnarray}
then $y_s(x)$ is an extreme point of the map $F(x).$ The second possible case is that there no exist a strictly positive solution  of the set of equations   (\ref{prodov8}) and inequalities  (\ref{prodov9}). Let us put
\begin{eqnarray*}           g(x,s) 
=\left \{ \begin{array}{ll}  y_s(x), & \textrm{if there exists  a strictly positive solution to (\ref{prodov8}), (\ref{prodov9}),}\\
0, & \textrm{if there  exist no a strictly positive solution to (\ref{prodov8}), (\ref{prodov9})}.
                                                    \end{array} \right. \end{eqnarray*}

It is obvious that  $g(x,s)$ is a continuous function of argument  $x \in X$  for every fixed $s, \ s=\overline{1, N},$ and satisfies the condition:
the map  $g(x,j)$ for every $x_1, x_2 \in X, \ 0 \leq \alpha \leq 1$ and arbitrary  $m_1, m_2 \in M_0$ is such that there exists a measure
$m_3 \in M_0$ that there holds equality
\begin{eqnarray*}
&   \alpha \sum\limits_{j=1}^Ng(x_1,j)m_1(j) + (1 -\alpha)\sum\limits_{j=1}^Ng(x_2,j)m_2(j) \\
& = \sum\limits_{j=1}^Ng(\alpha x_1+(1 -\alpha)x_2,j)m_3(j),\end{eqnarray*}
since it is a consequence of the convexity  down of the Leontieff technological map.

\chapter{Absence of arbitrage in  information model of economy}

\abstract* {On the basis  of random fields of consumer choice and decisions making by firms,  the main  objects of mathematical  information economics are constructed: demand and supply of the society  being constructed  from demand vectors of consumers, their  income functions  and supply of firms, correspondingly. The Theorem guaranteeing  the existence of equilibrium price vector in case of aggregated  description of economy system  is proved. The set of equations to that equilibrium
price vector satisfy are obtained.
In the case of availability of non-insatiable consumers in the economy system the Theorems giving algorithm of construction of all equilibrium states are proved. The examples  of stochastic  model of economy whose every industry is profitable in the equilibrium state are presented. Under the presence  of non-insatiable consumer, the set of equilibrium states is continual one.
 The Theorem of absence of  arbitrage  in the model of non-aggregated description of economy system is proved and the set of equations satisfied  by the Walras equilibrium state is found.}

In this Chapter, the  solution of the problem of the absence of  space  arbitrage\index{absence of  space  arbitrage} in the  information model of economy is given both in the case of aggregated description\index{aggregated description of the  economy system} and non-aggregated description of the  economy system  \cite{55, 71, 69, 92, 106}.\index{non-aggregated description of the  economy system} The absence of arbitrage in an economy system  means the existence with probability  one of an equilibrium price vector  under which  demand does not exceed supply.  Under such conditions  economic agents   have no possibility to buy goods for  low prices and to sell their  for  high  prices.

Here, the main mathematical model of economy system  is constructed in which the choice of consumers and decisions making by firms are described by random fields of consumer choice and  random fields of decisions making by firms.
The random fields of consumers choice and  random fields of decisions making by firms  are mutually dependent and contain information about the state of the economy system.
The investigation is carried  out in general form without special  definition of income pre-functions of  consumers and  a productive economic process.
The fundamental notion is the notion of income function of a consumer given  by Definition \ref{dl1} and the notion of random demand vector\index{ random demand vector} that has a simple  economic sense. Simple one-to-one correspondence  exists  between  the random demand vector and the random field of   consumer choice.
Demand  and supply  vectors of the  society,\index{demand  and supply  vectors of the  society} using  these
notions,  are constructed rather simply.
The structures of demand  and supply of the society are described on the basis of the   structure of random fields of  consumers choice  and decisions making
by firms   proposed in  Subsection 1. Notions of the Walras equilibrium state\index{Walras equilibrium state}
and   optimal  Walras equilibrium state\index{optimal  Walras equilibrium state}  are introduced. If an economy system is in the Walras equilibrium state, then   the  space arbitrage is absent  in it.
At first, the case of insatiable consumers is  considered. A key to  solve  the problem of the absence of arbitrage  in this case is the Theorem \ref{ch2l3}. In this Theorem, under sufficiently general assumptions on technological maps, the family of income pre-functions, and property vectors, the solvability of the set of equations  (\ref{al1}) with probability 1  in the unit simplex  of the nonnegative cone  is proved.
The proof of this Theorem is constructive   since  a   certain structure is presented that reduces the solution of this  set of equations to the Scarf algorithm\index{Scarf algorithm}  \cite{12}.
 In the Theorem \ref{doda1}, the conditions  are weakened under that there exists a solution to the set of equations. Under quite  general assumptions about technological maps of firms,  a family of income pre-functions, a productive economic process,  and property
vectors, the Theorem  \ref{ch2l1} of the absence of space arbitrage with probability 1 is proved, that is, the existence of the Walras  equilibrium  for all continuous realizations of random fields of consumers choice  and decisions making by firms relative to productive processes.  Under very general conditions,
 there  exists the  optimal Walras equilibrium state.
The next problem is a construction of a map,  whose fixed points  would be equilibrium states in the proposed model of economy. This problem is  solved by the Theorem \ref{ch2l2} in the case when all  consumers are insatiable. In this Theorem, sufficient conditions are formulated  under
the fulfilment of which  the Walras equilibrium state  satisfies the set of equations (\ref{g2l18}). The latter allow to use the Scarf algorithm \cite{12} to construct   equilibrium price vector that guarantees the absence of arbitrage in the constructed economic system.

The problem of construction of a map, whose fixed points (they and only they)  are fixed points of a certain family of maps in the case when not all consumers are insatiable  is solved by the Theorem \ref{ch2l4}.
The Theorem  \ref{ch2l5} gives an algorithm of construction  of all equilibrium states  when not all consumer are insatiable. The Arrow - Debreu model\index{Arrow - Debreu model} \cite{11, 13} is a partial case
of the economy model  considered, therefore, the problem of construction of a map
whose fixed point  (they and only they)  are the Walras equilibrium states is solved  for
the Arrow - Debreu model too.
In subsections 3.1.1 --- 3.1.3, examples of stochastic models of economy whose every branch   is profitable are presented. In these  models of economy, the specific examples of productive economic processes and demand vectors under which the equilibrium price vector  guarantees the profitability of every branch are constructed. The Theorem  \ref{ch2l10} generalizes the Arrow - Debreu Theorem  \cite{11, 13} and it is a consequence of the proved Theorems \ref{ch2l1} --- \ref{ch2l5}.

In subsection  3.2,  a model of non-aggregated description\index{model of non-aggregated description} of the  economy system is considered and the very important Theorem  \ref{1ch2l2} is proved  in that the existence of a solution  with probability one   to the set of equations (\ref{1g2l18}) is showed in a certain constructed  closed convex  subset of the unit simplex.

In the  Theorem \ref{1ch2l3}, very important for applications, the absence of arbitrage in the model of non-aggregated description of economy system is proved  and the set of equations   satisfied by the Walras equilibrium states  is found.
The proof of the existence of the Walras equilibrium state in this case is very important because  the strategies of consumers behavior are discontinuous on the unit simplex.
 This model is used in the next sections for the  investigation of real economy systems, in particular, the economy of Ukraine. The Theorem \ref{nen1} guarantees the absence of arbitrage  in the case when not all consumers are insatiable for the model of non-aggregated description of economy system.
From the  investigation carried  out,   the conclusion follows that, under the presence of
non-insatiable consumers in the economy system,  the set  of  Walras equilibrium states in the economy system  can be continual one. This  situation is of general  nature.
Therefore, we propose to consider, as the Walras equilibrium state  for such economy systems,
 the Walras equilibrium price vector that   minimizes the function of capital not used.\index{function of capital not used}
The notion of the  minimal and maximal degrees of possibility of a financial crisis\index{minimal and maximal degrees of possibility of a financial crisis} is introduced.
In the next sections we  carry on the search of models of economy in which the Walras equilibrium state is unique under conditions of insatiability of all consumers.
It will be reached by a certain claims relative to  the  agreement of  the structure of choice and the structure of supply that\index{agreement of  the structure of choice and the structure of supply } is natural.

\section{Absence of arbitrage in the economy  mo\-dels described aggregately }

We consider a model of economy system,  where  $m$ firms operate, whose production is described by  technological maps  $F_i(x), \ x \in X_i^1,$ $ \ i=\overline{1,m},$ belonging to the CTM or to the CTM class in a wide sense. We assume that, in the economy system, there are  $l$ consumers
that have the  property given by the property vectors $b_i(p,z), \ i=\overline{1,l},$ and there are $n$ kinds of goods  that are used by consumers.

In Chapter 1, we described the choice of consumers  by random fields of consumers choice.
In this section, we additionally assume  that  random fields of consumers choice
are continuous with probability 1 on the set of possible prices $\bar R_+^n,$ that is, every strategy of behavior of every consumer is  continuous if the opposite is not mentioned.
The last assumption restricts the possible spectrum of goods that are consumed.
Really, every consumer has to consume all goods available  in the  economy system, and this
 agrees with reality under aggregated description  of consumption.

Being guided by  the Theorems    \ref{tl4} and \ref{ttl3}, we introduce a fundamen\-tal notion:  a demand vector of the $i$-th consumer.
Since, for the random field of choice of the  $i$-th consumer, there hold representations  (\ref{pl7})
and   (\ref{ppl7}) let us define, on their basis, a random field that we call a random demand vector.
\begin{definition}\label{ch2d1}  Let
\begin{eqnarray*}         \eta_i^0(p,z,\omega_i)=
\{\eta_{ik}^0(p,z,\omega_i)\}_{k=1}^n, \quad  i=\overline{1,l},   \end{eqnarray*}
be  a random field of evaluation of information by  the  $i$-th consumer and
\begin{eqnarray*}         \zeta(p, \omega_0)=Q(p,\zeta_0(p, \omega_0))=\{ \zeta_i(p)\}_{i=1}^m\end{eqnarray*}
be a random field of decisions making by firms that satisfy the conditions of the Theorem  \ref{tl4}.
We call a random field
\begin{eqnarray*}         \gamma_i(p,\omega_0,\omega_i) =\{ \gamma_{ik}(p,\omega_0,\omega_i) \}_{k=1}^n,
\quad  i=\overline{1,l}, \end{eqnarray*}
 a random demand vector of the $i$-th  insatiable consumer,\index{random demand vector of   insatiable consumer} where
\begin{eqnarray*}          \gamma_{ik}(p,\omega_0,\omega_i)
=\frac{p_k\eta_{ik}^0(p, \zeta(p, \omega_0),\omega_i)}
{\sum\limits_{s=1}^n\eta_{is}^0(p, \zeta(p, \omega_0),\omega_i)p_s},
\quad  i=\overline{1,l}, \quad  k=\overline{1,n}.\end{eqnarray*}
\end{definition}
Take notations  that we shall use in further consideration.
We denote any realization of the random  demand vector of the $i$-th insatiable consumer  by
\begin{eqnarray*}          \gamma_i(p)=\{ \gamma_{ik}(p)\}_{k=1}^n,\end{eqnarray*}
where $ \gamma_{ik}(p)=\gamma_{ik}(p,\omega_0,\omega_i)$
 at  certain  $\omega_0 $ and $\omega_i,$
and we call it the demand vector of the  $i$-th insatiable consumer.
We denote any realization of random  field
\begin{eqnarray*}         \eta_i^0(p,\zeta(p, \omega_0),\omega_i)
=\{\eta_{ik}^0(p,\zeta(p, \omega_0),\omega_i)\}_{k=1}^n \end{eqnarray*}
 by
\begin{eqnarray*}          \mu_i(p)=\{\mu_{ki}(p)\}_{i=1}^n,\end{eqnarray*}
where
\begin{eqnarray*}          \mu_{ki}(p)=\eta_{ik}^0(p,\zeta(p, \omega_0),\omega_i)\end{eqnarray*}
at  certain  $\omega_0 $ and $\omega_i.$
Under such notations
\begin{eqnarray*}         \gamma_{ik}(p)=\frac{p_k\mu_{ki}(p)}{\sum\limits_{s=1}^n\mu_{si}(p)p_s},\quad  i=\overline{1,l},\quad  k=\overline{1,n}.\end{eqnarray*}

\begin{definition}\label{sh2d1} Let
\begin{eqnarray*}         \eta_i^1(p,z,\omega_i)
=\{\eta_{ik}^1(p, z,\omega_i)\}_{k=1}^n, \quad  i=\overline{1,l},  \end{eqnarray*}
and
\begin{eqnarray*}         \eta_i^0(p,z,\omega_i)
=\{\eta_{ik}^0(p, z,\omega_i)\}_{k=1}^n,
\quad  i=\overline{1,l},  \end{eqnarray*}
be correspondingly random fields of evaluation and overestimation information  by the  $i$-th non-insatiable consumer and  $\zeta(p, \omega_0)=Q(p,\zeta_0(p, \omega_0))$ be a random field of making decisions by firms that satisfy  conditions of the Theorem \ref{ttl3}.
We call a random field
\begin{eqnarray*}         \gamma_i(p,\omega_0,\omega_i) =\{ \gamma_{ik}(p,\omega_0,\omega_i)\}_{k=1}^n,
\quad  i=\overline{1,l}, \end{eqnarray*}            a random   vector of demand of the $i$-th non-insatiable consumer,\index{random   vector of demand of non-insatiable consumer}
where
\begin{eqnarray*}         \gamma_{ik}(p,\omega_0,\omega_i)
=\frac{p_k\eta_{ik}^0(p, \zeta(p, \omega_0),\omega_i)}
{\sum\limits_{s=1}^n\eta_{is}^1(p, \zeta(p, \omega_0),\omega_i)p_s},
\quad  i=\overline{1,l}.\end{eqnarray*}
\end{definition}
The random field of choice of the  $i$-th consumer\index{ random field of  consumer choice } is related to the  random   vector of demand of the  $i$-th consumer  by the formula
\begin{eqnarray}\label{pppl3}
\xi_i(p)=\left\{ \frac{K_i(p,\zeta_0(p,\omega_0))\gamma_{ik}(p,\omega_0,\omega_i)}{p_k}\right\}_{k=1}^n,
 \quad  i=\overline{1,l}.
\end{eqnarray}

We also denote  any realization of the random  demand  vector  of the  $i$-th consumer which is not insatiable by
\begin{eqnarray*}          \gamma_i(p)=\{ \gamma_{ik}(p)\}_{k=1}^n,\end{eqnarray*}
where $ \gamma_{ik}(p)=\gamma_{ik}(p,\omega_0,\omega_i)$
at  certain  $\omega_0 $ and $\omega_i,$ and we call it the demand vector of the   $i$-the
non-insatiable consumer.
We denote any realizations of random fields
\begin{eqnarray*}         \eta_i^0(p,\zeta(p, \omega_0),\omega_i)
=\{\eta_{ik}^0(p,\zeta(p, \omega_0),\omega_i)\}_{k=1}^n \end{eqnarray*}
and
\begin{eqnarray*}         \eta_i^1(p,\zeta(p, \omega_0),\omega_i)
=\{\eta_{ik}^1(p,\zeta(p, \omega_0),\omega_i)\}_{k=1}^n, \end{eqnarray*}
 correspondingly,
\begin{eqnarray*}          \mu_i^0(p)=\{\mu_{ki}^0(p)\}_{i=1}^n,\end{eqnarray*}
where
\begin{eqnarray*}          \mu_{ki}^0(p)=\eta_{ik}^0(p,\zeta(p, \omega_0),\omega_i),\end{eqnarray*}
and
\begin{eqnarray*}          \mu_i(p)=\{\mu_{ki}(p)\}_{i=1}^n,\end{eqnarray*}
where
\begin{eqnarray*}          \mu_{ki}(p)=\eta_{ik}^1(p,\zeta(p, \omega_0),\omega_i)\end{eqnarray*}
at  certain  $\omega_0 $ and $\omega_i.$

Under such notations
\begin{eqnarray*}         \gamma_{ik}(p)=\frac{p_k\mu_{ki}^0(p)}{\sum\limits_{s=1}^n\mu_{si}(p)p_s},\quad  i=\overline{1,l},\quad  k=\overline{1,n}.\end{eqnarray*}

Every realization of a random    demand vector of the  $i$-th consumer satisfies conditions:\\
1) $\gamma_i(p) $ is a continuous function of  $p \in K_+^n\  ;$\\
2) $\gamma_i(tp)=\gamma_i(p), \ t >0\  ;$  \\
3)
\begin{eqnarray}\label{pppl1}
\sum\limits_{k=1}^n\gamma_{ik}(p)=1, \quad p \in K_+^n,
\quad  i=\overline{1,l},
\end{eqnarray}
in the case of insatiable consumer and
\begin{eqnarray}\label{pppl2}
\sum\limits_{k=1}^n\gamma_{ik}(p) \leq 1, \quad  p \in K_+^n,
\quad  i=\overline{1,l},
\end{eqnarray}
in the case of non-insatiable consumer.

The component   $\gamma _{ik}(p)$ of demand vector  $\gamma_i(p)$  have such economic sense:  the part of income that the  $i$-th consumer spends to buy the  $k$-th goods.

We describe the  society  by a demand   matrix of  $l$ consumers\index{demand   matrix of  consumers}
\begin{eqnarray}\label{g2l4}
\gamma(p)=|| \gamma _{ij}(p)||_{i=1, j=1}^{l,n}.
\end{eqnarray}
From zero homogeneity  of random fields of consumers choice, it follows that $\gamma _{ik}(p)$ does not depend on the scale of prices, thus,
$\gamma _{ik}(p)$ are homogeneous functions of zero degree, that is,
\begin{eqnarray}\label{g2l5} \gamma
_{ik}(tp)=\gamma _{ik}(p), \quad \forall t > 0, \quad
i=\overline {1,l},\quad k=\overline {1,n}, \quad p \in K_+^n.
\end{eqnarray}
If  the $i$-th consumer is insatiable,  then for such
 $i$ there holds equality
\begin{eqnarray}\label{g2l6}
\sum\limits _{k=1}^n\gamma _{ik}(p)=1, \quad p \in K_+^n.
\end{eqnarray}
If the  $i$-th consumer is not insatiable, then for such $i$ the condition
\begin{eqnarray}\label{pog2l6}
\sum\limits _{k=1}^n\gamma _{ik}(p)\leq 1, \quad \gamma _{ik}(p)\geq 0,\quad p \in K_+^n,
\end{eqnarray}
is valid.
Indicate the relation  between realizations of random  demand  vectors
and realizations of random fields of consumers choice.

Remind that random fields of decisions making by firms are determined by the formula
$\zeta(p, \omega_0)=Q(p,\zeta_0(p, \omega_0))$
$=\{\zeta_i(p)\}_{i=1}^m.$
Let
$t_i(p)={\{t_{ki}(p)\}}_{k=1}^n$
be one of realizations  of the random field of choice of the $i$-th consumer\index{realization  of the random field of choice of  consumer}  $\xi _i(p)$, $i=\overline {1,l},$  and
$(x_s(p),y_s(p))=z^s(p)$ be a realization of the random field of decisions making by the $s$-th firm\index{realization of the random field of decisions making by firm} $\zeta _s(p),\ s=\overline {1,m}.$
We give the income of the  $i$-th consumer that corresponds to realization of firms behavior strategies   $z(p)$  by the formula
\begin{eqnarray}\label{pog2l7}
D_i(p)=K_i^0(p,z(p)),\quad i=\overline{1,l}, \quad p \in K_+^n,
\end{eqnarray}
\begin{eqnarray*}         \quad  z^i(p)=\{x_i(p), y_i(p)\}, \quad  i=\overline{1,m}, \quad z(p)=\{ z^1(p),\dots ,z^m(p)\},\end{eqnarray*}
where  $K_i^0(p,z),\ i=\overline{1,l},$ is income pre-functions of consumers.
Further, we assume that the random field
$\zeta_0(p, \omega_0)$ is  continuous with probability 1 and a productive economic process  $Q(p,z)$ is a continuous function of variables
$(p,z) \in  K_+^n\times \Gamma^m,$ therefore the random field of decisions making  by firms $\zeta(p, \omega_0)=Q(p,\zeta_0(p, \omega_0))$
is  continuous with probability  1.
Since  the random fields  $\zeta_s(p),\ s=\overline {1,m},$ are homogeneous of zero degree, we have $D_i(tp)=tD_i(p) ,\
i=\overline{1,l},\  p \in  K_+^n, \  \forall t>0.$
 We define a realization of a random  demand vector   of the  $i$-th consumer\index{realization of  random  demand vector   of  consumer} that corresponds to  a realization $t_i(p)$ of a random field of consumer choice  $\xi _i(p)$ by the formula
\begin{eqnarray*}         \gamma _i(p)={\left\{\gamma _{ik}(p)=\frac{t_{ki}(p)p_k}{
D_i(p)}\right\} }_{k=1}^n,\quad i=\overline {1,l}.\end{eqnarray*}          In this case, the demand matrix takes  the form  \begin{eqnarray*}         {||\gamma
_{ik}(p)||}_{i=1,k=1}^{l,n}={\left |\left|\frac{t_{ki}(p)p_k}{
D_i(p)}\right |\right |}_{i=1,k=1}^{l,n}.\end{eqnarray*}
If all consumers are insatiable, then the demand matrix satisfies conditions
(\ref{g2l5}) and (\ref{g2l6}).

When we say that firm chooses a behavior strategy   we understand to a certain extent  independence  of realization  of this behavior strategy from its will.

We assume that the state of an economy system is given in  the time interval $[0,T]$ of its operation
if we are given random fields of consumers choice  of all consumers, random fields of decisions making by firms of all firms, and an income function of every consumer. We prove  Theorems for continuous realizations of random fields, and income functions.
 If almost all realizations are continuous then we say that corresponding assertion is valid with probability 1.

 In what follows, we denote by $z(p)=\{z^i(p)\}_{i=1}^m$ any realization of random field of decisions making by firms  $\zeta(p, \omega_0)=Q(p,\zeta_0(p, \omega_0)),$
where $ z^i(p)$ is any realization of random field of decisions making by the $i$-th  firm.
\begin{definition}\label{chap2dl1} The description of an economy system is given in the time interval  $[0,T],$ if we  know:\\ 1)
a demand matrix $||\gamma _{ij}(p)||_{i=1~j=1}^{l\quad n}$
given on $K_+^n\  ;$ \\
2) a behavior strategy of every  firm  $z^j(p)=(x_j(p),y_j(p)),$ $j=\overline {1,m},$   defined on  $K_+^n\  ;$\\
3) an  income   $D_i(p) $ of the $i$-th consumer defined by the formula (\ref{pog2l7}), $ i=\overline {1,l}\  ;$\\
4) a  property vector  $b_i(p, z(p))$ that the  $i$-th consumer  has, $ i=\overline {1,l}.$
\end{definition}

Introduce vectors
\begin{eqnarray}\label{g2l11}
\phi (p)=\{\phi_k(p)\}_{k=1}^n, \quad \psi (p)=\{\psi_k(p)\}_{k=1}^n,
\end{eqnarray}
where
\begin{eqnarray*}         \phi _k(p)=\frac{1}{ p_k} \sum\limits
_{i=1}^l\gamma _{ik}(p)D_i(p),\quad k =\overline {1,n}, \end{eqnarray*}
   \begin{eqnarray*}      \psi_k(p)=\sum\limits_{i=1}^lb_{ki}(p,z(p))+
\sum\limits_{j=1}^m[y_{kj}(p) -x_{kj}(p)],
\quad k=\overline{1,n}. \end{eqnarray*}
\begin{eqnarray*} b_i(p,z)=\{b_{ki}(p,z)\}_{k=1}^n,\quad
   i=\overline{1,l},\end{eqnarray*}
\begin{eqnarray*} z(p)=\{ z^1(p),\dots ,z^m(p)\},\end{eqnarray*}
\begin{eqnarray*}  z^i(p)=\{x_i(p), y_i(p)\}, \quad  i=\overline{1,m},\end{eqnarray*}
\begin{eqnarray*} y_i(p)=\{y_{ki}(p)\}_{k=1}^n,\quad  x_i(p)=\{x_{ki}(p)\}_{k=1}^n.\end{eqnarray*}
We call the vector  $\phi (p)$ and the vector
$\psi (p),$  correspondingly, the  demand vector of the society\index{demand vector of the society }  and the final supply vector   of the society\index{final supply vector   of the society} under conditions that the choice of consumers describes the demand matrix $\gamma(p),$ firms realized the behavior strategies
 $z^j(p)=(x_j(p),y_j (p)),$
$~j=\overline {1,m},$ and $D_i(p)$ is the income of the  $i$-th consumer given by the formula
 (\ref{pog2l7}) via realized behavior  strategies of firms.  It is convenient to introduce  a notion  of  demand vectors of productive and unproductive  consumption of the society in the time interval
 $[0,T]$ of the economy operation.  We call also the vector $\phi (p)$
  the demand vector of  unproductive  consumption of the society,\index{demand vectors of unproductive  consumption of the society} we do the vector
$\sum\limits _{j=1}^mx_j(p)$  the demand vector of productive  consumption of  the society,\index{demand vectors of productive   consumption of the society} and we call the vector  $\sum\limits
_{j=1}^my_j(p)+\sum\limits_{i=1}^l b_i(p,z(p))$  the vector of supply of the society\index{vector of supply of the society}   in the time interval
 $[0,T]$ of the economy operation.
The  economic sense of the  vector  $\phi (p)$
is that it determines the  demand on goods that in the time interval $[0,T]$ will not be used for direct productive needs of firms  though its components  can be non zero  for goods destined for the productive consumption by firms or consumers  with the aim of its using in the next  period of the economy operation.
So, firms make store of raw materials, component parts, and so on.
As a consequence of that, just this type of the society  demand    determines the  productive demand of the society, then, in what follows,  we often shall  omit the word "unproductive", if it does not appear
the reason for confusion.

\begin{definition}\label{chapdl2}
The economy system is in the Walras equilibrium state,\index{Walras equilibrium state} in the period of its operation, if there exists  a price vector  $p^*, $
$~m$ productive processes
\begin{eqnarray*}         (x_i^* (p^* ),~y_i^* (p^* )),\quad  x_i^* (p^* )\in X_i,\quad y_i^* (p^*
)\in F_i(x_i^* (p^* )), \quad i=\overline
{1,m},  \end{eqnarray*}
such that there hold inequalities
\begin{eqnarray}\label{g2l13}
\phi (p^* )\leq\psi (p^* ),
\end{eqnarray}
\begin{eqnarray}\label{g2l14}
\left\langle\phi (p^*
),p^*\right\rangle \leq \left\langle\psi (p^* ),p^*\right\rangle ,
\end{eqnarray}
where $p^*\in K_+^n,$\quad $p^* =(p_1^* ,\dots
,p_n^* )$ is an equilibrium price vector, $F_i(x)$ is a technological map of the  $i$-th firm, $\ i=\overline{1,m}.$
\end{definition}
The inequality  (\ref{g2l13}) means that in the state of equilibrium there  exists a price vector  $p^* $  under which the demand of the society does not exceed the supply  and  the inequality  (\ref{g2l14}) means that the value of goods that the society wants to buy  does not exceed the value of goods proposed to  consume.
We call the price vector  $p^*$  that guarantees realization of inequalities  (\ref{g2l13}) and
(\ref{g2l14})  the equilibrium price vector.\index{ equilibrium price vector}
If all consumers in the economy system are insatiable, then inequality  (\ref{g2l14}) is turned into the equality and it is called the Walras law in the narrow sense.\index{ Walras law in the narrow sense}
\begin{definition}\label{chapdl3}
The Walras equilibrium state\index{optimal Walras equilibrium state } of the economy system is called  optimal if the productive process that corresponds to the  equilibrium price vector   $p^* $  is optimal, that is, there hold conditions
\begin{eqnarray*}
\left\langle
y_i^* (p^* )-x_i^* (p^* ),p^*\right\rangle = \sup\limits
_{x\in X_i}\sup\limits _{y\in F_i(x)} \left\langle y-x,p^*\right\rangle,
\end{eqnarray*}
\begin{eqnarray*}
 y_i^*(p^*) \in F_i(x_i^*(p^*)),\quad x_i^*(p^*) \in X_i, \quad i=\overline {1,m}, \end{eqnarray*}
where $X_i$ is an expenditure set of the $i$-th firm, $F_i(x)$ is its technological map.
\end{definition}

In the following  several Theorems, we assume that $K_+^n=\bar R_+^n$ and
\begin{eqnarray*}          P=\left\{ p=\{p_i\}_{i=1}^n  \in \bar R_+^n, \quad \sum\limits_{i=1}^np_i=1\right\}\end{eqnarray*}
is the unit simplex in $\bar R_+^n.$

In the next Theorem, we assume that the matrix
$||\gamma_{ik}(p)||_{i=1, k=1}^{l,n}$ is not   generated by the random fields of evaluation of information by consumers and it is arbitrary one that satisfies only the conditions of this Theorem.
\begin{theorem}\label{ch2l3}
Let technological maps $F_i(x), \  x \in X_i^1,
\ i=\overline{1,m},$ be  convex down, belong to the CTM class,  a productive economic process $Q(p,z),$  a family of income pre-functions $K_i^0(p,z), \ i=\overline{1,l},$  property vectors $b_i(p,z), \ i=\overline {1,l},$ be continuous maps of variables
$(p,z) \in \bar R_+^n\times \Gamma^m,$ and let random field of decisions making by firms satisfy conditions of the Theorem \ref{tl4}.
If, furthermore, there holds the condition
\begin{eqnarray}\label{g2l23}
R(p, Q(p,z)) > C e, \quad p \in P, \quad  z \in \Gamma^m,
\end{eqnarray}
\begin{eqnarray*}          R(p,z)= \sum\limits
_{i=1}^m[y_i-x_i]+ \sum\limits _{j=1}^l b_j(p,z), \quad e=\{1, 1, \ldots, 1\}, \quad C>0,\end{eqnarray*}
then for every continuous matrix  $||\gamma_{ik}(p)||_{i=1, k=1}^{l,\ n}$  on  $P,$
whose rows  satisfy conditions
\begin{eqnarray}\label{nn1s1}
\sum\limits_{k=1}^n\gamma_{ik}(p)=1, \quad i=\overline{1,l}, \quad p \in P,
\end{eqnarray}
and every continuous realization  $z(p)=\{z^i(p)\}_{i=1}^m$
of  random field of decisions making by firms
$\zeta(p),$ the set of equations
\begin{eqnarray}\label{al1} \sum\limits _{i=1}^l\gamma _{ik}(
p) D_i( p)
\end{eqnarray}
\begin{eqnarray*}         =p_k \left [\sum\limits _{i=1}^m[y_{ki}(p)-x_{ki}(p)]+
\sum\limits _{i=1}^l b_{ki}(p,z(p))\right ],\quad k=\overline{1,n},\end{eqnarray*}
has a solution  in the set $P,$ where
$D_i(p)=K_i^0(p,z(p)),  \ i=\overline{1,l}.$
\end{theorem}
\begin{proof}\smartqed
In accordance with the conditions of the Theorem \ref{ch2l3}, there holds inequality (\ref{g2l23}) from that it follows that on the simplex $P$ the inequalities
\begin{eqnarray}\label{al2}
\sum\limits _{i=1}^m y_{ki}(p)+ \sum\limits _{i=1}^l b_{ki}(p,z(p)) \geq
C>0,\qquad p\in P, \quad k=\overline{1,n},
\end{eqnarray}
are satisfied.
To prove the Theorem, let us construct on  the map
$f(p)=\{f_k(p)\}_{k=1}^n $ where
\begin{eqnarray*}         f_k(p)=\frac{\sum\limits _{i=1}^l\gamma _{ik}( p)D_i( p)+p_k
\sum\limits _{i=1}^mx_{ki}(p)}{\sum\limits _{i=1}^m y_{ki}(p)+
\sum\limits _{i=1}^l b_{ki}(p,z(p))},\quad p \in P, \quad k=\overline{1,n},\end{eqnarray*}
and the  map $\varphi(p)=\{\varphi_k(p)\}_{k=1}^n,$  where
\begin{eqnarray*}         \varphi_k(p)=\frac{f_k(p)}{\sum\limits_{k=1}^nf_k(p)},\quad p \in P, \quad
k=\overline{1,n}.\end{eqnarray*}
From  (\ref{al2}) and the conditions of the Theorem, it follows that  $f_k(p)$ is a continuous map on $P.$ Show that the map $\varphi(p)$ is also continuous on $P.$ For this, it is sufficient to prove that   $\sum\limits_{k=1}^nf_k(p) >0$ on $P.$
Really,
\begin{eqnarray*}         \sum\limits_{k=1}^nf_k(p) \geq \frac{1}{L}
\left[\sum\limits _{i=1}^m \left\langle y_i(p),p \right\rangle +\sum\limits
_{k=1}^l \left\langle b_k(p,z(p)), p \right\rangle \right] \geq \frac{C}{L}>0,\end{eqnarray*}          where
\begin{eqnarray*}         L=\max\limits_{k}\sup\limits_{p\in P}\left\{\sum\limits _{i=1}^m
y_{ki}(p)+\sum\limits _{i=1}^l b_{ki}(p,z(p))\right\}< \infty,\end{eqnarray*}
since technological maps belong to the CTM class.
On the basis of the Brouwer Theorem \cite{150},\index{Brouwer Theorem} there exists a fixed point   $\bar p=\{\bar p_k\}_{k=1}^n$
of the map $\varphi(p),$ that is,
\begin{eqnarray}\label{al3}
\bar p_k=\varphi_k(\bar p), \quad k=\overline{1,n}.
\end{eqnarray}
Show that $\bar p$  is a solution of the set of equations (\ref{al1}). We rewrite the set of equalities   (\ref{al3})  in the form
\begin{eqnarray}\label{nal3}
 f_k(\bar p)=\bar p_k \sum\limits_{k=1}^nf_k(\bar p), \quad
k=\overline{1,n}.
\end{eqnarray}
Further,
\begin{eqnarray*}         \sum\limits_{k=1}^nf_k(\bar p)\left\{\sum\limits _{i=1}^m
y_{ki}(\bar p)+ \sum\limits _{i=1}^l b_{ki}(\bar p,z(\bar p))\right\}
\end{eqnarray*}
\begin{eqnarray*}
 =\sum\limits _{i=1}^l\sum\limits_{k=1}^n\gamma _{ik}(\bar
p)D_i( \bar p)+\sum\limits_{k=1}^n \bar p_k \sum\limits
_{i=1}^mx_{ki}(\bar p) \end{eqnarray*}
\begin{eqnarray*}
 =\sum\limits _{i=1}^lD_i(
\bar p)+\sum\limits_{k=1}^n \bar p_k \sum\limits
_{i=1}^mx_{ki}(\bar p) \end{eqnarray*}
\begin{eqnarray*} =\sum\limits_{k=1}^n\left\{\sum\limits
_{i=1}^m y_{ki}(\bar p)+ \sum\limits _{i=1}^l b_{ki}(\bar p,z(\bar p))\right\}\bar
p_k>C>0.\end{eqnarray*}
Multiplying the left and the right parts of the $k$-th equality  (\ref{nal3}) by non zero multiplier
$\sum\limits _{i=1}^m y_{ki}(\bar p) + \sum\limits _{i=1}^l b_{ki}(\bar p,z(\bar p))$
and summing up over $k$ from  $1$ to $n,$
we have
\begin{eqnarray*}         \sum\limits_{k=1}^nf_k(\bar p)=1.\end{eqnarray*}
Hence, it follows that
\begin{eqnarray*}         \bar p_k=f_k(\bar p), \quad k=\overline{1,n},\end{eqnarray*}
and this is  equivalent to the set of equations  (\ref{al1}), since for every
$k$ the last equality   can be multiplied  by non zero multiplier
\begin{eqnarray*}         \sum\limits
_{i=1}^m y_{ki}(\bar p)+ \sum\limits _{i=1}^l b_{ki}(\bar p,z(\bar p)),
\quad k=\overline{1,n}.
\end{eqnarray*}
\qed
\end{proof}

The next Theorem generalizes  the Theorem \ref{ch2l3}.
\begin{theorem}\label{doda1}
Let technological maps  $F_i(x), \  x \in X_i^1,
\ i=\overline{1,m},$ be  convex down, belong to the CTM class,  a productive economic process  $Q(p,z),$   a family of income pre-functions  $K_i^0(p,z), \ i=\overline{1,l},$ property vectors  $b_i(p,z), \ i=\overline {1,l},$ be continuous maps of variables $(p,z) \in \bar R_+^n\times \Gamma^m,$ and let random fields of decisions making by firms satisfy the conditions of the Theorem  \ref{tl4}.
For every continuous matrix   $||\gamma_{ik}(p)||_{i=1, k=1}^{l,\ n}$
on  $P,$ whose rows  satisfy conditions
\begin{eqnarray}\label{dodat}
\sum\limits_{k=1}^n\gamma_{ik}(p)=1, \quad i=\overline{1,l},
\end{eqnarray}
and  every continuous realization  $z(p)=\{z^i(p)\}_{i=1}^m$
of random field of decisions making by firms
$\zeta(p),$ the set of equations  (\ref{al1})
has a solution in the set   $P.$
\end{theorem}
\begin{proof}\smartqed
In this Theorem, we do not require the  realization of the  condition  (\ref{g2l23}). In accordance with the Definition of  productive economic process,  the condition
\begin{eqnarray}\label{dodat1}
R(p, Q(p,z)) \geq 0, \quad p \in P, \quad  z \in \Gamma^m,
\end{eqnarray}
\begin{eqnarray*}          R(p,z)= \sum\limits
_{i=1}^m[y_i-x_i]+ \sum\limits _{j=1}^l b_j(p,z), \end{eqnarray*}
is always satisfied.
To prove the Theorem, we construct an auxiliary model of economy for that the condition (\ref{g2l23}) is already satisfied.

Let
$K_i^0(p,z), \ i=\overline{1,l},$ be the set of  income pre-functions that figures in the Theorem \ref{doda1}, and  $d$ be a positive number. Introduce a new family of income pre-functions
\begin{eqnarray*}         K_i^{0,d}(p,z)=K_i^0(p,z)+ d\sum\limits_{i=1}^np_i, \quad i=\overline{1,l},\quad  d >0.\end{eqnarray*}          We assume that, in this model of economy, the $i$-th consumer has a property vector
$b_i(p,z)+d e, \ i=\overline{1,l},$ where  $e=\{1, \ldots, 1 \}$ is the vector whose all component  equal unit and  $b_i(p,z), \ i=\overline{1,l},$ is the property vector of the  $i$-th consumer in the original model of economy. Introduce the set of income functions
$K_i^{d}(p,z)=K_i^{0,d}(p,Q(p,z)), $ where $Q(p,z)$ is the productive economic process that figures  in the Theorem \ref{doda1}. Such economy system satisfies the condition
\begin{eqnarray}\label{doda2}
R_d(p, Q(p,z)) > 0, \quad p \in P, \quad  z \in \Gamma^m,
\end{eqnarray}
\begin{eqnarray*}
  R_d(p,z)= \sum\limits
_{i=1}^m[y_i-x_i]+ \sum\limits _{j=1}^l [b_j(p,z)+d e].
\end{eqnarray*}
If $z(p)$ is a  realization of random field of decisions making by firms, then, in accordance with the Theorem  \ref{ch2l3}, there exists a price vector  $p^d$  that satisfies the set of equations
\begin{eqnarray}\label{doda3}
 \sum\limits_{i=1}^l \gamma_{ik}(p^{d})D_i^d( p^{d})
\end{eqnarray}
\begin{eqnarray*}
  = p_k^d\left[\sum\limits _{i=1}^m[y_{ki}(p^{d})-x_{ki}(p^{d})]+
\sum\limits _{i=1}^l [b_{ki}(p^{d},z(p^{d}))+ d]\right],\quad k=\overline{1,n}, \nonumber
\end{eqnarray*}
where $D_i^d( p^{d})=K_i^{d}(p,z(p))= D_i( p^{d})+d.$
We assume now that   $d$ belongs to a  certain set
$(0, d_0], \ d_0 >0.$ For every  $d \in (0, d_0]$ the vector $p^d,$ that is the solution to the set of equations  (\ref{doda3}),  belongs to the compact  $P.$
If  $d_s$ is a certain sequence that belongs to  $(0, d_0]$ and is such that  $d_s \to 0, $ as $s \to \infty,$ then, without loss of generality, we can assume the sequence of vectors  $p^{d_s}$ that is the solution to the set of equations  (\ref{doda3}) for  $d=d_s$ is   convergent  to  a certain vector  $p^0.$ The  vector  $p^{d_s}$ satisfies the set of equations
\begin{eqnarray}\label{doda5}
\sum\limits_{i=1}^l  \gamma_{ik}(p^{d_s}) D_i^{d_s}( p^{d_s})
\end{eqnarray}
\begin{eqnarray*}          = p_k^{d_s}\left[\sum\limits _{i=1}^m[y_{ki}(p^{d_s})-x_{ki}(p^{d_s})]+
\sum\limits _{i=1}^l [b_{ki}(p^{d_s},z(p^{d_s}))+ d_s]\right],\quad k=\overline{1,n},\end{eqnarray*}
where $D_i^{d_s}( p^{d_s})= D_i( p^{d_s})+d_s.$
Going to the limit in the set of equalities  (\ref{doda5}) as $s \to \infty,$
we obtain that the vector  $p^0$ satisfies the set of equations  (\ref{al1}).
\qed
\end{proof}

\begin{theorem}\label{ch2l1}
Let technological maps $F_i(x), \  x \in X_i^1,
\ i=\overline{1,m},$ be convex down, belong to the CTM class, a productive economic process  $Q(p,z),$  a family of income pre-functions  $K_i^0(p,z), \ i=\overline{1,l},$  and
property vectors  $b_i(p,z), \ i=\overline{1,l},$ be continuous maps of variables $(p,z) \in \bar R_+^n\times \Gamma^m,$  and let random fields of consumers choice and decisions making by firms satisfy conditions of the Theorem \ref{tl4}.
Then with probability one there exists the Walras equilibrium state, that is, for every realization  of random fields of consumers choice and decisions making by firms there exists
corresponding  to them equilibrium price vector    $p^*\in P$ such that the economy system is in the Walras equilibrium state. If, moreover, realizations of  random fields of decisions making by firms  are such that  among them there exist strategies of firms behavior arbitrarily close  to the optimal   in the sense of the Theorem   \ref{nnl1} of the  previous chapter, then with probability one there exists the optimal Walras equilibrium state.
\end{theorem}
\begin{proof}\smartqed
From the beginning, we assume  that the condition  (\ref{g2l23}) is carried out.
In accordance with the  introduced notations,  let any continuous realizations of random fields of consumers choice and decisions making by firms  generate continuous demand matrix on $P$ with matrix elements
\begin{eqnarray*}          \gamma_{ik}(p) =\frac{p_k\eta_{ik}^0(p, \zeta(p, \omega_0),\omega_i)}
{\sum\limits_{s=1}^n\eta_{is}^0(p, \zeta(p, \omega_0),\omega_i)p_s},\quad  k=\overline{1,n},
\quad  i=\overline{1,l},\end{eqnarray*}
for a certain  $\omega_0,$  $ \omega_i,$ and
\begin{eqnarray*}         \mu_{ki}(p)=\eta_{ik}^0(p, \zeta(p, \omega_0),\omega_i),
\quad  i=\overline{1,l}, \quad  k=\overline{1,n},\end{eqnarray*}
then
\begin{eqnarray*}          \gamma_{ik}(p)=\frac{p_k\mu_{ki}(p)}
{\sum\limits_{s=1}^n\mu_{si}(p)p_s},
\quad  i=\overline{1,l}, \quad  k=\overline{1,n}.\end{eqnarray*}
Construct an auxiliary matrix
$\gamma^{\delta}(p)= || \gamma_{ik}^{\delta}(p)||_{i,k=1}^{l,n},$ where
\begin{eqnarray*}          \gamma_{ik}^{\delta}(p)=\frac{p_k\mu_{ki}(p)+\delta}
{\sum\limits_{s=1}^n\mu_{si}(p)p_s + n \delta},
\quad  i=\overline{1,l}, \quad  k=\overline{1,n}, \quad \delta > 0,\end{eqnarray*}
and consider  the set of equations
\begin{eqnarray}\label{q1al1} \sum\limits _{i=1}^l\gamma _{ik}^{\delta}(
p) D_i( p)
\end{eqnarray}
\begin{eqnarray*}         =p_k \left [\sum\limits _{i=1}^m[y_{ki}(p)-x_{ki}(p)]+
\sum\limits _{i=1}^l b_{ki}(p,z(p))\right ],\quad k=\overline{1,n},\end{eqnarray*}
on the set $P.$
Quantities, entering the set of equations  (\ref{q1al1}), satisfy all conditions of the Theorem  \ref{ch2l3}.
Therefore there exists a solution  $p^{\delta}$ of the set of equations (\ref{q1al1}). Under assumptions  made in the Theorem,  every component of this solution is positive. Really, as a consequence of that  $p^{\delta}$  is a solution of the set of equations  (\ref{q1al1}), there hold such inequalities
\begin{eqnarray*}          p_k^{\delta} \geq  \frac{\delta}{(n\delta + \mu) \psi_k}\sum\limits _{i=1}^lD_i( p^{\delta}), \quad k=\overline{1,n},\end{eqnarray*}
where
\begin{eqnarray*}
   \mu=\max\limits_{i}\sup\limits_{p \in P}\sum\limits_{s=1}^n\mu_{si}(p)p_s< \infty,
\end{eqnarray*}
\begin{eqnarray*}  \psi_k=\sup\limits_{p \in P}\left[\sum\limits _{i=1}^m[y_{ki}(p)-x_{ki}(p)]+
\sum\limits _{i=1}^l b_{ki}(p,z(p))\right]< \infty.
\end{eqnarray*}
As a consequence of carrying out  the condition (\ref{g2l23}), there holds inequality
\begin{eqnarray*}         \sum\limits _{i=1}^lD_i( p^{\delta})=\left\langle p^{\delta}, \psi(p^{\delta})\right\rangle \geq C >0,\end{eqnarray*}
where $\psi(p)=\{\psi_k(p)\}_{k=1}^n, $
\begin{eqnarray*}         \psi_k(p)=\sum\limits _{i=1}^m[y_{ki}(p)-x_{ki}(p)]+
\sum\limits _{i=1}^l b_{ki}(p,z(p)).\end{eqnarray*}
As a consequence of that  $p^{\delta}$ is a solution of the set of equations  (\ref{q1al1}),
we obtain the set of inequalities
\begin{eqnarray}\label{q1al2}
\sum\limits_{i=1}^l \frac{\mu_{ki}(p^{\delta})}
{\sum\limits_{s=1}^n\mu_{si}(p^{\delta})p_s^{\delta} + n \delta}D_i( p^{\delta})
\end{eqnarray}
\begin{eqnarray*}          \leq \sum\limits _{i=1}^m[y_{ki}(p^{\delta})-x_{ki}(p^{\delta})]+
\sum\limits _{i=1}^l b_{ki}(p^{\delta},z(p^{\delta})),\quad k=\overline{1,n}.\end{eqnarray*}
Consider a sequence of positive numbers  $\delta_s$ that converges to zero, as $s \to \infty.$ Let   $p^{\delta_s}$ be a solution of the set of inequalities  (\ref{q1al2}) when as number
$\delta$ the number  $\delta_s$ is chosen.
The sequence  $p^{\delta_s}, $ as $\delta_s \to 0,$ is  compact since it belongs  to the set $P.$
Without loss of generality, we assume that the sequence  $p^{\delta_s}, $ as $\delta_s \to 0,$  converges to a  certain vector  $p^0 \in P.$
Due to the continuity of  $\sum\limits_{s=1}^n\mu_{si}(p)p_s$ on the set  $P$ and  the existence of  a positive  constant, that separates this quantity from zero, since the conditions of the Theorem \ref{tl4} are carried out, we can pass  to the limit in the set of inequalities (\ref{q1al2})  if we substitute  $\delta_s$ instead of  $\delta.$

Then  $p^0$ is a solution of the set of inequalities
\begin{eqnarray}\label{q1al3}
\sum\limits_{i=1}^l \frac{\mu_{ki}(p^{0})}
{\sum\limits_{s=1}^n\mu_{si}(p^{0})p_s^{0}} D_i( p^{0})
\end{eqnarray}
\begin{eqnarray*}
   \leq \sum\limits _{i=1}^m[y_{ki}(p^{0})-x_{ki}(p^{0})]+
\sum\limits _{i=1}^l b_{ki}(p^{0},z(p^{0})),\quad k=\overline{1,n}.\end{eqnarray*}
Prove now the correctness  of the  assertion of the Theorem  without carrying out the additional condition (\ref{g2l23}), that is, we assume that there holds natural condition (\ref{dodat1}).
Construct an auxiliary model of economy for that the condition (\ref{g2l23}) is valid.
We introduce a new set of income pre-functions
\begin{eqnarray*}         K_i^{0,d}(p,z)=K_i^0(p,z)+ d\sum\limits_{i=1}^np_i, \quad  i=\overline{1,l},\quad  d >0,\end{eqnarray*}
where  $K_i^0(p,z), \ i=\overline{1,l},$ is the set of income pre-functions that figures in the Theorem  \ref{ch2l1}.
We assume that, in this model, each consumer has the  property vector   $b_i(p,z)+d e, \ i=\overline{1,l},$ where the vector $e=\{1, \ldots, 1 \}$ has unit components and $b_i(p,z), \ i=\overline{1,l},$ is the property vector  of the  $i$-th consumer in the original model of economy. If to introduce the set of income functions
$K_i^{d}(p,z)=K_i^{0,d}(p,Q(p,z)), $ where $Q(p,z)$ is a productive economic process that figures in the Theorem, then in such economy system the condition
\begin{eqnarray}\label{dodl2}
R_d(p, Q(p,z)) > 0, \quad p \in P, \quad  z \in \Gamma^m,
\end{eqnarray}
\begin{eqnarray*}          R_d(p,z)= \sum\limits
_{i=1}^m[y_i-x_i]+ \sum\limits _{j=1}^l [b_j(p,z)+d e], \end{eqnarray*}
is already satisfied.
Let  $z(p)$ be a realization  of random field of decisions making by firms. In accordance with the proved assertion,  for any  $d>0$ there exists an equilibrium price vector  $p^d \in P$ such that
\begin{eqnarray}\label{dodl3}
\sum\limits_{i=1}^l \frac{\mu_{ki}(p^{d})}
{\sum\limits_{s=1}^n\mu_{si}(p^{d})p_s^{d}} D_i^d( p^{d})
\end{eqnarray}
\begin{eqnarray*}          \leq \sum\limits _{i=1}^m[y_{ki}(p^{d})-x_{ki}(p^{d})]+
\sum\limits _{i=1}^l [b_{ki}(p^{d},z(p^{d}))+ d],\quad k=\overline{1,n},\end{eqnarray*}
where $D_i^d( p^{d})=K_i^{d}(p,z(p))= D_i( p^{d})+d.$
The vector $p^d$ belongs to the compact $P$ for any positive  $d.$

Consider a sequence of positive numbers  $d_s$ that converges to zero, as $s \to \infty.$
Let   $p^{d_s}$ be a solution of the set of inequalities  (\ref{dodl3}) if instead of the number $d$ we  choose the number $d_s.$ The sequence  $p^{d_s}, $ as $d_s \to 0,$ is  compact, since it belongs to the set $P.$
Without loss of generality, we assume that the sequence  $p^{d_s} $ converges to a certain vector  $p^0 \in P,$ as $d_s \to 0.$
In the set of inequalities
(\ref{dodl3}) in that instead of the number $d$  the number $d_s$ is substituted,  we can pass  to the limit,
as $s \to \infty.$
As a results, we obtain that the vector $p^0$ satisfies the set of inequalities (\ref{q1al3}).

Prove the second part of the Theorem  \ref{ch2l1}, using the Kakutani continuity from above of technological maps.
Note that the existence of continuous realizations or continuous strategies of firms behavior arbitrarily close in income  to optimal one are  guaranteed by conditions of the Theorem  \ref{nnl1} of the  preceding Chapter.
The conditions of the Theorem \ref{ch2l1} guarantee only that among realizations of random fields of decisions making by firms  are such realizations    that are arbitrarily close to  optimal one in the sense of the Theorem  \ref{nnl1} of the Chapter 2.
Thus, for sufficiently small  $\varepsilon >0$
there exist  $m$ continuous realizations   $(x_i^\varepsilon
(p),y_i^\varepsilon (p)),$ $~i=\overline {1,m},$ of  random fields of decisions making by firms, where $x_i^\varepsilon (p)$ is the expenditure vector and
$y_i^\varepsilon (p)$ is the output vector of the  $i$-th firm,
$y_i^\varepsilon (p)\in F_i(x_i^\varepsilon (p))$ for that
\begin{eqnarray*}         \sup\limits _{p\in P}|\varphi _i(p)-\left\langle y_i^\varepsilon (p)-
x_i^\varepsilon (p),p\right\rangle |<\varepsilon, \quad
i=\overline{1,m}, \end{eqnarray*}
where
\begin{eqnarray*}         \varphi _i(p)=\sup\limits _{x\in
X_i}\sup\limits _{y\in F_i(x)} \left\langle y-x,p\right\rangle, \end{eqnarray*}
 $X_i \subset X_i^1,$ and every point  $x$ of the convex closed bounded set
 $ X_i$ is internal for $  X_i^1.$
Denote by  $\phi^{\varepsilon}(p)$ and  $ \psi^{\varepsilon}(p)$
the demand  and supply vectors of the society that correspond to chosen strategies of firms behavior
$(x_i^\varepsilon(p),y_i^\varepsilon (p)),$ $~i=\overline {1,m},$ where

\begin{eqnarray*}          \phi^{\varepsilon}(p) =\{\phi_k^{\varepsilon}(p)\}_{k=1}^n, \quad   \psi^{\varepsilon}(p) =\{\psi_k^{\varepsilon}(p)\}_{k=1}^n,
 \end{eqnarray*}
\begin{eqnarray*}
   \phi_k^{\varepsilon}(p)= \sum\limits_{i=1}^l \frac{\mu_{ki}^{\varepsilon}(p)}
{\sum\limits_{s=1}^n\mu_{si}^{\varepsilon}(p)p_s} D_i^{\varepsilon}( p), \end{eqnarray*}
\begin{eqnarray*}
     \mu_{ki}^{\varepsilon}(p)= \eta_{ik}^0(p, z^{\varepsilon}(p), \omega_i), \quad   D_i^{\varepsilon}( p)=K_i^0(p, z^{\varepsilon}(p)),
\end{eqnarray*}
\begin{eqnarray*}
& z^{\varepsilon}(p)=\{z^i_{\varepsilon}(p)\}_{i=1}^m, \quad
 z^i_{\varepsilon}(p)=(x_i^{\varepsilon}(p),  y_i^{\varepsilon}(p)),
 \end{eqnarray*}
\begin{eqnarray*}
\psi_k^{\varepsilon}(p)=\sum\limits _{i=1}^m[y_{ki}^{\varepsilon}(p)-x_{ki}^{\varepsilon}(p)]+
\sum\limits _{i=1}^l b_{ki}(p,z^{\varepsilon}(p)).
\end{eqnarray*}

Choose a positive sequence  $\varepsilon
_n$ that converges to zero.  For every $\varepsilon _n>0$
there exists equilibrium price vector  $p_n^* $ such that
\begin{eqnarray}\label{dodan1}
\phi ^{\varepsilon _n}(p_n^* )\leq \psi ^{\varepsilon _n}
(p_n^* )
\end{eqnarray}
and, moreover,
\begin{eqnarray}\label{g2l15}
\sup\limits _{p\in P}|\varphi _i(p)-\left\langle y_i^{\varepsilon
_n}(p)- x_i^{\varepsilon _n}(p),p\right\rangle |<\varepsilon _n, \quad i=\overline {1,m}.
\end{eqnarray}

Without loss of generality, we put that the sequences of vectors $x_i^{\varepsilon
_n}(p_n^* ),$ $~y_i^{\varepsilon _n}(p_n^* ),$  and $~p_n^* $
are convergent.

Really, if the last assumption is not valid, then, as a consequence of the  compactness  of  $P,X_i$   and the existence of general compact
$Y$ such that $F_i(x)\subseteq  Y,$  we can choose such subsequences for which this assumption will be already valid.
Denote by
\begin{eqnarray*}
     \lim\limits _{n\rightarrow\infty }p_n^* =p^*,\quad
  \lim\limits _{n\rightarrow\infty }y_i^{\varepsilon _n}(p_n^* )=
\bar y_i (p^* ),\quad
 \lim\limits _{n\rightarrow\infty}x_i^{\varepsilon _n}(p_n^* )=
\bar x_i(p^*),
\end{eqnarray*}
\begin{eqnarray*}
\bar x_i(p^*)=\{\bar x_{ki}(p^*)\}_{k=1}^n, \quad  \bar y_i(p^*)=\{\bar y_{ki}(p^*)\}_{k=1}^n.
 \end{eqnarray*}
From the Kakutani continuity from above  and (\ref{g2l15}), it follows that
\begin{eqnarray*}         \bar x_i(p^* )\in X_i,\quad\bar y_i(p^* )
\in F_i(\bar x_i(p^* )),\end{eqnarray*}
and
\begin{eqnarray*}         \varphi _i(p^* )=\left\langle\overline y_i(p^* )-
 \overline x_i(p^* ),p^*\right\rangle,\quad i=\overline {1,m}.\end{eqnarray*}
Passing to the limit in the  inequality (\ref{dodan1}), as  $\varepsilon _n \to 0,$ we obtain
\begin{eqnarray}\label{g2l16}
\phi (p^*)\leq\psi (p^* ),
\end{eqnarray}
where
\begin{eqnarray*}
 \phi (p^*)=\{\phi_k(p^*)\}_{k=1}^n, \quad   \psi (p^*)=\{\psi_k(p^*)\}_{k=1}^n,
\end{eqnarray*}
\begin{eqnarray*}
   \phi_k(p^*)= \sum\limits_{i=1}^l \frac{\mu_{ki}(p^*)}
{\sum\limits_{s=1}^n\mu_{si}(p^*)p_s^*} D_i(p^*),
\end{eqnarray*}
\begin{eqnarray*}
\mu_{ki}(p^*)= \eta_{ik}^0(p^*, \overline  z(p^*), \omega_i), \quad   D_i(p^*)=K_i^0(p^*, \overline z(p^*)),
\end{eqnarray*}
\begin{eqnarray*}
 \overline z(p^*)=\{\overline z^i(p^*)\}_{i=1}^m, \quad
 \overline z^i(p^*)=(\overline x_i(p^*),  \overline y_i(p^*)),
 \end{eqnarray*}
\begin{eqnarray*}
 \overline  x_i(p^*)=\{\overline  x_{ki}(p^*) \}_{k=1}^n, \quad  \overline  y_i(p^*)=\{\overline  y_{ki}(p^*) \}_{k=1}^n,
 \end{eqnarray*}
\begin{eqnarray*}
\psi_k(p^*)=\sum\limits _{i=1}^m[\overline y_{ki}(p^*)- \overline x_{ki}(p^*)]+
\sum\limits _{i=1}^l b_{ki}(p^*, \overline z(p^*)).
\end{eqnarray*}
\qed
\end{proof}

\begin{theorem}\label{ch2l2} Let technological maps   $F_i(x), \  x \in X_i^1, $
$i=\overline {1,m},$  a productive economic process  $Q(p,z),$ income pre-functions of consumers
$ K_i^0(p,z), \ i=\overline {1,l},$ property vectors  $b_i(p,z), \ i=\overline {1,l},$ satisfy conditions of the Theorem  \ref{ch2l1}.
Random fields of consumers choice and decisions making by firms satisfy the conditions of the Theorem  \ref{tl4}.
For every continuous  demand matrix
 $||\gamma_{ik}(p)||_{i=1, k=1}^{l,\ n}$ on  $P$  and continuous realization of random field of decisions making by firms  $z(p),$  every equilibrium price  vector $\bar p$ corresponding to them
   satisfies the set of equations
\begin{eqnarray*}
 \sum\limits
_{i=1}^l\gamma _{ik}(p)D_i(p)
\end{eqnarray*}
\begin{eqnarray}\label{g2l18}
 = p_k \left
[\sum\limits _{i=1}^m[y_{ki}(p)-x_{ki}(p)]+ \sum\limits
_{i=1}^l b_{ki}(p,z(p)) \right ], \quad  k=\overline{1,n}.
\end{eqnarray}
\end{theorem}
\begin{proof}\smartqed
Show that any equilibrium price vector satisfies the set of equations  (\ref{g2l18}).

Assume that the economy system is in the Walras equilibrium state, that is, there exists a vector  $\bar p$ such that there hold inequalities
\begin{eqnarray*}         \frac{1}{\bar p_k}\sum\limits _{i=1}^l
\gamma _{ik}(\bar p)D_i(\bar p)\leq  \sum\limits
_{i=1}^m [y_{ki}(\bar p)-x_{ki}(\bar p)]+ \sum\limits _{j=1}^l
b_{kj}(\bar p,z(\bar p)), \quad k=\overline{1,n}.\end{eqnarray*}
Multiplying the $k$-th inequality by  $\bar p_k,$ we obtain the set of inequalities
\begin{eqnarray*}         \sum\limits _{i=1}^l
\gamma _{ik}(\bar p)D_i(\bar p)\leq \bar p_k \left [ \sum\limits
_{i=1}^m [y_{ki}(\bar p)-x_{ki}(\bar p)]+ \sum\limits _{j=1}^l
b_{kj}(\bar p,z(\bar p))\right ], \quad k=\overline{1,n}.\end{eqnarray*}
The assumption that  there is at least one strict inequality in the last set of inequalities leads to the contradiction.
Really, if it is not so, then summing up over  $k$ from $1$ to $n$ the left and right parts of inequalities and using the property (\ref{g2l6}), we obtain
\begin{eqnarray*}         \sum\limits _{i=1}^n\psi_i(\bar p) \bar p_i <
\sum\limits _{i=1}^n\psi_i(\bar p)\bar p_i,\end{eqnarray*}
that is impossible. \qed
\end{proof}

The set of equations   (\ref{g2l18}) is very important since all equilibrium states of the economy system in the case of insatiable consumers are solutions of this set of equations.

The description of all equilibrium solutions of the set of equations  (\ref{g2l18}) is an important problem of mathematical economics.
\begin{note}
For finding the Walras equilibrium state  that is a solution to the set of equations  (\ref{g2l18}), one can  use the  Scarf method for finding a  solution of the set of equations  (\ref{al3}) and show whether the found solution satisfies the set of inequalities  (\ref{q1al3}).
\end{note}
Below, we  give the Theorem  \cite{55,71} that gives  an algorithm of construction of  solutions of the set of inequalities that are equilibrium states in the  economy system in the case when  not all consumers are insatiable. In this case a family of the set of equations is obtained  whose fixed points  are equilibrium states of the economy system.
\begin{theorem}\label{ch2l4}
Let technological maps  $F_i(x), \  x \in X_i^1,
\ i=\overline{1,m},$ be convex down and belong to the CTM class, a productive economic process $Q(p,z),$ a family of income pre-functions  $K_i^0(p,z), \ i=\overline{1,l},$ property vectors $b_i(p,z), \ i=\overline {1,l},$ be continuous maps of the variables
$(p,z) \in \bar R_+^n\times \Gamma^m,$ and let random fields of consumers choice and decisions making by firms  satisfy the conditions of the Theorem
\ref{ttl3} and a vector  $\bar p$
be an equilibrium price vector that satisfies the set of inequalities
 \begin{eqnarray*}       \frac{1}{\bar p_k}\sum\limits _{i=1}^l\gamma
_{ik}(\bar p)D_i(\bar p)\leq \left [\sum\limits _{i=1}^m[y_{ki}(\bar
p)-x_{ki}(\bar p)]+ \sum\limits _{i=1}^l b_{ki}(\bar p,z(\bar p))\right ], \quad
k=\overline{1,n},
\end{eqnarray*}
\begin{eqnarray*}
 D_i(\bar p) \not = 0, \quad i=\overline{1, l},\end{eqnarray*}
where $||\gamma_{ik}(p)||_{i,k=1}^{l,n}$ is a continuous demand matrix that describes  $1 \leq s \leq l$   non-insatiable consumers and
$l - s $  insatiable consumers.

Then there exists a matrix  $||\gamma_{ik}^*(p)||_{i,k=1}^{l,n}$ that is  continuous on  $P$
and such that   $\gamma_{ik}^*(p)=\gamma_{ik}(p), \  i=\overline{s+1, l}, \quad  k=\overline{1, n},$

\begin{eqnarray*}         \sum\limits_{k=1}^n\gamma^*_{ik}(p)=1, \quad  p \in P,
\quad i=\overline{1,l}, \end{eqnarray*}
the vector  $\bar
p$ is a solution of the set of equations \begin{eqnarray*}         \sum\limits
_{i=1}^l\gamma^* _{ik}(\bar p)D_i(\bar p)=\bar p_k \left
[\sum\limits _{i=1}^m[y_{ki}(\bar p)-x_{ki}(\bar p)]+ \sum\limits
_{i=1}^l b_{ki}(\bar p,z(\bar p))\right ], \quad k=\overline{1,n}.\end{eqnarray*}
\end{theorem}
\begin{proof}\smartqed
In accordance with the conditions of the Theorem, we assume that  $s$ realizations  $\gamma_i(p),\
i=\overline{1,s},$  of  random demand vectors are such that
 \begin{eqnarray*}         \sum\limits_{k=1}^n\gamma _{ik}(p)
\leq 1,\quad  p \in P, \quad i=\overline{1,s}. \end{eqnarray*}
For the rest realizations of demand vectors, there hold equalities  \begin{eqnarray*}         \sum\limits_{k=1}^n\gamma
_{ik}(p)=1,\quad  p \in P, \quad i=\overline{s+1,l}.\end{eqnarray*}
Let $\bar
p$ be a solution of the set of inequalities  \begin{eqnarray*}         \frac{1}{\bar p_k}\sum\limits
_{i=1}^l\gamma _{ik}(\bar p)D_i(\bar p)\leq  \sum\limits
_{i=1}^m[y_{ki}(\bar p)-x_{ki}(\bar p)]+ \sum\limits _{i=1}^l
b_{ki}(\bar p,z(\bar p)), \quad k=\overline{1,n},\end{eqnarray*}          then \begin{eqnarray*}         \sum\limits
_{i=1}^l\gamma _{ik}(\bar p)D_i(\bar p)\leq \bar p_k \left
[\sum\limits _{i=1}^m[y_{ki}(\bar p)-x_{ki}(\bar p)]+ \sum\limits
_{i=1}^l b_{ki}(\bar p,z(\bar p))\right ], \quad k=\overline{1,n}.\end{eqnarray*}
We prove the Theorem by induction. For  $s$ demand vectors the condition of insatiability is not satisfied on  the whole   set  $P,$ however, for  certain values
 $p \in P$ it can be valid. Consider the demand vector  $\gamma_s(p).$  There exist two possibilities: \\
1) for the vector $\bar p$ there holds the equality
\begin{eqnarray*}         \sum\limits_{k=1}^n\gamma _{sk}(\bar p)=1\  ;\end{eqnarray*}
\noindent 2)  for the vector $\bar p$  there holds the inequality
\begin{eqnarray*}         \sum\limits_{k=1}^n\gamma _{sk}(\bar p)<1.\end{eqnarray*}

If there holds the first condition, then since
\begin{eqnarray*}         \gamma_{sk}(p)=\frac{p_k\mu_{ks}^0(p)}{\sum\limits_{j=1}^n\mu_{js}(p)p_j},\quad    k=\overline{1,n},\end{eqnarray*}
we have
\begin{eqnarray*}         \sum\limits_{j=1}^n\mu_{js}(\bar p)\bar p_j=\sum\limits_{j=1}^n\mu_{js}^0(\bar p)\bar p_j.\end{eqnarray*}
Let us put
\begin{eqnarray*}         \mu_{ks}^1(p)=\mu_{ks}^0(p)+\sum\limits_{j=1}^n\mu_{js}( p) p_j-\sum\limits_{j=1}^n\mu_{js}^0(p)p_j.\end{eqnarray*}
On the set  $P,$ there holds the identity
\begin{eqnarray*}         \sum\limits_{j=1}^n\mu_{js}^1( p) p_j=\sum\limits_{j=1}^n\mu_{js}( p) p_j,\end{eqnarray*}
therefore, on the vector  $\bar p$
\begin{eqnarray*}         \frac{\bar p_k\mu_{ks}^1(\bar p)}{\sum\limits_{j=1}^n\mu_{js}^1(\bar p)\bar p_j}= \frac{\bar p_k\mu_{ks}^0(\bar p)}{\sum\limits_{j=1}^n\mu_{js}(\bar p)\bar p_j},\quad    k=\overline{1,n}.\end{eqnarray*}
Every component
\begin{eqnarray*}         \gamma_{sk}^1(p)=\frac{p_k\mu_{ks}^1( p)}{\sum\limits_{j=1}^n\mu_{js}^1(p)p_j},\quad    k=\overline{1,n},\end{eqnarray*}          is continuous on the set $P$ and there holds the condition
\begin{eqnarray*}         \sum\limits_{k=1}^n\gamma_{sk}^1(p)=1,\quad  p \in P.\end{eqnarray*}
Define in  this case
\begin{eqnarray*}         \gamma _{sk}^*(p)=\gamma _{sk}^1( p), \quad k=\overline{1,n},\quad
p \in P.\end{eqnarray*}

 Now, assume that there holds the second condition.
Consider the set of equations for the vector
$\alpha^s=\{\alpha_k^s\}_{k=1}^n$
\begin{eqnarray*}         \sum\limits_{i=s+1}^l\gamma
_{ik}(\bar p)D_i(\bar p)+ \left[\gamma _{sk}(\bar p)+
\alpha_k^s\left(1-\sum\limits _{k=1}^n\gamma_{sk}(\bar p)\right)\right]D_s(\bar
p)+ \sum\limits_{i=1}^{s-1}\gamma _{ik}(\bar p)D_i(\bar p)
\end{eqnarray*}
\begin{eqnarray*}
 =\bar p_k \left
[\sum\limits _{i=1}^m[y_{ki}(\bar p)-x_{ki}(\bar p)]+ \sum\limits
_{i=1}^l b_{ki}(\bar p,z(\bar p))\right ], \quad k=\overline{1,n}.
\end{eqnarray*}
A solution of this set of equations is a vector
$\alpha^s=\{\alpha_k^s\}_{k=1}^n,$ where
\begin{eqnarray*}
   \alpha_k^s=\frac{\bar p_k \left
[\sum\limits _{i=1}^m[y_{ki}(\bar p)-x_{ki}(\bar p)]+ \sum\limits
_{i=1}^l b_{ki}(\bar p,z(\bar p))\right ] - \sum\limits
_{i=1}^l\gamma_{ik}(\bar p)D_i(\bar p)}{\left[1-\sum\limits
_{k=1}^n\gamma_{sk}(\bar p)\right]D_s(\bar p)} \geq 0, \quad
  k=\overline{1,n}.
  \end{eqnarray*}
Then
\begin{eqnarray*}         \sum\limits _{k=1}^n\alpha_k^s=\frac{\sum\limits _{i=1}^s
\left[1-\sum\limits _{k=1}^n\gamma_{ik}(\bar p)\right]D_i(\bar p)}{
\left[1-\sum\limits _{k=1}^n\gamma_{sk}(\bar p)\right]
D_s(\bar p)} \geq 1,\end{eqnarray*}
in consequence of the equality
\begin{eqnarray*}         \sum\limits _{k=1}^n\bar p_k \left
[\sum\limits _{i=1}^m[y_{ki}(\bar p)-x_{ki}(\bar p)]+ \sum\limits
_{i=1}^l b_{ki}(\bar p,z(\bar p))\right ]=\sum\limits _{i=1}^lD_i(\bar p),\end{eqnarray*}
that follows from the Definition of the income function  $D_i(p)$
of the $i$-th consumer.
Consider a vector  $\bar \alpha^s=\{\bar \alpha^s_k\}_{k=1}^n,$ where
\begin{eqnarray*}         \bar
\alpha_k^s=\frac{\alpha_k^s}{\sum\limits_{k=1}^n\alpha_k^s},\quad
k=\overline{1,n}.\end{eqnarray*}          Then $\sum\limits_{k=1}^n \bar
\alpha_k^s=1.$
Let us put
\begin{eqnarray*}         \mu_{ks}^1(p)=p_k\mu_{ks}^0(p)+\bar \alpha_k^s \left[\sum\limits_{j=1}^n\mu_{js}( p) p_j-\sum\limits_{j=1}^n\mu_{js}^0(p)p_j\right].\end{eqnarray*}
Then on the set $P$ there holds  the identity
\begin{eqnarray*}         \sum\limits_{j=1}^n\mu_{js}^1( p)=\sum\limits_{j=1}^n\mu_{js}( p) p_j.\end{eqnarray*}
Now, determine  $\gamma _{sk}^*(p),$ putting
\begin{eqnarray*}         \gamma _{sk}^*(p)=\gamma _{sk}^1( p)=\frac{\mu_{ks}^1( p)}{\sum\limits_{j=1}^n\mu_{js}^1(p)}, \quad k=\overline{1,n},\quad
p \in P.\end{eqnarray*}
It is evident that the constructed vector
$\gamma_s^*(p)=\{\gamma^*_{sk}(p)\}_{k=1}^n$
both in the first and in the second cases is such that the equilibrium price vector
 $\bar p$ satisfies the set of inequalities
\begin{eqnarray*}
   \sum\limits_{i=s+1}^l\gamma
_{ik}(\bar p)D_i(\bar p)+ \gamma ^*_{sk}(\bar p)D_s(\bar p)+
\sum\limits_{i=1}^{s-1}\gamma _{ik}(\bar p)D_i(\bar p)
\end{eqnarray*}
\begin{eqnarray*}       \leq \bar p_k \left
[\sum\limits _{i=1}^m[y_{ki}(\bar p)-x_{ki}(\bar p)]+ \sum\limits
_{i=1}^l b_{ki}(\bar p,z(\bar p))\right ], \quad k=\overline{1,n}.
\end{eqnarray*}
Thus, in the economy system we diminished  the number of consumers  that are not insatiable  from  $s$ to $(s-1)$ and the equilibrium price vector  $\bar p$ is the equilibrium price vector  in such economy system. So, we can continue procedure carried out  further.
In consequence of finiteness  of $s,$  by the  finite number of steps we obtain the proof of the Theorem.
\qed
\end{proof}
It is evident  that there holds  inverse statement that gives an algorithm of the  construction of
  equilibrium price vectors  in the case when not all  consumers are  insatiable.

\begin{theorem}\label{ch2l5}
Let technological maps  $F_i(x), \  x \in X_i^1,
\ i=\overline{1,m},$ be convex down, belong to the CTM class, a productive  economic process  $Q(p,z),$ a family of income pre-functions  $K_i^0(p,z), \ i=\overline{1,l},$ property vectors $b_i(p,z), \ i=\overline {1,l},$  be continuous maps of variables
$(p,z) \in \bar R_+^n\times \Gamma^m,$ and let random fields of consumers choice and decisions making by firms satisfy the conditions of the Theorem
\ref{ttl3}.
For every continuous  demand matrix  $||\gamma_{ik}(p)||_{i=1, k=1}^{l,\ n}$  on  $P$ and continuous realization of random field of decisions making by firms  $z(p)$ there exists an equilibrium price vector  $ p^0=\{ p_k^0\}_{k=1}^n$ corresponding to them that is  a solution of the set of inequalities
\begin{eqnarray*}
     \frac{1}{ p_k}\sum\limits _{i=1}^l\gamma
_{ik}(p)D_i(p)
\end{eqnarray*}
\begin{eqnarray}\label{ndf2l18}
 \leq \sum\limits _{i=1}^m[y_{ki}(p)-x_{ki}(p)]+ \sum\limits _{i=1}^l b_{ki}(p,z(p)), \quad
k=\overline{1,n}.
\end{eqnarray}
There exist vectors    $\alpha_i(p)=\{\alpha_k^i(p)\}_{k=1}^n$ on  $P$ that satisfy the conditions
\begin{eqnarray*}          \sum\limits_{k=1}^n\alpha_k^i(p)=1, \quad i=\overline{1,s},\end{eqnarray*}
and such that the equilibrium price vector
$p^0$  satisfies the set of equalities
\begin{eqnarray*}
\sum\limits
_{i=1}^l\gamma _{ik}^*(p)D_i(p)
\end{eqnarray*}
\begin{eqnarray}\label{f2l18}
 = p_k \left
[\sum\limits _{i=1}^m[y_{ki}(p)-x_{ki}(p)]+ \sum\limits
_{i=1}^l b_{ki}(p,z(p)) \right ], \quad  k=\overline{1,n},
\end{eqnarray}
where
\begin{eqnarray*}
   \gamma _{ik}^*(p)=\frac{\mu_{ki}^1( p)}{\sum\limits_{j=1}^n\mu_{ji}^1(p)},\quad  p \in P, \quad k=\overline{1,n},\quad i=\overline{1,s},
\end{eqnarray*}
\begin{eqnarray*}       &  \mu_{ki}^1(p)=p_k\mu_{ki}^0(p)+\alpha_k^i(p) \left [\sum\limits_{j=1}^n\mu_{ji}( p) p_j-\sum\limits_{j=1}^n\mu_{ji}^0(p)p_j\right],
\end{eqnarray*}
\begin{eqnarray*}
 \gamma _{ik}(p)=\frac{p_k\mu_{ki}^0( p)}{\sum\limits_{j=1}^n\mu_{ji}(p)p_j}, \quad  p \in P, \quad k=\overline{1,n},\quad i=\overline{1,s},
\end{eqnarray*}
 \begin{eqnarray*}
 \gamma _{ik}^*(p)=\gamma _{ik}(p)=\frac{p_k\mu_{ki}( p)}{\sum\limits_{j=1}^n\mu_{ji}(p)p_j}, \quad  p \in P, \quad k=\overline{1,n},\quad i=\overline{s+1, l}.\end{eqnarray*}
\end{theorem}
\begin{proof}\smartqed
To prove the Theorem, consider an auxiliary model of economy.
Introduce a new set of income pre-functions
\begin{eqnarray*}         K_i^{0,d}(p,z)=K_i^0(p,z)+ d\sum\limits_{i=1}^np_i, \quad  i=\overline{1,l},\quad  d >0,\end{eqnarray*}
where  $K_i^0(p,z), \ i=\overline{1,l},$ is a set of income pre-functions that figures in the Theorem  \ref{ch2l5}.
We assume that in this model the $i$-th  consumer has a property vector   $b_i(p,z)+d e, \ i=\overline{1,l},$ where the vector $e=\{1, \ldots, 1 \}$ has unit components and $b_i(p,z), \ i=\overline{1,l},$ is a property vector of the  $i$-th consumer in the original model of economy. If we introduce a set of income functions
$K_i^{d}(p,z)=K_i^{0,d}(p,Q(p,z)), $ where $Q(p,z)$ is a productive economic process that figures in the Theorem \ref{ch2l5}, then such model of economy satisfies  the condition
\begin{eqnarray}\label{ndodl2}
 R_d(p, Q(p,z)) > 0, \quad p \in P, \quad  z \in \Gamma^m,
\end{eqnarray}
\begin{eqnarray*}
 R_d(p,z)= \sum\limits
_{i=1}^m[y_i-x_i]+ \sum\limits _{j=1}^l [b_j(p,z)+d e].
\end{eqnarray*}
Let $z(p)$ be a realization of random field of decisions making by firms  and
$\gamma_i(p)=\{\gamma_{ik}(p)\}_{k=1}^n, \ i=\overline{1,l},$ be a realization of random demand  vector of the $i$-th consumer.

We assume that $\alpha_i(p)=\{\alpha_k^i(p)\}_{k=1}^n, \ i=\overline{1,s},$ are continuous nonnegative vectors on the set $P,$   that satisfy  conditions
\begin{eqnarray*}          \sum\limits_{k=1}^n\alpha_k^i(p)=1, \quad i=\overline{1,s}.\end{eqnarray*}
Consider an auxiliary matrix   $\gamma^{* \delta}(p)=
||\gamma^{*\delta}_{ik}(p)||_{i=1, k=1}^{l, \ n},$ where
\begin{eqnarray*}         \gamma _{ik}^{*\delta}(p)=\frac{\mu_{ki}^1( p)+ \delta}{\sum\limits_{j=1}^n\mu_{ji}^1(p)+ n \delta}, \quad  p \in P, \quad k=\overline{1,n},\quad i=\overline{1,s}, \end{eqnarray*}
\begin{eqnarray*}         \gamma _{ik}^{*\delta}(p)=\frac{p_k\mu_{ki}( p)+ \delta}{\sum\limits_{j=1}^n\mu_{ji}(p)p_j+ n \delta}, \quad  p \in P, \quad k=\overline{1,n},\quad i=\overline{s+1, l}, \end{eqnarray*}
and  $\mu_{ki}^1( p), \  i=\overline{1,s}, \  k=\overline{1,n}, $ are defined in the condition of the Theorem.
In accordance with the Theorem \ref{ch2l3}, there exists a strictly positive solution   $p^{\delta,d}$ of the set of equations
\begin{eqnarray}\label{fq1al}
 \sum\limits _{i=1}^l\gamma _{ik}^{*\delta}(
p^{\delta,d})  D_i^d(p^{\delta,d})
\end{eqnarray}
\begin{eqnarray*}
& =p_k^{\delta,d} \left [\sum\limits _{i=1}^m[y_{ki}(p^{\delta,d})-x_{ki}(p^{\delta,d})]+
\sum\limits _{i=1}^l[ b_{ki}(p^{\delta,d},z(p^{\delta,d}))+d] \right ],\quad k=\overline{1,n}.
\end{eqnarray*}
Consider a certain sequence of numbers $\delta_s > 0$ that converges to zero as $s \to \infty.$
Let  $p^{\delta_s, d}$ be a solution of the set of equations  (\ref{fq1al}) if we put
 $ \delta= \delta_s.$
The set  of vectors  $p^{\delta_s, d}$ is  compact,  as $\delta_s \to 0,$ since it belongs to the unit simplex $P.$

If a vector $p^d=\{p_k^d\}_{k=1}^n$ is one of the limit points of the sequence  $p^{\delta_s, d},$
then it satisfies the set of inequalities
\begin{eqnarray}\label{1fq1al5}
\sum\limits_{i=1}^s \frac{\mu_{ki}^0(p^{d})}
{\sum\limits_{s=1}^n\mu_{si}^1(p^{d})} D_i^d( p^{d})+ \sum\limits_{i=s+1}^l \frac{\mu_{ki}(p^{d})}
{\sum\limits_{s=1}^n\mu_{si}(p^{d})p_s^{d}} D_i^d( p^{d})
\end{eqnarray}
\begin{eqnarray*}          \leq \sum\limits _{i=1}^m[y_{ki}(p^{d})-x_{ki}(p^{d})]+
\sum\limits _{i=1}^l[ b_{ki}(p^{d},z(p^{d}))+d],\quad k=\overline{1,n}.\end{eqnarray*}
It is evident that on the set $P$
\begin{eqnarray*}         \sum\limits_{j=1}^n\mu_{ji}^1( p)=\sum\limits_{j=1}^n\mu_{ji}( p) p_j,\quad i=\overline{1,s}.\end{eqnarray*}
Therefore,  $p^d=\{p_k^d\}_{k=1}^n$ is also a solution  of the set of inequalities
\begin{eqnarray}\label{fq1al5}
 \sum\limits_{i=1}^s \frac{\mu_{ki}^0(p^{d})}
{\sum\limits_{s=1}^n\mu_{si}(p^{d})p_s^{d}} D_i^d( p^{d})+ \sum\limits_{i=s+1}^l \frac{\mu_{ki}(p^{d})}
{\sum\limits_{s=1}^n\mu_{si}(p^{d})p_s^{d}} D_i^d( p^{d})
\end{eqnarray}
\begin{eqnarray*}
\leq \sum\limits _{i=1}^m[y_{ki}(p^{d})-x_{ki}(p^{d})]+
\sum\limits _{i=1}^l [ b_{ki}(p^{d},z(p^{d}))+d],\quad k=\overline{1,n}.
\end{eqnarray*}

For any  $d>0$ the vector  $p^d \in P$  satisfies the set of equations
\begin{eqnarray*}
& \sum\limits _{i=1}^l\gamma _{ik}^{*}(
p^{d})  D_i^d(p^{d})
\end{eqnarray*}
\begin{eqnarray}\label{ndodl3}
 =p_k^{d} \left [\sum\limits _{i=1}^m[y_{ki}(p^{d})-x_{ki}(p^{d})]+
\sum\limits _{i=1}^l[ b_{ki}(p^{d},z(p^{d}))+d] \right ],\quad k=\overline{1,n}.
\end{eqnarray}
The vector $p^d$ belongs to the compact  $P$ for every positive  $d.$

Consider a sequence of strictly positive numbers $d_s$ that converges to zero, as $s \to \infty.$
Let  $p^{d_s}$ be a solution of the set of equations  (\ref{ndodl3}) if to put  $d=d_s$ in the set of equations
 (\ref{ndodl3}.)
The sequence  $p^{d_s}, $  as  $d_s \to 0,$ is compact one since it belongs to the set $P.$
Without loss of generality, we assume that the sequence  $p^{d_s} $  converges to a certain vector $p^0 \in P,$ as $d_s \to 0.$
In the set of inequalities  (\ref{fq1al5}) and equalities
(\ref{ndodl3}) that are valid if  to substitute  $d_s$ instead of  $d,$  we can  pass to the limit, as  $s \to \infty.$
As a result, we obtain that the vector $p^0$ satisfies the set of inequalities  (\ref{ndf2l18}) and the set of equations  (\ref{f2l18}).

This is the needed set of inequalities  and   equalities   the Theorem said about.
\qed
\end{proof}

The last Theorem reduces the problem of  finding  equilibrium price vectors in the case  when not all consumers are insatiable  to the problem with insatiable consumers. Therefore, further we   investigate  economy systems with insatiable consumers only. From the last Theorem it follows  that the set of equilibrium price vectors can be continual in  economy systems with non-insatiable consumers

It is natural to consider as   optimal equilibrium price vector the vector that is a solution of the set of inequalities  (\ref{ndf2l18}) and which minimizes a function of capital not used\index{function of capital not used}
\begin{eqnarray*}         \sum\limits_{i=1}^s\left[1-\sum\limits_{k=1}^n\gamma _{ik}(\bar
p)\right]D_i(\bar p)\end{eqnarray*}
 on the set of all solutions of the set of inequalities (\ref{ndf2l18}). Such minimum exists because the set of vectors that are  solutions of the set of inequalities  (\ref{ndf2l18}) is closed  and
the function of capital  not used is a continuous function of the vector $\bar p.$
Denote the vector that minimizes the function of capital  not used by $p^{0}.$
One can introduce a notion of  norm of  indefinite behavior of consumers\index{norm of  indefinite behavior of consumers} that have free capital
\begin{eqnarray*}         n=\frac{\sum\limits_{i=1}^s\left[1-\sum\limits_{k=1}^n\gamma _{ik}(
p^{0})\right]D_i(p^{0})}{\sum\limits_{i=1}^lD_i(p^{0})}.\end{eqnarray*}
We can characterize by this norm the minimal degree  of financial crisis\index{norm of minimal degree  of financial crisis} that can arise
because  a large amount of goods  do not find   consumers in the economy system.
Similarly, one can introduce a notion of maximal degree of financial crisis\index{notion of maximal degree of financial crisis}  as ratio:
 \begin{eqnarray*}         N=\frac{\sum\limits_{i=1}^s\left[1-\sum\limits_{k=1}^n\gamma _{ik}(
p^1)\right]D_i(p^1)}{\sum\limits_{i=1}^lD_i(p^1)},\end{eqnarray*}
where  $p^1$ is a vector that maximizes the  function of  capital not used on the set of all solutions of the set of inequalities (\ref{ndf2l18}).

\subsection{Interindustry balance stochastic model}

Below, we construct example of the  economy model  satisfying  conditions of the Theorems proved.

We consider an economy system model with production described by technological maps generated by the Leontieff matrix: the economy has $n$ industries each of which produces a single kind of goods, production expenses for single goods unit are described by this matrix column. Therefore, let $A=||a_{ik}||_{i,k=1}^n$ be the Leontieff matrix, then matrix element $a_{ik}$ is the number of units of the $i$-th goods used to produce one unit of the $k$-th goods.
For  the non-negative matrix $A,$ we assume it to be productive and indecomposable.
Let $e_i=\{\delta_{ij}\}_{j=1}^n,\ i=\overline{1,n},$ be a vector whose  the $i$-th component equals 1 and the rest components are zero.
We define  technological maps in the model by the formulae
\begin{eqnarray}\label{prl1}
 F_i( x_i)=\{  y_i \in S, \  y_i=e_iu_i, \ 0 \leq  a_{ki}u_i \leq x_{ki}, \ k=\overline{1,n} \}, 
 \end{eqnarray}
\begin{eqnarray*}
   x_i=\{x_{ki}\}_{k=1}^n \in X_i,
\end{eqnarray*}
 \begin{eqnarray*}
 X_i=\{ x_i \in S, \ x_i=\{x_{ki}\}_{k=1}^n, \ 0 \leq x_{ki} \leq a_{ki}y_i^0, \ k=\overline{1,n}\}, \quad i=\overline{1,n},
\end{eqnarray*}
where $y_i^0$ is maximum available output of  the $i$-th industry, $i=\overline{1,n}.$
Let, further, the $i$-th consumer has property vector $b_i=\{b_{ki}\}_{k=1}^n, \ i=\overline{1,l}.$
Assume the condition $\sum\limits_{i=1}^lb_i \geq b_0$ holds, where the vector $b_0=\{b_i^0 \}_{i=1}^n$
has strictly positive components, i.e., $b_i^0 >0, \ i=\overline{1,n}.$
We give the set of possible  productive processes of the $i$-th industry by the formula
\begin{eqnarray*}         \Gamma_i=\{z=( x_i,  y_i), \  x_i \in X_i,\   y_i \in F_i(x_i)\}, \quad i=\overline{1,n},\end{eqnarray*}
and the set of possible productive  processes in the  economy system  by the formula
\begin{eqnarray*}         \Gamma^n=\prod\limits_{i=1}^n\Gamma_i. \end{eqnarray*}
In this Subsection we suppose that the set of possible  price vectors $K_+^n$ in the  economy system coincides with the set
 $\bar R_+^n.$

Introduce  continuous on $R^1$ function $\varphi_{[a_i, \infty)}(x)$ that equals 1 in the interval $[a_i, \infty),$ does zero in the interval $(-\infty, 0],$ and  is  positive less than 1 in the interval
$(0,a_i),$  continuous on $R^1$ function $u_{[d_i, \infty)}(x)$ that equals 1 in the interval $[d_i, \infty),$ does zero in the interval $(-\infty, 0],$ and is positive less than 1 in the interval
$(0,d_i),$ $a_i >0,\ d_i >0, \ i=\overline{1, n}.$
Denote
\begin{eqnarray*}          W_i(p)=\frac{p_i - \sum\limits_{s=1}^na_{si}p_s}{\sum\limits_{i=1}^np_i},  \quad p \in \bar R_+^n,   \quad i=\overline{1, n},  \end{eqnarray*}
\begin{eqnarray*}         a_i(p)=\varphi_{[a_i, \infty)}(W_i(p)), \quad   p \in \bar R_+^n, \quad i=\overline{1, n},\end{eqnarray*}
\begin{eqnarray*}         v_i(p,z)=u_{[d_i, \infty)}\left(a_i(p)u_i - \sum\limits_{s=1}^na_{is}a_s(p)u_s + \sum\limits_{k=1}^lb_{ik} - b_i^0\right), \quad i=\overline{1, n}. \end{eqnarray*}

The map
\begin{eqnarray}\label{prl2}
Q(p,z)=\{Q_i(p,z)\}_{i=1}^n, \quad
  (p,z) \in \bar R_+^n\times \Gamma^n,
\end{eqnarray}
where
\begin{eqnarray*}
     Q_i(p,z)=\{X_i(p,z), Y_i(p,z)\}, \quad i=\overline{1, n}, \end{eqnarray*}
\begin{eqnarray*}
   X_i(p,z)=\prod\limits_{s=1}^nv_s(p,z)\{a_{ki}a_i(p)u_i\}_{k=1}^n, \end{eqnarray*}          \begin{eqnarray*}
     Y_i(p,z)=\prod\limits_{s=1}^nv_s(p,z)a_i(p)u_ie_i, \quad i=\overline{1, n},
\end{eqnarray*}
is a productive economic process for which the production stops only in the detrimental industry.
Establish the inequality
\begin{eqnarray*}         \sum\limits_{i=1}^n[Y_i(p,z)- X_i(p,z)]+\sum\limits_{i=1}^lb_i\geq b_0, \quad (p,z) \in  \bar R_+^n\times\Gamma^n,\end{eqnarray*}
meaning that the set of values $Q(p,\Gamma^n)$ of the map
$Q(p,z)$ belongs to the set $G(p)$ built in the Lemma \ref{Ql1}.
Denote
\begin{eqnarray*}         y(p,z)=\{a_i(p)u_i \}_{i=1}^n, \quad a(p,z)=\prod\limits_{i=1}^nv_i(p,z).\end{eqnarray*}
Therefore,
\begin{eqnarray*}
 \sum\limits_{i=1}^n[Y_i(p,z)- X_i(p,z)]+\sum\limits_{i=1}^lb_i=a(p,z)[y(p,z) - Ay(p,z)]+ \sum\limits_{i=1}^lb_i
\end{eqnarray*}
\begin{eqnarray*}
    =\left\{\begin{array}{ll}
              [y(p,z) - Ay(p,z)]a(p,z) + \sum\limits_{i=1}^lb_i, & \textrm{if} \quad a(p,z)> 0,\\
               \sum\limits_{i=1}^lb_i, &      \textrm{if}  \quad   a(p,z)=0 \textrm{.}
                                                                                                                                                \end{array}
                                               \right.\end{eqnarray*}
However, for those $z$ for which $a(p,z)> 0,$ the set of inequalities
\begin{eqnarray*}          a_i(p)u_i - \sum\limits_{k=1}^na_{ik}a_k(p)u_k + \sum\limits_{k=1}^lb_{ik} >  b_i^0, \quad i=\overline{1,n},\end{eqnarray*}
holds. So, accounting for that $\sum\limits_{i=1}^lb_i  \geq b_0,$ we have
\begin{eqnarray*}         [y(p,z) - Ay(p,z)]a(p,z) + \sum\limits_{i=1}^lb_i \geq \sum\limits_{i=1}^lb_i (1 - a(p,z)) + a(p,z) b_0 \geq b_0.\end{eqnarray*}

Let $\zeta_0(p, \omega_0)=\{\zeta_0^i(p)\}_{i=1}^n,\  p \in \bar R_+^n,$ be a random field given on a probability space $\{\Omega_0, {\cal F}_0, P_0\},$
take values in the set $\prod\limits_{i=1}^n[0, y_i^0],$ be of zero homogeneity degree and   be continuous with probability 1 on $\bar R_+^n.$
Therefore,   a random field of decisions making  by industries
\begin{eqnarray}\label{ngb1}
 \zeta(p)=Q(p,\zeta_0(p, \omega_0))=\{\zeta_i(p) \}_{i=1}^n, \quad p \in \bar R_+^n,
\end{eqnarray}
 is continuous with probability 1
and
\begin{eqnarray*}
  \zeta_i(p)=Q_i(p,\zeta_0(p))=\{\zeta_i^1(p), \  \zeta_i^2(p)\},\quad  i=\overline{1,n},
  \end{eqnarray*}
where
\begin{eqnarray*}
 \zeta_i^1(p) =\{a_{ki}a_i(p)\zeta_{0}^i(p)\}_{k=1}^n \prod\limits_{i=1}^nv_i(p,\zeta_0(p, \omega_0)),\quad  i=\overline{1,n},
 \end{eqnarray*}
\begin{eqnarray*}
  \zeta_i^2(p) =a_i(p)\zeta_{0}^i(p)\prod\limits_{i=1}^nv_i(p,\zeta_0(p, \omega_0))e_i,\quad  i=\overline{1,n},
  \end{eqnarray*}
\begin{eqnarray*}
   v_i(p,\zeta_0(p, \omega_0))= u_{[d_i, \infty)}\left(a_i(p)\zeta_{0}^i(p) -
\sum\limits_{s=1}^na_{is}a_s(p)\zeta_{0}^s(p) + \sum\limits_{k=1}^lb_{ik} - b_i^0\right).
\end{eqnarray*}
Introduce a random field, namely,  the work intensity vector of industries $t(p)=\{t_i(p)\}_{i=1}^n,$ where
\begin{eqnarray*}         t_i(p)=a_i(p)\zeta_{0}^i(p) \prod\limits_{s=1}^nv_s(p,\zeta_0(p, \omega_0)).\end{eqnarray*}
It is obvious that $t(p)$ is homogeneous of zero degree and continuous on $\bar R_+^n$ if a realization of $\zeta_0(p, \omega_0)$ is such one.
Then
\begin{eqnarray*}         \zeta^1_i(p) =\{a_{ki}t_i(p)\}_{k=1}^n, \quad \zeta^2_i(p) =t_i(p)e_i, \quad i=\overline{1,n}.\end{eqnarray*}
We give  the final consumption vector $\psi(p)=\{\psi_k(p)\}_{k=1}^n$ by the formula
\begin{eqnarray*}
 \psi(p)=\sum\limits_{k=1}^n[Y_k(p, \zeta_0(p, \omega_0)) - X_k(p, \zeta_0(p, \omega_0))] +\sum\limits_{i=1}^lb_{i}
\end{eqnarray*}
 \begin{eqnarray*}
 =(E- A)t(p)+\sum\limits_{i=1}^lb_{i}, \quad  t(p)=\{t_i(p)\}_{i=1}^n.
 \end{eqnarray*}
Hence,
\begin{eqnarray*}
  \psi_k(p) =t_k(p)-\sum\limits_{i=1}^n
a_{ki}t_i(p)+\sum\limits_{i=1}^lb_{ki}, \quad k=\overline{1,n}.
\end{eqnarray*}
Then the  work intensity vector of  industries\index{work intensity vector of  industries} ~$t(p)=\{t_i(p)\}_{i=1}^n$~ is such that
\begin{eqnarray*}
  t_i(p)-\sum\limits_{k=1}^n
a_{ik}t_k(p)+\sum\limits_{k=1}^lb_{ik}\geq b_i^0, \quad i=\overline{1,n},
\end{eqnarray*}
with probability 1.

We give an income  pre-function of the $i$-th consumer by the formula
 \begin{eqnarray*}
  K_i^0(p,z)=\sum\limits _{j=1}^l\pi
_{ij}(p) \sum\limits _{k=1}^l r_{jk}(p) \left [\sum\limits
_{s=1}^n \alpha _{ks}(p)\left\langle  y_s -  x_s,p\right\rangle +\left\langle
b_k,p\right\rangle\right ], \quad  i=\overline{1,l},
\end{eqnarray*}
\begin{eqnarray}\label{al15}
 z^k=( x_k, y_k),\quad k=\overline {1,n},\quad  z=\{z^k\}_{k=1}^n,
\end{eqnarray}
where $b_k$ is an initial goods supply vector of the $k$-th
consumer at the beginning of the economy operation period. Then the $i$-th consumer income corresponding to a  realization of random field of decisions making by industries  we give  by the formula
\begin{eqnarray*}
   D_i(p)=K_i(p,\zeta_0(p, \omega_0)))=K_i^0(p,\zeta(p))
\end{eqnarray*}
\begin{eqnarray*}
     =\sum\limits _{j=1}^l
\pi _{ij}(p) \sum\limits _{k=1}^l
r_{jk}(p) \left [\sum\limits _{s=1}^n \alpha
_{ks}(p)t_s(p)\left[p_s-\sum\limits_{v=1}^na_{vs}p_v\right] +\left\langle
b_k,p\right\rangle\right ], \quad i=\overline{1,l},
\end{eqnarray*}
where $ \pi _{ij}(p),\ r_{jk}(p),  \alpha _{ks}(p) $ are continuous non-negative  functions on
$\bar R_+^n$ satisfying conditions (\ref{sl000}) and (\ref{sl0000}) when the cone $K_+^n=\bar R_+^n.$

\begin{theorem}\label{mgb1}
Let technological maps $F_i(x), \  x \in X_i,
\ i=\overline{1,n},$   have the form (\ref{prl1}),
a  productive economic process
 $Q(p,z)$ be given by the formulae (\ref{prl2}) and let a family of  income pre-functions  $K_i^0(p,z), \ i=\overline{1,l},$ be given by the formula (\ref{al15}).
If the  random field of decisions making  by  industries is defined by the formula (\ref{ngb1}) and the random fields of  consumers choice  under defined objects satisfy conditions of the Theorem \ref{tl4},
then with  probability 1 there exists the Walras  equilibrium state, i.e., for every realization of consumers choice and industries decisions
making  random fields there exists  corresponding to them a vector $p^*\in P$ such that the economy is in the Walras equilibrium state. If, additionally, realizations of random field of  decisions making by industries  are such that among them there are arbitrarily close to  optimum industries behavior strategy in the sense of the Theorem \ref{nnl1}, then with  probability 1
there exists the optimum Walras equilibrium state.
\end{theorem}

\subsection{Interindustry balance model with profitable pro\-duction}

In this Subsection, we consider the same economy system production model as in the previous Subsection.
As distinct from the  previous Subsection, here we suppose that the set of possible price vectors is the set
\begin{eqnarray*}         K_+^n=\{p=\{p_i\}_{i=1}^n  \in R_+^n, \ p_i >0, \ i=\overline{1,n}\}.\end{eqnarray*}
  We consider a productive economic  process given by the formula (\ref{prl2}) on the reduced set $ K_+^n\times \Gamma^n.$

To construct a  random field of choice of the $i$-th consumer, let us  build a probability space $\{\Omega_i, {\cal F}_i, P_i\}$
 and a random field of evaluation of information $ \eta_i^0(p, z, \omega_i), \ i=\overline{1,l},$ on this probability space.
Take, for $\Omega_i,$ the set of integer numbers $N=\{1,2, \ldots, \},$ $ i=\overline{1,l},$
and, for $\sigma$-algebra ${\cal F}_i,$ the set of all subsets of the set $N.$ We give a probability measure $P_i$ on ${\cal F}_i$ by the sequence $\pi_m^i, \ m=1,2,\ldots, \
\sum\limits_{m=1}^{\infty}\pi_m^i=1,$ putting $P_i(m)=\pi_m^i.$
Build a family of  maps
\begin{eqnarray*}         \eta_i^0(p,z, m)=\{\eta_{ik}^0(p,z, m)\}_{k=1}^n, \quad (p, z) \in
K_+^n\times\Gamma^n,   \quad m=1,2, \ldots, \ i=\overline{1,l}.\end{eqnarray*}
Put
\begin{eqnarray*}         \eta_{ik}^0(p,z, m)=\frac{1}{p_k}\left|u_k-
\sum\limits_{s=1}^{n}a_{ks}u_s+\sum\limits_{s=1}^{l}b_{ks}\right|
\sum\limits_{j=1}^{n}f_{kj}\varphi_{ji}(p,z, m),\end{eqnarray*}          \begin{eqnarray*}
 (p,z, m) \in K_+^n\times \Gamma^n\times\Omega_i, \quad i=\overline{1,l},  \quad k=\overline{1,n}.\end{eqnarray*}
Assume that random fields
$\varphi_{ji}(p,z, m), \ (p,z) \in K_+^n\times\Gamma^n,\  j=\overline{1,n},\ i=\overline{1,l}, $ given on a probability space $\{\Omega_i, {\cal F}_i, P_i\}$
and taking values in the set $R_+^1,$ are continuous with  probability 1 and such that for every realization the next inequalities and equalities hold:
\begin{eqnarray*}         \varphi_{ji}(p,z, m)> c(m),\quad  c(m) >0, \quad (p,z, m) \in P\times \Gamma^n\times\Omega_i,\end{eqnarray*}
\begin{eqnarray}\label{ngb2}
\varphi_{ji}(tp,z, m)=t\varphi_{ji}(p,z, m),\quad  t > 0, \quad (p,z, m) \in K_+^n\times \Gamma^n\times\Omega_i,
\end{eqnarray}
\begin{eqnarray*}          P=\left\{p \in K_+^n, \ \sum\limits_{i=1}^np_i=1\right\}. \end{eqnarray*}
  Random fields of consumers choice we define by the formula
\begin{eqnarray*}         \xi_i(p, m)
\end{eqnarray*}
\begin{eqnarray}\label{prl3}
=K_i^0(p,\zeta(p))\frac{\eta_i^0(p,\zeta(p), m)}
{\sum\limits_{k=1}^np_k\eta_{ik}^0(p,\zeta(p), m)}, \quad  i=\overline{1,l}, \quad  m=\overline{1,\infty},
\end{eqnarray}
and a random field of decisions making  by  firms we do by the formula $\zeta(p)=Q(p,\zeta_0(p, \omega_0)),$ where $\zeta_0(p, \omega_0),\  p \in K_+^n, $ is the reduction of the random field defined in the previous Subsection onto the set $K_+^n. $
Such defined random fields are continuous  with probability 1 and satisfy conditions of the Theorem \ref{tl4}.
We determine the set of vectors $f_1, \ldots,f_n$ as the set  of generatrices of the non-negative cone
\begin{eqnarray*}
 \bar T=\left\{  p \in K_+^n, \
p_i-\sum\limits_{s=1}^na_{si}p_s \geq 0, \ i=\overline{1,n}\right\}.
\end{eqnarray*}
   As the matrix $A$ is productive and indecomposable,
the set $\bar T$ is non-empty  because it contains,
e.g., the vector solving the set of equations
 \begin{eqnarray*}         rp_i=\sum\limits_{s=1}^na_{si}p_s, \quad
i=\overline{1,n},\end{eqnarray*}          where $r<1$ is the maximum proper value of the matrix $A.$ Let us put $f_j=(E-A^T)^{-1}s_j, \ j=\overline{1,n},$ where
$A^T$ is the matrix transposed to the matrix $A,$
the vector-column $s_j=\{\delta_{ij}\}_{i=1}^n$ is such that its
$j$-th component equals 1 and the rest components are zero.
It is obvious that the vectors $f_j, \ j=\overline{1,n},$ are linearly independent and strictly positive  because the matrix $A$ is productive and indecomposable  and \begin{eqnarray*}         \bar T=\left\{p \in K_+^n, \
p=\sum\limits_{i=1}^nc_if_i, \ c= \{c_i\}_{i=1}^n \in R_+^n, \  c \neq 0 \right \}.\end{eqnarray*}
Put $f_j=\{f_{kj}\}_{k=1}^n.$
 Random fields defined by the formula (\ref{prl3})  are continuous  with  probability 1 and satisfy the Theorem \ref{tl4} conditions.

Let $z(p), \ p \in K_+^n,$ be a continuous realization of the random field $\zeta(p).$
 The considered stochastic model of economy is such that all the industries are non-detrimental  with  probability 1.
 Demand vectors corresponding to  the constructed   random fields of evaluation of information by consumers  and  the random field of decisions making by industries   take the form
\begin{eqnarray}\label{al16}
\gamma
_i(p)=\{\gamma _{ik}(p)\}_{k=1}^n, \quad i=\overline{1,l},
\end{eqnarray}
where
\begin{eqnarray*}         \gamma_{ik}(p)=\frac{\psi_k(p)\sum\limits_{j=1}^n
f_{kj}\varphi_{ji}(p, z(p),m)}{\sum\limits_{j=1}^n\sum\limits_{k=1}^n
\psi_k(p)f_{kj}\varphi_{ji}(p,z(p),m)}, \quad k=\overline{1,n},  \quad
i=\overline{1,l},\end{eqnarray*}
for a certain $z(p)$ and $m,$
where $\varphi_{ji}(p,z(p),m)$ are continuous on
 $K_+^n$ functions,
\begin{eqnarray*}         \psi_k(p) =t_k(p)-\sum\limits_{i=1}^n
a_{ki}t_i(p)+\sum\limits_{i=1}^lb_{ik}, \quad k=\overline{1,n}.\end{eqnarray*}

\begin{theorem}\label{pr1} Assume that the production structure, in the economy system,  is described by technological maps (\ref{prl1}) and the matrix $A$ is productive and indecomposable.
 A productive economic process $Q(p,z)$ is expressed as (\ref{prl2}) and consumers income pre-functions are given by the formula (\ref{al15}), where
$\alpha _{ij}(p),$ $~r_{ij}(p),$ $~\pi _{ij}(p)$ are continuous on
$K_+^n$ and satisfy conditions (\ref{sl000}), (\ref{sl0000}).
If  random fields of consumers choice are defined by (\ref{prl3}) on the set
$K_+^n$
and   random field of decisions making by industries is given  by $\zeta(p)= Q(p,\zeta_0(p, \omega_0)),$ where the random field $\zeta_0(p, \omega_0)$ is continuous  with  probability 1,
then in such  economy system with  probability 1  all the industries are non-detrimental  in the Walras equilibrium state, i.e., there exists a strictly positive solution to the set of equations
\begin{eqnarray}\label{al17}
\sum\limits _{i=1}^l\gamma _{ik}( p)D_i(p)=p_k\psi_k(p), \quad k=\overline{1,n},
\end{eqnarray}
in the closed bounded convex set
\begin{eqnarray*}         T_1=\left\{p \in
K_+^n, \ p_i-\sum\limits_{s=1}^na_{si}p_s \geq 0, \
i=\overline{1,n}, \ \sum\limits_{i=1}^np_i=1\right\}.
\end{eqnarray*}
\end{theorem}
\begin{proof}\smartqed
It is obvious that $T_1$ is a closed bounded convex set. Prove that the set $T_1$ contains vectors with strictly positive components. If it is not the case, then the equalities
\begin{eqnarray*}         p_i -\sum\limits_{k=1}^na_{ki}p_k=0, \quad i=\overline{1,n},\end{eqnarray*}
 hold from which it follows that $p=0$ because the matrix $A$ is productive.
For all the vectors from $T_1$ in the set of inequalities
\begin{eqnarray*}         p_i -\sum\limits_{k=1}^na_{ki}p_k \geq 0, \quad i=\overline{1,n},\end{eqnarray*}
at least for one of $i$ the inequality
\begin{eqnarray*}         p_i -\sum\limits_{k=1}^na_{ki}p_k =\delta_i > 0 \end{eqnarray*}
holds. Therefore, the vector $p$ solves the set of equations
\begin{eqnarray*}         p_i -\sum\limits_{k=1}^na_{ki}p_k=\delta_i, \quad i=\overline{1,n},\end{eqnarray*}
where the vector $\delta= \{\delta_i\}_{i=1}^n$ has non-zero components. This solution has the representation $p=\sum\limits_{i=1}^nf_i\delta_i.$
From the indecomposability and productivity of $A$, it follows that $f_{kj} > 0,\ k,j=\overline{1,n},$
and, therefore, all the components of the vector $p$ are strictly positive.
Show the inequality
\begin{eqnarray*}         \sum\limits_{j=1}^n\sum\limits_{k=1}^n
\psi_k(p)f_{kj}\varphi_{ji}(p, z(p), m)> \delta >0, \quad p \in  T_1, \quad i=\overline{1,l},\end{eqnarray*}
holds, where $\delta$ does not depend on $ p \in T_1.$

Because the  inequalities (\ref{ngb2}) and the inequalities
\begin{eqnarray*} 
f_{kj} > 0,\quad  k,j=\overline{1,n}, \quad  \psi_k(p)> b_k^0 > 0, \quad  k=\overline{1,n}, \end{eqnarray*} 
hold, the needed inequality is valid.

From the structure of demand vectors, it follows that the  non-linear operator
\begin{eqnarray*}         \alpha(p)=\left\{\frac{\alpha_k(p)}{\sum\limits_{k=1}^n \alpha_k(p)}
\right\}_{k=1}^n,\end{eqnarray*}
where
\begin{eqnarray*}         \alpha_k(p)=\frac{1}{\psi_k(p)}\sum\limits
_{i=1}^l\gamma _{ik}( p)D_i( p), \quad k=\overline{1,n},\end{eqnarray*}
acts on the set $T_1,$
maps it into itself, and is continuous  on $T_1.$
Really, show that a positive number $\delta > 0$ exists such that the inequality
\begin{eqnarray*}         \sum\limits_{k=1}^n\alpha_k( p) > \delta\end{eqnarray*}
holds. As $\psi(p) > b_0 > 0, $ on the set $T_1$ the inequalities
\begin{eqnarray*}         \sum\limits_{i=1}^lD_i( p)=\left\langle \psi(p),  p \right\rangle  >  \left\langle  b_0, p \right\rangle  >  \min_{1 \leq i \leq n}b_i^0  \ >0\end{eqnarray*}
 hold. Therefore,
\begin{eqnarray*}         \sum\limits_{k=1}^n\alpha_k( p) >  \frac{\min\limits_{1 \leq i \leq n}b_i^0}{\max\limits_{1 \leq i \leq n}\sup\limits_{p \in P}\psi_i(p)}=\delta >0.\end{eqnarray*}
By the Schauder Theorem \cite{88}, there exists a fixed point of this map belonging to $T_1.$
By the same arguments as in the Theorem \ref{ch2l3}, this fixed point is that  for the set of equations (\ref{al17}).
\qed
\end{proof}

\subsection{ Equilibrium in the Stochastic Neumann Model of Production}

Construct the stochastic Neumann model of production, in which every firm is profitable  with  probability 1.

Consider the economy system containing $m$ firms, describe the production technology of
the $i$-th firm by the vector pair
$(a_i,b_i),\ i=\overline {1,m},$ where
$a_i=\{a_{ki}\}_{k=1}^n$ is an input vector and
$b_i=\{b_{ki}\}_{k=1}^n$ is an output vector corresponding to the input vector $a_i$.
 Suppose $m$ firms of the economy system produce $n$ kinds of goods for $l$ consumers.
We give technological maps, within the model, by the formulas
\begin{eqnarray}\label{pnrl1}
   F_i(x_i)=\{y_i \in S, \  y_i=\xi_i\{b_{ki}\}_{k=1}^n, \  a_{ki}\xi_i \leq x_{ki}, \ k=\overline{1,n} \}, 
\end{eqnarray}
\begin{eqnarray*}
 x_i \in    X_i=\{ x_i=\{x_{ki}\}_{k=1}^n \in S,  \ x_{ki} \leq  a_{ki}\xi_i^0, \ k=\overline{1,n}  \}, \quad i=\overline{1,m},
     \end{eqnarray*}
where $\xi_i^0$ is the maximum available work intensity of the $i$-th firm, $i=\overline{1,m}.$
If the $i$-th firm works with the intensity
 $\xi_i, \ i=\overline{1,m},$
then the added value under the price vector $p=\{p_i\}_{i=1}^n \in \bar R_+^n$ has the form
\begin{eqnarray}\label{gnnl1}
{\left\langle C_i(\xi),p\right\rangle} =\sum\limits _{k=1}^n
\xi_i (b_{ki}-a_{ki})p_k= \sum\limits _{k=1}^n C_{ki}(\xi)p_k,
\end{eqnarray}
\begin{eqnarray*}         C_{ki}(\xi)=\xi_i(b_{ki}-a_{ki}),\quad  k=\overline{1,n}, \quad
C_i(\xi)=\{C_{ki}(\xi)\}_{k=1}^n, \quad i=\overline{1,m}.\end{eqnarray*}
Let, further, $l$ consumers have property vectors
\begin{eqnarray*}         b_i^1=\{b_{ki}^1\}_{k=1}^n, \quad  i=\overline{1,l}.\end{eqnarray*}
Assume that  the condition $\sum\limits_{i=1}^lb_i^1 \geq b_0$ holds, where the vector $b_0=\{b_i^0 \}_{i=1}^n$
has strictly positive components.
We give the set of possible productive processes of the $i$-th firm by the formula
\begin{eqnarray*}         \Gamma_i=\{z^i=(x_i,y_i), \ x_i \in X_i,  \ y_i \in F_i(x_i)\}, \quad i=\overline{1,m}, \end{eqnarray*}
and the set of possible  productive processes in the  economy system by the formula
 $\Gamma^m=\prod\limits_{s=1}^m \Gamma_s.$

To every pair $z^i=(x_i,y_i) \in \Gamma_i,$ a certain   work intensity $\xi_i$ of the $i$-th firm  corresponds.
Therefore, the rule exists putting into correspondence  a certain work intensity $\xi_i $ of the $i$-th firm, lying within the interval $[0, \xi_i^0],$ to every pair $(x_i,y_i) \in \Gamma_i.$
To build a productive  economic process $ Q(p,z)$ given on $ \Gamma^m,$
it is sufficient to give a map of the set
\begin{eqnarray*}         \Sigma=\{\xi =\{\xi_i\}_{i=1}^m \in R_+^m, \ \xi_i \leq \xi_i^0,\ i=\overline{1,m}\}\end{eqnarray*}
 into the set $ \Gamma^m.$
Within this Subsection, we suppose  that the set of possible price vectors is
\begin{eqnarray*}         K_+^n=\{p=\{p_i\}_{i=1}^n  \in R_+^n, \ p_i >0, \ i=\overline{1,n}\}.\end{eqnarray*}
Introduce a continuous  function $\varphi_{[a_i, \infty)}(x),  \ a_i >0,\ i=\overline{1,m}, $ on $R^1$ that equals 1 on the interval $[a_i, \infty),$ does zero on the interval $(-\infty, 0],$ and is positive and less than 1 on the interval
$(0,a_i),$ and a continuous  function $u_{[d_s, \infty)}(x), \ d_s >0, \ s=\overline{1,n},$ on $R^1$ that equals 1 on the interval $[d_s, \infty),$  does zero on the interval $(-\infty, 0],$ and is positive  and less than 1 on the interval
$(0,d_s).$

Denote
\begin{eqnarray*}         a_i(p)=\varphi_{[a_i, \infty)}(W_i(p)),\quad      W_i(p)=\frac{\sum\limits_{s=1}^nb_{si}p_s - \sum\limits_{s=1}^na_{si}p_s}{\sum\limits_{i=1}^np_i}, \quad  i=\overline{1,m},\end{eqnarray*}
\begin{eqnarray*}         v_s(p,z)=u_{[d_s, \infty)}\left(\sum\limits_{i=1}^mb_{si}a_i(p)\xi_i - \sum\limits_{i=1}^ma_{si}a_i(p)\xi_i + \sum\limits_{i=1}^lb_{si}^1 - b_s^0\right), \quad  s=\overline{1,n}. \end{eqnarray*}
The map
\begin{eqnarray}\label{nrl2}
 Q(p,z)=\{ Q_i(p,z)\}_{i=1}^m, \quad (p,z) \in K_+^n\times \Gamma^m,
\end{eqnarray}
where
\begin{eqnarray*}          Q_i(p,z)=\{X_i(p,z), Y_i(p,z)\}, \quad  i=\overline{1,m},\end{eqnarray*}
and
\begin{eqnarray*}         X_i(p,z)=\prod\limits_{s=1}^nv_s(p,z)\{a_{ki}a_i(p)\xi_i\}_{k=1}^n,  \quad  i=\overline{1,m},\end{eqnarray*}          \begin{eqnarray*}          Y_i(p,z)=\prod\limits_{s=1}^nv_s(p,z)\{b_{ki}a_i(p)\xi_i\}_{k=1}^n, \quad  i=\overline{1,m},\end{eqnarray*}
is a productive economic  process,  within which a firm stops  production if it is detrimental.

Prove that the inequality
\begin{eqnarray*}         \sum\limits_{i=1}^m[Y_i(p,z)- X_i(p,z)]+\sum\limits_{i=1}^lb_i^1\geq b_0, \quad (p,z) \in  K_+^n\times\Gamma^m, \end{eqnarray*}
holds. It  means that the set of values $Q(p,\Gamma^m)$ of the map
$Q(p,z)$ belongs to the set $G(p)$ built in the Lemma \ref{Ql1}.
Denote
\begin{eqnarray*}         \xi(p,z)=\{a_i(p)\xi_i \}_{i=1}^m, \quad a(p,z)=\prod\limits_{s=1}^nv_s(p,z).\end{eqnarray*}
Then
\begin{eqnarray*}
    \sum\limits_{i=1}^m[Y_i(p,z)- X_i(p,z)]+\sum\limits_{i=1}^lb_i^1=a(p,z)[B\xi(p,z) - A\xi(p,z)]+ \sum\limits_{i=1}^lb_i^1
\end{eqnarray*}
\begin{eqnarray*}
       =\left\{\begin{array}{ll}
              [B\xi(p,z) - A\xi(p,z)]a(p,z) + \sum\limits_{i=1}^lb_i^1, & \textrm{if} \quad a(p,z) > 0,\\
               \sum\limits_{i=1}^lb_i^1, &      \textrm{if}  \quad   a(p,z)=0 \textrm{.}
                                                                                                                                                \end{array}
                                               \right.\end{eqnarray*}
However, for those $z$ for which $a(p,z) > 0,$ the set of inequalities
\begin{eqnarray*}          \sum\limits_{i=1}^mb_{ki}a_i(p)\xi_i - \sum\limits_{i=1}^ma_{ki}a_i(p)\xi_i + \sum\limits_{i=1}^lb_{ki}^1 >  b_k^0, \quad k=\overline{1,n},\end{eqnarray*}
holds. Therefore,
\begin{eqnarray*}         [B\xi(p,z) - A\xi(p,z)]a(p,z) + \sum\limits_{i=1}^lb_i^1 \geq \sum\limits_{i=1}^lb_i^1 (1 - a(p,z)) + a(p,z) b_0 \geq b_0.\end{eqnarray*}

From the fact  that $\sum\limits_{i=1}^lb_i^1  \geq b_0,$ when $a(p,z)=0,$ we establish the needed inequality.

Let
\begin{eqnarray*}         \psi^0(p,z)=\{\psi_k^0(p,z)\}_{k=1}^n, \quad \psi_k^0(p,z)=\sum\limits_{i=1}^m[y_{ki} - x_{ki}]+
\sum\limits_{i=1}^lb_{ki}^1, \quad k=\overline{1,n}.\end{eqnarray*}
Then the final supply vector
\begin{eqnarray*}          \psi(p,z)=\psi^0(p,Q(p,z))=\sum\limits_{i=1}^m[Y_i(p,z)- X_i(p,z)]+\sum\limits_{i=1}^lb_i^1.\end{eqnarray*}
Consider a probability space $\{\Omega, {\cal F}, \bar P\}$
being the direct product of $(l+1)$ probability  spaces $\{\Omega_i, {\cal F}_i, \bar P_i\}, \
i=\overline{0,l},$
where
\begin{eqnarray*}         \Omega=\prod\limits_{i=0}^{l}\Omega_i, \quad {\cal F}=\prod\limits_{i=0}^{l}
{\cal F}_i, \quad \bar P=\prod\limits_{i=0}^{l}\bar P_i.\end{eqnarray*}
 Random fields of  information evaluation by  consumers we give   by the formulae
\begin{eqnarray*}           \eta_i^0(p,z, \omega_i)=\{\eta_{is}^0(p,z, \omega_i)\}_{s=1}^n,\quad (p,z,\omega_i) \in K_+^n\times\Gamma^m\times\Omega_i, \quad i=\overline{1,l}.\end{eqnarray*}
Suppose they take values in the set $S,$ are continuous with  probability 1, and have components defined by formulae
\begin{eqnarray}\label{alpob1}
\eta_{ik}^0(p,z, \omega_i)=\frac{|\psi_k^0(p,z)|\sum\limits_{j=1}^nf_{kj}^1\varphi_{ij}(p,z, \omega_i)}{p_k},  \quad i=\overline{1,l},  \quad k=\overline{1,n},
\end{eqnarray}
where vectors $f_j^1=\{f_{kj}^1\}_{k=1}^n, \ j=\overline{1,n}, $ are such that $f_{kj}^1 > 0, \ k,\ j=\overline{1,n}.$

Assume that random fields $ \varphi_i(p,z, \omega_i)=\{\varphi_{ij}(p,z, \omega_i)\}_{j=1}^n$ are continuous  with probability 1 on a probability space
$\{\Omega_i, {\cal F}_i, \bar P_i\}, $ $(p,z) \in K_+^n \times \Gamma^m,$ $ \varphi_i(tp,z, \omega_i)= t\varphi_i(p,z, \omega_i), \ t > 0, \ i=\overline{1,l}.$
Let, additionally, random fields $ \varphi_i(p,z, \omega_i)$ satisfy the condition
\begin{eqnarray}\label{alpob2}
0 <  \varphi_0 < \varphi_{ij}(p,z, \omega_i) \leq   \varphi_1 < \infty,  \quad  (p, z, \omega_i) \in P\times\Gamma^m\times \Omega_i,
\end{eqnarray}
\begin{eqnarray*}
    P=\left\{ p \in K_+^n, \ \sum\limits_{i=1}^np_i=1\right\},
\end{eqnarray*}
where the strictly positive constants $\varphi_1, \varphi_0$ do not depend on  $ (p,z,\omega_i) \in P\times\Gamma^m\times\Omega_i $ and on $ j=\overline{1,n}$ and $ i=\overline{1,l}.$

Let $\zeta_0(p, \omega_0)$ be a random field given on a probability space $\{\Omega_0, {\cal F}_0, \bar P_0\}, \ $ $ p \in  K_+^n,$ and being continuous with probability 1, take values in the set $ \Gamma^m,$
and have the form
\begin{eqnarray*}
    \zeta_0(p, \omega_0)=\{(\zeta_i^{01}(p), \ \zeta_i^{02}(p))\}_{i=1}^m,
\end{eqnarray*}
where
\begin{eqnarray*}         \zeta_i^{01}(p) =\{a_{ki}\beta_i(p, \omega_0)\}_{k=1}^n,\quad  \zeta_i^{02}(p)= \{b_{ki}\beta_i(p, \omega_0)\}_{k=1}^n,  \end{eqnarray*}
and  let a random field $\beta(p, \omega_0)=\{\beta_i(p, \omega_0)\}_{i=1}^m$ be continuous with probability 1 on a probability space $\{\Omega_0, {\cal F}_0, \bar P_0\},\  p \in  K_+^n$, take values in the set $\Sigma$, and be zero-degree homogeneous.
Then random field of  decisions making by firms $ \zeta(p)=Q(p,\zeta_0(p, \omega_0)), p \in K_+^n,$ is continuous with probability 1 and zero-degree homogeneous.

If $\zeta(p)= \{\zeta_i(p)\}_{i=1}^m, \  \zeta_i(p)= (\zeta_i^1(p), \ \zeta_i^2(p)), $
then
\begin{eqnarray*}         \zeta_i^1(p) =\{a_{ki}a_i(p)\beta_i(p, \omega_0)\}_{k=1}^n \prod\limits_{s=1}^nv_s(p,\ \zeta_0(p, \ \omega_0)), \quad i=\overline{1,m},\end{eqnarray*}
\begin{eqnarray*}         \zeta_i^2(p) =\{b_{ki}a_i(p)\beta_i(p, \omega_0)\}_{k=1}^n \prod\limits_{s=1}^nv_s(p, \ \zeta_0(p, \omega_0)), \quad i=\overline{1,m},\end{eqnarray*}
\begin{eqnarray*}         v_s(p,\ \zeta_0(p, \omega_0))
\end{eqnarray*}
\begin{eqnarray*}
 =u_{[d_s, \infty)}\left(\sum\limits_{i=1}^mb_{si}a_i(p)\beta_i(p, \omega_0) -
\sum\limits_{i=1}^ma_{si}a_i(p)\beta_i(p, \omega_0) + \sum\limits_{i=1}^lb_{si}^1 - b_s^0\right).
\end{eqnarray*}
Introduce a random field of firms work intensity vector\index{random field of firms work intensity vector} $t(p)=\{t_i(p)\}_{i=1}^m,$ where
\begin{eqnarray*}         t_i(p)=a_i(p) \beta_i(p, \omega_0)\prod\limits_{s=1}^nv_s(p,\ \zeta_0(p, \omega_0)).\end{eqnarray*}
Then
\begin{eqnarray*}         \zeta^1_i(p) =\{a_{ki}t_i(p)\}_{k=1}^n, \quad \zeta^2_i(p) =\{b_{ki}t_i(p)\}_{k=1}^n, \quad i=\overline{1,m},\end{eqnarray*}
and the final consumption vector\index{final consumption vector} $\psi(p)=\{\psi_k(p)\}_{k=1}^n,$
where
\begin{eqnarray*}         \psi(p)=\sum\limits_{k=1}^m[Y_k(p, \zeta_0(p, \omega_0)) - X_k(p, \zeta_0(p, \omega_0))] +\sum\limits_{i=1}^lb_{i}^1
\end{eqnarray*}
\begin{eqnarray*}
   =[B - A]t(p)+\sum\limits_{i=1}^lb_{i}^1.
\end{eqnarray*}

We describe the $i$-th consumer by the income pre-function
\begin{eqnarray*}
    K_i^0(p,z)=\sum\limits _{j=1}^l\pi
_{ij}(p) \sum\limits _{k=1}^l r_{jk}(p) \left [\sum\limits
_{s=1}^m \alpha _{ks}(p)\left\langle y_s-x_s,p\right\rangle +\left\langle
b_k^1,p\right\rangle\right ], \quad i=\overline{1,l},\end{eqnarray*}
\begin{eqnarray}\label{nal15}
(p,z) \in K_+^n\times\Gamma^m,  \quad  z=\{z^k\}_{k=1}^m,\quad z^k=(x_k,y_k),\quad k=\overline {1,m},
\end{eqnarray} where $b_k^1$ is an  initial goods supply vector of the $k$-th
consumer at the beginning of the economy operation period. Then the $i$-th consumer income corresponding to  a realization of random field decisions making by firms has the form
\begin{eqnarray*}         D_i(p)=K_i(p,\zeta_0(p, \omega_0)))=K_i^0(p,\zeta(p)))\end{eqnarray*}
\begin{eqnarray*}         =\sum\limits _{j=1}^l
\pi _{ij}(p) \sum\limits _{k=1}^l
r_{jk}(p) \left [\sum\limits _{s=1}^n \alpha
_{ks}(p)t_s(p)\left[\sum\limits_{v=1}^nb_{vs}p_v -\sum\limits_{v=1}^na_{vs}p_v\right] +\left\langle
b_k^1,p\right\rangle\right ], \end{eqnarray*}
\begin{eqnarray*}          i=\overline{1,l},\end{eqnarray*}
where $ \pi _{ij}(p),\ r_{jk}(p),  \alpha _{ks}(p) $ are continuous on
$P$ and satisfy  conditions (\ref{sl000}) and (\ref{sl0000}). A demand vector of the $i$-th insatiable  consumer corresponding to a realization of  random field of consumer choice (\ref{alpob1}) is determined by the formula
\begin{eqnarray*}         \gamma
_i(p)=\{\gamma _{ik}(p)\}_{k=1}^n, \quad i=\overline{1,l},\end{eqnarray*}          satisfies conditions (\ref{g2l5}),
(\ref{g2l6}), and is continuous on $P,$
where
\begin{eqnarray*}         \gamma_{ik}(p)=\frac{\psi_k(p)\sum\limits_{j=1}^n
f_{kj}^1\varphi_{ji}(p)}{\sum\limits_{j=1}^n\sum\limits_{k=1}^n
\psi_k(p)f_{kj}^1\varphi_{ji}(p)}, \quad
i=\overline{1,l}, \quad  k=\overline{1,n},\end{eqnarray*}
\begin{eqnarray*}          \varphi_{ji}(p)=\varphi_{ji}(p, Q(p, \zeta_0(p, \omega_0),  \omega_i), \quad \psi_k(p)=\psi_k^0(p, Q(p,\zeta_0(p, \omega_0)).\end{eqnarray*}

According to the above given construction of the   work intensity vector of firms,\index{work intensity vector of firms} ~$t(p)=\{t_i(p)\}_{i=1}^m$~ are such that
\begin{eqnarray*}          \sum\limits_{k=1}^m
b_{ik}t_k(p)-\sum\limits_{k=1}^m
a_{ik}t_k(p)+\sum\limits_{k=1}^lb_{ik}^1\geq b_i^0,\quad  i=\overline{1,n}, \quad p \in K_+^n,\end{eqnarray*}
with probability 1.

Consider the set
\begin{eqnarray*}         T=\left\{p \in K_+^n,  \
\sum\limits_{s=1}^n(b_{si}-a_{si})p_s \geq 0, \
i=\overline{1,m}\right\}.\end{eqnarray*}

Let us give the sufficient conditions for the set $T$ to be non-empty one.
Here it is convenient to give the Definition for the productivity of the set of input-output vectors.
\begin{definition}\label{nbsv1}
Let $ m \leq n.$ The set of input-output vectors $\{a_i, b_i\}, \ i = \overline{1,m},$ is productive  of the rank $m$\index{productive  of the rank $m$} if  for the set of  vectors
  $\{b_i - a_i\}, \ i = \overline{1,m},$ of the rank $m$ there exists a set of biorthogonal vectors
\begin{eqnarray*}         f_j=\{f_{kj}\}_{k=1}^n, \quad  f_{kj} \geq 0, \quad  k = \overline{1,n}, \quad  j = \overline{1,m},\end{eqnarray*}
\begin{eqnarray*}         \quad \left\langle b_i - a_i, f_j \right\rangle=\delta_{ij}, \quad i = \overline{1,m}, \quad  j = \overline{1,m}, \end{eqnarray*}
 with non-negative components.
The set of  input-output vectors $\{a_i, b_i\}, $ $ i = \overline{1,m},$ is strictly productive  of the rank $m$\index{strictly productive  of the rank $m$} if
 for the  set of vectors $\{b_i - a_i\}, \ $  $i = \overline{1,m},$ there exists a set of biorthogonal vectors
\begin{eqnarray*}         f_j=\{f_{kj}\}_{k=1}^n, \quad  f_{kj} > 0, \quad  k = \overline{1,n}, \quad  j = \overline{1,m},\end{eqnarray*}
with strictly positive components.
\end{definition}
\begin{definition}\label{nbsv2}
The set of  input-output vectors $\{a_i, b_i\}, \  i =\overline{1,m},$ is (strictly) productive in a wide sense   of the rank $m_0$\index{productive in a wide sense   of the rank $m_0$} if a subset of  input-output vectors $\{b_{k_i}, a_{i_i}\},$  $ i = \overline{1,m_0},$  exists such that it is (strictly) productive of the rank $m_0$ in the sense of the Definition \ref{nbsv1},
and, for  the rest set  of  input-output vectors,  the representation
\begin{eqnarray*}         b_{k_s} - a_{k_s}=\sum\limits_{i=1}^{m_0}C_{k_i}^{k_s}[b_{k_i} - a_{k_i}], \quad C_{k_i}^{k_s} \geq 0, \quad s = \overline{m_0, m}, \quad i = \overline{1, m_0},\end{eqnarray*}
holds.
\end{definition}
\begin{definition}\label{albsv2}
Let a set of input-output vectors  $\{a_i, b_i\}, $ \ $ i=\overline{1,m},$ be productive  in a wide sense  of the rank $m_0.$ We say the set of input-output vectors $\{a_i, b_i\},\ i=\overline{1,m},$ generates the set
\begin{eqnarray*}         T=\left\{p \in K_+^n,  \
\sum\limits_{l=1}^n(b_{li}-a_{li})p_l \geq 0, \
i=\overline{1,m}\right\}\end{eqnarray*}
of the rank $n$ if a strictly productive set of input-output vectors $\{a_i^2, b_i^2\}, \ i=\overline{1,n},$ of the rank $n$ exists,   such  that the set $T$
contains the set
\begin{eqnarray*}         T_0=\left\{p \in K_+^n,  \
\sum\limits_{s=1}^n(b_{si}^2-a_{si}^2)p_s \geq 0, \ i=\overline{1,n}\right\}.\end{eqnarray*}
We call the set of input-output vectors $\{a_i^2, b_i^2\}, \ i=\overline{1,n},$ of the rank $n$ by subordinate  to the set of input-output vectors $\{a_i, b_i\},\ i=\overline{1,m}.$
\end{definition}

\begin{proposition}\label{alg1}
Let a set of input-output vectors $\{a_i, b_i\}, \ $ $i =\overline{1,m},$ be productive  in a wide sense  of the rank $m_0.$
Then the set $T$ is non-empty  and contains the set
	\begin{eqnarray*}          D_0=\left\{p \in R_+^n, \ p=\sum\limits_{i=1}^{m_0}c_if_i,\  c=\{c_i\}_{i=1}^{m_0} \in R_+^{m_0} \setminus \{0\}\right\},\end{eqnarray*}
where the set of vectors $f_i= \{f_{ki}\}_{k=1}^n \in R_+^n, \ i = \overline{1, m_0},$ is  the  set of   biorthogonal  vectors   to the  set of vectors  $\{b_{k_i} - a_{k_i}\}, \ i = \overline{1,m_0},$ and such that every component $f_{ki}\geq 0, \ i = \overline{1, m_0},\ k = \overline{1, n}.$
The set $T$ coincides with the set $D_0$ if $m_0=n.$
\end{proposition}

On the set
\begin{eqnarray*}         T_1=\left\{p \in
K_+^n,  \ \sum\limits_{l=1}^n(b_{li}-a_{li})p_l \geq 0, \
i=\overline{1,m}, \ \sum\limits_{i=1}^np_i=1\right\},\end{eqnarray*}
let us consider  demand vectors corresponding to realizations of random fields of  consumers choice and firms decisions making
 \begin{eqnarray}\label{nbsv3}
\gamma _i(p)=\{\gamma _{ik}(p)\}_{k=1}^n, \quad
i=\overline{1,l},
\end{eqnarray}
where
\begin{eqnarray*}         \gamma_{ik}(p)=\frac{\psi_k(p)\sum\limits_{j=1}^n
f_{kj}^1\varphi_{ji}(p)}{\sum\limits_{j=1}^n\sum\limits_{k=1}^n
\psi_k(p)f_{kj}^1\varphi_{ji}(p)}, \quad
i=\overline{1,l}, \quad  k=\overline{1,n},\end{eqnarray*}
\begin{eqnarray*}          \varphi_{ji}(p)=\varphi_{ji}(p, Q(p, \zeta_0(p, \omega_0),  \omega_i), \quad \psi_k(p)=\psi_k^0(p, Q(p,\zeta_0(p, \omega_0)),\end{eqnarray*}
and the set of equations
\begin{eqnarray}\label{nbsv4}
\frac{1}{\psi_k(p)}\sum\limits _{i=1}^l\gamma _{ik}( p)~D_i(
p)=p_k, \quad k=\overline{1,n}.
\end{eqnarray}

\begin{theorem}\label{alpob3}
Let technological maps (\ref{pnrl1}) describe  the production structure, in the considered economy system, a productive economic  process have the form (\ref{nrl2}), and  let
a set of input-output vectors
 $\{a_i, b_i\}, \ i = \overline{1,m},$
be  productive  in a wide sense  generating the set $T$ of the rank $n$ with a subordinate  set of input-output vectors $\{a_i^2, b_i^2\}, \ i = \overline{1,n},$ of the rank $n.$
If
$\alpha _{ij}(p),$ $~r_{ij}(p),$ $~\pi _{ij}(p)$ are continuous functions  on
$P$  satisfying conditions (\ref{sl000}) and (\ref{sl0000}), random fields of information evaluation by consumers  satisfy conditions (\ref{alpob1}) and (\ref{alpob2}), where the  vectors $f_j^1=\{f_{kj}^1\}_{k=1}^n, \ j=\overline{1,n},$ form the  set of  vectors being  biorthogonal to  the set of vectors $\{b_i^2 - a_i^2 \}, \ i = \overline{1,n},$ then the set of equations of economic equilibrium  (\ref{nbsv4})  has a solution in the set of strictly positive  price vectors with probability 1.
\end{theorem}
\begin{proof}\smartqed
On the set $T_1,$ let us  consider the non-linear map
\begin{eqnarray*}         \varphi(p)=\left\{\frac{f_k(p)}{\sum\limits_{k=1}^nf_k(p)}
\right\}_{k=1}^n,\end{eqnarray*}
where
\begin{eqnarray*}         f_k(p)=\frac{1}{\psi_k(p)}\sum\limits
_{i=1}^l\gamma _{ik}( p)D_i( p), \quad k=\overline{1,n},\end{eqnarray*}
and let the subset $T_1^0$ of the set $T_1$ be given by the rule
\begin{eqnarray*}         T_1^0=\left\{p \in
K_+^n,  \ \sum\limits_{s=1}^n(b_{si}^2-a_{si}^2)p_s \geq 0, \
i=\overline{1,m}, \ \sum\limits_{i=1}^np_i=1\right\}.\end{eqnarray*}
From the Theorem conditions, it follows that the non-linear map
$\varphi(p)$
maps the set $T_1^0$  into itself and is continuous  on $T_1^0.$
Really, as
\begin{eqnarray*}         f_k(p)=\frac{1}{\psi_k(p)}\sum\limits
_{i=1}^l\gamma _{ik}( p)D_i( p)
\end{eqnarray*}
\begin{eqnarray*}
= \sum\limits_{i=1}^l  \frac{\sum\limits_{j=1}^n
f_{kj}^1\varphi_{ji}(p)}{\sum\limits_{j=1}^n\sum\limits_{s=1}^n
\psi_s(p)f_{sj}^1\varphi_{ji}(p)}D_i( p)
\end{eqnarray*}
\begin{eqnarray*}
 = \sum\limits_{j=1}^n f_{kj}^1\sum\limits_{i=1}^l  \frac{
\varphi_{ji}(p)}{\sum\limits_{j=1}^n\sum\limits_{s=1}^n
\psi_s(p)f_{sj}^1\varphi_{ji}(p)}D_i( p), \quad k=\overline{1,n},
\end{eqnarray*}
the representation
\begin{eqnarray*}
\varphi(p)=\sum\limits_{j=1}^nf_{j}^1r_j(p)
\end{eqnarray*}
holds, where
\begin{eqnarray*}
 r_j(p)= \frac{1}{\sum\limits_{k=1}^nf_k(p)}\sum\limits_{i=1}^l  \frac{
\varphi_{ji}(p)}{\sum\limits_{j=1}^n\sum\limits_{k=1}^n
\psi_k(p)f_{kj}^1\varphi_{ji}(p)}D_i( p), \quad j=\overline{1,n},
\end{eqnarray*}
 are strictly positive.
Let us  find the upper and lower bounds for
\begin{eqnarray*}          \sum\limits_{k=1}^nf_k(p)=\sum\limits_{i=1}^l  \frac{ \sum\limits_{k=1}^n\sum\limits_{j=1}^n
f_{kj}^1\varphi_{ji}(p)}{\sum\limits_{j=1}^n\sum\limits_{s=1}^n
\psi_s(p)f_{sj}^1\varphi_{ji}(p)}D_i( p).\end{eqnarray*}
As
\begin{eqnarray*}         \max\limits_{1 \leq k \leq n} \sup\limits_{p \in P}\psi_k(p) \leq \max\limits_{1 \leq k \leq n}\sum\limits_{i=1}^n \left\{[b_{ki}+a_{ki}]\xi_i^0 +\sum\limits_{i=1}^l b_{ki}^1\right\}=\psi_0,\end{eqnarray*}
we obtain
\begin{eqnarray*}         \sum\limits_{k=1}^nf_k(p) \geq \frac{\sum\limits_{i=1}^lD_i( p)}{\psi_0} \geq
\frac{ \min\limits_{1 \leq i \leq n}b_i^0}{\psi_0} > 0,\end{eqnarray*}
because $\psi(p)> b_0>0, $  \ $\sum\limits_{k=1}^n\gamma _{ik}( p)=1,$ and
\begin{eqnarray*}         \sum\limits_{i=1}^lD_i( p)=\left\langle \psi(p),  p \right\rangle  >  \left\langle b_0, p \right\rangle  >  \min\limits_{1 \leq i \leq n}b_i^0  \ >0\end{eqnarray*}
on the set $T_1^0.$

Further,
\begin{eqnarray*}         \sum\limits_{k=1}^nf_k( p)\leq  \frac{\sum\limits_{i=1}^lD_i( p)}{\min\limits_{1 \leq i \leq n}b_i^0}= \frac{\left\langle \psi(p),  p\right\rangle }{\min\limits_{1 \leq i \leq n}b_i^0}\leq \frac{\psi_0 }{\min\limits_{1 \leq i \leq n}b_i^0}.\end{eqnarray*}
Therefrom,
\begin{eqnarray*}         r_j(p) \geq \frac{\varphi_0\left[ \min\limits_{1 \leq i \leq n}b_i^0\right]^2}{\varphi_1\psi_0^2\sum\limits_{j=1}^n\sum\limits_{s=1}^n
f_{sj}^1}>0, \quad j=\overline{1,n}.\end{eqnarray*}
By the Schauder Theorem\index{Schauder Theorem} \cite{88}, there exists a fixed point of this map belonging to $T_1^0.$
With the same arguments as in the Theorem \ref{ch2l3}, this fixed point is that  for the set of equations (\ref{nbsv4}).

Prove that an equilibrium price vector has strictly positive components.
 Really, it follows from the fact that the set $T_1^0$ contains vectors with strictly positive components.
\qed
\end{proof}

\begin{theorem}\label{nbsv5}
Let technological maps (\ref{pnrl1}) describe  the structure of  firms production  of the  economy system,  the formula (\ref{nrl2}) give a  productive economic process, and let a set of input-output vectors $\{a_i, b_i\}, \ i = \overline{1,m},$
be productive in a wide sense and generates the set $T$ of the rank $n$ with a subordinate  set of input-output vectors $\{a_i^2, b_i^2\}, \ i = \overline{1,n},$ of the rank $n.$
 If
$\alpha _{ij}(p),$ $~r_{ij}(p),$ $~\pi _{ij}(p)$ are continuous on
$P$ functions  and satisfy conditions (\ref{sl000}) and (\ref{sl0000}), demand and supply vectors are continuous  with probability 1 and satisfy the conditions
\begin{eqnarray*}         \sum\limits_{k=1}^n(b_{ki}^2-a_{ki}^2)\frac{1}{\psi_k(p)}\sum\limits _{i=1}^l\gamma _{ik}( p)~D_i(p) \geq 0, \quad  p \in T_1^0, \quad  i=\overline{1,n}, \end{eqnarray*}
then the  set  of equations of the economy equilibrium  (\ref{nbsv4}) has a solution in the set of strictly positive price vectors belonging to the set $T_1^0.$
\end{theorem}
\begin{proof}\smartqed
On the set $T_1,$ let us give  the non-linear map
\begin{eqnarray*}         \varphi(p)=\left\{\frac{f_k(p)}{\sum\limits_{k=1}^nf_k(p)}
\right\}_{k=1}^n,\end{eqnarray*}
where
\begin{eqnarray*}         f_k(p)=\frac{1}{\psi_k(p)}\sum\limits
_{i=1}^l\gamma _{ik}( p)D_i( p), \quad k=\overline{1,n}.\end{eqnarray*}
Consider a subset $T_1^0$ of the set $T_1$ given by the rule
\begin{eqnarray*}
 T_1^0=\left\{p \in
K_+^n,  \ \sum\limits_{l=1}^n(b_{li}^2-a_{li}^2)p_l \geq 0, \
i=\overline{1,m}, \ \sum\limits_{i=1}^np_i=1\right\}.
\end{eqnarray*}
From the conditions of the Theorem,  it follows that the non-linear map has the representation
\begin{eqnarray*}         \varphi(p)=\sum\limits_{j=1}^nf_{j}^1r_j(p)\end{eqnarray*}
on the set $T_1^0,$
where
\begin{eqnarray*}         r_j(p)= \frac{1}{\sum\limits_{k=1}^nf_k(p)}\sum\limits_{i=1}^l  \frac{
\varphi_{ji}(p)}{\sum\limits_{j=1}^n\sum\limits_{k=1}^n
\psi_k(p)f_{kj}^1\varphi_{ji}(p)}D_i( p), \quad j=\overline{1,n},\end{eqnarray*}
 are strictly positive.
Therefore, $\varphi(p)$
maps it into itself and is continuous  on $T_1^0.$
By the Schauder Theorem \cite{88}, there exists a fixed point of this map belonging to $T_1^0.$
With the same reasons as in the Theorem \ref{ch2l3}, this fixed point is that  for the set of equations (\ref{nbsv4}).

Prove an equilibrium price vector to have strictly positive components.
 Really, it follows from the fact that the set $T_1^0$ contains vectors with strictly positive components.
\qed
\end{proof}

\subsection{Economy model with optimum equilibrium}

\begin{theorem}\label{ch2l10} Let technological maps $F_i(x),\ x \in X_i, \  i=\overline{1,m},$ describing the structure of production of the economy system,   belong to the CTM class,
be strictly convex down, and satisfy the  conditions of the  Lemma \ref{ol1}. Suppose that the  consumers income pre-functions $K_i^0(p,z), \  i=\overline{1,l},$ are continuous functions of variables $(p,z) \in \bar R_+^n \times \Gamma^m,$ random fields of  consumers choice and  decisions making  by firms satisfy the  conditions of the Theorem \ref{tl4},  the next conditions
\begin{eqnarray*}         \sum\limits_{j=1}^m[ y_j- x_j]+ \sum\limits_{k=1}^lb_k(p,z) > 0,\quad
(p,z) \in K_+^n\times \Gamma^m, \quad z=\{z^i\}_{i=1}^m, \quad z^i=\{x_i,y_i\},
\end{eqnarray*}
hold, a  productive economic process is given by the rule
\begin{eqnarray*}         Q(p,z)=\{z^i\varphi^{\varepsilon}(h_i(p, z))\}_{i=1}^m,\end{eqnarray*}
 \begin{eqnarray*}         \varphi^{\varepsilon}(x)= \left\{\begin{array}{ll}
              1, & \textrm{if} \quad x \geq 0,\\
               1+\frac{x}{\varepsilon}, &      \textrm{if}  \quad - \varepsilon \leq x < 0,\\
               0,  & \textrm{if} \quad x < - \varepsilon  \textrm{,}
                                                                                                                                                \end{array}
                                               \right.\end{eqnarray*}
\begin{eqnarray*}         h_i(p, z)=\frac{\left\langle p, y_i - x_i\right\rangle }{\sum\limits_{i=1}^np_i},\end{eqnarray*}
the random field $\zeta_0(p, \omega_0)$ takes single value $z^0(p)=\{z_0^i(p)\}_{i=1}^m$ with probability 1, where $z_0^i(p)=\{x_i^0(p), y_i^0(p)\},$ and
\begin{eqnarray*}          \sup_{x \in X_i}\sup_{y \in F_i(x)}\left\langle y -x, p\right\rangle = \left\langle y_i^0(p)- x_i^0(p),p \right\rangle.\end{eqnarray*}
Then with probability 1 there exists optimum Walras equilibrium state,\index{optimum Walras equilibrium state} an equilibrium price vector solves the set of equations (\ref{tima1}), i.e., almost for every  realization of random fields of consumers choice  there exists a vector $\bar p \in P$ corresponding to it  such that  the economy system is in the Walras optimum equilibrium state and the vector $\bar p $  satisfies  the set of equations
\begin{eqnarray}\label{tima1}
\sum\limits
_{i=1}^l\gamma _{ik}( p)D_i( p)=\psi_k(p)p_k,\quad
k=\overline{1,n},
\end{eqnarray}
 where
\begin{eqnarray*}         \psi_k(p)=\sum\limits_{i=1}^lb_{ki}(p, z^0(p)) +
\sum\limits_{j=1}^m[ y_{kj}^0(p)- x_{kj}^0(p)] > 0,\quad
k=\overline{1,n}, \quad p \in P,\end{eqnarray*}          and the income function of the $i$-th
consumer $D_i(p)=K_i(p,z^0(p)).$
\end{theorem}
\begin{proof}\smartqed      Note that the Theorem does not require that  all the vectors $b_k(p,z), \ k=\overline{1,l},$  are strictly positive  despite a similar Theorem in \cite{13}.
Under the Theorem \ref{ch2l10} conditions, the Theorem \ref{ch2l1} conditions hold, i.e.,
\begin{eqnarray*}         R(p, Q(p,z)) >0, \quad p \in P, \ z \in \Gamma^m,\end{eqnarray*}
\begin{eqnarray*}          R(p,z)= \sum\limits
_{i=1}^m[y_i-x_i]+ \sum\limits _{j=1}^l b_j(p,z),\end{eqnarray*}
because just so the productive economic  process is built.
In view of the conditions of the Lemma \ref{ol1}, the optimum strategy of firms behavior   $z^0(p)$
is continuous  on the set $P$ and the equality
\begin{eqnarray*}         z(p)=Q(p,z^0(p))=z^0(p)\end{eqnarray*}
holds. The last equality holds because $\varphi^{\varepsilon}(\left\langle p, y_i^0(p) - x_i^0(p) \right\rangle)=1$
under condition that the inequality  $ \left\langle p, y_i^0(p) - x_i^0(p) \right\rangle \geq 0$ holds.
Therefore, the Theorem statement follows from the Theorem \ref{ch2l1}.
\qed
\end{proof}

The Theorem \ref{ch2l10}  generalizes the Arrow-Debreu Theorem\index{Arrow-Debreu Theorem} and gives the algorithm to find all equilibrium price vectors for competitive economy.

Give simple sufficient conditions under which equilibrium price vector is strictly positive (see also \cite{13}).
\begin{theorem}\label{ch2l21}
Assume all the  conditions of the Theorem \ref{ch2l1} hold. If also  numbers ~$0 \leq \alpha_i < 1,\ i=\overline{1,n},$~
and positive number $\rho$ exist such that
\begin{eqnarray*}         \min\limits_{k}\inf\limits_{p \in P}\frac{ \sum\limits _{i=1}^l
\gamma _{ik}(p)D_i( p)}{p_k^{\alpha_k}} > \rho>0,\end{eqnarray*}
then any solution to the set of equations (\ref{g2l18}) is strictly positive.
\end{theorem}
\begin{proof}\smartqed      The existence of at least one solution to the problem follows from the Theorem \ref{ch2l1}. From the condition of the  Theorem \ref{ch2l21},  the estimate for components $\bar p_k, \ k=\overline{1,n},$
of any vector $\bar p=\{\bar p_k\}_{k=1}^n$ solving the problem
(\ref{g2l18}) follows: \begin{eqnarray*}         \bar p_k^{(1-\alpha_k)} \geq
\frac{\rho}{\triangle} >0 ,\end{eqnarray*}          where \begin{eqnarray*}
\triangle=\max\limits_{k}\sup\limits_{p \in P}\left [ \sum\limits
_{i=1}^m [y_{ki}(p)-x_{ki}(p)]+ \sum\limits _{j=1}^l b_{kj}\right ].\end{eqnarray*}
From here, the needed follows.
\qed
\end{proof}

Further, we consider exchange model with constant consumption parts.

Assume that $\gamma_i=(\gamma_{ij})_{j=1}^n, \ i=\overline{1,l},$
are constant demand vectors, i.e., $\gamma_{ij}$ do not depend \hfill on \hfill $p \in
P.$ \hfill Let\hfill every \hfill $i$-th \hfill insatiable \hfill consumer \hfill have \hfill a \hfill goods \hfill vector \\ \hfill $b_i=(b_{1i}, \ldots, b_{ni}),\ $ $\left\langle p,b_i
\right\rangle=\sum\limits_{m=1}^nb_{mi}p_m, \ i=\overline{1,l},$\
 $\sum\limits_{i=1}^lb_{ki} > 0, \   k=\overline{1,n}.$
By the Theorem \ref{ch2l1}, the next Proposition holds.
\begin{proposition}
The set of equations
\begin{eqnarray*}         \sum\limits_{i=1}^l\gamma_{ik} \left\langle p,b_i
\right\rangle=p_k \sum\limits_{i=1}^lb_{ki}, \quad k=\overline{1,n},\end{eqnarray*}
is solvable in the set   $P.$
If  the matrix
$\left|\left|\sum\limits_{i=1}^l\gamma_{ik}b_{mi}\right|\right|_{k,m=1}^n$
is indecomposable
then, for exchange model with constant  parts of consumption,\index{exchange model with constant  parts of consumption} a non-negative vector $\bar p \in P$
 exists satisfying the above written set of equations,   this vector is an equilibrium price vector with strictly positive components and is unique up to a certain constant factor.
\end{proposition}

\section{Arbitrage absence in economy system models  described  non-aggregately}

Up to now, we supposed that\index{economy system models  described  non-aggregately}  behavior strategies of  consumers choice (realizations of random fields of consumers choice) are continuous functions on the simplex $P,$ however, for practically important situations it is not the case.
If the consumer consumes not all the goods the economy system produces, but some part of them, then  components of the  demand vector  corresponding to goods he does not consume equal zero. As a result, the demand vector of such consumer can not be given unambiguously on the whole simplex $P$  such that to be continuous  on this simplex.
To describe discontinuous  behavior strategies of consumers choice, define random fields of consumers choice  not on the whole cone $\bar R_+^n,$ but on a certain  cone
 $K_+^n \subset \bar R_+^n$ being a subcone of the cone $\bar R_+^n$
on which the conditions of  the Theorem \ref{tl4} and the Theorem \ref{ttl3}  on the existence of random fields of consumers choice  are valid.
Let $C=||c_{ik}||_{i,k=1}^{n,l}$ be a certain  $n \times l$-dimensional matrix satisfying conditions: $\sum\limits_{k=1}^n c_{ki} > 0, \ i=\overline{1,l},$
and $\min\limits_{k,s c_{ks} \neq 0}c_{ks}=1.$
Introduce the cone $K_+^n $ built by the rule
\begin{eqnarray}\label{kl1}
K_+^n=\left\{ p \in \bar R_+^n, \ \sum\limits_{k=1}^n c_{ki}p_k > 0, \ i=\overline{1,l}\right\}.
\end{eqnarray}
First, let us consider the case of all  insatiable consumers.
Suppose random fields of information evaluation by consumers    satisfy the condition:
for every $i$-th consumer  a random field of information evaluation by the $i$-th consumer  $\eta_i^0(p,z, \omega_i)=\{\eta_{ik}^0(p,z, \omega_i)\}_{k=1}^n$
on a probability space $\{\Omega_i, {\cal F}_i, P_i\}$
satisfies  the inequality
\begin{eqnarray*}
   \eta_i^0(p,z, \omega_i) \geq m_i C_i, \quad C_i=\{c_{ki}\}_{k=1}^n, \quad (p,z, \omega_i) \in  K_+^n\times \Gamma^m\times \Omega_i,  \quad i=\overline{1,l},
   \end{eqnarray*}
 where the real numbers $m_i >0,$ \ $ i=\overline{1,l},$
and  components
 $\eta_{ik}^0(p,z, \omega_i)=0$ if and only if $c_{ki}=0.$

Under these additional assumptions about the random fields $\eta_i^0(p,z, \omega_i),$ $  i=\overline{1,l},$
and assumptions about the matrix $C$ on the above built cone $K_+^n $ (with the rest  conditions of Theorems \ref{ptl3} and \ref{tl4} hold), there exist random fields of consumers choice and  decisions making by firms for insatiable consumers.

In what follows, we assume the conditions the above stated hold for the random fields on the above built cone.

Note that if a certain  components of the vector $C_i=\{c_{ki}\}_{k=1}^n$ equal zero, then the $i$-th consumer does not consume goods numbered by these components.

Let $z(p)$ be a certain  continuous realization of a random field $\zeta(p, \omega_0),$ and $\mu_i(p)=\{\mu_{ki}(p)\}_{k=1}^n$ be a continuous realization of a random field
\begin{eqnarray*}         \eta_i(p,\zeta_0(p, \omega_0),\omega_i)=
\eta_i^0(p,\zeta(p, \omega_0),\omega_i)
=\{\eta_{ik}^0(p,\zeta(p, \omega_0),\omega_i)\}_{k=1}^n. \end{eqnarray*}
Therefore, $\mu_{ki}(p)=\eta_{ik}^0(p,\zeta(p, \omega_0),\omega_i)$ for a certain $\omega_0$ and $\omega_i.$
Denote by
\begin{eqnarray*}          \gamma_{ik}(p)=\frac{\mu_{ki}(p)p_k}{\sum\limits_{j=1}^n
\mu_{ji}(p)p_j}, \quad k=\overline {1,n},\quad i=\overline {1,l},\end{eqnarray*}
\begin{eqnarray*}          \psi(p)=\{\psi_k(p)\}_{k=1}^n,\end{eqnarray*}
\begin{eqnarray*}         \psi_k(p)=\sum\limits _{i=1}^m[y_{ki}(p)-x_{ki}(p)]+ \sum\limits
_{i=1}^l b_{ki}(p,z(p)), \quad  k=\overline{1,n},\end{eqnarray*}
\begin{eqnarray*}         \Gamma_k(p)=\frac{1}{\psi_k(p)}\sum\limits_{i=1}^l
\frac{\mu_{ki}(p)p_kD_i(p)}{\sum\limits_{j=1}^n
\mu_{ji}(p)p_j}, \quad k=\overline {1,n},\end{eqnarray*}
\begin{eqnarray*}         \Gamma(p)=\{ \Gamma_k(p)\}_{k=1}^n. \end{eqnarray*}
Introduce the set
\begin{eqnarray*}         C_{\delta} =\left\{p \in R_+^n,\ \sum\limits_{s=1}^nc_{si}p_s \geq \delta ,\ i=\overline {1,l}, \ \sum\limits_{i=1}^np_i=1 \right\}\end{eqnarray*}
and the next notations
\begin{eqnarray*}          \sup_{k, p \in P\cap K_+^n} \psi_k(p)=R_1 , \quad  \inf_{k, p \in P\cap K_+^n} \psi_k(p)=R_0.\end{eqnarray*}

\begin{theorem}\label{1ch2l2}
Let technological maps $F_i(x), \  x \in X_i^1,
\ i=\overline{1,m},$ describing the economy system production be convex down, belong to  the CTM class, and let a productive economic  process $Q(p,z),$ a family of income pre-functions  $K_i^0(p,z), \ i=\overline{1,l},$ and property vectors $b_i(p,z), \ i=\overline {1,l},$ be continuous maps of variables
$(p,z) \in K_+^n\times \Gamma^m,$ where $K_+^n $ is a cone given by the formula (\ref{kl1}), and also let
 \begin{eqnarray*}          D_i(p)=K_i(p, \zeta_0(p, \omega_0))>a>0, \quad p \in P \cap K_+^n, \quad \omega_0 \in \Omega_0, \quad i=\overline{1,l},\end{eqnarray*}
 where $a$ does not depend on $(p, \omega_0).$
Assume that random fields of  information evaluation by consumers and  decisions making by firms satisfy the  conditions of the  Theorem \ref{tl4} and  the productive economic  process $Q(p,z)$ satisfies the condition
\begin{eqnarray}\label{1g2l23}
R(p, Q(p,z)) >r, \quad p \in K_+^n, \quad z \in \Gamma^m,
\end{eqnarray}
where
\begin{eqnarray*}          R(p,z)= \sum\limits
_{i=1}^m[y_i-x_i]+ \sum\limits _{j=1}^l b_j(p,z), \quad r=\{r_i\}_{i=1}^n, \quad r_i>0, \quad  i=\overline{1,n}.\end{eqnarray*}
If the set $C_{\delta}$
is non-empty  for some $\delta > 0,$ the inequality
\begin{eqnarray*}         \min_{\{k,s, f_{ks}=1\}}\inf_{p \in C_{\delta}}\frac{D_s(p)}{\psi_k(p)} \geq \frac{R_1}{R_0}\delta\end{eqnarray*}
holds, where
\begin{eqnarray*}         f_{ks}=\left\{\begin{array}{ll}
                 1, & \textrm{if} \quad c_{ks} \neq 0 \textrm{,}\\
                 0, & \textrm{if} \quad  c_{ks}  =0 \textrm{,}
                 \end{array} \right.\end{eqnarray*}
then for every continuous
demand matrix $||\gamma_{ik}(p)||_{i=1, k=1}^{l,\ n}$ and continuous realization of random field of decisions making by  firms $z(p)$ on $K_+^n,$ i.e., with probability 1, there exists  a price vector $\bar p$ corresponding to them
 satisfying the set of equations
\begin{eqnarray*}         \sum\limits
_{i=1}^l\gamma _{ik}(p)D_i(p)
\end{eqnarray*}
\begin{eqnarray}\label{1g2l18}
= p_k \left
[\sum\limits _{i=1}^m[y_{ki}(p)-x_{ki}(p)]+ \sum\limits
_{i=1}^l b_{ki}(p,z(p)) \right ], \quad  k=\overline{1,n}.
\end{eqnarray}
\end{theorem}
\begin{proof}\smartqed
 On the closed bounded convex set $C_\delta, $ let us consider the  non-linear operator
\begin{eqnarray*}
 f(p)=\{f_k(p)\}_{k=1}^n,
 \end{eqnarray*}
where
\begin{eqnarray*}
   f_k(p)=\frac{\Gamma_k(p)}{\sum\limits_{i=1}^n\Gamma_i(p)}, \quad
k=\overline {1,n}.
\end{eqnarray*}
Under the conditions of the  Theorem,  $R_0 > \min\limits_{1 \leq i \leq n} r_i > 0,\ $ and  $R_1 < \infty. $
The operator $f(p)$ transforms the set
$C_\delta $ into itself and is a continuous map. It is sufficient to check the inequalities
\begin{eqnarray*}         \sum\limits_{k=1}^nc_{ki}f_k(p) \geq \delta, \quad i=\overline {1,l},\end{eqnarray*}
and show the continuity of $f(p).$
First, prove the continuity of $f(p).$
In view of the continuity of $\mu_{ki}(p)$ for every realization and the  condition of the Theorem,
\begin{eqnarray*}         \sum\limits_{j=1}^n\mu_{ji}(p)p_j \geq m_i\sum\limits_{j=1}^nc_{ji}p_j > m_i\delta
> 0.\end{eqnarray*}
So, we have that $\sum\limits_{j=1}^n\mu_{ji}(p)p_j$ nowhere vanishes on the set $C_{\delta}.$ From the fact that $\psi_k(p) > r_k, \ k=\overline {1,n},$ we have that  $\psi_k(p)$ is a continuous function of $ p \in K_+^n $ and $\Gamma_k(p)$ is a continuous map on $C_{\delta}.$
Find lower bound for $\sum\limits_{k=1}^n\Gamma_k(p).$
We have
\begin{eqnarray*}          \sum\limits_{k=1}^n\Gamma_k(p) \geq \frac{1}{R_1}
\sum\limits_{i=1}^l
\frac{\sum\limits_{k=1}^n\mu_{ki}(p)p_k}{\sum\limits_{j=1}^n
\mu_{ji}(p)p_j}D_i(p)=\frac{1}{R_1}\sum\limits_{i=1}^n\psi_i(p)p_i \geq \frac{R_0}{R_1}.\end{eqnarray*}
Therefore, $f(p)$ is a continuous map on $C_{\delta}.$
For $\sum\limits_{k=1}^n\Gamma_k(p),$  the upper bound
\begin{eqnarray*}         \sup_{p \in P\cap K_+^n}\sum\limits_{r=1}^n\Gamma_r(p) \leq \frac{1}{R_0}\sup_{p \in P\cap K_+^n}\sum\limits_{i=1}^lD_i(p) = \frac{1}{R_0}\sup_{p \in P\cap K_+^n}\sum\limits_{i=1}^n\psi_i(p)p_i \leq \frac{R_1}{R_0}\end{eqnarray*}
is valid.
Finally, show that
\begin{eqnarray*}         \sum\limits_{k=1}^nc_{ki}f_k(p) \geq \delta, \quad i=\overline {1,l}.\end{eqnarray*}
Really,
\begin{eqnarray*}         \sum\limits_{k=1}^nc_{ki}f_k(p) =  \sum\limits_{k=1}^nc_{ki}
\frac{1}{\psi_k(p)\sum\limits_{r=1}^n\Gamma_r(p)}\sum\limits_{s=1}^l
\frac{\mu_{ks}(p)p_kD_s(p)}{\sum\limits_{j=1}^n
\mu_{js}(p)p_j}
\end{eqnarray*}
\begin{eqnarray*}          \geq \delta \sum\limits_{k=1}^nc_{ki}
\sum\limits_{s=1}^l
\frac{f_{ks}\mu_{ks}(p)p_k}{\sum\limits_{j=1}^n
\mu_{js}(p)p_j}=\delta\sum\limits_{s=1}^l
\frac{\sum\limits_{k=1}^nc_{ki}f_{ks}\mu_{ks}(p)p_k}{\sum\limits_{j=1}^n
\mu_{js}(p)p_j}.\end{eqnarray*}
Because of the assumptions for the matrix elements $c_{ki},$ for $s=i$
\begin{eqnarray*}         \frac{\sum\limits_{k=1}^nc_{ki}f_{ki}\mu_{ki}(p)p_k}{\sum\limits_{j=1}^n
\mu_{ji}(p)p_j} \geq \min_{\{k,i, f_{ki} =1\}}c_{ki}\frac{\sum\limits_{k=1}^n\mu_{ki}(p)p_k}{\sum\limits_{j=1}^n\mu_{ji}(p)p_j}=1.\end{eqnarray*}
So,
\begin{eqnarray*}         \sum\limits_{k=1}^nc_{ki}f_k(p) \geq \delta, \quad i=\overline{1,l}.\end{eqnarray*}
By the Schauder Theorem \cite{88}, there exists a fixed point of the map $f(p).$
This point is also a fixed point for the map $\Gamma(p).$
Really, as $p^*$ is a fixed point of the map $f(p),$
we have
\begin{eqnarray*}         \Gamma_k(p^*)=p_k^*\sum\limits_{s=1}^n\Gamma_s(p^*).\end{eqnarray*}
Multiplying by $\psi_k(p^*)$ the last equality and summing up over $k,$ we have
\begin{eqnarray*}         \sum\limits_{k=1}^n\psi_k(p^*)\Gamma_k(p^*)=\left\langle\psi(p^*), p^*\right\rangle\sum\limits_{s=1}^n\Gamma_s(p^*).\end{eqnarray*}
However,
\begin{eqnarray*}         \sum\limits_{k=1}^n\psi_k(p^*)\Gamma_k(p^*)=\sum\limits_{i=1}^lD_i(p^*)
=\left\langle\psi(p^*), p^*\right\rangle.\end{eqnarray*}
Reducing by $\left\langle\psi(p^*), p^*\right\rangle,$ in view of the inequality  $\left\langle\psi(p^*), p^*\right\rangle \neq 0,$
we obtain $ \sum\limits_{s=1}^n\Gamma_s(p^*)=1.$ The last means that $p^*$
solves the set of equations (\ref{1g2l18}). \qed
\end{proof}

Establish sufficient conditions for the previous Theorem conditions to hold.

 \begin{lemma}\label{l1ch2l1}
If the  matrix elements $c_{ki}$ satisfy the conditions stated above and the conditions of the  Theorem  \ref{1ch2l2} are valid and also
\begin{eqnarray*}         \min\limits_{1 \leq  i \leq l} \inf\limits_{p \in P \cap K_+^n}D_i(p)=d >0,\end{eqnarray*}
then for all
$\delta$ satisfying the inequalities
\begin{eqnarray*}         0 < \delta \leq \min\left\{ \frac{ R_0 d}{R_1^2 }, \ \frac{1}{n}\min_{1 \leq i \leq l}
 \sum\limits_{j=1}^nc_{ji}\right\} \end{eqnarray*}
the set $C_\delta $ is non-empty  and the inequality
\begin{eqnarray*}         \min_{\{k,s, f_{ks}=1\}}\inf_{p \in C_{\delta}}\frac{D_s(p)}{\psi_k(p)} \geq \frac{R_1}{R_0}\delta\end{eqnarray*}
holds.
\end{lemma}
\begin{proof}\smartqed
There hold bounds
\begin{eqnarray*}         \min_{\{k,s, f_{ks}=1\}}\inf_{p \in C_{\delta}}\frac{D_s(p)}{\psi_k(p)} \geq  \frac{d}{R_1} \geq \frac{R_1}{R_0}\delta. \end{eqnarray*}
Finally, the set $C_{\delta}$ is non-empty  because it contains the vector $\{1/n, \ldots, 1/n\}.$
\qed
\end{proof}

\begin{theorem}\label{rest1}
Let technological maps $F_i(x), \  x \in X_i^1,
\ i=\overline{1,m},$ describing the economy system production structure, be convex down, belong to the CTM class, and let  a  productive economic process $Q(p,z),$ a family of income pre-functions  $K_i^0(p,z), \ i=\overline{1,l},$ and property vectors $b_i(p,z), \ i=\overline {1,l},$ be continuous maps of variables
$(p,z) \in K_+^n\times \Gamma^m,$ where $K_+^n $ is a cone given by the formula (\ref{kl1}), and also let the conditions of the Lemma
 \ref{l1ch2l1}  hold.

Suppose that random fields of information evaluation by consumers   and  decisions making by firms satisfy the  conditions of the  Theorem \ref{tl4} and a productive economic  process $Q(p,z)$ satisfies the condition
\begin{eqnarray}\label{rest2}
R(p, Q(p,z)) >r, \quad p \in K_+^n, \ z \in \Gamma^m,
\end{eqnarray}
where
\begin{eqnarray*}          R(p,z)= \sum\limits
_{i=1}^m[y_i-x_i]+ \sum\limits _{j=1}^l b_j(p,z), \quad r=\{r_i\}_{i=1}^n, \quad r_i>0, \quad  i=\overline{1,n}.\end{eqnarray*}
If $\sum\limits_{i=1}^lf_{ki}>0, \  k=\overline{1,n},$ then
for every continuous
demand matrix $||\gamma_{ik}(p)||_{i=1, k=1}^{l,\ n}$ and continuous realization  of random field  of decisions making by firms  $z(p)$ on $K_+^n$  there exists a strictly positive price vector  $ p^{\varepsilon}$  corresponding to them and satisfying the set of equations
\begin{eqnarray*}         \sum\limits
_{i=1}^l\gamma _{ik}^{\varepsilon}(p)D_i(p) \end{eqnarray*}
\begin{eqnarray}\label{rest3}
= p_k \left
[\sum\limits _{i=1}^m[y_{ki}(p)-x_{ki}(p)]+ \sum\limits
_{i=1}^l b_{ki}(p,z(p)) \right ], \quad  k=\overline{1,n},
\end{eqnarray}
where
\begin{eqnarray*}         \gamma _{ik}^{\varepsilon}(p)=
\frac{\mu_{ki}(p)p_k+ \varepsilon f_{ki}}{\sum\limits_{j=1}^n
\mu_{ji}(p)p_j+ \varepsilon\sum\limits_{j=1}^nf_{ji}}, \quad k=\overline {1,n},\quad i=\overline {1,l}, \quad  0 < \varepsilon < 1. \end{eqnarray*}
\end{theorem}
\begin{proof}\smartqed
Denote
\begin{eqnarray*}         \Gamma_k^{\varepsilon}(p)=\frac{1}{\psi_k(p)}\sum\limits_{i=1}^l
\frac{(\mu_{ki}(p)p_k + \varepsilon f_{ki}) D_i(p)}{\sum\limits_{j=1}^n
\mu_{ji}(p)p_j +\varepsilon \sum\limits_{j=1}^nf_{ji}}, \quad k=\overline {1,n},\end{eqnarray*}
\begin{eqnarray*}         \Gamma^{\varepsilon}(p)=\{ \Gamma_k^{\varepsilon}(p)\}_{k=1}^n. \end{eqnarray*}
 On the closed bounded convex set $C_\delta, $ where $\delta>0$ satisfies  the  conditions of the Lemma \ref{l1ch2l1}, let us consider the non-linear map
\begin{eqnarray*}         f^{\varepsilon}(p)=\{f_k^{\varepsilon}(p)\}_{k=1}^n,\end{eqnarray*}
where
\begin{eqnarray*}         f_k^{\varepsilon}(p)=\frac{\Gamma_k^{\varepsilon}(p)}{\sum\limits_{i=1}^n\Gamma_i^{\varepsilon}(p)}, \quad
k=\overline {1,n}.\end{eqnarray*}          The map $f^{\varepsilon}(p)$ transforms the set
$C_\delta $ into itself and is a continuous map. It is sufficient to check the inequalities
\begin{eqnarray*}         \sum\limits_{k=1}^nc_{ki}f_k^{\varepsilon}(p) \geq \delta, \quad i=\overline {1,l},\end{eqnarray*}
and show the continuity of $f^{\varepsilon}(p).$
First, let us  prove the continuity of $f^{\varepsilon}(p).$
In view of the continuity of $\mu_{ki}(p)$ for every realization and the condition of the Theorem,
\begin{eqnarray*}         \sum\limits_{j=1}^n\mu_{ji}(p)p_j \geq m_i\sum\limits_{j=1}^nc_{ji}p_j > m_i\delta
> 0.\end{eqnarray*}
Thus, $\sum\limits_{j=1}^n\mu_{ji}(p)p_j$ nowhere vanishes on the set $C_{\delta}.$
Therefore, $\Gamma_k^{\varepsilon}(p)$ is a continuous map on $C_{\delta}.$

There hold the estimates
\begin{eqnarray*}          \sum\limits_{k=1}^n\Gamma_k^{\varepsilon}(p) \geq \frac{1}{R_1}
\sum\limits_{i=1}^l
D_i(p)=\frac{1}{R_1}\sum\limits_{i=1}^n\psi_i(p)p_i \geq \frac{R_0}{R_1},\end{eqnarray*}
\begin{eqnarray*}         \sup_{p \in P\cap K_+^n}\sum\limits_{r=1}^n\Gamma_r^{\varepsilon}(p) \leq \frac{1}{R_0}\sup_{p \in P\cap K_+^n}\sum\limits_{i=1}^lD_i(p) = \frac{1}{R_0}\sup_{p \in P\cap K_+^n}\sum\limits_{i=1}^n\psi_i(p)p_i \leq \frac{R_1}{R_0}.\end{eqnarray*}
Therefore, $f(p)$ is a continuous map on $C_{\delta}.$

Finally, show that
\begin{eqnarray*}         \sum\limits_{k=1}^nc_{ki}f_k^{\varepsilon}(p) \geq \delta, \quad i=\overline {1,l}.\end{eqnarray*}
Really,
\begin{eqnarray*}         \sum\limits_{k=1}^nc_{ki}f_k^{\varepsilon}(p) =  \sum\limits_{k=1}^nc_{ki}
\frac{1}{\psi_k(p)\sum\limits_{r=1}^n\Gamma_r^{\varepsilon}(p)}\sum\limits_{s=1}^l
\frac{(\mu_{ks}(p)p_k + \varepsilon f_{ks}) D_s(p)}{\sum\limits_{j=1}^n
\mu_{js}(p)p_j + \varepsilon \sum\limits_{j=1}^nf_{js}}
\end{eqnarray*}
\begin{eqnarray*}
\geq \delta \sum\limits_{k=1}^nc_{ki}
\sum\limits_{s=1}^l
\frac{f_{ks}(\mu_{ks}(p)p_k + \varepsilon f_{ks})}{\sum\limits_{j=1}^n
\mu_{js}(p)p_j + \varepsilon \sum\limits_{j=1}^nf_{js}}=\delta\sum\limits_{s=1}^l
\frac{\sum\limits_{k=1}^nc_{ki}f_{ks}(\mu_{ks}(p)p_k  + \varepsilon f_{ks})}{\sum\limits_{j=1}^n
\mu_{js}(p)p_j + \varepsilon \sum\limits_{j=1}^nf_{js}}.\end{eqnarray*}
In view of the assumptions about matrix elements $c_{ki},$ for $s=i$
\begin{eqnarray*}         \frac{\sum\limits_{k=1}^nc_{ki}f_{ki}(\mu_{ki}(p)p_k + \varepsilon f_{ki})}{\sum\limits_{j=1}^n
\mu_{ji}(p)p_j + \varepsilon \sum\limits_{j=1}^nf_{ji}} \geq \min_{\{k,i, f_{ki} =1\}}c_{ki}\frac{\sum\limits_{k=1}^n(\mu_{ki}(p)p_k + \varepsilon f_{ki})}{\sum\limits_{j=1}^n\mu_{ji}(p)p_j + \varepsilon \sum\limits_{j=1}^nf_{ji}}=1.\end{eqnarray*}
Therefore,
\begin{eqnarray*}         \sum\limits_{k=1}^nc_{ki}f_k^{\varepsilon}(p) \geq \delta, \quad i=\overline{1,l}.\end{eqnarray*}
By the Schauder Theorem \cite{88}, there exists a fixed point of the map $f^{\varepsilon}(p).$
This point is also a fixed point for the map $\Gamma^{\varepsilon}(p).$
Really, as $p^{\varepsilon}$ is the fixed point of the map $f^{\varepsilon}(p),$
we have
\begin{eqnarray*}         \Gamma_k(p^{\varepsilon})=p_k^{\varepsilon}\sum\limits_{s=1}^n\Gamma_s(p^{\varepsilon}).\end{eqnarray*}
Multiplying the last equality by $\psi_k(p^{\varepsilon})$  and summing up over $k,$ we have
\begin{eqnarray*}         \sum\limits_{k=1}^n\psi_k(p^{\varepsilon})\Gamma_k(p^{\varepsilon})=\left\langle\psi(p^{\varepsilon}), p^{\varepsilon}\right\rangle\sum\limits_{s=1}^n\Gamma_s(p^{\varepsilon}).\end{eqnarray*}
However,
\begin{eqnarray*}         \sum\limits_{k=1}^n\psi_k(p^{\varepsilon})\Gamma_k(p^{\varepsilon})=\sum\limits_{i=1}^lD_i(p^{\varepsilon})
=\left\langle\psi(p^{\varepsilon}), p^{\varepsilon}\right\rangle.\end{eqnarray*}
Reducing by $\left\langle\psi(p^{\varepsilon}), p^{\varepsilon}\right\rangle,$ in view of $\left\langle\psi(p^{\varepsilon}), p^{\varepsilon}\right\rangle \neq 0,$
we obtain $ \sum\limits_{s=1}^n\Gamma_s(p^{\varepsilon})=1.$ The latter means that $p^{\varepsilon}$
solves the set of equations (\ref{rest3}).
As the price vector $p^{\varepsilon}$
solves the set of equations (\ref{rest3}) and the conditions of the Lemma
 \ref{l1ch2l1}  hold, the inequalities  for components  $p_k^{\varepsilon}$
\begin{eqnarray*}          p_k^{\varepsilon} \geq  \frac{a \varepsilon \sum\limits_{i=1}^lf_{ki}}{R_1( \mu + \max\limits_{i} \sum\limits_{j=1}^nf_{ji} \varepsilon)} > 0, \quad k=\overline{1,n},\end{eqnarray*}
hold, where
\begin{eqnarray*}         \mu=\max\limits_{i}\sup\limits_{p \in P \cap K_+^n}\sum\limits_{s=1}^n\mu_{si}(p)p_s< \infty. \end{eqnarray*}
\qed
\end{proof}

\begin{theorem}\label{1ch2l3}
Let the conditions of the Theorem \ref{1ch2l2}, of the Lemma \ref{l1ch2l1}  and  the inequalities $\sum\limits_{i=1}^lf_{ki}>0, \  k=\overline{1,n},$ hold. Then there exists an equilibrium price vector $p^0 \in C_{\delta},$ under which the demand does not exceed the supply, i.e., the set of inequalities holds
\begin{eqnarray*}         \sum\limits_{i=1}^l\frac{\mu_{ki}(p^0)}{\sum\limits_{s=1}^n\mu_{si}(p^0)p_s^0}D_i(p^0) \end{eqnarray*}
\begin{eqnarray}\label{2g2l118}
\leq   \sum\limits _{i=1}^m[y_{ki}(p^0)-x_{ki}(p^0)]+ \sum\limits
_{i=1}^l b_{ki}(p^0,z(p^0)), \quad  k=\overline{1,n}.
\end{eqnarray}
Every equilibrium price vector satisfies  the set of equations (\ref{1g2l18}).
\end{theorem}
\begin{proof}\smartqed      Consider  the auxiliary set of equations
\begin{eqnarray}\label{11g2l119}
\sum\limits_{i=1}^l\frac{p_k\mu_{ki}(p)+\varepsilon f_{ki}}{\sum\limits_{s=1}^n\mu_{si}(p)p_s+  \varepsilon \sum\limits_{s=1}^nf_{si}}D_i(p)
\end{eqnarray}
\begin{eqnarray*}
= p_k\left[\sum\limits _{i=1}^m[y_{ki}(p)-x_{ki}(p)]+ \sum\limits
_{i=1}^l b_{ki}(p,z(p))\right], \quad  k=\overline{1,n}, \quad  0 < \varepsilon < 1,\end{eqnarray*}
built after the set of equations (\ref{1g2l18}).
Every component of the solution $p^{\varepsilon}= \{ p_k^{\varepsilon}\}_{k=1}^n$ to the set of equations (\ref{11g2l119}), by the Theorem \ref{rest1}, is strictly positive.

As $p^{\varepsilon}$ solves the set of equations (\ref{11g2l119}) and
$p_k^{\varepsilon} >0, \ k=\overline{1,n},$ we obtain the set of inequalities
\begin{eqnarray}\label{1g2l1110}
\sum\limits_{i=1}^l \frac{\mu_{ki}(p^{\varepsilon})}
{\sum\limits_{s=1}^n\mu_{si}(p^{\varepsilon})p_s^{\varepsilon} + \varepsilon \sum\limits_{s=1}^nf_{si}}D_i( p^{\varepsilon})
\end{eqnarray}
\begin{eqnarray*}
\leq \sum\limits _{i=1}^m[y_{ki}(p^{\varepsilon})-x_{ki}(p^{\varepsilon})]+
\sum\limits _{i=1}^l b_{ki}(p^{\varepsilon},z(p^{\varepsilon})),\quad k=\overline{1,n}.
\end{eqnarray*}
The sequence $p^{\varepsilon}, $ when $\varepsilon \to 0,$ is compact one, because it belongs to the set $C_{\delta}.$
Due to the continuity of $\sum\limits_{s=1}^n\mu_{si}(p)p_s$ on the set $P$ and the inequality
\begin{eqnarray*}         \sum\limits_{s=1}^n\mu_{si}(p)p_s > m_i \delta,\quad  p \in C_{\delta}, \end{eqnarray*}
one can go to the limit in the set of inequalities
(\ref{1g2l1110}). Denote one of the possible limit points of the sequence $p^{\varepsilon} $ by $p^0.$
Then $p^0$ solves the set of inequalities
\begin{eqnarray}\label{1g2l1111}
\sum\limits_{i=1}^l \frac{\mu_{ki}(p^{0})}
{\sum\limits_{s=1}^n\mu_{si}(p^{0})p_s^{0}} D_i( p^{0})
\end{eqnarray}
\begin{eqnarray*}          \leq \sum\limits _{i=1}^m[y_{ki}(p^{0})-x_{ki}(p^{0})]+
\sum\limits _{i=1}^l b_{ki}(p^{0},z(p^{0})),\quad k=\overline{1,n}.\end{eqnarray*}
It is obvious that the vector $p^0$ belongs to the set $C_{\delta}$ and solves the set of equations
(\ref{1g2l18}). \qed
\end{proof}

Now, consider the case where not all the consumers are insatiable. Suppose that assumptions about the matrix $C$ are the same as at the beginning of the Section.
Suppose random fields of  information evaluation by consumers  satisfy the condition:
for every $i$-th consumer a  random field of  information evaluation by  the $i$-th consumer $\eta_i^1(p,z, \omega_i)=\{\eta_{ik}^1(p,z, \omega_i)\}_{k=1}^n,$
on a probability space $\{\Omega_i, {\cal F}_i, P_i\},\ i=\overline{1,l},$
satisfies the inequality
\begin{eqnarray*}         \eta_i^1(p,z, \omega_i) \geq m_i C_i, \quad C_i=\{c_{ki}\}_{k=1}^n, \quad (p,z, \omega_i) \in  K_+^n\times \Gamma^m\times \Omega_i,  \quad i=\overline{1,l},\end{eqnarray*}
and components
$\eta_{ik}^1(p,z, \omega_i)=0$ if and only if $c_{ki}=0.$

Under these additional assumptions for random fields $\eta_i^1(p,z, \omega_i)$
and for the matrix $C$ on the above built cone $K_+^n $ (with the rest conditions of the Theorems \ref{tl4} and \ref{ttl3}), there exist the random fields of consumers choice and  decisions making by firms if not all consumers are insatiable.

Note that if certain components of the vector $C_i=\{c_{ki}\}_{k=1}^n$ equal zero, then  the $i$-th consumer does not consume goods numbered by them.

Let $z(p)$ be a certain  continuous realization of the random field $\zeta(p, \omega_0)$ and the random fields of information evaluation and  overestimation
 by  the $i$-th non-insatiable consumer
\begin{eqnarray*}         \eta_i^1(p,z,\omega_i)
=\{\eta_{ik}^1(p, z,\omega_i)\}_{k=1}^n, \quad  i=\overline{1,l},  \end{eqnarray*}
\begin{eqnarray*}         \eta_i^0(p,z,\omega_i)
=\{\eta_{ik}^0(p, z,\omega_i)\}_{k=1}^n,
\quad  i=\overline{1,l},  \end{eqnarray*}
satisfy the  conditions of  the Theorem \ref{ttl3}.
As earlier, we give the random demand vector of the  $i$-th non-insatiable consumer
by the formula
\begin{eqnarray*}         \gamma_i(p,\omega_0,\omega_i) =\{ \gamma_{ik}(p,\omega_0,\omega_i)\}_{k=1}^n,
\quad  i=\overline{1,l}, \end{eqnarray*}
where
\begin{eqnarray*}         \gamma_{ik}(p,\omega_0,\omega_i)
=\frac{p_k\eta_{ik}^0(p, \zeta(p, \omega_0),\omega_i)}
{\sum\limits_{s=1}^n\eta_{is}^1(p, \zeta(p, \omega_0),\omega_i)p_s},
\quad  i=\overline{1,l}.\end{eqnarray*}

We denote every realization of random demand vector  for the  $i$-th non-insatiable consumer by
\begin{eqnarray*}          \gamma_i(p)=\{ \gamma_{ik}(p)\}_{k=1}^n,\end{eqnarray*}
where $ \gamma_{ik}(p)=\gamma_{ik}(p,\omega_0,\omega_i)$
for certain $\omega_0 $ and $\omega_i$ and call it the demand vector of the $i$-th non-insatiable  consumer.
We denote any realizations of random fields
\begin{eqnarray*}         \eta_i^0(p,\zeta(p, \omega_0),\omega_i)
=\{\eta_{ik}^0(p,\zeta(p, \omega_0),\omega_i)\}_{k=1}^n,
\quad  i=\overline{1,l}, \end{eqnarray*}
and
\begin{eqnarray*}         \eta_i^1(p,\zeta(p, \omega_0),\omega_i)
=\{\eta_{ik}^1(p,\zeta(p, \omega_0),\omega_i)\}_{k=1}^n,
\quad  i=\overline{1,l}, \end{eqnarray*}
by
\begin{eqnarray*}          \mu_i^0(p)=\{\mu_{ki}^0(p)\}_{i=1}^n,\end{eqnarray*}
where
\begin{eqnarray*}          \mu_{ki}^0(p)=\eta_{ik}^0(p,\zeta(p, \omega_0),\omega_i),\end{eqnarray*}
and
\begin{eqnarray*}          \mu_i(p)=\{\mu_{ki}(p)\}_{i=1}^n,\end{eqnarray*}
where
\begin{eqnarray*}          \mu_{ki}(p)=\eta_{ik}(p,\zeta(p, \omega_0),\omega_i)\end{eqnarray*}
respectively, for certain $\omega_0 $ and $\omega_i.$

In these notations,
\begin{eqnarray*}         \gamma_{ik}(p)=\frac{p_k\mu_{ki}^0(p)}{\sum\limits_{s=1}^n\mu_{si}(p)p_s},\quad  i=\overline{1,l},\quad  k=\overline{1,n}.\end{eqnarray*}

\begin{theorem}\label{nen1}
Let technological maps $F_i(x), \  x \in X_i^1,
\ i=\overline{1,m},$ describing the  economy system production, be convex down, belong to the CTM class, and let a productive economic  process $Q(p,z),$ a family of income pre-functions  $K_i^0(p,z), \ i=\overline{1,l},$ and property vectors $b_i(p,z), \ i=\overline {1,l},$  be continuous maps of variables
$(p,z) \in K_+^n\times \Gamma^m,$ where $K_+^n $ is the cone given by the formula (\ref{kl1}),
 and also let
 \begin{eqnarray*}          D_i(p)=K_i(p, \zeta_0(p, \omega_0)> a>0, \quad p \in P \cap K_+^n, \quad \omega_0 \in \Omega_0, \quad i=\overline{1,l},\end{eqnarray*}
where $a$ does not depend on $(p, \omega_0).$
Assume that the  random fields of  information evaluation by consumers and  decisions making by firms  satisfy the  conditions of the Theorem \ref{ttl3} and  the productive economic process
 $Q(p,z)$ satisfies  the condition
\begin{eqnarray*}
   R(p, Q(p,z)) >r, \quad p \in K_+^n, \quad z \in \Gamma^m, \quad r=\{r_i\}_{i=1}^n, \quad r_i>0, \quad  i=\overline{1,n},
\end{eqnarray*}
\begin{eqnarray}\label{nen2}
R(p,z)= \sum\limits
_{i=1}^m[y_i-x_i]+ \sum\limits _{j=1}^l b_j(p,z).
\end{eqnarray}
If the  conditions of the Lemma \ref{l1ch2l1} and inequalities
$\sum\limits_{i=1}^lf_{ki}>0, \  k= \overline{1,n},$
hold, then for every continuous on $K_+^n$
demand matrix $||\gamma_{ik}(p)||_{i=1, k=1}^{l,\ n}$ and  continuous realization of random field of decisions making by firms  $z(p)$   there exists  a price vector
$\bar p$ corresponding to them and satisfying  the set of inequalities
\begin{eqnarray*}         \sum\limits
_{i=1}^l\frac{\mu_{ki}^0(p)}{\sum\limits_{s=1}^n\mu_{si}(p)p_s}D_i(p)
\end{eqnarray*}
\begin{eqnarray}\label{nen3}
\leq   \left
[\sum\limits _{i=1}^m[y_{ki}(p)-x_{ki}(p)]+ \sum\limits
_{i=1}^l b_{ki}(p,z(p)) \right ], \quad  k=\overline{1,n}.
\end{eqnarray}
This vector solves the set of equations
\begin{eqnarray*}         \sum\limits
_{i=1}^l\frac{p_k\mu_{ki}(p)}{\sum\limits_{s=1}^n\mu_{si}(p)p_s}D_i(p)
\end{eqnarray*}
\begin{eqnarray}\label{nen4}
= p_k \left
[\sum\limits _{i=1}^m[y_{ki}(p)-x_{ki}(p)]+ \sum\limits
_{i=1}^l b_{ki}(p,z(p)) \right ], \quad  k=\overline{1,n},
\end{eqnarray}
too.

\end{theorem}
\begin{proof}\smartqed      Under the conditions of the  Theorem \ref{nen1}, the set of equations (\ref{nen4}) has a solution because the  conditions of the  Theorems \ref{1ch2l2} and \ref{1ch2l3} hold guaranteeing the existence of such a solution. By the Theorem \ref{1ch2l3}, this solution solves the set of inequalities
(\ref{nen3}) too. \qed
\end{proof} 

\chapter{Foundations of economy system control}

\abstract*{The problem of finding conditions under that all firms, in the economy system, operate  in profitable regime in the state of equilibrium is solved. These conditions are very important  for economy system control. Among  parameters that permit to control economy system are: taxation system, consumption levels of economic subjects, subvention and credit policy.
The Lemma on the existence of a taxation system is proved. The notion of economic compatibility for the set of productive processes of firms and a notion of agreement of firm supply with consumers choice structure are introduced. Under conditions of economic compatibility of productive processes of firms,   agreement of firms supply with consumers choice structure, and  special taxation system the Theorem on the existence of equilibrium state under that firms operate  in profitable regime is proved. The Theorem giving the necessary and sufficient conditions of economic compatibility for the set of productive processes is proved. The examples of application of the proved Theorems are presented.}

The economy system models considered in the previous Chapters and the arbitrage absence conditions established  for them  are general. To control  the
 economy system,  one must have its detailed model  giving  the possibility to influence the output and consumption structure by changing the economic parameters.
 Among such parameters of the economy system,  there are, e.g., a taxation system,\index{taxation system} consumption levels of economic subjects,\index{consumption levels of economic subjects} a subvention policy\index{subvention policy} to give preferences for development of prospect economy directions, etc. Therefore, one must study those taxation systems that result in the income operation of the  economy system in the state of economy equilibrium. In this Chapter, for the economy system  whose structure of firms production  is described by technological maps, we clarify what   must be  these technological maps that the  economy system producing a certain set of the final consumption goods has no risk of firms default  \cite{92, 106}. For this, the goods produced in the economy system  must find their consumers, and firms, to have no default risk,\index{default risk} must be solvent, i.e, operate profitably. Therefore, the problem  is stated as follows: what must be technological maps describing firms and a taxation system for the economy system   to operate in a stable profitable regime and economic subjects  have no    arbitrage  opportunities.\index{arbitrage  opportunities} For this, we introduce significant notions of economic compatibility for firms productive processes,\index{economic compatibility for firms productive processes} agreement  of supply and choice structures\index{agreement  of supply and choice structures} giving the  possibility to build the set of taxation systems under which the product the economy system produces is consumed according to the choice structure.

This allows  to influence the economy system, choosing the needed taxation system, to stimulate the economic development and to do effectively managing the output structure  to satisfy consumers needs and to have no firms default risk.

In the Subsection "Foundations of economy system control", we prove the Lemma \ref{let1} named the  Lemma   on existence of a taxation system\index{ Lemma   on existence of a taxation system} that is fundamental  for the description of taxation systems in specific models of economy systems. We build a general economy system model whose firms production structure is described by convex down technological maps from the CTM class in a wide sense and a  general  productive economic process. For such an economy system, we introduce a notion of economic compatibility\index{ economic compatibility} for set of productive processes of firms with subvention  and initial goods supply vectors and a notion of  agreement of firms supply and consumers choice structures.\index{agreement of firms supply and consumers choice structures}
In the Subsection "Building taxation systems", in particular, in Lemmas \ref{opod11},
\ref{opcd1}, and \ref{opod111}
 we describe economically accepted solutions to the equation (\ref{opod1}). With the Lemma \ref{let1}
on the existence of a taxation system and the notion of agreement firms supply and consumers choice structures, Lemmas \ref{opod11},
\ref{opcd1}, and \ref{opod111} we establish the Theorem \ref{ptm1} guaranteeing the existence of corresponding taxation systems.

In the Subsection "Operation of economy systems without default", we prove the Theorem \ref{2ch2l2} giving an  algorithm to build the equilibrium state in which all the firms operate according to the productive economic process.

The Theorem \ref{3ch2l2} gives an algorithm to build equilibrium state in which all the firms operate according to the productive economic  process with stable taxation proportions within long time. The Theorem \ref{1uh2l2} generalizes Theorems \ref{2ch2l2} and \ref{3ch2l2}  onto the case when realizations of random fields  of decisions making  by firms  are regular.

In the Subsection "General results application", we prove the Theorem about the necessary and sufficient condition of economic compatibility for the set of productive processes with given sets of vectors of subventions   and initial goods supply, and do the Theorem \ref{sumi2} containing conditions for the existence of certain quality operation without default.

In the Subsection "Theory of interindustry equilibrium", we adapt the results obtained in the previous Sections to the case of the aggregated description of the economy.

\section{Economy systems without default}

In this Section, we study models of the economy system  applied to real economy systems. These models, of course, is a certain class of submodels of the earlier studied economy models. The main aim of the Subsection is not only to establish Theorems on the existence of the economy equilibrium in such models, but also to clarify what factors influence both the equilibrium state of the economy system and  the quality of this state to manage economic processes.
By the equilibrium state quality in the economy system, we understand the ability of the economy system to satisfy certain requirements to its work in certain regimes. Such regimes are determined by a productive economic  process, the present material balance between produced final product and consumers needs and a taxation system. Depending on a taken  productive economic  process and a redistribution system of incomes, we have one or another type of economy. These requirements to economy systems need to develop a new mathematical apparatus for  studies of such economy systems.

The first Subsection contains the construction of mathematical models of operation of economy systems  in certain regimes. We establish an auxiliary Lemma about the existence of a taxation system redistributing the product produced. We introduce the general class of productive economic  processes guaranteeing economy systems operation in profitable or subvention-profi\-table regime depending on the taken  productive economic  process. Such constructed  productive economic processes guarantee prescribed output levels in the economy system. For further investigation, the notions of economic compatibility and economic compatibility in a wide sense for productive processes of $m$ firms are important,  whose production structure is described by technological maps from the CTM class in a wide sense. This notion marks out a certain class of sets of technological maps after which we construct  models of the economy system  and which,  in what follows, are called  economically compatible. For such an economy system model, we construct a class of  taxation system  guaranteeing consumption in cost sense of the product produced within the economy system without arbitrage  opportunities  for economic agents. We describe available taxation systems depending on  the income part remaining in firms property. Under simple technical conditions, we establish Theorems about the existence of the given quality economic equilibrium for the economy systems economically compatible with the taken taxation system that guarantees the agreement of the   structure of  firms supply with the  structure of consumers choice.

In the second Subsection, we give the necessary and sufficient conditions for the economic compatibility of $m$ firms productive processes and prove the Theorem about the existence of prescribed quality economic equilibrium. Finally, in the third Subsection, we construct the theory of the given quality interindustry equilibrium.

\subsection{Foundations of economy systems control}

In this Subsection, we clarify the conditions for technological maps, the   structure of  firms
 supply, and the   structure of consumers choice under which the  economy system operates in certain regime.

Consider the economy system containing $m$ firms. We describe the  production  structure of  the $i$-th firm  by  a  convex   down  technological  map $F_i(x), $  $ x \in X_i^1,$  $   i =\overline{1,m},$ belonging to the  CTM class in a wide sense.
\begin{definition}
Let a  convex down technological map   $F_i(x),  \ x \in X_i^1,\ $ $ i =\overline{1,m},$ belong to the CTM class in a wide sense, and  $F_i^1(x),  \ x \in X_i^2,\  i =\overline{1,m},$ be its continuation to a convex down technological map from the  CTM class on the set
$ X_i^2,\  i =\overline{1,m},$  being the smallest linear convex span of the set   $ X_i^1,\  i =\overline{1,m},$ and the set containing  one point $0 \in S.$
If  a productive economic process  $Q_1(p,z)$  defined on the set $\Gamma_1^m = \prod\limits_{i=1}^m\Gamma^1_i, $ where   \begin{eqnarray*}   \Gamma_i^1=\{(x,y), \  x \in  X_i^2, \  y \in F_i^1(x)\}, \quad i=\overline{1,m},\end{eqnarray*}     admits a contraction  $Q(p,z)$ on the set $\Gamma^m = \prod\limits_{i=1}^m\Gamma_i, $ where
\begin{eqnarray*}   \Gamma_i=\{(x,y), \  x \in  X_i^1, \  y \in F_i(x)\}\cup(0,0),\quad   0 \in S,\end{eqnarray*}
and  $(0,0)$ is a zero productive process, then the  map
$Q(p,z)$  is called a  productive economic process for the set of convex down technological maps
 $F_i(x),  \ x \in X_i^1,\  i =\overline{1,m},$ belonging to the CTM class in a wide sense.
\end{definition}

Further, for the convex down technological maps $F_i(x),  \ x \in X_i^1,$ $  i =\overline{1,m},$ from the CTM class in a wide sense,  the set of possible  productive processes is  \begin{eqnarray*}   \Gamma^m=\prod\limits_{i=1}^m \Gamma_i,\end{eqnarray*}
 where now
\begin{eqnarray*}   \Gamma_i=\{(x,y), \  x \in  X_i^1, \  y \in F_i(x)\}\cup(0,0),\quad  0 \in S,  \quad i=\overline{1,m},\end{eqnarray*}
 $(0,0)$ is a zero productive process
and the set of possible  price vectors is the cone
\begin{eqnarray*}   K_+^n=\{p=\{p_i\}_{i=1}^n \in R_+^n, \ p_i > 0,\ i=\overline{1,n}\}.\end{eqnarray*}

For the considered set of technological maps, we give a productive economic  process
$Q(p,z)=\{Q_i(p,z)\}_{i=1}^m$ by the rule
\begin{eqnarray}   \label{001l0}
Q_i(p,z)=z_ia_i(p, z) \prod\limits_{s=1}^nv_s(p,z), \quad
  (p,z) \in K_+^n \times \Gamma^m,
\end{eqnarray}
\begin{eqnarray*}    a_i(p, z)=\chi_{[0, \infty)}(u_i(p,z)),   \quad u_i(p,z)= \left\langle p, y_i - x_i \right\rangle + \left\langle p, d_i + b_i \right\rangle, \end{eqnarray*}
\begin{eqnarray*}   v_s(p,z)=\chi_{[0, \infty)}(V_s(p,z)), \end{eqnarray*}
\begin{eqnarray*}    V_s(p,z) = \sum\limits_{i=1}^m \left[a_i(p,z)(y_{si} -x_{si}) +d_{si}\right] + \sum\limits_{i=1}^lb_{si} - c_s,\end{eqnarray*}
where $z^i=(x_i,y_i) \in \Gamma_i=\{(x,y), \  x \in X_i^1, \  y \in F_i(x)\}\cup(0,0),$  and
\begin{eqnarray*}    x_i=\{x_{ki}\}_{k=1}^n, \quad y_i=\{y_{ki}\}_{k=1}^n, \quad  i=\overline{1,m}, \end{eqnarray*}
\begin{eqnarray*}    b_i=\{b_{ki}\}_{k=1}^n, \quad i=\overline{1,l}, \quad d_i=\{d_{ki}\}_{k=1}^n, \quad  i=\overline{1,m}, \end{eqnarray*}
are vectors with non-negative components, $c=\{c_i\}_{i=1}^n, \ c_i>0,  \ i=\overline{1,n},$

\begin{eqnarray*}   \chi_{[0, \infty)}(x)= \left\{\begin{array}{ll}
            1, & \textrm{if} \quad x \geq 0,\\
            0  &         \textrm{if} \quad    x < 0  \textrm{.}
                                                                                                                                                                 \end{array}
                                               \right.\end{eqnarray*}
Further, we suppose the condition
\begin{eqnarray*}   \sum\limits_{i=1}^lb_i > c,\quad c=\{c_1, \ldots , c_n\},\quad  c_i>0, \quad i=\overline{1,n}, \end{eqnarray*}
holds.
Introduce the set
\begin{eqnarray*}   T=\{z \in \Gamma^m, \  W_s(z)>0,  \ s=\overline{1,n} \},\end{eqnarray*}
where
\begin{eqnarray}   \label{mg5}
W_s(z) = \sum\limits_{i=1}^m (y_{si} -x_{si} +d_{si}) + \sum\limits_{i=1}^lb_{si} - c_s , \quad s=\overline{1,n},
\end{eqnarray}
is the $s$-th component of the vector
\begin{eqnarray*}   W(z) = \sum\limits_{i=1}^m (y_i -x_i +d_i) + \sum\limits_{i=1}^lb_i - c.\end{eqnarray*}
In what follows, we consider only  sets of convex down technological maps
  $F_i(x),  \ x \in X_i^1,\  i =\overline{1,m},$ from the CTM class in a wide sense for which the set $T$ is non-empty.

Consider a probability space $\{\Omega, {\cal F}, \bar P\}$
being a direct product of $(l+1)$ probability spaces $\{\Omega_i, {\cal F}_i, \bar P_i\}, \
i=\overline{0,l},$
where
\begin{eqnarray*}   \Omega=\prod\limits_{i=0}^{l}\Omega_i, \quad {\cal F}=\prod\limits_{i=0}^{l}
{\cal F}_i, \quad \bar P=\prod\limits_{i=0}^{l}\bar P_i.\end{eqnarray*}
Denote, by $\{\tilde \Omega, \tilde {\cal F}, \tilde P\},$ a probability space with
\begin{eqnarray*}   \tilde \Omega=\prod\limits_{i=1}^{l}\Omega_i, \quad \tilde {\cal F}=\prod\limits_{i=1}^{l}
{\cal F}_i, \quad \tilde P=\prod\limits_{i=1}^{l}\bar P_i.\end{eqnarray*}
Then
\begin{eqnarray*}    \Omega=\Omega_0\times \tilde \Omega, \quad  {\cal F}={\cal F}_0\times \tilde {\cal F}, \quad \bar P=\bar P_0\times \tilde P.\end{eqnarray*}

Give the Theorem \ref{ptl3} version being applied in what follows.
\begin{theorem}\label{11tl4}
Let convex down technological maps $F_i(x),  \ x \in X_i^1,\  i =\overline{1,m},$ describing the structure of the  economy system production,  belong to the  CTM class in a wide sense,  a productive economic  process be given by the formula (\ref{001l0}),
 a  random field of information evaluation  by the $i$-th consumer
$\eta_i^0(p,z, \omega_i)=\eta_i^0(p,z), \  (p,z) \in  K_+^n\times \Gamma^m$ on a probability space $\{\Omega_i, {\cal F}_i, \bar P_i\}$  be a non-random continuous function of $(p,z) \in  K_+^n\times \Gamma^m, \ i=\overline{1,l},  $
with values in the set $S,$ and let  a random field
$\zeta_0(p, \omega_0), \  p \in K_+^n,$ on a probability space
$\{\Omega_0, {\cal F}_0, \bar P_0\}$
take values in the set $\Gamma^m$
and every its realization be a continuous function of ~$  p \in K_+^n.$ Let also
 $\eta_i^0(tp,z)=\eta_i^0(p,z), \ i=\overline{1,l}, \ t> 0,$
$\zeta_0(tp, \omega_0)=\zeta_0(p, \omega_0), \  p \in K_+^n,\ t > 0,$
and $K_i(p,z), (p,z) \in  K_+^n\times \Gamma^m, \ i=\overline{1,l}, $ be the set of consumers income functions satisfying  the conditions of the  Definition \ref{dl1}  with the productive economic process given by the formula (\ref{001l0}).
If
\begin{eqnarray*}   \left\langle\eta_i^0(p,z), p \right\rangle  > 0, \quad
 \eta_i(p,z)=\eta_i^0(p,Q(p,z)),	\end{eqnarray*}    \begin{eqnarray*}    (p,z)  \in
K_+^n\times \Gamma^m, \quad  i=\overline{1,l},   \end{eqnarray*}
 \begin{eqnarray*}    \eta_i^0(p,z) < \infty,		\quad (p,z)  \in
K_+^n\times \Gamma^m, \quad  i=\overline{1,l},\end{eqnarray*}
then random fields
\begin{eqnarray}   \label{11pl7}
\xi_i(p, \omega)
=\frac{K_i(p,\zeta_0(p, \omega_0)) \eta_i(p,\zeta_0(p, \omega_0))}
{\left\langle\eta_i(p,\zeta_0(p, \omega_0)),p\right\rangle }, \quad  i=\overline{1,l},
\end{eqnarray}
on the probability space $\{\Omega, {\cal F}, \bar P\}$ can be identified with the random fields  of choice of insatiable consumers if $ \zeta(p, \omega_0)=Q(p,\zeta_0(p, \omega_0))$
is identified with the random field of decisions making by firms  relative to productive processes.
\end{theorem}
Note that the conditions of the  Theorem \ref{ptl3}  hold if the  conditions of the Theorem \ref{11tl4} do because the productive economic process defined by the formula (\ref{001l0}) and the family  of income functions satisfy the  conditions of the Theorem \ref{ptl3}.
In addition, the set of values  of the productive economic  process $Q(p,z)$ is a Borel set.
Here random fields giving consumers choice by the formula (\ref{11pl7}) are not,  generally speaking, continuous  with probability 1 because  the consumers income functions are not such.
Suppose the first $m$ random fields of consumers choice describe the unproductive consumption\index{unproductive consumption} of $m$ firms each of which describes proper firms consumption for the accumulation, namely, creation of raw materials for the next economy operation period and goods for unproductive consumption within current term.

From now on, the exposition has the aim to build a family of income functions  from the condition of the Theorem \ref{11tl4}.
Prove an auxiliary Lemma that is  named  the  Lemma on the existence of  taxation system.\index{Lemma on the existence of  taxation system}
\begin{lemma}\label{let1}
Let $f_i(p,z), \ i=\overline{1,t}, $ be a set of non-negative functions given on the set $R \subseteq K_+^n\times \Gamma^m$ satisfying the condition: for every $(p,z) \in R $ there exist such numbers $1 \leq m \leq t $ and $1 \leq s \leq t $ that
\begin{eqnarray*}    f_{m}(p, z)+ f_s(p, z) > 0,  \quad (p,z) \in R, \quad m \neq s, \end{eqnarray*}
and a set of functions $g_j(p,z), \ j=\overline{1,r}, $ given on $R$ satisfies  the condition
\begin{eqnarray*}     \sum\limits _{i=1}^tf_i(p, z)= \sum\limits _{j=1}^rg_j(p, z), \quad (p,z) \in R.\end{eqnarray*}
Then a non-negative matrix $ ||\pi_{ij}(p,z)||_{i,j=1}^{t,r}$ exists such that
\begin{eqnarray*}     \sum\limits _{i=1}^t \pi_{ij}(p,z)=1 , \quad (p,z) \in R,  \quad  j=\overline{1,r}, \end{eqnarray*}
\begin{eqnarray*}    f_i(p,z)= \sum\limits _{j=1}^r \pi_{ij}(p,z)g_j(p,z),  \quad (p,z) \in R, \quad  i=\overline{1,t}.\end{eqnarray*}
\end{lemma}
\begin{proof}\smartqed   Without loss of generality, we suppose that $m=t-1$ and $s=t,$ i.e., the condition
\begin{eqnarray*}    f_{t-1}(p, z)+ f_t(p, z) > 0,  \quad (p,z) \in R, \end{eqnarray*}    holds. If it is not the case, one can reach this renumbering functions $f_i(p,z), \ i=\overline{1,t}.$
Let us give the constructive inductive method to build a taxation system.
We carry out the induction in the number $k$ of sets of functions $\varphi_1(p,z), \ldots \varphi_k(p,z)$ created from $t$ functions $f_i(p,z), \ i=\overline{1,t}, $ beginning from $k=1$ by the rule:
\begin{eqnarray*}   \varphi_1(p,z)= f_1(p,z), \ldots, \varphi_i(p,z)=f_i(p,z), \ldots, \varphi_{k-1}(p,z)=f_{k-1}(p,z),\end{eqnarray*}    \begin{eqnarray*}    \varphi_{k}(p,z)=\sum\limits_{i=k}^tf_{i}(p,z).\end{eqnarray*}
 For $k=1$ this set contains the function $\varphi_1(p,z)=\sum\limits _{i=1}^tf_i(p, z).$ By the Theorem condition,
\begin{eqnarray*}     \sum\limits _{i=1}^tf_i(p, z)= \sum\limits _{j=1}^r g_j(p,z),  \quad (p,z) \in R.\end{eqnarray*}
 In this case, one can take $ \pi_{1j}^1(p,z)=1, \ j=\overline{1,r}.$ Assume the Lemma is proved for the set containing $(k-1)$ functions
\begin{eqnarray*}   \varphi_1(p,z)= f_1(p,z), \ldots, \varphi_i(p,z)=f_i(p,z), \ldots, \varphi_{k-2}(p,z)=f_{k-2}(p,z),\end{eqnarray*}    \begin{eqnarray*}    \varphi_{k-1}(p,z)=\sum\limits_{i=k-1}^tf_{i}(p,z),\end{eqnarray*}
let us prove its validity for the set of  functions
\begin{eqnarray*}   \varphi_1(p,z)= f_1(p,z), \ldots, \varphi_i(p,z)=f_i(p,z), \ldots, \varphi_{k-1}(p,z)=f_{k-1}(p,z),\end{eqnarray*}    \begin{eqnarray*}    \varphi_{k}(p,z)=\sum\limits_{i=k}^tf_{i}(p,z).\end{eqnarray*}
From the latter,  the existence of a non-negative matrix
$ ||\pi_{ij}^{k-1}(p,z)||_{i,j=1}^{k-1,r}$ follows satisfying conditions
\begin{eqnarray*}     \sum\limits _{i=1}^{k-1} \pi_{ij}^{k-1}(p,z)=1 , \quad (p,z) \in R, \quad  j=\overline{1,r},\end{eqnarray*}
\begin{eqnarray*}    f_i(p,z)= \sum\limits _{j=1}^r \pi_{ij}^{k-1}(p,z)g_j(p,z),  \quad (p,z) \in R, \quad  i=\overline{1,k-2},\end{eqnarray*}
\begin{eqnarray*}     \sum\limits _{i=k-1}^tf_i(p, z)= \sum\limits _{j=1}^r \pi_{k-1,j}^{k-1}(p,z)g_j(p,z),  \quad (p,z) \in R.\end{eqnarray*}
Introduce the function
\begin{eqnarray*}   u_{k-1}(p,z)=\frac{f_{k-1}(p,z)}{\sum\limits _{i=k-1}^tf_i(p, z)}, \quad (p,z) \in R.\end{eqnarray*}
The function $u_{k-1}(p,z)$ is defined  on the set $R,$  it is  non-negative and  such  that
\begin{eqnarray*}   0 \leq u_{k-1}(p,z)\leq 1.\end{eqnarray*}
Put
\begin{eqnarray*}   \pi_{i,j}^{k}(p,z)=\pi_{i,j}^{k-1}(p,z),\quad (p,z) \in R, \quad i=\overline{1,k-2}, \quad j=\overline{1,r}, \end{eqnarray*}
\begin{eqnarray*}   \pi_{k-1,j}^{k}(p,z)= u_{k-1}(p,z)\pi_{k-1,j}^{k-1}(p,z),\quad (p,z) \in R,  \quad  j=\overline{1,r},  \end{eqnarray*}
\begin{eqnarray*}    \pi_{k,j}^{k}(p,z)= (1 -u_{k-1}(p,z))\pi_{k-1,j}^{k-1}(p,z), \quad (p,z) \in R,  \quad  j=\overline{1,r}.\end{eqnarray*}
Such constructed  matrix $||\pi_{i,j}^{k}(p,z)||_{i,j=1}^{k,r}$ satisfies the  conditions
\begin{eqnarray*}     \sum\limits _{i=1}^{k} \pi_{ij}^{k}(p,z)=1, \quad  j=\overline{1,r}, \quad (p,z) \in R, \end{eqnarray*}
\begin{eqnarray*}    f_i(p,z)= \sum\limits _{j=1}^r \pi_{ij}^{k}(p,z)g_j(p,z),  \quad (p,z) \in R, \quad  i=\overline{1,k-1},\end{eqnarray*}
\begin{eqnarray*}     \sum\limits _{i=k}^tf_i(p, z)= \sum\limits _{j=1}^r \pi_{k,j}^{k}(p,z)g_j(p,z),  \quad (p,z) \in R.\end{eqnarray*}
Continuing this process within the same scheme, we obtain the needed result.
\qed
\end{proof}
\begin{note} If one assumes  the continuity of functions
 $f_i(p,z), \ i=\overline{1,t}, \  g_j(p,z), \ j=\overline{1,r}, $
on the set $R,$ in the Lemma \ref{let1}
then the matrix $ ||\pi_{ij}(p,z)||_{i,j=1}^{t,r} $ is also continuous  on the set $R.$
\end{note}
The Lemma \ref{let1} holds under less strict assumptions.
\begin{lemma}\label{allalet1}
Let $f_i(p,z), \ i=\overline{1,t}, $ be a set of non-negative functions given on the set $R \subseteq K_+^n\times \Gamma^m$ satisfying the condition
\begin{eqnarray*}     \sum\limits _{i=1}^tf_i(p, z) > 0, \quad (p,z) \in R, \end{eqnarray*}
and let  a set of functions $g_j(p,z), \ j=\overline{1,r}, $ given on the set  $R$ satisfy the condition
\begin{eqnarray*}     \sum\limits _{i=1}^tf_i(p, z)= \sum\limits _{j=1}^rg_j(p, z), \quad (p,z) \in R.\end{eqnarray*}
Then  a non-negative matrix $ ||\pi_{ij}(p,z)||_{i,j=1}^{t,r}$ exists such  that
\begin{eqnarray*}     \sum\limits _{i=1}^t \pi_{ij}(p,z)=1 , \quad (p,z) \in R,  \quad  j=\overline{1,r}, \end{eqnarray*}
\begin{eqnarray*}    f_i(p,z)= \sum\limits _{j=1}^r \pi_{ij}(p,z)g_j(p,z),  \quad (p,z) \in R, \quad  i=\overline{1,t}.\end{eqnarray*}
Matrix elements of the matrix $ ||\pi_{ij}(p,z)||_{i,j=1}^{t,r}$  have the form
\begin{eqnarray*}   \pi_{ij}(p,z)= \frac{f_i(p, z)}{\sum\limits _{j=1}^rg_j(p, z)}, \quad i=\overline{1,t}, \quad j=\overline{1,r}.\end{eqnarray*}
\end{lemma}
Consider a set of vectors  $N_i\left(z,c^0\right)=\left[y_i -x_i\right] +d_i +b_i, \ i=\overline{1,m},$ where $z^i=\{x_i,y_i\} \in \Gamma_i$ is a productive process of the  $i$-th firm, $d_i$ is a vector of subvention\index{vector of subvention}  of the  $i$-th firm, and $b_i$ is a  vector of goods supply\index{ vector of goods supply}  of the $i$-th firm. Denote, by $c^0=\{d_i+b_i, i=\overline{1,m}\},$ the set of sum of vectors of subvention and goods supply  for every firm, respectively. We call the matrix
$N\left(z,c^0\right)=||N_{ki}\left(z,c^0\right)||_{k=1,i=1}^{n,m},$ where
\begin{eqnarray*}   N_{ki}\left(z,c^0\right)=y_{ki}- x_{ki} + d_{ki}+b_{ki},\quad z=\{z^i\}_{i=1}^m, \quad z^i=\{x_i, y_i\},\end{eqnarray*}
the matrix generated by the  set of vectors
\begin{eqnarray*}   N_i\left(z,c^0\right)=\left[y_i -x_i\right] +d_i +b_i, \quad i=\overline{1,m},\end{eqnarray*}
\begin{eqnarray*}   x_i=\{x_{ki}\}_{k=1}^n, \quad y_i=\{y_{ki}\}_{k=1}^n, \quad  d_i=\{d_{ki}\}_{k=1}^n, \quad b_i=\{b_{ki}\}_{k=1}^n.\end{eqnarray*}
In what follows, we denote by $N\left(z^0,c^0\right)^T$ the matrix transposed to the matrix $N\left(z^0,c^0\right).$
In the next Definition, we suppose $ m \leq n.$
\begin{definition}\label{oz1}
Let the structure of firms production  in the economy system be described by  technological maps $F_i(x),\  x \in X_i^1, \ i=\overline{1,m},$ from the  CTM class in a wide sense, and let  $z^0 \in T \subseteq  \Gamma^m$ be a certain  set of productive processes of $m$ firms the economy system contains. We call the set of vectors  \begin{eqnarray*}   N_i\left(z^0,c^0\right)=\left[y_i^0 -x_i^0\right] +d_i +b_i, \quad i=\overline{1,m},\end{eqnarray*}
constructed by the set of productive processes of $m$ firms $z^0=\left\{z_0^i\right\}_{i=1}^m,$  \ $ z_0^i=\left(x_i^0, y_i^0\right),$ a certain  set of vectors of  subvention and initial goods supply   $c^0=\{d_i+b_i\}_{i=1}^m,$ where $d_i$ is a  vector of  subvention of the $i$-th firm and $b_i$ is a  vector of  goods supply of the $i$-th firm,
economically compatible\index{economically compatible productive process} if there exists a non-negative matrix $D\left(z^0, c^0\right)=\left|\left|D_{ki}\left(z^0, c^0\right)\right|\right|_{k=1,i=1}^{n,m}$ with non-zero rows such that there exists the inverse matrix
$\left[N\left(z^0,c^0\right)^T D\left(z^0, c^0\right)\right]^{-1}$
with strictly positive matrix elements, and the inequality
\begin{eqnarray*}    \sum\limits_{i=1}^m\left[y_i^0 - x_i^0\right] + \sum\limits_{i=1}^m d_i > a \end{eqnarray*}
holds,  where $ a= \{a_i\}_{i=1}^n $ is a certain  vector with strictly positive components.

We call the set of just defined productive processes $z^0$ of $m$ firms economically compatible  with the set of vectors  subvention and initial goods supply  $c^0.$
\end{definition}
In the Definition \ref{ozn1}, we do not restrict the number $m$  of firms.
\begin{definition}\label{ozn1}
Let the structure of firms production  in the economy system be described by technological maps $F_i(x),\  x \in X_i^1, \ i=\overline{1,m},$ from the CTM class in a wide sense, and let  $z^0 \in T \subseteq  \Gamma^m$ be a certain  set of productive processes of $m$ firms the economy system contains. We call a set of vectors   \begin{eqnarray*}   N_i\left(z^0,c^0\right)=\left[y_i^0 -x_i^0\right] +d_i +b_i, \quad i=\overline{1,m},\end{eqnarray*}    constructed by the set of productive  processes of $m$ firms $z^0,$ a certain set of vectors of  subvention and initial goods supply   $c^0=\{d_i+b_i\}_{i=1}^m,$
economically compatible in a generalized sense\index{economically compatible  productive  processes in a generalized sense} if there exists a subset  of  vectors
\begin{eqnarray*}    N_i\left(z^0,c^0\right)=\left[y_i^0 -x_i^0\right] +d_i +b_i, \quad  i=\overline{1,m_0},\quad m_0 \leq n, \end{eqnarray*}    such \hfill that \hfill for \hfill the \hfill set \hfill of \hfill vectors \hfill of \hfill  subvention \hfill and  \hfill initial \hfill goods \hfill supply \hfill \\  $c^1=\{d_i+b_i\}_{i=1}^{m_0},$ where $d_i$ is a subvention vector of the $i$-th firm and $b_i$ is a goods supply vector of the $i$-th firm,
it is economically compatible  in the sense of the Definition \ref{oz1}, and the rest set of vectors
 \begin{eqnarray*}   N_i\left(z^0,c^0\right)=\left[y_i^0 -x_i^0\right] +d_i +b_i, \quad  i=\overline{m_0+1, m},\end{eqnarray*}
 constructed by the set of productive processes of the rest $m - m_0$ firms, a certain set of vectors of subvention and vectors of initial goods supply  $c^2=\{d_i+b_i\}_{i=m_0+1}^m,$ where $d_i$ is a vector of subvention  of the  $i$-th firm and $b_i$ is a vector of goods supply  of the $i$-th firm,
satisfies  the condition: there exists a non-negative matrix $L=||L_{ki}\left(z^0,c^0\right) ||_{k=m_0+1, i=1}^{m, m_0}$ such that
\begin{eqnarray*}    N_k\left(z^0,c^0\right) =\sum\limits_{i=1}^{m_0}L_{ki}\left(z^0,c^0\right)N_i\left(z^0,c^0\right), \quad k=\overline{m_0+1, m}.\end{eqnarray*}

We call the set of just defined productive processes $z^0$ of $m$ firms economically compatible in a generalized sense with a set of  vectors  of subvention and initial goods supply  $c^0=\{d_i+b_i\}_{i=1}^m.$
\end{definition}

We denote, by $M_0,$ the set of all productive processes $z^0$ of $m$ firms being  economically compatible (in a generalized sense) with a certain set of vectors  of subvention and initial goods supply  $c^0.$ It is obvious that $M_0 \subseteq T.$
\begin{lemma}
If a set of productive processes $z^0$ of $m$ firms is economically compatible with a   set of vectors of  subvention and initial goods supply   $c^0,$ then the set
\begin{eqnarray*}   T_0\left(z^0,c^0\right)=\left\{p=\{p_i\}_{i=1}^n \in K_+^n, \  a_i\left(p,z^0\right)=1, \ i=\overline{1,m}\right\},\end{eqnarray*}
is non-empty and  contains the set
\begin{eqnarray*}   T_1\left(z^0,c^0\right)
\end{eqnarray*}
\begin{eqnarray*} 
=\left\{p \in R_+^n, \  p=D\left(z^0, c^0\right)C\left(z^0,c^0\right)\delta, \  \delta =\{\delta_i\}_{i=1}^m \in R_+^m \setminus \{0, \ldots, 0\} \right\},\end{eqnarray*}
where we introduced the notation $C\left(z^0,c^0\right)=\left[N\left(z^0,c^0\right)^T D\left(z^0, c^0\right)\right]^{-1}.$
\end{lemma}
\begin{proof}\smartqed
Let the vector $p \in T_1\left(z^0,c^0\right).$ Prove that  it  belongs to the set $T_0\left(z^0,c^0\right).$
For this it is sufficient to prove that
$u_i\left(p, z^0\right)  \geq 0, \ i=\overline{1,m}.$ However, at the vector $p=D\left(z^0, c^0\right)C\left(z^0,c^0\right)\delta,$ the function $u_i\left(p, z^0\right)=\delta_i,$ ~$i=\overline{1,m}.$ As $ \delta =\{\delta_i\}_{i=1}^m \in R_+^m \setminus \{0, \ldots, 0\},$ we have $u_i\left(p, z^0\right) \geq 0.$ From here it follows that $a_i\left(p,z^0\right)=1, \ i=\overline{1,m}.$
Also, it is obvious that the vector $ p=D\left(z^0, c^0\right)C\left(z^0,c^0\right)\delta$ has strictly positive components for any $\delta =\{\delta_i\}_{i=1}^m \in R_+^m \setminus \{0, \ldots, 0\} \},$ therefore $p \in K_+^n.$
\qed
\end{proof}
\begin{note}
Sets $T_0\left(z^0,c^0\right)$ and $T_1\left(z^0,c^0\right)$ are the same if $m=n.$
\end{note}
Really, one must prove in this case only the next imbedding
\begin{eqnarray*}   T_0\left(z^0,c^0\right) \subseteq T_1\left(z^0,c^0\right).\end{eqnarray*}
Let $p \in  T_0\left(z^0,c^0\right).$ Then $u_i\left(p, z^0\right)  \geq 0, \ i=\overline{1,m}.$ Denote
\begin{eqnarray*}    u_i\left(p, z^0\right)=\delta_i^0, \quad  i=\overline{1,m}.\end{eqnarray*}    It is obvious that the vector $\delta^0 =\{\delta_i^0\}_{i=1}^m$ does not equal zero because the matrix $ N\left(z^0,c^0\right) $ is non-degenerate. From the same consideration of non-degeneracy of the matrix $ N\left(z^0,c^0\right), $ it follows that the vector $p$ has the representation
  \begin{eqnarray*}    p=D\left(z^0, c^0\right)C\left(z^0,c^0\right)\delta^0.\end{eqnarray*}

\begin{lemma}\label{stone1}
If a set of productive processes $z^0$ of $m$ firms is economically compatible  in a generalized sense with a set of vectors  subvention and initial goods supply  $c^0,$ then the set
\begin{eqnarray*}   T_0\left(z^0,c^0\right)=\left\{p=\{p_i\}_{i=1}^n \in K_+^n, \  a_i\left(p,z^0\right)=1, \ i=\overline{1,m}\right\}
\end{eqnarray*}
\begin{eqnarray*}   =\left\{p \in K_+^n, \  a_i\left(p,z^0\right)=1, \ i=\overline{1,m_0}\right\}\end{eqnarray*}
is non-empty and contains the set
\begin{eqnarray*}   T_1\left(z^0,c^0\right)
\end{eqnarray*}
\begin{eqnarray*} 
=\left\{p \in R_+^n, \  p=D\left(z^0, c^0\right)C\left(z^0,c^0\right)\delta, \  \delta =\{\delta_i\}_{i=1}^{m_0} \in R_+^{m_0} \setminus \{0, \ldots, 0\} \right\},
\end{eqnarray*}
where we introduced the notation $C\left(z^0,c^0\right)=\left[N\left(z^0,c^0\right)^T D\left(z^0, c^0\right)\right]^{-1}.$
\end{lemma}
Note that, in the Lemma \ref{stone1}, we build the matrix $ N\left(z^0,c^0\right)$ only after the  set of vectors
$N_i\left(z^0,c^0\right), \ i=\overline{1, m_0},$ therefore, the matrix $D\left(z^0, c^0\right)$ is non-negative  of the dimension $n \times m_0.$

 Let a set of productive processes of $m$ firms $z^0 \in T \subseteq  \Gamma^m$ be economically compatible  with a set of vectors of subvention  and initial goods supply  $c^0.$ Let the matrix $D\left(z^0, c^0\right)$ guarantee this economic compatibility.
We denote, by $\Gamma_0^m\left(z^0, c^0\right),$ a subset of productive processes of $m$ firms $z \in T \subseteq  \Gamma^m$ for which the set of  vectors  $N_i\left(z,c^0\right)=\left[y_i -x_i\right] +d_i +b_i, \ i=\overline{1,m},$ is economically compatible  with the set of vectors of  subvention  and initial goods supply   $c^0$ and the same matrix $D\left(z^0, c^0\right)$ and we call it the set of economically compatible productive processes of $m$ firms corresponding to the set of vectors of subvention  and initial goods supply   $c^0.$ It is obvious that if the point $z^0$ is an internal point of the set $T \subseteq  \Gamma^m,$ then
$\Gamma_0^m\left(z^0, c^0\right)$ is an open set, if $z^0$ is a limit point of the set $T \subseteq \Gamma^m,$ then we put by the definition
$\Gamma_0^m\left(z^0, c^0\right)=z^0.$ Build the sets
\begin{eqnarray*}   R=\bigcup\limits_{z^0 \in A_0 \cup A_1}T_0\left(z^0,c^0\right)\times z^0,\end{eqnarray*}
\begin{eqnarray*}   R_2=\bigcup\limits_{z^0 \in A_0 \cup A_1}T_1\left(z^0,c^0\right)\times z^0,\end{eqnarray*}
where $A_0$ is the set of internal points of economically compatible sets of productive processes of $m$ firms belonging to the set $T$ and $A_1$ is not more than a countable set of limit points of economically compatible sets of productive processes of $m$ firms belonging to the set $T.$
Then $R$ and $R_2$ are Borel subsets from ${\cal B}(K_+^n)\times{\cal B}(\Gamma^m).$
Introduce the Borel set
\begin{eqnarray*}   R_1 = \left\{( p,z) \in K_+^n\times \Gamma^m, \ u_i(p,z) \geq 0, \ i=\overline{1,m},\  \sum\limits_{i=1}^m\left[y_i - x_i\right] + \sum\limits_{i=1}^m d_i > a \right\}, \end{eqnarray*}
where $a$ is a certain strictly positive vector.
The next inclusions  hold:
\begin{eqnarray*}   R_1 \supseteq R \supseteq R_2.\end{eqnarray*}

The subsequent Definition is fundamental  to all what follows.
\begin{definition}\label{oz2} Let the economy system be such as in the Theorem \ref{11tl4}.
In the economy system,  the  structure of firms supply agrees with the  structure of consumers choice\index{structure of firms supply agrees with the  structure of consumers choice} on the set $B \subseteq K_+^n\times \Gamma^m $ if there exist $l$ fields
\begin{eqnarray*}   y_i(p,z)> t_0 >0, \quad (p,z) \in B,  \quad i=\overline{1,l}, \end{eqnarray*}    where a  number $t_0$ is independent of
$ (p,z) \in B,$ such that the equalities
\begin{eqnarray*}   \sum\limits_{i=1}^m\left[y_{ki} - x_{ki} + d_{ki}\right]+\sum\limits_{i=1}^lb_{ki}
\end{eqnarray*}
\begin{eqnarray}   \label{2dile1}
=\sum\limits_{i=1}^l\eta_{ik}^0(p,z) y_i(p,z),  \quad (p,z) \in B, \quad  k=\overline{1,n},
\end{eqnarray}
hold.
Here $b_i=\{b_{ki} \}_{k=1}^n, \ i=\overline{1,l},$ is a property vector of the $i$-th consumer at the beginning of the economy operation period.
\end{definition}

\begin{note}
If the structure of firms supply  agrees with the structure of consumers choice  in the  economy system,  then, under the  productive economic  process $Q(p,z)$  having the form (\ref{001l0}), the equalities
\begin{eqnarray*}   \sum\limits_{i=1}^m\left[Y_{ki}(p,z) - X_{ki}(p,z)+ d_{ki} \right]+\sum\limits_{i=1}^lb_{ki}
\end{eqnarray*}
\begin{eqnarray}   \label{2ole1}
=\sum\limits_{i=1}^l\eta_{ik}(p,z) y_i(p, Q(p,z)), \quad
(p,z) \in  R_1, \quad  k=\overline{1,n}, 
\end{eqnarray}
 hold, where
\begin{eqnarray*}    Q(p,z)=\{X_i(p,z),Y_i(p,z)\}_{i=1}^m,\end{eqnarray*}
\begin{eqnarray*}   X_i(p,z)=\{X_{ki}(p,z) \}_{k=1}^n, \quad  Y_i(p,z)=\{Y_{ki}(p,z) \}_{k=1}^n.\end{eqnarray*}
\end{note}

The latter holds because if the equalities (\ref{2dile1}) hold, then, obviously, the equalities (\ref{2ole1}) holds too as
\begin{eqnarray*}    X_i(p,z)=x_i, \quad  Y_i(p,z)=y_i,  \quad (p,z) \in R_1, \quad  i=\overline{1,m}.\end{eqnarray*}

\subsection{Taxation systems construction}

Consider, on the set $R_0$ being one of sets $R_1, R,$ or $ R_2,$ the family  of functions
\begin{eqnarray*}   f_i(p,z)= \pi_i(p,z )\left[\left\langle p, y_i - x_i + d_i \right\rangle  + \left\langle b_i, p \right\rangle\right], \quad i=\overline{1,m},\end{eqnarray*}
\begin{eqnarray}   \label{rev5}
f_i(p,z)=y_i(p,z)\sum\limits_{k=1}^n\eta_{ik}^0(p, z)p_k ,  \quad i=\overline{m+1,l},
\end{eqnarray}
and the family of functions
\begin{eqnarray*}   g_i(p,z)= \left\langle p, y_i - x_i + d_i\right\rangle  + \left\langle b_i, p\right\rangle , \quad i=\overline{1,m},\end{eqnarray*}
\begin{eqnarray}   \label{rev4}
 g_i(p,z)= \left\langle p, b_i \right\rangle,  \quad i=\overline{m+1,l},
\end{eqnarray}
where $0 < \pi_i(p,z), \ i=\overline{1,m}, $ and that  are continuous functions of $(p,z) \in R_1.$
The set of functions  $g_i(p,z), \ f_i(p,z), \ i=\overline{1,l},$ on the set $R_0 $
satisfy  the conditions of the Lemma \ref{let1}  about the existence of taxation system  if there hold conditions:
\begin{eqnarray*}     \sum\limits _{i=1}^lf_i(p, z)= \sum\limits _{j=1}^lg_j(p, z), \quad (p,z) \in R_0,\end{eqnarray*}
there exist such numbers $k$ and $s$ that $f_{k}(p,z)+f_s(p,z) > 0$ on the set $R_0,$  or
\begin{eqnarray*}     \sum\limits _{i=1}^lf_i(p, z)> 0, \quad (p,z) \in R_0.\end{eqnarray*}

As  fields of information evaluation by  consumers  $ \eta_{i}^0(p, z),$ $ i=\overline{1,l},$
 satisfy the  conditions of the Theorem \ref{11tl4}, the inequalities follow
\begin{eqnarray}   \label{arevla1}
\sum\limits_{k=1}^n\eta_{ik}^0(p, z)p_k > 0, \quad (p,z) \in K_+^n\times \Gamma^m ,  \quad  i=\overline{1,l}.
\end{eqnarray}
The last equality takes the form
\begin{eqnarray*}    \sum\limits _{i=1}^m\left\langle p, y_i - x_i + d_i\right\rangle  +\sum\limits _{i=1}^l\left\langle b_i, p\right\rangle
\end{eqnarray*}
\begin{eqnarray*}    =\sum\limits _{i=1}^m\pi_i(p,z)\left\langle p, y_i - x_i + d_i +b_i\right\rangle  +\sum\limits _{i=m+1}^ly_i(p,z)\sum\limits_{k=1}^n\eta_{ik}^0(p, z)p_k.\end{eqnarray*}

Now assume  the structure of firms supply  agrees with  the  structure of consumers choice on the set $R_1.$
Then, accounting for (\ref{2dile1}), this equality takes the form
\begin{eqnarray*}    \sum\limits _{i=1}^m\pi_i(p,z)\left[\left\langle p, y_i - x_i + d_i \right\rangle + \left\langle b_i, p \right\rangle\right]\end{eqnarray*}
\begin{eqnarray}   \label{opod1}
=\sum\limits _{i=1}^my_i(p,z)\sum\limits_{k=1}^n\eta_{ik}^0(p, z)p_k.
\end{eqnarray}

Let us give the sufficient conditions for the existence of a solution  to the equation (\ref{opod1}) for the vector $\pi(p,z)=\{\pi_i(p,z)\}_{i=1}^m.$

\begin{lemma}\label{opod11}
Let, in the  economy system, the structure of firms  supply  agree with the structure of consumers choice  on the set $R_1,$ let fields $y_i(p,z), \ i=\overline{1,l},$ from the set of equalities (\ref{2dile1})
be continuous functions of $(p,z) \in R_1,$  and let  fields of information evaluation by  consumers $\eta_i^0(p,z), \ i=\overline{1,l}, $ be continuous functions of $(p,z) \in K_+^n\times \Gamma^m$ and satisfy conditions
\begin{eqnarray*}    \sum\limits_{i=1}^m\eta_{is}^0(p,z)> q>0, \quad (p,z) \in K_+^n\times \Gamma^m, \quad  s=\overline{1,n},  \end{eqnarray*}
where $q$ is a certain  positive number.
Then there exists the infinite number of solutions to the equation
(\ref{opod1}) for the vector $\pi(p,z)=\{\pi_i(p,z)\}_{i=1}^m$ whose every component $\pi_i(p,z),  \  i=\overline{1,m},$ is a continuous function of $(p,z) \in R_1$ such that \begin{eqnarray*}   \pi_i(p,z) > h >0, \quad  (p,z) \in R_1, \quad   i=\overline{1,m},  \end{eqnarray*}    where $h$ is a certain positive number.
\end{lemma}
\begin{proof}\smartqed
 The solution satisfying all the conditions of the Lemma is, e.g., the vector  $\pi(p,z)=\{\pi_i(p,z)\}_{i=1}^m,$ where
\begin{eqnarray*}   \pi_i(p,z)=\pi^0(p,z)=\frac{ \sum\limits_{i=1}^my_i(p,z)\sum\limits_{s=1}^n\eta_{is}^0(p,z)p_s}
{\sum\limits_{i=1}^m \left\langle p, y_i - x_i +d_i + b_i \right\rangle},\quad  i=\overline{1,m}.\end{eqnarray*}
The denominator in this formula nowhere vanishes because the lower bound holds
\begin{eqnarray*}   \sum\limits_{i=1}^m \left\langle p, y_i - x_i +d_i + b_i \right\rangle \geq \left\langle p,a \right\rangle  >  0, \end{eqnarray*}
where $p$ and $a$ are strictly positive vectors. The numerator and the denominator are continuous functions of $(p,z) \in R_1.$

Establish the necessary lower bound.
The set  of vectors
\begin{eqnarray*}   N_i\left(z,c^0\right)=\left[y_i -x_i\right] +d_i +b_i, \quad  i=\overline{1,m},\end{eqnarray*}
satisfies  the condition
\begin{eqnarray*}   0 < a < \sum\limits_{i=1}^m\left[y_i - x_i\right] + \sum\limits_{i=1}^m (d_i+b_i) \leq w, \quad w=\{w_i\}_{i=1}^n, \quad w_i < \infty, \quad i=\overline{1,n},\end{eqnarray*}
where the vector $w=\{w_i\}_{i=1}^n $ has strictly positive components and does not depend on $z \in \Gamma^m$ because all technological maps belong to the CTM class in a wide sense. Therefore, the upper bound holds:
\begin{eqnarray*}    \sum\limits_{i=1}^m \left\langle p, y_i - x_i +d_i + b_i \right\rangle \leq \max\limits_{1 \leq i \leq n}w_i \sum\limits_{i=1}^np_i,\quad (p,z) \in R_1.\end{eqnarray*}
Further,
\begin{eqnarray*}   \sum\limits_{i=1}^my_i(p,z)\sum\limits_{s=1}^n\eta_{is}^0(p,z)p_s \geq  q\min\limits_{1 \leq i \leq n}y_i(p,z) \sum\limits_{i=1}^np_i,\quad (p,z) \in R_1.\end{eqnarray*}
In view of assumptions valid in the case, we have
\begin{eqnarray*}    \frac{ \sum\limits_{i=1}^my_i(p,z)\sum\limits_{s=1}^n\eta_{is}^0(p,z)p_s}
{\sum\limits_{i=1}^m \left\langle p, y_i - x_i +d_i + b_i \right\rangle} \geq \frac{q\min\limits_{1 \leq i \leq n}y_i(p,z) }{\max\limits_{1 \leq i \leq n}w_i}.\end{eqnarray*}
However, there exists  the constant $t_0 > 0 $ such  that
\begin{eqnarray*}   \min\limits_{1 \leq i \leq m}y_i(p,z) \geq t_0 > 0, \quad (p,z) \in R_1.\end{eqnarray*}
The latter means that
\begin{eqnarray*}    \frac{ \sum\limits_{i=1}^my_i(p,z)\sum\limits_{s=1}^n\eta_{is}^0(p,z)p_s}
{\sum\limits_{i=1}^m \left\langle p, y_i - x_i +d_i + b_i \right\rangle} \geq \frac{t_0q }{\max\limits_{1 \leq i \leq n}w_i}=h > 0, \quad  (p,z) \in R_1.\end{eqnarray*}
Therefore, $\pi_i(p,z)=\pi^0(p,z), \  i=\overline{1,m}, $ is a continuous function of $ (p,z) \in R_1 $ and the vector $\pi(p,z)=\{\pi_i(p,z)\}_{i=1}^m$ is a continuous solution to the equation
(\ref{opod1}).
As a rule, within economy systems, the condition
\begin{eqnarray}   \label{opodd1}
\sum\limits _{i=m+1}^l\left\langle b_i, p\right\rangle \leq \sum\limits _{i=m+1}^ly_i(p,z)\sum\limits_{k=1}^n\eta_{ik}^0(p, z)p_k, \quad  (p,z) \in R_1,
\end{eqnarray}
 holds meaning that a part of the product produced in the economy  is redistributed by taxing to guarantee the economy system operation.

In this case, for the built solution the inequality holds
 $\pi^0(p,z) \leq 1$
as a result of the condition (\ref{opodd1}).

Let us prove the existence of the infinity number of solutions to the equation
(\ref{opod1}) for the vector $\pi(p,z)=\{\pi_i(p,z)\}_{i=1}^m$ whose every component $\pi_i(p,z),  \  i=\overline{1,m},$ is a continuous function of $(p,z) \in R_1$ such that $\pi_i(p,z) > h >0, \  i=\overline{1,m},$ where $h$ is a certain  positive number.

 Take, for the solution to the equation (\ref{opod1}), the vector $\pi(p,z)=\{\pi_i(p,z)\}_{i=1}^m,$ where
\begin{eqnarray*}    \pi_i(p,z)=\pi_0(p,z)\pi_i(p), \quad  0< \pi_i(p) < \infty, \quad  i=\overline{1,m},\end{eqnarray*}

\begin{eqnarray}   \label{opod22}
\pi_0(p,z)=\frac{ \sum\limits_{i=1}^my_i(p,z)\sum\limits_{s=1}^m\eta_{is}^0(p,z)p_s}
{\sum\limits_{i=1}^m \pi_i(p)\left\langle p, y_i - x_i +d_i + b_i \right\rangle},\quad  i=\overline{1,m}.
\end{eqnarray}
Suppose that $\pi_i(p),\  i=\overline{1,m},$ are continuous bounded functions of $p \in K_+^n,$
$\pi_i=\sup\limits_{p \in K_+^n}\pi_i(p) < \infty,\  i=\overline{1,m},$ $\pi_i^0=\inf\limits_{p \in K_+^n}\pi_i(p) >\pi_0 > 0,\  i=\overline{1,m}.$
As
\begin{eqnarray*}   \sum\limits_{i=1}^m \pi_i(p)\left\langle p, y_i - x_i +d_i + b_i \right\rangle \geq   \pi_0 \left\langle  p,a \right\rangle,\quad  (p,z) \in R_1,\end{eqnarray*}
\begin{eqnarray*}   \frac{ \sum\limits_{i=1}^my_i(p,z)\sum\limits_{s=1}^m\eta_{is}^0(p,z)p_s}
{\sum\limits_{i=1}^m \pi_i(p)\left\langle p, y_i - x_i +d_i + b_i \right\rangle} \geq \end{eqnarray*}    \begin{eqnarray*}   \geq \frac{t_0q }{\max\limits_{1 \leq i \leq n}\pi_i \max\limits_{1 \leq j \leq n}w_j}=h > 0, \quad (p,z) \in R_1, \quad  i=\overline{1,m},\end{eqnarray*}
$ \pi_i(p,z)$ is a continuous function of $(p, z) \in R_1.$
The last inequalities hold because  the inequalities
\begin{eqnarray*}   \left\langle p, y_i - x_i +d_i + b_i \right\rangle  \geq 0, \quad i=\overline{1,m},\end{eqnarray*}    hold on the set $R_1.$

\begin{note}\label{zau1}
The solution just built does not satisfy, generally speaking, the conditions
$\pi_i(p,z) \leq 1, \ i=\overline{1,m}.$ However, such a solution can be useful to be applied because, under certain conditions, the need can appear to redistribute incomes of some firms for the use of other ones in view of the need to  modernize supported firms production.
If it is not the case, additional conditions occur to put $\pi_i(p),\ i=\overline{1,m},$
\begin{eqnarray*}   \frac{\pi_i(p)\sum\limits_{i=1}^my_i(p,z)\sum\limits_{s=1}^n\eta_{is}^0(p,z)p_s}{
\sum\limits_{i=1}^m \pi_i(p)\left\langle p, y_i - x_i +d_i + b_i \right\rangle} \leq 1, \quad \ (p,z) \in R_1, \quad  i=\overline{1,m}.\end{eqnarray*}
\end{note}
\qed
\end{proof}

Let us formulate the Lemma guaranteeing the existence of a continuous solution to the equation (\ref{opod1}) for the vector $\pi(p,z)=\{\pi_i(p,z)\}_{i=1}^m$ under significantly less strict assumptions. In this case, one can not guarantee uniform lower bounds for the solution. Just such conditions guaranteeing the existence of a continuous solution  are acceptable for applications.
\begin{lemma}\label{opcd1}
Let, in  the  economy system, the structure of firms supply  agree with the  structure of consumers choice  on the set $R_1,$ let fields $y_i(p,z), \ i=\overline{1,m},$ from the set of equalities (\ref{2dile1}) be continuous functions of
$(p,z) \in R_1,$ and let fields of  information evaluation by  consumers $\eta_i^0(p,z), \ i=\overline{1,l}, $ satisfy the  conditions of the Theorem \ref{11tl4}, i.e., be continuous functions of $ (p,z) \in K_+^n \times \Gamma^m $ satisfying the condition
\begin{eqnarray*}      \sum\limits_{i=1}^m\sum\limits_{s=1}^n\eta_{is}^0(p,z)p_s  >0,  \quad (p,z) \in K_+^n\times \Gamma^m, \end{eqnarray*}
then there exists the infinite number of solutions to the equation (\ref{opod1}) for the vector $\pi(p,z)=\{\pi_i(p,z)\}_{i=1}^m,$ whose every component $\pi_i(p,z),  \  i=\overline{1,m},$ is a continuous function of $(p,z) \in R_1$ such that $\pi_i(p,z) > 0, \  i=\overline{1,m}, $ $ (p,z) \in R_1. $
\end{lemma}
The Proof of this Lemma is a very simplified repetition of the Proof of the previous Lemma because one must prove that the marked solution is continuous and, for every $(p,z) \in R_1,$ every component of the  solution is strictly positive. The mentioned follows from the conditions formulated in the Lemma. The previous Lemma is useful because it gives a lower bound for a part of the income to remain in producer's property.

\begin{lemma}\label{opod111}
Let, in the economy system, the  structure of firms supply  agree with the structure of consumers choice  on the set $R_1,$ let fields $y_i(p,z), \ i=\overline{1,m},$ from the set of equalities (\ref{2dile1}) be
continuous functions of $(p,z) \in R_1,$ and let  fields of information evaluation by  consumers $\eta_i^0(p,z), \ i=\overline{1,l}, $ are continuous functions of $ (p,z) \in K_+^n \times \Gamma^m $
satisfying  the condition
\begin{eqnarray*}     \sum\limits_{i=1}^m \sum\limits_{s=1}^n\eta_{is}^0(p,z)> q>0,\quad  (p,z) \in  K_+^n\times \Gamma^m, \end{eqnarray*}
where $q$ is a certain positive number, then there exists a solution to the equation
(\ref{opod1}) for the vector $\pi(p,z)=\{\pi_i(p,z)\}_{i=1}^m$ whose every component $\pi_i(p,z),  \  i=\overline{1,m},$ is a continuous function of $p $ on the set
\begin{eqnarray*}   T_1\left(z,c^0\right)=\left\{p \in K_+^n, \  p=D\left(z,c^0\right)C\left(z,c^0\right)\delta, \ \delta \in R_+^m\setminus \{0,\ldots,0\}\right\}\end{eqnarray*}    for every fixed $z  \in M_0$ such that $\pi_i(p,z) > h >0, \  i=\overline{1,m}.$
\end{lemma}
\begin{proof}\smartqed
Find a solution to the equation (\ref{opod1}) on the set $T_1\left(z,c^0\right).$
If $z \in M_0,$ then, in view of the economic compatibility of vectors
\begin{eqnarray*}   N_i\left(z,c^0\right)=\left[y_i -x_i\right] +d_i +b_i, \quad  i=\overline{1,m},\end{eqnarray*}
a  matrix $D\left(z, c^0\right)=||D_{ki}\left(z, c^0\right)||_{k,i=1}^{n,m}$ exists with non-negative elements and non-zero rows such that the inverse matrix $\left[N\left(z,c^0\right)^TD\left(z, c^0\right)\right]^{-1}$ exists with strictly positive elements. Denote
\begin{eqnarray*}   \left[N\left(z,c^0\right)^TD\left(z, c^0\right)\right]^{-1}=||C_{ki}\left(z,c^0\right) ||_{k,i=1}^m.\end{eqnarray*}
On the set $T_1\left(z,c^0\right),$ this equation turns into the equation
\begin{eqnarray}   \label{opod14}
 \sum\limits_{j=1}^m \pi_j(p,z)\delta_j=\sum\limits_{j=1}^m \sum\limits_{i=1}^m  \sum\limits_{s=1}^ny_i(p,z)\eta_{is}^0(p,z)\sum\limits_{k=1}^mD_{sk}\left(z, c^0\right)C_{kj}\left(z,c^0\right)\delta_j.
\end{eqnarray}
Here we use the equality $\delta_j=\left\langle p, N_j\left(z,c^0\right)\right\rangle$
if  one takes $p=D\left(z, c^0\right)C\left(z, c^0\right)\delta$ in (\ref{opod1}).

The solution to the equation (\ref{opod14}) is the vector $ \pi(p,z)=\{  \pi_j(p,z)\}_{j=1}^n,$ where
\begin{eqnarray}\label{opod13}
\pi_j(p,z)=\sum\limits_{i=1}^m  \sum\limits_{s=1}^ny_i(p,z)\eta_{is}^0(p,z)\sum\limits_{k=1}^mD_{sk}\left(z, c^0\right)C_{kj}\left(z,c^0\right), \quad j=\overline{1,m}.
\end{eqnarray}
Matrix elements $\sum\limits_{k=1}^mD_{sk}\left(z, c^0\right)C_{kj}\left(z,c^0\right)$ are strictly positive. Denote
\begin{eqnarray*}   g_0(z)= \min\limits_{1\leq  k \leq m,1 \leq s \leq n } \sum\limits_{i=1}^m C_{ik}\left(z,c^0\right)D_{si}\left(z, c^0\right).\end{eqnarray*}
Therefore, the lower bound
\begin{eqnarray*}    \pi_k(p,z)\geq g_0(z) t_0 \sum\limits_{j=1}^m\sum\limits_{s=1}^n\eta_{js}^0(p,z) \geq g_0(z) t_0 q =h > 0, \quad k=\overline{1,m},\end{eqnarray*}
holds.
\qed
\end{proof}

\begin{theorem}\label{ptm1}
If the conditions of one of the  Lemmas \ref{opod11}, \ref{opcd1}, or \ref{opod111} hold,
then a taxation system
$||\pi_{ij}^0(p,z)||_{ i,j=1}^l$ exists on the set $R_0$ such  that
\begin{eqnarray*}   f_i(p,z)=\sum\limits _{j=1}^l\pi_{ij}^0(p,z)g_j(p,z), \quad (p,z) \in R_0.\end{eqnarray*}
\end{theorem}

Let $\pi_{ij}^1(p,z), \ i,j=\overline{1,l}, \ (p,z) \in K_+^n\times \Gamma^m \setminus R_0$ be a certain  taxation system on the set $K_+^n\times \Gamma^m \setminus R_0.$
Introduce the taxation system
\begin{eqnarray*}   \pi_{ij}(p,z)= \pi_{ij}^0(p,z)\chi_{R_0}(p,z)+ \pi_{ij}^1(p,z)\left[1 - \chi_{R_0}(p,z)\right],\end{eqnarray*}    \begin{eqnarray*}  (p,z) \in K_+^n\times \Gamma^m,  \quad    i,j=\overline{1,l}. \end{eqnarray*}
We give the  family of income pre-functions of consumers  by the formula
\begin{eqnarray*}   K_i^0(p,z)=\sum\limits _{j=1}^l\pi_{ij}(p,z)g_j(p,z), \quad (p,z) \in K_+^n\times \Gamma^m,   \quad  i=\overline{1,l}. \end{eqnarray*}
Within these conditions, the family of income functions  has the form
\begin{eqnarray}\label{001l1}
K_i(p,z)= K_i^0(p, Q(p,z))
\end{eqnarray}
 \begin{eqnarray*}   =\sum\limits _{j=1}^l\pi_{ij}(p,Q(p,z))g_j(p,Q(p,z)), \quad  (p,z) \in K_+^n\times \Gamma^m, \quad  i=\overline{1,l}, 
  \end{eqnarray*}
where $Q(p,z)$ is the productive economic process given by the formula (\ref{001l0}).

\subsection{Non-default operation of economy systems}

In what follows, we suppose the set of income functions from the condition of the Theorem \ref{11tl4}  is given by formulae  (\ref{001l1}).
Suppose that  $\zeta_0(p, \omega_0)=z^0, \ p \in K_+^n, $ with probability 1, where $ z^0   \in M_0.$
\begin{theorem}\label{2ch2l2}
 Let convex down technological maps $F_i(x),\ x \in X_i^1,$
$~i=\overline {1,m},$ belong to the CTM class in a wide sense,
the set $R$ be non-empty, the set of income functions be given by formulae (\ref{001l1}),
and let the structure of firms supply  agree with the structure of consumers choice  on the set $R_1,$
 fields $y_i(p,z),\ i=\overline{1,l}, $ be continuous functions of $(p,z) \in R_1,$ the conditions of the  Lemma \ref{opcd1} or of the  Lemma \ref{opod111}  hold.

If   fields of information evaluation by consumers $\eta_{i}^0(p,z),\  i=\overline{1,l},$ and a random field $\zeta_0(p,\omega_0)$ satisfy the  conditions of the Theorem \ref{11tl4}, the condition (\ref{arevla1}) holds, then the set of equations of the economy equilibrium
\begin{eqnarray*}   \sum\limits
_{i=1}^l \frac{\mu_{ki}(p)D_i(p)}{\sum\limits_{j=1}^n
\mu_{ji}(p)p_j}
\end{eqnarray*}
\begin{eqnarray}   \label{1g2p18}
=  \sum\limits _{i=1}^m\left[Y_{ki}\left(p, z^0\right)-X_{ki}\left(p, z^0\right)+ d_{ki}\right]+ \sum\limits
_{i=1}^l b_{ki}, \quad  k=\overline{1,n},
\end{eqnarray}
has a solution in the set of   strictly positive  price vectors $p^0=\{p_i^0\}_{i=1}^n$
for every $ z^0 \in
 M_0.$
Such an economy system can operate profitably in the Walras equilibrium state if $\left\langle y_i^0 -x_i^0, p^0 \right\rangle  > 0, \ i=\overline{1,m},$ or subvention-profitably if
\begin{eqnarray*}   \left\langle y_i^0 -x_i^0+ d_i , p^0 \right\rangle  > 0, \quad i=\overline{1,m},\end{eqnarray*}
for $b_i \neq 0.$
\end{theorem}
\begin{proof}\smartqed
The net income of the $i$-th consumer-producer, under the condition that  $\zeta_0(p, \omega_0)=z^0, \ z^0 \in M_0,$ with probability 1, takes the form
\begin{eqnarray*}   D_i(p)= K_i\left(p,z^0\right)= K_i^0\left(p,Q\left(p,z^0\right)\right)
\end{eqnarray*}
\begin{eqnarray*}   =\pi_i\left(p,Q\left(p,z^0\right)\right)\left[\left\langle p, y_i^0 - x_i^0\right\rangle a_i\left(p,z^0\right)\prod\limits_{i=1}^nv_i\left(p,z^0\right) +\left\langle d_i + b_i, p\right\rangle \right]
\end{eqnarray*}    \begin{eqnarray*}
=\pi_i\left(p,z^0\right)\left[\left\langle p, y_i^0 - x_i^0+ d_i\right\rangle +\left\langle b_i, p\right\rangle \right],\quad  \left(p,z^0\right)\in  R,
 \quad  i=\overline{1,m}.
\end{eqnarray*}
Introduce, as earlier, quantities
\begin{eqnarray*}    N_i\left(z^0, c^0\right)=\left[y_i^0 -x_i^0\right] +d_i +b_i, \quad C\left(z^0,c^0\right)=\left[N\left(z^0, c^0\right)^T D\left(z^0, c^0\right)\right]^{-1}.\end{eqnarray*}
Then, at the vector $p= D\left(z^0, c^0\right)C\left(z^0,c^0\right)\delta,$ where $\delta=\{\delta_i\}_{i=1}^m \in R_+^m,$
\begin{eqnarray*}   \tilde D_i(\delta)= \left\langle N_i\left(z^0, c^0\right), p \right\rangle = \left\langle N_i\left(z^0, c^0\right), D\left(z^0, c^0\right)C\left(z^0,c^0\right)\delta\right\rangle
\end{eqnarray*}
 \begin{eqnarray*}   =
\left[N\left(z^0, c^0\right)^T D\left(z^0, c^0\right)C\left(z^0,c^0\right)\delta\right]_i=\delta_i, \quad i=\overline{1,m},\end{eqnarray*}
\begin{eqnarray*}   N\left(z^0, c^0\right)^T D\left(z^0, c^0\right)=\left|\left|\sum\limits_{s=1}^nN_{si}\left(z^0, c^0\right)D_{sk}\left(z^0, c^0\right)\right|\right|_{i,k=1}^m,\end{eqnarray*}
\begin{eqnarray*}    N\left(z^0, c^0\right)=\left|\left|N_{si}\left(z^0, c^0\right)\right|\right|_{s=1,i=1}^{n,m}, \quad
D\left(z^0, c^0\right)=\left|\left|D_{sk}\left(z^0, c^0\right)\right|\right|_{s,k=1}^{n,m},
\end{eqnarray*}
\begin{eqnarray*}    C\left(z^0, c^0\right)=||C_{sk}\left(z^0, c^0\right)||_{s,k=1}^m,
\end{eqnarray*}
and $\left[N\left(z^0, c^0\right)^T D\left(z^0, c^0\right)C\left(z^0,c^0\right)\delta\right]_i$ is the $i$-th component of the vector
\begin{eqnarray*}   N\left(z^0, c^0\right)^TD\left(z^0, c^0\right)C\left(z^0,c^0\right)\delta, \quad \delta=\{\delta_i\}_{i=1}^m \in R_+^m.\end{eqnarray*}
Consider the set of equations
\begin{eqnarray*}   \delta_i=\frac{y_i\left(D\left(z^0, c^0\right)C\left(z^0,c^0\right)\delta, z^0\right)}{\pi_i\left(D\left(z^0, c^0\right)C\left(z^0,c^0\right)\delta, z^0\right)}\sum\limits_{s=1}^n
\eta_{is}^0\left(D\left(z^0,c^0\right)C\left(z^0,c^0\right)\delta, z^0\right)
 \end{eqnarray*}
\begin{eqnarray*}    \times \sum\limits_{k=1}^mD_{sk}\left(z^0, c^0\right)\sum\limits_{j=1}^mC_{kj}\left(z^0, c^0\right)\delta_j,\quad i=\overline{1,m},\quad \delta=\{\delta_i\}_{i=1}^m \in R_+^m, \quad z^0 \in
M_0,\end{eqnarray*}
for the vector $\delta=\{\delta_i\}_{i=1}^m.$
In the notations given, the above introduced set of equations has operator form
\begin{eqnarray*}   \delta = M\left(D\left(z^0, c^0\right)C\left(z^0,c^0\right)\delta, z_0\right)D\left(z^0, c^0\right)C\left(z^0,c^0\right)\delta, \end{eqnarray*}
where
\begin{eqnarray*}    M\left(p,z^0\right)=\left|\left |\frac{y_i\left(p,z^0\right)}{\pi_i\left(p,z^0\right)}\eta_{is}^0\left(p,z^0\right)\right|\right|_{i=1, s=1}^{m, n}.\end{eqnarray*}
Introduce the matrix
$ U\left(p, z_0\right), \ p \in T_0\left(z^0,c^0\right),$
and built then the $m \times m$-dimensional matrix
\begin{eqnarray*}   \tilde U(\delta)= U\left(D\left(z^0, c^0\right)C\left(z^0,c^0\right)\delta, z^0\right), \quad  \tilde U(\delta)=||\tilde U_{ik}(\delta)||_{i,k=1}^m, \end{eqnarray*}
where $p=D\left(z^0, c^0\right)C\left(z^0,c^0\right)\delta$ and $ U\left(p,z^0\right)$ is a matrix of the form
\begin{eqnarray*}   U\left(p,z^0\right)= M\left(p,z^0\right)D\left(z^0,c^0\right)C\left(z^0,c^0\right), \quad  \left(p,z^0\right) \in T_0\left(z^0,c^0\right)\times z^0,\end{eqnarray*}
being strictly positive  because the condition (\ref{arevla1}) holds. Establish the existence of a strictly positive solution to the set of equations
\begin{eqnarray}   \label{1o2p20}
\delta_i=\sum\limits_{k=1}^m\tilde U_{ik}(\delta)\delta_k,   \quad i=\overline{1,m}.
\end{eqnarray}
For this, build the auxiliary map $H(\delta)=\{H_i(\delta)_i\}_{i=1}^m$ given on the unit simplex
\begin{eqnarray*}   P_1=\left\{ \delta=\{\delta_i\}_{i=1}^m \in R_+^m, \ \sum\limits_{i=1}^m\delta_i=1\right\}\end{eqnarray*}
by the rule
 \begin{eqnarray*}   H_i(\delta)=\frac{\delta_i+\sum\limits_{k=1}^m\tilde U_{ik}(\delta)\delta_k}{
1+\sum\limits_{i=1}^m\sum\limits_{k=1}^m\tilde U_{ik}(\delta)\delta_k}, \quad i=\overline{1,m}. \end{eqnarray*}
The map $H(\delta)$ continuously maps the unit simplex $P_1$ into itself, therefore, there exists a fixed point of this map belonging to $P_1,$ i.e.,
\begin{eqnarray*}   \delta_i^0 =H_i\left(\delta^0\right), \quad i=\overline{1,m}.\end{eqnarray*}
Or
\begin{eqnarray*}    \delta_i^0\sum\limits_{i=1}^m\sum\limits_{k=1}^m\tilde U_{ik}\left(\delta^0\right)\delta_k^0 =
 \sum\limits_{k=1}^m\tilde U_{ik}\left(\delta^0\right)\delta_k^0,  \quad i=\overline{1,m}.\end{eqnarray*}
As the matrix $\tilde U(\delta)$ has strictly positive elements, the vector
$\delta^0$ has strictly positive components.
Introduce notations
\begin{eqnarray*}   \lambda=\sum\limits_{i=1}^m\sum\limits_{k=1}^m\tilde U_{ik}\left(\delta^0\right)\delta_k^0,
\quad p^0=D\left(z^0, c^0\right)C\left(z^0,c^0\right)\delta^0.\end{eqnarray*}
It is obvious that the number $\lambda >0.$
The vector $p^0 \in  T_1\left(z^0,c^0\right)\subseteq T_0\left(z^0,c^0\right)$ has strictly positive components and satisfies the set of equations
\begin{eqnarray*}   \lambda\pi_i\left(p^0,z^0\right) \left\langle N_i\left(z^0, c^0\right), p^0 \right\rangle
\end{eqnarray*}
\begin{eqnarray}   \label{1o2p91}
= y_i\left(p^0,z^0\right)\sum\limits_{s=1}^n\eta_{is}\left(p^0,z^0\right)p_s^0,
\quad i=\overline{1,m}.
\end{eqnarray}
From the last, it follows that the vector $p^0$ solves the set of equations
\begin{eqnarray}\label{1o2p99}
\lambda D_i(p)=y_i\left(p,z^0\right)\sum\limits_{s=1}^n\eta_{is}\left(p, z^0\right)p_s, \quad i=\overline{1,m},
\end{eqnarray}
because the equality holds
\begin{eqnarray*}   D_i(p)= K_i\left(p,z^0\right)= K_i^0\left(p,Q\left(p,z^0\right)\right)
\end{eqnarray*}
\begin{eqnarray*}   =\pi_i\left(p,Q\left(p,z^0\right)\right)\left[\left\langle p, y_i^0 - x_i^0\right\rangle a_i\left(p,z^0\right)\prod\limits_{i=1}^nv_i\left(p,z^0\right) +\left\langle d_i + b_i, p\right\rangle \right]
\end{eqnarray*}
\begin{eqnarray*}
=\pi_i\left(p,z^0\right)\left[\left\langle p, y_i^0 - x_i^0+ d_i\right\rangle +\left\langle b_i, p\right\rangle \right],\quad  \left(p,z^0\right)\in  R,  \quad  i=\overline{1,m},
\end{eqnarray*}
in view of the taxation system built.
Summing up  equalities (\ref{1o2p99}) over $i$ from $1$ to $m,$ we obtain
\begin{eqnarray*}    \lambda \sum\limits_{i=1}^m\pi_i\left(p^0,z^0\right)\left[\left\langle p^0, y_i^0 - x_i^0+ d_i\right\rangle  +\left\langle b_i, p^0\right\rangle \right]
\end{eqnarray*}
\begin{eqnarray*}   =\sum\limits_{i=1}^m
y_i\left(p^0,z^0\right)\sum\limits_{s=1}^n\eta_{is}\left(p^0,z^0\right)p_s^0.
\end{eqnarray*}
As the vector $\pi\left(p,z^0\right)=\{\pi_i\left(p,z^0\right)\}_{i=1}^m$ satisfies  the equation
\begin{eqnarray*}    \sum\limits_{i=1}^m\pi_i\left(p,z^0\right)\left[\left\langle p, y_i^0 - x_i^0+ d_i\right\rangle  +\left\langle b_i, p\right\rangle \right]
\end{eqnarray*}
\begin{eqnarray*}   =\sum\limits_{i=1}^m
y_i\left(p,z^0\right)\sum\limits_{s=1}^n\eta_{is}\left(p,z^0\right)p_s, \quad \left(p,z^0\right) \in R_1,
\end{eqnarray*}
we obtain $ \lambda=1.$
Therefore, the vector $p^0$ solves the set of equations
\begin{eqnarray}\label{allamoja}
D_i(p)=y_i\left(p,z^0\right)\sum\limits_{s=1}^n\eta_{is}\left(p, z^0\right)p_s, \quad i=\overline{1,m},
\end{eqnarray}
and, owing to the taxation system built, there hold equalities
\begin{eqnarray*}   D_i(p)= K_i\left(p,z^0\right)= K_i^0\left(p,Q\left(p,z^0\right)\right)\end{eqnarray*}
\begin{eqnarray}   \label{0z1l1}
= y_i\left(p,z^0\right)\sum\limits_{s=1}^n \eta_{is}\left(p,z^0\right)p_s, \quad \left(p,z^0\right)\in  R, \quad i=\overline{m+1,l}.
\end{eqnarray}
Due to the conditions of the Theorem, the equalities
\begin{eqnarray*}   \sum\limits_{i=1}^m\left[Y_{ki}(p,z) - X_{ki}(p,z)+ d_{ki} \right]+\sum\limits_{i=1}^lb_{ki}
\end{eqnarray*}
\begin{eqnarray}\label{2oke12}
=\sum\limits_{i=1}^l\eta_{ik}(p,z) y_i(p, Q(p,z)), \quad  (p,z) \in R, \quad  k=\overline{1,m},
\end{eqnarray}
hold or, inserting   $z =\zeta_0(p, \omega_0)=z_0$ and $p=p_0$ into the equality (\ref{2oke12}), we obtain
\begin{eqnarray*}   \sum\limits_{i=1}^m\left[Y_{ki}\left(p^0,z^0\right) - X_{ki}\left(p^0,z^0\right)+ d_{ki} \right]+\sum\limits_{i=1}^lb_{ki}
\end{eqnarray*}
\begin{eqnarray*}   =\sum\limits_{i=1}^l\eta_{ik}\left(p^0,z^0\right) y_i\left(p^0,z^0\right), \quad
\left(p^0, z^0\right) \in R, \quad  k=\overline{1,m}.
\end{eqnarray*}
Accounting for that equalities (\ref{0z1l1}) hold and inserting the expression for $y_i\left(p^0,z^0\right)$
from the equalities (\ref{allamoja}) and  (\ref{0z1l1}) into the last equalities,
we obtain
\begin{eqnarray*}    \sum\limits_{i=1}^m\left[Y_{ki}\left(p^0,z^0\right) - X_{ki}\left(p^0,z^0\right)+ d_{ki} \right] +\sum\limits_{i=1}^lb_{ki}
\end{eqnarray*}
\begin{eqnarray}   \label{k61l}
= \sum\limits_{i=1}^l\frac{\mu_{ki}\left(p^0\right)D_i\left(p^0\right)}{\sum\limits_{j=1}^n\mu_{ji}\left(p^0\right)p_j^0},\quad
\left(p^0 ,z^0\right) \in R, \quad  k=\overline{1,m}.
\end{eqnarray}
The latter means that the vector $p^0$ is an equilibrium price vector.
\qed
\end{proof}

In practice, it is important to have a  system of taxation with stable proportions\index{taxation  system with stable proportions} during a long time. The next Theorem shows that such a taxation system  exists.

\begin{theorem}\label{3ch2l2}
Let  convex down technological maps $F_i(x),\ x \in X_i^1,$
$i=\overline {1,m},$  belong to the CTM class in a wide sense,
the set $R$ be non-empty, the   structure of firms supply agree with the structure of consumers choice on the set $R_1,$
and let fields $y_i(p,z),\ i=\overline{1,l}, $ be continuous functions of $(p,z) \in R_1,$ the conditions of the Lemma \ref{opcd1}  hold. The solution to the equation (\ref{opod1}) with $z=z^0$ for the vector $\pi\left(p,z^0\right)=\{\pi_i\left(p,z^0\right)\}_{i=1}^m$ has the form
$\pi\left(p,z^0\right)=\{\pi_i\left(p,z^0\right)\}_{i=1}^m,$ where
\begin{eqnarray*}    \pi_i\left(p,z^0\right)=\pi_0\left(p,z^0\right)\pi_i, \quad  0< \pi_i, \quad  i=\overline{1,m},\end{eqnarray*}
\begin{eqnarray}   \label{3pod1}
\pi_0\left(p,z^0\right)=\frac{ \sum\limits_{i=1}^my_i\left(p,z^0\right)\sum\limits_{s=1}^n\eta_{is}^0\left(p,z^0\right)p_s}
{\sum\limits_{i=1}^m \pi_i\left\langle  p, y_i^0 - x_i^0 +d_i + b_i \right\rangle},\quad  i=\overline{1,m},
\end{eqnarray}
and $\pi=\{\pi_i\}_{i=1}^m$ is a certain  strictly positive vector.
If  fields of information evaluation by  consumers $\eta_{i}^0(p,z),\  i=\overline{1,l},$ and random field   $\zeta_0(p,\omega_0)$ satisfy the conditions of the  Theorem \ref{11tl4}  with the set of income functions  given by the formulae (\ref{001l1}),  then the  set of equations
\begin{eqnarray*}   \sum\limits
_{i=1}^l \frac{\mu_{ki}(p)D_i(p)}{\sum\limits_{j=1}^n
\mu_{ji}(p)p_j}
\end{eqnarray*}
\begin{eqnarray}\label{3g2p18}
= \sum\limits _{i=1}^m\left[Y_{ki}\left(p, z^0\right)-X_{ki}\left(p, z^0\right)+ d_{ki}\right]+ \sum\limits
_{i=1}^l b_{ki}, \quad  k=\overline{1,n},
\end{eqnarray}
has a solution $p^0=\{p_i^0\}_{i=1}^n$ in the set of strictly positive  price vectors
for every $ z^0 \in
 M_0.$
Such an  economy system can operate profitably in the Walras equilibrium state if $\left\langle y_i^0 -x_i^0, p^0 \right\rangle  > 0, \ i=\overline{1,m},$ or subvention-profitably if
\begin{eqnarray*}   \left\langle y_i^0 -x_i^0+ d_i , p^0 \right\rangle  > 0, \quad i=\overline{1,m},\end{eqnarray*}
for $b_i \neq 0.$

The net income of the $i$-th firm in the state of the economy equilibrium
for every $ z^0 \in
 M_0$ is given by the formula
\begin{eqnarray*}   D_i\left(p^0\right)= \frac{\pi_i}{\lambda} \left\langle y_i^0 -x_i^0 +d_i +b_i, p^0\right\rangle , \quad i=\overline{1,m}.\end{eqnarray*}
The equilibrium vector $p^0=\{p_i^0\}_{i=1}^n$ solves the set of equations
\begin{eqnarray*}    \pi_i \left\langle y_i^0 -x_i^0 +d_i +b_i, p\right\rangle = \lambda y_i\left(p,z^0\right)\sum\limits_{s=1}^n\eta_{is}^0\left(p,z^0\right)p_s,\quad i=\overline{1,m}, \end{eqnarray*}
where $\lambda$ is a certain strictly positive number.
\end{theorem}
\begin{proof}\smartqed
The set of equations
\begin{eqnarray*}   \pi_i\left(p, z^0\right) \left\langle y_i^0 -x_i^0 +d_i +b_i, p\right\rangle
\end{eqnarray*}
\begin{eqnarray}\label{novs1}
=y_i\left(p,z^0\right)\sum\limits_{s=1}^n\eta_{is}^0\left(p,z^0\right)p_s, \quad i=\overline{1,m},
\end{eqnarray}
for the vector $p=\{p_i\}_{i=1}^n$ with the taxation system considered takes the form
\begin{eqnarray*}   \frac{\pi_i \left\langle y_i^0 -x_i^0 +d_i +b_i, p\right\rangle }{\sum\limits_{i=1}^m \pi_i \left\langle p, y_i^0 - x_i^0 +d_i + b_i \right\rangle}=\frac{y_i\left(p,z^0\right)\sum\limits_{s=1}^n\eta_{is}^0\left(p,z^0\right)p_s}{ \sum\limits_{i=1}^my_i\left(p,z^0\right)\sum\limits_{s=1}^n\eta_{is}^0\left(p,z^0\right)p_s}, \quad i=\overline{1,m}.\end{eqnarray*}
Consider the set of equations
\begin{eqnarray}   \label{novs2}
 \pi_i \left\langle y_i^0 -x_i^0 +d_i +b_i, p\right\rangle = \lambda y_i\left(p,z^0\right)\sum\limits_{s=1}^n\eta_{is}^0\left(p,z^0\right)p_s,\quad i=\overline{1,m},
\end{eqnarray}
for the vector $p=\{p_i\}_{i=1}^n.$
As in the  Proof of the  Theorem \ref{2ch2l2}, we establish the existence of a strictly positive vector
$p^0=\{p_i^0\}_{i=1}^n$
for every $ z^0 \in
 M_0$ and a strictly positive number $\lambda$ such that $p^0=\{p_i^0\}_{i=1}^n$ solves the set of equations
(\ref{novs2}). The equality
\begin{eqnarray*}   \sum\limits_{i=1}^m \pi_i \left\langle  p^0, y_i^0 - x_i^0 +d_i + b_i \right\rangle = \lambda \sum\limits_{i=1}^my_i\left(p,z^0\right)\sum\limits_{s=1}^n\eta_{is}^0\left(p^0,z^0\right)p_s^0 \end{eqnarray*}
holds,  from which it follows that this vector solves the set of equations (\ref{novs1}) too.
At this solution
\begin{eqnarray*}   D_i\left(p^0\right)=\frac{ \pi_i}{\lambda} \left\langle y_i^0 -x_i^0 +d_i +b_i, p^0\right\rangle , \quad i=\overline{1,m}.\end{eqnarray*}
\qed
\end{proof}

Let us  generalize the results of the  Theorem \ref{2ch2l2}   onto the case when realizations of a random field  of   decisions making by firms\index{regular random field  of   decisions making by firms}  are regular.
In what follows, we suppose $ m \leq n.$
\begin{definition}
In the economy system, we describe  the structure of firms production  by technological maps $F_i(x),\  x \in X_i^1, \ i=\overline{1,m},$ from the CTM class in a wide sense.
We call the random field $\zeta_0(p,\omega_0),$ $p \in K_+^n,$ taking values in the set $M_0,$ regular  if, for every realization $z(p)$ of the random field $\zeta_0(p,\omega_0),$ there exists a set of productive processes of $m$ firms $z^0 \in M_0$ such that, for any $p \in K_+^n,$ the set of  vectors
\begin{eqnarray*}   N_i\left(z(p),c^0\right)=\left[y_i(p) -x_i(p)\right] +d_i +b_i, \quad i=\overline{1,m},\end{eqnarray*}
built after the set of productive processes of $m$ firms $z(p)=\{z^i(p)\}_{i=1}^m,$  \ $ z^i(p)=(x_i(p), y_i(p)),$ a certain set of vectors of subvention and initial goods supply   $c^0=\{d_i+b_i\}_{i=1}^m,$ where $d_i$ is a subvention vector of the $i$-th firm and $b_i$ is a  vector of goods supply of the  $i$-th firm, is economically compatible  with a single strictly positive matrix $D\left(z^0, c^0\right)=||D_{ki}\left(z^0, c^0\right)||_{k=1,i=1}^{n,m},$ i.e., there exists the inverse matrix
\begin{eqnarray*}   \left[N\left(z(p),c^0\right)^T D\left(z^0, c^0\right)\right]^{-1},  \quad p \in K_+^n,\end{eqnarray*}
with strictly positive matrix elements, and the conditions
\begin{eqnarray*}    \left\langle y_i(p) - x_i(p)+ d_i+b_i, p \right\rangle \  \geq 0, \quad p \in K_+^n,   \quad i=\overline{1,m},\end{eqnarray*}
\begin{eqnarray*}    \sum\limits_{i=1}^m\left[y_i(p) - x_i(p)\right] + \sum\limits_{i=1}^m d_i > a > 0, \quad p \in K_+^n, \end{eqnarray*}
 hold,  where $ a= \{a_i\}_{i=1}^n $ is a certain vector with strictly positive components.
\end{definition}

Consider the case of a  regular random field\index{regular random field} $\zeta_0(p,\omega_0)$ being continuous  with probability 1. Let  $z(p)$ be a certain regular continuous realization, that is,  $\zeta_0(p,\omega_0)=z(p)$ at a certain $\omega_0.$

For a continuous realization $z(p)$ of the regular random field $\zeta_0(p,\omega_0)$, there exist a  set of productive processes $z^0$ and a strictly positive matrix $D\left(z^0, c^0\right)$ such that, for any $p \in K_+^n,$  the set of  vectors  $N_i\left(z(p),c^0\right), \ i=\overline{1,m},$ is  economically compatible  with the single matrix
$D\left(z^0, c^0\right).$ Denote
\begin{eqnarray*}   G\left(z^0\right)=\left\{p \in K_+^n,\  p= D\left(z^0, c^0\right)\delta, \  \delta \in  P_{0}\right\},\end{eqnarray*}
where
\begin{eqnarray*} 
 P_{0}=\{\delta=\left\{\delta_i \}_{i=1}^m \in R_+^m, \ \sum\limits_{i=1}^m\delta_i=1\right\}.\end{eqnarray*}
Let  $z\left(G\left(z^0\right)\right)$ be the set of values for $z(p), \ p \in G\left(z^0\right).$ Then $z\left(G\left(z^0\right)\right)$ is a closed set in
$M_0.$ The matrix
\begin{eqnarray*} S(z)=\left[N\left(z,c^0\right)^T D\left(z^0, c^0\right)\right]^{-1}
\end{eqnarray*}
is a continuous function of $z$ on the set $z\left(G\left(z^0\right)\right)$  such that  every element $S_{ij}(z)$ of the matrix $S(z)$ is majorized from below by a strictly positive number independent of $z \in z\left(G\left(z^0\right)\right).$

\begin{lemma}\label{rev1}
Let  fields of information evaluation by consumers $\eta_{i}^0(p,z),\  $ $i=\overline{1,l},$ satisfy the  conditions of the Theorem \ref{11tl4}, the random field $\zeta_0(p,\omega_0)$ be continuous with probability 1 and regular,
the structure of firms  supply  agree with the  structure of consumers  choice on the set $K_+^n\times M_0.$
For a continuous realization $z(p)$ of the regular random field $\zeta_0(p,\omega_0)$, there exists a solution to the equation
\begin{eqnarray*}
 \sum\limits _{i=1}^m\pi_i^0(p)\left\langle p, y_i(p) - x_i(p) + d_i +b_i \right\rangle
 \end{eqnarray*}
  \begin{eqnarray}\label{opod15}
=\sum\limits _{i=1}^my_i(p,z(p))\sum\limits_{k=1}^n\eta_{ik}^0(p, z(p))p_k
\end{eqnarray}
for the vector $\pi^0(p)=\{\pi_i^0(p)\}_{i=1}^n$
being a continuous function of $ p \in K_+^n.$
\end{lemma}
\begin{proof}\smartqed
 A solution to the equation (\ref{opod15}) for the vector $\pi^0(p)=\{\pi_i^0(p)\}_{i=1}^m$ has, e.g., the form
$\pi^0(p)=\{\pi_i^0(p)\}_{i=1}^m,$ where
\begin{eqnarray*}    \pi_i^0(p)=\pi_0(p), \quad   i=\overline{1,m},\end{eqnarray*}
\begin{eqnarray}   \label{rev2}
\pi_0(p)=\frac{ \sum\limits_{i=1}^my_i(p,z(p))\sum\limits_{s=1}^n\eta_{is}^0(p,z(p))p_s}
{\sum\limits_{i=1}^m \left\langle p, y_i(p) - x_i(p) +d_i + b_i \right\rangle}.
\end{eqnarray}
This solution is a continuous function of $ p \in  K_+^n.$

One can construct other solutions to the equation (\ref{opod15}) for the vector $\pi^0(p)=\{\pi_i^0(p)\}_{i=1}^m$ putting
$\pi^0(p)=\{\pi_i^0(p)\}_{i=1}^m,$ where
\begin{eqnarray*}    \pi_i^0(p)=\pi_0(p)\pi_i, \quad  0< \pi_i, \quad  i=\overline{1,m},\end{eqnarray*}
\begin{eqnarray}   \label{reh2}
\pi_0(p)=\frac{ \sum\limits_{i=1}^my_i(p,z(p))\sum\limits_{s=1}^n\eta_{is}^0(p,z(p))p_s}
{\sum\limits_{i=1}^m \pi_i\left\langle  p, y_i(p) - x_i(p) +d_i + b_i \right\rangle},
\end{eqnarray}
and $\pi=\{\pi_i\}_{i=1}^m$ is a certain  strictly positive vector.
\qed
\end{proof}

On the set $ K_+^n,$ consider  the  family of functions
\begin{eqnarray*}   f_i(p)= \pi_i^0(p )\left[\left\langle p, y_i(p) - x_i(p) + d_i \right\rangle + \left\langle b_i, p \right\rangle\right], \quad i=\overline{1,m},\end{eqnarray*}
\begin{eqnarray}   \label{rev7}
f_i(p)=y_i(p,z(p))\sum\limits_{k=1}^n\eta_{ik}^0(p, z(p))p_k ,  \quad i=\overline{m+1,l},
\end{eqnarray}
and the  family of functions
\begin{eqnarray*}   g_i(p)= \left\langle p, y_i(p) - x_i(p) + d_i\right\rangle  + \left\langle b_i, p\right\rangle , \quad i=\overline{1,m},\end{eqnarray*}
\begin{eqnarray}   \label{rev6}
 g_i(p)= \left\langle p, b_i \right\rangle,  \quad i=\overline{m+1,l},
\end{eqnarray}
where $\pi^0(p)=\{\pi_i^0(p)\}_{i=1}^m$ is a continuous solution to the equation (\ref{opod15}).

\begin{theorem}\label{reh3}
Let the  fields of information evaluation by  consumers $\eta_{i}^0(p,z),$  $i=\overline{1,l},$ satisfy the  conditions of the Theorem \ref{11tl4}, the random field $\zeta_0(p,\omega_0)$ be continuous with probability 1 and regular,
 let the structure  of firms supply agree with the structure of consumers  choice  on the set $K_+^n\times M_0.$
For every regular realization $z(p)$ of the random field $\zeta_0(p,\omega_0)$ on the set $ K_+^n,$ a taxation system     $||\pi_{ij}^0(p)||_{ i,j=1}^l$ exists  such that
\begin{eqnarray*}   f_i(p)=\sum\limits _{j=1}^l\pi_{ij}^0(p)g_i(p), \quad  p \in  K_+^n,  \quad  i=\overline{1,l}, \end{eqnarray*}
if $\pi^0(p)=\{\pi_i^0(p)\}_{i=1}^m$ is a continuous solution to the equation (\ref{opod15}).
\end{theorem}

In the next Theorem, we suppose the  family of income functions is given by the formulae
\begin{eqnarray}   \label{rev8}
D_i(p)= \sum\limits _{j=1}^l\pi_{ij}^0(p))g_i(p)), \quad  p \in  K_+^n,  \quad  i=\overline{1,l}.
 \end{eqnarray}

\begin{theorem}\label{1uh2l2}
 Let   convex down technological maps $F_i(x),\ x \in X_i^1,$
$~i=\overline {1,m},$  belong to the CTM class in a wide sense,
the set $M_0$ be non-empty, and  let  the structure  of firms  supply agree with the structure of consumers choice  on the set $K_+^n\times M_0.$
If the fields of information evaluation by  consumers $\eta_{i}^0(p,z),\  i=\overline{1,l},$ satisfy the  conditions of the  Theorem \ref{11tl4}, the random field $\zeta_0(p,\omega_0)$ is continuous with probability 1 and regular,
 fields $y_i(p,z),\ i=\overline{1,l}, $ are continuous functions of $(p, z) \in K_+^n \times M_0,$ the condition (\ref{arevla1}) hold, then the set of equations  of the economy equilibrium
\begin{eqnarray*}   \sum\limits
_{i=1}^l \frac{p_k\mu_{ki}(p)D_i(p)}{\sum\limits_{j=1}^n
\mu_{ji}(p)p_j}
\end{eqnarray*}
\begin{eqnarray}\label{1g9q18}
= p_k 
\left[\sum\limits _{i=1}^m\left[Y_{ki}(p, z(p))-X_{ki}(p, z(p))+ d_{ki}\right]+ \sum\limits
_{i=1}^l b_{ki} \right], \quad  k=\overline{1,n},
\end{eqnarray}
has  a strictly positive solution being an equilibrium price vector $p^0=\{p_i^0\}_{i=1}^n$
for a continuous realization $ z(p)$ of the random field $\zeta_0(p,\omega_0).$
Such an economy system can operate profitably in the state of the Walras equilibrium if \begin{eqnarray*}   \left\langle y_i\left(p^0\right) -x_i\left(p^0\right),  p^0 \right\rangle  > 0, \quad i=\overline{1,m},\end{eqnarray*}
or subvention-profitably if
\begin{eqnarray*}   \left\langle y_i\left(p^0\right) -x_i\left(p^0\right)+ d_i , p^0 \right\rangle  > 0, \quad i=\overline{1,m},\end{eqnarray*}
where  $\mu_{ki}(p)=\eta_{ik}(p, z(p)), \ k=\overline{1,n},  \ i=\overline{1,l}.$
\end{theorem}
\begin{proof}\smartqed
Consider the set of equations
\begin{eqnarray}   \label{opod16}
 \left\langle N_i\left(z(p),c^0\right),p\right\rangle = \frac{y_i(p,z(p))}{\pi_i^0(p)}\sum\limits_{s=1}^n\eta_{is}^0(p, z(p))p_s, \quad i=\overline{1,m},
\end{eqnarray}
for the price vector $p=\{p_i\}_{i=1}^n.$
Introduce the notations
\begin{eqnarray*}   y_i^*(\delta)=y_i\left(D\left(z^0, c^0\right)\delta,z\left(D\left(z^0, c^0\right)\delta\right)\right), \quad \pi_i^*(\delta)=\pi_i^0\left(D\left(z^0, c^0\right)\delta\right),\end{eqnarray*}
\begin{eqnarray*}   \quad \eta_{is}^*(\delta)=\eta_{is}^0\left(D\left(z^0, c^0\right)\delta, z\left(D\left(z^0, c^0\right)\delta\right)\right),\end{eqnarray*}
\begin{eqnarray}   \label{opod17}
N^*(\delta)= N\left(z\left(D\left(z^0, c^0\right)\delta\right),c^0\right).
\end{eqnarray}
We find a solution to the set of equations (\ref{opod16}) in the form
\begin{eqnarray*}   p=D\left(z^0, c^0\right)\delta, \quad  \delta \in P_0.\end{eqnarray*}
 In the notations given, the set of equations (\ref{opod16}) can be written as follows
\begin{eqnarray*}   \left[N^*(\delta)^T D\left(z^0, c^0\right)\delta\right]_i
\end{eqnarray*}
\begin{eqnarray}\label{opod18}
=\frac{y_i^*(\delta)}{\pi_i^*(\delta)}\sum\limits_{s=1}^n\eta_{is}^*(\delta)\sum\limits_{k=1}^mD_{sk}\left(z^0, c^0\right)\delta_k, \quad i=\overline{1,m}.
\end{eqnarray}
Denote  $\left[N^*(\delta)^T D\left(z^0, c^0\right)\right]^{-1}=||Q_{ij}(\delta) ||_{i,j=1}^m.$ Then the set of equations
(\ref{opod18}) is equivalent to the set of equations
\begin{eqnarray}   \label{opod19}
\delta_k=\sum\limits_{i=1}^mQ_{ki}(\delta)\frac{y_i^*(\delta)}{\pi_i^*(\delta)}\sum\limits_{s=1}^n\eta_{is}^*(\delta)\sum\limits_{k=1}^mD_{sk}\left(z^0, c^0\right)\delta_k, \quad k=\overline{1,m}.
\end{eqnarray}
As in the Proof of the Theorem \ref{2ch2l2}, we establish the existence of a strictly positive solution $\delta_0=\{\delta_i^0\}_{i=1}^m \in P_0$
to the set of equations (\ref{opod19}). Similarly to the Proof of the Theorem \ref{2ch2l2}, we establish that   $p^0=D\left(z^0, c^0\right)\delta^0$ solves the set of equations (\ref{opod16}).
However, for this solution, the following equality holds:
\begin{eqnarray*}   \pi_i^0\left(p^0\right) \left\langle N_i\left(z\left(p^0\right),c^0\right),p^0\right\rangle =D_i\left(p^0\right), \quad i=\overline{1,m}.\end{eqnarray*}
Therefore, the vector $p^0$ solves the set of equations
\begin{eqnarray}   \label{opod20}
D_i(p)= y_i(p,z(p))\sum\limits_{s=1}^n\eta_{is}^0(p, z(p))p_s, \quad i=\overline{1,m}.
\end{eqnarray}
With the same arguments as in the Theorem \ref{2ch2l2}, we establish that $p^0$ solves the set of equations (\ref{1g9q18}).
\qed
\end{proof}

Let us generalize the results onto the case when the firms number in the economy system $m > n,$
where $n$ is the number of  goods produced within the economy system.

Suppose that  $\zeta_0(p, \omega_0)=z^0, \ p \in K_+^n, $ with probability 1, where $ z^0 \in M_0.$ In the next Theorem, we suppose the income functions are given by formulae (\ref{001l1}).

\begin{theorem}\label{9nl2}
 Let   convex down technological maps $F_i(x),\ x \in X_i^1,$
$i=\overline {1,m},$ belong to the CTM class in a wide sense,
the set $R$ be  non-empty such that its every point is economically compatible  in a generalized sense, and let  the solution to the equation (\ref{opod1}) for $\pi(p,z)=\{\pi_i(p,z)\}_{i=1}^m$ satisfy  the conditions of the Lemma \ref{opcd1} or of the Lemma \ref{opod111}.

If the  fields of information evaluation by  consumers  $\eta_{i}^0(p,z),\  i=\overline{1,l},$ and the random field $\zeta_0(p,\omega_0)$ satisfy the  conditions of the Theorem \ref{11tl4}, the structure of firms  supply  agrees with the structure of consumers  choice  on the set $R_1,$
 and the fields $y_i\left(p,z^0\right),\ i=\overline{1,l}, $ are continuous functions of the vector $p$ on the set $ K_+^n$ for every $ z^0 \in
 M_0, $ and also
\begin{eqnarray*}   y_j\left(p,z^0\right)=\frac{\sum\limits_{i=1}^{m_0}L_{ji}(p)\sum\limits_{s=1}^{n}\mu_{si}(p)p_s y_i\left(p,z^0\right)}{\sum\limits_{s=1}^{n}\mu_{sj}(p)p_s},\quad p \in T_0\left(z^0, c^0\right),\quad  j=\overline{m_0, m},   \end{eqnarray*}    \begin{eqnarray*}    L_{ji}(p)=\frac{\pi_j\left(p,z^0\right)L_{ji}\left(z^0,c^0\right)}{\pi_i\left(p,z^0\right)}, \quad  j=\overline{m_0, m}, \quad  i=\overline{1, m_0},\end{eqnarray*}
then the set of  equations for the economy equilibrium
\begin{eqnarray*}   \sum\limits
_{i=1}^l \frac{p_k\mu_{ki}(p)D_i(p)}{\sum\limits_{j=1}^n
\mu_{ji}(p)p_j}
\end{eqnarray*}
\begin{eqnarray}\label{9nl3}
= p_k 
\left[\sum\limits _{i=1}^m\left[Y_{ki}\left(p, z^0\right)-X_{ki}\left(p, z^0\right)+ d_{ki}\right]+ \sum\limits
_{i=1}^l b_{ki} \right ], \quad  k=\overline{1,n},
\end{eqnarray}
has a  strictly positive solution $p^0=\{p_i^0\}_{i=1}^n$ being an equilibrium price vector
for every $ z^0 \in
 M_0.$
Such an economy system can operate profitably in the Walras equilibrium state if $\left\langle y_i^0 -x_i^0, p^0 \right\rangle > 0, \ i=\overline{1,m},$ or subvention-profitably if
\begin{eqnarray*}   \left\langle y_i^0 -x_i^0+ d_i , p^0 \right\rangle  > 0, \quad i=\overline{1,m},\end{eqnarray*}
for $b_i \neq 0.$
\end{theorem}

\subsection{Application of general results}

Let us give a series of definitions to be applied in what follows.
Let $A$ and $B$ be  two square matrices.
\begin{definition}
The pair of non-negative matrices $\{A, B\}$ of the dimension $d\times d$ belongs to the class $\Pi_0$ if
there exists the inverse matrix $(B - A)^{-1}$ whose all elements are strictly positive.
\end{definition}
Let us give features of matrices belonging to the class $\Pi_0.$
\begin{proposition}
Let $\{ A_0, B_0\}$ and $\{A_1, B_1\}$ be two pairs of non-negative matrices of the dimension $d\times d$ belonging to the class $\Pi_0.$ Then pairs of matrices 
\begin{eqnarray*}
\{B_0B_1+ A_0A_1, A_0B_1 + B_0A_1\}, \quad \{B_1B_0+ A_1A_0, A_1B_0 + B_1A_0\}
\end{eqnarray*}
 belong to the class $\Pi_0$ too.
\end{proposition}
This feature obviously follows from the evident fact that if $\{ A_0, B_0\}$ and $\{A_1, B_1\}$ are two pairs of $d\times d-$dimensional matrices belonging to the class $\Pi_0,$ then the matrix product
\begin{eqnarray*}   (B_1 - A_1)^{-1}(B_0 - A_0)^{-1}=\left[(B_0 - A_0)(B_1 - A_1)\right]^{-1}\end{eqnarray*}
has strictly positive matrix elements. However,
\begin{eqnarray*}   (B_0 - A_0)(B_1 - A_1)= (B_0B_1+ A_0A_1) -(A_0B_1 + B_0A_1).\end{eqnarray*}
Similarly, one can establish another feature.
\begin{proposition}
Let $\{ A_0, B_0\}$ be a matrix pair belonging to the class $\Pi_0.$
Then the matrix pair $\{ A_0+H, B_0+H\}$ belongs to the class $\Pi_0,$ where $H$
is any non-negative matrix.
\end{proposition}

\begin{proposition}
Let \hfill $\{ A, B\}$  \hfill be  \hfill a  \hfill matrix  \hfill pair  \hfill of  \hfill the  \hfill dimension  \hfill $d\times d$  \hfill  such  \hfill that \\ $(B - A)^{-1}$ exists.
If there exists a non-degenerate non-negative matrix $C$ such that the pair $\{ AC, BC\} $ or the pair
$\{ CA, CB\} $ belongs to the class $\Pi_0,$
then the pair $\{ A, B\}$ belongs to the class $\Pi_0.$
\end{proposition}
This feature follows from that if, e.g., the pair $\{ AC, BC\}$ belongs to the class $\Pi_0,$ then there exists such a pair
$\{ A_0, B_0\}$ from the class $\Pi_0$ that
\begin{eqnarray*}   BC - AC=B_0 - A_0. \end{eqnarray*}
 From here, $(B - A)^{-1}=C(B_0 -A_0)^{-1}.$ The last matrix is the product of a non-degenerate non-negative matrix and the matrix with strictly positive elements, therefore, it is the matrix with strictly positive elements.

Consider a generalization for the notion of the belonging to the class $\Pi_0$ if the matrix pair $\{A, B\}$ has the dimension $m\times n, \ m\neq n.$
\begin{definition}
The pair of rectangular non-negative matrices $\{A,B \}$ of certain dimension $m \times n$ belongs to the class $\Pi_0$ if there exists a non-negative matrix $D$  of the dimension
$n \times m$  having no zero columns  such that there exists the inverse matrix $\left[(B - A)D\right]^{-1}$ whose all the matrix elements are strictly positive.
\end{definition}
Let $D=||D_{ki} ||_{k,i=1}^{n, m}$ be a matrix  of the dimension  $n \times m,$
$ m \leq n,$ and the rank $m.$ Relate  to the  matrix $D$  $ m$ vector-columns $d_i =\{D_{k,i}\}_{k=1}^n, \ i=\overline{1,m}.$
\begin{definition}
 A \hfill matrix \hfill pair \hfill  $\{A,B\}$ \hfill  is \hfill built \hfill after \hfill a \hfill non-negative \hfill matrix \\ $D=||D_{ki}||_{k,i=1}^{n,m}$ of  the rank $m$ if elements of the matrices $A=||a_{ki}||_{k,i=1}^{m,n}$ and $B=||b_{ki}||_{k,i=1}^{m,n}$ are given by the formulae
\begin{eqnarray*}    a_{ki}= \left\{ \begin{array}{cl}
                     - m_{ki},& \textrm{if} \quad  m_{ki} < 0,\\
                      0,    & \textrm{if } \quad m_{ki} \geq 0\textrm{,}
                      \end{array}
                      \right.\end{eqnarray*}
\begin{eqnarray*}    b_{ki}= \left\{ \begin{array}{cl}
                      m_{ki},& \textrm{if} \quad  m_{ki} > 0,\\
                      0,    & \textrm{if } \quad m_{ki} \leq 0\textrm{,}
                      \end{array}
                    \right.\end{eqnarray*}
where $m_i=\{m_{ij}\}_{j=1}^n, i=\overline{1,m},$ is   set of vectors being biorthogonal  to the set of  vectors  $d_i=\{D_{k,i}\}_{k=1}^n, \ i=\overline{1,m}.$
\end{definition}
\begin{note}
After a non-negative matrix $D=||D_{ki}||_{k,i=1}^{n,m}$ of the rank $m,\ m \leq n, $ one can build an
$(n - m) $-parameter family of matrix pairs $\{A,B\}.$
\end{note}
\begin{lemma}
Let $C$ be a non-degenerate matrix of the dimension $m \times m$ with strictly positive elements, and the matrix pair $\{A,B\}$ be built after a non-negative matrix $D$ of the rank $m.$ If \begin{eqnarray*}   C^{-1}A \geq 0, \quad  C^{-1}B \geq 0,\end{eqnarray*}
then the matrix pair $\{C^{-1}A, C^{-1}B\}$ belongs to the class $\Pi_0.$
\end{lemma}
\begin{proof}\smartqed  Proof  follows from that the matrix $B - A $ is left inverse to the matrix $D,$ i.e., $(B - A)D = E,$ and the equality $ C^{-1}(B - A)D = C^{-1},$ where $E$ is the unit matrix of the dimension $m \times m.$
\qed
\end{proof}
Now let us  give the necessary and sufficient conditions for the economic compatibility of  a set of vectors $N_i\left(z^0,c^0\right)=\left[y_i^0 -x_i^0\right] +d_i +b_i, \ i=\overline{1,m},$ built after a  set of productive processes $m$ firms $z^0$ and a certain set of vectors of  subvention and initial goods supply  $c^0=\{d_i+b_i\}_{i=1}^m.$
\begin{theorem}\label{sumi1}
Let $ m \leq n,$ and let there exist a vector $ a= \{a_i\}_{i=1}^n $ with strictly positive components such that
\begin{eqnarray*}    \sum\limits_{i=1}^m\left[y_i^0 - x_i^0\right] + \sum\limits_{i=1}^m d_i > a > 0. \end{eqnarray*}
The necessary and sufficient condition of the economic compatibility for the set of vectors
$N_i\left(z^0,c^0\right)=\left[y_i^0 -x_i^0\right] +d_i +b_i, \ i=\overline{1,m},$ built after a set of productive processes  of $m$ firms $z^0,$ a certain set of vectors  subvention and goods supply  $c^0=\{d_i+b_i\}_{i=1}^m$ is the existence of a set of vectors $L_i\left(z^0,c^0\right),$  $ \ i=\overline{1,m},$ being  biorthogonal to the  set  of vectors $N_i\left(z^0,c^0\right),$  $  \ i=\overline{1,m},$ such that every vector  $L_i\left(z^0,c^0\right), \ i=\overline{1,m},$ has strictly positive components.
\end{theorem}
\begin{proof}\smartqed    Necessity. If the set of vectors
\begin{eqnarray*}   N_i\left(z^0,c^0\right)=\left[y_i^0 -x_i^0\right] +d_i +b_i, \quad  i=\overline{1,m},\end{eqnarray*}
 is economically compatible, then there exists a non-negative matrix
 $D\left(z^0, c^0\right)$ $=||D_{ki}\left(z^0, c^0\right)||_{k,i=1}^{n,m}$ with non-zero  rows such that there exists the  inverse matrix
$\left[N\left(z^0,c^0\right)^T D\left(z^0, c^0\right)\right]^{-1}$ with strictly positive matrix  elements or
\begin{eqnarray*} 
C^{-1} =\left[N\left(z^0,c^0\right)^T D\left(z^0, c^0\right)\right], 
\end{eqnarray*}
where a  non-degenerate matrix $C$ has strictly positive matrix elements. Consider the matrix $D^1\left(z^0, c^0\right)=D\left(z^0, c^0\right)C.$ The matrix $D^1\left(z^0, c^0\right)$ has strictly positive elements and the equality $N\left(z^0, c^0\right)^T D^1\left(z^0, c^0\right)=E$ holds, where
$E$ is the unit matrix of the dimension $m \times m.$ Denote by
$L_i\left(z^0,c^0\right)=\{D_{ki}^1\left(z^0, c^0\right)\}_{k=1}^n, \ i=\overline{1,m},$ columns of the matrix
$D^1\left(z^0, c^0\right)=||D_{ki}^1\left(z^0, c^0\right)||_{k,i=1}^{n,m}.$ Then, from the last equality,  we have
$\left\langle N_j\left(z^0,c^0\right), L_i\left(z^0,c^0\right) \right\rangle  =\delta_{ij},  \ i=\overline{1,m}.$ The last means the existence of a  set of  vectors with strictly positive components being  biorthogonal to the set of vectors $N_i\left(z^0,c^0\right)=\left[y_i^0 -x_i^0\right] +d_i +b_i, \  i=\overline{1,m}.$ The necessity is established.

 Sufficiency. Under the conditions of the  Theorem, there exists a  set of vectors $L_i\left(z^0,c^0\right)=\{D_{ki}^1\left(z^0, c^0\right)\}_{k=1}^n, \ i=\overline{1,m},$ with strictly positive components being  biorthogonal to the set of vectors  $N_i\left(z^0,c^0\right), \ i=\overline{1,m}.$
In the  matrix form, it can be written as $N\left(z^0,c^0\right)^T D^1\left(z^0, c^0\right)=E,$ where $D^1\left(z^0, c^0\right)=||D_{ki}^1\left(z^0, c^0\right)||_{k,i=1}^{n,m}.$ Show the existence of a non-negative matrix $ D\left(z^0, c^0\right)$ such that $\left[N\left(z^0,c^0\right)^T D\left(z^0, c^0\right)\right]^{-1}$ has strictly positive elements. Consider the matrix $D\left(z^0, c^0\right)=D^1\left(z^0, c^0\right)(E - R(\varepsilon)), $ where $R(\varepsilon)=||R_{ij}(\varepsilon) ||_{i,j=1}^m, $ $ R_{ij}(\varepsilon)=(1- \delta_{ij})\varepsilon,$ and the number $ \varepsilon > 0.$ Matrix  elements of the matrix
\begin{eqnarray*}   D\left(z^0, c^0\right)=D^1\left(z^0, c^0\right)(E - R(\varepsilon)) \end{eqnarray*}
 are given by the formulae
\begin{eqnarray*}   D_{ki}\left(z^0, c^0\right)=D_{ki}^1\left(z^0, c^0\right) - \sum\limits_{j=1}^mD_{kj}^1\left(z^0, c^0\right)(1 - \delta_{ji})\varepsilon
\end{eqnarray*}
\begin{eqnarray*}   =D_{ki}^1\left(z^0, c^0\right) - \varepsilon
\sum\limits_{j=1, \ j\neq i}^mD_{kj}^1\left(z^0, c^0\right), \quad k=\overline{1,n}, \quad i=\overline{1,m}.
\end{eqnarray*}
Choose  $\varepsilon > 0$ such that the  inequality
\begin{eqnarray*}    \varepsilon < \min\left\{\max\limits_{1 \leq k \leq n, \ 1 \leq i \leq m}\frac{D_{ki}^1}{\sum\limits_{j =1}^m D_{kj}^1}, \ \frac{1}{n}\right\}\end{eqnarray*}
holds.

Then the inequalities
\begin{eqnarray*}   D_{ki}^1\left(z^0, c^0\right) - \varepsilon \sum\limits_{j=1, \ j \neq i}^m D_{kj}^1\left(z^0, c^0\right) > 0,\quad  k =\overline{1,n}, \quad  i =\overline{1,m},\end{eqnarray*}    hold.

Therefore, the matrix $D\left(z^0, c^0\right)$ has strictly positive elements, and the equality $N\left(z^0,c^0\right)^T D\left(z^0, c^0\right)=E - R(\varepsilon)$ holds. Under assumptions on the number  $\varepsilon$,
the matrix $(E - R(\varepsilon))^{-1}$ exists and has strictly positive elements. Finally, show that the set $T_0\left(z^0, c^0\right)$ is non-empty. For this, it is sufficient to establish the existence of a set of vectors $p \in K_+^n $ for which the inequalities
$ a_i\left(p,z^0\right) > 0, \ i =\overline{1,m}, $ hold. Such a set is the set of vectors $p$ of the form $p=D\left(z^0, c^0\right)(E - R(\varepsilon))^{-1}\delta, \ \delta \in R_+^m, \  \delta  > 0,$ because $ a_i\left(p,z^0\right)=1, \  \left\langle N_i\left(z^0, c^0\right), p\right\rangle  =\delta_i > 0.$ With the condition of the  Theorem
\begin{eqnarray*}    \sum\limits_{i=1}^m\left[y_i^0 - x_i^0\right] + \sum\limits_{i=1}^m d_i > a > 0 \end{eqnarray*}
we obtain the economic compatibility of the set of  vectors  $N_i\left(z^0, c^0\right), \ i=\overline{1,m}.$
\qed
\end{proof}

In practice, simple models of economic equilibrium are significant that satisfy simple sufficient conditions, under which it is possible to manage economic processes without invasion into the price formation properly, however, regulating by taxing, credit policy, and investment. It is particularly important in the case of the  necessity of economic reforms accounting for global trends in the today world. Consider a simple model describing the choice of consumers. Suppose the fields of information evaluation by consumers  are non-random  and
\begin{eqnarray}   \label{sim1}
\eta_i^0(p,z, \omega_i)=C_i(z)=\{C_{ki}(z)\}_{k=1}^n, \quad i=\overline{1,l},
\end{eqnarray}
with probability 1, where $\sum\limits_{k=1}^nC_{ki}(z) > \ 0, \  C_{ki}(z) \geq 0, \ k=\overline{1,n},\ i=\overline{1,l}.$
Matrix elements $ C_{ki}(z) $ have simple economic treatment:
the number of the $k$-th goods units the $i$-th consumer wants to consume within the  economy operation period if  firms  realize productive processes $z$ in  the economy system.  We assume the final supply vector belongs to the interior of the cone created by vector-columns $C_i(z)=\{C_{ki}(z)\}_{k=1}^n$ of the matrix $C(z)=||C_{ki}(z) ||_{k=1,i=1}^{n,l}.$
From mathematical point of view, the last means that
\begin{eqnarray}   \label{sim2}
\psi(z)=\sum\limits_{i=1}^lC_i(z)y_i(z), \quad y(z)=\{y_i(z)\}_{i=1}^l, \quad y_i(z) >0, \quad i=\overline{1,l},
\end{eqnarray}
where
\begin{eqnarray*}   \psi(z)=\{\psi_k(z)\}_{k=1}^n, \quad \psi_k(z)=\sum\limits_{i=1}^m\left[y_{ki} - x_{ki} + d_{ki}+ b_{ki}\right]+ \sum\limits_{i=m+1}^lb_{ki}.\end{eqnarray*}
In the next Theorem, we suppose the family of  income functions  is given by the formulae
(\ref{001l1}) and the random field $\zeta_0(p, \omega_0)=z^0.$

\begin{theorem}\label{sumi2}
Let  convex down technological maps $F_i(x),\ x \in X_i^1,$
$i=\overline {1,m},$  belong to the CTM class in a wide sense.
Let the fields of information evaluation by consumers  $\eta_{i}^0(p,z, \omega_i),\  i=\overline{1,l},$ be given  by the formulae (\ref{sim1}), a  set of productive processes of $m$ firms $z^0$ satisfy the conditions:
there exists a set of vectors $L_i\left(z^0,c^0\right),$ $ \ i=\overline{1,m},$ being biorthogonal to the set vectors  $N_i\left(z^0,c^0\right)=\left[y_i^0 -x_i^0\right] +d_i +b_i,$ $  \ i=\overline{1,m},$ such  that every vector  $L_i\left(z^0,c^0\right), \ i=\overline{1,m},$ has strictly positive components, the inequality
 \begin{eqnarray*}    \sum\limits_{i=1}^m\left[y_i^0 - x_i^0\right] + \sum\limits_{i=1}^m d_i > a > 0 \end{eqnarray*}
holds  for a certain strictly positive vector $ a= \{a_i\}_{i=1}^n. $ And finally, let the final supply vector $\psi\left(z^0\right)$ belong to the interior of the cone created by vector-columns $C_i\left(z^0\right)=\{C_{ki}\left(z^0\right)\}_{k=1}^n, \ i=\overline{1,l,}$ of the matrix $C\left(z^0\right)=||C_{ki}\left(z^0\right) ||_{k=1,i=1}^{n,l}.$
Then there exist a taxation system and an equilibrium price vector such  that every consumer satisfies his needs according to his choice structure. The equilibrium price vector under which  the needs of consumers are satisfied solves the  set of equations of the economy equilibrium
\begin{eqnarray}\label{1z2p1u}
\sum\limits_{i=1}^l \frac{C_{ki}\left(z^0\right)D_i(p)}{\sum\limits_{j=1}^n
C_{ji}\left(z^0\right)p_j}=  \psi_k\left(z^0\right),  \quad  k=\overline{1,n}.
\end{eqnarray}
\end{theorem}
\begin{proof}\smartqed   Under the conditions of the Theorem \ref{sumi2}, the set of productive processes  $z^0$ of $m$ firms is economically compatible  according to the Theorem \ref{sumi1}. The rest conditions of the Theorem \ref{sumi2} guarantee the fulfilment of conditions of the  Theorem \ref{2ch2l2}. According to this Theorem, the taxation system exists, for which there exists the equilibrium price vector solving the set of equations (\ref{1z2p1u}).
\qed
\end{proof}

\subsection{Theory of interindustry equilibrium}

Here, we use the  results of the  previous Subsection to construct the theory of interindustry equilibrium.
Consider an economy model containing $n$ net industries whose production is described by technological maps $F_i(x_i),\  x_i \in X_i, \ i=\overline{1,n},$
where
\begin{eqnarray}\label{g2l27}   F_i(x_i)=\{  y_i=\{\delta_{ik} u_i\}_{k=1}^n \in S, \  0 \leq a_{ki}u_i \leq x_{ki}, \ k=\overline{1,n}\},\end{eqnarray}
\begin{eqnarray*}
x_i=\{x_{ki}\}_{k=1}^n \in X_i,
\end{eqnarray*}
and $F_i(x_i)$ is a set of plans  for an input vector $x_i.$
If $y_i^{0}$ is the maximum available output of the $i$-th industry, then
\begin{eqnarray*}   X_i=\left\{ x_i=\{x_{ki}\}_{k=1}^n \in S, \  x_{ki} \leq a_{ki}y_i^0 ,\ i=\overline{1,n}\right\}.\end{eqnarray*}
Consider a productive economic  process given by the formula
\begin{eqnarray}   \label{g2l28}
Q(p,z)=\{Q_i(p,z)\}_{i=1}^n,
\end{eqnarray}
\begin{eqnarray*}   Q_i(p,z)=\{X_i(p,z), Y_i(p,z)\}, \end{eqnarray*}
\begin{eqnarray*}   X_i(p,z)=a_i(p,z)\prod\limits_{i=1}^nv_i(p,z)\left \{a_{ki}u_i\right\}_{k=1}^n, \end{eqnarray*}
\begin{eqnarray*}   Y_i(p,z)=a_i(p,z)\prod\limits_{i=1}^nv_i(p,z)u_ie_i, \quad e_i=\{\delta_{ij}\}_{j=1}^n,\end{eqnarray*}
where
\begin{eqnarray*}    a_i(p,z)=\chi_{[0, \infty)}\left(u_ip_i - \sum\limits_{s=1}^n a_{si}p_su_i +\left\langle d_i, p\right\rangle \right),\end{eqnarray*}
\begin{eqnarray*}    v_i(p,z)=\chi_{[0, \infty)}\left(a_i(p,z)u_i - \sum\limits_{s=1}^n a_{is}a_s(p,z)u_s +\sum\limits_{k=1}^l b_{ik} -c_i\right),\end{eqnarray*}
and the vector $c=\{c_i\}_{i=1}^n$ is strictly positive, i.e., $c_i >0, \  i=\overline{1,n}.$

Consider a set of productive processes  $( x_i,  y_i), \  i=\overline{1,n},$
where
\begin{eqnarray*}    x_i=\{  a_{ki}u_i\}_{k=1}^n,\quad   y_i=\{  \delta_{ki}u_i\}_{k=1}^n, \quad u_i \leq y_i^0,  \quad   i=\overline{1,n}.\end{eqnarray*}
Let also $N_i\left(z,c^0\right)=\{\delta_{ki}u_i - a_{ki}u_i +d_{ki} +b_{ki} \}_{k=1}^n,\  i=\overline{1,n},$ be the set of vectors  built in  the previous Subsection.
The set of productive processes  $( x_i,  y_i), \  i=\overline{1,n},$ is economically compatible  if the  conditions
\begin{eqnarray}\label{ec1}
 u_k - \sum\limits_{i=1}^n a_{ki}u_i + \sum\limits_{i=1}^n d_{ki} > 0, \quad k=\overline{1,n},
\end{eqnarray}
hold, and there exists the inverse matrix $C\left(z,c^0\right)$ to the matrix \begin{eqnarray*}   N\left(z,c^0\right)=||\delta_{ki}u_i - a_{ki}u_i +d_{ki} +b_{ki}||_{k,i=1}^n,\end{eqnarray*}    whose elements are strictly positive. We call the vector
$d_i=\{d_{ki}\}_{k=1}^n$ the subvention vector and the vector $b_i=\{b_{ki}\}_{k=1}^n$ we do the initial goods supply vector of the $i$-th industry.

Let us  give the sufficient conditions guaranteeing the economic compatibility of the  productive processes.

\begin{lemma}\label{ecs1}
Let $A$ be a productive and indecomposable matrix, and let  the strictly positive output vector
$ y=\{u_i\}_{i=1}^n$ in the economy system
 satisfy  the condition (\ref{ec1}) and inequalities
$a_{ki}u_i - d_{ki} - b_{ki} > 0$  for those $k, i=\overline{1,n},$ for which $ \quad a_{ki} > 0.$
If $ a_{ki} = 0,$ then suppose $d_{ki} + b_{ki}=0.$
Under these conditions, the set of productive processes  $( x_i,  y_i), \  i=\overline{1,n},$ is economically compatible.
\end{lemma}
The Proof of the Lemma is obvious  because the conditions of the Definition of compatibility of productive processes  hold, and the requirements applied guarantee the existence of the matrix inverse to the matrix $N\left(z,c^0\right)$ having strictly positive elements. Really,
if the conditions of the  Lemma  hold, then there exists the inverse matrix to the matrix
 \begin{eqnarray*}   N\left(z,c^0\right)=||\delta_{ki}u_i - a_{ki}u_i +d_{ki} +b_{ki}||_{k,i=1}^n\end{eqnarray*}    because the spectral radius of the matrix $A(y)=|| a_{ki} - d_{ki}/u_i - b_{ki}/u_i||_{k,i=1}^n$ is strictly less than 1.  It is non-negative and indecomposable, and
\begin{eqnarray*}   N\left(z,c^0\right)^{-1}=f(y)\left[E - A(y)\right]^{-1}, \quad f(y)=\frac{1}{\prod\limits_{i=1}^nu_i }.\end{eqnarray*}
The matrix $\left[E - A(y)\right]^{-1}$ has strictly positive elements because $A(y)$ is indecomposable as well as $A.$

\begin{theorem}\label{2ch002}
Let technological maps describing net industries be given by formulae (\ref{g2l27}) and the conditions of the  Lemma \ref{ecs1}  hold. A set of productive processes  $M_0$ is determined by the  conditions of the  Lemma \ref{ecs1}.
A solution to the equation (\ref{opod1}) for $\pi(p,z)=\{\pi_i(p,z)\}_{i=1}^m$ satisfies the  conditions of the  Lemma \ref{opcd1} or of the Lemma  \ref{opod111} if $m=n,$ and the productive economic process is given by the formulae (\ref{g2l28}).

If the fields  of information evaluation by  consumers $\eta_{i}^0(p,z),\  i=\overline{1,l},$ and the random field $\zeta_0(p,\omega_0)$ satisfy the  conditions of the Theorem \ref{11tl4}, the  structure  of firms  supply agrees with the structure of consumers  choice
 and the fields $y_i\left(p,z^0\right),\ i=\overline{1,l}, $ are continuous functions of the vector $p$ on the set $ K_+^n,$ for every $ z^0 \in
 M_0,$ then for $\zeta_0(p,\omega_0)= z^0 \in
 M_0$ the   set of equations of the economy equilibrium
\begin{eqnarray*}   \sum\limits
_{i=1}^l \frac{p_k\mu_{ki}(p)D_i(p)}{\sum\limits_{j=1}^n
\mu_{ji}(p)p_j}
\end{eqnarray*}
\begin{eqnarray}\label{1qt1p18}
= p_k \left[
\sum\limits _{i=1}^m\left[Y_{ki}\left(p, z^0\right)-X_{ki}\left(p, z^0\right)+ d_{ki}\right]+ \sum\limits
_{i=1}^l b_{ki} \right ], \quad  k=\overline{1,n},
\end{eqnarray}
has a solution in the set of strictly positive price vectors $p^0=\{p_i^0\}_{i=1}^n$
for every $ z^0 \in
 M_0.$
Such an economy system can operate profitably in the Walras equilibrium state if $\left\langle  y_i^0 - x_i^0, p^0 \right\rangle \ > 0, \ i=\overline{1,m},$ or subvention-profitably if
\begin{eqnarray*}   \left\langle  y_i^0 -  x_i^0+ d_i , p^0 \right\rangle  > 0, \quad i=\overline{1,n},\end{eqnarray*}
for $b_i \neq 0.$
\end{theorem}

\chapter{Equilibrium States Constructing Algorithms}

\abstract*{For an exchange economy  model with constant elasticity some Theorems of the existence of equilibrium states are proved. In a few Theorems, algorithms of construction of the equilibrium state for an exchange economy model with proportional consumption are presented.
The necessary and sufficient conditions of the existence of strictly positive equilibrium price vector for exchange  models with proportional consumption are given. The problem of supply of goods on a market of an economy system that does not change equilibrium price vector, levels of consumption is solved. It is proved that in some class of equivalence the equilibrium price vector is not changed.
Several Theorems giving the sufficient conditions of the existence  of equilibria for the economy model with fixed gains are found. Algorithm of construction of equilibrium states for the economy models with constant part of consumption  are presented. }

In this Chapter, we establish  algorithms of constructing  equilibrium price vectors  for various  models of consumers choice  \cite{55, 71, 92, 106}. The class of exchange models considered is important in view of their interpretation and possible using to describe the reality. Such models are exchange economy models with constant consumption elasticity,\index{exchange economy models with constant consumption elasticity} exchange economy models with proportional consumption,\index{exchange economy models with proportional consumption} and exchange economy models with fixed gain levels and proportional consumption.\index{exchange economy models with fixed gain levels and proportional consumption} The existence of strictly positive equilibrium price vector for which space arbitrage opportunities\index{space arbitrage opportunities}  are not possible in the economy system guarantees optimal possibility to form the needed goods supply for economic activity of every  subject of economic activity.

In the Section "Exchange economy with constant elasticity", we consider a class of models in which consumers choice are described by demand vectors from the class $C^{\alpha}.$  Such a class of models contains, as a particular case, a consumers choice model with constant demand elasticity.\index{consumers choice model with constant demand elasticity} From here the name of the  class of models  follows. In the Theorem \ref{fas2}, we establish the  conditions for the set of equations (\ref{fas1}) to be solved in the set $C_{\delta},$ and in the Lemma, we give the conditions for the set $C_{\delta}$  to be non-empty.
 This Theorem is used to establish the Theorem  \ref{mod0} guaranteeing the existence of a  strictly positive  price vector under which economic agents  whose choice is described by demand vectors with constant   elasticity\index{demand vectors with constant   elasticity}  have no arbitrage opportunities.\index{arbitrage opportunities}

 In the Section "Exchange economy model with proportional consumption",\index{exchange economy model with proportional consumption} we establish the conditions for the existence of a strictly positive equilibrium price vector in models with
  fields of information  evaluation  by  consumers  independent of the price vector. Such exchange economy model has a simple economic interpretation.  The matrix $C$ constructed after  fields of information evaluation   by  consumers we called the matrix of unproductive consumption.\index{matrix of unproductive consumption}
In the Theorem \ref{mod3}, we propose an explicit construction of the matrix $B$ builded after  property vectors of economic agents for which an equilibrium price vector in the model considered  is strictly positive  and we also give an algorithm of construction of  equilibrium states. Under a certain additional assumptions, a simplified construction of the matrix $B$
is proposed in the Theorem \ref{alla7}.

For  further detailed   study of the  structure of the matrix $B,$
under which  there exists a strictly positive equilibrium price vector, the Theorem \ref{alla8}  is proved  in which  the necessary and sufficient conditions of the existence of  a strictly positive equilibrium price vector are established.

 Using the Theorem \ref{alla8}, in the Theorem \ref{alla16}, we give the necessary conditions for the elements of the matrix $B$  under which a strictly positive equilibrium price vector exists.
For a class of particular models, in the Theorem \ref{alla4}, we establish the  necessary and sufficient conditions for the matrix $B$ under which a strictly positive equilibrium price vector exists.

In the Section "Economy model with fixed gains", we study a class of models with fixed gains\index{class of models with fixed gains} and arbitrary structure of  fields of information evaluation   by  consumers. Before establishing the conditions under which a strictly positive equilibrium price vector exists in such models of economic systems, we prove the Theorem \ref{algl2} that  establishes the existence of a solution to the set of equations (\ref{algl4}) under rather general assumptions. By the Theorem \ref{algl2}, we establish the Theorem \ref{algl5} on the existence of equilibrium price vector in such economic models. The Theorem \ref{Pas3} gives an algorithm to construct equilibrium price vectors in the economy models with fixed gains and proportional consumption. Finally, in the Theorem \ref{Pas4}, we give the sufficient conditions for the existence of a strictly positive equilibrium price vector, and in the Theorem \ref{Pas6}, we propose an algorithm to construct equilibrium price vectors.

In the Section "On economic interpretation of the economy model with proportional consumption",\index{economy model with proportional consumption} we economically interpret matrix elements of the
fields of evaluation of information by  consumers for the case of proportional consumption.
The Theorem \ref{Pas5} gives the necessary and sufficient conditions for the existence of equilibrium price vector in the model with fixed gains and proportional consumption.\index{model with fixed gains and proportional consumption} Also it gives an algorithm to construct this vector.

Finally, in the Section "Economy model with constant parts of the consumption" in the Theorem \ref{alla17}, we give the necessary and sufficient conditions under which the considered production model operates without default if demand vectors describe models with constant parts of the consumption.\index{model with constant parts of the consumption}

\section{Exchange Models}

\subsection{Exchange economy with constant elasticity}

In the considered model of the  economy, there are $l$ insatiable consumers having goods supplies that they exchange with each other.
We suppose the $i$-th consumer to have a vector of goods supply  $b_i=\{b_{ki}\}_{k=1}^n, \ i=\overline{1,l}.$ The  presence at the market of the $i$-th consumer means that  $\sum\limits_{k=1}^nb_{ki}> 0, \ i=\overline{1,l}.$ In such a model, we assume that the net income function of the  $i$-th consumer is given by the formula
\begin{eqnarray*}  K_i(p)=D_i(p)=\sum\limits_{k=1}^nb_{ki}p_k,  \quad   p \in R_+^n, \quad  i=\overline{1,l}.\end{eqnarray*}
The supply vector in such a model has the form
\begin{eqnarray*}  \psi=\{\psi_k\}_{k=1}^n,\quad  \psi_k=\sum\limits_{i=1}^lb_{ki}> 0, \quad  k=\overline{1,n}.\end{eqnarray*}
 We describe the choice of the $i$-th insatiable consumer by a random demand vector of the $i$-th insatiable consumer
\begin{eqnarray*} \gamma_i(p,\omega_i) =\{ \gamma_{ik}(p,\omega_i) \}_{k=1}^n,
\quad  i=\overline{1,l}, \end{eqnarray*}
on a probability space $\{\Omega, {\cal F}, \bar P_0\},$ where
\begin{eqnarray*}   \Omega=\prod\limits_{i=1}^l\Omega_i, \quad {\cal F}=\prod\limits_{i=1}^l {\cal F}_i,\quad \bar P_0= \prod\limits_{i=1}^l\bar P_i,\end{eqnarray*}
and
\begin{eqnarray*}  \gamma_{ik}(p,\omega_i)
=\frac{p_k\eta_{ik}^0(p, \omega_i)}
{\sum\limits_{s=1}^n\eta_{is}^0(p, \omega_i)p_s},
\quad  i=\overline{1,l}, \quad  k=\overline{1,n}.\end{eqnarray*}

 A certain realization of a random demand vector of the $i$-th insatiable consumer we denote
\begin{eqnarray*}  \gamma_i(p)=\{ \gamma_{ik}(p)\}_{k=1}^n,\end{eqnarray*}
where $ \gamma_{ik}(p)=\gamma_{ik}(p,\omega_i)$
for a certain  $\omega_i,$
and we call it the demand vector of the $i$-th insatiable consumer.
 Any realization of a  random  field  of information evaluation by the $i$-th consumer
\begin{eqnarray*} \eta_i^0(p,\omega_i)
=\{\eta_{ik}^0(p, \omega_i)\}_{k=1}^n, \quad  i=\overline{1,l}, \end{eqnarray*}
we denote
\begin{eqnarray*}  \mu_i(p)=\{\mu_{ki}(p)\}_{i=1}^n,\end{eqnarray*}
where
\begin{eqnarray*}  \mu_{ki}(p)=\eta_{ik}^0(p, \omega_i), \quad  k=\overline{1,n}, \quad  i=\overline{1,l},\end{eqnarray*}
for a certain   $\omega_i.$
With such notations
\begin{eqnarray*} \gamma_{ik}(p)=\frac{p_k\mu_{ki}(p)}{\sum\limits_{s=1}^n\mu_{si}(p)p_s},\quad  i=\overline{1,l},\quad  k=\overline{1,n}.\end{eqnarray*}

\begin{definition} A demand vector
\begin{eqnarray*}  \gamma_i(p)=\{ \gamma_{ik}(p)\}_{k=1}^n, \quad  i=\overline{1,l},\end{eqnarray*}
belongs to the class
$C^{\alpha},$  if a vector $\alpha_i=\{\alpha_{ik}\}_{k=1}^n, \   0 \leq \alpha_{ik}\leq 1,$ and a vector $C_i=\{c_{ki}\}_{k=1}^n, \  c_{ki} \geq 0, \ \sum\limits_{k=1}^n
c_{ki}>0, \  i=\overline{1,l}$ exist such that for the components $\gamma_{ik}(p)$ of the demand vector $ \gamma_i(p)$
the representation
\begin{eqnarray*} \gamma_{ik}(p)=\frac{c_{ki}p_i^{\alpha_{ik}}}{\sum\limits_{k=1}^n
c_{ki}p_k^{\alpha_{ik}}}, \quad  p \in P,
 \quad k=\overline{1,n}, \quad i=\overline{1,l},\end{eqnarray*}
holds, where
\begin{eqnarray*} P=\left\{p=\{p_i\}_{i=1}^n \in R_+^n, \  \sum\limits_{i=1}^np_i=1\right\}.\end{eqnarray*}
\end{definition}
\begin{theorem}\label{fas2}
Let demand vectors describing $l$ consumers belong to the class
$C^{\alpha}.$
If for a certain $\delta >0$ the set
\begin{eqnarray*} C_\delta =\left\{p\in P,\ \sum\limits
_{s=1}^nc_{si}p_s^{\alpha_{is}} \geq \delta ,\
i=\overline {1,l}\right\} \end{eqnarray*}  is not empty,
$\psi_k=\sum\limits_{i=1}^lb_{ki} >0, \ k=\overline {1,n},$
and functions
\begin{eqnarray*} D_i(p)=\sum\limits_{k=1}^nb_{ki}p_k,  \quad p \in P, \quad i=\overline {1,l}, \end{eqnarray*}
satisfy the condition
\begin{eqnarray*} \inf\limits _{p\in C_\delta }{c_{ki}D_i(p)\over\psi _k
\sum\limits_{k=1}^n\Gamma_k(p)}\geq\delta, \end{eqnarray*}
for those $k,i,$ for which $c_{ki}\not= 0,$ where
\begin{eqnarray*} \Gamma_k(p)=\frac{1}{\psi_k}\sum\limits_{i=1}^l
\frac{c_{ki}p_k^{\alpha_{ik}}D_i(p)}{\sum\limits_{j=1}^n
c_{ji}p_j^{\alpha_{ij}}}, \quad k=\overline {1,n},\end{eqnarray*}
\begin{eqnarray*} \Gamma(p)=\{ \Gamma_k(p)\}_{k=1}^n, \end{eqnarray*}
then the set of equations
\begin{eqnarray} \label{fas1}
p=\Gamma (p)
\end{eqnarray}
is solvable  in the set $C_\delta.$
\end{theorem}
\begin{proof}\smartqed
On the set $C_\delta, $ let us consider  a non-linear operator
\begin{eqnarray*} f(p)=\{f_k(p)\}_{k=1}^n,\end{eqnarray*}
where
\begin{eqnarray*} f_k(p)=\frac{\Gamma_k(p)}{\sum\limits_{i=1}^n\Gamma_i(p)}, \quad
k=\overline {1,n}.\end{eqnarray*}  The operator $f(p)$ maps the convex set
$C_\delta $ into itself. Indeed, it is sufficient to check the inequalities
\begin{eqnarray*} \sum\limits_{k=1}^nc_{ki}f_k^{\alpha_{ik}}(p) \geq \delta,
\quad i=\overline {1,l}.\end{eqnarray*}
We have
\begin{eqnarray*} \sum\limits_{k=1}^nc_{ki}f_k^{\alpha_{ik}}(p)\geq
\sum\limits_{k=1}^nc_{ki}f_k(p)=
\sum\limits_{k=1}^n\frac{c_{ki}}{\psi_k
\sum\limits_{s=1}^n\Gamma_s(p)}\sum\limits_{s=1}^l
\frac{c_{ks}p_k^{\alpha_{sk}}D_s(p)}{\sum\limits_{j=1}^n
c_{js}p_j^{\alpha_{sj}}}
\end{eqnarray*}
\begin{eqnarray*} \geq \delta \sum\limits_{k=1}^nc_{ki}
\sum\limits_{s=1}^l
\frac{f_{ks}p_k^{\alpha_{sk}}}{\sum\limits_{j=1}^n
c_{js}p_j^{\alpha_{sj}}}=\delta\sum\limits_{s=1}^l
\frac{\sum\limits_{k=1}^nc_{ki}f_{ks}p_k^{\alpha_{sk}}}{\sum\limits_{j=1}^n
c_{js}p_j^{\alpha_{sj}}}\geq \delta,\quad i=\overline {1,l}, \end{eqnarray*}
where $f_{ks}=1,$ if $c_{ks}>0,$  and  $f_{ks}=0,$ if $c_{ks}=0.$
The last inequality holds because on the set $C_\delta, $
 for $s=i,$
\begin{eqnarray*} \frac{\sum\limits_{k=1}^nc_{ki}f_{ki}p_k^{\alpha_{ik}}}{\sum\limits_{j=1}^n
c_{ji}p_j^{\alpha_{ij}}}=1.\end{eqnarray*}

On the set $C_\delta, $ the operator $f(p)$ is continuous. By the Schauder theorem
 \cite{88}, there exists a fixed point of this map belonging to $C_\delta.$
This point is fixed one for the map $\Gamma(p)$ too.
Really, from the fact that $p^*$ is a fixed point of the map $f(p),$
we have
\begin{eqnarray*} \Gamma_k(p^*)=p_k^*\sum\limits_{s=1}^n\Gamma_s(p^*).\end{eqnarray*}
Multiplying by $\psi_k$ the last equality and summing up over $k$ from $1$ to $n,$ we have
\begin{eqnarray*} \sum\limits_{k=1}^n\psi_k\Gamma_k(p^*)=\langle \psi, p^* \rangle \sum\limits_{s=1}^n\Gamma_s(p^*).\end{eqnarray*}
However,
\begin{eqnarray*} \sum\limits_{k=1}^n\psi_k\Gamma_k(p^*)=\sum\limits_{i=1}^lD_i(p^*)
=\langle \psi, p^* \rangle.\end{eqnarray*}
In view of $\langle \psi, p^* \rangle \neq 0,$ dividing by $\langle \psi, p^* \rangle,$
we obtain $ \sum\limits_{s=1}^n\Gamma_s(p^*)=1.$ The last means that $p^*$
solves the set of equations (\ref{fas1}).
\qed
\end{proof}

\begin{lemma}\label{cod1}
Let the components $\psi_k, \ k=\overline{1,n},$ of the vector of the final supply\index{vector of the final supply}  $\psi=\{\psi_k\}_{k=1}^n$ be strictly positive, and $ D_i(p), \
i=\overline{1,l},$ be such that
\begin{eqnarray} \label{alla1}
 \min\limits_{1 \leq i \leq l}\inf\limits_{ p \in P}D_i(p)=d>0.
\end{eqnarray}
If
\begin{eqnarray*}  \sum\limits_{s=1}^nc_{si}>0,\quad i=\overline{1,l},\end{eqnarray*}
then the conditions of the Theorem \ref{fas2}  hold.
\end{lemma}
\begin{proof}\smartqed  Let us introduce notations
\begin{eqnarray*}  \min_{k}\psi_k=a > 0, \quad \max_{k}\psi_k=A<\infty, \end{eqnarray*}
\begin{eqnarray*}  c=\min_{k,i, c_{ki} \not = 0}c_{ki}, \quad
 \sup\limits_{1 \leq i \leq l, \ p \in P}D_i(p)=D.\end{eqnarray*}
If $\delta>0$ satisfies the inequality
\begin{eqnarray*} \delta \leq
\min\left\{\frac{cda}{ADl}, \
\min_{i}\sum\limits_{j=1}^nc_{ji}n^{-\alpha_{ij}}\right\}=\delta_0,\end{eqnarray*}  then
$C_\delta$ is non-empty since it contains  the vector
$\{\frac{1}{n}, \ldots, \frac{1}{n}\} \in P.$ It is obvious that
$\delta_0 > 0,$ because
\begin{eqnarray*} \min_{i}\sum\limits_{j=1}^nc_{ji}n^{-\alpha_{ij}}>0,\end{eqnarray*}
as a result of that
\begin{eqnarray*} \min_{i}\sum\limits_{j=1}^nc_{ji}>0.\end{eqnarray*}
Let us prove that  for those $k,i,$ for which $c_{ki}\not= 0,$ the inequality
\begin{eqnarray*} \inf\limits _{p\in
C_\delta }{c_{ki}D_i(p)\over\psi _k
\sum\limits_{k=1}^n\Gamma_k(p)}\geq\delta\end{eqnarray*}
holds.

Really, in view of
\begin{eqnarray*} \sum\limits_{k=1}^n\Gamma_k(p) \leq \frac{1}{a}
\sum\limits_{i=1}^lD_i(p)\leq\frac{lD}{a},\end{eqnarray*}
we obtain that  for those $k,i,$ for which $c_{ki}\not= 0,$
\begin{eqnarray*} \inf\limits _{p\in
C_\delta }{c_{ki}D_i(p)\over\psi _k
\sum\limits_{k=1}^n\Gamma_k(p)}\geq  \frac{cda}{ADl} \geq \delta. \end{eqnarray*}
\qed
\end{proof}

The condition (\ref{alla1}), e.g., holds if every consumer has a basket of all $n$ kinds of goods or some funds and some goods basket.

From the Theorem \ref{fas2} and the Lemma \ref{cod1} the next Theorem follows.
\begin{theorem}\label{alla2}
Let demand vectors, describing $l$ consumers, belong to the class
$C^{\alpha}.$
If $\delta>0$ satisfies the inequality
\begin{eqnarray*} \delta \leq
\min\left\{\frac{cda}{ADl}, \
\min_{i}\sum\limits_{j=1}^nc_{ji}n^{-\alpha_{ij}}\right\}=\delta_0,\end{eqnarray*}
and functions
\begin{eqnarray*} D_i(p)=\sum\limits_{k=1}^nb_{ki}p_k,  \quad p \in P, \quad i=\overline {1,l}, \end{eqnarray*}
are such  that the condition (\ref{alla1})  holds,
then the set of equations (\ref{fas1})
has a solution in the set $C_\delta .$
\end{theorem}
The next Theorem does not need the condition (\ref{alla1}).
\begin{theorem}\label{mod0}
Let demand vectors, describing the choice of $l$ consumers, belong to the class
$C^{\alpha},$
where vectors $\alpha_i=\{\alpha_{ik}\}_{k=1}^n, \ i=\overline{1,l},$ are such that $0 \leq \alpha_{ik} < 1, \  k=\overline{1,n}, \   i=\overline{1,l},$  and let conditions
\begin{eqnarray*} \sum\limits_{i=1}^lc_{ki}b_{si} > 0, \quad  k,s=\overline{1,n}, \quad \sum\limits_{i=1}^nc_{ik} > 0,  \quad  k=\overline{1,l},\end{eqnarray*}
\begin{eqnarray*} \sum\limits_{i=1}^lb_{si} > 0,\quad  s=\overline{1,n},\end{eqnarray*}
hold.

Then there exists a strictly positive equilibrium price vector solving the problem
\begin{eqnarray} \label{mod1}
p=\Gamma (p),
\end{eqnarray}
where
\begin{eqnarray*} \Gamma(p)=\{ \Gamma_k(p)\}_{k=1}^n, \end{eqnarray*}
\begin{eqnarray*} \Gamma_k(p)=\frac{1}{\psi_k}\sum\limits_{i=1}^l
\frac{c_{ki}p_k^{\alpha_{ik}}D_i(p)}{\sum\limits_{j=1}^n
c_{ji}p_j^{\alpha_{ij}}}, \quad k=\overline {1,n},\end{eqnarray*}
\begin{eqnarray*} D_i(p)=\langle b_i, p \rangle=\sum\limits_{s=1}^nb_{si}p_s, \quad \psi_k=\sum\limits_{i=1}^lb_{ki}.\end{eqnarray*}
\end{theorem}
\begin{proof}\smartqed  Let a certain number $\varepsilon > 0.$ Consider the auxiliary problem
\begin{eqnarray} \label{mod2}
 p=\Gamma^{\varepsilon} (p),
\end{eqnarray}
\begin{eqnarray*} \Gamma^{\varepsilon}(p)=\{ \Gamma^{\varepsilon}_k(p)\}_{k=1}^n, \end{eqnarray*}
where
\begin{eqnarray*} \Gamma^{\varepsilon}_k(p)=\frac{1}{\psi_k^{\varepsilon}}\sum\limits_{i=1}^l
\frac{c_{ki}p_k^{\alpha_{ik}}D_i^{\varepsilon}(p)}{\sum\limits_{j=1}^n
c_{ji}p_j^{\alpha_{ij}}}, \quad k=\overline {1,n},\end{eqnarray*}
\begin{eqnarray*}  \psi_k^{\varepsilon}=\psi_k + l\varepsilon, \quad D_i^{\varepsilon}(p)=D_i(p)+ \varepsilon.\end{eqnarray*}
The problem (\ref{mod2}) has a solution in $C_{\delta}$ in view of the conditions of the  Lemma \ref{cod1}.  Let us  prove the strict positivity for this solution denoting it as $p^{\varepsilon}=\{p_k^{\varepsilon}\}_{k=1}^n.$ Let $\varepsilon$  be such that the inequality $l \varepsilon \leq \psi_k, \ k=\overline {1,n},$ holds. Then we have the estimate
\begin{eqnarray*}  p_k^{\varepsilon} \geq \frac{[p_k^{\varepsilon}]^{\alpha_0}}{2 \max\limits_{k} \psi_k \max\limits_{i}\sum\limits_{s=1}^nc_{si}} \sum\limits_{i=1}^lc_{ki}D_i^{\varepsilon}(p^{\varepsilon})  \end{eqnarray*}
\begin{eqnarray*}  \geq  \frac{[p_k^{\varepsilon}]^{\alpha_0}}{2 \max\limits_{k} \psi_k \max\limits_{i}\sum\limits_{s=1}^nc_{si}} \sum\limits_{i=1}^lc_{ki}D_i(p^{\varepsilon}), \end{eqnarray*}
where $\alpha_0= \max\limits_{i,k}\alpha_{ik} < 1.$
The inequality
\begin{eqnarray*} [p_k^{\varepsilon}]^{(1- \alpha_0)} \geq \frac{\min\limits_{k,s} \sum\limits_{i=1}^lc_{ki}b_{si}}{2\psi C}\end{eqnarray*}
holds, where
\begin{eqnarray*}  \psi= \max\limits_{k} \psi_k, \quad C=\max\limits_{i}\sum\limits_{s=1}^nc_{si}.\end{eqnarray*}
Let $D_1=\min\limits_{k,s} \sum\limits_{i=1}^lc_{ki}b_{si}.$
Then
\begin{eqnarray*}  p_k^{\varepsilon} \geq \left[\frac{D_1}{2 \psi C}\right]^{\frac{1}{1 - \alpha_0}}= \gamma.\end{eqnarray*}
Each component of the  solution  is estimated  from below by $\varepsilon$-independent constant. As the vector $p^{\varepsilon}$ belongs to the unit compact simplex, there exists a converging subsequence $\varepsilon_k \to 0$ such that $p^{\varepsilon_k}$ converges to a vector
$p^0.$ For components of  the vector $p^0$  there holds the inequality $p_k^0 \geq \gamma > 0.$ Going to the limit in the  equalities (\ref{mod2}), we obtain the existence of a  strictly positive solution for the problem (\ref{mod1}).
\qed
\end{proof}

\subsection{Exchange economy model with proportional con\-sumption}

In this Subsection, we consider an exchange model\index{exchange economy model with proportional consumption} for the case when the  choice of every consumer is described by a demand vector belonging to the class $C^{\alpha}$ under the condition that each vector $\alpha_i=\{\alpha_{ik}\}_{k=1}^n$ is such that $ \alpha_{ik}=1,$ $  k=\overline{1,n},\  i=\overline{1,l}.$

\begin{theorem}\label{mod3}
If a rectangular matrix
$\tau=|| \tau_{ki}||_{k=1, i=1}^{l,n}$ is such that $\tau C$ is a non-negative indecomposable matrix\index{indecomposable matrix} and $\tau C \tau$ is a non-zero non-negative matrix, then for the matrix $B=C B_1,$
where
\begin{eqnarray*} B_1=||b_{ks}^1 ||_{k,s=1}^l, \quad  b_{ks}^1=b_s^1\sum\limits_{i=1}^n\tau_{ki}C_{is},\end{eqnarray*}
and $b^1=\{b_s^1\}_{s=1}^l$ is the right Frobenius vector\index{right Frobenius vector} of the matrix $\tau C,$ there exists a non-negative solution to the set of equations
\begin{eqnarray} \label{mod4}
\sum\limits_{i=1}^lC_{ki}\frac{\langle b_i, p \rangle}{\langle C_i, p \rangle}=\psi_k,\quad  \psi_k=\sum\limits_{i=1}^lb_{ki}, \quad k=\overline{1,n}.
\end{eqnarray}
The solution to the problem (\ref{mod4}) is given by the formula
\begin{eqnarray} \label{mod5}
p^0=\{p_i^0\}_{i=1}^n, \quad p_i^0=\frac{1}{\lambda}\sum\limits_{k=1}^l\delta_k^0\tau_{ki},\quad i=\overline{1,n},
\end{eqnarray}
where $\delta^0=\{\delta_s^0\}_{s=1}^l$ is the strictly positive left Frobenius vector\index{left Frobenius vector} of the problem
\begin{eqnarray} \label{mod6}
\sum\limits_{k=1}^l\delta_k^0\sum\limits_{i=1}^n\tau_{ki}C_{is}= \lambda\delta_s^0, \quad s=\overline{1,l},
\end{eqnarray}
$\lambda$ is the largest proper value of the problem (\ref{mod6}).

If the matrix $\tau C \tau$ does not contain zero columns, then the problem (\ref{mod4})  has a strictly positive solution.
\end{theorem}
\begin{proof}\smartqed  From the conditions of the Theorem,  it follows that the  solution of the  problem (\ref{mod6})  always exists in view of the Perron-Frobenius Theorem\index{Perron-Frobenius Theorem} and is strictly positive. Also from the Perron-Frobenius theorem, the existence of a strictly positive solution to the conjugate problem
\begin{eqnarray*}  \sum\limits_{s=1}^l\sum\limits_{i=1}^n\tau_{ki}C_{is}b_s^1= \lambda b_k^1, \quad k=\overline{1,l},\end{eqnarray*}
follows, where the vector $b^1=\{b_s^1\}_{s=1}^l$ is the proper vector of the matrix $\tau C$ corresponding to the largest proper value $\lambda.$

Multiplying the $s$-th  equality from (\ref{mod6}) by $b_s^1$ and dividing by $\delta_s^0,$ we obtain
\begin{eqnarray*} \frac{\sum\limits_{k=1}^l\delta_k^0\sum\limits_{i=1}^n\tau_{ki}C_{is}b_s^1}{\delta_s^0}= \lambda b_s^1, \quad s=\overline{1,l},\end{eqnarray*}
or
\begin{eqnarray} \label{mod7}
 \frac{\sum\limits_{k=1}^l\delta_k^0b_{ks}^1}{\delta_s^0}= \lambda b_s^1, \quad s=\overline{1,l}.
\end{eqnarray}
Introduce the vector
\begin{eqnarray*} p^0=\{p_i^0\}_{i=1}^n, \quad p_i^0=\frac{1}{\lambda}\sum\limits_{k=1}^l\delta_k^0\tau_{ki},\quad i=\overline{1,n}. \end{eqnarray*}
Let us show this vector to be non-negative.
Inserting into the formula (\ref{mod5}) the expression for components of $\delta_k^0$
and using the set of equations (\ref{mod6}) satisfied by these components, we have
\begin{eqnarray*}  p_j^0= \frac{1}{\lambda^2}\sum\limits_{s=1}^l\delta_s^0\sum\limits_{i=1}^n\tau_{si}\sum\limits_{k=1}^l
C_{ik}\tau_{kj}=\frac{1}{\lambda^2}\sum\limits_{s=1}^l\delta_s^0[\tau C \tau]_{sj},\end{eqnarray*}
where
\begin{eqnarray*} [\tau C \tau]_{sj}=\sum\limits_{i=1}^n\tau_{si}\sum\limits_{k=1}^l
C_{ik}\tau_{kj}.\end{eqnarray*}
Under the  conditions of the  Theorem, matrix elements $[\tau C \tau]_{sj}$ of the matrix $\tau C \tau$ are non-negative.
This fact proves non-negativity for components of the vector $p^0=\{p_i^0\}_{i=1}^n.$

From the set of equalities (\ref{mod6}), we obtain
\begin{eqnarray*} \delta_s^0= \sum\limits_{i=1}^np_i^0C_{is}.\end{eqnarray*}
In view of the last,
\begin{eqnarray*} \frac{\sum\limits_{k=1}^l\delta_k^0b_{ks}^1}{\delta_s^0}=\frac{\sum\limits_{k=1}^l\sum\limits_{i=1}^np_i^0C_{ik}b_{ks}^1}{\sum\limits_{i=1}^np_i^0C_{ik}}=\frac{\sum\limits_{i=1}^np_i^0\sum\limits_{k=1}^lC_{ik}b_{ks}^1}{\sum\limits_{i=1}^np_i^0C_{ik}}, \quad s=\overline{1,n}.\end{eqnarray*}
Introduce the matrix
 $B=C B_1,$
where $B=||b_{ks} ||_{k,s=1}^l, $ $ b_{is}=\sum\limits_{k=1}^lC_{ik}b_{ks}^1.$
Then, accounting for (\ref{mod7}), we have
\begin{eqnarray} \label{1mod8}
\frac{\langle b_s, p^0 \rangle}{\langle  C_s, p^0 \rangle}=\lambda b_s^1,    \quad s=\overline{1,l},
\end{eqnarray}
where we introduced the notations
\begin{eqnarray*} \langle b_s, p^0 \rangle=\sum\limits_{i=1}^np_i^0b_{is}, \quad \langle C_s, p^0 \rangle=\sum\limits_{i=1}^np_i^0C_{is},  \quad s=\overline{1,l}.\end{eqnarray*}
Multiplying by $C_{ks}$ and summing up over $s$ from $1$ to $l,$ we obtain
\begin{eqnarray} \label{mod8}
 \sum\limits_{s=1}^lC_{ks}\frac{\langle b_s, p^0 \rangle}{\langle C_s, p^0\rangle }=\lambda\sum\limits_{s=1}^lC_{ks}b_s^1, \quad k=\overline{1,n}.
\end{eqnarray}
Let us establish now that the equality
\begin{eqnarray} \label{mod9}
 \lambda\sum\limits_{s=1}^lC_{ks}b_s^1=\sum\limits_{s=1}^lb_{ks}, \quad k=\overline{1,n},
\end{eqnarray}
holds.

Really,
\begin{eqnarray} \label{mod10}
 b_{ks}=\sum\limits_{m=1}^lC_{km}b_{ms}^1=\sum\limits_{m=1}^lC_{km}
 \sum\limits_{i=1}^n\tau_{mi}C_{is}b_s^1, \quad k=\overline{1,n}, \quad s=\overline{1,l}.
\end{eqnarray}
From (\ref{mod10}) the equality
\begin{eqnarray*} \sum\limits_{s=1}^lb_{ks}=\sum\limits_{m=1}^lC_{km}\sum\limits_{s=1}^l\left[\sum\limits_{i=1}^n\tau_{mi}C_{is}\right]b_s^1=\sum\limits_{m=1}^lC_{km}\lambda b_m^1\end{eqnarray*}
follows.
The last equality holds because
\begin{eqnarray*} \sum\limits_{s=1}^l\left[\sum\limits_{i=1}^n\tau_{mi}C_{is}\right]b_s^1=\lambda b_m^1.\end{eqnarray*}
From (\ref{mod8}) and (\ref{mod9}) we obtain
\begin{eqnarray*} \sum\limits_{s=1}^lC_{ks}\frac{\langle b_s, p^0 \rangle }{\langle C_s, p^0 \rangle }=\sum\limits_{s=1}^lb_{ks},  \quad k=\overline{1,n}.\end{eqnarray*}
\qed
\end{proof}

The next Theorem is a Corollary of the last one. It simplifies the method to construct the matrix $B$ if the additional condition holds
\begin{eqnarray*}  \sum\limits_{s=1}^lC_{ks}> 0, \quad  k=\overline{1,n},\end{eqnarray*}
and the matrix $\tau$ is non-negative.
\begin{theorem}\label{alla7}
Let conditions
\begin{eqnarray*} \sum\limits_{s=1}^lC_{is}> 0, \quad i=\overline{1,n}, \quad \sum\limits_{k=1}^nC_{ks}> 0,\quad s=\overline{1,l},\end{eqnarray*}
hold and a non-negative matrix
$ \tau_1=||\tau_{ki}^1 ||_{k=1, i=1}^{l,n}$
is such that the matrix $\tau_1 C$ is indecomposable  and
\begin{eqnarray*} \sum\limits_{i=1}^n\tau_{ki}^1 =a > 0, \quad k=\overline{1,l}.\end{eqnarray*}
If $B=C B_1=||b_{ki} ||_{k=1, i=1}^{n,l},$ where
\begin{eqnarray*} B_1=||b_{km}^1 ||_{k,m=1}^l,\quad
b_{km}^1=\sum\limits_{i=1}^n\tau_{ki}^1\frac{C_{im}}{\sum\limits_{s=1}^lC_{is}},\end{eqnarray*}
then there exists a strictly positive solution to the set of equations  (\ref{mod4}).
The solution to the problem (\ref{mod4}) is given by the formula
\begin{eqnarray} \label{mod11}
p^0=\{p_i^0\}_{i=1}^n, \quad p_i^0=\frac{1}{a}\sum\limits_{k=1}^l\delta_k^0\tau_{ki},\quad i=\overline{1,n},
\end{eqnarray}
where $\delta^0=\{\delta_s^0\}_{s=1}^l$ is strictly positive proper vector\index{strictly positive proper vector} of the problem
\begin{eqnarray} \label{mod12}
\sum\limits_{k=1}^l\delta_k^0\sum\limits_{i=1}^n\tau_{ki}C_{is}= a \delta_s^0, \quad s=\overline{1,l},
\end{eqnarray}
$a$ is the largest proper value of the problem (\ref{mod12}),
\begin{eqnarray*} \tau = \left|\left| \tau_{ki}^1\frac{1}{\sum\limits_{s=1}^lC_{is}}\right|\right|_{k=1, i=1}^{l,n}.\end{eqnarray*}
\end{theorem}
\begin{proof}\smartqed  To prove the Theorem, it is sufficient to check the validity of conditions of the previous Theorem
\ref{mod3}. The matrix
$\tau$ is such that $\tau C$ is indecomposable because $\tau_1 C$ is such one. The matrix
$\tau C$ has proper vector $e=\{1, 1,  \ldots, 1\}$ with proper value $a.$ This proper value is the largest one. Hence, all the conditions of the Theorem \ref{mod3} hold.
\qed
\end{proof}

Let vectors $C_i=\{C_{ki}\}_{k=1}^n, \ i=\overline{1,l}, $ satisfy the conditions
\begin{eqnarray*}   \sum\limits_{k=1}^n
C_{ki}>0, \quad   C_{ki} \geq 0, \quad  k=\overline{1,n},  \quad  i=\overline{1,l},\end{eqnarray*}
and $C=|| C_{ik}||_{i=1, k=1}^{n,l}$ be a  non-negative rectangular matrix constructed after vectors $C_i=\{C_{ki}\}_{k=1}^n, \ i=\overline{1,l}, $ and let $B=|| b_{ik}||_{i=1, k=1}^{n,l}$ be a matrix constructed after property vectors $b_i=\{b_{ki}\}_{k=1}^n, \ i=\overline{1,l}, $
where
\begin{eqnarray*}   \sum\limits_{k=1}^n
b_{ki}>0, \quad  b_{ki} \geq 0, \quad   k=\overline{1,n}, \quad   i=\overline{1,l}.\end{eqnarray*}
 Let us  use as previously the notations
\begin{eqnarray*}   \langle b_i, p \rangle=\sum\limits_{k=1}^nb_{ki}p_k,\quad  \langle C_i, p \rangle=\sum\limits_{k=1}^nC_{ki}p_k, \quad i=\overline{1,l},\end{eqnarray*}
where $ p=\{p_i\}_{i=1}^n $ is a certain vector from $R_+^n. $

\begin{theorem}\label{alla8}
Let $\psi=\{\psi_k\}_{k=1}^n$ be a certain  strictly positive vector.
The necessary conditions for solvability in the set of strictly positive vectors of the set of equations
 \begin{eqnarray} \label{alla9}
\sum\limits_{i=1}^lC_{ki}\frac{\langle b_i, p \rangle}{\langle C_i, p \rangle}=\psi_k, \quad k=\overline{1,n},
\end{eqnarray}
with respect to the vector $p \in R_+^n$ are the following:\\
1) the vector $\psi=\{\psi_k\}_{k=1}^n$ belongs to the interior of a positive cone created by vectors\index{interior of a positive cone created by vectors} $C_i=\{C_{ki}\}_{k=1}^n, \ i=\overline{1,l}, $ i.e.
\begin{eqnarray} \label{alla10}
\psi=\sum\limits_{i=1}^ly_iC_i, \quad y_i >0, \quad   i=\overline{1,l}\  ;
\end{eqnarray}
2) the rank $r$ of the  set of the  vectors $b_i - y_i C_i, \ i=\overline{1,l}, $ does not exceed $n-1\  ;$\\
3) every subset of $r$ linearly independent vectors
\begin{eqnarray*} b_{i_k} - y_{i_k}C_{i_k}, \quad  k=\overline{1,r}, \end{eqnarray*}
is such that any vector
$d_s=\{b_{s i_k} - y_{i_k} C_{s i_k} \}_{k=1}^r, \ s=\overline{1,n},$
belongs to the interior of the cone created by vectors
 \begin{eqnarray*} -d_m=\{- b_{m i_k} + y_{i_k} C_{m i_k} \}_{k=1}^r, \quad  m=\overline{1,n}, \quad  m \neq s.\end{eqnarray*}
The sufficient conditions for solvability of  the set of equations (\ref{alla9}) in the set of strictly positive vectors
are conditions 1) -- 2) and the condition\\
4) at least for one subset of $r$ linearly independent vectors
\begin{eqnarray*} b_{i_k} - y_{i_k}C_{i_k}, \quad  k=\overline{1,r}, \end{eqnarray*}
 the vector
$d_1=\{b_{1 i_k} - y_{i_k} C_{1 i_k} \}_{k=1}^r$
belongs to the interior of the cone created by vectors
 \begin{eqnarray*} -d_m=\{- b_{m i_k} + y_{i_k} C_{m i_k} \}_{k=1}^r, \quad  m=\overline{2,n}.\end{eqnarray*}
\end{theorem}
\begin{proof}\smartqed
 Necessity. Let a strictly positive vector $p_0$ solving the set of equations (\ref{alla9})  exist. Introduce  notations
\begin{eqnarray} \label{alla11}
 y_i= \frac{ \langle b_i, p_0 \rangle}{ \langle C_i, p_0 \rangle}, \quad i=\overline{1,l}.
\end{eqnarray}
In view of the assumptions before the Theorem \ref{alla8} about the vectors $C_i=\{C_{ki}\}_{k=1}^n,$
$ b_i=\{b_{ki}\}_{k=1}^n, \ i=\overline{1,l}, $ we have
$y_i > 0, \ i=\overline{1,l},$
and the equality (\ref{alla10}) holds.
From the equalities (\ref{alla11}), we obtain the set of equalities
\begin{eqnarray} \label{alla12}
\langle b_i - y_i C_i, p_0 \rangle =0, \quad  i=\overline{1,l},
\end{eqnarray}
from which the conditions 2) and 3) of the Theorem \ref{alla8} follow because of strict positivity of the vector $p_0.$

 Sufficiency.
If the conditions 1), 2), and 4) of the Theorem \ref{alla8} hold, then there exist strictly positive numbers $\alpha_i >0, \ i=\overline{2,n}, $
such that
\begin{eqnarray} \label{alla13}
 - d_1= \sum\limits_{i=2}^n\alpha_id_i.
\end{eqnarray}
Let us put
$p_0=\{p_i^0\}_{i=1}^n,$ where $p_i^0=\alpha_i, \ i=\overline{2,n}, \ p_1^0=1. $
Then the set of equations (\ref{alla13}) turns into the set of equalities
\begin{eqnarray} \label{alla14}
\langle b_{i_k} - y_{i_k} C_{i_k}, p_0 \rangle  =0, \quad  k=\overline{1,r}.
\end{eqnarray}
The  set of vectors
 $b_{i_k} - y_{i_k}C_{i_k}, \  k=\overline{1,r}, $ contains maximum number of linearly independent vectors from the vector set $b_i - y_i C_i, \ i=\overline{1,l}. $ From here it follows that the set of equalities
(\ref{alla12}) holds for the vector $p_0$ introduced. From this fact and the condition 1) of the Theorem we have that $p_0$ solves the set of equations (\ref{alla9}).
\qed
\end{proof}

Consider a non-negative matrix $B$ given by the formula $B=CB_1,$ where
\begin{eqnarray*} B=|| b_{ik}||_{i=1, k=1}^{n,l}, \quad B_1=|| b_{ik}^1||_{i=1, k=1}^{l}.\end{eqnarray*}
The Theorem \ref{alla8} has a Corollary.

\begin{theorem}\label{alla16}
Let strictly positive components of the supply vector $\psi=\{\psi_k\}_{k=1}^n$ have the form $ \psi_k=\sum\limits_{i=1}^lb_{ki},\ k =\overline{1,n},$ and positive components of the vector $y_0=\{y_i^0\}_{i=1}^l$ have the form $y_i^0=\sum\limits_{k=1}^lb_{ik}^1, \ i=\overline{1,l}.$

The sufficient conditions of solvability for the set of equations (\ref{alla9}) in the set of strictly positive vectors
are the following:\\
1) the rank $r$ of the set of the vectors  $b_i - y_i^0 C_i, \ i=\overline{1,l}, $ does not exceed $n-1\  ;$\\
2) at least for one subset of $r$ linearly independent vectors \begin{eqnarray*} b_{i_k} - y_{i_k}^0C_{i_k}, \quad  k=\overline{1,r}, \end{eqnarray*}
 the vector
$d_1=\{b_{1 i_k} - y_{i_k}^0 C_{1 i_k} \}_{k=1}^r$
belongs to the interior of the cone created by vectors
 \begin{eqnarray*} -d_m=\left\{- b_{m i_k} + y_{i_k}^0 C_{m i_k} \right\}_{k=1}^r, \quad  m=\overline{2,n}.\end{eqnarray*}
\end{theorem}

As consequence of the previous investigation we obtain the next theorem.
\begin{theorem}\label{liuda1} 
Let  $\psi=\{\psi_k\}_{k=1}^n$ be a strictly positive vector belonging to the interior of the cone generated by the set of vectors $C_i=\{C_{ki}\}_{k=1}^n, \ i=\overline{1,l},$ being the column of the matrix $C=|C_{ki}|_{k=1, i=1}^{n, l},$
and also let the inequalities
\begin{eqnarray}\label{marijka1} \sum\limits_{s=1}^lC_{ks} >0,\quad  k=\overline{1,n},\quad  \sum\limits_{k=1}^nC_{ki} >0,  \quad i=\overline{1,l},\end{eqnarray} 
hold.
Suppose that there exist a vector $y=\{y_i\}_{i=1}^l, \ y_i > 0, \ i=\overline{1,l},$ such that
\begin{eqnarray*} \psi =\sum\limits_{i=1}^lC_iy_i  
\end{eqnarray*}
and a matrix $ \tau_1=|| \tau_{ki}^1||_{k=1, i=1}^{l,n}$ satisfying conditions:
  $\sum\limits_{i=1}^n \tau_{ki}^1=1, \  k=\overline{1,l}.$
If the matrix  elements  of the matrix $\tau=|| \tau_{ki}||_{k=1, i=1}^{l,n}$  constructed by  the rule
\begin{eqnarray*}
 \tau_{ki}= \frac{y_k  \tau_{ki}^1}{\psi_i}, \quad  k=\overline{1,l}, \quad i=\overline{1,n}, 
\end{eqnarray*} 
are such that the matrix  $\tau C$ is nonnegative and indecomposable, the matrix  $\tau C \tau$ is nonnegative and does not contain zero column, then  for the matrix $B=|| b_{ik}||_{i=1, k=1}^{n,l}$ having  the representation $B= CB_1, $ where
\begin{eqnarray*}  B_1=|| b_{ik}^1||_{i=1, k=1}^{l},\quad 
 b_{ks}^1= y_s  \sum\limits_{i=1}^n\tau_{ki} C_{is},\quad  k,s=\overline{1,l},
\end{eqnarray*} 
the set of equations (\ref{alla9}) has a strictly positive solution. If the rank of the set of the vectors $b_i -y_iC_i, \ i=\overline{1,l},$ equals $n-1,$ then a strictly positive solution is unique up to constant factor.
\end{theorem}
\begin{proof}\smartqed 
It is easy to verify  that $ \psi_k= \sum\limits_{i=1}^lb_{ki},\ k=\overline{1,n}.$  From the conditions of the Theorem \ref{liuda1},
the vector $y$ is the right Frobenius vector of the matrix $\tau C$ with proper value 1. The rest conditions of the Theorem \ref{mod3} are also valid. 
\qed
\end{proof}
\begin{theorem}\label{liuda2} Let the inequalities  (\ref{marijka1}) hold.
The set of equations  (\ref{alla9}) has a strictly positive solution if and only if there exist
a strictly positive vector $y=\{y_i\}_{i=1}^l, \ y_i > 0, \ i=\overline{1,l},$ and a matrix
\begin{eqnarray*}
 \tau_1=|| \tau_{ki}^1||_{k=1, i=1}^{l,n}, \quad
  \sum\limits_{i=1}^n \tau_{ki}^1=1, \quad  k=\overline{1,l},
\end{eqnarray*}
such that:\\
1) for a strictly positive  vector $\psi$ the representation
\begin{eqnarray*}
 \psi =\sum\limits_{i=1}^lC_iy_i
\end{eqnarray*}
 holds;\\ 
2) a matrix
\begin{eqnarray*}
\tau=|| \tau_{ki}||_{k=1, i=1}^{l,n}, \quad \tau_{ki}= \frac{y_k  \tau_{ki}^1}{\psi_i}, \quad  k=\overline{1,l}, \quad i=\overline{1,n},
\end{eqnarray*}
satisfying conditions: the matrix $\tau C$ is non-negative and indecomposable, the matrix  $\tau C \tau$ is non-negative and does not contain zero column;\\
3) the conditions
\begin{eqnarray*}
\langle b_i, p^0\rangle=\langle \bar b_i, p^0\rangle,  \quad  i=\overline{1,l}, 
\end{eqnarray*}
hold,  where
\begin{eqnarray*} 
\bar B=|| \bar b_{ik}||_{i=1, k=1}^{n,l}, \quad \bar B= CB_1,  \quad B_1=|| b_{ik}^1||_{i=1, k=1}^{l},
\end{eqnarray*}
\begin{eqnarray*}
 b_{ks}^1= y_s  \sum\limits_{i=1}^n\tau_{ki} C_{is},\quad  k,s=\overline{1,l}, \quad b_i=\{b_{ki}\}_{k=1}^n, \quad \bar b_i=\{\bar b_{ki}\}_{k=1}^n,
\end{eqnarray*} 
$p^0$ is a strictly positive vector constructed in the Theorem \ref{liuda1} for the matrix $\bar B$ instead of matrix $B.$ 
\end{theorem}
\begin{proof}\smartqed Necessity. Let there exist a strictly positive solution $p^0$ to the set of equations (\ref{alla9}). Denote
\begin{eqnarray*}
y_i=\frac{\langle b_i, p^0 \rangle}{\langle C_i, p^0 \rangle}, \quad  i=\overline{1,l},
\end{eqnarray*}
and introduce the matrix
\begin{eqnarray*}
\tau_1=|| \tau_{ki}^1||_{k=1, i=1}^{l,n}, \quad
   \tau_{ki}^1=\frac{p_i^0 \psi_i}{\langle \psi, p^0 \rangle},  \quad  k=\overline{1,l}, \quad  i=\overline{1,n}.
\end{eqnarray*}
It is evident that  $\sum\limits_{i=1}^n \tau_{ki}^1=1, \  k=\overline{1,l}.$
A matrix $\tau=|| \tau_{ki}||_{k=1, i=1}^{l,n}$  with matrix elements
\begin{eqnarray*}
 \tau_{ki}= \frac{y_k  \tau_{ki}^1}{\psi_i}=\frac{y_k p_i^0}{\langle \psi, p^0 \rangle}, \quad  k=\overline{1,l}, \quad i=\overline{1,n}, 
\end{eqnarray*} 
is such that the matrix
\begin{eqnarray*}
\tau C=\left | \frac{y_k  \langle C_s, p^0 \rangle }{\langle \psi, p^0 \rangle}   \right |_{k,s=1}^l
\end{eqnarray*}
is nonnegative and indecomposible, a matrix 
\begin{eqnarray*}
\tau  C \tau=\left |  \frac{y_k p_i^0}{\langle \psi, p^0 \rangle} \right|_{k=1, i=1}^{l, n}
\end{eqnarray*}
is nonnegative  and does not contain zero rows.
 It is not difficult to show that the vector $y$
 is a strictly positive proper vector of the matrix $ \tau C$ with a proper value 1, that is,
\begin{eqnarray}\label{liuda3}
\tau  C  y= y.
\end{eqnarray}
Let us construct the matrix
\begin{eqnarray*} 
\bar B=|| \bar b_{ik}||_{i=1, k=1}^{n,l}, \quad \bar B= CB_1,  \quad B_1=|| b_{ks}^1||_{k=1, s=1}^{l}, 
\end{eqnarray*}
where
\begin{eqnarray*}
 b_{ks}^1=y_s \sum\limits_{i=1}^n \tau_{ki}C_{is}=\frac{y_s y_k \sum\limits_{i=1}^n p_i^0 C_{is}}{\langle \psi, p^0 \rangle}, \quad  k,s =\overline{1,l},
\end{eqnarray*}
\begin{eqnarray*}
 \bar b_{ki}= \sum\limits_{j=1}^lC_{kj}b_{ji}^1,\quad  k=\overline{1,n}, \quad i=\overline{1,l}.
\end{eqnarray*} 
There hold equalities
\begin{eqnarray*}
\sum\limits_{i=1}^l\bar b_{ki}=\psi_k, \quad  k=\overline{1,n}.
\end{eqnarray*}
The conjugate problem
\begin{eqnarray*}
\sum\limits_{k=1}^l\delta_k^0\sum\limits_{i=1}^n \tau_{ki}C_{is} = \delta_s^0, \quad s=\overline{1,l},
\end{eqnarray*}
to the problem (\ref{liuda3}) has strictly positive solution $  \delta^0=\{\delta^0_s \}_{s=1}^l, $  $ \delta^0_s=\sum\limits_{i=1}^np_i^0 C_{is}.$
In correspondence with the conditions of the Theorem \ref{mod3},
let us construct the vector $ \bar p^0 =\{\bar p_i^0 \}_{i=1}^n.$  We have
\begin{eqnarray*}
\bar p_i^0=\sum\limits_{k=1}^l \delta^0_k \tau_{ki}=\sum\limits_{k=1}^l \sum\limits_{i=1}^np_i^0 C_{ik} \frac{y_k p_i^0}{\langle \psi, p^0 \rangle}= p_i^0, \quad i=\overline{1,n}.
\end{eqnarray*}
Since 
\begin{eqnarray*}
 \bar b_{ki}= \sum\limits_{j=1}^lC_{kj}b_{ji}^1,= \frac{\psi_k y_i \sum\limits_{m=1}^n p_m^0 C_{mi}}{\langle \psi, p^0 \rangle}, \quad  k=\overline{1,n}, \quad i=\overline{1,l},
\end{eqnarray*} 
we have 
\begin{eqnarray*}
\langle \bar b_i, p^0 \rangle =\sum\limits_{k=1}^n \bar b_{ki} p_k^0=y_i  \sum\limits_{m=1}^n p_m^0 C_{mi}= \sum\limits_{k=1}^n  b_{ki} p_k^0= \langle b_i, p^0 \rangle.
\end{eqnarray*}
The necessity is proved.

Sufficiency. If sufficient conditions of the Theorem \ref{liuda2}  are valid,  then we are in the  condions of the Theorem \ref{liuda1}.   From the Theorem \ref{liuda1} and the conditions of the Theorem \ref{liuda2}  we obtain that the set of equations 
\begin{eqnarray*}
\sum\limits_{i=1}^l  C_{ki} \frac{\langle\bar b_i, p \rangle }{\langle\bar C_i, p \rangle}= \psi_k, \quad k=\overline{1,n},
\end{eqnarray*}
has a strictly positive solution $p^0$  satisfying conditions
\begin{eqnarray*}
\langle\bar b_i, p^0 \rangle =\langle b_i, p^0 \rangle, \quad i=\overline{1,l}.
\end{eqnarray*}
This proves the Theorem. 
\qed
\end{proof}

\begin{corollary}
If $p^0$ is a strictly positive solution to the set of equations  (\ref{alla9}), then 
for the vectors $b_i$ the representations 
\begin{eqnarray*}
b_i=\bar b_i +d_i, \quad i=\overline{1,l},
\end{eqnarray*}
hold, where
\begin{eqnarray*}
  \sum\limits_{i=1}^ld_i=0, \quad \langle p^0, d_i \rangle=0, \quad \bar b_i=\frac{\psi y_i \langle C_i, p^0  \rangle }{ \langle \psi, p^0  \rangle}, \quad i=\overline{1,l}. 
\end{eqnarray*}
\end{corollary}
\begin{corollary} For given strictly positive vectors $\psi, y, p^0$ and matrix $ C$ satisfying conditions  (\ref{marijka1}) and the vector $\psi$ having the represantation $\psi=\sum\limits_{i=1}^lC_iy_i, \ C_i=\{c_{ki}\}_{k=1}^n,\ i=\overline{1,l},$ the set of equations  (\ref{alla9}) has strictly positive solution $p^0$ if and only if there exists vectors $d_i, \ i=\overline{1,l},$ such that
\begin{eqnarray*}
b_i=\bar b_i +d_i, \quad \langle p^0, d_i \rangle=0,   \quad i=\overline{1,l}, \quad \sum\limits_{i=1}^ld_i=0,
\end{eqnarray*}
where
\begin{eqnarray*}
\bar b_i=\frac{\psi y_i \langle C_i, p^0  \rangle }{ \langle \psi, p^0  \rangle}, \quad  \quad i=\overline{1,l}.
\end{eqnarray*}
\end{corollary}
Partial case of the model considered is the model with the number of consumers  that equal to the number of goods  that consumers exchange with each other and every consumer has only single goods type.
In this case $D_i(p)=b_ip_i, \  b_i > 0, \ i=\overline {1,n}.$

Let us establish the necessary and sufficient conditions for the existence of a strictly positive solution to the set of equations
\begin{eqnarray} \label{g2l34} \sum\limits
_{i=1}^n\frac{p_kC_{ki}b_ip_i}{
\sum\limits _{s=1}^nC_{si}p_s}=p_k b_k, \quad k=\overline {1,n}.
\end{eqnarray}
Consider the set of inequalities
\begin{eqnarray} \label{g2l35}
\sum\limits
_{i=1}^n\frac{C_{ki}b_ip_i}{
\sum\limits _{s=1}^nC_{si}p_s} \leq b_k, \quad k=\overline {1,n}.
\end{eqnarray}
The main problem  is that  to describe the set of all the solutions to the set of inequalities (\ref{g2l35}).

Let $p^0 $ be a price vector satisfying the set of inequalities
\begin{eqnarray*} \sum\limits
_{i=1}^n\frac{C_{ki}b_ip_i}{\sum\limits _{s=1}^n C_{si}p_s}
-b_k=0,\quad k\in I,\end{eqnarray*}
\begin{eqnarray} \label{g4l36}
\sum\limits _{i=1}^n\frac{C_{ki}b_ip_i}{\sum\limits _{s=1}^n C_{si}p_s}
-b_k<0,\quad k\in J,
\end{eqnarray}
where $I$ and $J$ are subsets of the set
$N=\{1,2,\dots n\},$ $~I\bigcap J=\emptyset,$ and $\emptyset$ is empty set,
$~J\bigcup I=N.$
Let us describe all the solutions to the set of inequalities (\ref{g4l36}).
\begin{lemma}\label{Pas1} Let the vector $p^0 =(p_1^0
,\ldots p_n^0 )\in R_+^n$ be a certain solution to the set of inequalities (\ref{g4l36}), then components of the vector $p^0$
whose indices belong to the set $J$ equal zero.
\end{lemma}
\begin{proof}\smartqed
 If $p^0 $ solves (\ref{g4l36}), then it solves (\ref{g2l34}).  Multiplying the $k$-th  inequality of (\ref{g4l36}) by $p_k^0 $ and summing up over all $k,$ we obtain
\begin{eqnarray*} 0=\sum\limits _{k=1}^n\left(\sum\limits
_{i=1}^n\frac{C_{ki} b_ip_i^0}{\sum\limits
_{s=1}^nC_{si}p_s^0} -b_k\right) p_k^0 =\sum\limits _
{k\in J}\left(\sum\limits _{i=1}^n\frac{C_{ki}b_ip_i^0}
{\sum\limits _
{s=1}^n C_{si}p_s^0}-b_k\right) p_k^0 .\end{eqnarray*}
If at least a certain component
$p_k^{0}>0$, $k\in J,$ then rhs of this equality is strictly negative  and lhs equals zero identically. This contradiction proves the Lemma.
\qed
\end{proof}
The fact that $p^0 $ solves (\ref{g4l36}) means that
$\sum\limits _{s=1}^n
C_{si}p_s^0 >0,$ $\ i=\overline {1,n}.$ Further we will use the notation
\begin{eqnarray*} y_i^0 =\frac {p_i^0}{\sum\limits _{s=1}^n
C_{si}p_s^0 }, \quad i=\overline {1,n}.\end{eqnarray*}

\begin{theorem} Let the vector $y^0 =(y_1^0 ,\dots
,y_n^0 )$ be a certain solution to the set of inequalities
\begin{eqnarray*} \sum\limits
_{i=1}^nC_{ki}b_iy_i= b_k,\quad k\in I,\end{eqnarray*}
\begin{eqnarray} \label{g4l37}
\sum\limits _{i=1}^nC_{ki}b_iy_i< b_k,\quad k\in J,
\end{eqnarray}
and satisfies  conditions $y_i^0\geq 0,$ $~i=\overline{1,n},$
$~I\bigcup J=N,$ $~y_i^0=0,$ $~i\in J,$ then there exists a non-negative solution $p^0 =(p_1^0 ,\dots ,p_n^0 )$
to the problem \begin{eqnarray*} p_i^0 =y_i^0\sum\limits _{l=1}^n C_{li}p_l^0
, \quad i=\overline {1,n}.\end{eqnarray*}  If this solution is such that
\begin{eqnarray*} \sum\limits _{s=1}^n C_{si}
p_s^0\not= 0,\quad  i=\overline {1,n},\end{eqnarray*}  then the vector $p^0 $
solves the problem (\ref{g4l36}).
 \end{theorem}
\begin{proof}\smartqed  Let $I$ be non-empty set, then those components of the vector $y^0 $ that solve the set of inequalities (\ref{g4l37}) and belong to the set $I$ are determined  by the set of equations
\begin{eqnarray} \label{g4l38}
\sum\limits _{i\in I}C_{ki}b_iy_i^0 =b_k,\quad k\in
I.  \end{eqnarray}
Denote by $C_{y^0}^I=||C_{ki}y_i^0 ||_{k,i \in I}$ the matrix of the dimension $|I|\times |I|$
whose matrix elements are given by the formula $C_{ki}y_i^0, \quad k,i\in I,$
where $|I|$ is the number of elements of the set
$I.$ The operator
$C_{y^0 }^I,$
constructed after the matrix
$||C_{ki}y_i^0 ||_{k,i \in I}$ in the space of the dimension $|I|$ has the norm that does not exceed 1 in the norm
$||x||=\max\limits _{i\in I}\frac{|x_i|}{b_i}, \ x \in R^{|I|}.$

 From (\ref{g4l38}) it follows that 1 is proper value of the problem
\begin{eqnarray*} C_{y^0}^I\tilde x=\tilde x \end{eqnarray*}
with proper vector $\tilde x=\{b_i\}_{i \in I}.$
The matrix conjugate to the matrix $C_{y^0 }^I$ is the transposed one. As proper values of the transposed matrix are the same as  of the matrix $C_{y^0 }^I,$  then 1 is maximum proper value of the matrix transposed to the matrix $\bar C_{y^0 }^I.$
By the Perron-Frobenius Theorem\index{Perron-Frobenius Theorem} there exists a non-negative vector
$\tilde p^0 $  of the conjugate problem
\begin{eqnarray} \label{g4l39}
y_i^0\sum\limits _{k\in I}C_{ki}\tilde p_k^0 =
\tilde p_i^0 ,\quad i\in I.
\end{eqnarray}

Define the vector
$p^0 =(p_1^0 ,\dots ,p_n^0 )$
by the rule: if indices of components of the vector
$p^0 $ belong to the set
$J,$ then they equal zero; those components of the vector
$p^0$ whose indices belong to the set
$I$ are given by the formula
\begin{eqnarray*} p_i^0 =\tilde p_i^0,\quad i\in I.\end{eqnarray*}
The constructed vector solves the problem
\begin{eqnarray} \label{g4l40}
y_i^0\sum\limits _{k=1}^nC_{ki}p_k^0 =p_i^0
, \quad i=\overline {1,n}.
\end{eqnarray}
If this solution is such that
$\sum\limits _{k=1}^n
C_{ki}p_k^0\not= 0,$ $\  i={\overline {1,n}},$ then
\begin{eqnarray*} y_i^0 =\frac{p_i^0}{\sum\limits _{k=1}^n
C_{ki}p_k^0}, \quad i=\overline {1,n}.\end{eqnarray*}
Inserting the expression for $y_i^0 $ into (\ref{g4l37}), we obtain the proof of the Theorem.
\qed
\end{proof}
\begin{theorem}\label{alla3} If the matrix $C$ is indecomposable and the vector $y^0 $
is such that $y_i^0 >0, \ i=\overline {1,n},$ then the vector $p^0
=(p_1^0 ,\dots ,p_n^0 )$ solving the problem
(\ref{g4l36}) is such that $p_i^0> 0,\ i=\overline {1,n}.$
\end{theorem}

\begin{theorem}\label{alla4}
Let $C$ be an indecomposable matrix.\index{indecomposable matrix} The necessary and sufficient condition for solvability of the set of equations (\ref{g2l34}) in the set of strictly positive vectors is belonging of the vector $b=\{b_i\}_{i=1}^n$ to the interior of the cone created by vectors-columns of the matrix\index{cone created by vectors-columns of the matrix} $C.$
\end{theorem}
\begin{proof}\smartqed  Necessity. If there exists a strictly positive solution $p^0=\{p_i^0\}_{i=1}^n$ to the set of equations (\ref{g2l34}), then the vector $p^0$ satisfies also the set of equations
\begin{eqnarray} \label{alla5} \sum\limits
_{i=1}^n\frac{C_{ki}b_ip_i^0}{
\sum\limits _{s=1}^nC_{si}p_s^0}= b_k, \quad k=\overline {1,n}.
\end{eqnarray}
Using introduced notations
\begin{eqnarray*} y_i^0 =\frac{p_i^0}{\sum\limits _{k=1}^n
C_{ki}p_k^0} > 0, \quad i=\overline {1,n},\end{eqnarray*}
we can express (\ref{alla5}) as
\begin{eqnarray} \label{alla6} \sum\limits
_{i=1}^nC_{ki}b_i y_i^0= b_k, \quad k=\overline {1,n}.
\end{eqnarray}
The last means that the vector $b=\{b_i\}_{i=1}^n$ belongs to the interior of the cone created by vectors-columns of the matrix\index{interior of the cone created by vectors-columns of the matrix} $C.$

 Sufficiency. If the vector $b=\{b_i\}_{i=1}^n$ belongs to the interior of the cone created by vectors-columns of the matrix $C,$ then there exists such strictly positive vector $\xi=\{\xi_i\}_{i=1}^n$ that $b=C\xi.$ From the last it follows that the set of equations
\begin{eqnarray} \label{alla26}
\sum\limits _{i=1}^nC_{ki}b_iy_i^0 =b_k,\quad k=\overline{1,n}.
\end{eqnarray}
has a strictly positive solution $y^0=\{y_i^0\}_{i=1}^n,$ where $y_i^0=\xi_i/b_i, \
i=\overline{1,n}.$ By the Theorem \ref{alla3} the set of equations (\ref{g2l34}) has a strictly positive solution.
\qed
\end{proof}

\section{Consumption economy model with fixed gains}

 	We introduced a model of consumption economy  with fixed gains\index{model of consumption economy  with fixed gains} and studied it in the papers and monographs \cite{17}, \cite{16},   \cite{55}, \cite{71}. Under rather simple restrictions on the consumption structure,\index{consumption structure} supply vector,\index{supply vector} and consumers gains,\index{consumers gains} we proved the existence Theorem for equilibrium price vector.
Suppose that in a certain economic system there are $n$ kinds of goods and $l$ consumers.
Consider that the $i$-th consumer has fund $D_i>0,\
 i=\overline{1,l}.$ On the economic system market, goods supply vector\index{goods supply vector} has the form \begin{eqnarray*} \psi=\{\psi _i\}_{i=1}^n, \quad  \psi _i > 0, \quad  i=\overline{1,n}. \end{eqnarray*}

The set of possible price vectors is a cone $K_+^n \subset  \bar R_+^n,$ being a subcone of the cone $\bar R_+^n.$
Let us build the cone $K_+^n.$
Let $C=||c_{ik}||_{i,k=1}^{n,l}$ be a certain  non-negative matrix of the dimension $n \times l$
satisfying conditions
\begin{eqnarray*} \sum\limits_{k=1}^n c_{ki} > 0, \quad  i=\overline{1,l}, \quad
 \min\limits_{k,  s \ c_{ks} \neq 0}c_{ks}=1.\end{eqnarray*}
  The matrix $C$ determines the cone $K_+^n $ by the rule
\begin{eqnarray} \label{algl1}
K_+^n=\left\{ p \in \bar R_+^n, \ \sum\limits_{k=1}^n c_{ki}p_k > 0, \ i=\overline{1,l}\right\}.
\end{eqnarray}
Consider the case of insatiable consumers.
Suppose that random   fields of information evaluation by consumers satisfy the condition:
for every $i$-th consumer the random   field of information evaluation $\eta_i^0(p,\omega_i)$ by the $i$-th consumer on a  probability space
 $\{\Omega_i, {\cal F}_i, P_i\},\ i=\overline{1,l},$
satisfies the inequality
\begin{eqnarray} \label{algl0}
\eta_i^0(p, \omega_i) \geq m_i C_i, \quad  (p, \omega_i) \in  K_+^n\times  \Omega_i,  \quad m_i > 0, \quad  i=\overline{1,l},
\end{eqnarray}
where $C_i=\{c_{ki}\}_{k=1}^n,$
and components $\eta_{ik}^0(p,\omega_i)$ of the random field of information evaluation by the $i$-th consumer
 \begin{eqnarray*} \eta_i^0(p,\omega_i)=\{\eta_{ik}^0(p, \omega_i)\}_{k=1}^n\end{eqnarray*}
satisfy the condition: $\eta_{ik}^0(p,\omega_i)=0$ if and only if $c_{ki}=0.$

Suppose that consumers operate independently and their random  fields of choice  have the form
\begin{eqnarray*}  \xi_i(p)= \frac{D_i \eta_i^0(p,\omega_i)}{\langle p,  \eta_i^0(p,\omega_i) \rangle}, \quad i=\overline{1,l}.\end{eqnarray*}

In the next Theorem, we assume that the above formulated conditions hold for random fields of consumers choice    on the cone built above.

Note that if certain components of the vector $C_i=\{c_{ki}\}_{k=1}^n$ equal zero, then the $i$-th consumer does not consume goods numbered by these components.

Let $\mu_i(p)=\{\mu_{ki}(p)\}_{k=1}^n$ be a continuous realization of the random field of information evaluation by consumer
\begin{eqnarray*} \eta_i(p,\omega_i)=
\eta_i^0(p,\omega_i)
=\{\eta_{ik}^0(p,\omega_i)\}_{k=1}^n. \end{eqnarray*}
Therefore, $\mu_{ki}(p)=\eta_{ik}^0(p,\omega_i)$ for some $\omega_0.$
As earlier, denote
\begin{eqnarray*}  \gamma_{ik}(p)=\frac{\mu_{ki}(p)p_k}{\sum\limits_{j=1}^n
\mu_{ji}(p)p_j}, \quad k=\overline {1,n},\quad i=\overline {1,l},\end{eqnarray*}
components of the demand vector
\begin{eqnarray*}  \gamma_i(p)=\{ \gamma_{ik}(p)\}_{k=1}^n, \quad  i=\overline {1,l}.\end{eqnarray*}
Introduce into  consideration a set
\begin{eqnarray*} C_{\delta} =\left\{p \in K_+^n,\ \sum\limits_{s=1}^nc_{si}p_s \geq \delta ,\
i=\overline {1,l}, \ \sum\limits_{i=1}^np_i\psi_i=\sum\limits_{i=1}^lD_i \right\}\end{eqnarray*}
and notations $R_1=\max\limits_{k}\psi_k,\ R_0=\min\limits_{k}\psi_k, \ D_0=\sum\limits_{i=1}^lD_i. $

\begin{theorem}\label{algl2}
Suppose that random fields of information evaluation by  consumers  on the cone $K_+^n,$ given by the formula (\ref{algl1}), are continuous  with  probability 1, satisfy the condition (\ref{algl0}), and a number $\delta$ satisfies  the inequality
\begin{eqnarray*}  0 < \delta \leq \min \left\{\min\limits_{i,k}\frac{D_i}{\psi_k}, \ \frac{D_0}{n R_1} \min\limits_{i}\sum\limits_{k=1}^n c_{ki}\right\}.\end{eqnarray*}
Then for every continuous on $K_+^n$ demand matrix $||\gamma_{ik}(p)||_{i=1, k=1}^{l,\ n},$ i.e., with probability 1, there exists a corresponding price vector $\bar p$ satisfying the set of equations
\begin{eqnarray} \label{algl4}
\sum\limits
_{i=1}^l\frac{\mu_{ki}(p)p_k}{\sum\limits_{j=1}^n
\mu_{ji}(p)p_j}D_i = p_k \psi_k, \quad  k=\overline{1,n}.
\end{eqnarray}
\end{theorem}
\begin{proof}\smartqed   Introduce into consideration a map
\begin{eqnarray*} \Gamma(p)=\{ \Gamma_k(p)\}_{k=1}^n, \end{eqnarray*}
\begin{eqnarray*} \Gamma_k(p)=\frac{1}{\psi_k}\sum\limits_{i=1}^l
\frac{\mu_{ki}(p)p_kD_i}{\sum\limits_{j=1}^n
\mu_{ji}(p)p_j}, \quad k=\overline {1,n},\end{eqnarray*}
and show that  it  maps the non-empty set $C_{\delta}$ into itself.

If $\delta$ satisfies the conditions of the  Theorem, then the set $C_{\delta}$
is not empty  and the inequality
\begin{eqnarray*} \min_{\{k,s, f_{ks}=1\}}\frac{D_s}{\psi_k} \geq \delta\end{eqnarray*}
holds, where
\begin{eqnarray*} f_{ks}=\left\{\begin{array}{ll}
                 1, & \textrm{if} \quad c_{ks} \neq 0 \textrm{,}\\
                 0, & \textrm{if} \quad  c_{ks}  =0 \textrm{.}
                 \end{array} \right.\end{eqnarray*}
Check the validity of the inequalities
\begin{eqnarray*} \sum\limits_{k=1}^nc_{ki} \Gamma_k(p) \geq \delta, \quad i=\overline {1,l}.\end{eqnarray*}
Really,
\begin{eqnarray*} \sum\limits_{k=1}^nc_{ki}\Gamma_k(p) =  \sum\limits_{k=1}^nc_{ki}
\frac{1}{\psi_k}\sum\limits_{s=1}^l
\frac{\mu_{ks}(p)p_kD_s}{\sum\limits_{j=1}^n
\mu_{js}(p)p_j}
\end{eqnarray*}
\begin{eqnarray*}  \geq \delta \sum\limits_{k=1}^nc_{ki}
\sum\limits_{s=1}^l
\frac{f_{ks}\mu_{ks}(p)p_k}{\sum\limits_{j=1}^n
\mu_{js}(p)p_j}=\delta\sum\limits_{s=1}^l
\frac{\sum\limits_{k=1}^nc_{ki}f_{ks}\mu_{ks}(p)p_k}{\sum\limits_{j=1}^n
\mu_{js}(p)p_j}.\end{eqnarray*}
Because of assumptions about the matrix elements $c_{ki},$ for  $s=i$
\begin{eqnarray*} \frac{\sum\limits_{k=1}^nc_{ki}f_{ki}\mu_{ki}(p)p_k}{\sum\limits_{j=1}^n
\mu_{ji}(p)p_j} \geq \min_{\{k,i,f_{ki}=1\}}c_{ki}\frac{\sum\limits_{k=1}^n\mu_{ki}(p)p_k}{\sum\limits_{j=1}^n\mu_{ji}(p)p_j}=1.\end{eqnarray*}
Therefore,
\begin{eqnarray*} \sum\limits_{k=1}^nc_{ki}\Gamma_k(p) \geq \delta, \quad i=\overline{1,l}.\end{eqnarray*}
The map $\Gamma(p)$ is a continuous map of the convex compact set $C_{\delta}$ into itself.
By the Schauder Theorem\index{Schauder Theorem} \cite{88}, there exists a fixed point of the map $\Gamma(p).$
\qed
\end{proof}

\begin{theorem}\label{rect1}
Assume that random fields of information evaluation by  consumers on the cone $K_+^n$ given by the formula (\ref{algl1}) are continuous  with probability 1, satisfy the condition (\ref{algl0}), and a number $\delta$ satisfies conditions of the Theorem \ref{algl2}.
If $\sum\limits_{i=1}^lf_{ki}>0, \ k=\overline{1,n},$
then for every continuous  demand matrix $||\gamma_{ik}(p)||_{i=1, k=1}^{l,\ n}$ on $K_+^n,$ i.e., with  probability 1, there exists corresponding it a strictly positive price vector
$p^{\varepsilon} \in C_{\delta},$ solving the set of equations
\begin{eqnarray} \label{rect2}
\sum\limits
_{i=1}^l\gamma _{ik}^{\varepsilon}(p)D_i = p_k \psi_k, \quad  k=\overline{1,n},
\end{eqnarray}
where
\begin{eqnarray*} \gamma _{ik}^{\varepsilon}(p)=
\frac{\mu_{ki}(p)p_k+ \varepsilon f_{ki}}{\sum\limits_{j=1}^n
\mu_{ji}(p)p_j+ \varepsilon\sum\limits_{j=1}^nf_{ji}}, \quad k=\overline {1,n},\quad i=\overline {1,l}, \quad  0 < \varepsilon < 1. \end{eqnarray*}
\end{theorem}
\begin{proof}\smartqed
 Introduce into consideration a map
\begin{eqnarray*} \Gamma^{\varepsilon}(p)=\{ \Gamma_k^{\varepsilon}(p)\}_{k=1}^n. \end{eqnarray*}
\begin{eqnarray*} \Gamma_k^{\varepsilon}(p)=\frac{1}{\psi_k}\sum\limits_{i=1}^l
\frac{(\mu_{ki}(p)p_k + \varepsilon f_{ki})D_i}{\sum\limits_{j=1}^n
\mu_{ji}(p)p_j + \varepsilon \sum\limits_{j=1}^nf_{ji}}, \quad k=\overline {1,n},\end{eqnarray*}
and show that it  maps the non-empty set $C_{\delta}$ into itself.

If $\delta$ satisfies the  conditions of the Theorem, then the set $C_{\delta}$
is not empty and the inequality
\begin{eqnarray*} \min_{\{k,s, f_{ks}=1\}}\frac{D_s}{\psi_k} \geq \delta,\end{eqnarray*}
holds.
Check the validity of the inequalities
\begin{eqnarray*} \sum\limits_{k=1}^nc_{ki} \Gamma_k^{\varepsilon}(p) \geq \delta, \quad i=\overline {1,l}.\end{eqnarray*}
Really,
\begin{eqnarray*} \sum\limits_{k=1}^nc_{ki}\Gamma_k^{\varepsilon}(p) =  \sum\limits_{k=1}^nc_{ki}
\frac{1}{\psi_k}\sum\limits_{s=1}^l
\frac{(\mu_{ks}(p)p_k + \varepsilon f_{ks})D_s}{\sum\limits_{j=1}^n
\mu_{js}(p)p_j + \varepsilon\sum\limits_{j=1}^nf_{ji}}
\end{eqnarray*}
\begin{eqnarray*}  \geq \delta \sum\limits_{k=1}^nc_{ki}
\sum\limits_{s=1}^l
\frac{f_{ks}(\mu_{ks}(p)p_k +  \varepsilon f_{ks})}{\sum\limits_{j=1}^n
\mu_{js}(p)p_j +  \varepsilon \sum\limits_{j=1}^nf_{js}}=\delta\sum\limits_{s=1}^l
\frac{\sum\limits_{k=1}^nc_{ki}f_{ks}(\mu_{ks}(p)p_k + \varepsilon  f_{ks}) }{\sum\limits_{j=1}^n
\mu_{js}(p)p_j+ \varepsilon  \sum\limits_{j=1}^nf_{js}}.\end{eqnarray*}
Due to assumptions about the matrix elements $c_{ki},$ for $s=i$
\begin{eqnarray*} \frac{\sum\limits_{k=1}^nc_{ki}f_{ki}(\mu_{ki}(p)p_k + \varepsilon  f_{ki})}{\sum\limits_{j=1}^n
\mu_{ji}(p)p_j + \varepsilon  \sum\limits_{j=1}^nf_{ji}} \geq \min_{\{k,i,f_{ki}=1\}}c_{ki}\frac{\sum\limits_{k=1}^n(\mu_{ki}(p)p_k + \varepsilon f_{ki})}{\sum\limits_{j=1}^n\mu_{ji}(p)p_j+ \varepsilon\sum\limits_{j=1}^nf_{ji}}=1.\end{eqnarray*}
Therefore,
\begin{eqnarray*} \sum\limits_{k=1}^nc_{ki}\Gamma_k^{\varepsilon}(p) \geq \delta, \quad i=\overline{1,l}.\end{eqnarray*}
The map $\Gamma^{\varepsilon}(p)$ is a continuous map of the convex compact set $C_{\delta}$ into itself.
By the Schauder Theorem \cite{88}, there exists a fixed point $p^{\varepsilon}=\{p^{\varepsilon}_k\}_{k=1}^n$ of the map $\Gamma^{\varepsilon}(p)$ whose components satisfy inequalities
\begin{eqnarray*}  p_k^{\varepsilon} \geq  \frac{ \varepsilon  \min\limits_{i}D_i \sum\limits_{i=1}^lf_{ki}}{R_1( \mu + \varepsilon \max\limits_{i} \sum\limits_{j=1}^nf_{ji} )}> 0, \quad k=\overline{1,n},\end{eqnarray*}
where
\begin{eqnarray*} \mu=\max\limits_{i}\sup\limits_{p \in P\cap K_+^n}\sum\limits_{s=1}^n\mu_{si}(p)p_s< \infty.
\end{eqnarray*}
\qed
\end{proof}

\begin{theorem}\label{algl5}
Let the conditions of the Theorem \ref{algl2}  and the inequalities
\begin{eqnarray*} \sum\limits_{i=1}^lf_{ki}>0, \quad  k=\overline{1,n},\end{eqnarray*}
 hold. Then there exists an equilibrium price vector $p^0 \in C_{\delta}$ for which the demand does not exceed the supply, i.e., the set of inequalities
\begin{eqnarray} \label{algl6}
\sum\limits_{i=1}^l\frac{\mu_{ki}(p^0)}{\sum\limits_{s=1}^n\mu_{si}(p^0)p_s^0}D_i\leq   \psi_k, \quad  k=\overline{1,n},
\end{eqnarray}
hold.
Every equilibrium price vector satisfies the set of equations (\ref{algl4}).
\end{theorem}
\begin{proof}\smartqed  Consider the auxiliary set of equations
\begin{eqnarray} \label{algl7}
\sum\limits
_{i=1}^l\gamma _{ik}^{\varepsilon}(p)D_i = p_k \psi_k, \quad  k=\overline{1,n},
\end{eqnarray}
built after the set of equations (\ref{algl4}), where
\begin{eqnarray*} \gamma _{ik}^{\varepsilon}(p)=
\frac{\mu_{ki}(p)p_k+ \varepsilon f_{ki}}{\sum\limits_{j=1}^n
\mu_{ji}(p)p_j+ \varepsilon\sum\limits_{j=1}^nf_{ji}}, \quad k=\overline {1,n},\quad i=\overline {1,l}, \quad  0 < \varepsilon < 1. \end{eqnarray*}

Every component of a solution $p^{\varepsilon}= \{ p_k^{\varepsilon}\}_{k=1}^n$ for the set of equations (\ref{algl7}) is strictly positive  by the Theorem \ref{rect1}.
From the fact that the strictly positive vector $p^{\varepsilon}$ satisfies the set of equations (\ref{algl7}), the validity of the set of inequalities
\begin{eqnarray} \label{algl9}
\sum\limits_{i=1}^l\frac{\mu_{ki}(p^{\varepsilon})}{\sum\limits_{s=1}^n\mu_{si}(p^{\varepsilon})p_s^{\varepsilon}+ \varepsilon \sum\limits_{s=1}^nf_{si} }D_i\leq   \psi_k, \quad  k=\overline{1,n},
\end{eqnarray}
follows.
The sequence $p^{\varepsilon} $ for $\varepsilon \to 0$ is compact  because it belongs to the compact set $C_{\delta}.$
Due to the continuity of $\sum\limits_{s=1}^n\mu_{si}(p)p_s$ on the set $C_{\delta}$ and the inequality
\begin{eqnarray*} \sum\limits_{s=1}^n\mu_{si}(p)p_s > m_i\delta,\quad  p \in C_{\delta},\end{eqnarray*}
one can go to the limit in the set of inequalities
(\ref{algl9}), as $\varepsilon \to 0.$ Denote one of the possible limit points of the sequence $p^{\varepsilon}$ by  $p^0.$
Then $p^0$ solves the set of inequalities
\begin{eqnarray} \label{algl10}
\sum\limits_{i=1}^l \frac{\mu_{ki}(p^{0})}
{\sum\limits_{s=1}^n\mu_{si}(p^{0})p_s^{0}} D_i\leq \psi_k,\quad k=\overline{1,n}.
\end{eqnarray}
It is obvious that the vector $p^0$ belongs to the set $C_{\delta}.$
\qed
\end{proof}

Further, we study in detail a particular case when the net income\index{net income} $D_i, \ i=\overline{1,l},$~
a structure matrix of consumption\index{structure matrix of consumption}   ${C_0=||C_{ki}||}_{k=1,i=1}^{n,l}$ and a supply vector $\psi=\{\psi_k\}_{k=1}^n$ do not depend on the price vector. Therefore, further we suppose that $\mu_{ki}(p)=C_{ki}, \ k=\overline{1,n}, \  i=\overline{1,l},$ \
$ \sum\limits_{k=1}^nC_{ki} > 0, \  i=\overline{1,l}.$

\begin{lemma}\label{Pas2} If the vector $p^0 =(p_1^0 ,\dots
,p_n^0 ) \in B$ solves the set of inequalities
\begin{eqnarray*} \sum\limits _{i=1}^l\frac
{C_{ki}D_i}{\sum\limits _{s=1}^n C_{si}p_s}=\psi _k, \quad k\in
I,\end{eqnarray*}
\begin{eqnarray} \label{gonl27} \sum\limits _{i=1}^l\frac
{C_{ki}D_i}{\sum\limits _{s=1}^nC_{si}p_s} <\psi _k,\quad k\in J,
 \end{eqnarray}
then $p_k^0 =0,$ $k\in J,$ where $I\cup J=\{1,\dots ,n\},$
$I\cap J=\emptyset,$~  $D_i>0,$~ $i=\overline{1,l},$ ~$\psi _i>0, \
i=\overline{1,n},$
\begin{eqnarray*} B=\left\{p=\{p_i\}_{i=1}^n \in R_+^n,\ \sum\limits _{i=1}^n
p_i\psi _i=\sum\limits _{i=1}^lD_i\right\}.\end{eqnarray*}  \end{lemma}

The Proof of the Lemma \ref{Pas2} is similar to the Proof of the Lemma  \ref{Pas1}.

\begin{theorem}\label{Pas3} Let the vector $y=\{y_1,\dots, y_l\}$ be a solution of the set of inequalities
\begin{eqnarray*} \sum\limits _{i=1}^lC_{ki}y_i=\psi _k,\quad k\in I,\end{eqnarray*}
\begin{eqnarray} \label{gonl28}
   \sum\limits _{i=1}^lC_{ki}y_i<\psi _k,\quad k\in J,
\end{eqnarray}
where $I$ and $J$ are the same as in the Lemma \ref{Pas2}, and also $y_i>0,~
i=\overline {1,l}.$

If there exists a solution $ \tilde p=\{\tilde p_s\}_{s=1}^n$ to the set of equations \begin{eqnarray} \label{gonl29} \sum\limits _{s\in
I}C_{si}\tilde {p_s} =\frac {D_i}{ y_i}, \quad i=\overline {1,l},
\end{eqnarray}
then the vector $p^0 ={\{p^0 _s\}}_{s=1}^n$
solves the set of inequalities
\begin{eqnarray} \label{gonl30}
\sum\limits _{i=1}^l\frac {C_{ki}D_i}{\sum\limits
_{s=1}^nC_{si}p^0 _s}\leq\psi _k, \quad k=\overline{1,n},
\end{eqnarray}
where
$p_s^0 =\tilde p_s$ for those  $s$, for which  $\tilde p_s>0,$
and $p^0 _s=0,$ for $\tilde p_s\leq 0$  or $s \in J.$
\end{theorem}
\begin{proof}\smartqed
 From the fact that the set of equations (\ref{gonl29}) has a solution
and $\frac {D_i} {y_i}>0,$ it follows that the inequalities
\begin{eqnarray*} \frac {D_i} {y_i}=\sum\limits _{s\in I}C_{si}\tilde p_s\leq
\sum\limits _{s=1}^nC_{si}p_s^0, \quad i=\overline{1,l},\end{eqnarray*}
or the inequalities
\begin{eqnarray} \label{gonl31}
y_i\geq\frac {D_i}{\sum\limits _{s=1}^nC_{si}p_s^0 }
, \quad i=\overline{1,l},
\end{eqnarray}
hold.

Inserting (\ref{gonl31}) into (\ref{gonl28}), we obtain the Proof of the Theorem
\ref{Pas3}.
\qed
\end{proof}
Let us give the sufficient conditions for the existence of a strictly positive solution to the set of inequalities (\ref{gonl27}).

\begin{theorem}\label{Pas4} Let $y=\{y_1,\dots ,y_l\}$ be a certain strictly positive solution to the set of equations
 \begin{eqnarray} \label{gonl32}
\sum\limits _{i=1}^lC_{ki}y_i=\psi _k, \quad k=\overline{1,n},
\end{eqnarray}
and $\tilde p={\{\tilde p_s\}}_{s=1}^n$
solves the set of equations
\begin{eqnarray} \label{gonl33}
\sum\limits _{s=1}^nC_{si}\tilde p_s=\frac {D_i}{ y_i},\quad
i=\overline {1,l}.
\end{eqnarray}
If there exists a subset of $n$ linearly independent vectors $C_{i_1},\dots
,C_{i_n},$  where $C_{i_k}=\{C_{si_k}\}_{s=1}^n,$  such  that the vector
$b_{i_1\dots i_n}=(\frac {D_{i_1}} {y_{i_1}},\dots ,\frac
{D_{i_n}} {y_{i_n}})$ belongs to the interior of the non-negative cone created by vectors\index{interior of the non-negative cone created by vectors} $C_{i_1},\dots ,C_{i_n},$ then the vector
$\tilde p=\{\tilde p_s\}_{s=1}^n,$ solving the problem (\ref{gonl33}) is such that $\tilde p_s>0,\
s=\overline {1,n}.$
\end{theorem}

\begin{proof}\smartqed  From the condition that vectors
$C_{i_k},~ k= \overline {1,n},$ are linearly independent and
$b_{i_1\dots i_n}$ belongs to the interior of the non-negative polyhedral cone created by vectors\index{interior of the non-negative polyhedral cone created by vectors}
$C_{i_1},\dots ,C_{i_n},$ it follows that there exists unique solution to the set of equations

\begin{eqnarray} \label{gonl34} \sum\limits
_{s=1}^nC_{s_{i_k}}p_s^0 =\frac {D_{i_k}}{y_{i_k}}, \quad
k=\overline {1,n}.
 \end{eqnarray}

 The consequence of the fact that the solution to
(\ref{gonl34}) is unique  is $p_s^0 = \tilde p_s, \ s=\overline{1,n}.$
From here it follows that the set of equations
 (\ref{gonl33}) has a solution with positive components.
\qed
\end{proof}

 From the Theorem \ref{Pas4} it follows that if its conditions hold, then every solution $y=\{y_1,\dots, y_l\}$ of the set of equations
(\ref{gonl32}) has corresponding unique strictly positive equilibrium price vector. To find it, one need to select the basis of linearly independent vectors
$C_{i_1},\dots, C_{i_n}$ and interior vector
 $b_{{i_1},\dots, i_n}$ in polyhedral cone and to solve the set of equations (\ref{gonl34}).
The rest $D_i,~i\not= i_k, \ k=\overline {1,n},$
are given by \begin{eqnarray*} D_i=y_i\sum\limits _{s=1}^nC_{si} p_s^0.\end{eqnarray*}
  The next Theorem gives an algorithm to find equilibrium price vector. The Theorems \ref{jant39} and \ref{allupotka4} contain an algorithm to find  a positive solution $y=\{y_1,\dots, y_l\}.$

\begin{theorem}\label{Pas6} A certain equilibrium price vector corresponds to every strictly positive solution $y=\{y_1,\dots, y_l\}$
of the set of inequalities (\ref{gonl28})  if and only if, when the ranks of the matrix ${||c_{ki}||}_{k\in I,~ i= \overline {1,l}}$
and  the extended matrix of the set of equations (\ref{gonl29}) are the same.
\end{theorem}

\begin{proof}\smartqed  Necessity is obvious. Prove the sufficiency. If $r$ is the rank of the matrix ${||c_{ki}||}_{k\in I,~
i=\overline {1,l}}$ and the extended matrix, then it is obvious that $1\leq r\leq |I|.$ From here it follows that the set of equations
(\ref{gonl29}) has a solution. Therefore, we obtain as a result of the Theorem \ref{Pas3} the statement of the Theorem \ref{Pas6}. If  the rank
$r<|I|$, then there exist $|I|-r$ linearly independent solutions  corresponding to homogeneous set of equations
(\ref{gonl29}). Therefore the set of equations (\ref{gonl29}) has
$(|I|-r)$-parametric family of solutions. By the Theorem
\ref{Pas3}, the set of inequalities (\ref{gonl27}) has
$(|I|-r)$-parametric family of equilibrium price vectors.
\qed
\end{proof}
Finally, let us describe an algorithm to solve the set of inequalities
(\ref{gonl28}). For that one needs to solve the set of equations
\begin{eqnarray*} \sum\limits _{i=1}^lC_{ki}y_i=\psi _k,\quad k\in I,\end{eqnarray*}
using the Theorems \ref{jant39}, \ref{allupotka4}. One must insert solutions obtained into the set of inequalities
\begin{eqnarray*} \sum\limits
_{i=1}^lC_{ki}y_i<\psi _k,\quad k\in J.\end{eqnarray*}
If all the inequalities are satisfied, then the vector
$y=\{y_1,\dots ,y_l\}$ solves the set of inequalities
(\ref{gonl28}).

\subsection{On economic interpretation of the economy model with proportional consumption and fixed gains}

Let us give possible applications of the considered model. The first application is budget formation,\index{budget formation} more exactly formation of expenses of  budget.
In the economy system there are $l$ social groups. One must share the final consumption vector\index{final consumption vector} $\psi$ among $l$ social groups. If we know equilibrium price vector and  the matrix elements $C_{ki},$ then solving the set of equations (\ref{gonl46}) with respect to budget financing\index{budget financing} $D_i$ of the $i$-th social group,  we can find distribution of budget costs\index{distribution of budget costs}  without break of economic equilibrium in the economy  system.

The next application  of the model is that  to determine equilibrium prices at a commodity exchange if we know supply vector $\psi,$ funds  $D_i$ of every $i$-th consumer and the matrix elements $C_{ki}.$

Finally, we can use the model in the scope of all the economy to determine proportions between productive, unproductive, and social consumptions and accumulation\index{proportions between productive, unproductive, and social consumptions and accumulation} or to determine equilibrium prices at known productive, unproductive, and social consumptions\index{productive consumptions} and accumulation.\index{unproductive consumptions} In this case, as the vector\index{social consumptions} $\psi$ we must take gross output vector\index{gross output vector}  in the economy system.

The set of equations
\begin{eqnarray} \label{gonl46}
\sum\limits _{i=1}^l{C_{ki}}\frac {D_i} {\sum\limits
_{k=1}^{n}C_{ki}p_k}= \psi _k,\quad k=\overline
 {1,n},
\end{eqnarray}
can be solved using an algorithm proposed in the Theorem \ref{Pas3}. Let clarify economic sense of introduced variables
\begin{eqnarray} \label{gonl47}
y_i=\frac
{D_i}{\sum\limits _{k=1}^nC_{ki}p_k}, \quad i=\overline{1,l}.
\end{eqnarray}

Let us give economic interpretation of the matrix elements for the vectors $C_i=\{C_{k i}\}_{k=1}^n,
i=\overline {1,l}.$ The matrix element $C_{ki}$ is a number of units of the $k$-th goods that the $i$-th consumer wants to consume in a certain  period of the economy operation.
If $D_i$ is the gain of the $i$-th consumer during this period, then suppose him to expense his gain completely to buy goods vector ${\{C_{ki}y_i\}}_{k=1}^n,$~ i.e., the consumer buys goods vector ${\{C_{ki}y_i\}}_{k=1}^n,$ where $y_i$ has the form (\ref{gonl47}). The equality (\ref{gonl46}) means that the quantity of the $k$-th goods consumed by all $l$ consumers equals the quantity of this goods $\psi _k$ before consumption. If
$y_i=1,$ then we say that needs of  the $i$-th consumer  are satisfied completely.\index{needs of a  consumer  are satisfied completely} If $y_i<1,$ then  needs of the consumer are satisfied partially.\index{needs of a consumer are satisfied partially} Therefore, we can interpret $y_i$ as  a level of satisfaction of needs of the $i$-th consumer.\index{level of satisfaction of needs of a consumer}

From the previous consideration, necessary and sufficient conditions follow for the existence of strictly positive price vector  solving the set of equations
\begin{eqnarray} \label{gonl48}
\sum\limits_{i=1}^lC_{ki}\frac
 {D_i} {\sum\limits _{s=1}^nC_{si}p_s}= \psi_k,\quad k=\overline
 {1,n}.
\end{eqnarray}

\begin{theorem}\label{Pas5}
Let the  vector of the final consumption\index{vector of the final consumption} $\psi=\{\psi_k \}_{k=1}^n $ belong to the interior of the positive cone created by vectors\index{interior of the positive cone created by vectors} $C_i={\{C_{ki}\}}_{k=1}^n,$
$i=\overline{1,l},$ the ranks of matrices
$C={||C_{ki}||}_{k=1,i=1}^{n,l}$ and $\tilde C$ equal $n,$
where $\tilde C$ is the matrix extended to the matrix $C,$ i.e., first $n $ rows of the matrix $\tilde C$ equal to rows of the matrix $C,$ and $(n+1)$-th row of the matrix $\tilde C$ equals ${\{{D_i\over y_i}\}} _{i=1}^l,$ and let
the vector $y={\{y_i\}}_{i=1}^l$ be a strictly positive solution to the set of equations
\begin{eqnarray*} \sum\limits
_{i=1}^lC_{ki}y_i=\psi _k,\quad k=\overline {1,n}.\end{eqnarray*}
The necessary and sufficient condition for the existence of a strictly positive solution to the set of equations
(\ref{gonl48}) is the existence of a subset of linearly independent vectors $C_{i_1},\dots ,C_{i_n}$ from the  set of vectors
$C_i, \ i=\overline {1,l},$ such that the inequalities $\langle f_k,  b_{i_1 \dots i_n}\rangle >0,$\ $
k=\overline {1,n},$ hold, where \begin{eqnarray*} b_{i_1 \dots i_n}=\left \{ {D_{i_1}\over
y_{i_1}},\dots , {D_{i_n}\over y_{i_n}}\right\},\end{eqnarray*}  $\{f_k,\ k=\overline {1,n}\}$ is  biorthogonal  set of  vectors to the set of vectors
\begin{eqnarray*} d_s={\{C_{si_k}\}}_{k=1}^n,\quad s=\overline {1,n}. \end{eqnarray*}
Moreover, there hold equalities
\begin{eqnarray*} p_k=\langle f_k, b_{i_1 \dots i_n} \rangle ,\quad k=\overline {1,n}.\end{eqnarray*}
\end{theorem}

\begin{proof}\smartqed   Necessity.
The existence of a strictly positive solution to the set of equations
(\ref{gonl48}) means that
\begin{eqnarray*} \sum\limits _{i=1}^lC_{ki}y_i=\psi _k,\quad k=\overline
 {1,n},\end{eqnarray*}
where
\begin{eqnarray*} y_i=\frac {D_i} {\sum\limits_{s=1}^nC_{si}p_s},\quad i=\overline
 {1,l},\end{eqnarray*}
$p=\{p_1,\dots ,p_n\}$ is  a strictly positive solution to the set of equations
(\ref{gonl48}). As the rank of the matrix
$||C_{ki} ||_{k,l=1}^{n,l}$ equals $n,$ it follows that there exists a subset of indices  $i_k,$ $k=\overline {1, n},$~
such that $C_{i_k}=\{C_{si_k}\}_{s=1}^n$ are linearly independent. Therefore, the set of vectors  $d_s={\{C_{si_k}\}}_{k=1}^n, \
s=\overline{1,n}$ is linearly independent too. Hence, the vector $p=(p_1,\dots ,p_n)$
solves the set of linear equations \begin{eqnarray*} \sum\limits
_{s=1}^nd_sp_s=b_{i_1 \dots i_n}\end{eqnarray*}  written in the operator form. From here \begin{eqnarray*} p_s=\langle f_s,b_{i_1 \dots i_n} \rangle \ >0,\quad s=\overline
{1,n}.\end{eqnarray*}   Sufficiency. The consequence of equality of ranks for matrices $C$ and
$\tilde C$ is the existence of a solution to the set of equations
\begin{eqnarray*} \sum\limits _{s=1}^n C_{si}p_s=\frac {D_i}{y_i},\quad
i=\overline {1,l}.\end{eqnarray*}
 From the Theorem \ref{Pas5}, it follows the existence of a set of linearly independent vectors $C_{i_1},\dots ,C_{i_n}$ such that
\begin{eqnarray*} \langle f_k, b_{i_1 \dots i_n} \rangle   > 0,\quad k=\overline {1,n}.\end{eqnarray*}
However, $p_s=\langle f_s, b_{i_1 \dots i_n}\rangle,\ s=\overline {1,n},$
because \begin{eqnarray*} \sum\limits _{s=1}^nd_sp_s=b_{i_1 \dots i_n},\end{eqnarray*}
where $\{f_s, s=\overline{1,n},\}$ is biorthogonal  set of  vectors to the set of vectors
 $\{d_s, ~ s=\overline {1,n}\}.$
\qed
\end{proof}

Economic sense of the $i$-th component of the vector
$y={\{y_i\}}_{i=1}^l$ is a level of satisfaction  of needs of
the $i$-th consumer\index{level of satisfaction  of needs of
a consumer} during a certain period of the economy operation ($y_i$ is a number of baskets
$C_i=\{C_{si}\}_{s=1}^n$ a consumer can buy within his gain $D_i$).

In the next Theorem, we assume that the vectors
$C_i=\{C_{ki}\}_{k=1}^n$ and
$ b_i=\{b_{ki}\}_{k=1}^n, \ i=\overline{1,l}, $ satisfy conditions formulated before the Theorem \ref{alla8}.
\begin{theorem}\label{luda8}
Let $\psi=\{\psi_k\}_{k=1}^n$ be a strictly positive vector and the $i$-th consumer has a non-negative  vector of goods supply\index{non-negative  vector of goods supply} $b_i=\{b_{ki}\}_{k=1}^n$ and funds $D_i > 0, \ i=\overline{1, l}.$
The necessary and sufficient conditions for the existence of a strictly positive solution to the set of equations
 \begin{eqnarray} \label{luda9}
\sum\limits_{i=1}^lC_{ki}\frac{(D_i+ \langle b_i, p \rangle)}{\langle C_i, p \rangle }=\psi_k, \quad k=\overline{1,n},
\end{eqnarray}
with respect to the vector $p \in R_+^n$ are the conditions:\\
1) the vector $\psi=\{\psi_k\}_{k=1}^n$ belongs to the interior of a non-negative cone created by vectors $C_i=\{C_{ki}\}_{k=1}^n, \ i=\overline{1,l}, $ i.e.,
\begin{eqnarray} \label{luda10}
\psi=\sum\limits_{i=1}^ly_iC_i, \quad y_i >0, \quad   i=\overline{1,l}\  ;
\end{eqnarray}
2) the vector $D=\{D_i\}_{i=1}^l$ belongs to the interior of a non-negative cone created by vectors\index{interior of a non-negative cone created by vectors}
$d_k=\{ - b_{ki} + y_i C_{ki} \}_{i=1}^l, \ k=\overline{1,n}.$
\end{theorem}
\begin{proof}\smartqed
 Necessity. Let there exists a strictly positive vector $p_0$ solving the set of equations (\ref{luda9}). Introduce the notation
\begin{eqnarray} \label{luda11}
 y_i= \frac{D_i+ \langle b_i, p_0 \rangle }{ \langle C_i, p_0 \rangle }, \quad i=\overline{1,l}.
\end{eqnarray}
Due to assumptions before the Theorem \ref{alla8} about vectors $C_i=\{C_{ki}\}_{k=1}^n,$
$ b_i=\{b_{ki}\}_{k=1}^n, \ i=\overline{1,l}, $ we have
$y_i > 0, \ i=\overline{1,l},$
and the equality
(\ref{luda10}) holds.
From equalities (\ref{luda11}), we obtain the set of equalities
\begin{eqnarray} \label{luda12}
\langle - b_i + y_i C_i, p_0 \rangle =D_i, \quad  i=\overline{1,l},
\end{eqnarray}
from which the condition 2) of the Theorem \ref{luda8} follows due to strict positivity of the vector $p_0.$

 Sufficiency.
If conditions 1) --- 2) of the Theorem \ref{luda8} hold, then there exist strictly positive numbers $\alpha_i >0, \ i=\overline{1,n}, $
such that
\begin{eqnarray} \label{luda13}
 D= \sum\limits_{i=1}^n\alpha_kd_k.
\end{eqnarray}
Let us put
$p_0=\{p_i^0\}_{i=1}^n,$ where $p_i^0=\alpha_i, \ i=\overline{1,n}. $
Then the set of equations (\ref{luda13}) turns into the set of equalities
\begin{eqnarray} \label{luda14}
y_i= \frac{D_i+ \langle b_i, p_0 \rangle }{ \langle C_i, p_0 \rangle }, \quad i=\overline{1,l}.
\end{eqnarray}
Inserting the expression for $y_i$ into (\ref{luda10}), we obtain that built vector $p_0=\{p_i^0\}_{i=1}^n$ is a strictly positive solution to the set of equations (\ref{luda9}).
\qed
\end{proof}

\section{Economy model with constant parts of consumption}
\begin{lemma}
Let $A$ be an indecomposable matrix\index{indecomposable matrix} and $t=\{t_1, \ldots , t_n\}$ be a vector with strictly positive components.\index{economy model with constant parts of consumption} Then the set of equations
\begin{eqnarray} \label{ans1}
\frac{t_i}{\sum\limits_{k=1}^na_{ki}t_k}=\frac{y_i}{\sum\limits_{k=1}^na_{ik}y_k}, \quad i=\overline{1,n},
\end{eqnarray}
is solvable in the set of strictly positive vectors $y=\{y_1, \ldots , y_n\}.$
\end{lemma}
\begin{proof}\smartqed   The set of equations (\ref{ans1}) can be expressed as follows
\begin{eqnarray} \label{ans2}
y_i = \sum\limits_{k=1}^na_{ik}^*(t)y_k, \quad i=\overline{1,n},
\end{eqnarray}
where
\begin{eqnarray*} a_{ik}^*(t)=\frac{a_{ik}t_i}{\sum\limits_{k=1}^na_{ki}t_k}.\end{eqnarray*}
Consider the problem conjugate to the problem (\ref{ans2})
\begin{eqnarray} \label{ans3}
\tau_k = \sum\limits_{i=1}^na_{ik}^*(t)\tau_i, \quad k=\overline{1,n},
\end{eqnarray}
or
\begin{eqnarray} \label{ans4}
\tau_k = \sum\limits_{i=1}^n\frac{a_{ik}t_i\tau_i}{\sum\limits_{k=1}^na_{ki}t_k}, \quad k=\overline{1,n}.
\end{eqnarray}
The solution of the problem (\ref{ans4})  is the vector $\tau=\{\tau_i\}_{i=1}^n$ with components
\begin{eqnarray*}  \tau_i=\sum\limits_{k=1}^na_{ki}t_k, \quad i=\overline{1,n},\end{eqnarray*}
each of which is strictly positive.
Now it is easy to establish that the unit is the largest proper value of the problem\index{largest proper value of the problem} (\ref{ans3}). From this and the Perron-Frobenius Theorem\index{Perron-Frobenius Theorem} it follows that the problem (\ref{ans2}) has a strictly positive solution.
\qed
\end{proof}

Let $\gamma=||\gamma_{ik}||_{i,k=1}^n$ be a non-negative matrix satisfying conditions
\begin{eqnarray*}  \sum\limits_{k=1}^n\gamma_{ik}=1, \quad i=\overline{1,n}.\end{eqnarray*}

\begin{theorem}\label{alla17}
Let the matrix $A$ be productive\index{productive matrix}  and do not contain zero rows and columns, $\gamma(E -A)^{-1} $ be indecomposable. The set of equations of the economy equilibrium
\begin{eqnarray} \label{ahs1}
\frac{1}{p_k}\sum\limits_{i=1}^n\gamma_{ik}x_i\left(p_i - \sum\limits_{s=1}^na_{si}p_s\right)= x_k - \sum\limits_{k=1}^na_{ki}x_i, \quad k=\overline{1,n},
\end{eqnarray}
is solvable in the set
\begin{eqnarray*}  T=\left\{p=\{p_i\}_{i=1}^n  \in R_+^n, \  p_k - \sum\limits_{s=1}^na_{sk}p_s > 0, \ k=\overline{1,n}\right\} \end{eqnarray*}
if and only if, when there exists a certain  strictly positive vector
\begin{eqnarray*} z=\{z_1, \ldots , z_n\}, \quad z_i > 0,\quad i=\overline{1,n},\end{eqnarray*}
such that the vector of the final supply
\begin{eqnarray*} \psi=\{\psi_i\}_{i=1}^n, \quad  \psi_i=x_i - \sum\limits_{k=1}^na_{ik}x_k, \quad i=\overline{1,n}, \end{eqnarray*}
solves the set of equations
\begin{eqnarray} \label{ahs2}
\frac{\psi_k}{\sum\limits_{i=1}^n\gamma_{ik}z_i \sum\limits_{s=1}^na_{is}^{-1}\psi_s}=
\frac{1}{\sum\limits_{i=1}^na_{ik}^{-1}z_i}, \quad k=\overline{1,n}.
\end{eqnarray}
At that
\begin{eqnarray*} p=(E - A^T)^{-1}z, \quad  a_{ki}^{-1}=(E - A)^{-1}_{ki}, \quad  x=(E - A)^{-1}\psi.\end{eqnarray*}
\end{theorem}
\begin{proof}\smartqed   Necessity. Let the  vector of the final supply\index{vector of the final supply}
\begin{eqnarray*} \psi=\{\psi_k\}_{k=1}^n, \quad  \psi_k=x_k - \sum\limits_{i=1}^na_{ki}x_i, \quad k=\overline{1,n}, \end{eqnarray*}
have strictly positive components. From here it follows that output vector $x=\{x_1, \ldots , x_n\}$
also has strictly positive components. By the Theorem condition, there exists a vector $p=\{p_1, \ldots , p_n\},$
solving the problem (\ref{ahs1}) and belonging to the set $T.$
Denote $z_i=p_i - \sum\limits_{s=1}^na_{si}p_s, \ i=\overline{1,n},$ and introduce the vector
$z=\{z_1, \ldots , z_n\}.$ Then the set of equations (\ref{ahs1}) can be written in the form
\begin{eqnarray*} x_k= \sum\limits_{k=1}^na_{ki}x_i + \frac{1}{[(E-A^T)^{-1}z]_k}\sum\limits_{i=1}^n\gamma_{ik}x_iz_i, \quad k=\overline{1,n}.\end{eqnarray*}
Accounting for the notations introduced, the last set of equations can be written as
\begin{eqnarray*} \frac{\psi_k}{\sum\limits_{i=1}^n\gamma_{ik}z_i \sum\limits_{s=1}^na_{is}^{-1}\psi_s}=
\frac{1}{[(E - A^T)^{-1}z]_k}, \quad   \quad k=\overline{1,n},\end{eqnarray*}
that is identic to (\ref{ahs2}).

 Sufficiency. Let there exist a certain  vector $z=\{z_1, \ldots , z_n\}$ with strictly positive components such that the problem (\ref{ahs2}) be solvable with respect to the vector $\psi=\{\psi_k\}_{k=1}^n.$

Show that the vector $\psi=\{\psi_i\}_{i=1}^n$ solving the problem (\ref{ahs2}) has strictly positive components. Introduce the matrix
$ C=||C_{ki}||_{i,k=1}^n, $ where $ C_{ki}=\gamma_{ik}z_i.$ Then
$[C^Te]_i =\sum\limits_{k=1}^n\gamma_{ik}z_i=z_i, \ e=\{1, \ldots, 1\}.$ Accounting for these notations the set of equations
(\ref{ahs2}) can be written as follows
\begin{eqnarray} \label{ahs3}
\frac{\psi_k}{[C(E - A)^{-1}\psi]_k }=
\frac{1}{[E - A^T)^{-1}C^T e]_k }, \quad k=\overline{1,n}.
\end{eqnarray}
The matrix $C(E - A)^{-1}$ is indecomposable  after the Theorem condition. On the ground of the previous Lemma, there exists a strictly positive solution to the set of equations (\ref{ahs3}) and, consequently, to (\ref{ahs2}).
If to introduce output vector $x=(E - A)^{-1}\psi,$ then (\ref{ahs2}) can be written as follows
\begin{eqnarray} \label{ahs4}
x_i= \sum\limits_{k=1}^na_{ik}x_k + \frac{1}{\sum\limits_{k=1}^na_{ki}^{-1}z_k}\sum\limits_{k=1}^n\gamma_{ik}x_kz_k, \quad i=\overline{1,n}.
\end{eqnarray}
Introduce the vector $p=\{p_1, \ldots , p_n\},$ supposing $ p_i= \sum\limits_{k=1}^na_{ki}^{-1}z_k, \quad i=\overline{1,n}.$ Then, obviously, $  p_i - \sum\limits_{s=1}^na_{si}p_s =z_i > 0, \ i=\overline{1,n},$ and (\ref{ahs4}) takes the form (\ref{ahs1}).
\qed
\end{proof}

Consider an exchange model, in which $l$ consumers have property vectors $b_i=\{b_{ki}\}_{k=1}^n, \ i=\overline{1,l}, $ and funds $D_i, \ i=\overline{1,l}.$
Let $\gamma_i=\{\gamma_{ik}\}_{k=1}^{n},  \ i=\overline{1,l},$ be demand vector of the $i$-th consumer satisfying condition
\begin{eqnarray*}  \sum\limits_{k=1}^n\gamma_{ik}=1, \quad  i=\overline{1,l}.\end{eqnarray*}
Denote
$ \langle b_i, p\rangle =  \sum\limits_{k=1}^nb_{ki}p_k.$
\begin{theorem}\label{allunja4}
Let conditions
$ \sum\limits_{i=1}^lD_i>0, \ D_i \geq 0 $ hold.
The set of  equations of economic equilibrium
\begin{eqnarray} \label{allunja5}
\frac{1}{p_k}\sum\limits_{i=1}^l\gamma_{ik}(D_i +\langle b_i, p\rangle )= \psi_k, \quad k=\overline{1,n},
\end{eqnarray}
is solvable in the set of strictly positive price vectors if the matrix
\begin{eqnarray*} \left|\left|\sum\limits_{i=1}^l\gamma_{ik}b_{si}\right|\right|_{k,s=1}^n\end{eqnarray*}
is indecomposable, the inequalities
\begin{eqnarray*}  \psi_k \geq \sum\limits_{i=1}^l b_{ki}, \quad \psi_k > 0,\quad   k=\overline{1,n},\end{eqnarray*}
hold, there exist $k$ and $s,$ $1\leq k, s \leq n,$ such that $ \psi_k > \sum\limits_{i=1}^l b_{ki}, \ $ $\sum\limits_{i=1}^l\gamma_{is}D_i>0.$
\end{theorem}
\begin{proof}\smartqed  The set of equations (\ref{allunja5}) is equivalent to the set of equations
\begin{eqnarray} \label{allunja6}
p_k=\frac{1}{\psi_k}\sum\limits_{i=1}^l\gamma_{ik}(D_i + \langle b_i, p\rangle ), \quad k=\overline{1,n}.
\end{eqnarray}
Let us prove that the spectral radius of the matrix
\begin{eqnarray*} \left|\left|\frac{1}{\psi_k}\sum\limits_{i=1}^l\gamma_{ik}b_{si}\right|\right|_{k,s=1}^n\end{eqnarray*}
is strictly smaller than unit. Consider the set of equations
\begin{eqnarray} \label{allunja7}
\sum\limits_{s=1}^n\frac{1}{\psi_k}\sum\limits_{i=1}^l\gamma_{ik}b_{si}x_s=\lambda x_k, \quad k=\overline{1,n},
\end{eqnarray}
with respect to the vector $x=\{x_i\}_{i=1}^n.$
By the Perron-Frobenius Theorem, there exists a strictly positive vector $x^0=\{x_i^0\}_{i=1}^n$ with the largest proper value $\lambda_0 > 0$
corresponding to it and solving the set of equations
\begin{eqnarray} \label{allunja8}
\sum\limits_{s=1}^n\frac{1}{\psi_k}\sum\limits_{i=1}^l\gamma_{ik}b_{si}x_s^0=\lambda_0 x_k^0, \quad k=\overline{1,n}.
\end{eqnarray}
Multiplying the $k$-th equality in (\ref{allunja8}) by $\psi_k$ and summing up over $k$ from $1$ to $n$, we obtain
\begin{eqnarray*} \sum\limits_{s=1}^n\sum\limits_{i=1}^lb_{si}x_s^0= \lambda_0\sum\limits_{k=1}^n\psi_k  x_k^0.\end{eqnarray*}
From here
\begin{eqnarray*}   \lambda_0=\frac{\sum\limits_{s=1}^n\sum\limits_{i=1}^lb_{si}x_s^0}{\sum\limits_{k=1}^n\psi_k  x_k^0}< 1.\end{eqnarray*}
\qed
\end{proof}

Introduce matrices
\begin{eqnarray*} \gamma=||\gamma_{ik}||_{i=1, k=1}^{l,\ n}, \quad
\bar \gamma=||\gamma_{ik}||_{i, k=1}^{n}, \quad \tilde \gamma=||\gamma_{ik}||_{i=n+1, k=1}^{l, n},\end{eqnarray*}
\begin{eqnarray*} B=||b_{si}||_{s=1, i=1}^{n,\ l},\quad \psi^{-1}=||[\psi_k]^{-1} \delta_{ik}||_{i,k=1}^n, \end{eqnarray*}
and vectors $D=\{ D_i\}_{i=1}^l, \ \bar D=\{ D_i\}_{i=1}^n, \ \tilde D=\{ D_i\}_{i=n+1}^l.$
Let us  write the set of equations (\ref{allunja6}) with respect to the vector $p=\{p_i\}_{i=1}^n$ in operator form. We have
\begin{eqnarray} \label{allunja9}
p=\psi^{-1}\gamma^TD+\psi^{-1}\gamma^TB^Tp.
\end{eqnarray}
By $\gamma^T,\  B^T$ we denote matrices transposed  to matrices $\gamma,\  B$ correspondingly.
From the Theorem \ref{allunja4} it follows that the spectral radius of the matrix\index{spectral radius of the matrix} $\psi^{-1}\gamma^TB^T$ is strictly smaller than unit.
Then the unique strictly positive solution to the set of equations (\ref{allunja9}) has the form
\begin{eqnarray} \label{allunja10}
p= \left[E - \psi^{-1}\gamma^TB^T\right]^{-1}\psi^{-1}\gamma^TD.
\end{eqnarray}

Let the vector $\pi=\{\pi_i\}_{i=1}^n$ have strictly positive components and satisfy conditions $0 < \pi_i < 1, \ i=\overline{1,n},$ and let the output vector $x^0=\{x_i^0\}_{i=1}^n$  in the economy system  be such that the conditions
\begin{eqnarray*}  x_i^0 - \sum\limits_{k=1}^n a_{ik}x_k^0 > 0,  \quad  i=\overline{1,n},\end{eqnarray*}
 hold.
Consider the set of equations of  economic equilibrium
\begin{eqnarray*} \frac{1}{p_k}\sum\limits_{i=1}^n\gamma_{ik}\left[\pi_ix_i^0\left(p_i - \sum\limits_{s=1}^na_{si}p_s\right)+\langle b_i,p\rangle \right]+ \frac{1}{p_k}\sum\limits_{i=n+1}^l\gamma_{ik}(D_i + \langle b_i, p\rangle )\end{eqnarray*}
\begin{eqnarray} \label{allunja11}
 = x_k^0 - \sum\limits_{k=1}^na_{ki}x_i^0+ \sum\limits_{i=1}^lb_{ki}, \quad k=\overline{1,n},
\end{eqnarray}
with respect to the vector $p=\{p_i\}_{i=1}^n.$

\begin{theorem}\label{allunja12}
Let the matrix $A=|| a_{ik}||_{i,k=1}^n$ be productive and do not contain zero columns, there hold inequalities $  D_i \geq 0, \ i=\overline{n+1,l}, $
 the output vector in the economic system $x^0=\{x_i^0\}_{i=1}^n$ be such that the conditions
\begin{eqnarray*}  x_i^0 - \sum\limits_{k=1}^n a_{ik}x_k^0 > 0,  \quad  i=\overline{1,n},\end{eqnarray*}
 hold, the matrix
\begin{eqnarray*} \left|\left|\sum\limits_{i=1}^l\gamma_{ik}b_{si}\right|\right|_{k,s=1}^n\end{eqnarray*}
is indecomposable.
The set of equations of economic equilibrium  (\ref{allunja11})
is solvable in the set
\begin{eqnarray*} T=\left\{p \in \bar R_+^n, \  p_i -\sum\limits_{s=1}^na_{si}p_s > 0, \ i=\overline{1,n}\right\}\end{eqnarray*}
if and only if, when the vector $ \left[E - \psi^{-1}\gamma^TB^T\right]^{-1}\psi^{-1}\tilde \gamma^T \tilde D$
belongs to the interior of the cone created by vectors-columns of the matrix
\begin{eqnarray*} \left[E - A^T\right]^{-1}\pi^{-1}X_0^{-1} -  \left[E - \psi^{-1}\gamma^TB^T\right]^{-1}\psi^{-1}\bar \gamma^T,\end{eqnarray*}
where $\pi^{-1}=||\delta_{ij}[\pi_i]^{-1}||_{i,j=1}^n,\ X_0^{-1}=||\delta_{ij}[x_i^0]^{-1}||_{i,j=1}^n.$
\end{theorem}
\begin{proof}\smartqed   Necessity. Let a strictly positive price vector $p^0=\{p_i^0\}_{i=1}^n$   belonging to the set $T$ exist and solve the set of equations (\ref{allunja11}).
Introduce notations
\begin{eqnarray} \label{allunja13}
D_i=\pi_ix_i^0\left(p_i^0 - \sum\limits_{s=1}^na_{si}p_s^0\right),\quad  i=\overline{1,n}.
\end{eqnarray}
Within such notations, the vector $p^0=\{p_i^0\}_{i=1}^n$ solves the set of equations
\begin{eqnarray} \label{allunja14}
p_k^0=\frac{1}{\psi_k}\sum\limits_{i=1}^l\gamma_{ik}(D_i + \langle b_i, p^0\rangle ), \quad k=\overline{1,n},
\end{eqnarray}
where $ \psi_k=x_k^0 - \sum\limits_{k=1}^na_{ki}x_i^0+ \sum\limits_{i=1}^lb_{ki}, \ k=\overline{1,n}.$
The unique solution to the set of equations (\ref{allunja14}) is the vector
\begin{eqnarray*} p^0=\left[E - \psi^{-1}\gamma^TB^T\right]^{-1}\psi^{-1}\gamma^TD=\left[E - \psi^{-1}\gamma^TB^T\right]^{-1}\psi^{-1}\bar \gamma^T\bar D \end{eqnarray*}
\begin{eqnarray} \label{allunja15}
+\left[E - \psi^{-1}\gamma^TB^T\right]^{-1}\psi^{-1}\tilde \gamma^T \tilde D.
\end{eqnarray}
From other side, with (\ref{allunja13}), for $p^0$ we obtain the formula
\begin{eqnarray} \label{allunja16}
p^0= \left[E - A^T\right]^{-1}\pi^{-1}X_0^{-1}\bar D.
\end{eqnarray}
Inserting the expression for $p^0$ into
(\ref{allunja15}), we obtain
\begin{eqnarray*} [E - A^T]^{-1}\pi^{-1}X_0^{-1}\bar D - [E - \psi^{-1}\gamma^TB^T]^{-1}\psi^{-1}\bar \gamma^T\bar D \end{eqnarray*}
\begin{eqnarray} \label{allunja17}
=[E - \psi^{-1}\gamma^TB^T]^{-1}\psi^{-1}\tilde \gamma^T \tilde D.
\end{eqnarray}
The last proves the necessity.

 Sufficiency. If the Theorem conditions hold, then there exists a strictly positive vector $\bar D$ such that there hold equalities (\ref{allunja17}).
Define strictly positive vector $p^0$ by the formula (\ref{allunja16}). The last is equivalent that there hold equalities
\begin{eqnarray} \label{allunja18}
D_i=\pi_ix_i^0\left(p_i^0 - \sum\limits_{s=1}^na_{si}p_s^0\right),\quad  i=\overline{1,n}.
\end{eqnarray}
From the formula (\ref{allunja17}), we have for the vector $p^0$ the representation
\begin{eqnarray*} p^0=\left[E - \psi^{-1}\gamma^TB^T\right]^{-1}\psi^{-1} \gamma^T  D.\end{eqnarray*}
However, the last vector is the  unique solution to the set of equations
\begin{eqnarray} \label{allunja19}
p_k^0=\frac{1}{\psi_k}\sum\limits_{i=1}^l\gamma_{ik}(D_i + \langle b_i, p^0\rangle ), \quad k=\overline{1,n}.
\end{eqnarray}
Finally, inserting $D_i, \ i=\overline{1,n},$ given by the formulae (\ref{allunja18}) into the formula (\ref{allunja19}), we obtain the proof of the sufficiency.
\qed
\end{proof}

\chapter{Output distribution in a society with fixed consumption levels }

\abstract*{ The problem of distribution of the output created in the economy system according to the consumption structure determined by non-random fields of information evaluation by consumers  that are independent of the price vector is solved. Consumers choice is described by non-random fields of information evaluation  that are independent of price vector.  Particular taxation systems and a family of income functions of consumers are constructed. Using the introduced taxation systems and a family of income functions of consumers the Theorems guaranteeing the existence of equilibrium price vector for which every industry  is profitable and every consumer satisfies his needs according to the demand structure  are established. The conditions guaranteeing industries profitability are necessary under simple technical conditions.}

In this Chapter, we solve the important problem of distributing  the output created in the economic system according to the consumption structure determined by non-random fields  of information evaluation by consumers  independent of the price vector \cite{55, 71, 92, 106}.
The main task of this Chapter is to construct a family of income functions of consumers for which there exists  an equilibrium price vector such that the  consumers satisfy their needs with fixed consumption levels.\index{needs with fixed consumption levels} The solution  of this problem is related to the formation of taxation system\index{taxation system} and establishing  its dependence on the output vector,\index{output vector} the production structure, and consumption levels\index{consumption levels} being  of  practical importance.

We suppose that the structure   of  firms supply\index{structure   of  firms supply} is compatible with the structure of consumers   choice\index{structure of consumers   choice} that means a certain set of material balance equations\index{material balance equations} holds in the economy system. Under condition that the structure of firms supply  is compatible with the structure of consumers choice, we construct a taxation system and corresponding  a family of income functions of consumers\index{family of income functions of consumers} for which there exists an equilibrium price vector such that every consumer can buy the wanted quantity of goods in a fixed proportion.

In the Section "Economy model with particular taxation system",\index{economy model with particular taxation system} we introduce the economy model studied. The character of the model considered is that we describe consumers by
non-random fields  of information evaluation by consumers\index{non-random fields  of information evaluation by consumers} independent of price vector.
In the Lemma \ref{ant4}, we establish the sufficient conditions for the equation (\ref{ant5}) to have   independent of the price vector solution having strictly positive components smaller than unit.
 With the solution built in the Lemma \ref{ant4}, in the Theorem \ref{hbant1} we construct a particular taxation system.\index{particular taxation system} The specific character of such taxation system is that taxation levels of  industries\index{taxation levels of  industries} do not depend on the price vector. The solution built in the Lemma \ref{ant4} we   name the taxation vector. Economic sense of the $i$-th component $\pi_i(z^0)$ of the taxation vector\index{taxation vector} is the part of the value created remaining in the $i$-th industry under productive process $z^0$ realized in the economy system.

For the built family of income functions of consumers,\index{family of income functions of consumers} using the introduced taxation system in the Theorem \ref{bant1} we establish the existence of equilibrium price vector for which every industry is profitable  and every $i$-th consumer satisfies  his needs proportionally to his  field of information evaluation with proportionality coefficient $y_i(z^0).$

In the Section "Economic systems with fixed consumption levels", we construct economic system models by a non-negative matrix $C,$  named unproductive consumption matrix,\index{unproductive consumption matrix}  a vector of  levels of  satisfaction of consumers needs\index{vector of  levels of  satisfaction of consumers needs} $y=\{y_i\}_{i=1}^l,$ a  matrix $B$ built by vectors of the initial goods supply.\index{vectors of the initial goods supply} The peculiarity of such model systems are conditions which the output vector\index{output vector } and taxation system\index{taxation system} have to  satisfy  in order that the product created in the economic system does find its buyer. Therefore, in this Section, the primary quantities are a certain  non-negative matrix $C,$ a vector of  levels of  satisfaction of consumers needs\index{vector of  levels of  satisfaction of consumers needs} $y=\{y_i\}_{i=1}^l,$ an input-output  matrix\index{input-output  matrix} $A,$ an initial goods supply matrix\index{initial goods supply matrix} $B.$ For the above quantities the output vector solves the set of equations and its components have special representation.

In the Theorem \ref{ant11}, we establish the necessary and sufficient conditions for the existence of a strictly positive solution to the set of equations (\ref{ant10}) belonging to the set $T_0.$ The  conditions of the  Lemma \ref{jant1} guarantee that the set
 $T_0$ is non-empty. The conditions of the Lemma \ref{jant4} guarantee the existence of a strictly positive solution to the equation (\ref{janm5}) that is independent of the price vector and has strictly positive, smaller than unit, components.
The Theorem \ref{jant6} guarantees the existence of peculiar taxation system\index{peculiar taxation system} built after strictly positive tax vector\index{tax vector} whose existence the Lemma
 \ref{jant4} establishes. By the constructed taxation system, we have built a family of income functions of consumers and established the Theorem \ref{bant2} guaranteeing the existence of a strictly positive solution  to the set of equations of economic equilibrium (\ref{gok1}) in the set $T_0.$

In the Section "Consumers needs satisfaction under arbitrary tax vector", we clarify conditions under which equilibrium price vector exists such  that created in the  economy system output will be consumed according to the consumption structure. Unlike the previous Section where a certain equation determines the tax vector, in this Section we suppose the tax vector to be known. The main task is to clarify the dependence of the output vector of  tax levels\index{tax levels} and consumption levels\index{consumption levels} for which  equilibrium price vector exists such that consumers satisfy their needs according to the demand structure.\index{demand structure} For that the Theorem \ref{jant12} gives the sufficient conditions under which a solution to the equation (\ref{jant13})  for the given tax vector exists in the set $T_0.$ This solution depends on the price vector in the set $T_0$ and on the tax vector. In the Theorem \ref{jant22}, we build a taxation system for the economy system model considered after this solution. The Theorem \ref{jant25} gives the sufficient conditions under which a taxation system determined by the known tax vector guarantees the existence of equilibrium price vector for which consumers satisfy their needs according to the demand structure.

In the Section "Necessary conditions of industries profitability at economic equilibrium", we establish that the previously stated sufficient conditions for the existence of equilibrium price vector under which every consumer can satisfy his needs according to the demand structure are also necessary.
The Theorem \ref{jant39} describes the structure of strictly positive solutions to the set of equations (\ref{gonl40}).

\section{Economy model with proportional con\-sumption }

\subsection{Economy model with peculiar taxation system}\index{economy model with peculiar taxation system}

Consider the model of economic system containing $n$ net industries.
Let $A=||a_{ki} ||_{k=1,i=1}^n$ be a non-negative productive matrix.

We describe  the structure of the $i$-th industry production  by the technological map
\begin{eqnarray*} F_i( x_i)=\{  y_i \in S,\   y_i= u_ie_i, \ 0 \leq a_{ki}u_i \leq x_{ki}, \ k=\overline{1,n} \}, \quad e_i=\{\delta_{ik}\}_{k=1}^n,\end{eqnarray*}
\begin{eqnarray*}   x_i=\left\{x_{ki}\}_{k=1}^n \in  X_i=\{  x_i=\{x_{ki}\}_{k=1}^n  \in S, \ 0 \leq x_{ki} \leq a_{ki}u_i^0, \ k=\overline{1,n}\right\},\end{eqnarray*}
where $u_i^0$ is a maximum possible output of the $i$-th industry,\index{maximum possible output of an industry}  $i=\overline{1,n},$
and $\delta_{ik}$ is the Kronecker symbol.
Further, let $l$ consumers have property vectors $b_i=\{b_{ki}\}_{k=1}^n, \ i=\overline{1,l}.$
The set of possible productive processes\index{set of possible productive processes} in the model considered is
 \begin{eqnarray*} \Gamma^n=\prod\limits_{i=1}^n\Gamma_i, \quad
\Gamma_i=\{( x_i,  y_i), \ x_i \in X_i, \  y_i \in F_i(  x_i)\}, \quad z=\{z^i\}_{i=1}^n, \quad  z^i=( x_i,  y_i),\end{eqnarray*}
and the set of possible price vectors\index{set of possible price vectors} is the cone
\begin{eqnarray*} K_+^n=\{p \in R_+^n, \ p_i >0,  \  i=\overline{1,n}\}.\end{eqnarray*}
The map $Q(p,z)=\{Q_i(p,z))\}_{i=1}^n$ that maps $\{\Gamma^n,{\cal B}(\Gamma^n)\}$
into itself for every $p \in K_+^n $
and is given as follows
\begin{eqnarray*} Q_i(p,z)=\prod\limits_{s=1}^n a_s(p)v_s(p,z) \bar Q_i(z),\end{eqnarray*}
where
\begin{eqnarray*} \bar Q_i(z)=\{\{a_{ki}u_i\}_{k=1}^n, \   u_ie_i \},\end{eqnarray*}
\begin{eqnarray*} a_i(p)=\chi_{[0, \infty)}\left(p_i - \sum\limits_{s=1}^na_{si}p_s\right), \quad
v_i(p,z)=\chi_{[0, \infty)}\left(u_i - \sum\limits_{k=1}^na_{ik}u_k + \sum\limits_{k=1}^lb_{ik}\right),\end{eqnarray*}
is a productive economic process.\index{productive economic process}
The structure of the productive economic process is such that industries operate without loss.

 The equality $Q(p,Q(p,z))=Q(p,z), \ (p, z) \in K_+^n \times \Gamma^n$ takes place.
The vector $u=\{u_i\}_{i=1}^n$ is called  a possible output vector in the economy system\index{possible output vector in the economy system} corresponding to a productive process $  z \in \Gamma^n.$
 Let us consider a probability space  $\{\Omega, {\cal F}, \bar P\},$
where
\begin{eqnarray*} \Omega= \prod\limits_{i=0}^l\Omega_i, \quad {\cal F}=\prod\limits_{i=0}^l{\cal F}_i, \quad \bar P=\prod\limits_{i=0}^l\bar P_i.\end{eqnarray*}
We suppose that in the economy system  random fields of information evaluation by consumers\index{random fields of information evaluation by consumers}  on  a probability space $\{\Omega, {\cal F}, \bar P\}$
are non-random and have the form
\begin{eqnarray*} \eta_i^0(p,z, \omega_i)=\{\eta_{ik}^0(p,z, \omega_i)\}_{k=1}^n, \quad i=\overline{1,l},\end{eqnarray*}
where
\begin{eqnarray*} \eta_{ik}^0(p,z, \omega_i)=c_{ki}(z),  \quad (p,z) \in K_+^n \times \Gamma^n, \quad i=\overline{1,l}, \quad k=\overline{1,n},\end{eqnarray*}
and a random field $\zeta_0(p, \omega_0), \ (p, \omega_0)  \in K_+^n \times \Omega_0, $ takes values in $ \Gamma^n.$
Therefore, in this Section, we suppose that random fields of information evaluation by consumers depend only on the set of productive processes $z \in \Gamma ^n.$

Introduce into  consideration the set
 \begin{eqnarray*} T =\left\{z \in \Gamma^n, \ z=\{z^i\}_{i=1}^n, \  z^i=( x_i,  y_i), \ u_k -  \sum\limits_{s=1}^na_{ks}u_s +\sum\limits_{i=1}^lb_{ki} \geq 0, \  k=\overline{1,n}\right\}\end{eqnarray*}
being a subset of the set $\Gamma ^n.$
\begin{definition}\label{ant3}
The structure of supply   agrees with the structure of choice\index{structure of supply   agrees with the structure of choice}  on a certain set of productive processes $B_0 \subseteq T$ if for every productive process $ z^0=\{z_0^i\}_{i=1}^n \in B_0 \subseteq T,$ $ \ z_0^i= \left\{\left\{x^0_ia_{ki}\right\}_{k=1}^n, x_i^0e_i\right\}, $  the  corresponding vector of the final consumption\index{vector of the final consumption}
\begin{eqnarray*}  \psi\left(z^0\right)=\{\psi_k\left(z^0\right)\}_{k=1}^n, \quad \psi_k\left(z^0\right)=x_k^0 -  \sum\limits_{s=1}^na_{ks}x_s^0 +\sum\limits_{i=1}^lb_{ki}, \quad  k=\overline{1,n},\end{eqnarray*}
has the following representation
\begin{eqnarray} \label{jant26}
\psi_k\left(z^0\right)=\sum\limits_{i=1}^l c_{ki}\left(z^0\right)y_i\left(z^0\right), \quad z^0 \in B_0, \quad  k=\overline{1,n},
\end{eqnarray}
 where the vector $y\left(z^0\right)=\{y_i\left(z^0\right)\}_{i=1}^l$ has strictly positive components.

We call the vector $y\left(z^0\right)$   the vector of levels of satisfaction of consumers needs\index{vector of levels of satisfaction of consumers needs}  and the vector $x^0=\{x_i^0\}_{i=1}^n$ the  gross output vector\index{gross output vector} in  the economy system corresponding to the productive process $z^0 \in B_0.$
\end{definition}
Further we suppose that technological maps describing production of industries are such that the set $B_0$ is non-empty.

Introduce matrices
\begin{eqnarray*} X_0=||\delta_{ij}x_i^0||_{i,j=1}^n, \quad  A=||a_{ij}||_{i,j=1}^n,\quad B=||b_{ij}||_{i=1,j=1}^{n,\ l}, \end{eqnarray*}  \begin{eqnarray*}  C\left(z^0\right)=||c_{ij}\left(z^0\right)||_{i=1,j=1}^{n,\ l},\quad \tilde  B=||b_{ij}||_{i=1,j=n+1}^{n,\ l},\quad \bar B=||b_{ij}||_{i=1,j=1}^{n},\end{eqnarray*}
\begin{eqnarray*} \bar C\left(z^0\right)=||c_{ij}\left(z^0\right)||_{i=1,j=1}^{n},\quad \tilde  C\left(z^0\right)=||c_{ij}\left(z^0\right)||_{i=1,j=n+1}^{n,\ l},\end{eqnarray*}
where $\delta_{ij}$ is the Kronecker symbol and vectors
\begin{eqnarray*}    y\left(z^0\right)=\{y_i\left(z^0\right)\}_{i=1}^l, \quad \tilde y\left(z^0\right)=\{y_i\left(z^0\right)\}_{i=n+1}^l, \quad \bar y\left(z^0\right)=\{y_i\left(z^0\right)\}_{i=1}^n, \end{eqnarray*}
\begin{eqnarray*}   e=\{e_i\}_{i=1}^l, \quad  \bar e=\{e_i\}_{i=1}^n, \quad \tilde e=\{e_i\}_{i=n+1}^l,\quad e_i=1, \quad i=\overline{1,l}.\end{eqnarray*}

Let us give a series of definitions useful for what follows.
\begin{definition}\label{mant0}
By a polyhedral non-negative cone created by a set of  vectors\index{polyhedral non-negative cone created by a set of  vectors} $\{a_i,\
i=\overline {1,t}\}$ of $n$-dimensional space $R^n$ we understand the set of vectors of the form
\begin{eqnarray*}  d = \sum\limits_{i=1}^t\alpha_i a_i,\end{eqnarray*}
where $\alpha=\{\alpha_i\}_{i=1}^t$ runs over the set $R_+^t.$
\end{definition}
\begin{definition}\label{mant1}
The dimension of a non-negative polyhedral cone created by a set of vectors $\{a_i,\
i=\overline {1,t}\}$ in $n$-dimensional space $R^n$ is maximum number of linearly independent vectors from the set of vectors $\{a_i,\
i=\overline {1,t}\}.$
\end{definition}

\begin{definition}\label{1mant2}
The vector $b $ belongs to the interior of the non-negative polyhedral $r$-dimensional cone,\index{interior of the non-negative polyhedral  cone created by the set of vectors} $r \leq n,$ created by the set of vectors
 $\{a_1,\dots ,a_t\}$ in $n$-dimensional vector space $R^n$
 if   a strictly positive vector $\alpha=\{\alpha_i\}_{i=1}^t \in R_+^t$ exists such that
\begin{eqnarray*} b =\sum\limits _{s=1}^t a_s\alpha_s,\end{eqnarray*}
where $\alpha_s>0, \ s=\overline{1,t}.$
\end{definition}

Let us give the necessary and sufficient conditions under which a certain  vector belongs to the interior of the polyhedral cone.

\begin{theorem}\label{mant2} Let $\{a_1,\dots ,a_m\},$ $1\leq m\leq n,$ be the set of linearly independent vectors in $R_+^n.$ The necessary and sufficient conditions for the vector $b $ to belong to the interior of the non-negative cone $K_a^+$ created by vectors
$\{a_i,\ i=\overline {1,m}\}$ are the conditions
\begin{eqnarray} \label{mant3}
\langle f_i, b \rangle  >0, \quad i=\overline {1,m}, \quad \langle f_i, b \rangle=
0,\quad i=\overline {m+1,n},
\end{eqnarray}
where  $f_i,\
i=\overline {1,n},$   is a set of vectors being  biorthogonal to the set of linearly independent vectors $\bar a_i, \ i=\overline {1,n},$ and
$\bar a_i=a_i, \ i=\overline {1,m}.$
\end{theorem}
\begin{proof}\smartqed  It is obvious that the set of  biorthogonal  vector  exists. Really, the vector $f_j, \ j=\overline {1,n},$ exists that solves the set of equations
\begin{eqnarray} \label{mant4}
\langle \bar a_i, f_j \rangle = \delta _{ij},\quad i=\overline {1,n},
\end{eqnarray}
due to linear independency of vectors
$\bar a_i,\ i=\overline {1,n}.$

 Necessity. If the vector $b $ belongs to the interior of the non-negative cone $K_a^+,$ then there exist  numbers
 $\alpha_i>0,$~ $i=\overline {1,m},$ such that
\begin{eqnarray*} b=\sum\limits_{i=1}^m a_i\alpha_i.\end{eqnarray*}  From here \begin{eqnarray*} \alpha_i=\langle b ,f_i\rangle  >0,\quad
i=\overline{1,m}, \quad \langle f_i, b \rangle =
0,\quad i=\overline {m+1,n}.\end{eqnarray*}

 Sufficiency is obvious because $\alpha_i$
is determined by the formula $\alpha_i=\langle b ,f_i\rangle $
unambiguously from the representation $b =\sum\limits _{i=1}^n\alpha_i \bar a_i.$
\qed
\end{proof}

The next statement is obvious.
\begin{theorem}\label{allapotka1} Let $\{a_1,\dots ,a_m\},$ $1\leq m\leq n,$ be the set of linearly independent vectors in $R_+^n.$ The necessary and sufficient conditions for the vector $b $ to belong to  the non-negative cone $K_a^+$ created by the set of  vectors
$\{a_i,\ i=\overline {1,m}\}$ are the conditions
\begin{eqnarray} \label{allapotka2}
\langle f_i, b \rangle  \geq 0, \quad i=\overline {1,m}, \quad \langle f_i, b \rangle=
0,\quad i=\overline {m+1,n},
\end{eqnarray}
where  $f_i,\
i=\overline {1,n},$   is a set of vectors being  biorthogonal to the set of linearly independent vectors $\bar a_i, \ i=\overline {1,n},$ and
$\bar a_i=a_i, \ i=\overline {1,m}.$
\end{theorem}
From the Theorem \ref{allapotka1} the representation for the vector $b$ belonging to the cone
$K_a^+$
\begin{eqnarray} \label{allapotka3}
 b=\sum\limits_{i=1}^m\alpha_ia_i, \quad \alpha_i \geq 0, \quad i=\overline{1,m},
\end{eqnarray}
is valid.

Therefore, to check belonging of the vector $b$
to the interior of the non-negative cone $K_a^+$ created by vectors
$\{a_i,\  i=\overline {1,m}\}$
one must enlarge the set of $m$ linearly independent vectors
$\{a_i,\ i=\overline {1,m}\}$~ up to the set of $n$ linearly independent vectors in $R^n,$ then to build the biorthogonal set of vectors $\{f_i,\ i=\overline
{1,n}\}$ for the enlarged set of vectors and to check the conditions of the Theorem.

Describe now an algorithm of constructing strictly positive solutions to the set of equations
\begin{eqnarray} \label{luda05}
 \psi=\sum\limits_{i=1}^lC_i y_i, \quad y_i>0, \quad i=\overline{1,l},
\end{eqnarray}
with respect to the vector $y=\{y_i\}_{i=1}^l$
or the same set of equations in coordinate form
\begin{eqnarray} \label{gonl40}
\sum\limits _{i=1}^l
c_{ki}y_i =\psi _k,\quad k=\overline{1,n},
\end{eqnarray}
for the vector
$\psi =\{\psi _1,\dots ,\psi _n\}$ belonging to the interior of the polyhedral cone created by vectors $\{C_i=\{c_{ki}\}_{k=1}^n, \ i=\overline {1,l}\}.$

\begin{theorem}\label{jant39}
If a certain  vector
$\psi$ belonging to the interior of a non-negative $r$-dimensional polyhedral cone created by vectors $\{C_i=\{c_{ki}\}_{k=1}^n,\ i=\overline {1,l}\},$
is such that there exists a subset of $r$ linearly independent vectors of the set of  vectors
$\{C_i, \ i=\overline {1,l}\},$ such that the vector $\psi$ belongs to the interior of the cone created by this  subset of vectors,
then there exist $l-r+1$ linearly independent non-negative solutions $z_i$ to the set of equations (\ref{gonl40})  such that the set of strictly positive solutions to the set of equations (\ref{gonl40})
is given by the formula
\begin{eqnarray} \label{gonl41} y=\sum\limits
_{i=r}^l\gamma _iz_i , \end{eqnarray}
where \begin{eqnarray*} z_i=\{\langle\psi
,f_1\rangle - \langle C_i,f_1\rangle y_i^*,\dots , \langle\psi
,f_r\rangle - \langle C_i,f_r\rangle y_i^*,0,\dots ,y_i^*, 0,\dots
,0\},\quad i=\overline {r+1,l},\end{eqnarray*}
\begin{eqnarray*} z_r=\{\langle\psi ,f_1\rangle ,\dots
,\langle\psi ,f_r\rangle ,0,\dots ,0\},\end{eqnarray*}
 \begin{eqnarray*} y_i^*=\left\{\begin{array}{ll}
  \min\limits_{s\in K_i}\frac {\langle \psi
,f_s \rangle} {\langle  C_i,f_s \rangle },& K_i=\{s,\langle
C_i,f_s \rangle >0\},\\ 1, & {\langle C_i,f_s \rangle}\leq 0,
~\forall \ s=\overline{1,r}, \end{array} \right.
\end{eqnarray*}
and the components of the vector
$\{\gamma_i\}_{i=r}^l$  satisfy the set of inequalities
\begin{eqnarray*} \sum\limits _{i=r}^l\gamma _i=1, \quad
\gamma _i>0, \quad i=\overline {r+1,l}, \end{eqnarray*}
\begin{eqnarray} \label{eq2}
\sum\limits_{i=r+1}^l{\langle C_i,f_k \rangle} y_i^*\gamma_i<
{\langle\psi,f_k\rangle}, \quad k=\overline{1,r}.
\end{eqnarray}
\end{theorem}
\begin{proof}\smartqed  The vector $\psi $ belongs to the interior of
$r $-dimensional polyhedral cone, $r \leq n,$ and there exist
$r$ linearly independent vectors from the set of  vectors
$C_1,\dots ,C_l,$~ ~$l\geq n,$ such that $\psi $ is interior  for this set of
 $r$ linearly independent vectors.
Without loss of generality, suppose these vectors are $C_1,\dots ,C_r.$ However, if it is not the case, then one can get it by renumbering vectors $C_i$ and components of the vector $y=\{y_1,\dots ,y_l\},$  respectively.
Therefore, the vector $ \psi $ has the representation
\begin{eqnarray} \label{luda45}
 \psi=\sum\limits_{i=1}^r\alpha_i C_i, \quad \alpha_i>0, \quad i=\overline{1,r}.
\end{eqnarray}
Consider the set of equations
\begin{eqnarray} \label{gonl42}
\sum\limits _{i=1}^lC_iy_i=\psi
\end{eqnarray}
for the vector $y=\{y_1,\dots ,y_l\}.$

Build the set of vectors $f_1,\dots ,f_n$ being the set of biorthogonal  vectors to the set of linearly independent vectors
$C_1,\dots ,C_r$ and satisfying the conditions
\begin{eqnarray*} \langle f_i, C_j\rangle =\delta _{ij},\quad i,j=\overline{1,r},
\quad \langle f_i,C_j \rangle =0,\quad j=\overline{1, r}, \quad i=\overline{r+1,n}, \end{eqnarray*}  where
$\langle x,y \rangle $ is  the scalar product of the vectors $x,y$ in $R^n.$
The set of equations (\ref{gonl42}) is equivalent to the set of equations
\begin{eqnarray} \label{gonl43}
\sum\limits
_{i=r+1}^l\langle C_i,f_j \rangle y_i+y_j= \langle \psi ,f_j \rangle,\quad j=\overline{1,r},
\end{eqnarray}
where $ \langle \psi ,f_i \rangle  >0, \ i=\overline{1,r}.$

This equivalence holds because there hold the next equalities $ \langle f_i, C_j\rangle =0,\ j=\overline{1, l}, \ i=\overline{r+1, n}, $
$ \langle \psi, f_i \rangle =0, \ i=\overline{r+1, n}. $ The last equalities are the consequence of the representation (\ref{luda45}) for the vector $\psi.$
Note that the general solution to the set of equations (\ref{gonl43})
has the form
\begin{eqnarray*}  y=\left\{\langle \psi ,f_1 \rangle -\sum\limits _{i=r+1}^l\langle C_i, f_1\rangle y_i, \dots\right. \end{eqnarray*}
 \begin{eqnarray} \label{gonl44}
\left.\dots ,\langle \psi ,f_r >- \sum\limits _{i=r+1}^l\langle C_i, f_r\rangle y_i, y_{r+1},\dots
,y_l\right\},
\end{eqnarray}
where the vector $\tilde y=\{y_{r+1},\dots ,y_l\}$ runs over the set  $R^{l-r}.$

The vectors
$z_i,\ i=\overline {r,l},$ defined in the Theorem solve the set of equations
(\ref{gonl43}), their components are non-negative, and they themselves are linearly independent.
Build after vectors
$z_i$ the vector
\begin{eqnarray*} \bar y=\sum\limits
_{i=r}^l{\gamma _i}{z_i},\end{eqnarray*}
where
\begin{eqnarray*} \sum\limits _ {i=r}^l\gamma
_i=1.\end{eqnarray*}
Then
\begin{eqnarray} \label{gonl45}
\bar y=\left\{ \langle \psi
,f_1 \rangle  - \sum\limits _{i=1+r}^l\langle C_i, f_1\rangle \gamma _iy_i^*,\right.
\end{eqnarray}  \begin{eqnarray*} \left.\dots ,\langle \psi ,f_r \rangle
-\sum\limits_{i=r+1}^l\langle C_i, f_r\rangle \gamma _iy_i^*, \gamma
_{r+1}y_{r+1}^*,\dots ,\gamma _ly_l^*\right\}\end{eqnarray*}
satisfies  the  conditions of the  Theorem.
Show that there is reciprocal one-to-one correspondence between vectors $\bar y$ and $\tilde y$ determined by formulae (\ref{gonl45}) and
(\ref{gonl44}), respectively.
To prove this, it is sufficient to prove that every vector
~$\tilde y={\{y_{r+1},\dots
,y_l\}}\in R^{l-r}$ has one and only one corresponding set
$\gamma _i, ~ i=\overline {r,l},$ such that
$\sum\limits _{i=r}^l\gamma _i =1.$
Really, from the linear independence of vectors
$z_r,z_{r+1},\dots ,z_l$ it follows that the set of equations
\begin{eqnarray*}  \gamma _{r}z_r+\dots +\gamma
_{l}z_l=y\end{eqnarray*}  is equivalent to the set of equations
\begin{eqnarray*} \gamma _{r+1}(z_{r+1}-z_r)+\dots +\gamma
_l{(z_l-z_r)}=y-z_r.\end{eqnarray*}
The vectors $z_{r+1}-z_r,\dots
,z_l-z_r$ are linearly independent, therefore we can determine $\gamma
_{r+1},\dots ,\gamma _l$ for every vector
$\tilde y =\{y_{r+1},\dots ,y_l\}$ unambiguously. From the equality
$\sum\limits _{i=r}^l \gamma _i=1 $ we determine the number $  \gamma _r$
unambiguously too. It is easy to see that $y_i=\gamma _iy_i^*, \ i=\overline{r+1,l}.$ The solution $\bar
y$ is strictly positive  if $\gamma _i,~ i=\overline
{r,l},$ satisfy the set of inequalities defined in the Theorem \ref{jant39}.
\qed
\end{proof}

\begin{definition}\label{allunja1}
We call a subset of vectors $\{b_1, \ldots, b_m\}$   from the set of vectors
$\{a_1, \ldots, a_t\}$
the  generating set of  vectors of the $r$-dimensional cone created by the vectors\index{generating set of  vectors of a  cone created by the vectors}
$\{a_1, \ldots, a_t\}$
 if it satisfies  the conditions: \\
1) every vector $b_i, \ i=\overline{1,m},$ from the set of vectors $\{b_1, \ldots, b_m\}$ does not belong to  the cone created by the set of  vectors $\{b_1, \ldots, b_m\}\setminus \{b_i\}\  ;$ \\
2) the cone created by the set of vectors  $\{a_1, \ldots, a_t\}$ coincides with the cone created by the vectors $\{b_1, \ldots, b_m\}.$
\end{definition}
\begin{proposition}\label{nastya2}
The  generating set of vectors of  $r$-dimensional cone created by a set of vectors  $\{a_1, \ldots, a_t\}$ exists and contains no less than $r$ vectors.
\end{proposition}
\begin{proof}\smartqed The proof we carry out by  induction on the number of vectors. Denote   $\{b_1,\ldots ,b_m\}$  the generating set of vectors for the cone created by the set of vectors
$\{a_1,\ldots, a_k\}.$
Let  $K_{a_1, \ldots, a_k}^+$ and  $K_{b_1,\ldots ,b_m}^+$ be the cones created by the set of vectors
$\{a_1,\ldots, a_k \}$ and  $\{b_1,\ldots ,b_m\},$
correspondingly. In accordance with the Definition \ref{allunja1}, for  the generating set of vectors the equality  $K_{a_1, \ldots, a_k}^+=K_{b_1,\ldots ,b_m}^+$ holds.
Let $a_{k+1}$  be a certain vector not belonging to the cone $K_{a_1, \ldots, a_k}^+.$
The equality $K_{a_1, \ldots, a_{k+1}}^+= K_{b_1,\ldots ,b_m, a_{k+1}}^+$  takes place.
For the vector $b_i, \ i=\overline{1,m}, $ consider the cone $ K_{b_1,\ldots b_{i-1}, b_{i+1}, \ldots ,b_m, a_{k+1}}^+$ created by the set of vectors $\{a_{k+1}\}\cup\{b_1,\ldots ,b_m\} \setminus \{b_i\}.$
Two cases are possible: the vector $b_i$  belongs to the cone $ K_{b_1,\ldots b_{i-1}, b_{i+1}, \ldots ,b_m, a_{k+1}}^+$ and  it does not belong to this one. If $b_i$ belongs to this cone, then  we  throw out it from the set of vectors $\{b_1,\ldots ,b_m\}\cup \{a_{k+1}\}.$
It is obvious that $ K_{b_1,\ldots, b_m, a_{k+1}}^+=K_{b_1,\ldots b_{i-1}, b_{i+1}, \ldots ,b_m, a_{k+1}}^+.$
Further we act analogously. If the vector $b_j$
belongs to the cone  $K_{b_1,\ldots b_{i-1}, b_{i+1}, \ldots ,  b_{j-1}, b_{j+1}, \ldots, b_m, a_{k+1}}^+,$ then  we  throw out it from the set of vectors $\{a_{k+1}\} \cup\{b_1,\ldots ,b_m\} \setminus \{b_i\}.$ Then the  following equalities
\begin{eqnarray*}  K_{b_1,\ldots, b_m, a_{k+1}}^+=K_{b_1,\ldots b_{i-1}, b_{i+1}, \ldots ,b_m, a_{k+1}}^+ \end{eqnarray*}   \begin{eqnarray*} =K_{b_1,\ldots b_{i-1}, b_{i+1}, \ldots ,  b_{j-1}, b_{j+1}, \ldots, b_m, a_{k+1}}^+\end{eqnarray*}
hold.
Having carried out the finite number of steps, we come to the set of vectors $\{a_{k+1}\}\cup\{b_{1}^1,\ldots  b_{m_1}^1\}$
that does  not contain  none  vector $b_j^1, \ j=\overline{1, m_1},$   belonging to   the corresponding cone $ K_{b_1^1,\ldots,   b_{j-1}^1, b_{j+1}^1, \ldots, b_{m_1}^1, a_{k+1}}^+.$
It is obvious that the  equality  $K_{b_1^1,\ldots ,b_{m_1}^1, a_{k+1}}^+= K_{b_1,\ldots ,b_m, a_{k+1}}^+$ takes place.   To finish the proof one has to note that as the  basis of induction we can choose  any set of $r$ linearly independent  vectors from the set of vectors
$\{a_i,\  i=\overline {1,t}\}.$
\qed
\end{proof}
\begin{proposition}\label{allunja2}
A certain vector $d$ belongs to the interior of  $r$-dimen\-sional cone created by vectors $\{a_i,\  i=\overline {1,t}\},$
if and only if for the vector $d$  the representation
\begin{eqnarray} \label{allunja3}
d=\sum\limits_{i=1}^m \alpha_i b_i
\end{eqnarray}
holds, where
$\{b_1, \ldots, b_m \}$ is the generating set of vectors of the considered cone,
$\alpha_i \geq 0, $ and among coefficients  $\alpha_i,\ i=\overline {1,m}, $ there are no less than $r$ strictly positive numbers.
\end{proposition}
\begin{proof}\smartqed  Necessity.
The cone created by vectors  $\{a_i,\  i=\overline {1,t}\}$
is the union of the cones  created by all  $r$ linearly independent subsets of vectors $\{b_{i_1}, \ldots, b_{i_r} \}$ belonging to the set $ \{b_{1}, \ldots, b_{m} \}.$
Therefore, there exists a maximal number of subsets $\{b_{i_1^s}, \ldots, b_{i_r^s} \},\  s=\overline{1,w},$ of $r$ linearly independent  vectors such that  the vector $d$ belongs to every cone created by subset of vectors $\{b_{i_1^s}, \ldots, b_{i_r^s} \},\  s=\overline{1,w}.$ In accordance with the Theorem \ref{allapotka1}, for the vector $d$ belonging to the cone  created by vectors  $\{a_i,\  i=\overline {1,t}\}$ there holds the representations
\begin{eqnarray*}  d=\sum\limits_{k=1}^r \alpha_{i_k^s} b_{i_k^s}, \quad \alpha_{i_k^s} \geq 0, \quad  k=\overline{1,r}, \quad  s=\overline{1,w}.\end{eqnarray*}
From here, it follows the representation
\begin{eqnarray*}  d=\frac{1}{w}\sum\limits_{s=1}^w \sum\limits_{k=1}^r \alpha_{i_k^s} b_{i_k^s}.\end{eqnarray*}
In this representation the number of different generating vectors entering with strictly positive coefficients  is  not less  than $r. $
Because if it is not the case  then the vector $d$ do not belong to the interior of the cone
created by vectors  $\{a_i,\  i=\overline {1,t}\}.$

 Sufficiency. If  conditions of the  Proposition \ref{allunja2} hold, then the vector $d$ is the sum of two vectors, namely, a certain  vector $x$ that gets into the interior of the cone created by a subset of $r$ linearly independent vectors of the generating set of vectors  and a vector $y$ from the cone created by vectors
$\{a_i,\  i=\overline {1,t}\}.$
Let us apply the Theorem \ref{jant39} to the vector $x$ getting into the interior of the cone created by the subset of
$r$ linearly independent vectors from the generating set of vectors, taking for the vector set  $C_i$ the vector set $a_i$ under the condition that $l=t.$
Therefore, the vector $x$ has the representation
\begin{eqnarray*} x =\sum\limits _{s=1}^t \gamma_s a_s,\end{eqnarray*}
where $\gamma_s>0, \ s=\overline{1,t}.$
From here it immediately follows that the vector $d$ has the representation
\begin{eqnarray*} d =\sum\limits _{s=1}^t \gamma_s^1 a_s,
\end{eqnarray*}
where $\gamma_s^1>0, \ s=\overline{1,t}.$
\qed
\end{proof}

The algorithm of checking whether the vector $d$ belongs to the interior of the cone created by the vectors
$\{a_i,\  i=\overline {1,t}\}$ consists of building the generating  set of vectors  $\{b_i,\  i=\overline {1,m}\}\  ;$
for every  subset of $r$  linearly independent vectors from the generating set of vectors one must  use the Theorem \ref{allapotka1} to choose only those subsets of $r$ linearly independent vectors $\{b_{i_k^s}, \ k=\overline{1,r}\}$ from the  generating set of vectors for which  the representations
\begin{eqnarray} \label{allochka1000}
d=\sum\limits_{k=1}^r\delta_{i_k^s}b_{i_k^s}, \quad \delta_{i_k^s} \geq 0, \quad k=\overline{1,r},
\end{eqnarray}
are valid.

Two cases are possible:

\noindent 1) there not exists a subset of $r$ linearly independent vectors from the generating set of vectors such that the representation  (\ref{allochka1000}) is valid.

\noindent 2)  there exists a certain set of  subsets of  $r$ linearly independent vectors from the generating set of vectors such that for every subset $r$ linearly independent vectors belonging to this set all coefficients
of the  expansion  (\ref{allochka1000})  are non-negative.

In the first case, the vector $d$ does not belong to the interior of the cone created by  the vectors
$\{a_i,\  i=\overline {1,t}\}.$
In the   second one, if the number of different generating vectors that  appeared in expansions (\ref{allochka1000}) for the vector $d$ with  positive  coefficients in expansions  is not less  than  $r$
then the vector $d$ belongs to the interior of the cone created by the set of  vectors
$\{a_i,\  i=\overline {1,t}\}.$
In the opposite case the vector $d$ does not belong to the interior of the cone created by the vectors
$\{a_i,\  i=\overline {1,t}\}.$

In the next Theorem, we solve the problem of constructing  the strictly positive  solutions to the set of equations  (\ref{gonl40}) without additional  assumption figuring in the Theorem \ref{jant39}.  The rank of the set of the vectors  $\{C_i=\{c_{ki}\}_{k=1}^n,\ i=\overline {1,l}\}$ we denote  $r.$

\begin{theorem}\label{allupotka4}
Let   a   vector   of  final  consumption\index{ vector   of  final  consumption }    $ \psi$  belong  to  the   interior    of   the   cone   created    by    the    set    of    vectors   $\{C_i=\{c_{ki}\}_{k=1}^n,\ i=\overline {1,l}\}.$
 Then there exists  a  set of vectors  $ \psi_s=\{\psi_k^s\}_{k=1}^n, \ s=\overline{1, 2},$ and a real number $0 \leq \alpha \leq 1$
 satisfying conditions:\\
1) every vector $ \psi_s=\{\psi_k^s\}_{k=1}^n, \ s=\overline{1, 2},$ belongs to the interior of a cone created by a set of $r$ linearly independent vectors $\{C_{i_1^s}, \ldots C_{i_r^s}\}.$\\
2) there holds the representation
$ \psi= \alpha \psi_1+ (1-\alpha)\psi_2.$ \\
A strictly positive solution  to the set of equations (\ref{gonl40})
can be represented in the form
\begin{eqnarray*}  y=\alpha y_1+ (1-\alpha)y_2, \quad y_s= \{y_i^s \}_{i=1}^l,\end{eqnarray*}
where $y_s$ is a strictly positive solution  to the set of equations
\begin{eqnarray} \label{allupotka5}
\sum\limits _{i=1}^l
c_{ki}y_i^s =\psi _k^s,\quad k=\overline{1,n}, \quad s=\overline{1,2},
\end{eqnarray}
constructed in the Theorem \ref{jant39}.
\end{theorem}
\begin{proof}\smartqed Without loss of generality, denote the generating set of vectors for the cone $K_{C_1, \ldots, C_l}^+$  by $C_1, \ldots,C_m,$  since  it is not the case we can renumber the set of vectors ${C_1, \ldots, C_l}.$ Since the cone $K_{C_1, \ldots, C_l}^+$  has dimension $r,$ let us consider all subsets  of vectors $\{C_{i_1^s}, \ldots C_{i_r^s}\},\ s=\overline{1,w},\ $ from the set $C_1, \ldots,C_m,$
being lenear independent.  It is evident that 
\begin{eqnarray*} K_{C_1, \ldots, C_l}^+ =\bigcup\limits_{s=1}^wK_{C_{i_1^s}, \ldots C_{i_r^s}}^+, \end{eqnarray*}
where $K_{C_{i_1^s}, \ldots C_{i_r^s}}^+$ is a nonnegative cone created by the vectors $C_{i_1^s}, \ldots C_{i_r^s}.$
Consider two subcones $K_{C_{i_1^{s_1}}, \ldots C_{i_r^{s_1}}}^+$ and $K_{C_{i_1^{s_2}}, \ldots C_{i_r^{s_2}}}^+$ from the cone $ K_{C_1, \ldots, C_l}^+$
such that
\begin{eqnarray*} K_{C_{i_1^{s_1}}, \ldots C_{i_r^{s_1}}}^+ \cap K_{C_{i_1^{s_2}}, \ldots C_{i_r^{s_2}}}^+\neq \emptyset \end{eqnarray*}
and
\begin{eqnarray*} K_{C_{i_1^{s_1}}, \ldots C_{i_r^{s_1}}}^+ \neq K_{C_{i_1^{s_2}}, \ldots C_{i_r^{s_2}}}^+. \end{eqnarray*}
If the vector $\psi$  belongs to set
\begin{eqnarray*} K_{C_{i_1^{s_1}}, \ldots C_{i_r^{s_1}}}^+ \cap K_{C_{i_1^{s_2}}, \ldots C_{i_r^{s_2}}}^+ \end{eqnarray*}
and is internal vectors for the cone $ K_{C_1, \ldots, C_l}^+,$ then there exist two internal vectors $\psi_1$ and $\psi_2$ belonging correspondingly  to cones $K_{C_{i_1^{s_1}}, \ldots C_{i_r^{s_1}}}^+$ and $K_{C_{i_1^{s_2}}, \ldots C_{i_r^{s_2}}}^+$ and a real number $ 0< \alpha< 1$ such that
\begin{eqnarray*} \psi=\alpha \psi_1+ (1-\alpha)\psi_2 \end{eqnarray*}
due to convexity of the cone  $ K_{C_1, \ldots, C_l}^+.$ 
\qed
\end{proof}

In this Chapter, we suppose that all subvention vectors $ d_i, \ i=\overline{1,n},$ are equal zero.
As in Chapter 4, denote  $c^0$ the set of  vectors of  property   $b_i, \ i=\overline{1,n}.$

\begin{lemma}\label{ant4}
Let the structure of supply  agree with the structure of  choice\index{structure of supply  agree with the structure of  choice}  on the set $B_0,$ and
let an output vector\index{output vector} $x^0=\{x_i^0\}_{i=1}^n$ corresponding to the  productive process $z^0 \in B_0,$ matrices $A$ and $\bar B$ be such that the couple of matrices $\{AX_0, \bar B + X_0\}$ belong to the class $\Pi_0,$ the matrix  $ \bar C\left(z^0\right)$ does not contain zero rows.

If
$ \sum\limits_{s=1}^nc_{si}\left(z^0\right) > 0, \ i=\overline{1,l}, $
the vector $\tilde C\left(z^0\right) \tilde y\left(z^0\right) - \tilde B \tilde e$ belongs to the interior of a non-negative cone created by the vectors-columns of the matrix  $\bar B + X_0- AX_0,$ then for any $p$ from the set
\begin{eqnarray*} T_1\left(z^0,c^0\right)=\left\{p \in K_+^n, \  p=H^T\left(z^0,c^0\right)\delta, \ \delta \in R_+^n\setminus \{0,\ldots,0\}\right\},\end{eqnarray*}
where
\begin{eqnarray*}  H\left(z^0,c^0\right)=[\bar B + X_0 - AX_0]^{-1},\end{eqnarray*}
and $H^T\left(z^0,c^0\right)$ is a matrix transposed to the matrix $H\left(z^0,c^0\right),$
there exists a solution to the equation
\begin{eqnarray*}  \sum\limits _{i=1}^n\pi_i \left(z^0\right)\left[x_i^0\left(p_i - \sum\limits_{s=1}^na_{si}p_s\right) + \langle p,b_i\rangle \right] \end{eqnarray*}
\begin{eqnarray} \label{ant5}
=\sum\limits _{i=1}^ny_i\left(z^0\right)\sum\limits_{k=1}^nc_{ki}\left(z^0\right)p_k
\end{eqnarray}
with respect to the vector $\pi\left(z^0\right)=\{\pi_i\left(z^0\right)\}_{i=1}^n$
 such that the components $\pi_i\left(z^0\right),  \  i=\overline{1,n},$ of the solution
$\pi\left(z^0\right)=[\bar B + X_0 - AX_0]^{-1}\bar C\left(z^0\right)\bar y\left(z^0\right)$
to the equation (\ref{ant5}) satisfy inequalities $0 < \pi_i(z^0) < 1,  \  i=\overline{1,n}.$
\end{lemma}
\begin{proof}\smartqed
If the vector $\pi\left(z^0\right)=\{\pi_i\left(z^0\right)\}_{i=1}^n$ solves the set of equations
\begin{eqnarray*}   \sum\limits _{j=1}^n\pi_j\left(z^0\right)[\delta_{ij}x_j^0 - a_{ij}x_j^0 +b_{ij}]=\sum\limits_{k=1}^nc_{ik}\left(z^0\right)y_k\left(z^0\right), \quad i=\overline{1,n},\end{eqnarray*}
then, obviously, it solves the equation (\ref{ant5}). The solution to the last set of equations is unique and can be expressed as follows
\begin{eqnarray*} \pi\left(z^0\right)=[\bar B + X_0 - AX_0]^{-1}\bar C\left(z^0\right)\bar y\left(z^0\right).\end{eqnarray*}
 From the conditions of the Theorem, it follows that the components of  this solution  are strictly positive  because matrix  elements of the matrix $[\bar B + X_0 - AX_0]^{-1}$  are strictly positive and the vector $\bar C\left(z^0\right)\bar y\left(z^0\right)$ does not equal zero.
The vector $ \bar e - \pi\left(z^0\right)=\bar \pi\left(z^0\right)$ solves the set of equations
 \begin{eqnarray*} [\bar B + X_0 - AX_0]\bar \pi\left(z^0\right)=\tilde C\left(z^0\right) \tilde y\left(z^0\right) - \tilde B \tilde e\end{eqnarray*}
and every its component   is strictly positive. Really, from the fact that the vector
$\tilde C\left(z^0\right) \tilde y\left(z^0\right) - \tilde B \tilde e$ belongs to the interior of the non-negative cone created by the vectors-columns of the matrix $\bar B + X_0- AX_0$ the representation
\begin{eqnarray*} \tilde C\left(z^0\right) \tilde y\left(z^0\right) - \tilde B \tilde e= [\bar B + X_0 - AX_0]\alpha\end{eqnarray*}
is valid, where  the components  of the vector $\alpha=\{\alpha_i\}_{i=1}^n$ are strictly positive. From the non-degeneracy of the matrix
$[\bar B + X_0 - AX_0],$ we obtain the equality $ \bar e - \pi\left(z^0\right)=\alpha.$ From here it follows that components of the solution  satisfy
 inequalities
\begin{eqnarray*} 0 < \pi_i\left(z^0\right) < 1,  \quad   i=\overline{1,n}.\end{eqnarray*}
\qed
\end{proof}
\begin{theorem}\label{hbant1}
If the  conditions of the  Lemma \ref{ant4} hold, then on the set
$T_1\left(z^0,c^0\right)\times z^0$
a taxation system\index{taxation system}
\begin{eqnarray*} ||\pi_{ij}^0\left(p,z^0\right)||_{i,j=1}^l, \quad  \left(p,z^0\right) \in T_1\left(z^0,c^0\right)\times z^0,\end{eqnarray*}
exists satisfying conditions
\begin{eqnarray*} \pi_{ij}^0\left(p,z^0\right)=\pi_i\left(z^0\right)\delta_{ij},  \quad \left(p,z^0\right) \in T_1\left(z^0,c^0\right)\times z^0,\quad  i =\overline{1,n}, \quad j =\overline{1,l},\end{eqnarray*}
\begin{eqnarray*}
 \pi_j\left(z^0\right)+ \sum\limits_{i=n+1}^l\pi_{ij}^0\left(p,z^0\right)=1,  \quad \left(p,z^0\right) \in T_1\left(z^0,c^0\right)\times z^0, \quad  j =\overline{1,n}, \end{eqnarray*}
\begin{eqnarray} \label{qqant1}
\sum\limits_{i=n+1}^l\pi_{ij}^0\left(p,z^0\right)=1,  \quad  \left(p,z^0\right) \in T_1\left(z^0,c^0\right)\times z^0, \quad j =\overline{n+1,l},
\end{eqnarray}
where the vector $\pi\left(z^0\right)=\{\pi_i\left(z^0\right)\}_{i=1}^n$ solves the equation
(\ref{ant5}).
\end{theorem}
\begin{proof}\smartqed
Consider on the set $T_1\left(z^0,c^0\right)\times z^0$ two sets of functions
\begin{eqnarray*}  f_i\left(p,z^0\right)= y_i\left(z^0\right)\sum\limits_{k=1}^nc_{ki}\left(z^0\right)p_k, \quad i=\overline{n+1,l},\end{eqnarray*}
\begin{eqnarray*}  g_i\left(p,z^0\right)=\left(1 - \pi_i\left(z^0\right)\right)\left[x_i^0\left(p_i - \sum\limits_{s=1}^na_{si}p_s\right)+ \langle b_i, p\rangle \right], \quad i=\overline{1,n}, \end{eqnarray*}  \begin{eqnarray*}  g_i\left(p,z^0\right)=\langle b_i, p \rangle, \quad  i=\overline{n+1,l},\end{eqnarray*}
where the vector $\pi\left(z^0\right)=\{\pi_i\left(z^0\right)\}_{i=1}^n$ solves the  equation
(\ref{ant5}).
Then on the set $T_1\left(z^0,c^0\right)\times z^0$ the equality
\begin{eqnarray*}  \sum\limits_{i=n+1}^ly_i\left(z^0\right)\sum\limits_{k=1}^nc_{ki}\left(z^0\right)p_k \end{eqnarray*}
\begin{eqnarray} \label{dant2}
 =\sum\limits_{i=1}^n\left(1 - \pi_i\left(z^0\right)\right)\left[x_i^0\left(p_i - \sum\limits_{s=1}^na_{si}p_s\right)+ \langle b_i, p\rangle \right]+  \sum\limits_{i=n+1}^l \langle b_i, p\rangle ,
\end{eqnarray}
holds because with the conditions of the Theorem  it turns into the equation (\ref{ant5}) for the vector
$\pi\left(z^0\right)=\{\pi_i\left(z^0\right)\}_{i=1}^n.$ However, the equality (\ref{dant2}) in terms of the sets of  functions  $f_i\left(p,z^0\right), \ i=\overline{n+1,l},$ and $g_i\left(p,z^0\right), \ i=\overline{1,l},$ has the form
\begin{eqnarray*} \sum\limits_{i=n+1}^lf_i\left(p,z^0\right)= \sum\limits_{i=1}^lg_i\left(p,z^0\right),\quad \left(p,z^0\right) \in T_1\left(z^0,c^0\right)\times z^0,\end{eqnarray*}
that is valid on the set $T_1\left(z^0,c^0\right)\times z^0.$ Due to the Lemma \ref{let1} about the existence of a  taxation system,\index{Lemma about the existence of   taxation system} whose the rest of conditions are valid  because the conditions  of the  Theorem  \ref{hbant1} hold,
  a taxation system $||\bar \pi_{ij}\left(p,z^0\right)||_{i,j=n+1}^l$ exists such  that
\begin{eqnarray*}  f_i\left(p,z^0\right)=\sum\limits_{j=1}^l\bar \pi_{ij}\left(p, z^0\right)g_j\left(p,z^0\right), \quad \left(p,z^0\right) \in T_1\left(z^0,c^0\right)\times z^0, \quad  i=\overline{n+1,l}.\end{eqnarray*}
After the taxation system just built, introduce on the set $T_1\left(z^0,c^0\right)\times z^0$ a new taxation system supposing
\begin{eqnarray*}    \pi_{ij}^0\left(p,z^0\right)=\delta_{ij}\pi_i\left(z^0\right), \quad i=\overline{1,n}, \quad j=\overline{1,l},\end{eqnarray*}  \begin{eqnarray*}  \pi_{ij}^0\left(p,z^0\right)=\left(1 - \pi_j\left(z^0\right)\right)\bar \pi_{ij}\left(p,z^0\right), \quad i=\overline{n+1,l},  \quad j=\overline{1,n},\end{eqnarray*}  \begin{eqnarray*}
\pi_{ij}^0\left(p,z^0\right)=\bar \pi_{ij}\left(p,z^0\right), \quad i=\overline{n+1,l},  \quad j=\overline{n+1,l}.\end{eqnarray*}
The taxation system built in such a way satisfies  the  conditions of the Theorem \ref{hbant1}.
\qed
\end{proof}

Let us construct a taxation system
$||\pi_{ij}(p,z)||_{i,j=1}^l, \ (p,z) \in K_+^n \times \Gamma^n,$ satisfying conditions
\begin{eqnarray*} \pi_{ij}(p,z)=\pi_i(z)\delta_{ij}, \quad   0< \pi_i(z) <1, \quad (p,z) \in K_+^n \times \Gamma^n,\quad  i =\overline{1,n}, \quad j =\overline{1,l},\end{eqnarray*}
\begin{eqnarray*}
 \pi_j(z)+ \sum\limits_{i=n+1}^l\pi_{ij}(p,z)=1,  \quad (p,z) \in K_+^n \times \Gamma^n, \quad  j =\overline{1,n}, \end{eqnarray*}
\begin{eqnarray} \label{ant1}
\sum\limits_{i=n+1}^l\pi_{ij}(p,z)=1,  \quad (p,z) \in K_+^n \times \Gamma^n, \quad j =\overline{n+1,l}.
\end{eqnarray}

 Let a productive process $z^0$ and the corresponding output vector $x^0$ satisfy conditions of the Lemma \ref{ant4} and the Theorem \ref{hbant1}. Suppose that on the set $T_1\left(z^0,c^0\right)\times z^0$ a taxation system is determined by the Theorem \ref{hbant1}. On the set $[K_+^n \times \Gamma^n \setminus  T_1\left(z^0,c^0\right)]\times z^0,$ one takes it arbitrary provided that equalities (\ref{ant1}) hold.

The  taxation system built above we call  the taxation system\index{taxation system} determined by the Lemma \ref{ant4} and the Theorem \ref{hbant1}.

Consider a family  of income pre-functions\index{family  of income pre-functions}
\begin{eqnarray*}  K^0_i\left(p,z\right)=\pi_i\left(z\right)[\langle y_i -  x_i, p \rangle + \langle b_i,p \rangle],  \quad \left(p,z\right) \in K_+^n \times \Gamma^n, \quad i=\overline{1,n},\end{eqnarray*}
 \begin{eqnarray*}  K^0_i\left(p,z\right)=\sum\limits_{j=1}^n\pi_{ij}\left(p,z\right)[\langle y_j -  x_j, p \rangle + \langle b_j,p \rangle] \end{eqnarray*}
\begin{eqnarray} \label{ant2}
+ \sum\limits_{j=n+1}^l\pi_{ij}\left(p,z\right)  \langle b_j,p \rangle,  \quad \left(p,z\right) \in K_+^n \times \Gamma^n, \quad  i=\overline{n+1,l},
\end{eqnarray}
and a family of income functions\index{family of income functions} built after it
\begin{eqnarray*}  K_i\left(p,z\right)= K^0_i\left(p,Q\left(p,z\right)\right) \end{eqnarray*}
 \begin{eqnarray*} = \pi_i(Q(p,z))\left[u_i\left(p_i - \sum\limits_{s=1}^na_{si}p_s\right)\prod\limits_{s=1}^na_s(p)b_s(p,z)+ \langle b_i,p \rangle\right], \quad i=\overline{1,n},\end{eqnarray*}

\begin{eqnarray*}  K_i(p,z)= K^0_i(p,Q(p,z)) \end{eqnarray*}  \begin{eqnarray*} =\sum\limits_{j=1}^n\pi_{ij}(p,Q(p,z))\left[u_i\left(p_i - \sum\limits_{s=1}^na_{si}p_s\right)\prod\limits_{s=1}^na_s(p)b_s(p,z)+ \langle b_i, p \rangle\right] \end{eqnarray*}
\begin{eqnarray} \label{qant2}
+ \sum\limits_{j=n+1}^l\pi_{ij}(p,Q(p,z))  \langle b_j,p \rangle,  \quad  i=\overline{n+1,l},
\end{eqnarray}
\begin{eqnarray*}   z \in \Gamma^n, \quad z=\{z^i\}_{i=1}^n, \quad  z^i=( x_i,  y_i), \quad i=\overline{1,n}.\end{eqnarray*}
Here the vector $u=\{u_i\}_{i=1}^n$ is a gross output in the economy system corresponding to a productive process $ z \in \Gamma^n.$

Since for $z^0 \in B_0 \subseteq T$ there hold equalities
\begin{eqnarray*} v_i(p,\zeta_0(p, \omega_0))=\chi_{[0, \infty)}\left(x_i^0 - \sum\limits_{k=1}^na_{ik}x_k^0 + \sum\limits_{k=1}^lb_{ik}\right)=1, \quad  i=\overline{1,n},\end{eqnarray*}
the reduction of the considered family of income functions  onto the set $T_1\left(z^0,c^0\right)\times z^0$ takes the form
\begin{eqnarray*}
K_i\left(p,z^0\right)=\pi_i\left(d(p)z^0\right)\left[x_i^0d(p)\left(p_i - \sum\limits_{s=1}^na_{si}p_s\right)+ \langle b_i,p \rangle\right], \quad   i=\overline{1,n},
\end{eqnarray*}
\begin{eqnarray*} K_i\left(p, z^0\right)=\sum\limits_{j=1}^n\pi_{ij}\left(p,d(p)z^0\right)\left[x_j^0d(p)\left(p_j - \sum\limits_{s=1}^na_{sj}p_s\right)+ \langle b_j,p \rangle \right] \end{eqnarray*}  \begin{eqnarray} \label{dant101} + \sum\limits_{j=n+1}^l\pi_{ij}\left(p, d(p)z^0\right)  \langle b_j,p \rangle, \quad
i=\overline{n+1,l},\end{eqnarray}
where 
\begin{eqnarray*}
d(p)=\prod\limits_{s=1}^na_s(p).
\end{eqnarray*}
For those $p$ for which  $d(p)=1,$ 
the reduction of the considered family of income functions  onto the set $T_1\left(z^0,c^0\right)\times z^0$ takes the form
\begin{eqnarray*}
K_i\left(p,z^0\right)=\pi_i\left(z^0\right)\left[x_i^0\left(p_i - \sum\limits_{s=1}^na_{si}p_s\right)+ \langle b_i,p \rangle\right], \quad   i=\overline{1,n},
\end{eqnarray*}
\begin{eqnarray*} K_i\left(p, z^0\right)=\sum\limits_{j=1}^n\pi_{ij}^0\left(p,z^0\right)\left[x_j^0\left(p_j - \sum\limits_{s=1}^na_{sj}p_s\right)+ \langle b_j,p \rangle \right] \end{eqnarray*}  \begin{eqnarray*} + \sum\limits_{j=n+1}^l\pi_{ij}^0\left(p, z^0\right)  \langle b_j,p \rangle, 
\end{eqnarray*}
\begin{eqnarray} \label{dant1}
=y_i\left(z^0\right)\sum\limits_{k=1}^nc_{ki}\left(z^0\right)p_k,  \quad  i=\overline{n+1,l}.\end{eqnarray}

Consider the case of a random field $\zeta_0(p, \omega_0)$ taking single value $z^0$ in the set $B_0 $ with probability 1 and does not depend on $p.$

Therefore, we suppose
\begin{eqnarray*}  \zeta_0(p, \omega_0)=\left\{\zeta_0^i(p)\right\}_{i=1}^n=z^0, \end{eqnarray*}  \begin{eqnarray*}  \zeta_0^i(p)=\left\{\left\{a_{ki}x_i^0\right\}_{k=1}^n, \   x_i^0e_i\right\}=z^i_0, \quad i=\overline{1,n}.\end{eqnarray*}
Then
\begin{eqnarray*} \zeta(p)=Q(p,\zeta_0(p, \omega_0))= Q(p,z^0)=\{\zeta_i(p)\}_{i=1}^n=d(p)z^0, \end{eqnarray*}  \begin{eqnarray*}  \zeta_i(p)=\prod\limits_{s=1}^n a_s(p) v_s(p,\zeta_0(p, \omega_0))\left\{\left\{ a_{ki}x_i^0 \right\}_{k=1}^n, \ x_i^0e_i\right\}
\end{eqnarray*}
\begin{eqnarray*}
=\{\zeta_i^{1}(p), \zeta_i^{ 2}(p)\}=d(p)z_0^i,\end{eqnarray*}
where
\begin{eqnarray*} \zeta_i^{1}(p)=d(p)\left\{ a_{ki}x_i^0 \right\}_{k=1}^n, \quad  \zeta_i^{ 2}(p)=d(p)x_i^0e_i. \end{eqnarray*}
Under these conditions,
\begin{eqnarray*} D_i(p)= K_i^0\left(p,\zeta(p)\right)=\pi_i(\zeta(p))\left[ \langle p, \zeta_i^{2}(p) - \zeta_i^{1}(p) \rangle + \langle p,b_i \rangle\right] \end{eqnarray*}  \begin{eqnarray*} =\pi_i(\zeta(p))\left[d(p)  x_i^0\left(p_i - \sum\limits_{s=1}^na_{si}p_s\right) + \langle p,b_i \rangle \right], \quad  i=\overline{1,n},\end{eqnarray*}
\begin{eqnarray*} D_i(p)= K_i^0(p,\zeta(p))=\end{eqnarray*}  \begin{eqnarray*} = \sum\limits_{j=1}^n\pi_{ij}(p,\zeta(p))\left[d(p) x_j^0\left(p_j - \sum\limits_{s=1}^na_{sj}p_s\right) +  \langle p,b_j \rangle \right] \end{eqnarray*}  \begin{eqnarray*} + \sum\limits_{j=n+1}^l\pi_{ij}(p,\zeta(p))\langle p, b_j \rangle, \quad i=\overline{n+1,l}.\end{eqnarray*}
  The set of equations of the economy equilibrium takes the form
\begin{eqnarray*} \sum\limits_{i=1}^l\frac{p_kc_{ki}(z^0)D_i(p)}{\sum\limits_{s=1}^np_sc_{si}(z^0)} \end{eqnarray*}
\begin{eqnarray*} =
p_k\left[d(p)\left(x_k^0 -  \sum\limits_{s=1}^na_{ks}x_s^0\right) +\sum\limits_{i=1}^lb_{ki}\right], \quad  k=\overline{1,n}, \end{eqnarray*}
\begin{eqnarray*} D_i(p)=\pi_i(\zeta(p))\left[d(p)x_i^0\left(p_i - \sum\limits_{s=1}^na_{si}p_s\right) + \langle p,b_i \rangle \right], \quad  i=\overline{1,n},\end{eqnarray*}
\begin{eqnarray*} D_i(p)=\sum\limits_{j=1}^n\pi_{ij}(p, \zeta(p))\left[d(p)x_j^0\left(p_j - \sum\limits_{s=1}^na_{sj}p_s\right) + \langle p,b_j\rangle \right] \end{eqnarray*}
\begin{eqnarray*} + \sum\limits_{j=1}^n\pi_{ij}(p, \zeta(p))\langle p, b_j \rangle, \quad  i=\overline{n+1,l}.\end{eqnarray*}

If $z^0$ satisfies  conditions of the Lemma \ref{ant4} and the Theorem \ref{hbant1}, $d(p)=1,$ then on the set
 $T_1\left(z^0,c^0\right)\times z^0$ the considered above set of equations of the economy equilibrium  turns into the set of equations and inequalities
\begin{eqnarray*} \sum\limits_{i=1}^l\frac{p_kc_{ki}\left(z^0\right)D_i(p)}{\sum\limits_{s=1}^np_sc_{si}\left(z^0\right)} \end{eqnarray*}  \begin{eqnarray*} =
p_k\left[x_k^0 -  \sum\limits_{s=1}^na_{ks}x_s^0 +\sum\limits_{i=1}^lb_{ki}\right], \quad  k=\overline{1,n}.\end{eqnarray*}
\begin{eqnarray*} p_i - \sum\limits_{s=1}^na_{si}p_s \geq 0, \quad  i=\overline{1,n},\end{eqnarray*}
\begin{eqnarray*} x_k^0 -  \sum\limits_{s=1}^na_{ks}x_s^0 +\sum\limits_{i=1}^lb_{ki} \geq 0, \quad k=\overline{1,n}.\end{eqnarray*}
\begin{eqnarray*} D_i(p)=\pi_i\left(z^0\right)\left[x_i^0\left(p_i - \sum\limits_{s=1}^na_{si}p_s\right) + \langle p,b_i \rangle \right], \quad  i=\overline{1,n}, \end{eqnarray*}
\begin{eqnarray*} D_i(p)=\sum\limits_{j=1}^n\pi_{ij}^0\left(p, z^0\right)\left[x_j^0\left(p_j - \sum\limits_{s=1}^na_{sj}p_s\right) + \langle p,b_j \rangle \right] \end{eqnarray*}  \begin{eqnarray*} + \sum\limits_{j=n+1}^l\pi_{ij}^0\left(p, z^0\right)\langle p,b_j\rangle  =y_i\left(z^0\right)\sum\limits_{k=1}^nc_{ki}\left(z^0\right)p_k, \quad  i=\overline{n+1,l}.\end{eqnarray*}
The determination of the conditions of  the solvability  of the last problem  represents important problem for mathematical economics.

\begin{theorem}\label{bant1}
Let a gross output vector\index{gross output vector} $x^0=\left\{x_i^0\right\}_{i=1}^n$ and the corresponding productive process $z^0 \in B_0$ satisfy conditions of the Lemma \ref{ant4} and the Theorem \ref{hbant1},
and let  a taxation system\index{taxation system} be determined by the Lemma \ref{ant4} and the Theorem \ref{hbant1}, the matrix  $ \bar C\left(z^0\right)$ do not contain zero rows. 
If there exists a non-negative vector $v_0=\{v_i\}_{i=1}^n$ such that
\begin{eqnarray*}  y_i\left(z^0\right) > \pi_i\left(z^0\right) v_i, \quad   i =\overline{1,n},\end{eqnarray*}
and the non-negative matrix $\bar C(v_0) - \bar B$ has no zero rows or columns, where $\bar C(v_0)=||c_{ki}\left(z^0\right)v_i||_{k,i=1}^n,$
then the set of equations of the  economy equilibrium
\begin{eqnarray} \label{rok1} \sum\limits_{i=1}^l\frac{p_kc_{ki}\left(z^0\right)D_i(p)  }{\sum\limits_{s=1}^np_sc_{si}\left(z^0\right)}
\end{eqnarray}
 \begin{eqnarray*} =p_k\left[x_k^0 -  \sum\limits_{s=1}^na_{ks}x_s^0 +\sum\limits_{i=1}^lb_{ki}\right], \quad  k=\overline{1,n},\end{eqnarray*}
has a solution in the set of strictly positive price vectors belonging to the set
$T_1\left(z^0,c^0\right),$
every industry  is profitable  in the state of the economy equilibrium.
 \end{theorem}
\begin{proof}\smartqed  A set of productive processes  of $n$ industries $ (x_i, y_i), \ i=\overline{1,n},$ where
$x_i=\left\{a_{ki}x_i^0\right\}_{k=1}^n, \  y_i=x_i^0e_i,\  i =\overline{1,n},$
is economically compatible\index{economically compatible} in the sense of Chapter 4 if the couple of matrices $\{AX_0, B + X_0\}$ belongs to the class $\Pi_0.$ Due to the  agreement of the structure of supply  with the  structure of choice\index{ agreement of the structure of supply  with the  structure of choice} on the set $B_0$ and $\ z^0 \in B_0,$ there holds the representation
\begin{eqnarray*} \sum\limits_{i=1}^lc_{ki}\left(z^0\right) y_i\left(z^0\right) = x_k^0 -  \sum\limits_{s=1}^na_{ks}x_s^0 +\sum\limits_{i=1}^lb_{ki}, \quad y_i\left(z^0\right) > 0, \quad i=\overline{1,l},\quad k=\overline{1,n}.\end{eqnarray*}
For the vector $p=\{p_i\}_{i=1}^n$, consider the set of equations
\begin{eqnarray*}  \lambda \pi_i\left(z^0\right)\left[x_i^0\left(p_i-\sum\limits_{s=1}^n
a_{si}p_s\right)+ \sum\limits_{k=1}^nb_{ki}p_k\right] \end{eqnarray*}
\begin{eqnarray} \label{ant7}
 =y_i\left(z^0\right)\sum\limits_{s=1}^n c_{si}\left(z^0\right)p_s, \quad
i=\overline{1,n}.
\end{eqnarray}
Under the  assumptions of the Theorem, this set of equations is equivalent to the set of equations
\begin{eqnarray} \label{gant7}
\lambda p = \left\{\left[\bar B +X_0 - A X_0\right]^T\right\}^{-1}\pi^{-1}\left(z^0\right)\left[\bar C\left(\bar y\left(z^0\right)\right)\right]^Tp,
\end{eqnarray}
where $[\bar B +X_0 - A X_0]^T$ is the matrix transposed to the matrix
$\bar B +X_0 - A X_0,$
 \begin{eqnarray*}  \pi^{-1}\left(z^0\right)=\left|\left|\frac{1}{\pi_i\left(z^0\right)}\delta_{ij}\right|\right|_{i,j=1}^n,\end{eqnarray*}
and $\left[\bar C\left(\bar y\left(z^0\right)\right)\right]^T$ is the matrix transposed to the matrix
\begin{eqnarray*} \bar C\left(\bar y\left(z^0\right)\right)=||c_{ki}\left(z^0\right)y_i\left(z^0\right) ||_{k,i=1}^n.\end{eqnarray*}

Due to the Perron-Frobenius Theorem\index{Perron-Frobenius Theorem} whose conditions hold because elements  of the matrix
\begin{eqnarray*} \left\{\left[\bar B +X_0 - A X_0\right]^T\right\}^{-1}\left[\bar C\left(\bar y\left(z^0\right)\right)\right]^T\end{eqnarray*}
are strictly positive, there exists a positive number $\lambda$ and a strictly positive price vector $p_0=\{p_i^0\}_{i=1}^n$ solving the set of equations (\ref{gant7}). The same vector $p_0=\{p_i^0\}_{i=1}^n$ solves the set of equations (\ref{ant7}). Show that under the conditions of the  Theorem  $\lambda=1.$
For the taxation system built and $p_0$ solving the set of equations (\ref{ant7}), we obtain the series of equalities
\begin{eqnarray*}  D_i(p_0)=y_i\left(z^0\right)\sum\limits_{s=1}^n c_{si}\left(z^0\right)p_s^0, \quad  i=\overline{n+1,l}, \end{eqnarray*}
\begin{eqnarray} \label{ant8}
 \lambda \pi_i\left(z^0\right)\left[x_i^0\left(p_i^0-\sum\limits_{s=1}^n
 a_{si}p_s^0\right)+ \sum\limits_{k=1}^nb_{ki}p_k^0\right] 
\end{eqnarray}
\begin{eqnarray*} 
 =y_i\left(z^0\right)\sum\limits_{s=1}^n c_{si}\left(z^0\right)p_s^0, \quad  i=\overline{1,n}.
\end{eqnarray*}
Summing up over $i$ from $1$ to $n$ equalities (\ref{ant8}), we obtain
\begin{eqnarray*} \lambda \sum\limits_{i=1}^n \pi_i\left(z^0\right)\left[x_i^0\left(p_i^0-\sum\limits_{s=1}^n
a_{si}p_s^0\right)+ \sum\limits_{k=1}^nb_{ki}p_k^0\right] \end{eqnarray*}
\begin{eqnarray} \label{allalove}
= \sum\limits_{i=1}^n y_i\left(z^0\right)\sum\limits_{s=1}^n c_{si}\left(z^0\right)p_s^0, \quad
i=\overline{1,n}.
\end{eqnarray}
The vector $\pi\left(z^0\right)=\{\pi_i\left(z^0\right) \}_{i=1}^n$ solves the set of equations
(\ref{ant5}), therefore,  $\lambda=1.$
To establish  profitability  of every industry, one must use the fact that the vector $p_0=\{p_i^0 \}_{i=1}^n$ solves the set of equations (\ref{allalove}) with $\lambda=1$
 and use the  conditions of the Theorem.
 Hence, a strictly positive equilibrium price vector $p_0=\{p_i^0 \}_{i=1}^n$ exists solving the set of equations (\ref{rok1}).
\qed
\end{proof}

\subsection{Economic systems with fixed consumption levels}

In this Subsection, relying on the statements obtained, we construct the economy system model with known levels of  consumption  determined by a strictly positive  vector of levels of satisfaction of consumers needs  $y_0=\{y_i^0\}_{i=1}^l$ and by matrix $C=||c_{ij}||_{i=1,j=1}^{n,\ l}$ that is independent of the gross output vector $x^0=\{x_i^0\}_{i=1}^n,$  vector of levels of satisfaction of consumers needs\index{vector of levels of satisfaction of consumers needs}  $y_0=\{y_i^0\}_{i=1}^l,$  after which we construct fields of evaluation of information  by consumers.\index{fields of evaluation of information  by consumers}
Further, we suppose that a strictly positive vector $x^0=\{x_i^0\}_{i=1}^n$ is the gross output vector in the  economy system\index{gross output vector in the  economy system } whose production we describe by technological maps introduced at the beginning of the Chapter
and $z^0=\{z_i^0\}_{i=1}^n \in T, \ $
$z_i^0=\{\{a_{ki}x_i^0\}_{k=1}^n, \   x_i^0e_i\}, \ i=\overline{1,n},$ is a productive process corresponding to the gross output vector $x^0.$

Consider  non-negative matrices
\begin{eqnarray*} X_0=||\delta_{ij}x_i^0||_{i,j=1}^n, \quad  A=||a_{ij}||_{i,j=1}^n,\quad B=||b_{ij}||_{i=1,j=1}^{n,\ l}, \end{eqnarray*}  \begin{eqnarray*}  C=||c_{ij}||_{i=1,j=1}^{n,\ l},\quad \tilde  B=||b_{ij}||_{i=1,j=n+1}^{n,\ l},\quad \bar B=||b_{ij}||_{i=1,j=1}^{n},\end{eqnarray*}
\begin{eqnarray*} \bar C=||c_{ij}||_{i=1,j=1}^{n},\quad \tilde  C=||c_{ij}||_{i=1,j=n+1}^{n,\ l},\end{eqnarray*}
where $\delta_{ij}$ is the Kronecker symbol and vectors
\begin{eqnarray*}    y_0=\{y_i^0\}_{i=1}^l, \quad \tilde y_0=\{y_i^0\}_{i=n+1}^l, \quad \bar y_0=\{y_i^0\}_{i=1}^n, \end{eqnarray*}
\begin{eqnarray*}  e=\{e_i\}_{i=1}^l, \quad \bar e=\{e_i\}_{i=1}^n, \quad \tilde e=\{e_i\}_{i=n+1}^l,\quad e_i=1, \quad i=\overline{1,l}.\end{eqnarray*}
 As earlier, the matrices introduced have the same economic sense.
Suppose that the strictly positive  vector of gross outputs $x^0$ solves the set of equations
\begin{eqnarray} \label{qant9}
x_k^0 - \sum\limits_{i=1}^na_{ki}x_i^0 + \sum\limits_{i=1}^lb_{ki}=\sum\limits_{i=1}^l c_{ki}y_i^0,  \quad  k=\overline{1,n},
\end{eqnarray}
 where the vector $y_0=\{y_i^0\}_{i=1}^l$ has strictly positive components.

The simple sufficient conditions of this are: the vector $Cy_0 -Be$ belongs to the interior of the cone created by the  vectors-columns of the matrix $E - A.$

The vector $y_0$ is called the vector of levels of satisfaction of consumers needs\index{vector of levels of satisfaction of consumers needs}  and the vector $x^0=\{x_i^0\}_{i=1}^n$ is the vector  of gross outputs in the economy system\index{vector  of gross outputs in the economy system} corresponding to the productive process $z^0.$

Establish a series of auxiliary statements.

Let $\varphi=\{\varphi_i\}_{i=1}^n$ be a non-negative vector. Consider the set of equations
\begin{eqnarray} \label{ant10} \sum\limits_{i=1}^n
\frac{c_{ki}\left[x_i^0\left(p_i-\sum\limits_{s=1}^n
a_{si}p_s\right)+ \sum\limits_{k=1}^nb_{ki}p_k\right]}{\sum\limits_{s=1}^n c_{si}p_s}=\varphi_k, \quad
k=\overline{1,n},
\end{eqnarray}
for the vector $p=\{p_i\}_{i=1}^n.$

\begin{theorem}\label{ant11} Let $ \sum\limits_{s=1}^n c_{si}>0, \ i=\overline{1,n},$ there exists a non-negative vector
$v_0=\{v_i\}_{i=1}^n, \ v_i \geq 0, \ i=\overline{1,n},$ such that $\bar C(v_0) - \bar B $ is a non-negative matrix having no zero rows or      columns and
$ A + \bar C(v_0) - \bar B$ is an indecomposable matrix. If
$x_i^0>0,$~ $i=\overline{1,n},$ the spectral radius of the matrix $A$ is less than 1,
then the set of equations (\ref{ant10}) has a solution in the set
\begin{eqnarray*} T_0=\left\{ p \in K_+^n, \
p_i-\sum\limits_{s=1}^n\left[a_{si} + \frac{1}{x_i^0}(c_{si}v_i - b_{si})\right]p_s > 0, \ i=\overline{1,n}\right\}\end{eqnarray*}
if and only if when the vector $\varphi $ belongs to the non-negative cone created by vectors $C_i=\{c_{ki}\}_{k=1}^n ,\ i=\overline{1,n},$ i.e.,
\begin{eqnarray*} \varphi=\sum\limits_{i=1}^n C_iy_i,\quad
y_i >v_i, \quad i=\overline{1,n},  \quad \bar y=\{y_i\}_{i=1}^n,\end{eqnarray*}
there exists a strictly positive vector
$g=\{g_i\}_{i=1}^n,$ \  $ g_i>0,$ \ $   i=\overline{1,n},$  such that for components
 $x_i^0, \  i=\overline{1,n},$ of the vector $x^0=\{x_i^0\}_{i=1}^n$ there hold the representations
\begin{eqnarray*} x_i^0=\frac{[(E-A)^{-1}[\bar C(\bar y)- \bar B]g]_i}{
g_i}, \ \quad i=\overline{1,n},\end{eqnarray*}
where $\bar C(v_0)=||c_{si}v_i||_{s,i=1}^{n},$
$\bar C(\bar y)=||c_{si}y_i||_{s,i=1}^{n}.$
\end{theorem}
\begin{proof}\smartqed
 Necessity. Let  a solution $p_{0}=\{p_i^{0}\}_{i=1}^n$ to the problem
(\ref{ant10}) belonging to the set $T_0$ exist. Denote
\begin{eqnarray} \label{ant12}
y_i=\frac{x_i^0\left(p_i^{0}-\sum\limits_{s=1}^n
a_{si}p_s^{0}\right)+ \sum\limits_{s=1}^n b_{si}p_s^0}{\sum\limits_{s=1}^n c_{si}p_s^{0}}, \quad
i=\overline{1,n}.
\end{eqnarray}
Then, obviously,
\begin{eqnarray*} \varphi =\sum\limits_{i=1}^nC_iy_i, \quad i=\overline{1,n}, \end{eqnarray*}
and from the fact that the vector $p_{0}$ belongs to the set $T_0$ the next inequalities follow: $y_i >v_i, \ i=\overline{1,n}. $
Prove the representation for the vector $x^0.$
From (\ref{ant12}) it follows that the vector
$p_{0}=\{p_i^{0}\}_{i=1}^n$ solves the set of equations
\begin{eqnarray} \label{ant13}
p_i^{0}=\sum\limits_{s=1}^na_{si}p_s^{0}+\frac{1}{x_i^0}
\sum\limits_{s=1}^n[c_{si}y_i- b_{si}] p_s^{0},
\quad i=\overline{1,n}.
\end{eqnarray}
Because $p_{0} \in K_+^n,$ i.e., all the components of $p_{0}$ are strictly positive,
the conjugate problem
\begin{eqnarray} \label{ant14}
z_k=\sum\limits_{i=1}^na_{ki}z_i+
\sum\limits_{i=1}^n[c_{ki}y_i- b_{ki}]\frac{z_i}{x_i^0}, \quad
k=\overline{1,n},
\end{eqnarray}
to the problem (\ref{ant13}) has a solution in the set of strictly positive vectors. The latter follows from the indecomposability of the  matrix $A+ \bar C(v_0) - \bar B$  and the Perron-Frobenius Theorem. From here, it follows that for the vector
$z=\{z_i\}_{i=1}^n$ being strictly positive solution to the problem (\ref{ant14}) there holds the representation
\begin{eqnarray*} z=(E-A)^{-1}[\bar C(\bar y) - \bar B]g,\end{eqnarray*}
where components of the vector $g$  are strictly positive  and determined by the formula \begin{eqnarray*} g_i=\frac{z_i}{x_i^0},\quad
i=\overline{1,n}.\end{eqnarray*}  From here, we obtain
\begin{eqnarray*} x_i^0=\frac{z_i}{g_i}=\frac{[(E-A)^{-1}
[\bar C(\bar y)- \bar B]g]_i}{g_i}, \quad i=\overline{1,n}.\end{eqnarray*}
 Sufficiency. Assume that the vector $\varphi $ has the representation
\begin{eqnarray*} \varphi =\sum\limits_{i=1}^n C_iy_i, \quad y_i> v_i, \quad i=\overline{1,n},\end{eqnarray*}
and there exists  a strictly positive vector
$g=\{g_i\}_{i=1}^n, \ g_i>0,$ \ $i=\overline{1,n},$
such  that the components $x_i^0$ of the vector $x^0=\{x_i^0\}_{i=1}^n$
have the representation \begin{eqnarray*} x_i^0=\frac{[(E-A)^{-1}[\bar C(\bar y) - \bar B]g]_i}{ g_i}, \quad i=\overline{1,n}.\end{eqnarray*}  Consider the set of equations
 \begin{eqnarray*} z_k=\sum\limits_{i=1}^na_{ki}z_i+
\sum\limits_{i=1}^n[c_{ki}y_i- b_{ki}]\frac{z_i}{x_i^0}, \quad
k=\overline{1,n}.
\end{eqnarray*}
From the conditions of the  Theorem \ref{ant11}  on validity of the representation for the components of the vector $x^0=\{x_i^0\}_{i=1}^n,$  the last set of equations has a strictly  positive solution $z=(E-A)^{-1}[\bar C(\bar y) - \bar B]g.$ From here and the conditions of the Theorem,  it follows that there exists a strictly positive solution to the conjugate problem \begin{eqnarray} \label{ant15}
p_i^{0}=\sum\limits_{s=1}^na_{si}p_s^{0}+\frac{1}{x_i^0}
\sum\limits_{s=1}^n[c_{si}y_i- b_{si}] p_s^{0},
\quad i=\overline{1,n},
\end{eqnarray}
obviously belonging to $T_0,$ since $\sum\limits_{s=1}^nc_{si}y_i > \sum\limits_{s=1}^nc_{si}v_i, \ i=\overline{1,n}.$
Having determined from this set of equations components of the vector
$y=\{y_i\}_{i=1}^n$ and inserting them into the representation for the vector $\varphi,$ we obtain  the needed.
\qed
\end{proof}

\begin{note}
If the conditions of the Theorem \ref{ant11}  hold, then every industry is profitable. Really,
then there exists a solution to the set of equations (\ref{ant15}) belonging to the set $T_0.$
This means that components of the strictly positive vector $p^0$ solving the set of equations (\ref{ant15}) satisfy the inequalities
\begin{eqnarray*} p_i^0-\sum\limits_{s=1}^n\left[a_{si} + \frac{1}{x_i^0}(c_{si}v_i - b_{si})\right]p_s^0 > 0, \quad i=\overline{1,n}.\end{eqnarray*}
As the matrix $\bar C(\bar y) - \bar B$ is non-negative and has no zero columns,
therefore the vector $[\bar C(\bar y) -\bar B]^{T}p^0$ has strictly positive components. From  here we obtain that  the vector
$(E - A)^Tp^0$ has strictly positive components.
\end{note}
\begin{lemma}\label{jant1}
Let $ \sum\limits_{k=1}^nc_{ki}> 0,\  i=\overline{1,n}, $ and let  there exist a non-negative vector
$v_0=\{v_i\}_{i=1}^n, \ v_i \geq 0, \ i=\overline{1,n},$ such  that $\bar C(v_0) - \bar B $ is a non-negative matrix having no zero rows or columns and a
 matrix $ A + \bar C(v_0) - \bar B$ is indecomposable. If the spectral radius of the matrix\index{spectral radius of the matrix} $A$ is less than 1,
there exists  a  strictly positive vector
$g=\{g_i\}_{i=1}^n,$  $ g_i>0,$ $   i=\overline{1,n},$ such that for components $x_i^0, \  i=\overline{1,n},$ of the vector $x^0=\{x_i^0\}_{i=1}^n$   the representations
\begin{eqnarray*} x_i^0=\frac{[(E-A)^{-1}[\bar C(\bar y)- \bar B]g]_i}{
g_i}, \ \quad i=\overline{1,n},\end{eqnarray*}
 hold, where
\begin{eqnarray*} \bar C(v_0)=||c_{si}v_i||_{s,i=1}^{n},\quad \bar C(\bar y)=||c_{si}y_i||_{s,i=1}^{n},\end{eqnarray*}
and $ y_i > v_i, \  i=\overline{1,n}, $ $\ \bar y=\{y_i\}_{i=1}^n, $
then the spectral radius of the matrix
\begin{eqnarray*} A+ [\bar C(v_0) - \bar B]X_0^{-1}\end{eqnarray*}
 is strictly less than 1.
\end{lemma}
\begin{proof}\smartqed
As components $x_i^0$ of the vector $x^0=\{x_i^0\}_{i=1}^n$
have the representation \begin{eqnarray*} x_i^0=\frac{[(E-A)^{-1}[\bar C(\bar y) - \bar B]g]_i}{ g_i}, \quad i=\overline{1,n},\end{eqnarray*}  it follows that the set of equations
\begin{eqnarray} \label{jant2}
z_k=\sum\limits_{i=1}^na_{ki}z_i+
\sum\limits_{i=1}^n[c_{ki}y_i- b_{ki}]\frac{z_i}{x_i^0}, \quad
k=\overline{1,n},
\end{eqnarray}
has a solution $z=(E-A)^{-1}[\bar C(\bar y) - \bar B]g$
being strictly positive  because $z_k=g_k x_k^0, \ k=\overline{1,n}.$
From here and the conditions of the  Theorem,  it follows that there exists a strictly positive solution $p_0=\{p_i^0\}_{i=1}^n$ to the conjugate problem \begin{eqnarray} \label{jant3}
p_i^{0}=\sum\limits_{s=1}^na_{si}p_s^{0}+\frac{1}{x_i^0}
\sum\limits_{s=1}^n[c_{si}y_i- b_{si}] p_s^{0},
\quad i=\overline{1,n},
\end{eqnarray}
obviously belonging to the set  $T_0.$

 Write the set of equations (\ref{jant3}) in the form

\begin{eqnarray*}  p_i^{0}=\sum\limits_{s=1}^na_{si}p_s^{0} \end{eqnarray*}
\begin{eqnarray} \label{jant31}
+\frac{1}{x_i^0}
\sum\limits_{s=1}^n[c_{si}v_i- b_{si}] p_s^{0}+\frac{1}{x_i^0}(y_i - v_i)
\sum\limits_{s=1}^nc_{si}p_s^{0},
\quad i=\overline{1,n}.
\end{eqnarray}
It is obvious that numbers $m_i,$ where
\begin{eqnarray*} m_i =\frac{1}{x_i^0}(y_i - v_i)\sum\limits_{s=1}^nc_{si}p_s^{0}, \quad i=\overline{1,n},\end{eqnarray*}
are strictly positive and $p_k^0 > m_k, \ k=\overline{1,n}.$ The last follows from the indecomposability of the matrix
$A+ \bar C(v_0) - \bar B $ and strict positivity of the vector $p_0.$
Introduce into the space $R^n$ the norm determining  it on the vectors $t=\{t_k\}_{k=1}^n \in R^n$ by the rule
\begin{eqnarray*} ||t||=\max\limits_{k}\frac{|t_k|}{p_k^0}.\end{eqnarray*}
For the  norm of the  matrix $[A+ [\bar C(v_0) - \bar B]X_0^{-1}]^T,$ the estimate holds
\begin{eqnarray*} ||[A+ [\bar C(v_0) - \bar B]X_0^{-1}]^T|| \leq \max\limits_{k}(1 - m_k/p_k^0) < 1.\end{eqnarray*}
\qed
\end{proof}

\begin{corollary}
If the conditions  of the Lemma \ref{jant1} hold, then the set
\begin{eqnarray*} T_0=\left\{ p \in K_+^n, \
p_i-\sum\limits_{s=1}^n\left[a_{si} + \frac{1}{x_i^0}(c_{si}v_i - b_{si})\right]p_s > 0, \ i=\overline{1,n} \right \}\end{eqnarray*}
is non-empty.
\end{corollary}
For example, the set $T_0$ contains a strictly positive proper vector of the problem
\begin{eqnarray*} \lambda p_i=\sum\limits_{s=1}^n\left[a_{si} + \frac{1}{x_i^0}(c_{si}v_i - b_{si})\right]p_s, \quad i=\overline{1,n},\end{eqnarray*}
corresponding to the largest proper value $\lambda <1 $ existing due to the Perron-Frobenius Theorem.

\begin{lemma}\label{jant4}
If the vector $x^0$ is a strictly positive solution to the set of equations (\ref{qant9}), the
vectors $\tilde C \tilde y_0 - \tilde B \tilde e$ and $\bar C \bar y_0 $ belong to the interior of the non-negative cone\index{ interior of the non-negative cone} created by vectors-columns of the non-degenerate matrix $\bar B + X_0- AX_0,$ then for any $p \in K_+^n$
there exists a solution to the equation
\begin{eqnarray} \label{janm5}
 \sum\limits _{i=1}^n\pi_i \left[x_i^0\left(p_i - \sum\limits_{s=1}^na_{si}p_s\right) + \langle p,b_i\rangle \right]
=\sum\limits _{i=1}^ny_i^0\sum\limits_{k=1}^nc_{ki}p_k
\end{eqnarray}
with respect to the vector $\pi=\{\pi_i\}_{i=1}^n$
such  that components $\pi_i,  \  i=\overline{1,n},$ of the solution to the equation
 (\ref{janm5}) satisfy inequalities $0 < \pi_i < 1,  \  i=\overline{1,n}.$
\end{lemma}
\begin{proof}\smartqed  From the conditions of the  Lemma,  it follows that the equation (\ref{janm5})
has a solution for any $p \in K_+^n$ if the vector $\pi=\{\pi_i\}_{i=1}^n$ solves the set of equations
\begin{eqnarray*}   \sum\limits _{j=1}^n\pi_j[\delta_{ij}x_j^0 - a_{ij}x_j^0 +b_{ij}]=\sum\limits_{k=1}^nc_{ik}y_k^0, \quad i=\overline{1,n}.\end{eqnarray*}
The solution to the last set of equations is unique and expressed as follows
\begin{eqnarray*} \pi=[\bar B + X_0 - AX_0]^{-1}\bar C\bar y_0.\end{eqnarray*}
 From the conditions of the  Theorem,  it follows that components of this solution are strictly positive  because the vector $\bar C\bar y_0$ belongs to the interior of a non-negative cone created by the  vectors-columns of the non-degenerate matrix $\bar B + X_0 - AX_0.$
The vector $ \bar e - \pi=\bar \pi$ solves the set of equations
 \begin{eqnarray*} [\bar B + X_0 - AX_0]\bar \pi=\tilde C \tilde y_0 - \tilde B \tilde e,\end{eqnarray*}
  having unique solution whose every component is strictly positive  because the vector
$\tilde C \tilde y_0 - \tilde B \tilde e$ belongs to the interior of a  non-negative cone created by the vectors-columns of the matrix $\bar B + X_0- AX_0.$ From here, it follows that components  of the  solution satisfy inequalities
\begin{eqnarray*} 0 < \pi_i < 1,  \quad   i=\overline{1,n}.\end{eqnarray*}
\qed
\end{proof}
Introduce into  consideration the matrix $\pi=||\pi_i\delta_{ki}||_{k,i=1}^n,$
where $\pi=\{\pi_i\}_{i=1}^n$ solves the equation (\ref{janm5}). Further, by $\pi^{-1}$ we denote the matrix inverse to the matrix $\pi.$ From the context one can  understand if the vector $\pi=\{\pi_i\}_{i=1}^n$ or the matrix $\pi$ is considered because the text contains only the inverse matrix $\pi^{-1}$ built after the vector $\pi=\{\pi_i\}_{i=1}^n.$
\begin{theorem}\label{jant6}
Let $ \sum\limits_{s=1}^nc_{si}>0,  \ i=\overline{1,l},$ the conditions of the  Lemma \ref{jant4}  hold,
 and let there exist a vector
$v_0=\{v_i\}_{i=1}^n, \ v_i \geq 0, \ i=\overline{1,n}$ such that $\bar C(v_0) - \bar B $ is a non-negative matrix having no zero rows or columns and a matrix
 $ A + \bar C(v_0) - \bar B$ is indecomposable, and also let there hold inequalities $ y_i^0 > \pi_i v_i, \ i=\overline{1,n}.$
If the spectral radius of the matrix $A$ is less than 1,
there exists a  strictly positive vector
$g=\{g_i\}_{i=1}^n,$  $ g_i>0,$ $   i=\overline{1,n},$ such that for
components $x_i^0,  \ i=\overline{1,n},$ of the vector $x^0=\{x_i^0\}_{i=1}^n$ there hold the representations
\begin{eqnarray*} x_i^0=\frac{[(E-A)^{-1}[\bar C(\pi^{-1}\bar y_0)-  \bar B]g]_i}{
g_i}, \ \quad i=\overline{1,n},\end{eqnarray*}
where
\begin{eqnarray*} \bar C(\pi^{-1}\bar y_0)=\left|\left|c_{si}\frac{y_i^0}{\pi_i}\right|\right|_{s,i=1}^{n},\end{eqnarray*}
then on the set $T_0\times z^0$
there exists a taxation system
\begin{eqnarray*} ||\pi_{ij}^0\left(p,z^0\right)||_{i,j=1}^l, \quad  \left(p,z^0\right) \in T_0\times z^0,\end{eqnarray*}
satisfying conditions
\begin{eqnarray*} \pi_{ij}^0\left(p,z^0\right)=\pi_i\delta_{ij},  \quad \left(p,z^0\right) \in T_0\times z^0,\quad  i =\overline{1,n}, \quad j =\overline{1,l},\end{eqnarray*}
\begin{eqnarray*}
 \pi_j+ \sum\limits_{i=n+1}^l\pi_{ij}^0\left(p,z^0\right)=1,  \quad \left(p,z^0\right) \in T_0\times z^0, \quad  j =\overline{1,n}, \end{eqnarray*}
\begin{eqnarray} \label{qyant1}
\sum\limits_{i=n+1}^l\pi_{ij}^0\left(p,z^0\right)=1,  \quad  \left(p,z^0\right) \in T_0\times z^0, \quad j =\overline{n+1,l},
\end{eqnarray}
where the vector $\pi=\{\pi_i\}_{i=1}^n$ solves the set of equations
(\ref{janm5}).
\end{theorem}
\begin{proof}\smartqed
Under the  conditions  of the Theorem \ref{jant6}   the  conditions of the Lemma \ref{jant1} hold if one takes for the vector $\bar y$ the vector $\pi^{-1}\bar y_0,$  therefore, the set $T_0$ is non-empty.
Consider on the set $T_0$ two sets of functions
\begin{eqnarray*}  f_i\left(p,z^0\right)= y_i^0\sum\limits_{k=1}^nc_{ki}p_k, \quad i=\overline{n+1,l},\end{eqnarray*}
\begin{eqnarray*}  g_i\left(p,z^0\right)=(1 - \pi_i)\left[x_i^0\left(p_i - \sum\limits_{s=1}^na_{si}p_s\right)+ \langle b_i, p\rangle \right], \quad i=\overline{1,n}, \end{eqnarray*}  \begin{eqnarray*}  g_i\left(p,z^0\right)=\langle b_i, p\rangle , \quad  i=\overline{n+1,l}.\end{eqnarray*}
On the set $T_0$ there holds the condition
\begin{eqnarray*}  \sum\limits_{i=n+1}^ly_i^0\sum\limits_{k=1}^nc_{ki}p_k \end{eqnarray*}
\begin{eqnarray} \label{jant8}
 =\sum\limits_{i=1}^n(1 - \pi_i)\left[x_i^0\left(p_i - \sum\limits_{s=1}^na_{si}p_s\right)+ \langle b_i, p\rangle \right ]+  \sum\limits_{i=n+1}^l \langle b_i, p\rangle ,
\end{eqnarray}
because \hfill due \hfill to \hfill the \hfill Theorem \hfill it \hfill turns \hfill into \hfill the \hfill equation \hfill (\ref{janm5}) \hfill for \hfill the \hfill vector \\
$\pi=\{\pi_i\}_{i=1}^n.$ However, the equality (\ref{jant8}) in terms of the family of functions  $f_i\left(p,z^0\right), \ i=\overline{n+1,l},$ and  the family of functions  $g_i\left(p,z^0\right),  i=\overline{1,l},$ has the form
\begin{eqnarray*} \sum\limits_{i=n+1}^lf_i\left(p,z^0\right)= \sum\limits_{i=1}^lg_i\left(p,z^0\right)\end{eqnarray*}
on the set $T_0\times z^0 .$ Due to the Lemma on the existence of taxation system whose conditions hold on the set $T_0\times z^0,$ there exists a  taxation system $||\bar \pi_{ij}\left(p,z^0\right)||_{i=n+1, j=1}^l$ such that
\begin{eqnarray*}  f_i\left(p,z^0\right)=\sum\limits_{j=1}^l\bar \pi_{ij}\left(p,z^0\right)g_j\left(p,z^0\right),  \quad  i=\overline{n+1,l}.\end{eqnarray*}
After just built taxation system, introduce a new taxation system supposing
\begin{eqnarray*}    \pi_{ij}^0\left(p, z^0\right)=\delta_{ij}\pi_i, \quad i,j=\overline{1,n}, \end{eqnarray*}  \begin{eqnarray*}  \pi_{ij}^0\left(p, z^0\right)=(1 - \pi_j)\bar \pi_{ij}\left(p, z^0\right), \quad i=\overline{n+1,l},  \quad j=\overline{1,n},\end{eqnarray*}  \begin{eqnarray*}
\pi_{ij}^0\left(p,z^0\right)=\bar \pi_{ij}\left(p, z^0\right), \quad i=\overline{n+1,l},  \quad j=\overline{n+1,l}.\end{eqnarray*}
The taxation system built in such a way satisfies the  conditions of the  Theorem.
\qed
\end{proof}

 Construct the taxation system
$||\pi_{ij}(p,z)||_{i,j=1}^l, \ (p,z) \in K_+^n \times \Gamma^n,$ satisfying conditions
\begin{eqnarray*} \pi_{ij}(p,z)=\pi_i(z)\delta_{ij}, \quad  0< \pi_i(z) <1, \quad (p,z) \in K_+^n \times \Gamma^n,\quad  i =\overline{1,n}, \quad j =\overline{1,l},\end{eqnarray*}
\begin{eqnarray*}
 \pi_j(z)+ \sum\limits_{i=n+1}^l\pi_{ij}(p,z)=1,  \quad (p,z) \in K_+^n \times \Gamma^n, \quad  j =\overline{1,n}, \end{eqnarray*}
\begin{eqnarray} \label{qwant6}
\sum\limits_{i=n+1}^l\pi_{ij}(p,z)=1,  \quad (p,z) \in K_+^n \times \Gamma^n, \quad j =\overline{n+1,l}.
\end{eqnarray}

 Suppose that on the set $T_0\times z^0$ the taxation system is determined by the Theorem \ref{jant6}. On the set $K_+^n \times \Gamma^n \setminus  T_0\times z^0$, we take it arbitrary but require that there hold equalities (\ref{qwant6}).

Consider the family of  income pre-functions
\begin{eqnarray*}  K^0_i(p,z)=\pi_i(z)[\langle y_i -  x_i, p \rangle + \langle b_i,p \rangle ],  \quad (p,z) \in K_+^n \times \Gamma^n, \quad i=\overline{1,n},\end{eqnarray*}
 \begin{eqnarray*}  K^0_i(p,z)=\sum\limits_{j=1}^n\pi_{ij}(p,z)\left[\langle y_j -  x_j, p \rangle + \langle b_j, p \rangle\right] \end{eqnarray*}
\begin{eqnarray} \label{qwant1}
+ \sum\limits_{j=n+1}^l\pi_{ij}(p,z)  \langle b_j,p\rangle,  \quad (p,z) \in K_+^n \times \Gamma^n, \quad  i=\overline{n+1,l},
\end{eqnarray}
and built after it the family of income functions
\begin{eqnarray*}  K_i(p,z)= K^0_i(p,Q(p,z)) \end{eqnarray*}  \begin{eqnarray*} = \pi_i(Q(p,z))\left[u_i\left(p_i - \sum\limits_{s=1}^na_{si}p_s\right)\prod\limits_{s=1}^na_s(p)b_s(p,z)+ \langle b_i,p\rangle \right], \quad i=\overline{1,n},\end{eqnarray*}

\begin{eqnarray*}  K_i(p,z)= K^0_i(p,Q(p,z)) \end{eqnarray*}  \begin{eqnarray*} =\sum\limits_{j=1}^n\pi_{ij}(p,Q(p,z))\left[u_i\left(p_i - \sum\limits_{s=1}^na_{si}p_s\right)\prod\limits_{s=1}^na_s(p)b_s(p,z)+ \langle b_i,p\rangle \right] \end{eqnarray*}
\begin{eqnarray} \label{qwant2}
+ \sum\limits_{j=n+1}^l\pi_{ij}(p,Q(p,z))  \langle  b_j,p \rangle ,  \quad  i=\overline{n+1,l}.
\end{eqnarray}
\begin{eqnarray*}   z \in \Gamma^n, \quad z=\{z^i\}_{i=1}^n, \quad  z^i=( x_i,  y_i), \quad i=\overline{1,n}.\end{eqnarray*}
Here the vector $u=\{u_i\}_{i=1}^n$ is a gross output in the economy system corresponding to a productive process $  z \in \Gamma^n.$

The reduction onto the set $T_0\times z^0$ of the  considered family of income functions  takes the form
\begin{eqnarray} \label{qwant3}
K_i\left(p, z^0\right)=\pi_i\left\{x_i^0\left[p_i - \sum\limits_{s=1}^na_{si}p_s\right]+ \langle b_i,p\rangle \right\}, \quad i=\overline{1,n},
\end{eqnarray}
\begin{eqnarray*} K_i\left(p, z^0\right)=\sum\limits_{j=1}^n\pi_{ij}^0\left(p,z^0\right)\left\{x_j^0\left[p_j - \sum\limits_{s=1}^na_{sj}p_s\right]+ \langle b_j,p\rangle \right\} \end{eqnarray*}  \begin{eqnarray*} + \sum\limits_{j=n+1}^l\pi_{ij}^0\left(p, z^0\right)  \langle b_j,p\rangle =y_i^0\sum\limits_{k=1}^nc_{ki}p_k,  \quad  i=\overline{n+1,l}.\end{eqnarray*}
Therefore, on the random field realization $\zeta_0\left(p, \omega_0\right)=z^0$ the net income functions takes the form
\begin{eqnarray} \label{jant7}
 D_i\left(p\right)=\pi_i\left[x_i^0\left(p_i - \sum\limits_{s=1}^na_{si}p_s\right)+ \langle b_i,p\rangle \right], \quad i=\overline{1,n},
\end{eqnarray}
\begin{eqnarray*}  D_i\left(p\right)=\sum\limits_{j=1}^n\pi_{ij}^0\left(p, z^0\right)\left[x_i^0\left(p_i - \sum\limits_{s=1}^na_{si}p_s\right)+ \langle b_i, p\rangle \right] \end{eqnarray*}
\begin{eqnarray*} + \sum\limits_{j=n+1}^l\pi_{ij}^0\left(p, z^0\right)  \langle b_j, p\rangle =y_i^0\sum\limits_{k=1}^nc_{ki}p_k,  \quad  i=\overline{n+1,l}.
\end{eqnarray*}

\begin{theorem}\label{bant2}
Let $ \sum\limits_{s=1}^nc_{si}>0,  \ i=\overline{1,l},$
there exist a non-negative vector
$v_0=\left\{v_i\right\}_{i=1}^n, $ $  v_i \geq 0, \ i=\overline{1,n},$ such that $\bar C\left(v_0\right) - \bar B $ is a non-negative matrix having no zero rows or columns and a matrix
$ A + \bar C\left(v_0\right) - \bar B$ is indecomposable, and let
the vector $x^0$ be strictly positive solution to the set of equations (\ref{qant9}),
 there hold inequalities $ y_i^0 > \pi_i v_i, \ i=\overline{1,n},$ and a taxation system be determined by the Lemma \ref{jant4} and the Theorem \ref{jant6}.
If the spectral radius of the matrix $A$ is less than 1,
there exists a strictly positive vector
$g=\left\{g_i\right\}_{i=1}^n,$ \  $ g_i>0,$ \ $   i=\overline{1,n},$ such that for components
$x_i^0, \  i=\overline{1,n},$ of the vector $x^0=\left\{x_i^0\right\}_{i=1}^n$ the representations
\begin{eqnarray*} x_i^0=\frac{\left[\left(E-A\right)^{-1}\left[\bar C\left(\pi^{-1}\bar y_0\right)-  \bar B\right]g\right]_i}{
g_i}, \ \quad i=\overline{1,n},\end{eqnarray*}
hold, where
\begin{eqnarray*} \bar C\left(\pi^{-1}\bar y_0\right)=\left|\left|c_{si}\frac{y_i^0}{\pi_i}\right|\right|_{s,i=1}^{n},\end{eqnarray*}
then there exists a strictly positive solution to the set of equations
\begin{eqnarray} \label{gok1} \sum\limits_{i=1}^l\frac{c_{ki}D_i\left(p\right)  }{\sum\limits_{s=1}^np_sc_{si}}
\end{eqnarray}
 \begin{eqnarray*} =x_k^0 -  \sum\limits_{s=1}^na_{ks}x_s^0 +\sum\limits_{i=1}^lb_{ki}, \quad  k=\overline{1,n},\end{eqnarray*}
in the set $T_0$ for which every industry is profitable.
\end{theorem}
\begin{proof}\smartqed  Consider the auxiliary set of equations
\begin{eqnarray} \label{jant9} \sum\limits_{i=1}^n
\frac{c_{ki}\left[x_i^0\left(p_i-\sum\limits_{s=1}^n
a_{si}p_s\right)+ \sum\limits_{k=1}^nb_{ki}p_k\right]}{\sum\limits_{s=1}^n c_{si}p_s}=\sum\limits_{i=1}^n c_{ki}\frac{y_i^0}{\pi_i}, \quad
k=\overline{1,n}.
\end{eqnarray}
 For such rhs in the set of equations (\ref{jant9}) and the conditions of the  Theorem \ref{bant2},  there hold all the conditions of the Theorem \ref{ant11}, so there exists a strictly positive vector $p_0=\left\{p_i^0\right\}_{i=1}^n$ solving the set of equations (\ref{jant9}) and belonging to the set $T_0.$ From the Proof of the Theorem \ref{ant11}, it follows that the same vector solves the set of equations
 \begin{eqnarray} \label{jant10}
p_i=\sum\limits_{s=1}^na_{si}p_s+\frac{1}{x_i^0}
\sum\limits_{s=1}^n\left[c_{si}\frac{y_i^0}{\pi_i}- b_{si}\right] p_s,
\quad i=\overline{1,n},
\end{eqnarray}
and, therefore, also the set of equations
\begin{eqnarray} \label{jant11} \sum\limits_{i=1}^n
\frac{c_{ki}\pi_i\left[x_i^0\left(p_i-\sum\limits_{s=1}^n
a_{si}p_s\right)+ \sum\limits_{k=1}^nb_{ki}p_k\right]}{\sum\limits_{s=1}^n c_{si}p_s}=\sum\limits_{i=1}^n c_{ki}y_i^0, \quad
k=\overline{1,n}.
\end{eqnarray}
 From the last and the conditions of the Theorem, it follows that the vector $p_0=\left\{p_i^0\right\}_{i=1}^n$ solves the set of equations (\ref{gok1}) and belongs to the set $T_0.$ Such a solution guarantees the profitability to every industry.
\qed
\end{proof}
\subsection{Consumers needs satisfaction for arbitrary taxation vector }

Consider the case when a taxation vector\index{taxation vector} $\pi=\left\{\pi_i\right\}_{i=1}^n$ is strictly positive  and the taxation system depends on taxation  and price vectors that is important for various possible situations in real economic systems.

\begin{lemma}\label{jant18}
Let $\sum\limits_{k=1}^nc_{ki}>0, \ i=\overline{1,l},$  there exist a non-negative vector
$v_0=\left\{v_i\right\}_{i=1}^n, $ $ \ v_i \geq 0, \ i=\overline{1,n},$ such   that $\bar C\left(v_0\right) - \bar B $ is a non-negative matrix having no zero rows or columns and
$ A + \bar C\left(v_0\right) - \bar B$ is an indecomposable matrix.\index{indecomposable matrix} If
the spectral radius of the matrix\index{ spectral radius of the matrix} $A$ is less than 1,
 there exists a strictly positive vector
$g=\left\{g_i\right\}_{i=1}^n,$  $ g_i>0,$ $   i=\overline{1,n},$ such that for  components $x_i^0, \ i=\overline{1,n},$ of the vector $x^0=\left\{x_i^0\right\}_{i=1}^n$  the representations
\begin{eqnarray*} x_i^0=\frac{\left[\left(E-A\right)^{-1}\left[\bar C\left(\pi^{-1}\bar y\right)- \bar B\right]g\right]_i}{
g_i}, \ \quad i=\overline{1,n},\end{eqnarray*}
hold, where
\begin{eqnarray*} \bar C\left(v_0\right)=||c_{si}v_i||_{s,i=1}^{n}, \quad \bar C\left(\pi^{-1}\bar y\right)=\left|\left|c_{si}\frac{y_i}{\pi_i}\right|\right|_{s,i=1}^{n},\end{eqnarray*}
and $ \pi=\left\{\pi_i\right\}_{i=1}^n, \  \pi_i >0, \  i=\overline{1,n},$ is a certain strictly positive vector satisfying inequalities $y_i > \pi_i v_i, \  i=\overline{1,n}, $ $\bar y=\left\{y_i\right\}_{i=1}^n, $
then the spectral radius of the matrix $A+ \left[\bar C\left(v_0\right) - \bar B\right]X_0^{-1}$ is strictly less than 1 and the set $T_0$ is non-empty.
\end{lemma}
\begin{proof}\smartqed
The Proof of this Lemma is the same as that of the Lemma \ref{jant1}.
\qed
\end{proof}

\begin{theorem}\label{jant12}
Let $ \sum\limits_{s=1}^nc_{si}>0,  \ i=\overline{1,l},$
there exist  a  non-negative vector
$v_0=\left\{v_i\right\}_{i=1}^n, \ v_i \geq 0, \ i=\overline{1,n},$ such  that $\bar C\left(v_0\right) - \bar B $ is a non-negative matrix having no zero rows or columns and
$ A + \bar C\left(v_0\right) - \bar B$ is an indecomposable matrix, and let the vector $x^0$ be a strictly positive solution to the set of equations (\ref{qant9}), the spectral radius of the matrix $A$ be less than 1.  If there exists a strictly positive vector
$g=\left\{g_i\right\}_{i=1}^n,$  $ g_i>0,$ $   i=\overline{1,n},$ such  that for components $x_i^0, \ i=\overline{1,n},$ of the vector $x^0=\left\{x_i^0\right\}_{i=1}^n$ the representations
\begin{eqnarray*} x_i^0=\frac{\left[\left(E-A\right)^{-1}\left[\bar C\left(\pi^{-1}\bar y_0\right)- \bar B\right]g\right]_i}{
g_i}, \ \quad i=\overline{1,n},\end{eqnarray*}
hold, where
\begin{eqnarray*} \bar C\left(v_0\right)=||c_{si}v_i||_{s,i=1}^{n}, \quad \bar C\left(\pi^{-1}\bar y_0\right)=\left|\left|c_{si}\frac{y_i^0}{\pi_i}\right|\right|_{s,i=1}^{n},\end{eqnarray*}
and $ \pi=\left\{\pi_i\right\}_{i=1}^n, \ \pi_i >0, \  i=\overline{1,n},$ is a certain strictly positive vector satisfying inequalities $ y_i^0 > \pi_i v_i, \  i=\overline{1,n}, $ $\bar y_0=\left\{y_i^0\right\}_{i=1}^n, $
 then on the set $T_0$ there exists a strictly positive solution to the equation
\begin{eqnarray} \label{jant13}
 \sum\limits_{i=1}^n\pi_i\left(p,z^0\right)\left[x_i^0\left(p_i- \sum\limits_{s=1}^na_{si}p_s \right)+ \sum\limits_{s=1}^nb_{si}p_s\right]=\sum\limits_{i=1}^n y_i^0\sum\limits_{s=1}^nc_{si}p_s,
\end{eqnarray}
having the form
$\pi\left(p,z^0\right)=\left\{  \pi_i \left(p, z^0\right)\right\}_{i=1}^n,$
where
\begin{eqnarray*}    \pi_i\left(p, z^0\right)=\pi_i\pi\left(p, z^0\right), \quad \pi\left(p, z^0\right)=\frac{\sum\limits_{i=1}^n y_i^0\sum\limits_{s=1}^nc_{si}p_s}{\sum\limits_{i=1}^n\pi_i\left[x_i^0\left(p_i- \sum\limits_{s=1}^na_{si}p_s\right) + \sum\limits_{s=1}^nb_{si}p_s\right]}.\end{eqnarray*}
\end{theorem}
\begin{proof}\smartqed
From the conditions of the Theorem, it follows that the set $T_0$ is non-empty  because there hold the  conditions of the Lemma \ref{jant18}.
To prove the Theorem, it is sufficient to establish that on the set $T_0$ the expression
\begin{eqnarray*} \sum\limits_{i=1}^n\pi_i\left[x_i^0\left(p_i- \sum\limits_{s=1}^na_{si}p_s\right) + \sum\limits_{s=1}^nb_{si}p_s\right]\end{eqnarray*}
does not equal zero.
From the definition of the set $T_0$, it follows that any vector $p \in T_0$ has the estimate
\begin{eqnarray*}  x_i^0\left(p_i- \sum\limits_{s=1}^na_{si}p_s\right) + \sum\limits_{s=1}^nb_{si}p_s > 0,  \quad i=\overline{1,n},\end{eqnarray*}
therefore,
\begin{eqnarray*} \sum\limits_{i=1}^n\pi_i\left[x_i^0\left(p_i- \sum\limits_{s=1}^na_{si}p_s\right) + \sum\limits_{s=1}^nb_{si}p_s\right]  \end{eqnarray*}
\begin{eqnarray*}  \geq \min\limits_{i}\pi_i
\sum\limits_{i=1}^n\left[x_i^0\left(p_i- \sum\limits_{s=1}^na_{si}p_s\right) + \sum\limits_{s=1}^nb_{si}p_s\right]>0, \quad p \in T_0.\end{eqnarray*}
\qed
\end{proof}

\begin{theorem}\label{jant22}
Let there hold the  conditions of  Theorem \ref{jant12}.
On the set
$T_0 \times z^0$, there exists a  taxation system $||\pi_{ij}\left(p, z^0\right)||_{i,j=1}^l$ for which
\begin{eqnarray} \label{jant23}
f_i\left(p,z^0\right)=\sum\limits_{j=1}^l\pi_{ij}\left(p,z^0\right)g_j\left(p,z^0\right), \quad i=\overline{1,l},
\end{eqnarray}
where
\begin{eqnarray*} f_i\left(p,z^0\right)=\pi_i\left(p,z^0\right)\left[x_i^0\left(p_i - \sum\limits_{s=1}^na_{si}p_s\right)+ \langle b_i, p\rangle \right],  \quad i=\overline{1,n},\end{eqnarray*}
\begin{eqnarray*} f_i\left(p,z^0\right)=y_i^0\sum\limits_{k=1}^nc_{ki}p_k,  \quad  i=\overline{n+1,l},\end{eqnarray*}
\begin{eqnarray*}  g_j\left(p,z^0\right)=x_j^0\left[\left(p_j - \sum\limits_{s=1}^na_{sj}p_s\right)+ \langle b_j, p\rangle \right],  \quad j=\overline{1,n},\end{eqnarray*}  \begin{eqnarray*}  g_j\left(p,z^0\right)=\langle b_j, p\rangle, \quad j=\overline{n+1,l},\end{eqnarray*}
and the vector $\pi\left(p,z^0\right)=\left\{  \pi_i \left(p, z^0\right)\right\}_{i=1}^n$ is built in the Theorem \ref{jant12} as a solution to the equation (\ref{jant13}).
\end{theorem}
\begin{proof}\smartqed  For two sets of functions  $f_i\left(p,z^0\right)$ and $ \ g_i\left(p,z^0\right), \ i=\overline{1,l},$ on the set $T_0 \times z^0,$ the equality \begin{eqnarray*} \sum\limits_{i=n+1}^ly_i^0\sum\limits_{k=1}^nc_{ki}p_k+\sum\limits_{i=1}^n\pi_i\left(p,z^0\right)\left[x_i^0\left(p_i - \sum\limits_{s=1}^na_{si}p_s\right)+ \langle b_i, p\rangle \right]  \end{eqnarray*}
\begin{eqnarray} \label{jant24}
 =\sum\limits_{i=1}^n\left[x_i^0\left(p_i - \sum\limits_{s=1}^na_{si}p_s\right)+ \langle b_i, p\rangle \right]+  \sum\limits_{i=n+1}^l \langle b_i, p\rangle
\end{eqnarray}
holds because it turns into the equation (\ref{jant13}) for the vector
\begin{eqnarray*} \pi\left(p,z^0\right)=\left\{\pi_i\left(p,z^0\right)\right\}_{i=1}^n\end{eqnarray*}
on the set $T_0\times z^0$ if the vector $x^0$ solves the set of equations (\ref{qant9}).
However, the equality (\ref{jant24}) in terms of the families of functions  $f_i\left(p,z^0\right)$ and  $g_i\left(p,z^0\right), \ i=\overline{1,l},$ has the form of the equality
\begin{eqnarray*} \sum\limits_{i=1}^lf_i\left(p,z^0\right)= \sum\limits_{i=1}^lg_i\left(p,z^0\right), \quad \left(p,z^0\right) \in T_0\times z^0.\end{eqnarray*}
 Because of the Lemma on the existence of a taxation system  whose conditions hold on the set $T_0\times z^0,$ there exists a  taxation system $||\pi_{ij}\left(p,z^0\right)||_{i,j=1}^l$ such that
\begin{eqnarray*}  f_i\left(p,z^0\right)=\sum\limits_{j=1}^l \pi_{ij}\left(p, z^0\right)g_i\left(p,z^0\right),  \quad \left(p,z^0\right) \in T_0\times z^0, \quad  i=\overline{1,l}.\end{eqnarray*}
\qed
\end{proof}

\begin{theorem}\label{jant14}
 Let $\sum\limits_{k=1}^nc_{ki}>0, \ i=\overline{1,n},$ there exist a non-negative vector
$v_0=\left\{v_i\right\}_{i=1}^n, \ v_i \geq 0, \ i=\overline{1,n},$   such that $\bar C\left(v_0\right) - \bar B $ is a non-negative matrix having no zero rows or columns and
$ A + \bar C\left(v_0\right) - \bar B$ is an indecomposable matrix,\index{indecomposable matrix} and let the vector $x^0=\left\{x_i^0\right\}_{i=1}^n$ have strictly positive components, and the spectral radius of the matrix $A$ be strictly less than 1.

For the known strictly positive vector $\pi=\left\{\pi_i\right\}_{i=1}^n,\  \pi_i > 0, \ i=\overline{1,n}, $ and the  strictly positive vector
$\bar y=\left\{y_i\right\}_{i=1}^n$ whose components satisfy inequalities $  \ y_i  > \pi_iv_i, \ i=\overline{1,n}, $
a solution to the set of equations
\begin{eqnarray} \label{jant15}
p_i= \sum\limits_{s=1}^na_{si}p_s +  \frac{1}{x_i^0}\sum\limits_{s=1}^n\left[c_{si} \frac{y_i}{\pi_i} - b_{si}\right]p_s, \quad   i=\overline{1,n},
\end{eqnarray}
in the set $T_0$ exists if and only if components of the  vector $x^0=\left\{x_i^0\right\}_{i=1}^n$  have the representation
\begin{eqnarray*} x_i^0=\frac{\left[\left(E-A\right)^{-1}\left[\bar C\left(\pi^{-1}\bar y\right)-  \bar B\right]g\right]_i}{
g_i}, \ \quad i=\overline{1,n},\end{eqnarray*}
where
\begin{eqnarray*} \bar C\left(\pi^{-1}\bar y\right)=\left|\left|c_{si}\frac{y_i}{\pi_i}\right|\right|_{s,i=1}^{n}, \quad
g=\left\{g_i\right\}_{i=1}^n, \quad g_i>0, \quad   i=\overline{1,n}.\end{eqnarray*}
\end{theorem}

\begin{proof}\smartqed
 Necessity. Let there exist a  solution $p_{0}=\left\{p_i^{0}\right\}_{i=1}^n$ to the problem
(\ref{jant15}) belonging to $T_0.$
Prove the representation for the vector $x^0.$
Because $p_{0} \in K_+^n,$ i.e., all the components of $p_{0}$ are strictly positive, the conditions  of the  Theorem \ref{jant14} hold,
the problem
\begin{eqnarray} \label{ant16}
z_k=\sum\limits_{i=1}^na_{ki}z_i+
\sum\limits_{i=1}^n\left[c_{ki}\frac{y_i}{\pi_i}- b_{ki}\right]\frac{z_i}{x_i^0}, \quad
k=\overline{1,n},
\end{eqnarray}
conjugate to the problem (\ref{jant15}) has a solution in the set of strictly positive vectors. The latter follows from indecomposability of  $A+ \bar C\left(v_0\right) - \bar B$ and the Perron-Frobenius Theorem.\index{Perron-Frobenius Theorem} From here, it follows that the vector
$z=\left\{z_i\right\}_{i=1}^n$ that is strictly positive solution to the problem (\ref{ant16}) has the representation
\begin{eqnarray*} z=\left(E-A\right)^{-1}\left[\bar C\left(\pi^{-1}\bar y\right) - \bar B\right]g,\end{eqnarray*}
where components of the vector $g$ are strictly positive and given by the formula \begin{eqnarray*} g_i=\frac{z_i}{x_i^0},\quad
i=\overline{1,n}.\end{eqnarray*}  From here, we obtain
\begin{eqnarray*} x_i^0=\frac{z_i}{g_i}=\frac{\left[\left(E-A\right)^{-1}
\left[\bar C\left(\pi^{-1}\bar y\right)- \bar B\right]g\right]_i}{g_i}, \quad i=\overline{1,n}.\end{eqnarray*}
 Sufficiency. Assume a certain vector
$g=\left\{g_i\right\}_{i=1}^n, \ g_i>0,$ \ $i=\overline{1,n},$
exists such that for components $x_i^0, \ i=\overline{1,n},$ of the vector $x^0=\left\{x_i^0\right\}_{i=1}^n$
 the representations
\begin{eqnarray*} x_i^0=\frac{\left[\left(E-A\right)^{-1}\left[\bar C\left(\pi^{-1}\bar y\right) - \bar B\right]g\right]_i}{ g_i}, \quad i=\overline{1,n},\end{eqnarray*}   hold.  Consider the set of equations \begin{eqnarray*} z_k=\sum\limits_{i=1}^na_{ki}z_i+
\sum\limits_{i=1}^n\left[c_{ki}\frac{y_i}{\pi_i}- b_{ki}\right]\frac{z_i}{x_i^0}, \quad
k=\overline{1,n}.
\end{eqnarray*}
This set of equations has a strictly positive solution
\begin{eqnarray*} z=\left(E-A\right)^{-1}\left[\bar C\left(\pi^{-1}\bar y\right) - \bar B\right]g.\end{eqnarray*}
From here and the  conditions of the Theorem, it follows that there exists a strictly positive solution to the conjugate problem \begin{eqnarray} \label{ant17}
p_i^{0}=\sum\limits_{s=1}^na_{si}p_s^{0}+\frac{1}{x_i^0}
\sum\limits_{s=1}^n\left[c_{si}\frac{y_i}{\pi_i}- b_{si}\right] p_s^{0},
\quad i=\overline{1,n},
\end{eqnarray}
which, obviously, belongs to $T_0$ because
\begin{eqnarray*} \frac{1}{x_i^0}\left[\frac{y_i}{\pi_i}- v_i\right]\sum\limits_{s=1}^nc_{si} p_s^{0} > 0,
\quad i=\overline{1,n}.\end{eqnarray*}
\qed
\end{proof}

Let us  use the Theorem \ref{jant14}
and construct on the set $ K_+^n \times \Gamma^n$ a taxation system
$||\pi_{ij}\left(p,z\right)||_{i,j=1}^l$
given on the set $T_0 \times z^0$ by the Theorem \ref{jant22} and on the set $K_+^n \times \Gamma^n \setminus  T_0\times z^0$ being arbitrary
provided that
\begin{eqnarray} \label{gint1}
\sum\limits_{i=1}^l\pi_{ij}\left(p,z\right)=1,  \quad \left(p,z\right) \in K_+^n \times \Gamma^n  \setminus  T_0\times z^0, \quad j =\overline{1,l}.
\end{eqnarray}
Let now give a family of income pre-functions\index{family of income pre-functions} by the formula
\begin{eqnarray} \label{gint2}
K_i^0\left(p, z\right)=\sum\limits_{j=1}^l\pi_{ij}\left(p,z\right)g_j\left(p,z\right),   \quad \left(p,z\right) \in K_+^n \times \Gamma^n, \quad i=\overline{1,l},
\end{eqnarray}
where
\begin{eqnarray*} g_i\left(p,z\right)=\left[\langle y_i -  x_i, p \rangle + \langle b_i,p \rangle\right],   \quad \left(p,z\right) \in K_+^n \times \Gamma^n, \quad i=\overline{1,n},\end{eqnarray*}
\begin{eqnarray*}  g_i\left(p,z\right)=\langle b_i, p\rangle,  \quad \left(p,z\right) \in K_+^n \times \Gamma^n, \quad  i=\overline{n+1,l}.\end{eqnarray*}
and define a family of  income functions by formulae
\begin{eqnarray*}  K_i\left(p,z\right)= K^0_i\left(p,Q\left(p,z\right)\right),  \quad \left(p,z\right) \in K_+^n \times \Gamma^n, \quad i=\overline{1,l}.\end{eqnarray*}

The reduction onto the set $T_0\times z^0$ of the above family of income functions takes the form
\begin{eqnarray} \label{gint3}
K_i\left(p, z^0\right)=\pi_i\left(p,z^0\right)\left\{x_i^0\left[p_i - \sum\limits_{s=1}^na_{si}p_s\right]+ \langle b_i,p\rangle \right\}, \quad i=\overline{1,n},
\end{eqnarray}
\begin{eqnarray*} K_i\left(p, z^0\right)=y_i^0\sum\limits_{k=1}^nc_{ki}p_k,  \quad  i=\overline{n+1,l}.\end{eqnarray*}
Therefore, on the realization of random field $\zeta_0\left(p\right)=z^0$ net income functions take the form
\begin{eqnarray*}  D_i\left(p\right)=\pi_i\left(p,z^0\right)\left[x_i^0\left(p_i - \sum\limits_{s=1}^na_{si}p_s\right)+ \langle b_i,p\rangle \right],   \quad p \in T_0, \quad i=\overline{1,n},\end{eqnarray*}
\begin{eqnarray} \label{gint4}
 D_i\left(p\right)=y_i^0\sum\limits_{k=1}^nc_{ki}p_k, \quad p \in T_0, \quad  i=\overline{n+1,l}.
\end{eqnarray}

In the next Theorem, we suppose that consumers net income corresponding to the realization of random field $\zeta_0\left(p\right)=z^0$ with probability  1 has, on the set $T_0,$ the form (\ref{gint4}).
\begin{theorem}\label{jant25}
Let $\sum\limits_{k=1}^nc_{ki}>0, \ i=\overline{1,l},$ there exists a non-negative vector
$v_0=\left\{v_i\right\}_{i=1}^n, \ v_i \geq 0, \ i=\overline{1,n},$  such  that $\bar C\left(v_0\right) - \bar B $ is a non-negative matrix having no zero rows or columns and
$ A + \bar C\left(v_0\right) - \bar B$ is an indecomposable matrix, and let the strictly positive vector $x^0$ be a solution to the set of equations (\ref{qant9}).
If the spectral radius of the matrix $A$ is less than 1, there exists a strictly positive vector
$g=\left\{g_i\right\}_{i=1}^n,$ \ $ g_i>0,$  \ $   i=\overline{1,n},$ such that for components $x_i^0, \  i=\overline{1,n},$ of the vector $x^0=\left\{x_i^0\right\}_{i=1}^n$   the representations
\begin{eqnarray*} x_i^0=\frac{\left[\left(E-A\right)^{-1}\left[\bar C\left(\pi^{-1}\bar y_0\right)- \bar B\right]g\right]_i}{
g_i}, \ \quad i=\overline{1,n},\end{eqnarray*}
 hold, where
\begin{eqnarray*} \bar C\left(v_0\right)=||c_{si}v_i||_{s,i=1}^{n}, \quad \bar C\left(\pi^{-1}\bar y_0\right)=\left|\left|c_{si}\frac{y_i^0}{\pi_i}\right|\right|_{s,i=1}^{n}, \quad  y_i^0 >  \pi_i v_i, \quad  i=\overline{1,n}, \end{eqnarray*}
and $ \pi=\left\{\pi_i\right\}_{i=1}^n, \ \pi_i >0, \  i=\overline{1,n},$ is a certain strictly positive vector,
then there exists a strictly positive equilibrium price vector $p_0=\left\{p_i^0\right\}_{i=1}^n \in T_0 $ solving the set of equations
\begin{eqnarray} \label{jant27} \sum\limits_{i=1}^l\frac{c_{ki}D_i\left(p\right)  }{\sum\limits_{s=1}^np_sc_{si}}=x_k^0 -  \sum\limits_{s=1}^na_{ks}x_s^0 +\sum\limits_{i=1}^lb_{ki}, \quad  k=\overline{1,n},
\end{eqnarray}
for which every industry is profitable, net income of productive industries at the state of economic equilibrium is given by formulae
\begin{eqnarray} \label{jant29}
D_i\left(p_0\right)=\pi_i\left[x_i^0\left(p_i^0- \sum\limits_{s=1}^na_{si}p_s^0\right) + \sum\limits_{s=1}^nb_{si}p_s^0\right], \quad   i=\overline{1,n},
\end{eqnarray}
and consumers net income is expressed as follows
\begin{eqnarray*} D_i\left(p_0\right)=y_i^0\sum\limits_{s=1}^nc_{si}p_s^0, \quad  i=\overline{n+1, l}.  \end{eqnarray*}

This equilibrium price vector solves the set of equations
\begin{eqnarray} \label{jant28}
\pi_i\left[x_i^0\left(p_i- \sum\limits_{s=1}^na_{si}p_s\right) + \sum\limits_{s=1}^nb_{si}p_s\right]= y_i^0\sum\limits_{s=1}^nc_{si}p_s, \quad   i=\overline{1,n}.
\end{eqnarray}
\end{theorem}
\begin{proof}\smartqed  For the  considered taxation system, the  net income of the $i$-th industry\index{net income of an industry} on the set $T_0$ is given by the formula
\begin{eqnarray*} D_i\left(p\right)=\pi_i\left(p,z^0\right)\left[x_i^0\left(p_i - \sum\limits_{s=1}^na_{si}p_s\right)+ \langle b_i,p\rangle \right],  \quad p \in T_0, \quad i=\overline{1,n}.\end{eqnarray*}
Consider the set of equations
\begin{eqnarray} \label{jant33}
\pi_i\left(p,z^0\right)\left[x_i^0\left(p_i - \sum\limits_{s=1}^na_{si}p_s\right)+ \langle b_i,p\rangle \right]=y_i^0\sum\limits_{k=1}^nc_{ki}p_k,  \quad i=\overline{1,n},
\end{eqnarray}
for the vector $p=\left\{p_i\right\}_{i=1}^n.$
Under the condition that the taxation system on the set $T_0\times z^0$ is given by Theorems \ref{jant12}, \ref{jant22}, the considered set of equations (\ref{jant33}) is equivalent to the set of equations
\begin{eqnarray} \label{jant34}
\frac{\pi_i\left[x_i^0\left(p_i-\sum\limits_{s=1}^na_{si}p_s\right)+\langle b_i, p\rangle \right]}{\sum\limits_{i=1}^n\pi_i\left[x_i^0\left(p_i- \sum\limits_{s=1}^na_{si}p_s\right) + \sum\limits_{s=1}^nb_{si}p_s\right]}=
\frac{y_i^0\sum\limits_{k=1}^nc_{ki}p_k}{\sum\limits_{i=1}^n y_i^0\sum\limits_{s=1}^nc_{si}p_s},  \quad i=\overline{1,n}.
\end{eqnarray}
Under the  conditions of the  Theorem \ref{jant25}, the conditions of the Theorem \ref{jant14} hold guaranteeing the existence of a strictly positive solution $p_0=\left\{p_i^0\right\}_{i=1}^n \in T_0 $ to the set of equations (\ref{jant28}). This solution also solves the set of equations (\ref{jant34}) because the equality
\begin{eqnarray*}  \sum\limits_{i=1}^n\pi_i\left[x_i^0\left(p_i^0- \sum\limits_{s=1}^na_{si}p_s^0\right) + \sum\limits_{s=1}^nb_{si}p_s^0\right]=\sum\limits_{i=1}^n y_i^0\sum\limits_{s=1}^nc_{si}p_s^0\end{eqnarray*}  holds.
From this equality it follows that for the solution $p_0=\left\{p_i^0\right\}_{i=1}^n $ to the set of equations (\ref{jant28})
there holds the equality
\begin{eqnarray*} D_i\left(p_0\right)=\pi_i\left[x_i^0\left(p_i^0 - \sum\limits_{s=1}^na_{si}p_s^0\right)+ \langle b_i,p_0\rangle \right],  \quad i=\overline{1,n},\end{eqnarray*}
because
\begin{eqnarray*} \pi_i\left(p_0,z^0\right)=\pi_i\frac{\sum\limits_{i=1}^n y_i^0\sum\limits_{s=1}^nc_{si}p_s^0}{\sum\limits_{i=1}^n\pi_i\left[x_i^0\left(p_i^0- \sum\limits_{s=1}^na_{si}p_s^0\right) + \sum\limits_{s=1}^nb_{si}p_s^0\right]}=\pi_i.\end{eqnarray*}
\qed
\end{proof}

\subsection{Necessary conditions of industries profitability at economy equilibrium}

In this Subsection, we establish that conditions guaranteeing industries profitability in Theorems proven in previous Sections are necessary  under simple technical conditions.
\begin{theorem}\label{jant35}
Let $\sum\limits_{k=1}^nc_{ki}>0, \ i=\overline{1,l},$ there exist a non-negative vector
$v_0=\left\{v_i\right\}_{i=1}^n$ such that $\bar C\left(v_0\right) - \bar B $ is a non-negative matrix having no zero rows or columns and a matrix
$ A + \bar C\left(v_0\right) - \bar B$ is indecomposable, and let  the set $T_0$ be not empty,  $D_i\left(p\right) >0, \  i=\overline{n+1, l}, \ p \in  T_0.$ If the vector $\pi=\left\{\pi_i\right\}_{i=1}^n$ has strictly positive components,
the spectral radius of the matrix\index{ spectral radius of the matrix} $A$ is less than 1, the  strictly positive vector $x^0=\left\{x^0_i\right\}_{i=1}^n$
is such that $ \ x_k^0 -  \sum\limits_{s=1}^na_{ks}x_s^0 +\sum\limits_{i=1}^lb_{ki} > 0,\  k=\overline{1,n},\ $ there exists a strictly positive vector $p_0=\left\{p_i^0\right\}_{i=1}^n \in  T_0 $ being a solution to the set of equations
\begin{eqnarray} \label{jant36} \sum\limits_{i=1}^l\frac{c_{ki}D_i\left(p\right)  }{\sum\limits_{s=1}^np_sc_{si}}=x_k^0 -  \sum\limits_{s=1}^na_{ks}x_s^0 +\sum\limits_{i=1}^lb_{ki}, \quad  k=\overline{1,n},
\end{eqnarray}
where
\begin{eqnarray} \label{ojont4}
 D_i\left(p\right)=\pi_i\left[x_i^0\left(p_i - \sum\limits_{s=1}^na_{si}p_s\right)+ \langle b_i,p\rangle \right], \quad i=\overline{1,n},
 \end{eqnarray}
then the vector
\begin{eqnarray*} \psi=\left\{\psi_k\right\}_{k=1}^n, \quad  \psi_k=x_k^0 -  \sum\limits_{s=1}^na_{ks}x_s^0 +\sum\limits_{i=1}^lb_{ki},\quad  k=\overline{1,n}, \end{eqnarray*}
belongs to the interior of the cone created by the vectors-columns of the matrix\index{interior of the cone created by the vectors-columns of the matrix} $C=||c_{ki} ||_{k=1, i=1}^{n, \ l},$
 components $x_i^0$ of the vector $x^0=\left\{x_i^0\right\}_{i=1}^n$ have the representation
\begin{eqnarray*} x_i^0=\frac{\left[\left(E-A\right)^{-1}\left[\bar C\left(\pi^{-1}\bar y\right)- \bar B\right]g\right]_i}{
g_i}, \quad i=\overline{1,n},\end{eqnarray*}
where
$g=\left\{g_i\right\}_{i=1}^n,$  $ g_i>0,$ $ i=\overline{1,n},$
is a certain strictly positive vector,
\begin{eqnarray*} \bar y=\left\{y_i\right\}_{i=1}^n, \quad  y_i=\frac{D_i\left(p_0\right)}{\sum\limits_{s=1}^nc_{si}p_s^0},\quad  i=\overline{1,l},\end{eqnarray*}
\begin{eqnarray*} \bar C\left(v_0\right)=||c_{si}v_i||_{s,i=1}^{n}, \quad \bar C\left(\pi^{-1}\bar y\right)=\left|\left|c_{si}\frac{y_i}{\pi_i}\right|\right|_{s,i=1}^{n},\end{eqnarray*}
there hold inequalities
\begin{eqnarray*}  y_i > \pi_i v_i, \quad  i=\overline{1,n}.\end{eqnarray*}
\end{theorem}
\begin{proof}\smartqed  Because the solution $p_0=\left\{p_i^0\right\}_{i=1}^n \in  T_0 $ to the set of equations (\ref{jant36}) exists, denoting
\begin{eqnarray*}  y=\left\{y_i\right\}_{i=1}^l, \quad  y_i=\frac{D_i\left(p_0\right)}{\sum\limits_{s=1}^nc_{si}p_s^0},\quad  i=\overline{1,l},\end{eqnarray*}
we have that
\begin{eqnarray*}  \psi = \sum\limits_{i=1}^lC_i y_i, \quad y_i > 0, \quad C_i=\left\{c_{ki}\right\}_{i=1}^n, \quad   i=\overline{1,l}.\end{eqnarray*}
The latter means that the vector of the final consumption $\psi=\left\{\psi_k\right\}_{k=1}^n$ belongs to the interior of the cone created by the vectors-columns of the matrix $C.$
It is obvious that the strictly positive vector $p_0=\left\{p_i^0\right\}_{i=1}^n \in  T_0 $
is a solution to the set of equations
\begin{eqnarray} \label{jant37}
p_i= \sum\limits_{s=1}^na_{si}p_s + \frac{1}{x_i^0}\sum\limits_{s=1}^nc_{si}\left[\frac{y_i}{\pi_i} - b_{si}\right] p_s, \quad   i=\overline{1,n},
\end{eqnarray}
and there hold inequalities
\begin{eqnarray*} \frac{y_i}{\pi_i}=\frac{D_i^0\left(p_0\right)}{\sum\limits_{s=1}^nc_{si}p_s^0} > v_i , \quad i=\overline{1,n}.\end{eqnarray*}
As earlier, we establish that there exists a strictly positive solution to the conjugate problem
\begin{eqnarray} \label{jant38}
z_k=\sum\limits_{i=1}^na_{ki}z_i+
\sum\limits_{i=1}^n\left[c_{ki}\frac{y_i}{\pi_i}- b_{ki}\right]\frac{z_i}{x_i^0}, \quad
k=\overline{1,n},
\end{eqnarray}
from which the needed representations for components $x_i^0, \  i=\overline{1,n},$ of the gross output vector $x^0=\left\{x_i^0\right\}_{i=1}^n$ follow.
\qed
\end{proof}

Let give sufficient conditions for the set $T_0$ to be non-empty one.
\begin{lemma}\label{lat1}
Let $\sum\limits_{k=1}^nc_{ki}>0, \ i=\overline{1,l},$ the matrix $A$ be productive and indecomposable,
there exist a non-negative vector
 $v_0=\left\{v_i\right\}_{i=1}^n, \ v_i \geq 0, \ i=\overline{1,n},$ such
that the inequality $\bar C\left(v_0\right) - \bar B  \geq 0$ holds, and let the gross output vector\index{} $x^0=\left\{x_i^0\right\}_{i=1}^n$ be strictly positive. If the inequalities
\begin{eqnarray*} v_i < \frac{\left(1 -\lambda\right)x_i^0  p_i^0}{\sum\limits_{k=1}^nc_{ki}p_k^0}, \quad i=\overline{1,n},\end{eqnarray*}
hold,  where $p^0=\left\{p_i^0\right\}_{i=1}^n$ is left Frobenius vector\index{left Frobenius vector} of the matrix $A$ and $\lambda$ is the Frobenius number,\index{Frobenius number} then the set $T_0$ is not empty.
\end{lemma}
\begin{proof}\smartqed  The set $T_0$ contains the left Frobenius vector of the matrix $A.$
\qed
\end{proof}

The  proven  Theorem \ref{jant35} means that earlier conditions  only sufficient for the existence of economic equilibrium at which all industries have profits are also necessary. Therefore, it is worth  considering only those economy system models whose the structure of supply  agrees with the structure  of choice or the final consumption vector\index{final consumption vector} belongs to the interior of the  non-negative cone created by vectors-columns of the matrix $C,$ the income functions of consumers-non-producers\index{income functions of consumers-non-producers} have the form $D_i(p)=y_i \sum\limits_{s=1}^nc_{si}p_s, \ i=\overline{n+1, l}.$
As the solution to the set of equations (\ref{jant28} in the Theorem \ref{jant25} is unique up to constant factor that one must take   accounting  for the face-value of the monetary unit passing in the economic system, then proven Theorems  mean that having chosen industries taxation vector\index{industries taxation vector} and consumers needs satisfaction levels,\index{consumers needs satisfaction levels} we determine equilibrium price vector unambiguously.
This fact opens wide possibilities to apply considered models to real economic systems for clarifying  possible imbalances in economy development resulting from taxation system deformations, to construct economic transformations theory, to study monopolistic influence onto the economy.\index{monopolistic influence onto the economy} All that will be the subject of subsequent Chapters.

\chapter{Economy transformations theory}

\abstract*{The theory of economic transformations based on the principle of  openness of economic system to its environment is constructed. A notion of agreement of taxation vector with the structure of consumption and foreign economic relations is introduced.
Two types of transformations of characteristic  of the economy system not going outside the given class of models and conserving the equilibrium price vector are studied. These transformations are transformations enlarging the part of the value added of industry in the industry output and transformations
 changing levels of taxation. Explicit formulae for  structure matrix of production  and  matrix of unproductive consumption  are obtained under which the equilibrium price vector is the same one. The results obtained are applied to analyze real economy systems.}

In this Chapter, we construct the theory of  economic transformations  based on the principle of openness of economic systems  \cite{ 92, 106, 19, 74, 75, 200, 102}.\index{ principle of openness of economic systems}
In the Section "Open economic system", we describe open economic systems basing on the general concept of description of economic systems. We introduce a notion of agreement of the structure of  supply  with the  structures of  choice and  foreign economic relations.\index{agreement of the structure of  supply  with the  structures of  choice and  foreign economic relations} We construct a family of income functions\index{family of income functions} after a productive economic process introduced and a family of income pre-functions built. We establish the Theorem \ref{gat22} about the absence of arbitrage\index{absence of arbitrage} and the ability of the considered economic system  to operate without default.\index{operate without default} Grounding on a specific form for the field of information evaluation by an economic agent\index{field of information evaluation by an economic agent} that consumes goods exported by the economic system and the  vector of  supply of goods\index{vector of  supply of goods} that he has to import into the economy system, the Theorem \ref{gat24} gives the  conditions of absence of arbitrage   for such an open economy system.

In the Section "Aggregated open economic system", we construct the basic model studied in what follows. We give the crucial definition of agreement of  the structure of supply  with the structures of choice  and foreign economic relations.\index{agreement of  the structure of supply  with the structures of choice  and foreign economic relations} The Lemma \ref{vjant18} and the Theorem \ref{vjant12} establish the sufficient conditions for the  vector of  gross outputs under which there exists a certain form of taxation system given by  a certain strictly positive vector called the taxation vector. The Theorem \ref{vjant22} gives the  conditions of the existence of a taxation system with which we construct a family of income pre-functions  for the economic system model studied in what follows. The Theorem \ref{vjant14} establishes the necessary and sufficient condition of the existence  of a solution to the set of equations (\ref{vjant15}) belonging to the set $T_0$ for the known  strictly positive taxation vector.
This condition is equivalent to the solvability of the set of equations with respect to the  vector of  gross outputs.\index{vector of  gross outputs} The Theorem \ref{vjant25} gives a  method to construct equilibrium price vectors in the considered economy model.

 In the Section "Economy systems transformations", we reformulate the earlier given definition of the agreement of the structure of  supply with the structure of  choice\index{agreement of the structure of  supply with the structure of  choice} in the  more convenient  definition of the agreement of a certain  strictly positive vector called the  taxation vector with  the structures of  consumption  and foreign economic relations  in the economy system determined by  fields of information evaluation by consumers.

In the Theorem \ref{gat9}, we adapt the results of the Section "Aggregated open economic system"  to the form convenient to apply it in what follows taking into account the definition of the agreement of the taxation vector with the  structures of consumption and foreign economic relations.\index{ agreement of the taxation vector with the  structures of consumption and foreign economic relations}

In the Section "Transformations conserving equilibrium price vector", within the models with proportional consumption, we study transformations of characteristics of the economy system  that do not go outside the given class of models and conserve the equilibrium price vector.
Transformations of the first kind result in changes of technological coefficients, changes of unproductive consumption matrix, and  changes of a vector of gross outputs  under a constant taxation system, levels of consumption,  vectors of export and import. In cost parameters, such transformations enlarge the part of the  value added  of industry in  the industry output.\index{part of   value added  of industry in  the industry output}
Transformations of the second kind result in change of technological coefficients, change of a matrix of unproductive consumption, change of a vector of gross outputs  and levels of taxation,\index{levels of taxation} and also do  changes of levels of consumption and   vectors of export and import.
In this Section, we construct four types of   model  economic systems. We describe the model economy system $E_1$ by a certain structure matrix of production,\index{structure matrix of production} a certain  matrix of  unproductive consumption,\index{matrix of  unproductive consumption} and a matrix of an  initial goods supply\index{matrix of an  initial goods supply} at the beginning of economy operation period, given levels of satisfaction of consumers needs,\index{ levels of satisfaction of consumers needs} given  taxation system\index{ taxation system}  and  system of foreign economic relations,\index{system of foreign economic relations} a vector of  gross outputs.\index{vector of  gross outputs} We suppose that an equilibrium price vector in  the economy system is the vector $p^0=\{p_i^0\}_{i=1}^n.$ The model economic system $E_2$ differs from the model economy system $E_1$  by the improved  structure matrix of production,\index{structure matrix of production}  the improved matrix of unproductive consumption,\index{improved matrix of unproductive consumption} and higher levels of industry gross outputs.\index{higher levels of industry gross outputs} We prove the Theorem \ref{fa1} establishing that the equilibrium price vector in the model  economic system $E_2$ is the same vector $p^0=\{p_i^0\}_{i=1}^n.$ In macroeconomic sense, the transition from the model economic system $E_1$ to the model economic system $E_2$ results in enlarging of the part of GDP (gross domestic product)\index{gross domestic product} in the gross national product.\index{gross national product}  The explicit formulae for  matrix  elements of unproductive consumption\index{ matrix  elements of unproductive consumption} in the model  economy  system $E_2$ in terms of matrix elements of unproductive consumption  in the model  economic system $E_1,$ levels of enlarging of value added in every industry,\index{levels of enlarging of value added in every industry} industry gross outputs,\index{industry gross outputs} and the equilibrium price vector are obtained.
The model economy system $E_3$ are constructed so that all characteristics of the economic system change. The most significant change is change of  levels  of taxation. We obtain explicit formulae for the structure matrix of production\index{ structure matrix of production}  and the  matrix of unproductive consumption\index{matrix of unproductive consumption} under which the equilibrium price vector in the model economy system is the same vector $p^0=\{p_i^0\}_{i=1}^n.$
The  construction of the  model economy system $E_4$ contains elements of construction  of the model economy system $E_2$ as well as elements of construction  of the model economy system $E_3,$ i.e.,  the part of GDP in the  gross national  product\index{ gross national  product} enlarges and  levels of  taxation are changed.

To every model economy system constructed, we put into correspondence the cost parameters description and establish the relations between quantities introduced. As the consequence of these relations, we obtain a certain set of balance equations\index{set of balance equations} and the set of relations between matrix elements of unproductive consumption\index{set of relations between matrix elements of unproductive consumption}  and other aggregated characteristics. In the case of the model  economy system $E_1$, these relations are (\ref{gurl1}).
We postulate the validity of these relations for the description of real economic systems. On this ground,  formulae for technological coefficients of production,\index{technological coefficients of production} the vector of gross outputs,\index{vector of gross outputs} matrix elements of unproductive  consumption,\index{matrix elements of unproductive  consumption} vectors of  export and import\index{vectors of  export and import} corresponding to a hypothetical economy of social agreement\index{hypothetical economy of social agreement} are obtained.
In the Subsection "Mathematical foundations of social agreement economy", we use the relations obtained between cost parameters to describe real economic systems.

We set up the correspondence between aggregated description of the  economy system\index{aggregated description of the  economy system} and non-aggregated one. The Theorem \ref{ss0s9} establishes that constructed non-aggregated description\index{non-aggregated description} is equivalent to aggregated one. The last opens wide possibilities to apply obtained results to real economic systems.

 We compare a real economic system with a certain social agreement economy\index{social agreement economy} to reveal deformations existing in the real economy and negative influences they do onto the real economy. In this framework, we analyze the state of Ukrainian economy.

\section{Problem statement}

To understand processes in an  economic system and reasons why one or another industry is detrimental, balance of foreign economic relations\index{balance of foreign economic relations } is negative, the taxation system does not favor investment, one needs adequate mathematical model to answer the questions above mentioned. We suppose that the equilibrium price vector formed in an open economy reflects the structure of production, demand and supply structures,
the structure of foreign economic relations, the current taxation system.\index{current taxation system} If the production is energy and material expensive, the  part of the added value in the industry gross output\index{part of the added value in the industry gross output} is less than that in the corresponding industry of other country with less energy and material expensive production.\index{energy and material expensive production} The taxation system under energy and material expensive production is deformed  because for the open operation of the economy system there exist privileges for ones and discriminations for other ones.
The main task is to construct the mathematical model of the description of the economy  to reveal negative influences onto the economy and propose changes to remove negative influences.

Let us give heuristic reasons to solve the problem stated.

Let two economy systems $E_1$ and $E_2$ interact with each other exchanging goods and have the same structures of industries producing the same goods.
The economy system $E_2$ has more perfect production technologies than the economy system
$E_1.$  Let $E_2$ be perfectly competitive. Because of openness, the prices in economy systems are the same. The taxation system in $E_2$
is equitable and in $E_1$ is deformed. To clarify which industries in $E_1$
have backward technologies,\index{backward technologies} one must go in $E_1$ to the same  parts of added values in  industries outputs\index{parts of added values in  industries outputs} as in the economy system $E_2$ and then to taxation system\index{taxation system} in $E_2$   such that the price vector  in $E_1$ remain previous.
Having made such transition, we find to what measure one must change production technologies or, in other words, input structure,\index{input structure} raise wages, expand investment, change foreign economic relations, gross outputs and so on in $E_1$ to reach the desired level of  economy development.\index{desired level of  economy development} Therefore, in the Section "Transformations conserving equilibrium price vector", we study the above transformations.
This will help to establish changes of structures of production and consumption  needed to raise  level of economy development.\index{level of economy development}

For the economy system $E_2$ one can take a hypothetical economy system to compare it with the real economy system $E_1.$ This hypothetical economy system we call the economy of social agreement.\index{economy of social agreement}

\section{Open economy system and its description}

Show  that  the description of an open economy system  is a partial case of the description considered in previous Sections.
Let, as earlier, in the economy system $m$ firms operate, the structure of production of the $i$-th industry is described by a convex down technological map $F_i(x),  \ x \in X_i^1,\  i =\overline{1,m},$ belonging to the CTM class in a wide sense.
In this Subsection, we suppose that the set of possible price vectors\index{ set of possible price vectors} is the cone
\begin{eqnarray*} K_+^n=\{p=\{p_i\}_{i=1}^n \in \bar R_+^n, \ p_i > 0,\ i=\overline{1,n}\}.\end{eqnarray*}
We give a productive economic process\index{productive economic process}
$Q(p,z)=\{Q_i(p,z)\}_{i=1}^m$ by the rule
\begin{eqnarray*} Q_i(p,z)=z_ia_i(p, z) \prod\limits_{s=1}^nv_s(p,z), \quad
a_i(p, z)=\chi_{[0, \infty)}(u_i(p,z)),\end{eqnarray*}
\begin{eqnarray*}    u_i(p,z)= \left\langle p, y_i - x_i \right\rangle + \left\langle p, d_i + b_i \right\rangle, \end{eqnarray*}
\begin{eqnarray*} v_s(p,z)=\chi_{[0, \infty)}(V_s(p,z)), \end{eqnarray*}
\begin{eqnarray} \label{gat15}
V_s(p,z) = \sum\limits_{i=1}^m \{a_i(p,z)[y_{si} -x_{si}] +d_{si}\} + \sum\limits_{i=1}^{l+1}b_{si} - c_s,
\end{eqnarray}
where $z^i=(x_i,y_i) \in \Gamma_i=\{(x,y), \  x \in X_i, \  y \in F_i(x)\}\cup (0,0), \ 0 \in S,$ and
\begin{eqnarray*}  b_i=\{b_{ki}\}_{k=1}^n, \quad i=\overline{1,l+1}, \quad d_i=\{d_{ki}\}_{k=1}^n, \quad  i=\overline{1,m}, \end{eqnarray*}
\begin{eqnarray*}  x_i=\{x_{ki}\}_{k=1}^n,  \quad y_i=\{y_{ki}\}_{k=1}^n, \quad  i=\overline{1,n}, \end{eqnarray*}
are vectors with non-negative components $c=\{c_i\}_{i=1}^n, \ c_i \geq 0,  \ i=\overline{1,n},$

\begin{eqnarray*} \chi_{[0, \infty)}(x)= \left\{\begin{array}{ll}
            1, & \textrm{if} \quad x \geq 0,\\
            0  &         \textrm{if} \quad    x < 0  \textrm{.}
                                                                                                                                                                 \end{array}
                                               \right.\end{eqnarray*}
Further, we suppose that the condition
\begin{eqnarray*} \sum\limits_{i=1}^{l+1}b_i \geq c,\quad c=\{c_1, \ldots , c_n\},\quad  c_i \geq 0, \quad i=\overline{1,n}, \end{eqnarray*}
holds.
Consider the set
\begin{eqnarray*} T=\{z \in \Gamma^m, \ W_s(z)>0,  \ s=\overline{1,n} \},\end{eqnarray*}
where
\begin{eqnarray} \label{gat16}
W_s(z) = \sum\limits_{i=1}^m (y_{si} -x_{si} +d_{si}) + \sum\limits_{i=1}^{l+1}b_{si} - c_s, \quad s=\overline{1,n},
\end{eqnarray}
is the $s$-th component of the vector
\begin{eqnarray*} W(z) = \sum\limits_{i=1}^m (y_i -x_i +d_i) + \sum\limits_{i=1}^{l+1}b_i - c.\end{eqnarray*}
Let $K_i^0(p,z), \ i=\overline{1,l+1},$ be a certain family  of  income pre-functions\index{family  of  income pre-functions}   in the economy system that will be specified later.
Further, we consider only those  technological maps
  $F_i(x),  \ x \in X_i^1,\  i =\overline{1,m},$ from the  CTM class in a wide sense that are convex down for which the set $T$ is non-empty.

On a probability space $\{\Omega, {\cal F}, \bar P\},$
that is the direct product of $(l+2)$ measurable spaces $\{\Omega_i, {\cal F}_i, \bar P_i\}, \
i=\overline{0,l+1},$
where
\begin{eqnarray*} \Omega=\prod\limits_{i=0}^{l+1}\Omega_i, \quad {\cal F}=\prod\limits_{i=0}^{l+1}
{\cal F}_i, \quad \bar P=\prod\limits_{i=0}^{l+1}\bar P_i,\end{eqnarray*}
consider random fields of information evaluation  by $(l+1)$ consumers\index{random fields of information evaluation  by  consumers}
\begin{eqnarray*}   \eta_i^0(p,z, \omega_i)=\{\eta_{is}^0(p,z, \omega_i)\}_{s=1}^n,\quad (p,z,\omega_i) \in K_+^n\times\Gamma^m\times\Omega_i, \quad i=\overline{1,l+1},\end{eqnarray*}
that take values in the set $S$ and are continuous with probability 1
 and random field $\zeta_0(p,\omega_0)$ taking values in the set $\Gamma^m$ and being continuous with probability 1.
We suppose that  fields of information evaluation by consumers\index{fields of information evaluation by consumers }  are not random, i.e.,
\begin{eqnarray*}   \eta_i^0(p,z, \omega_i)= \eta_i^0(p,z)=\{\eta_{is}^0(p,z)\}_{s=1}^n,\quad (p,z) \in K_+^n\times\Gamma^m, \quad i=\overline{1,l+1}.\end{eqnarray*}
Assume that they satisfy the  conditions of the  Theorem \ref{11tl4}.

First $m$  random fields of consumers choice,  built after first $m$ random fields of information evaluation by consumers, describe unproductive consumption of $m$ firms each of which contains proper firm consumption forming accumulation and raw materials for future period of operation and goods for unproductive consumption.
 Fields of  information evaluation by  consumers $ \eta_{i}^0(p, z),\ i=\overline{1,l+1},$ satisfy the conditions
\begin{eqnarray*} \sum\limits_{k=1}^n\eta_{ik}^0(p, z)p_k > 0, \quad (p,z) \in K_+^n\times \Gamma^m ,  \quad  i=\overline{1,l+1}.\end{eqnarray*}
Introduce the vectors $e^0=\{e_i^0\}_{i=1}^n$ and $i^0=\{i_k^0\}_{k=1}^n,$
 the first vector we call the export vector\index{export vector} and the second one we call the import vector.\index{import vector}

To describe an open economy system,\index{open economy system} we suppose that the $(l+1)$-th consumer has a vector of goods  $b_{l+1}=\{b_{k, l+1}\}_{k=1}^n$ where $ \ b_{k, l+1}= i_k^0,\ k=\overline{1,n},$ and the consumption of the  $(l+1)$-th consumer  is given by a field of information evaluation
$\eta_{l+1}^0(p,z)=\{\eta_{l+1,k}^0(p,z)\}_{k=1}^n,$ where $ \ \eta_{l+1,k}^0(p,z)=e_k^0, \ k=\overline{1,n},$ and we suppose that his income pre-function is given by the formula
\begin{eqnarray*}  K_{l+1}^0(p,z)=\sum\limits_{s=1}^ne_s^0p_s.\end{eqnarray*}
Further, in the definition of productive economic process given by the formula (\ref{gat15}), we
assume that $c_k \geq e_k^0, \ k=\overline{1,n}.$
\begin{definition}\label{gat17}  The structure of  supply  agrees with the structures of choice and foreign economic relations\index{structure of  supply  agrees with the structures of choice and foreign economic relations}  on the set $B_0 \subseteq K_+^n\times \Gamma^m $ if there exist  fields $y_i(p,z),  \ i=\overline{1,l},$ satisfying conditions
\begin{eqnarray*} y_i(p,z)> t_0 >0,  \quad i=\overline{1,l}, \quad (p,z) \in B_0, \end{eqnarray*}  where the number $t_0$ does not depend on
$ (p,z) \in B_0$ such that the equalities hold
\begin{eqnarray*} \sum\limits_{i=1}^m[y_{ki} - x_{ki} + d_{ki}]+\sum\limits_{i=1}^lb_{ki}+ i_k^0 \end{eqnarray*}
\begin{eqnarray} \label{gat18}
=\sum\limits_{i=1}^l\eta_{ik}^0(p,z) y_i(p,z)+ e_k^0, \quad  k=\overline{1,n}, \quad (p,z) \in B_0.
\end{eqnarray}
Here $b_i, \ i=\overline{1,l},$ is the property vector of the $i$-th consumer at the beginning of the economy operation period, $e^0=\{e_i^0\}_{i=1}^n$ is the export vector and $i^0=\{i_k^0\}_{k=1}^n$ is the import vector.
\end{definition}
The above  definition of the agreement of the structure of  supply  with the structures of  choice and foreign economic relations for such $(l+1)$-th consumer  is the definition of agreement of the structure of supply  with the structure  of  choice in the  economy system with $m$ firms and $(l+1)$ consumers. To complete assignment of the economy system with $m$ firms and $(l+1)$ consumers, it is sufficient to give a family  of income pre-functions of the rest $l$ consumers.

Definitions of notations used below are given in the Chapter 4.

Consider on the set $R_0$ that is one of sets $R_1, R,$ or $ R_2$ defined in the Chapter 4  a family of functions
\begin{eqnarray*} f_i(p,z)= \pi_i(p,z )[\left\langle p, y_i - x_i + d_i \right\rangle + \left\langle b_i, p\right\rangle ], \quad i=\overline{1,m},\end{eqnarray*}
\begin{eqnarray} \label{gat20}
f_i(p,z)=y_i(p,z)\sum\limits_{k=1}^n\eta_{ik}^0(p, z)p_k ,  \quad i=\overline{m+1,l+1},
\end{eqnarray}
where
\begin{eqnarray*} y_{l+1}(p,z)=1, \quad 0 < \pi_i(p,z), \quad i=\overline{1,m}, \end{eqnarray*}
and a family of functions
\begin{eqnarray*} g_i(p,z)= \left\langle p, y_i - x_i + d_i\right\rangle  + \left\langle b_i, p\right\rangle , \quad i=\overline{1,m},\end{eqnarray*}
\begin{eqnarray} \label{gat19}
 g_i(p,z)= \left\langle p, b_i \right\rangle,  \quad i=\overline{m+1,l+1},
\end{eqnarray}
that are continuous functions of $(p,z) \in R_1.$
 On the set $R_0, $ the  set of functions $g_i(p,z), \ f_i(p,z), \ i=\overline{1,l+1},$
satisfies  the conditions of the Lemma \ref{let1}  about the existence of a taxation system\index{Lemma   about the existence of a taxation system }  if conditions hold
\begin{eqnarray*}   \sum\limits _{i=1}^{l+1}f_i(p, z)= \sum\limits _{j=1}^{l+1}g_j(p, z), \quad (p,z) \in R_0,\end{eqnarray*}
there exist  numbers $k$ and $s$ such that $f_{k}(p,z)+f_s(p,z) > 0$ on the set $R_0.$

The last equality takes the form
\begin{eqnarray*}  \sum\limits _{i=1}^m\left\langle p, y_i - x_i + d_i\right\rangle  +\sum\limits _{i=m+1}^{l+1}\left\langle b_i, p\right\rangle \end{eqnarray*}
\begin{eqnarray*}  =\sum\limits _{i=1}^m\pi_i(p,z)\left\langle p, y_i - x_i + d_i +b_i\right\rangle  +\sum\limits _{i=m+1}^{l+1}y_i(p,z)\sum\limits_{k=1}^n\eta_{ik}^0(p, z)p_k.\end{eqnarray*}

Assume that the structure of supply  agrees with structures of choice and foreign economic relations\index{ structure of supply  agrees with structures of choice and foreign economic relations}  on the set $R_0.$

With (\ref{gat18}) the equality turns into
\begin{eqnarray*}  \sum\limits _{i=1}^m\pi_i(p,z)[\left\langle p, y_i - x_i + d_i \right\rangle + \left\langle b_i, p\right\rangle ]
\end{eqnarray*}
\begin{eqnarray} \label{gat21}
=\sum\limits _{i=1}^my_i(p,z)\sum\limits_{k=1}^n\eta_{ik}^0(p, z)p_k.
\end{eqnarray}

\begin{theorem}\label{gat26}
If there hold conditions of  one of the Lemmas \ref{opod11}, \ref{opcd1},   \ref{opod111},
then on the set $R_0$ there exists a taxation system
$||\pi_{ij}^0(p,z)||_{ i,j=1}^{l+1}$ such  that
\begin{eqnarray*} f_i(p,z)=\sum\limits _{j=1}^{l+1}\pi_{ij}^0(p,z)g_j(p,z), \quad (p,z) \in R_0.\end{eqnarray*}
\end{theorem}

Let $\pi_{ij}^1(p,z), \ i,j=\overline{1,l+1}, \ (p,z) \in K_+^n\times \Gamma^m \setminus R_0$ be a certain  taxation system on the set $K_+^n\times \Gamma^m \setminus R_0.$
Introduce into  consideration a taxation system
\begin{eqnarray*} \pi_{ij}(p,z)= \pi_{ij}^0(p,z)\chi_{R_0}(p,z)+ \pi_{ij}^1(p,z)[1 - \chi_{R_0}(p,z)],\end{eqnarray*}  \begin{eqnarray*}   (p,z) \in K_+^n\times \Gamma^m, \quad i,j=\overline{1,l+1}. \end{eqnarray*}
Give  a family of income pre-functions of consumers by the formulae
\begin{eqnarray*} K_i^0(p,z)=\sum\limits _{j=1}^{l+1}\pi_{ij}(p,z)g_j(p,z), \quad (p,z) \in K_+^n\times \Gamma^m,  \quad  i=\overline{1,l+1}. \end{eqnarray*}
Under so defined   family of income pre-functions,  a family of income functions  takes the form
\begin{eqnarray} \label{gat27}
K_i(p,z)= K_i^0(p, Q(p,z))
\end{eqnarray}
 \begin{eqnarray*} =\sum\limits _{j=1}^{l+1}\pi_{ij}(p,Q(p,z))g_j(p,Q(p,z)), \quad (p,z) \in K_+^n\times \Gamma^m, \quad  i=\overline{1,l+1},   \end{eqnarray*}
where $Q(p,z)$ is a productive economic process given by the formula (\ref{gat15}).

Suppose, as in the Chapter 4, that with probability 1 $\zeta_0(p, \omega_0)=z^0, \ p \in K_+^n, $ where $ z^0 \in M_0.$

The next Theorem follows from Theorems of Chapter 4.

Introduce the next notations
\begin{eqnarray*} \mu_{ki}(p)=\eta_{ik}^0\left(p,z^0\right), \quad k=\overline{1,n}, \quad i=\overline{1,l+1}, \end{eqnarray*}
\begin{eqnarray*}  D_i(p)=f_i\left(p,z^0\right), \quad  \left(p,z^0\right) \in R_0,  \quad  i=\overline{1,l+1}. \end{eqnarray*}

\begin{theorem}\label{gat22}
 Let technological maps $F_i(x),\ x \in X_i^1,$
$i=\overline {1,m},$ be convex down and belong to the CTM class in a wide sense,
the set $R$ be non-empty, the structure of supply  agree with the structure of choice  on the set $R_1$
 and let  fields $y_i(p,z),\ i=\overline{1,l}, $ be continuous functions of $(p,z) \in R_1,$ there hold conditions of the Lemma \ref{opcd1} or the Lemma \ref{opod111}.

If  fields of   information evaluation by consumers $\eta_{i}^0(p,z),\  i=\overline{1,l+1},$ and a random field $\zeta_0(p,\omega_0)$ satisfy the  conditions of the Theorem \ref{11tl4}, then the set of equations of the  economy equilibrium
\begin{eqnarray*} \sum\limits
_{i=1}^{l+1} \frac{\mu_{ki}(p)D_i(p)}{\sum\limits_{j=1}^n
\mu_{ji}(p)p_j} \end{eqnarray*}
\begin{eqnarray} \label{gat23}
=  \sum\limits _{i=1}^m\left[Y_{ki}\left(p, z^0\right)-X_{ki}\left(p, z^0\right)+ d_{ki}\right]+ \sum\limits
_{i=1}^{l+1} b_{ki}, \quad  k=\overline{1,n},
\end{eqnarray}
has a solution  $p^0=\{p_i^0\}_{i=1}^n$ in the set of strictly positive  price vectors
for every $ z^0 \in M_0.$
Such economy system can operate profitably in the Walras equilibrium state under the condition $\left\langle y_i^0 -x_i^0, p^0\right\rangle  \ > 0, \ i=\overline{1,m},$ or subvention-profitably in the Walras equilibrium state if
\begin{eqnarray*} \left\langle y_i^0 -x_i^0+ d_i , p^0 \right\rangle   > 0, \quad i=\overline{1,m},\end{eqnarray*}
for $b_i \neq 0.$
\end{theorem}
Taking into account  the structure of  property vector of the $(l+1)$-th consumer and the structure of his consumption, the Theorem \ref{gat22} takes the form
\begin{theorem}\label{gat24}
 Let technological maps $F_i(x),\ x \in X_i^1,$
$i=\overline {1,m},$ be convex down and belong to the  CTM class in a wide sense,
the set $R$ be non-empty, the structure of supply  agree with the structures of choice and foreign economic relations  on the set $R_1,$
 and let fields $y_i(p,z),\ i=\overline{1,l}, $ be continuous functions of $(p,z) \in R_1,$ there hold conditions of the Lemma \ref{opcd1} or the Lemma \ref{opod111}.

If  fields of information evaluation by consumers  $\eta_{i}^0(p,z),\  i=\overline{1,l},$ and a random field $\zeta_0(p,\omega_0)$ satisfy the  conditions of the  Theorem \ref{11tl4}, then the set of equations of the economy equilibrium
\begin{eqnarray*} \sum\limits
_{i=1}^{l} \frac{\mu_{ki}(p)D_i(p)}{\sum\limits_{j=1}^n
\mu_{ji}(p)p_j} \end{eqnarray*}
\begin{eqnarray} \label{gat25}
=  \sum\limits _{i=1}^m\left[Y_{ki}\left(p, z^0\right)-X_{ki}\left(p, z^0\right)+ d_{ki}\right]+ \sum\limits
_{i=1}^{l} b_{ik}+i_k^0 -e_k^0, \quad  k=\overline{1,n},
\end{eqnarray}
has a solution $p^0=\{p_i^0\}_{i=1}^n$ in the set of strictly positive  price vectors
for every $ z^0 \in M_0.$
Such economy system can operate profitably in the Walras equilibrium state\index{operate profitably in the Walras equilibrium state} under the condition $\left\langle y_i^0 -x_i^0, p^0\right\rangle  > 0, \ i=\overline{1,m},$ or subvention-profitably in the Walras equilibrium state\index{operate subvention-profitably in the Walras equilibrium state} if
\begin{eqnarray*} \left\langle y_i^0 -x_i^0+ d_i , p^0\right\rangle  > 0, \quad i=\overline{1,m},\end{eqnarray*}
for $b_i \neq 0.$
\end{theorem}
The division of consumers onto external and internal leads to the classification of economy equilibrium states.
If in the equilibrium state $ \left\langle p^0, i^0 -e^0 \right\rangle   > 0,$ then the economy system ticks, and if $ \left\langle p^0, i^0 - e^0 \right\rangle   < 0,$ then it credits other economies, and it is in equilibrium with environment if $ \left\langle p^0, i^0 - e^0 \right\rangle = 0.$
\section{Aggregated open economy system }

Consider an  open economy system\index{open economy system} containing $n$ net industries that are producers and consumers simultaneously and $l-n$ consumers $l > n.$ The economy system operates during certain time interval, e.g., a year.
Formulate the model that we apply to describe the economy of a  state in the aggregated form.

Let $A=||a_{ki}||_{k, i=1}^n$ be a non-negative productive matrix of direct production inputs, and $R=||r_{ki}||_{k, i=1}^n$ be a non-negative matrix of constant production inputs.

In the considered economy system, we describe the $i$-th industry production structure by the technological map
\begin{eqnarray*} F_i( x_i)=\{  y_i \in S,\    y_i= u_ie_i, \ a_{ki}(u_i)u_i \leq x_{ki}, \ k=\overline{1,n} \},\end{eqnarray*}
\begin{eqnarray*}    e_i=\{\delta_{ik}\}_{k=1}^n, \quad   x_i=\{x_{ki}\}_{k=1}^n \in X_i,\end{eqnarray*}
\begin{eqnarray*}  X_i=\{ x_i=\{x_{ki}\}_{k=1}^n  \in S, \ r_{ki} \leq x_{ki} \leq a_{ki}\left(u_i^0\right)u_i^0, \ k=\overline{1,n}\}, \quad i=\overline{1,n},\end{eqnarray*}
where $u_i^0$ is a maximum possible output of the $i$-th industry,  $i=\overline{1,n},$
 $\delta_{ik}$ is the Kronecker symbol,
\begin{eqnarray*} a_{ki}(u_i)=a_{ki}+ \frac{r_{ki}}{u_i}, \quad k,i=\overline{1,n}.\end{eqnarray*}
Let $l$ consumers have property vectors $b_i=\{b_{ki}\}_{k=1}^n, \ i=\overline{1,l}.$
We call the vector $e^0=\{e_i^0\}_{i=1}^n$  the export vector and the vector $i^0=\{i_k^0\}_{k=1}^n$  the import vector.

In this Section, we assume that the set of possible price vectors is the cone
\begin{eqnarray*} K_+^n=\{p=\{p_i\}_{i=1}^n \in R_+^n , \  p_i >0, \  i=\overline{1,n}\}.\end{eqnarray*}

For technological maps considered, the set of  possible productive processes is given by the formula
 \begin{eqnarray*} \Gamma^n=\prod\limits_{i=1}^n\Gamma_i, \quad
\Gamma_i=\{(  x_i,  y_i), \  x_i \in X_i, \  y_i \in F_i(  x_i)\}\cup (0,0), \quad 0 \in S,\end{eqnarray*}
\begin{eqnarray*}   z \in \Gamma^n, \quad z=\{z^i\}_{i=1}^n, \quad  z^i=( x_i,  y_i), \quad i=\overline{1,n}.\end{eqnarray*}

The map $Q(p,z)=\{Q_i(p,z))\}_{i=1}^n$ reflecting $\{\Gamma^n,{\cal B}(\Gamma^n)\}$
into itself for every fixed $p \in K_+^n$ whose components $Q_i(p,z)$ are given by the formula
\begin{eqnarray*} Q_i(p,z)=\prod\limits_{s=1}^n a_s(p,z)v_s(p,z) \bar Q_i(z),\end{eqnarray*}
where
\begin{eqnarray*} \bar Q_i(z)=\{\{a_{ki}(u_i)u_i\}_{k=1}^n, \   u_ie_i \},\end{eqnarray*}
\begin{eqnarray*} a_i(p,z)=\chi_{[0, \infty)}\left(p_i - \sum\limits_{s=1}^na_{si}(u_i)p_s\right), \end{eqnarray*}  \begin{eqnarray*}
v_i(p,z)=\chi_{[0, \infty)}\left(u_i - \sum\limits_{k=1}^na_{ik}(u_k)u_k + \sum\limits_{k=1}^lb_{ik}   +i_i^0 - c_i\right), \quad c_i \geq e_i^0, \quad i=\overline{1,n},\end{eqnarray*}
is a productive economic process.
  The vector $u=\{u_i\}_{i=1}^n$ is called  a possible vector of gross outputs  in the economy system.

Let us connect with  the possible vector of  gross outputs\index{possible vector of  gross outputs}  $x^0=\{x^0_i\}_{i=1}^n$  the productive process
\begin{eqnarray*} z^0=\{z_0^i\}_{i=1}^n, \quad  z_0^i= \{\{x^0_ia_{ki}\left(x_i^0\right)\}_{k=1}^n,\ x_i^0e_i\}.\end{eqnarray*}
In what follows, we describe the economy system by three matrices, namely, a matrix of initial goods supply\index{ matrix of initial goods supply} $B=||b_{ij}||_{i,j=1}^{n,\ l},$  a matrix of technology coefficients\index{matrix of technology coefficients}
 $A\left(x^0\right)=||a_{ij}\left(x_j^0\right)||_{i,j=1}^n,$ \ $ x^0 \in X \subseteq  R_+^n $
and an unproductive consumption matrix\index{unproductive consumption matrix}
$C=||c_{ij}||_{i,j=1}^{n,\ l}$ whose the $i$-th column $C_i=\{c_{ki}\}_{k=1}^n, \ i=\overline{1,l},$ is the  field  of information evaluation by the $i$-th consumer.\index{field  of information evaluation by  consumer}
Economic sense of
 $a_{ij}\left(x_j^0\right)$ is the number of units of the $i$-th goods needed to produce one unit of the $j$-th goods if the economy system produces the vector of goods
 $x^0=\{x_i^0\}_{k=1}^n,$  $c_{ij}$ is the number of units of the $i$-th goods of the  final consumption which the $j$-th consumer wants to buy in the economy operation period.

First $n$  fields  of information evaluation $C_i$,\ $i=\overline{1,n},$ determine the structure of unproductive consumption  for producers\index{structure of unproductive consumption  for producers} whose number we suppose to be $n.$ The rest $l-n$
consumers have consumption structure characterized by fields  of information evaluation
$C_i=\{c_{ki}\}_{k=1}^n$, $i=\overline{n+1,l}$.

 The presence  of a consumer in the economy system means that he consumes at least one kind of goods, therefore, we suppose that conditions
\begin{eqnarray} \label{allochka}
\sum\limits_{s=1}^n c_{si} > 0, \quad i=\overline{1,l},
\end{eqnarray}
 hold.
We suppose that in the economy system random  fields  of information evaluation by consumers\index{random  fields  of information evaluation by consumers}   on a probability space $\{\Omega, {\cal F}, \bar P\}$ have the form
\begin{eqnarray*} \eta_i^0(p,z, \omega_i)=\{\eta_{ik}^0(p,z, \omega_i)\}_{k=1}^n, \quad \eta_{ik}^0(p,z, \omega_i)=c_{ki}, \quad i=\overline{1,l}, \quad k=\overline{1,n},\end{eqnarray*}
and random field $\zeta_0(p, \omega_0)$ with probability 1 takes one value $\zeta_0(p, \omega_0)=z^0, \ p \in K_+^n, $ $ z^0   \in M_0,$
where, as earlier,
\begin{eqnarray*} \Omega= \prod\limits_{i=0}^l\Omega_i, \quad {\cal F}=\prod\limits_{i=0}^l{\cal F}_i, \quad \bar P=\prod\limits_{i=0}^l\bar P_i.\end{eqnarray*}

Determine conditions under which the economy system with $l$ consumers whose demand is described by non-random demand vectors
\begin{eqnarray*}  \gamma_i(p,z)=\left\{\frac{p_kc_{ki}}{\sum\limits_{s=1}^nc_{si}p_s}\right\}_{k=1}^n, \quad i=\overline{1,l},\end{eqnarray*}
can operate profitably.

 The set of equations of the economy equilibrium has the form
\begin{eqnarray} \label{vjant31} \sum\limits_{i=1}^{l}\frac{c_{ki}D_i(p)  }{\sum\limits_{s=1}^np_sc_{si}}
\end{eqnarray}
 \begin{eqnarray*} =x_k^0 -  \sum\limits_{s=1}^na_{ks}\left(x_s^0\right)x_s^0 +\sum\limits_{i=1}^{l}b_{ki} + i_k^0 - e_k^0, \quad  k=\overline{1,n}.\end{eqnarray*}
In this set of equations, all the consumers  are  already internal.
The division of consumers onto external and internal leads to the classification of economy equilibrium states.
If in the equilibrium state $ \left\langle p^0, i^0 -e^0 \right\rangle   > 0,$ then the economy system ticks, and if $ \left\langle p^0, i^0 - e^0 \right\rangle   < 0,$ then it credits other economies, and it is in equilibrium with environment if $ \left\langle p^0, i^0 - e^0 \right\rangle = 0.$
 \begin{definition}\label{vant3}
The structure of supply  agrees with the structures of choice and foreign economic relations\index{structure of supply  agrees with the structures of choice and foreign economic relations}  if for the vector of final consumption\index{vector of final consumption}
\begin{eqnarray*} \psi =\{\psi_k\}_{k=1}^n, \quad  \psi_k= x_k^0 -  \sum\limits_{s=1}^na_{ks}\left(x_s^0\right)x_s^0  +\sum\limits_{i=1}^lb_{ki} - e_k^0 + i_k^0, \quad  k=\overline{1,n},\end{eqnarray*}
corresponding to the strictly positive  vector of gross outputs $x^0=\{x_i^0\}_{k=1}^n,$
 the representation
\begin{eqnarray} \label{vjant26}
x_k^0 -  \sum\limits_{s=1}^na_{ks}\left(x_s^0\right)x_s^0  +\sum\limits_{i=1}^lb_{ki} - e_k^0 + i_k^0 =\sum\limits_{i=1}^l c_{ki}y_i^0,  \quad  k=\overline{1,n},
\end{eqnarray}
is valid,  where the vector $y_0=\{y_i^0\}_{i=1}^l$ has strictly positive components.
The  vector $y_0$ is called      the  vector of levels of satisfaction of consumers needs.\index{vector of levels of satisfaction of consumers needs}
\end{definition}

Introduce into  consideration matrices
\begin{eqnarray*} X_0=||\delta_{ij}x_i^0||_{i,j=1}^n, \quad  A\left(x^0\right)=||a_{ij}\left(x_j^0\right)||_{i,j=1}^n, \end{eqnarray*}  \begin{eqnarray*}  B=||b_{ij}||_{i=1,j=1}^{n,\ l}, \quad \bar B=||b_{ij}||_{i=1,j=1}^{n},  \quad \tilde  B=||b_{ij}||_{i=1,j=n+1}^{n,\ l},\end{eqnarray*}
\begin{eqnarray*}  C=||c_{ij}||_{i=1,j=1}^{n,\ l}, \quad \bar C=||c_{ij}||_{i=1,j=1}^{n},\quad \tilde  C=||c_{ij}||_{i=1,j=n+1}^{n,\ l},\end{eqnarray*}
where $\delta_{ij}$ is the Kronecker symbol, and vectors
\begin{eqnarray*}    y_0=\{y_i^0\}_{i=1}^l,  \quad \bar y_0=\{y_i^0\}_{i=1}^n, \quad \tilde y_0=\{y_i^0\}_{i=n+1}^l, \end{eqnarray*}
\begin{eqnarray*}  e=\{e_i\}_{i=1}^l, \quad \bar e=\{e_i\}_{i=1}^n, \quad \tilde e=\{e_i\}_{i=n+1}^l,\quad e_i=1, \quad i=\overline{1,l}.\end{eqnarray*}

Let $\pi=\{\pi_i\}_{i=1}^n$ be an  arbitrary strictly positive vector.

\begin{lemma}\label{vjant18}
Let conditions (\ref{allochka}) hold,  there  exist  a non-negative vector
$v_0=\{v_i\}_{i=1}^n, \ v_i \geq 0, \ i=\overline{1,n},$ such that $\bar C(v_0) - \bar B $ is a non-negative matrix having no zero rows or columns and
$ A\left(x^0\right) + \bar C(v_0) - \bar B$ is an indecomposable matrix.

If the spectral radius of the matrix\index{spectral radius of the matrix} $A\left(x^0\right)$ is less than 1,
 there exists a strictly positive vector
$g=\{g_i\}_{i=1}^n,$  $ g_i>0,$ $   i=\overline{1,n},$ such that components $x_i^0, \   i=\overline{1,n}, $ of the strictly positive  vector of gross outputs  $x^0=\{x_i^0\}_{i=1}^n$ satisfy the set of equations
\begin{eqnarray*} x_i^0=\frac{\left[\left(E-A\left(x^0\right)\right)^{-1}\left[\bar C(\pi^{-1}\bar y)- \bar B\right]g\right]_i}{
g_i}, \ \quad i=\overline{1,n},\end{eqnarray*}
where
\begin{eqnarray*} \bar C(v_0)=||c_{si}v_i||_{s,i=1}^{n}, \quad \bar C(\pi^{-1}\bar y)=\left|\left|c_{si}\frac{y_i}{\pi_i}\right|\right|_{s,i=1}^{n},\end{eqnarray*}
and $ \pi=\{\pi_i\}_{i=1}^n, \  \pi_i >0, \  i=\overline{1,n},$ is a certain  strictly positive vector satisfying inequalities $  y_i > \pi_i v_i, \  i=\overline{1,n}, $ $\bar y=\{y_i\}_{i=1}^n, $
then the spectral radius of the matrix $A\left(x^0\right)+ [\bar C(v_0) - \bar B]X_0^{-1}$ is strictly less than 1 and the set
\begin{eqnarray*} T_0=\left\{ p \in K_+^n, \
p_i-\sum\limits_{s=1}^n\left[a_{si}\left(x_i^0\right) + \frac{1}{x_i^0}(c_{si}v_i - b_{si})\right]p_s > 0, \ i=\overline{1,n} \ \right\}\end{eqnarray*}
is non-empty.
\end{lemma}
The Proof of this Lemma is the same as that of the Lemma \ref{jant18}.

Introduce the set
\begin{eqnarray*} T=\{(p,z) \in K_+^n\times \Gamma^n, \ \prod\limits_{s=1}^n a_s(p,z)v_s(p,z)=1\}.\end{eqnarray*}
\begin{lemma}\label{gat5}
If the conditions of the  Lemma \ref{vjant18}  hold, then the set $T$ is non-empty and contains the set $T_0\times z^0.$
\end{lemma}

\begin{theorem}\label{vjant12}
Let the structure  of supply agrees with the  structures of choice and foreign economic relations,\index{structure  of supply agrees with the  structures of choice and foreign economic relations}
conditions  (\ref{allochka}) hold,
a non-negative vector
$v_0=\{v_i\}_{i=1}^n, \ v_i \geq 0, \ i=\overline{1,n},$  exist  such that $\bar C(v_0) - \bar B $ is a non-negative matrix having no zero rows or columns and
$ A\left(x^0\right) + \bar C(v_0) - \bar B$ is an indecomposable matrix, and let the spectral radius of the matrix $A\left(x^0\right)$ be less than 1.
If there exists a strictly positive vector
$g=\{g_i\}_{i=1}^n,$  $ g_i>0,$ $   i=\overline{1,n},$ such  that components $x_i^0$ of the possible  vector of gross outputs $x^0=\{x_i^0\}_{i=1}^n$ satisfy the set of equations
\begin{eqnarray*} x_i^0=\frac{\left[\left(E-A\left(x^0\right)\right)^{-1}\left[\bar C(\pi^{-1}\bar y_0)- \bar B\right]g\right]_i}{
g_i}, \ \quad i=\overline{1,n},\end{eqnarray*}
where
\begin{eqnarray*} \bar C(v_0)=||c_{si}v_i||_{s,i=1}^{n}, \quad \bar C(\pi^{-1}\bar y_0)=\left|\left|c_{si}\frac{y_i^0}{\pi_i}\right|\right|_{s,i=1}^{n},\end{eqnarray*}
and $ \pi=\{\pi_i\}_{i=1}^n, \  \pi_i >0, \  i=\overline{1,n},$ is a certain positive vector satisfying inequalities $ y_i^0 > \pi_i v_i, \  i=\overline{1,n}, $ $\bar y_0=\{y_i^0\}_{i=1}^n, $
 then in the set $T_0$ there exists a strictly positive solution to the equation
\begin{eqnarray} \label{vjant100}
 \sum\limits_{i=1}^n\pi_i\left(p,z^0\right)\left[x_i^0\left(p_i- \sum\limits_{s=1}^na_{si}p_s \right)+ \sum\limits_{s=1}^nb_{si}p_s\right]=\sum\limits_{i=1}^n y_i^0\sum\limits_{s=1}^nc_{si}p_s
\end{eqnarray}
with respect to the vector $\pi\left(p,z^0\right)=\{  \pi_i \left(p, z^0\right)\}_{i=1}^n,$
whose components are given by the formula
\begin{eqnarray*}    \pi_i\left(p, z^0\right)=\pi_i\pi\left(p, z^0\right),
\end{eqnarray*}
\begin{eqnarray*}
  \pi\left(p, z^0\right)=\frac{\sum\limits_{i=1}^n y_i^0\sum\limits_{s=1}^nc_{si}p_s}{\sum\limits_{i=1}^n\pi_i\left[x_i^0\left(p_i- \sum\limits_{s=1}^na_{si}\left(x_i^0\right)p_s\right) + \sum\limits_{s=1}^nb_{si}p_s\right]}.\end{eqnarray*}
\end{theorem}

On the set $T_0\times z^0$, consider two sets of functions
 \begin{eqnarray*} f_i\left(p,z^0\right)=\pi_i\left(p,z^0\right)\left[x_i^0\left(p_i - \sum\limits_{s=1}^na_{si}\left(x_i^0\right)p_s\right)+ \left\langle b_i, p \right\rangle \right],  \quad i=\overline{1,n},\end{eqnarray*}
\begin{eqnarray} \label{gat2}
f_i\left(p,z^0\right)=y_i^0\sum\limits_{k=1}^n c_{ki}p_k,  \quad i=\overline{n+1,l},
\end{eqnarray}
\begin{eqnarray*}  g_i\left(p,z^0\right)=x_i^0\left(p_i - \sum\limits_{s=1}^na_{si}\left(x_i^0\right)p_s\right)+ \left\langle b_i, p\right\rangle,  \quad i=\overline{1,n},\end{eqnarray*}
\begin{eqnarray} \label{gat1}
g_i\left(p,z^0\right)= \left\langle b_i,p\right\rangle, \quad i=\overline{n+1,l},
\end{eqnarray}
where $\pi\left(p,z^0\right)=\{  \pi_i \left(p, z^0\right)\}_{i=1}^n$ is a solution to the equation (\ref{vjant100}) built in the Theorem \ref{vjant12}.

\begin{theorem}\label{vjant22}
If the  conditions of the Lemma \ref{vjant18}, the Theorem \ref{vjant12}, and also  conditions
 (\ref{allochka}) hold, then on the set
$T_0\times z^0$ there exists a taxation system $||\pi_{ij}^0\left(p,z^0\right)||_{i,j=1}^l$ for which
\begin{eqnarray} \label{vjant23}
f_i\left(p,z^0\right)=\sum\limits_{j=1}^l\pi_{ij}^0\left(p,z^0\right)g_j\left(p,z^0\right), \quad i=\overline{1,l}.
\end{eqnarray}
\end{theorem}

Apply the Theorem \ref{vjant22}
and construct on the set $ K_+^n \times \Gamma^n$ a taxation system
$||\pi_{ij}(p,z)||_{i,j=1}^l$
which on the set $T_0 \times z^0$ we determine by the Theorem \ref{vjant22} and on the set $K_+^n \times \Gamma^n \setminus  T_0\times z^0$  we take it arbitrary if only equalities
\begin{eqnarray} \label{vgint1}
\sum\limits_{i=1}^l\pi_{ij}(p,z)=1,  \quad (p,z) \in K_+^n \times \Gamma^n  \setminus  T_0\times z^0, \quad j =\overline{1,l},
\end{eqnarray}
hold. We give a family of  income pre-functions by formulae
\begin{eqnarray} \label{vgint2}
K_i^0(p, z)=\sum\limits_{j=1}^l\pi_{ij}(p,z)g_j(p,z),   \quad (p,z) \in K_+^n \times \Gamma^n, \quad i=\overline{1,l},
\end{eqnarray}
where
\begin{eqnarray*} g_i(p,z)=[\left\langle y_i -  x_i, p \right\rangle + \left\langle b_i,p \right\rangle ],   \quad (p,z) \in K_+^n \times \Gamma^n, \quad i=\overline{1,n},\end{eqnarray*}
\begin{eqnarray*}  g_i(p,z)=<b_i,p>,  \quad (p,z) \in K_+^n \times \Gamma^n, \quad  i=\overline{n+1,l},\end{eqnarray*}
and define a family of  income functions  by the formula
\begin{eqnarray*}  K_i(p,z)= K^0_i(p,Q(p,z)),  \quad (p,z) \in K_+^n \times \Gamma^n, \quad i=\overline{1,l}.\end{eqnarray*}

The reduction on the set $T_0\times z^0$ of the considered family of income functions  takes the form
\begin{eqnarray*}  K_i\left(p, z^0\right)=\pi_i\left(p,z^0\right)\left[x_i^0\left(p_i - \sum\limits_{s=1}^na_{si}\left(x_i^0\right)p_s\right) + \left\langle b_i, p\right\rangle \right],\quad p \in T_0, \quad i=\overline{1,n}, \end{eqnarray*}
\begin{eqnarray} \label{vgint3}
K_i\left(p, z^0\right)=y_i^0\sum\limits_{k=1}^nc_{ki}p_k,\quad p \in T_0,  \quad  i=\overline{n+1,l}.
\end{eqnarray}
On the realization of random field $\zeta_0(p, \omega_0)=z^0,$  functions of the  net income have the form
\begin{eqnarray*}  D_i(p)=\pi_i\left(p,z^0\right)\left[x_i^0\left(p_i - \sum\limits_{s=1}^na_{si}\left(x_i^0\right)p_s\right)+ \left\langle b_i, p\right\rangle \right],   \quad p \in T_0, \quad i=\overline{1,n},\end{eqnarray*}
\begin{eqnarray} \label{vgint4}
 D_i(p)=y_i^0\sum\limits_{k=1}^nc_{ki}p_k, \quad p \in T_0, \quad  i=\overline{n+1,l}.
\end{eqnarray}

Consider the case when the random field $\zeta_0(p, \omega_0)$ takes a single value in the set $\Gamma^n$ with probability 1 and does not depend on $p.$

Suppose
\begin{eqnarray*}  \zeta_0(p, \omega_0)=\{\zeta_0^i(p)\}_{i=1}^n, \end{eqnarray*}  \begin{eqnarray*}  \zeta_0^i(p)=\{\{a_{ki}\left(x_i^0\right)x_i^0\}_{k=1}^n, \   x_i^0e_i\}, \quad i=\overline{1,n},\end{eqnarray*}
where the output vector $x^0$ satisfies  conditions of the    Lemma \ref{vjant18}, the Theorem \ref{vjant12} and \ref{vjant22}.
Then
\begin{eqnarray*} \zeta(p)=Q(p,\zeta_0(p, \omega_0))=\{\zeta_i(p)\}_{i=1}^n, \end{eqnarray*}  \begin{eqnarray*}  \zeta_i(p)=\prod\limits_{s=1}^n a_s(p,\zeta_0(p, \omega_0)) v_s(p,\zeta_0(p, \omega_0))\{\{ a_{ki}\left(x_i^0\right)x_i^0 \}_{k=1}^n, \ x_i^0e_i\} \end{eqnarray*}  \begin{eqnarray*} =\{\zeta_i^{1}(p), \zeta_i^{2}(p)\},\end{eqnarray*}
where
\begin{eqnarray*} v_i(p,\zeta_0(p, \omega_0))=\chi_{[0, \infty)}\left(x_i^0 - \sum\limits_{k=1}^na_{ik}\left(x_k^0\right)x_k^0 + \sum\limits_{k=1}^lb_{ik} + i_i^0 - c_i\right),\end{eqnarray*}
\begin{eqnarray*} a_i(p,\zeta_0(p, \omega_0))=\chi_{[0, \infty)}\left(p_i - \sum\limits_{s=1}^na_{si}\left(x_i^0\right)p_s\right). \end{eqnarray*}
Under these conditions,
\begin{eqnarray*} D_i(p)= K_i^0(p,\zeta(p))=\pi_i(p,\zeta(p))[ \left\langle p, \zeta_i^{2}(p) - \zeta_i^{1}(p) \right\rangle  + \left\langle p,b_i\right\rangle ] \end{eqnarray*}
 \begin{eqnarray*} =\pi_i(p,\zeta(p))\prod\limits_{s=1}^n a_s(p, \zeta_0(p, \omega_0))  v_s(p,\zeta_0(p, \omega_0))x_i^0\left[p_i - \sum\limits_{s=1}^na_{si}\left(x_i^0\right)p_s\right] \end{eqnarray*}
\begin{eqnarray*} +\pi_i(p,\zeta(p)) \left\langle p,b_i\right\rangle, \quad  i=\overline{1,n},\end{eqnarray*}
\begin{eqnarray*} D_i(p)= K_i^0(p,\zeta(p)) \end{eqnarray*}
 \begin{eqnarray*} = \sum\limits_{j=1}^n\pi_{ij}(p,\zeta(p)) \prod\limits_{s=1}^n a_s(p,\zeta_0(p, \omega_0))  v_s(p,\zeta_0(p, \omega_0))x_j^0\left[p_j - \sum\limits_{s=1}^na_{sj}\left(x_j^0\right)p_s\right]  \end{eqnarray*}  \begin{eqnarray*} + \sum\limits_{j=1}^n\pi_{ij}(p,\zeta(p)) <p,b_j>+ \sum\limits_{j=n+1}^l\pi_{ij}(p, \zeta(p))\left\langle p,b_j\right\rangle, \quad i=\overline{n+1,l},\end{eqnarray*}
\begin{eqnarray*}  \zeta_i^{2}(p)= \prod\limits_{s=1}^n a_s(p, \zeta_0(p, \omega_0))  v_s(p,\zeta_0(p, \omega_0))
x_i^0 e_i,\end{eqnarray*}
\begin{eqnarray*} \zeta_i^{1}(p)=\prod\limits_{s=1}^n a_s(p, \zeta_0(p, \omega_0)) v_s(p,\zeta_0(p, \omega_0))
\{a_{ki}\left(x_i^0\right)x_i^0\}_{k=1}^n.\end{eqnarray*}
The set of equations of the economy equilibrium turns into the form
\begin{eqnarray*} \sum\limits_{i=1}^l\frac{p_kc_{ki}D_i(p)}{\sum\limits_{s=1}^np_sc_{si}} \end{eqnarray*}  \begin{eqnarray*} =
p_k\left[d(p)\left(x_k^0 -  \sum\limits_{s=1}^na_{ks}\left(x_s^0\right)x_s^0\right) +\sum\limits_{i=1}^lb_{ki} - e_k^0 +i_k^0\right], \quad  k=\overline{1,n}, \end{eqnarray*}
\begin{eqnarray*} D_i(p)=\pi_i(p,\zeta(p))\left[d(p)x_i^0\left(p_i - \sum\limits_{s=1}^na_{si}\left(x_i^0\right)p_s\right) + \left\langle p,b_i\right\rangle\right], \quad  i=\overline{1,n},\end{eqnarray*}
\begin{eqnarray*} D_i(p)=\sum\limits_{j=1}^n\pi_{ij}(p,\zeta(p))\left[d(p)x_j^0\left(p_j - \sum\limits_{s=1}^na_{sj}\left(x_j^0\right)p_s\right) + \left\langle p,b_j\right\rangle\right] \end{eqnarray*}  \begin{eqnarray*} + \sum\limits_{j=1}^n\pi_{ij}(p,\zeta(p))\left\langle p,b_j\right\rangle, \quad  i=\overline{n+1,l},\end{eqnarray*}
where
\begin{eqnarray*}  d(p)=\prod\limits_{s=1}^n a_s(p, \zeta_0(p, \omega_0))  v_s(p,\zeta_0(p, \omega_0)).\end{eqnarray*}

The reduction onto the set $T_0\times z^0$ of the  above set of equations of the economy equilibrium  takes the form
\begin{eqnarray*} \sum\limits_{i=1}^l\frac{p_kc_{ki}D_i(p)}{\sum\limits_{s=1}^np_sc_{si}} \end{eqnarray*}  \begin{eqnarray*} =
p_k\left[x_k^0 -  \sum\limits_{s=1}^na_{ks}\left(x_s^0\right)x_s^0 +\sum\limits_{i=1}^lb_{ki} - e_k^0 +i_k^0\right], \quad  k=\overline{1,n},\end{eqnarray*}
\begin{eqnarray*} p_i - \sum\limits_{s=1}^na_{si}\left(x_i^0\right)p_s \geq 0, \quad  i=\overline{1,n},\end{eqnarray*}
\begin{eqnarray*} x_k^0 -  \sum\limits_{s=1}^na_{ks}\left(x_s^0\right)x_s^0 +\sum\limits_{i=1}^lb_{ki} - e_k^0 +i_k^0 \geq 0, \quad k=\overline{1,n},\end{eqnarray*}
\begin{eqnarray*} D_i(p)=\pi_i\left(p, z^0\right)\left[x_i^0\left(p_i - \sum\limits_{s=1}^na_{si}\left(x_i^0\right)p_s\right) + \left\langle p,b_i\right\rangle \right], \quad  i=\overline{1,n}, \end{eqnarray*}
\begin{eqnarray*} D_i(p)=\sum\limits_{j=1}^n\pi_{ij}\left(p, z^0\right)\left[x_j^0\left(p_j - \sum\limits_{s=1}^na_{sj}\left(x_j^0\right)p_s\right) + \left\langle p, b_j\right\rangle \right] \end{eqnarray*}  \begin{eqnarray*} + \sum\limits_{j=n+1}^l\pi_{ij}\left(p, z^0\right)\left\langle p,b_j\right\rangle =y_i^0\sum\limits_{s=1}^nc_{si}p_s , \quad  i=\overline{n+1,l}.\end{eqnarray*}

\begin{theorem}\label{vjant14}
 Let conditions (\ref{allochka}) hold, there   exist a non-negative vector
$v_0=\{v_i\}_{i=1}^n, \ v_i \geq 0, \ i=\overline{1,n},$  such that $\bar C(v_0) - \bar B $ is a non-negative matrix having no zero rows or columns and
$ A\left(x^0\right) + \bar C(v_0) - \bar B$ is an  indecomposable matrix, and let the vector $x^0=\{x_i^0\}_{i=1}^n$ have strictly positive components, the spectral radius of the matrix $A\left(x^0\right)$ be less than 1.

For the known strictly positive vector $\pi=\{\pi_i\}_{i=1}^n,\  \pi_i > 0, \ i= \overline{1,n}, $ and the strictly positive vector
$\bar y=\{y_i\}_{i=1}^n$ whose components satisfy inequalities $  \ y_i  > \pi_iv_i, \ i=\overline{1,n}, $
there exists a solution to the set of equations
\begin{eqnarray} \label{vjant15}
p_i= \sum\limits_{s=1}^na_{si}\left(x_i^0\right)p_s +  \frac{1}{x_i^0}\sum\limits_{s=1}^n\left[c_{si} \frac{y_i}{\pi_i} - b_{si}\right]p_s, \quad   i=\overline{1,n},
\end{eqnarray}
belonging to the set $T_0$ if and only if when the  components of the vector $x^0=\{x_i^0\}_{i=1}^n$ satisfy the set of equations
 \begin{eqnarray*} x_i^0=\frac{\left[\left(E-A\left(x^0\right)\right)^{-1}\left[\bar C(\pi^{-1}\bar y)-  \bar B\right]g\right]_i}{
g_i}, \ \quad i=\overline{1,n},\end{eqnarray*}
for a certain strictly positive vector $ g=\{g_i\}_{i=1}^n, \ g_i>0, \   i=\overline{1,n},$
where
\begin{eqnarray*} \bar C(\pi^{-1}\bar y)=\left|\left|c_{si}\frac{y_i}{\pi_i}\right|\right|_{s,i=1}^{n}.
\end{eqnarray*}
\end{theorem}
The Theorem \ref{vjant14} is similar to the Theorem \ref{jant14}. Therefore, it has the same Proof as the Theorem
\ref{jant14}.

In the next Theorem, we suppose that the net income of consumers  corresponding to the realization of random field $\zeta_0(p, \omega_0)=z^0$ with probability 1 on the set $T_0$ is given by formulae (\ref{vgint4}).
\begin{theorem}\label{vjant25}
Let the structure of  supply  agree with the  structures of  choice and foreign economic relations,
conditions (\ref{allochka}) hold,   a non-negative vector
$v_0=\{v_i\}_{i=1}^n, \ v_i \geq 0, \ i=\overline{1,n},$ exist such that $\bar C(v_0) - \bar B $ is a non-negative matrix having no zero rows or columns and
$ A\left(x^0\right) + \bar C(v_0) - \bar B$ is an indecomposable matrix.\index{indecomposable matrix}
If the spectral radius of the matrix\index{spectral radius of the matrix} $A\left(x^0\right)$ is less than 1, there exists a strictly positive vector
$g=\{g_i\}_{i=1}^n,$ \ $ g_i>0,$ $ \  i=\overline{1,n},$ such that components
$x_i^0, \  i=\overline{1,n},$ of the vector $x^0=\{x_i^0\}_{i=1}^n$ satisfy the set of equations
\begin{eqnarray*} x_i^0=\frac{\left[\left(E-A\left(x^0\right)\right)^{-1}\left[\bar C(\pi^{-1}\bar y_0)- \bar B\right]g\right]_i}{
g_i}, \ \quad i=\overline{1,n},\end{eqnarray*}
where
\begin{eqnarray*} \bar C(v_0)=||c_{si}v_i||_{s,i=1}^{n}, \quad \bar C(\pi^{-1}\bar y_0)=\left|\left|c_{si}\frac{y_i^0}{\pi_i}\right|\right|_{s,i=1}^{n}, \quad  y_i^0 >  \pi_i v_i, \quad  i=\overline{1,n}, \end{eqnarray*}
and $ \pi=\{\pi_i\}_{i=1}^n, \  \pi_i >0, \  i=\overline{1,n},$ is a certain  strictly positive vector, then there exists a strictly positive price vector $p_0=\{p_i^0\}_{i=1}^n \in T_0 $ being a solution to  the set of equations
\begin{eqnarray} \label{vjant28}
\pi_i\left[x_i^0\left(p_i- \sum\limits_{s=1}^na_{si}\left(x_i^0\right)p_s\right) + \sum\limits_{s=1}^nb_{si}p_s\right]= y_i^0\sum\limits_{s=1}^nc_{si}p_s, \quad   i=\overline{1,n}.
\end{eqnarray}
This price vector is a solution to the set of equations
\begin{eqnarray} \label{vjant47} \sum\limits_{i=1}^l\frac{c_{ki}D_i(p)  }{\sum\limits_{s=1}^np_sc_{si}}=x_k^0 -  \sum\limits_{s=1}^na_{ks}\left(x_s^0\right)x_s^0 +\sum\limits_{i=1}^lb_{ki} -e_k^0 + i_k^0, \quad  k=\overline{1,n},
\end{eqnarray}
and, therefore, is an equilibrium price vector under which every industry is profitable,  net income of  productive industries in the economy equilibrium state\index{net income of  productive industries in the economy equilibrium state} is given by formulae
\begin{eqnarray} \label{vjant29}
D_i(p_0)=\pi_i\left[x_i^0\left(p_i^0- \sum\limits_{s=1}^na_{si}\left(x_i^0\right)p_s^0\right) + \sum\limits_{s=1}^nb_{si}p_s^0\right], \quad   i=\overline{1,n},
\end{eqnarray}
and net income of the $i$-th consumer  is given by the formula
\begin{eqnarray} \label{seq6}
D_i(p_0)=y_i^0\sum\limits_{s=1}^np_s^0c_{si},
 \quad   i=\overline{n+1, l}.
\end{eqnarray}

\end{theorem}

\section{Economy systems transformations}

To have the possibility to apply results of the  previous Section  to the theory of economic transformations, one must look at the results of the formulated theorems  from another point of view having done somewhat different accents.

 As earlier, we describe the economy system by three matrices, namely, a matrix of initial goods supply\index{ matrix of initial goods supply}  $B=||b_{ij}||_{i,j=1}^{n,\ l},$ a  matrix of technological coefficients\index{matrix of technological coefficients}
 $A(x)=||a_{ij}(x_j)||_{i,j=1}^n,$ \ $ x \in X \subseteq  R_+^n $
and a matrix of unproductive consumption\index{matrix of unproductive consumption}
$C=||c_{ij}||_{i,j=1}^{n,\ l}$ which have the same, as before,  economic sense.

Further, we suppose that the considered economy system is open to external environment, exchanges goods, and is in the economy equilibrium with it.

In these terms, the set of equations of the economy equilibrium takes the form
\begin{eqnarray}\label{gat13} \sum\limits_{i=1}^{l}\frac{c_{ki}D_i(p)  }{\sum\limits_{s=1}^np_sc_{si}}
\end{eqnarray}
 \begin{eqnarray*}=x_k^0 -  \sum\limits_{s=1}^na_{ks}\left(x_s^0\right)x_s^0 +\sum\limits_{i=1}^{l}b_{ki} + i_k^0 - e_k^0, \quad  k=\overline{1,n}.\end{eqnarray*}
where the vector $e^0=\{e_i^0\}_{i=1}^n$ is the export vector and the vector $i^0=\{i_k^0\}_{k=1}^n$ is the import vector.

We consider the agreement of the structure of  supply  with the structures of choice and foreign economic relations\index{agreement of the structure of  supply  with the structures of choice and foreign economic relations}  as the set of equations for the strictly positive vector of gross outputs\index{vector of gross outputs} $x^0=\{x_i^0\}_{k=1}^n.$
Therefore, we suppose that the strictly positive  vector of  gross outputs\index{strictly positive  vector of  gross outputs} $x^0=\{x_i^0\}_{k=1}^n$ is a solution to  the set of equations
\begin{eqnarray}\label{gat8}
x_k^0 -  \sum\limits_{s=1}^na_{ks}\left(x_s^0\right)x_s^0 +\sum\limits_{i=1}^lb_{ki} - e_k^0 + i_k^0 =\sum\limits_{i=1}^l c_{ki}y_i,  \quad  k=\overline{1,n},
\end{eqnarray}
where the vector $y=\{y_i\}_{i=1}^l$ has strictly positive components. In what follows, we suppose  that the vector of levels of satisfaction of  consumers needs\index{vector of levels of satisfaction of  consumers needs}  $y=\{y_i\}_{i=1}^l$ is known and  just it  determines levels of consumption of  consumers in the economy system\index{levels of consumption of  consumers in the economy system}   and a strictly positive vector of gross outputs  $x^0=\{x_i^0\}_{k=1}^n $ guarantees these consumption levels and is a solution to the set of equations (\ref{gat8}).

\begin{definition}\label{1gat13}
Let there  exist  a vector
$v_0=\{v_i\}_{i=1}^n, \ v_i \geq 0, \ i=\overline{1,n},$  such that $\bar C(v_0) - \bar B $ is a non-negative matrix having no zero  rows or columns and
$ A\left(x^0\right) + \bar C(v_0) - \bar B$ is an  indecomposable matrix,\index{indecomposable matrix} and let  the spectral radius of the matrix $A\left(x^0\right)$ be less than 1.
 In the economy system, a taxation vector $\pi=\{\pi_i\}_{i=1}^n$ agrees with the structure of consumption  given by a vector of levels of satisfaction of  consumers needs\index{ taxation vector  agrees with the structure of consumption  given by a vector of levels of satisfaction of  consumers needs} $y=\{y_i\}_{i=1}^l$ if the  inequalities
$ y_i >  \pi_i v_i, \  i=\overline{1,n}, $
hold and there  exists a strictly positive vector
$g=\{g_i\}_{i=1}^n,$  $ g_i>0,$ $   i=\overline{1,n},$ such that components $g_i$ of the vector $g$ satisfy the set of equations
\begin{eqnarray*}g_i=\frac{\left[\left(E-A\left(x^0\right)\right)^{-1}\left[\bar C(\pi^{-1}\bar y)- \bar B\right]g\right]_i}{x_i^0
},  \quad i=\overline{1,n},  \quad \bar y=\{y_i\}_{i=1}^n,\end{eqnarray*}
where
\begin{eqnarray*}\bar C(v_0)=||c_{si}v_i||_{s,i=1}^{n}, \quad \bar C(\pi^{-1}\bar y)=\left|\left|c_{si}\frac{y_i}{\pi_i}\right|\right|_{s,i=1}^{n},\end{eqnarray*}
and  $\bar y=\{y_i\}_{i=1}^n.$
\end{definition}
Under these assumptions, we can reformulate the Theorem \ref{vjant25} in the form needed to further applications.
\begin{theorem}\label{gat9}
Let a strictly positive  vector of  gross outputs $x^0=\{x_i^0\}_{k=1}^n$ be a solution to the set of equations (\ref{gat8}),
conditions (\ref{allochka}) hold,
the  taxation vector $\pi=\{\pi_i\}_{i=1}^n$ agrees with the   structure of  consumption in the  economy system.

Then there exists a strictly positive price vector $p_0=\{p_i^0\}_{i=1}^n $ being a solution to  the set of equations
\begin{eqnarray*}\pi_i\left[x_i^0\left(p_i- \sum\limits_{s=1}^na_{si}\left(x_i^0\right)p_s\right) + \sum\limits_{s=1}^nb_{si}p_s\right]
\end{eqnarray*}
\begin{eqnarray}\label{gat12}
 =y_i\sum\limits_{s=1}^nc_{si}p_s, \quad   i=\overline{1,n}.
\end{eqnarray}
This price vector $p_0=\{p_i^0\}_{i=1}^n $ is a solution to  the set of equations
\begin{eqnarray}\label{gat10} \sum\limits_{i=1}^l\frac{c_{ki}D_i(p)  }{\sum\limits_{s=1}^np_sc_{si}}
\end{eqnarray}
 \begin{eqnarray*}=x_k^0 -  \sum\limits_{s=1}^na_{ks}\left(x_s^0\right)x_s^0 +\sum\limits_{i=1}^lb_{ki} - e_k^0 + i_k^0, \quad  k=\overline{1,n},\end{eqnarray*}
 therefore, it is an equilibrium price vector with which every industry is profitable in the  state of the  economy equilibrium, where the net income of  industry\index{net income of  industry} is given by the formula
\begin{eqnarray}\label{11}
D_i(p_0)=\pi_i\left[x_i^0\left(p_i^0- \sum\limits_{s=1}^na_{si}\left(x_i^0\right)p_s^0\right) + \sum\limits_{s=1}^nb_{si}p_s^0\right], \quad   i=\overline{1,n},
\end{eqnarray}
and the net profit of the  $i$-th consumer\index{ net profit of consumer}  is given by the formula
\begin{eqnarray}\label{gat6}
D_i(p_0)=y_i\sum\limits_{s=1}^np_s^0c_{si},
 \quad   i=\overline{n+1, l}.
\end{eqnarray}
\end{theorem}
Further, we describe the  state  of the  economy system in terms of the set of equations (\ref{gat8}), (\ref{gat12}), and  functions  of the net profit of the $i$-th
consumer $D_i(p), \ 1 \leq i \leq  l,$ from the Theorem \ref{gat9}.

\subsection{Transformations conserving equilibrium price \\vector}

We proceed from globalization principle of economy systems\index{globalization principle of economy systems}  or openness principle\index{ openness principle} for every economy system with respect to its foreign economic environment. According to this principle, the openness means that the economy system exchanges goods, capital, and labour force with its environment. Such exchange equalizes prices for goods and services  in the economy system and its environment. There  exist no  economy systems identical in  level of  development. They all differ one from another by natural sources,\index{natural sources} by levels of development of  productive forces,\index{levels of development of  productive forces} by  the structure of economy industries.
To integrate itself into the world economy under existing  difference in levels of development of   productive forces   inside and outside, the economy system must have different from environment  levels of  consumption  and taxation.
Just these differences guarantee the economy system openness to its environment. Mathematical formulation of the openness principle is that the  economy system can not influence, at this stage of  development,  on levels of  prices  for  goods and services, therefore, it accept them. Therefore, we suppose that in the considered economy system equilibrium price vector guarantees open operation of the  economy system\index{open operation of the  economy system}  with respect to its environment. Transitions from one level of  development  to another one is accompanied by changes of such characteristics of the   economy system  as level of  development of productive forces,\index{level of development of productive forces}   level of  consumption,\index{level of  consumption} levels of  gross outputs,\index{levels of  gross outputs} levels of taxation, and custom restrictions\index{custom restrictions} and export and import structures. The most significant among these changes are those leading to economic growth related to growing the  part of the value added  in the industry output.\index{ part of the value added  in the industry output} The latter  relates to new technologies introduced.
Another significant change is  change of levels of taxation\index{change of levels of taxation}  leading to equal right non-discriminative integral taxation.\index{equal right non-discriminative integral taxation} In reality, simultaneously there happen changes of  structures  of production and consumption and  levels of taxation. Therefore, it is important to consider also the changes that simultaneously change structures of production and consumption  as well as  levels of taxation. These changes must happen  such  that equilibrium price vector remains invariant  with respect to these changes in a certain period of the economy operation and every industry has the profit.

Therefore, in this Section we construct three types of transformations for characteristics of the considered economy model, namely, those improving production technologies and consumption levels; those changing taxation system; those simultaneously improving production technologies and consumption levels and changing  levels of  taxation.

Consider an open economy system $E_1$ containing
$n$ net industries and one social group of consumers. We describe the structure of production in the economy system   by a productive input matrix\index{productive input matrix}
\begin{eqnarray*} A_1(x_1)=||a_{ki}^1\left(x_i^1\right)||_{k,i=1}^{n}, \quad a_{ki}^1\left(x_i^1\right)=a_{ki}^1+ \frac{r_{ki}^1}{x_i^1},\quad x_1=\{x_i^1\}_{i=1}^n,\end{eqnarray*}
where $R_1=|r_{ki}^1 |_{k,i=1}^n$ is a non-negative  matrix of  constant inputs of production,\index{matrix of  constant inputs of production}
$A_{1}=||a^{1}_{ki}||_{k,i=1}^{n}$ is a non-negative productive matrix of direct inputs of production.\index{ productive matrix of direct inputs of production}
  We describe the structure of unproductive consumption\index{ structure of unproductive consumption}  by a non-negative matrix of unproductive  consumption\index{ matrix of unproductive  consumption}  $C_1=||c_{ki}^1||_{k,i=1}^{n,n+1}$ whose columns are fields  of information evaluation by consumers  and by non-negative matrix $B_1=||b_{ki}^1||_{k,i=1}^{n,n+1}$ built after property vectors of $(n+1)$ consumers.

 Suppose conditions
\begin{eqnarray}\label{1alla1}
 \sum\limits_{k=1}^n c_{ki}^1>0, \quad  i=\overline{1, n+1},
\end{eqnarray}
 hold.

 Consider only those vectors of gross outputs  $x_1$ for which the matrix $ A_1(x_1)$ is productive.
The vector of  gross outputs\index{vector of  gross outputs }
$x_1=\{x_i^1\}_{i=1}^{n}$
 corresponding to the given consumption levels in the economy system $E_1$
is a strictly positive solution to the set of equations
\begin{eqnarray}\label{bl20}
x_i^1 - \sum\limits_{k=1}^{n} a_{ik}^1\left(x_k^1\right)x_k^1 - e_i^1 +i_i^1 +\sum\limits_{k=1}^{n+1} b_{ik}^1=
 \sum\limits_{k=1}^{n+1} c_{ik}^1y_k^1,
\quad  i=\overline{1,n},
\end{eqnarray}
where
$e_1=\{e_i^1\}_{i=1}^{n},
\ i_1=\{i_k^1\}_{k=1}^{n} $ are export and import vectors in natural parameters,
$y_1=\{y_k^1\}_{k=1}^{n+1} $ is a  vector of levels of satisfaction of consumers needs.\index{vector of levels of satisfaction of consumers needs}

Assume a taxation vector $\pi_1=\{\pi_i^1\}_{i=1}^{n}$ agrees with the structure of  consumption\index{taxation vector  agrees with the structure of  consumption}  determined by the vector of levels of satisfaction of consumers needs\index{vector of levels of satisfaction of consumers needs}  $y_1=\{y_i^1\}_{i=1}^{n+1}.$
The latter means that there exist a non-negative vector
 $v_1=\{v_i^1\}_{i=1}^n, \ v_i^1 \geq 0, \ i=\overline{1,n},$ such that $\bar C_1(v_1) - \bar B_1 $ is a non-negative matrix having no zero rows or columns, the spectral radius of the matrix $A_1(x_1)$ is less than 1 and the matrix
$ A_1(x_1) + \bar C_1(v_1) - \bar B_1$ is indecomposable,
there exists a strictly positive solution
$g_1=\{g_i^1\}_{i=1}^{n}$
to the set of equations
\begin{eqnarray*}g_i^1=\frac{[(E-A_1(x_1))^{-1}[\bar C_1(\pi_1^{-1}\bar y_1)- \bar B_1]g_1]_i}{
x_i^1},  \quad i=\overline{1,n},  \end{eqnarray*}
where
\begin{eqnarray*}\bar C_1(v_1)=||c_{si}^1v_i^1||_{s,i=1}^{n}, \quad \bar C_1(\pi_1^{-1}\bar y_1)=\left|\left|c_{si}^1\frac{y_i^1}{\pi_i^1}\right|\right|_{s,i=1}^{n}, \quad
\bar B_1= ||b_{ki}^1||_{k,i=1}^{n,n},\end{eqnarray*}
\begin{eqnarray*} y_i^1 >  \pi_i^1 v_i^1, \quad  i=\overline{1,n},\quad \bar y_1=\{y_i^1\}_{i=1}^{n}.\end{eqnarray*}
Under these conditions, unique up to constant factor equilibrium price vector
$p^0=\{ p_i^0\}_{i=1}^{n},$ in the economy system $E_1,$ exists solving the set of equations
\begin{eqnarray}\label{bl19}
p_i^0=\sum\limits_{k=1}^{n}p_k^0 a_{ki}^1\left(x_i^1\right) +
\frac{1}{x_i^1}\sum\limits_{k=1}^{n} \left[c_{ki}^1\frac{y_i^1}{\pi_i^1} - b_{ki}^1\right]p_k^0, \quad
i=\overline{1,n}.
\end{eqnarray}
We give the income function for a social group of consumers\index{income function for a social group of consumers} by the formula
\begin{eqnarray*} D_{n+1}^1\left(p^0\right)=y_{n+1}^1\sum\limits_{k=1}^{n}c_{k,n+1}^1p_k^0.\end{eqnarray*}

Consider an open economy system $E_2$ containing
$n$ net industries and a single consumer social group. We describe the structure of production in the  economy system $E_2$
by an input  matrix
\begin{eqnarray*} A_{2}(x_{2})=||a^{2}_{ki}\left(x_i^{2}\right)||_{k,i=1}^{n},\quad a_{ki}^2\left(x_i^2\right)=a_{ki}^2+ \frac{r_{ki}^2}{x_i^2},\end{eqnarray*}
where $R_2=|r_{ki}^2 |_{k,i=1}^n$ is a non-negative matrix of constant inputs of  production\index{matrix of constant inputs of  production}  in $E_2,$
$A_{2}=||a^{2}_{ki}||_{k,i=1}^{n}$ is a non-negative  matrix of direct production inputs.\index{ matrix of direct production inputs}
We describe the structure of  consumption  in the economy system $E_2$ by a non-negative  matrix of  unproductive consumption\index{matrix of  unproductive consumption}
$C_2=||c_{ki}^2||_{k,i=1}^{n,n+1}$ whose columns are  fields of information evaluation by consumers\index{fields of information evaluation by consumers}  and by a non-negative matrix $B_2=||b_{ki}^2||_{k,i=1}^{n,n+1}$ built after property vectors of $(n+1)$ consumers.
Let the strictly positive  vector of gross outputs\index{vector of gross outputs} $x_{2}=\{x_i^{2}\}_{i=1}^{n}$ corresponding to the given levels of  consumption  in the economy system\index{ given levels of  consumption  in the economy system} $E_2$ solve the set of equations
\begin{eqnarray*} x_i^{2} - \sum\limits_{k=1}^{n} a_{ik}^{2}\left(x_k^{2}\right)x_k^{2} - e_i^2 +i_i^2 +\sum\limits_{k=1}^{n+1} b_{ik}^2  \end{eqnarray*}
\begin{eqnarray}\label{bl25}
=\sum\limits_{k=1}^{n+1} c_{ik}^{2}y_k^2 ,
\quad  i=\overline{1,n},
\end{eqnarray}
where  $y_2=\{y_k^2\}_{k=1}^{n+1}$ is a  vector of levels of satisfaction of consumers needs in the economy system $E_2.$

Assume that  in the  considered economy systems $E_1$ and $E_2$
matrix elements
$a_{ij}^1\left(x_j^1\right)$ and $ a^{2}_{ij}\left(x_j^{2}\right),$ components of vectors  gross outputs   $x_1=\{x_i^{1}\}_{i=1}^{n},$
$x_2=\{x_i^{2}\}_{i=1}^{n},$ and equilibrium price vector components $p^0=\{ p_i^0\}_{i=1}^{n}$ relate with each other by equalities
\begin{eqnarray*}\sum\limits_{j=1}^{n}
p_i^0a_{ij}^1\left(x_j^1\right)x_j^1=
\sum\limits_{j=1}^{n}
p_i^0a^{2}_{ij}\left(x_j^{2}\right)x_j^{2},\end{eqnarray*}
\begin{eqnarray}\label{dl131}
\sum\limits_{i=1}^{n}
p_i^0a_{ij}^1\left(x_j^1\right)x_j^1=
\sum\limits_{i=1}^{n}
p_i^0a^{2}_{ij}\left(x_j^{2}\right)x_j^{2}, \quad
i,j=\overline{1,n},
\end{eqnarray}
\begin{eqnarray*} x_j^{2} \geq x_j^1, \quad
p_j^0(x_j^{2} - x_j^1)=\sigma_j,  \quad \sigma_j \geq 0, \quad
j=\overline{1,n}, \quad  \sum\limits_{j=1}^{n}\sigma_j > 0. \end{eqnarray*}
Matrix elements of unproductive consumption  are related to each other by the equalities
\begin{eqnarray}\label{dl23}
c_{kj}^{2}=c_{kj}^1+ \frac{h_{kj}}{p_k^0y_j^1}, \quad h_{kj} \geq 0, \quad
k=\overline{1,n}, \quad
j=\overline{1,n+1},
\end{eqnarray}
where matrix elements of the matrix
$ H=||h_{ki}||_{k,i=1}^{n, n+1} $  solve the set of equations

\begin{eqnarray*}\sum\limits_{j=1}^{n+1} h_{kj}=\sigma_k, \quad k=\overline{1,n}, \quad
\sum\limits_{k=1}^{n} h_{kj}=\sigma_j\pi_j^1, \quad j=\overline{1,n}, \end{eqnarray*}
\begin{eqnarray}\label{el23}
\sum\limits_{k=1}^{n} h_{k,n+1}=\sum\limits_{k=1}^{n}\sigma_k\left(1 - \pi_k^1\right).
\end{eqnarray}
Introduce the notation  $\bar C_2=||c_{ki}^{2}||_{k,i=1}^{n,n},$ $\bar B_2=||b_{ki}^{2}||_{k,i=1}^{n,n}.$

Suppose the matrix
$ A_2\left(x_2\right) + \bar C_2(v_2) - \bar B_2$ is indecomposable and the equalities
\begin{eqnarray}\label{babp1}
 B_1=B_2, \quad  y_1=y_2, \quad \pi_1=\pi_2, \quad e_1=e_2, \quad i_1=i_2, \quad v_1=v_2.
\end{eqnarray}
\begin{eqnarray*} y_1=\{y_i^1\}_{i=1}^{n+1}, \quad  y_2=\{y_i^2\}_{i=1}^{n+1}, \quad  \pi_1=\{\pi_i^1\}_{i=1}^n, \quad  \pi_2=\{\pi_i^2\}_{i=1}^n,\end{eqnarray*} \begin{eqnarray*}  e_1=\{e_i^1\}_{i=1}^n, \quad  e_2=\{e_i^2\}_{i=1}^n, \quad  v_1=\{v_i^1\}_{i=1}^n, \quad  v_2=\{v_i^2\}_{i=1}^n,\end{eqnarray*}
hold.
We assume that the equilibrium price vector in the economy system $E_2$ within a certain period of the economy operation is given by the solution to the set of equations
\begin{eqnarray}\label{bl24}
p_i=
\sum\limits_{k=1}^{n}p_k a_{ki}^{2}\left(x_i^{2}\right) +
\frac{1}{x_i^{2}}\sum\limits_{k=1}^{n}\left [c_{ki}^{2}\frac{y_i^2}{\pi_i^2} -b_{ki}^2\right]p_k, \quad
i=\overline{1,n},
\end{eqnarray}
for the price vector $p=\{p_k\}_{k=1}^{n}.$
We give the income function of social consumers group as follows
\begin{eqnarray*} D_{n+1}^{2}\left(p^0\right)=
y_{n+1}^2\sum\limits_{k=1}^{n}c_{k,n+1}^{2}p_k^0.\end{eqnarray*}

\begin{lemma}\label{ssos1}
If the following conditions
\begin{eqnarray*}\sum\limits_{j=1}^{n+1}\beta_j=\sum\limits_{k=1}^{n}\sigma_k, \quad \sum\limits_{k=1}^{n}\sigma_k> 0,\quad
\beta_j \geq 0,\quad  j=\overline{1, n+1}, \quad  \sigma_k \geq 0, \quad k=\overline{1, n},\end{eqnarray*}
hold, then the set of equations
\begin{eqnarray}\label{ssos2}
\sum\limits_{j=1}^{n+1} h_{kj}=\sigma_k, \quad k=\overline{1,n}, \quad
\sum\limits_{k=1}^{n} h_{kj}=\beta_j, \quad j=\overline{1,n+1},
\end{eqnarray}
for the matrix $H=||h_{ki} ||_{k=1, j=1}^{n, \ n+1}$ has a solution in the set of non-negative matrices one of which can be given as
\begin{eqnarray*}
h_{kj}=\frac{\sigma_k\beta_j}{\sum\limits_{k=1}^{n}\sigma_k}, \quad k,j=\overline{1,n}, \end{eqnarray*}
\begin{eqnarray}\label{ssos3}
h_{k, n+1}=
\frac{\sigma_k \left(\sum\limits_{k=1}^n\sigma_k- \sum\limits_{j=1}^{n}\beta_j\right)}{\sum\limits_{k=1}^n\sigma_k}, \quad k=\overline{1, n}.
\end{eqnarray}
\end{lemma}
The Proof of the Lemma  is obvious.

\begin{theorem}\label{fa1}
Let the  conditions
\begin{eqnarray*} \sum\limits_{k=1}^{n}\sigma_k> 0,\quad
  \sigma_k \geq 0, \quad k=\overline{1, n}, \quad  \sum\limits_{k=1}^{n}\sigma_k -  \sum\limits_{k=1}^{n}\pi_k^1\sigma_k \geq 0,\end{eqnarray*}
hold. The equilibrium price vector in the economy system $E_2$
is the same vector
$p^0=\{p_k^0\}_{k=1}^{n}$
as in the economy system $E_1,$
i.e., $p^0$ is a solution to  the set of equations
(\ref{bl24}) and the vector
$x_{2}=\{x_k^{2}\}_{k=1}^{n}$ solves the set of equations (\ref{bl25}). The vector $\pi_2=\{\pi_i^2\}_{i=1}^n$ agrees with the structure of consumption  described by a vector of levels of satisfaction of consumers needs    $y_2=\{y_k^2\}_{k=1}^{n+1}. $
\end{theorem}
\begin{proof}\smartqed  Show that the vector $p^0$ solves the set of equations
(\ref{bl24}). Multiplying the expression
\begin{eqnarray*} \sum\limits_{k=1}^{n}p_k^0 a_{ki}^{2}\left(x_i^{2}\right) +
\frac{1}{x_i^{2}}\sum\limits_{k=1}^{n} \left[c_{ki}^{2}\frac{y_i^2}{\pi_i^2} - b_{ki}^2\right]p_k^0, \quad
i=\overline{1,n},\end{eqnarray*}
by $x_i^{2}$ and using relations (\ref{dl131}), (\ref{babp1})
and the expression (\ref{dl23})
for matrix elements $c_{ki}^{2},$
we obtain
\begin{eqnarray*}
\sum\limits_{k=1}^{n}p_k^0 a_{ki}^{2}\left(x_i^{2}\right)x_i^{2} +
\sum\limits_{k=1}^{n} \left[c_{ki}^{2}\frac{y_i^2}{\pi_i^2} - b_{ki}^2\right]p_k^0
 \end{eqnarray*} \begin{eqnarray*}=\sum\limits_{k=1}^{n}p_k^0 a_{ki}\left(x_i^1\right)x_i^1 +
\sum\limits_{k=1}^{n} \left[c_{ki}^1  \frac{y_i^1}{\pi_i^1} - b_{ki}^1\right]p_k^0 +
\frac{1}{\pi_i^1}\sum\limits_{k=1}^{n} h_{ki},
\quad i=\overline{1,n}.\end{eqnarray*}
Multiplying  $x_i^{2}$  by $p_i^0$
and  using the equality
$p_i^0\left(x_i^{2} - x_i^1\right)=\sigma_i, $
we obtain
\begin{eqnarray*} p_i^0x_i^{2}= x_i^1p_i^0+\sigma_i.\end{eqnarray*}
However, the equality
\begin{eqnarray*}x_i^1p_i^0+\sigma_i=\end{eqnarray*} \begin{eqnarray*}=
\sum\limits_{k=1}^{n}p_k^0 a_{ki}\left(x_i^1\right)x_i^1 +
\sum\limits_{k=1}^{n} \left[c_{ki}^1\frac{y_i^1}{\pi_i^1} - b_{ki}^1\right]p_k^0 +
\frac{1}{\pi_i^1}\sum\limits_{k=1}^{n} h_{ki},
\quad i=\overline{1,n},\end{eqnarray*}
holds because the vector $p^0$ solves the set of equations (\ref{bl19})
and matrix  elements of the matrix
$ H=||h_{ki}||_{k,i=1}^{n, n+1} $ solve the set of equations (\ref{el23}).

Prove that the vector
$x_{2}=\{x_i^{2}\}_{i=1}^{n}$ is a solution to  the set of equations (\ref{bl25}).
Multiplying by $p_i^0$ the expression
\begin{eqnarray*}x_i^{2} - \sum\limits_{k=1}^{n} a_{ik}^{2}\left(x_k^{2}\right)x_k^{2} - e_i^2 +i_i^2 +\sum\limits_{k=1}^{n+1}b_{ik}^2 , \quad
i=\overline{1,n},\end{eqnarray*}
 and using relations (\ref{dl131}), (\ref{babp1}) and  the expression (\ref{dl23})
for matrix elements $c_{ki}^{2},$
we obtain
\begin{eqnarray*}p_i^0x_i^{2} - \sum\limits_{k=1}^{n}p_i^0 a_{ik}^{2}\left(x_k^2\right)x_k^{2} -
p_i^0e_i^2 +p_i^0i_i^2 +p_i^0\sum\limits_{k=1}^{n+1}b_{ik}^2
 \end{eqnarray*}
 \begin{eqnarray*}=
p_i^0x_i^1+\sigma_i - \sum\limits_{k=1}^{n}p_i^0 a_{ik}^1\left(x_k^1\right)x_k^1 -
p_i^0e_i^1 +p_i^0i_i^1 + p_i^0\sum\limits_{k=1}^{n}b_{ik}^1, \quad i=\overline{1,n}.\end{eqnarray*}
Multiplying by $p_i^0$ the expression
\begin{eqnarray*}\sum\limits_{k=1}^{n+1} c_{ik}^{2}y_k^2,
\quad  i=\overline{1,n},\end{eqnarray*}
we obtain
\begin{eqnarray*}\sum\limits_{k=1}^{n+1}p_i^0 c_{ik}^{2}y_k^2 =\end{eqnarray*} \begin{eqnarray*}=\sum\limits_{k=1}^{n+1}p_i^0 c_{ik}^1y_k^1 +
\sum\limits_{k=1}^{n+1}h_{ik},
\quad  i=\overline{1,n}.\end{eqnarray*}
However, the equality
\begin{eqnarray*}p_i^0x_i^1+\sigma_i - \sum\limits_{k=1}^{n}p_i^0 a_{ik}^1\left(x_k^1\right)x_k^1 -
p_i^0e_i^1 +p_i^0i_i^1 + p_i^0\sum\limits_{k=1}^{n+1}b_{ik}^1
 \end{eqnarray*}
 \begin{eqnarray*}=
\sum\limits_{k=1}^{n+1}p_i^0 c_{ik}^1y_k^1 +
\sum\limits_{k=1}^{n+1}h_{ik}, \quad i=\overline{1,n},\end{eqnarray*}
holds because the vector $x_1=\{x_i^1\}_{i=1}^{n}$ solves the set of equations (\ref{bl20})
and matrix  elements of the matrix
$ H=||h_{ki}||_{k,i=1}^{n, n+1} $ solve the set of equations
(\ref{el23}).

To finish the Proof of the Theorem, one must prove that there exists at least one solution to the set of equations (\ref{el23}). Due to the Lemma \ref{ssos1}, the solution to the set of equations
 (\ref{el23}) is the matrix
$H$ with elements
\begin{eqnarray*} h_{kj}=\frac{\sigma_k\sigma_j\pi_j^1}
{\sum\limits_{j=1}^{n}\sigma_j},
\quad k,j=\overline{1, n}, \end{eqnarray*} \begin{eqnarray*}
 h_{k, n+1}=\frac{\sigma_k
\sum\limits_{j=1}^{n}\sigma_j\left(1-\pi_j^1\right)}
{\sum\limits_{j=1}^{n}\sigma_j},
\quad k= \overline{1, n}.\end{eqnarray*}
Prove the agreement of the taxation vector
$\pi_2=\{\pi_i^2\}_{i=1}^n$ with the structure of  consumption  described by a vector of levels of satisfaction of consumers needs  $y_2=\{y_k^2\}_{k=1}^{n+1}. $
It is obvious that
$\sum\limits_{k=1}^n c_{ki}^2 \geq \sum\limits_{k=1}^n c_{ki}^1>0, \ i=\overline{1, n+1},$
the matrix $\bar C_2(v_2) - \bar B_2 \geq  \bar C_1(v_1) - \bar B_1 $ is non-negative  having no zero rows or columns.
From the solvability of the set of equations (\ref{bl24}) in the set
\begin{eqnarray*}T_0=\left\{p \in \bar R_+^n, \   p_i > \sum\limits_{s=1}^na^2_{si}\left(x^2_i\right)p_s+ \frac{1}{x_i^2}\sum\limits_{s=1}^n[c_{si}^2v_i^2 -b_{si}]p_s,   \ i=\overline{1,n}\right\},\end{eqnarray*}
and the Theorem \ref{vjant14} it follows the needed.
\qed
\end{proof}

As earlier, let the vector $p^0=\{p_i^0\}_{i=1}^{n}$ be an equilibrium price vector in economy system $E_1.$

Consider a non-negative productive matrix
$ A_3=||a_{ki}^3||_{k,i=1}^{n}$ and non-negative matrices
\begin{eqnarray*}c_3=||c_{ki}^3||_{k,i=1}^{n, n+1}, \quad   b_3=||b_{ki}^3||_{k=1,i=1}^{n, n+1}.\end{eqnarray*}
Assume the validity of inequalities $\sum\limits_{k=1}^nc_{ki}^3 >0, \ i=\overline{1, n+1}.$
After these matrices, build the matrices
\begin{eqnarray*} \bar A_3=||\bar a_{ij}^3||_{i,j=1}^{n}, \quad
  C_3=||C_{ki}^3||_{k,i=1}^{n,\ n+1}, \quad
\bar C_3=||C_{ki}^3||_{k=1,i=1}^{n},\end{eqnarray*}
\begin{eqnarray*} B_3=||B_{ki}^3||_{k=1,i=1}^{n, \ n+1}, \quad  \bar B_3=||B_{ki}^3||_{k=1,i=1}^{n,\ n},\quad \bar a_{ij}^3=
\frac{p_i^0 a_{ij}^3}{p_j^0},
\quad  i,j=\overline{1,n},\end{eqnarray*} \begin{eqnarray*}
C_{ki}^3= p_k^0c_{ki}^3,\quad  B_{ki}^3= p_k^0b_{ki}^3,
\quad  k=\overline{1,n},
\quad  i=\overline{1,n+1}.\end{eqnarray*}
Introduce non-negative vectors
\begin{eqnarray*}X_3=\{X_i^3\}_{i=1}^{n}, \quad X_i^3 > 0, \quad i=\overline{1,n}, \quad
{\cal E}_3=\{{\cal E}_i^3\}_{i=1}^{n},
\quad I_3=\{I_k^3\}_{k=1}^{n}, \end{eqnarray*} \begin{eqnarray*}
Y_3=\{Y_i^3\}_{i=1}^{n+1}, \quad  Y_i^3 > 0,
\quad  i=\overline{1,n+1}.\end{eqnarray*}
Assume that  there exists a  non-negative vector $v_3=\{v_i^3\}_{i=1}^n$   such that $ \bar C_3(v_3) - \bar B_3$ is a non-negative  matrix having no zero rows or columns and
$ \bar A_3  + \bar C_3(v_3) - B_3$ is an  indecomposable  matrix,
where the matrix $\bar C_3(v_3)=||C_{ki}^3v_i^3||_{k,i=1}^{n}, \  v_3=\{v_i^3\}_{i=1}^{n}, \ v_i^3 \geq 0, \ i=\overline{1, n}.$
Let a strictly positive vector
$X_3=\{X_i^3\}_{i=1}^{n}, \ X_i^3 > 0, \ i=\overline{1, n},$ solve the set of equations
\begin{eqnarray}\label{bl21}
X_i^3 - \sum\limits_{k=1}^{n} \bar a_{ik}^3X_k^3 +\sum\limits_{k=1}^{n+1} B_{ik}^3- {\cal E}_i^3 +I_i^3 =
 \sum\limits_{k=1}^{n+1} C_{ik}^3Y_k^3,
\quad  i=\overline{1,n},
\end{eqnarray}
where
\begin{eqnarray*}Y_3= \{Y_k^3\}_{k=1}^{n+1}, \quad   Y_k^3 > 0, \quad k=\overline{1,n+1}, \quad Y_k^3 > \pi_k^3 v_k^3, \quad  k=\overline{1,n}.\end{eqnarray*}
We suppose the vector $\pi_3=\{\pi_i^3\}_{i=1}^{n}$ is such  that there exists a strictly positive solution
$g_3=\{g_i^3\}_{i=1}^{n}$
to the set of equations
\begin{eqnarray}\label{el1}
\frac{[(E- \bar A_3)^{-1}[\bar C_3(\pi_3^{-1}\bar Y_3) - \bar  B_3]g_3]_i}{X_i^3} = g_i^3,
\quad  i=\overline{1,n},
\end{eqnarray}
\begin{eqnarray*} \bar C_3(\pi_3^{-1}\bar Y_3)=\left|\left|C_{ki}^3  \frac{Y_i^3}{\pi_i^3} \right|\right|_{k,i=1}^{n}.\end{eqnarray*}
Consider the set of equations for the vector $p_3=\{p^3_i\}_{i=1}^{n}$
\begin{eqnarray}\label{dl19}
p_i^3=\sum\limits_{k=1}^{n}\bar a_{ki}^3p_k^3  +
\frac{1}{X_i^3}\sum\limits_{k=1}^{n}\left[C_{ki}^3 \frac{Y_i^3}{\pi_i^3} - B_{ki}^3\right]p_k^3, \quad
i=\overline{1,n}.
\end{eqnarray}
From the  agreement  of  taxation vector with the  structure of  consumption\index{agreement  of  taxation vector with the  structure of  consumption} determined by  a vector of levels of satisfaction of consumers needs\index{vector of levels of satisfaction of consumers needs} $Y_3= \{Y_k^3\}_{k=1}^{n+1},$ the existence  of unique up to constant factor a strictly positive solution $p_3=\{p_i^3\}_{i=1}^n$
to the set of equations (\ref{dl19}) follows.

Construct an  open economy system $E_3$ containing
$n$ net industries and a single  social group of consumers. We describe the  structures of production and consumption in this economy system by an input matrix
$  A(p_3)=||a_{ki}(p_3)||_{k,i=1}^{n}$
and by a  matrix of  unproductive consumption
 $ C(p_3)=||c_{ki}(p_3)||_{k=1,i=1}^{n,\ n+1},$
and also by  a matrix of initial goods supply  $B(p_3)=||b_{ki}(p_3)||_{k=1,i=1}^{n,\ n+1}, $
where
\begin{eqnarray*}a_{ki}(p_3)=\frac{p_k^3 a_{ki}^3}{p_i^3},\quad  \ k, i =\overline{1, n}, \quad b_{ki}(p_3)=p_k^3b_{ki}^3,\quad  k=\overline{1,n}, \quad i =\overline{1, n+1},   \end{eqnarray*} \begin{eqnarray*}
c_{ki}(p_3)=p_k^3c_{ki}^3, \quad  k=\overline{1,n}, \quad i =\overline{1, n+1}.
  \end{eqnarray*}
 An equilibrium price vector solving the set of equations
\begin{eqnarray}\label{bl22}
p_i=\sum\limits_{k=1}^{n}p_k a_{ki}( p_3) +
\frac{1}{x_i^3}\sum\limits_{k=1}^{n} \left[c_{ki}(p_3)\frac{y_i^3}{\pi_i^3} - b_{ki}(p_3)\right]p_k, \quad
i=\overline{1,n},
\end{eqnarray}
relative to the  price vector
$p=\{ p_i\}_{i=1}^{n},$
where $y_i^3= Y_i^3, \ i=\overline{1, n+1}, $ and a strictly positive solution to the set of equations
\begin{eqnarray*}x_k^3 - \sum\limits_{i=1}^{n}a_{ki}(p_3)x_i^3 -  e_k^{3}(p_3) +
i_k^{3}(p_3)+\sum\limits_{i=1}^{n+1}b_{ki}(p_3) \end{eqnarray*}
\begin{eqnarray}\label{bl23}
 = \sum\limits_{k=1}^{n+1} c_{ki}(p_3)y_i^3,
\quad  k=\overline{1,n},
\end{eqnarray}
for the vector of  gross outputs\index{vector of  gross outputs}
 $x_3=\{x_i^3\}_{i=1}^{n},$ where
\begin{eqnarray*}
x_i^3=\frac{p_i^3X_i^3}{p_i^0}, \quad
i=\overline{1,n}, \quad
 e_k(p_3)=\frac{p_k^3{\cal E}_k^3}{p_k^0}, \quad
i_k\left(p^3\right)=\frac{p_k^3I_k^3}{p_k^0}, \quad
k=\overline{1,n}, \end{eqnarray*}
guarantee the equilibrium in the economy system $E_3.$
In such a manner built the economy system $E_3,$ the taxation vector $\pi_3=\{\pi_i^3\}_{i=1}^{n}$ agrees with the  structure of  consumption\index{ taxation vector  agrees with the  structure of  consumption}, i.e., there exists a strictly positive solution to the set of equations
\begin{eqnarray*}\frac{[(E-  A(p_3))^{-1} [C_3(p_3, \pi_3^{-1}\bar y_3) - \bar B_3(p_3)]g_3]_i}{x_i^3} = g_i^3,
\quad  i=\overline{1,n}, \end{eqnarray*}
\begin{eqnarray*} C_3(p_3,  \pi_3^{-1}\bar y_3)= \left|\left|c_{ki}(p_3)\frac{y_i^3}{\pi_i^3}\right|\right|_{k,i=1}^{n}, \quad \bar y_3 = \{y_i^3\}_{i=1}^{n},\end{eqnarray*}
\begin{eqnarray*}\bar B_3(p_3)=||b_{ki}(p_3)||_{k,i=1}^{n},\quad y_i^3 > \pi_i^3 v_i^3, \quad  i=\overline{1,n}, \end{eqnarray*}
for the vector $g_3=\{g_i^3\}_{i=1}^{n}.$
We give an  income function of social group of  consumers\index{income function of social group of  consumers}  as follows
\begin{eqnarray*} D_{n+1}\left(p_3, p^0\right)=y_{n+1}^3\sum\limits_{k=1}^{n}c_{k,n+1}(p_3)p_k^0.\end{eqnarray*}
\begin{theorem}
The equilibrium price vector in the economy system $E_3$ exists and coincides
with the equilibrium price  vector  $p^0=\{p_i^0\}_{i=1}^{n}$ being such as in the  economy system $E_1$
if the vector $p_3=\{p_i^3\}_{i=1}^{n}$ solves the set of equations
(\ref{dl19}) and the vector $X_3=\{X_i^3\}_{i=1}^{n}$ solves the set of equations (\ref{bl21}).
\end{theorem}
\begin{proof}\smartqed  From the productivity  of the  matrix\index{productivity  of a  matrix}
$ A_3$  the productivity of the matrix
$ \bar A_3$ follows. The existence of a strictly positive solution to the set of equations
(\ref{el1}) means the existence of a strictly positive solution to the set of equations
(\ref{dl19}).
Using solvability of the sets of equations
(\ref{dl19}),  (\ref{bl21}), by elementary transformations we prove the validity of the  equalities
\begin{eqnarray}\label{el2}
p_i^0=\sum\limits_{k=1}^{n}p_k^0 a_{ki}( p_3) +
\frac{1}{x_i^3}\sum\limits_{k=1}^{n}\left[ c_{ki}(p_3)\frac{\pi_i^3}{y_i^3} - b_{ki}(p_3)\right]p_k^0, \quad
i=\overline{1,n},
\end{eqnarray}
\begin{eqnarray*} x_k^3 - \sum\limits_{i=1}^{n} a_{ki}( p_3)x_i^3 -  e_k(p_3) +
i_k(p_3)+\sum\limits_{i=1}^{n+1} b_{ki}( p_3) \end{eqnarray*}
\begin{eqnarray}\label{el3}
 =\sum\limits_{i=1}^{n+1} c_{ki}(p_3)y_i^3,
\quad  k=\overline{1,n},
\end{eqnarray}
where the vector $p^0=\{p_i^0\}_{i=1}^{n}$ is the equilibrium price vector in the economy system $E_1$
and
\begin{eqnarray*}x_3=\{x_i^3\}_{i=1}^{n}, \quad  x_i^3=\frac{p_i^3X_i^3}{p_i^0}, \quad  i=\overline{1,n}.\end{eqnarray*}
\qed
\end{proof}

Describe each of three introduced economy systems in cost parameters.

In cost parameters, we determine the economy system $E_1$ by
\\ 1) a vector of  gross outputs \index{vector of  gross outputs }
$X_1=\{X_i^1\}_{i=1}^{n}, \ X_i^1=p_i^0x_i^1,
\ i=\overline{1, n}\ ;$
\\ 2) matrix of  financial flows\index{ matrix of  financial flows}
 \begin{eqnarray*}||X_{ij}^1||_{i,k=1}^{n}, \quad
X_{ij}^1= p_i^0a_{ij}\left(x_j^1\right)x_j^1, \quad i,j=\overline{1, n}\ ;\end{eqnarray*}
\\ 3) an input  matrix\index{ input  matrix}
\begin{eqnarray*} \bar A_1=||\bar a_{ij}(X_1)||_{i,j=1}^{n},
\quad \bar a_{ij}(X_1)=\frac{X_{ij}^1}{X_j^1}=\frac{p_i^0a_{ij}\left(x_j^1\right)}{p_j^0},
\quad i,j=\overline{1, n} \ ;\end{eqnarray*}
4) a  matrix of unproductive consumption\index{matrix of unproductive consumption}
\begin{eqnarray*} C_1=||C_{ki}^1||_{k,i=1}^{n, n+1},\quad
 C_{ki}^1=p_k^0c_{ki}^1, \quad k=\overline{1, n}, \quad i=\overline{1, n+1}, \end{eqnarray*}
and a matrix of  initial goods supply\index{matrix of  initial goods supply}
\begin{eqnarray*} B_1=||B_{ki}^1||_{k,i=1}^{n, n+1},\quad
 B_{ki}^1=p_k^0b_{ki}^1, \quad k=\overline{1, n}, \quad i=\overline{1, n+1}\ ; \end{eqnarray*}
5) vectors of export and import\index{vectors of export and import}
\begin{eqnarray*}{\cal E}_1=\{{\cal E}_k^1\}_{k=1}^{n}, \quad
I_1=\{I_k^1\}_{k=1}^{n}, \quad I_k^1=i_k^1p_k^0, \quad
{\cal E}_k^1=e_k^1p_k^0, \quad k=\overline{1, n}\ ;\end{eqnarray*}
 6) vectors of final consumption\index{ vectors of final consumption and gross accumulation and supply change} $C_1^f=\{C_i^1\}_{i=1}^{n}$
and gross accumulation and supply change $N_1=\{N_i^1\}_{i=1}^{n}$
such that \begin{eqnarray*} N_i^1+C_i^1=\sum\limits_{k=1}^{n+1}C_{ik}^1Y_k^1 - \sum\limits_{k=1}^{n+1}B_{ik}^1,
\quad i=\overline{1, n}\ ; \end{eqnarray*}
\\ 7) a value added of the $i$-th industry\index{ value added of the $i$-th industry}
$\Delta_i^1$ and an  income of   social  group of  consumers $\bar D_{n+1}^1$
\begin{eqnarray*}\Delta_i^1=\sum\limits_{k=1}^{n}\left[C_{ki}^1\frac{Y_i^1}{\pi_i^1}  - B_{ki}^1\right],
 \quad i=\overline{1, n}, \end{eqnarray*} \begin{eqnarray*}
\bar D_{n+1}^1=Y^1_{n+1}\sum\limits_{k=1}^{n}C_{k,n+1}^1\ ;\end{eqnarray*}
8) a taxation vector\index{taxation vector} $\pi_1=\{\pi_i^1\}_{i=1}^{n}\ ;$
\\ 9) a  vector of levels  of satisfaction of   consumers needs\index{vector of levels  of satisfaction of   consumers needs}
 \begin{eqnarray*}Y_1=\{Y_i^1\}_{i=1}^{n+1}, \quad  Y_k^1=y_k^1,
\quad k=\overline{1, n+1}. \end{eqnarray*}
As a result of the set of equations (\ref{bl20}), (\ref{bl19}),
the  introduced quantities satisfy  the set of  balance  equations
\begin{eqnarray*} X_i^1 - \sum\limits_{k=1}^{n} X_{ik}^1 - {\cal E}_i^1
+ I_i^1 =N_i^1+C_i^1, \quad i=\overline{1,n}, \end{eqnarray*}
\begin{eqnarray*} \sum\limits_{i=1}^{n} X_{ik}^1 + \Delta_k^1=X_k^1,
\quad  k=\overline{1,n}, \end{eqnarray*}
holds.
 \begin{corollary}
Matrix elements $C_{ki}^1$ of the  matrix  of unproductive consumption in cost parameters\index{ matrix  of unproductive consumption in cost parameters} satisfy the sets of inequalities and equations
\begin{eqnarray*} C_{ki}^1 \geq 0, \quad k=\overline{1,n}, \quad
i=\overline{1,n+1}, \end{eqnarray*}
\begin{eqnarray*} \sum\limits_{i=1}^{n+1}C_{ki}^1Y_i^1=C_k^1+N_k^1 +\sum\limits_{i=1}^{n+1}B_{ki}^1 , \quad
k=\overline{1,n}, \end{eqnarray*}
\begin{eqnarray*} Y_i^1\sum\limits_{k=1}^{n}C_{ki}^1=\pi_i^1\Delta_i^1 + \pi_i^1\sum\limits_{k=1}^{n}B_{ki}^1,
\quad i=\overline{1,n},\end{eqnarray*}
\begin{eqnarray*}Y_{n+1}^1\sum\limits_{k=1}^{n}C_{k,n+1}^1 \end{eqnarray*}
\begin{eqnarray}\label{gurl1}
=\sum\limits_{i=1}^{n}\left(1- \pi_i^1\right)\Delta_i^1 - {\cal E}_1^0 +I_1^0 + \sum\limits_{i=1}^{n}\sum\limits_{s=1}^{n}\left(1 - \pi_s^1\right)B_{is}^1+ \sum\limits_{i=1}^{n}B_{i, n+1}^1,
\end{eqnarray}
where ${\cal E}_1^0= \sum\limits_{i=1}^{n}{\cal E}_i^1, \
I_1^0= \sum\limits_{i=1}^{n}I_i^1.$
\end{corollary}
First two  set of  equations follow from definitions introduced and the third one follows from the set of balance  equations.

Really, from the set of   balance equations  the equality follows
\begin{eqnarray}\label{ssos5}
\sum\limits_{i=1}^{n}(N_i^1+C_i^1)+\sum\limits_{i=1}^{n}{\cal E}_i^1 - \sum\limits_{i=1}^{n}I_i^1=\sum\limits_{i=1}^{n}\Delta_i^1.
\end{eqnarray}
From the definition, we have
\begin{eqnarray*} N_i^1+C_i^1=\sum\limits_{k=1}^{n+1}C_{ik}^1Y_k^1 - \sum\limits_{k=1}^{n+1}B_{ik}^1,
\quad i=\overline{1, n},\end{eqnarray*}
from here
\begin{eqnarray*}\sum\limits_{i=1}^{n}(N_i^1+C_i^1)=\sum\limits_{k=1}^{n+1}\sum\limits_{i=1}^{n}C_{ik}^1Y_k^1 -
\sum\limits_{i=1}^{n}\sum\limits_{k=1}^{n+1}B_{ik}^1.\end{eqnarray*}
In view of
\begin{eqnarray*}\sum\limits_{i=1}^{n}C_{ik}^1Y_k^1=\pi_k^1\Delta_k^1+ \pi_k^1 \sum\limits_{i=1}^{n} B_{ik}^1,
\quad k=\overline{1, n}, \end{eqnarray*}
we obtain
\begin{eqnarray*}\sum\limits_{i=1}^{n}(N_i^1+C_i^1) \end{eqnarray*} \begin{eqnarray*}=\sum\limits_{i=1}^{n}C_{i,n+1}^1Y_{n+1}^1 + \sum\limits_{k=1}^{n}\pi_k^1\Delta_k^1 - \sum\limits_{k=1}^{n}\left(1 - \pi_k^1\right)\sum\limits_{i=1}^{n}B_{ik}^1 - \sum\limits_{i=1}^{n}B_{i,n+1}^1.\end{eqnarray*}
Inserting the last equality into (\ref{ssos5}), we obtain (\ref{gurl1}).

In cost parameters, we describe economy system $E_2$ by
\\ 1) a vector of gross outputs\index{vector of gross outputs}
\begin{eqnarray*}X_2=\{X_i^2\}_{i=1}^{n}, \quad X_i^2=p_i^0x_i^{2}=X_i^1 + \sigma_i, \quad
i=\overline{1, n}\ ;\end{eqnarray*}
\\ 2)  a matrix of financial flows\index{matrix of financial flows}
 \begin{eqnarray*}||X_{ij}^{2}||_{i,k=1}^{n}, \quad
X_{ij}^{2}= p_i^0a^{2}_{ij}\left(x_j^{2}\right)x_j^{2},
\quad i,j=\overline{1, n}\ ;\end{eqnarray*}\\
3) an input  matrix\index{input  matrix}
\begin{eqnarray*} \bar A_{2}=||\bar a_{ij}^{2}||_{i,j=1}^{n},
\quad \bar a_{ij}^{2}=\frac{X_{ij}^{2}}{X_j^2}=\frac{p_i^0a^{2}_{ij}\left(x_j^{2}\right)}{p_j^0}=\frac{X_{ij}^{2}}{X_j^1+\sigma_j},
\quad i,j=\overline{1, n}\ ;\end{eqnarray*}
4) a matrix of unproductive consumption\index{matrix of unproductive consumption}
\begin{eqnarray*} C_{2}=||C_{ki}^{2}||_{k,i=1}^{n, n+1},\quad
 C_{ki}^{2}=p_k^0c_{ki}^{2}, \quad  k=\overline{1, n},
\quad i=\overline{1, n+1}, \end{eqnarray*}
and a matrix of  initial goods supply\index{matrix of  initial goods supply}
\begin{eqnarray*} B_{2}=||B_{ki}^{2}||_{k,i=1}^{n, n+1},\quad
 B_{ki}^{2}=p_k^0b_{ki}^{2}, \quad b_{ki}^{2}=b_{ki}^{1}, \quad   k=\overline{1, n},
\quad i=\overline{1, n+1}\ ; \end{eqnarray*}
5)  vectors of  export and import\index{vectors of  export and import}
\begin{eqnarray*}{\cal E}_2=\{{\cal E}_k^2\}_{k=1}^{n},  \quad
I_2=\{I_k^2\}_{k=1}^{n}, \quad
{\cal E}_k^2=e_k^2p_k^0, \quad \end{eqnarray*} \begin{eqnarray*}I_k^2=i_k^2p_k^0, \quad  e_k^2=e_k^1, \quad  i_k^2=i_k^1,
\quad k=\overline{1, n}\ ;\end{eqnarray*}
 6)  vectors of final consumption\index{vectors of final consumption and gross accumulation and supply change} $C_2^f=\{C_i^2\}_{i=1}^{n}$
and gross accumulation and supply change $N_2=\{N_i^2\}_{i=1}^{n}$
such that
\begin{eqnarray*} N_i^2+C_i^2= \sum\limits_{k=1}^{n+1}C_{ik}^{2}Y_k^2 -  \sum\limits_{k=1}^{n+1}B_{ik}^2=
N_i^1+C_i^1+\sigma_i, \quad  i=\overline{1, n}\ ; \end{eqnarray*}
\\ 7) a value added of  the $i$-th industry\index{value added of  the $i$-th industry}
$\Delta_i^2$ and an income of consumers social group $\bar D_{n+1}^2$
\begin{eqnarray*}\Delta_i^2 =\sum\limits_{k=1}^{n}\left[C_{ki}^{2}\frac{Y_i^2}{\pi_i^2}  -B_{ki}^2\right]=
 \Delta_i^1 +\sigma_i,  \quad  i=\overline{1, n},\end{eqnarray*}
 \begin{eqnarray*} \bar D_{n+1}^2=Y^2_{n+1}\sum\limits_{k=1}^{n}C^{2}_{k,n+1}\ ;\end{eqnarray*}
 8) a  taxation vector\index{ taxation vector} $\pi_2=\{\pi_i^2\}_{i=1}^{n}, \ \pi_i^2=\pi_i^1, \ i=\overline{1, n}\ ;$ \\
9)  a  vector of levels  of satisfaction of   consumers needs\index{vector of levels  of satisfaction of   consumers needs}
 \begin{eqnarray*}Y_2=\{Y_i^2\}_{i=1}^{n+1}, \quad Y_k^2=y_k^2=y_k^1,
\quad k=\overline{1, n+1}. \end{eqnarray*}
As in the previous case, for the  introduced quantities  the  set of  balance equations\index{set of  balance equations}
\begin{eqnarray*} X_i^2 - \sum\limits_{k=1}^{n} X_{ik}^{2} - {\cal E}_i^2
+ I_i^2 =N_i^2+C_i^2, \quad i=\overline{1,n}, \end{eqnarray*}
\begin{eqnarray*} \sum\limits_{i=1}^{n} X_{ik}^{2} + \Delta_k^2=X_k^2,
\quad  k=\overline{1,n}, \end{eqnarray*}
 holds.
\begin{corollary}
In cost parameters, matrix  elements $C_{ki}^{2}$ of the  matrix of  unproductive consumption\index{ matrix of  unproductive consumption} satisfy the set of equations
\begin{eqnarray*} \sum\limits_{i=1}^{n+1}C_{ki}^{2}Y_i^2=C_k^2+N_k^2+ \sum\limits_{i=1}^{n+1}B_{ki}^2, \quad
k=\overline{1,n}, \end{eqnarray*} \begin{eqnarray*}
Y_i^2\sum\limits_{k=1}^{n}C_{ki}^{2}=\pi_i^2\Delta_i^2 + \pi_i^2\sum\limits_{k=1}^{n}B_{ki}^2,
\quad i=\overline{1,n}, \end{eqnarray*}
\begin{eqnarray*}Y_{n+1}^2\sum\limits_{k=1}^{n}C_{k,n+1}^{2}=\sum\limits_{i=1}^{n}\left(1-\pi_i^2\right)\Delta_i^2 \end{eqnarray*}
\begin{eqnarray}\label{gosl1}
+ \sum\limits_{i=1}^{n}\sum\limits_{s=1}^{n}\left(1 - \pi_s^2\right)B_{is}^2 + \sum\limits_{i=1}^{n}B_{i, n+1}^2 - {\cal E}_2^0 +I_2^0,
\end{eqnarray}
where ${\cal E}_2^0= \sum\limits_{i=1}^{n}{\cal E}_i^2, \
I_2^0= \sum\limits_{i=1}^{n}I_i^2,$
\end{corollary}
With points 4) and 5) of the description of the economy system $E_2$, we have
\begin{corollary}
In cost parameters, matrix elements $C_{ki}^{2}$ of the  matrix of  unproductive consumption satisfy the set of equations
\begin{eqnarray*} \sum\limits_{i=1}^{n+1}C_{ki}^{2}Y_i^2=C_k^1+N_k^1+ \sum\limits_{i=1}^{n+1}B_{ki}^1 +\sigma_k, \quad
k=\overline{1,n}, \end{eqnarray*} \begin{eqnarray*}
Y_i^2\sum\limits_{k=1}^{n}C_{ki}^{2}=\pi_i^1\left(\Delta_i^1+\sigma_i\right) + \pi_i^1\sum\limits_{k=1}^{n}B_{ki}^1,
\quad i=\overline{1,n}, \end{eqnarray*}
\begin{eqnarray*}Y_{n+1}^2\sum\limits_{k=1}^{n}C_{k,n+1}^{2} \end{eqnarray*}
\begin{eqnarray}\label{pgosl1}
=\sum\limits_{i=1}^{n}\left(1-\pi_i^1\right)\left(\Delta_i^1+\sigma_i\right)+
\sum\limits_{i=1}^{n}\sum\limits_{s=1}^{n}\left(1 - \pi_s^1\right)B_{is}^1 + \sum\limits_{i=1}^{n}B_{i, n+1}^1 - {\cal E}_1^0 +I_1^0.
\end{eqnarray}

\end{corollary}

In cost parameters, we determine the economy system $E_3$ by
\\ 1) a vector of gross outputs\index{vector of gross outputs}
\begin{eqnarray*}X_3(p_3)=\{X_i(p_3)\}_{i=1}^{n}, \quad X_i(p_3)=p_i^0x_i^3=p_i^3X_i^3,
\quad i=\overline{1, n}\ ;\end{eqnarray*}
\\ 2) a matrix of financial flows\index{matrix of financial flows}
 \begin{eqnarray*}||X_{ki}(p_3)||_{k,i=1}^{n}, \
X_{ki}(p_3)= p_k^0a_{ki}(p_3)x_i^3,
\quad k,i=\overline{1, n}\ ;\end{eqnarray*}
\\ 3) an input  matrix\index{input  matrix}
\begin{eqnarray*}  A(p_3)=||\bar a_{ki}(p_3)||_{i,j=1}^{n},
 \end{eqnarray*}
\begin{eqnarray*}
\bar a_{ki}(p_3)=\frac{X_{ki}(p_3)}{X_i(p_3)}
=\frac{p_k^0 a_{ki}(p_3)}{p_i^0}=\frac{p_k^3\bar a_{ki}^3}{p_i^3}, \quad k,i=\overline{1, n}\ ;\end{eqnarray*}
4) a matrix  of unproductive consumption\index{matrix  of unproductive consumption}
\begin{eqnarray*} C(p_3)=||C_{ki}(p_3)||_{k=1,i=1}^{n,\ n+1},\end{eqnarray*} \begin{eqnarray*}
 C_{ki}(p_3)=p_k^3C_{ki}^3, \quad k=\overline{1, n},
\quad i=\overline{1, n+1}, \end{eqnarray*}
and a matrix of  initial goods supply\index{ matrix of  initial goods supply}
\begin{eqnarray*} B(p_3)=||B_{ki}(p_3)||_{k=1,i=1}^{n,\ n+1}, \quad B_{ki}(p_3)=B_{ki}^3p_k^3, \quad k=\overline{1, n}, \quad i=\overline{1, n+1}\ ;\end{eqnarray*}
 5)  vectors of  export and import\index{vectors of  export and import}
\begin{eqnarray*}{\cal E}(p_3)=\{{\cal E}_k(p_3)\}_{k=1}^{n}, \quad
I(p_3)=\{I_k(p_3)\}_{k=1}^{n},\end{eqnarray*} \begin{eqnarray*}
{\cal E}_k(p_3)=p_k^3{\cal E}_k^3, \quad
 I_k(p_3)=p_k^3 I_k^3, \quad k=\overline{1, n}\ ; \end{eqnarray*}
 6) vectors of final consumption\index{ vectors of final consumption and gross accumulation and supply change} $C^f(p_3)=\{C_i(p_3)\}_{i=1}^{n}$
and gross accumulation and supply change $N(p_3)=\{N_i(p_3)\}_{i=1}^{n}$
 such  that \begin{eqnarray*} N_i(p_3)+C_i(p_3)=\sum\limits_{k=1}^{n+1}C_{ik}(p_3)Y_k^3 - \sum\limits_{k=1}^{n+1}B_{ik}(p_3),
\quad i=\overline{1, n}\ ; \end{eqnarray*}
7) a value added  of the $i$-th industry\index{value added  of the $i$-th industry}
$\Delta_i(p_3)$ and an income of social  group of  consumers $\bar D_{n+1}(p_3)$
\begin{eqnarray*}\Delta_i(p_3) = \sum\limits_{k=1}^{n}\left[C_{ki}(p_3)\frac{Y_i^3}{\pi_i^3} - B_{ki}(p_3)\right],
 \quad i=\overline{1, n},\end{eqnarray*} \begin{eqnarray*}
\bar D_{n+1}(p_3)= Y_{n+1}^3\sum\limits_{k=1}^{n}C_{k,n+1}(p_3)\ ;\end{eqnarray*}
8) a taxation vector\index{taxation vector} $\pi_3=\{\pi_i^3\}_{i=1}^{n}\ ;$ \\
9) a  vector of levels  of satisfaction of   consumers needs\index{vector of levels  of satisfaction of   consumers needs}
 \begin{eqnarray*} Y_3=\{ Y_i^3\}_{i=1}^{n+1}, \quad   Y_k^3> 0,
\quad k=\overline{1, n+1}. \end{eqnarray*}
For the quantities introduced,  the  set of balance  equations\index{set of balance  equations}
\begin{eqnarray*} X_i(p_3) - \sum\limits_{k=1}^{n} X_{ik}(p_3) - {\cal E}_i(p_3)
+ I_i(p_3) =N_i(p_3)+C_i(p_3), \quad i=\overline{1,n}, \end{eqnarray*}
\begin{eqnarray*}X_k(p_3)=\sum\limits_{i=1}^{n} X_{ik}(p_3) + \Delta_k(p_3),
\quad  k=\overline{1,n}, \end{eqnarray*}
holds.

The vector
$p_3=\{p_i^3\}_{i=1}^{n}$ solves the set of equations
(\ref{dl19}), the vector $X_3=\{X_i^3\}_{i=1}^{n}$ solves the set of equations
(\ref{bl21}), and the vector
$\pi_3=\{\pi_i^3\}_{i=1}^{n}$ is such that there exists a strictly positive solution to the set of equations (\ref{el1}).

In cost parameters, we describe a real economy system with the same notations as in the case of the economy system $E_1.$ Some of these parameters can be unsatisfactory. It may be an  input  matrix $ \bar A_1=||\bar a_{ki}(X_1)||_{i,j=1}^{n}$
 the part of value added  of the  $i$-th  industry
$\Delta_i^1$
in  the  industry output $X_i^1,$ a taxation system inadequate to the  economy development needs. Therefore, to enlarge, e.g., the  part of the value added   in the industry output,\index{ part of the value added   in the industry output} one must go from the industry output
$X_i^1$ to $X_i^1+\sigma_i$ enlarging  the industry value added\index{industry value added} from
$\Delta_i^1$ to $\Delta_i^1+ \sigma_i, \  \sigma_i \geq 0, \ i=\overline{1, n}.$
Such enlargement of the industry value added   can relate to  need of the  industry to reach the  given profitability, to increase  wages. One can reach it improving production technologies  such that the equilibrium price vector is the same because for the open economy system it is determined by demand and supply structures as well as by competing environment which does not allow to domestic firms to do prices exceeding external ones.

Therefore, we accept the main postulate that equilibrium prices in an open economy system determine whether  the economy is competitive  or not;
with advanced technologies \index{advanced technologies} or energy and materials expensive; with high levels of  consumption\index{high levels of  consumption}  or low; with equitable taxation system\index{equitable taxation system} or discriminative one. The discriminative taxation system\index{discriminative taxation system } only favors energy and material expensive production\index{energy and material expensive production} transferring taxation pressure on consumers that does not favor increase of   savings of  consumers  as domestic source of investment.\index{domestic source of investment}

Construct  a model economy system $E_4$ after the pattern of the  model economy system $E_3$ taking  natural parameters in the  economy system $E_3$ such  as in the economy system $E_2,$ i.e., suppose
\begin{eqnarray*}a_{ki}^4=a^{2}_{ki}\left(x_i^2\right), \quad k=\overline{1, n}, \quad i=\overline{1, n}, \quad c_{ki}^4= c_{ki}^2, \quad  k=\overline{1, n}, \quad i=\overline{1, n+1}.\end{eqnarray*}

Under such conditions, in cost parameters, we describe economy system $E_4$ by\\
1) a vector of  gross outputs\index{ vector of  gross outputs}
\begin{eqnarray*}X(p_4)=\{X_i(p_4)\}_{i=1}^{n}, \ X_i(p_4)=p_i^0x_i^4= p_i^4X_i^4,
\quad i=\overline{1, n}\ ;\end{eqnarray*}
\\ 2) a  matrix of financial flows\index{matrix of  financial flows}
 \begin{eqnarray*}||X_{ki}(p_4)||_{k,i=1}^{n}, \quad
X_{ki}(p_4)= p_k^0  a_{ki}(p_4)x_i^4,
\quad  a_{ki}(p_4)=\frac{p_k^4 a^{2}_{ki}\left(x_i^2\right)}{p_i^4},
\quad k,i=\overline{1, n}\ ;\end{eqnarray*}
\\ 3) an input  matrix\index{input  matrix}
\begin{eqnarray*} A(p_4)=||\bar a_{ki}(p_4)||_{k,i=1}^{n},  \end{eqnarray*}
\begin{eqnarray*} \bar a_{ki}(p_4)=\frac{X_{ki}(p_4)}{X_i(p_4)}
=\frac{p_k^0 a_{ki}(p_4)}{p_i^0}=\frac{p_k^4 \bar a_{ki}^2}{p_i^4}
=\frac{p_k^4X_{ki}^{2}}{\left(X_i^1+\sigma_i\right)p_i^4}, \quad k,i=\overline{1, n}\ ;\end{eqnarray*}
4) a matrix of unproductive consumption\index{ matrix of unproductive consumption}
\begin{eqnarray*}C(p_4)=||C_{ki}(p_4)||_{k,i=1}^{n, n+1},\end{eqnarray*} \begin{eqnarray*}
 C_{ki}(p_4)=p_k^4C_{ki}^{2}, \quad C_{ki}^{2} = p_k^0c_{ki}^{2}, \quad k=\overline{1, n},
\quad i=\overline{1, n+1}, \end{eqnarray*}
and a  matrix of  initial goods supply\index{ matrix of  initial goods supply}
\begin{eqnarray*}B(p_4)=||B_{ki}(p_4)||_{k,i=1}^{n, n+1},\end{eqnarray*} \begin{eqnarray*}
 B_{ki}(p_4)=p_k^4B_{ki}^{4}, \quad k=\overline{1, n},
\quad i=\overline{1, n+1}\ ; \end{eqnarray*}
 5)  vectors of  export and import\index{vectors of  export and import}
\begin{eqnarray*}{\cal E}(p_4)=\{{\cal E}_k(p_4)\}_{k=1}^{n}, \
I(p_4)=\{I_k(p_4)\}_{k=1}^{n},\end{eqnarray*} \begin{eqnarray*}
{\cal E}_k(p_4)=p_k^4{\cal E}_k^4, \quad
 I_k(p_4)=p_k^4 I_k^4,  \quad k=\overline{1, n}\ ; \end{eqnarray*}
 6) vectors of final consumtion\index{vectors of final consumption and gross accumulation and supply change} $C^f(p_4)=\{C_i(p_4)\}_{i=1}^{n}$
and gross accumulation and supply change $N(p_4)=\{N_i(p_4)\}_{i=1}^{n}$
 such that
\begin{eqnarray*} N_i(p_4)+C_i(p_4)=\sum\limits_{k=1}^{n+1}C_{ik}(p_4) Y_k^4 - \sum\limits_{k=1}^{n+1}B_{ik}(p_4),
\quad i=\overline{1, n}\ ; \end{eqnarray*}
7) a value added of the $i$-th industry\index{ value added of the $i$-th industry}
$\Delta_i(p_4)$ and an income of  social group of  consumers  $ D_{n+1}(p_4)$
\begin{eqnarray*}\Delta_i(p_4) = \sum\limits_{k=1}^{n}\left[C_{ki}(p_4)\frac{ Y_i^4}{\pi_i^4} - B_{ki}(p_4)\right],
 \quad i=\overline{1, n}, \end{eqnarray*} \begin{eqnarray*}
D_{n+1}(p_4)= Y_{n+1}^4\sum\limits_{k=1}^{n}C_{k,n+1}(p_4)\ ;\end{eqnarray*}
8) a taxation vector\index{taxation vector} $\pi_4=\{\pi_i^4\}_{i=1}^{n}\ ;$ \\
 9) a vector of levels  of satisfaction of   consumers needs\index{vector of levels  of satisfaction of   consumers needs}
 \begin{eqnarray*} Y_4=\{Y_i^4\}_{i=1}^{n+1},
\quad k=\overline{1, n+1}. \end{eqnarray*}
For quantities introduced, the  set of balance equations
\begin{eqnarray*} X_i(p_4) - \sum\limits_{k=1}^{n} X_{ik}(p_4) - {\cal E}_i(p_4)
+ I_i(p_4) =N_i(p_4)+C_i(p_4), \quad i=\overline{1,n}, \end{eqnarray*}
\begin{eqnarray*}X_k(p_4)=\sum\limits_{i=1}^{n} X_{ik}(p_4) + \Delta_k(p_4),
\quad  k=\overline{1,n}, \end{eqnarray*}
 holds.
The strictly positive vector
$X^4=\{X_i^4\}_{i=1}^{n}$ solves the set of equations
\begin{eqnarray}\label{boml21}
X_i^4 - \sum\limits_{k=1}^{n} \bar a_{ik}^{2}X_k^4 - {\cal E}_i^4 +I_i^4 + \sum\limits_{k=1}^{n+1}B_{ik}^4 =
 \sum\limits_{k=1}^{n+1} C_{ik}^{2} Y_k^4,
\quad  i=\overline{1,n},
 \end{eqnarray}
where
\begin{eqnarray*}  \bar a_{ij}^{2}=\frac{X_{ij}^{2}}{X_j^2}=\frac{p_i^0a^{2}_{ij}\left(x_j^{2}\right)}{p_j^0}=\frac{X_{ij}^{2}}{X_j^1+\sigma_j},
\quad i,j=\overline{1, n}.\end{eqnarray*}
Suppose \hfill  a \hfill non-negative \hfill vector \hfill $v_4=\{v_i^4\}_{i=1}^n$ \hfill exists \hfill such \hfill that \hfill the \hfill  matrix \\ $\bar C_2(v_4) - \bar B_4 $  is non-negative  having no zero rows or columns and the matrix $\bar A_2+ \bar C_2(v_4) - \bar B_4 $ is indecomposable. Assume the vector $\pi_4=\{\pi_i^4\}_{i=1}^{n}$ to be such that there exists a  strictly positive solution $g_4=\{g_i^4\}_{i=1}^{n}$
to the set of equations
\begin{eqnarray}\label{eoml1}
\frac{[(E- \bar A_2)^{-1}[\bar C_{2}(\pi_4^{-1}\bar Y_4) - \bar B_4]g_4]_i}{X_i^4} = g_i^4,
\quad  Y_i ^4> \pi_i^4v_i^4,   \quad i=\overline{1,n},
\end{eqnarray}
\begin{eqnarray*} \bar A_2=|| \bar a_{ij}^{2} ||_{i=1,j=1}^n, \quad   \bar C_2(\pi_4^{-1}\bar Y_4)=\left|\left|C_{ik}^{2}\frac{ Y_k^4}{\pi_k^4}\right|\right|_{i,k=1}^{n},\quad \bar B_4=||B_{ki}^4||_{k,i=1}^{n},\end{eqnarray*}
and a strictly positive vector $p_4=\{p_i^4\}_{i=1}^{n}$ solves the set of equations
\begin{eqnarray}\label{doml19}
p_i^4=\sum\limits_{k=1}^{n}p_k^4 \bar a_{ki}^{2} +
\frac{1}{X_i^4}\sum\limits_{k=1}^{n} \left[C_{ki}^{2}\frac{Y_i^4}{\pi_i^4}  - B_{ki}^4\right]p_k^4, \quad
i=\overline{1,n}.
\end{eqnarray}
Further, we suppose that in cost parameters the matrix elements $C_{ki}^2$ of the matrix $C_2=||C_{ki}^2||_{k=1,i=1}^{n, \ n+1}$ are determined by the solution to the set of equations (\ref{pgosl1}).

\section{Mathematical foundations of social agreement economy}

In this Section, we use relations between characteristics of the economy system in cost parameters obtained for the  model economy system with proportional consumption to describe real economy systems.
Principally, we accept the hypothesis that in a real economy system relations between characteristics of the  economy system  are the same  as in the economy with proportional consumption. We support the validity of this hypothesis introducing such notions as weighted mean price vector\index{weighted mean price vector} and net industries used to construct interindustry balance.
We construct aggregated  matrix of financial flows of unproductive consumption\index{aggregated  matrix of financial flows of unproductive consumption} in the real economy system as a solution to the set of equations (\ref{pgosl1}).

We accept the hypothesis that the transition to another taxation system\index{taxation system} happens in the same manner as in the economy with proportional consumption.\index{economy with proportional consumption}
The Theorem \ref{ss0s9} states that every aggregated description\index{aggregated description} corresponds to non-aggregated\index{non-aggregated description } one with the same integral taxation system\index{integral taxation system} as in the real economy system.
This result and hypotheses accepted clarify what structure changes must happen in technologies, gross outputs, consumptions, and foreign economic relations to reach the desired economy state.

\begin{definition}\label{ss0s10} We say the economy system is described aggregately if the information about the economy system is given by\\
1) a vector of  gross outputs\index{vector of  gross outputs} $X=\{X_i\}_{i=1}^{n}, \ X_i >0, \ i=\overline{1,n}\ ;$  \\
2) a matrix of financial flows\index{matrix of financial flows}
$||X_{ik}||_{i,k=1 }^{n}$ describing input structure of $n$
net industries;\\
3) an input  matrix\index{input  matrix} related to financial flows and gross outputs and expressed as follows
\begin{eqnarray*}  \bar A=||\bar a_{ij}||_{i,j=1}^{n},  \quad
\bar a_{ij}=\frac{X_{ij}}{X_j}\ ;\end{eqnarray*}
4)  vectors of export and import\index{vectors of export and import}
${\cal E}=\{ {\cal E}_i\}_{i=1}^{n},$  $I=\{ I_i\}_{i=1}^{n}\ ;$ \\
5) vectors of final consumption\index{vectors of final consumption and gross accumulation and supply change}
$C^f=\{C_i\}_{i=1}^{n}$
and gross accumulation and supply change  $N=\{N_i\}_{i=1}^{n}\ ;$ \\
6) a value added\index{value added}  $\Delta_i, \  i=\overline{1, n},$ of the  $i$-th industry.
 \\
7) initial goods supply matrix\index{initial goods supply matrix} $ B=||B_{ki}||_{k=1, i=1}^{n, n+1}$ at the beginning of the economy operation period; \\
8) a vector  of integral taxation\index{vector  of integral taxation} $\pi=\{\pi_i\}_{i=1}^{n}.$ \\
All these data generate interindustry balance\index{interindustry balance}
\begin{eqnarray*}\sum\limits_{k=1}^{n}X_{ik} +C_i+N_i+ {\cal
E}_i -I_i= X_i, \quad i=\overline{1,n},\end{eqnarray*}
\begin{eqnarray*}\sum\limits_{i=1}^{n} X_{ik} +\Delta_k= X_k, \quad
k=\overline{1,n}.\end{eqnarray*}
\end{definition}
By the characteristics of the  aggregated description of the economy system,\index{ characteristics of the  aggregated description of the economy system} let us construct a non-negative rectangular matrix
$ C=||C_{ki}||_{k=1, i=1}^{n, n+1}$
whose matrix  elements satisfy the set of equations
\begin{eqnarray*} \sum\limits_{i=1}^{n+1}C_{ki}Y_i=C_k+N_k+ \sum\limits_{i=1}^{n+1}B_{ki}, \quad
k=\overline{1,n}, \end{eqnarray*} \begin{eqnarray*}
Y_i\sum\limits_{k=1}^{n}C_{ki}=\pi_i\Delta_i + \pi_i\sum\limits_{k=1}^{n}B_{ki},
\quad i=\overline{1,n}, \end{eqnarray*}
\begin{eqnarray*}Y_{n+1}\sum\limits_{k=1}^{n}C_{k,n+1}=\end{eqnarray*}
\begin{eqnarray}\label{gorl60}
=\sum\limits_{i=1}^{n}(1-\pi_i)\Delta_i+
\sum\limits_{i=1}^{n}\sum\limits_{s=1}^{n}(1 - \pi_s)B_{is} + \sum\limits_{i=1}^{n}B_{i, n+1} - {\cal E} +I,
\end{eqnarray}
where ${\cal E}= \sum\limits_{i=1}^{n}{\cal E}_i, \
I= \sum\limits_{i=1}^{n}I_i,$ and $Y= \{Y_i\}_{i=1}^{n+1}$ is a certain strictly positive vector.

\begin{theorem}\label {ss0s9} Let the economy system be described aggregately and for aggregated data\index{aggregated data } the inequalities
 \begin{eqnarray*}C_k+N_k+ \sum\limits_{i=1}^{n+1}B_{ki}> 0, \quad  k=\overline{1,n},\end{eqnarray*}
\begin{eqnarray}\label{ssos8}
\sum\limits_{i=1}^{n}(1-\pi_i)\Delta_i+
\sum\limits_{i=1}^{n}\sum\limits_{s=1}^{n}(1 - \pi_s)B_{is} + \sum\limits_{i=1}^{n}B_{i, n+1} - {\cal E} +I > 0,
\end{eqnarray}
hold.
Then there exists a  matrix of  unproductive consumption\index{ matrix of  unproductive consumption}  $\bar C=||C_{ki}||_{k=1, i=1}^{n, n+1}$  being a solution to  the set of equations (\ref{gorl60}) and a non-negative vector $v=\{v_i\}_{i=1}^n$ for which the matrix
 $\bar C(v) - \bar B$ is non-negative  having no zero rows or columns and the matrix $A + \bar C(v) - \bar B$ is indecomposable. If the inequalities
\begin{eqnarray*}Y_i > \pi_i v_i, \quad  i=\overline{1,n},\end{eqnarray*}
hold then the integral taxation vector\index{integral taxation vector agrees with  structure of consumption}
\begin{eqnarray*}\pi=\{\pi_i\}_{i=1}^{n}, \quad  0  < \pi_i <1, \quad
i=\overline{1,n},\end{eqnarray*}
  agrees with  structure of consumption in the  economy system  given by a vector of levels of  satisfaction of  consumers needs
$Y= \{Y_i\}_{i=1}^{n+1}.$
\end{theorem}
\begin{proof}\smartqed
Prove the existence of a solution to the set of equations
(\ref{gorl60}).
The equality
\begin{eqnarray*}\sum\limits_{k=1}^{n}\left(C_k+N_k+ \sum\limits_{i=1}^{n+1}B_{ki}\right) - \sum\limits_{k=1}^{n}
\left(\pi_k\Delta_k + \pi_k\sum\limits_{i=1}^{n}B_{ik}\right)
\end{eqnarray*}
\begin{eqnarray*}=
\sum\limits_{i=1}^{n}(1-\pi_i)\Delta_i+
\sum\limits_{i=1}^{n}\sum\limits_{s=1}^{n}(1 - \pi_s)B_{is} + \sum\limits_{i=1}^{n}B_{i, n+1} - {\cal E} +I\end{eqnarray*}
 holds which is nothing else than the equality
\begin{eqnarray}\label{ssos7}
 \sum\limits_{j=1}^{n}(C_j +N_j)=\sum\limits_{j=1}^{n}\Delta_j - {\cal E}+ I.
\end{eqnarray}
The equality (\ref{ssos7}) holds because it follows from the  set of balance equations.
In view of conditions (\ref{ssos8}) of the Theorem \ref{ss0s9}, the conditions of the  Lemma \ref{ssos1}  hold.
Therefore, according to the Lemma \ref{ssos1}, the solution to the set of equations
(\ref{gorl60})
is the matrix
$ C=||C_{ki}||_{k=1, i=1}^{n, n+1},$
where
\begin{eqnarray*}
C_{ki}=\frac{\left(C_k+N_k + \sum\limits_{j=1}^{n+1} B_{kj}\right)\left(\pi_i\Delta_i+ \pi_i \sum\limits_{j=1}^{n}B_{ji}\right)}{Y_i \sum\limits_{j=1}^{n}\left(C_j +N_j +\sum\limits_{m=1}^{n+1}B_{jm}\right)},
\quad  k,i=\overline{1,n},\end{eqnarray*}
\begin{eqnarray*}
C_{k,n+1}\end{eqnarray*}
\begin{eqnarray*}
=\frac{\left(C_k+N_k + \sum\limits_{j=1}^{n+1} B_{kj}\right)\left[\sum\limits_{j=1}^{n}\left(C_j +N_j +\sum\limits_{m=1}^{n+1}B_{jm}\right)  -\sum\limits_{i=1}^{n} \left(\pi_i\Delta_i+ \pi_i \sum\limits_{j=1}^{n}B_{ji}\right) \right]}{Y_{n+1}\sum\limits_{j=1}^{n}\left(C_j +N_j +\sum\limits_{m=1}^{n+1}B_{jm}\right)}, \end{eqnarray*} \begin{eqnarray*}  k=\overline{1,n}.\end{eqnarray*}
From the  conditions  of the  Theorem \ref{ss0s9} it follows that matrix  elements of the matrix
$ C$
are strictly positive, therefore, a non-negative vector
$v=\{v_i\}_{i=1}^n$
exists for which strict inequality $\bar C(v) - \bar B  >  0$ holds and the matrix $A + \bar C(v) - \bar B$ is indecomposable.

From  the  conditions   of   the   Theorem \ref{ss0s9},  it  follows  that  the  vector  $p=\{p_i\}_{i=1}^{n},$  $ p_i=1,
\ i=\overline{1,n},$ is
a solution to the set of equations
\begin{eqnarray}\label{bl1}
p_i=\sum\limits_{k=1}^{n}\bar a_{ki}p_k +
\frac{1}{X_i}\sum\limits_{k=1}^{n}\left[C_{ki}\frac{Y_i}{\pi_i} - B_{ki}\right] p_k,
\quad i=\overline{1,n}.
\end{eqnarray}
Really, inserting the vector $p$ to the set of equations (\ref{bl1}), we have
\begin{eqnarray*} 1 -\sum\limits_{k=1}^{n}\bar a_{ki} = \frac{1}{X_i}\sum\limits_{k=1}^{n}\left[C_{ki}\frac{Y_i}{\pi_i} - B_{ki}\right]=
\frac{\Delta_i}{X_i},
\quad i=\overline{1,n},
\end{eqnarray*}
that is the same as
\begin{eqnarray*}X_i =\sum\limits_{k=1}^{n}\bar a_{ki}X_i+\Delta_i,
\quad i=\overline{1,n}.
\end{eqnarray*}
Let a vector  of levels  of satisfaction of consumers needs  is such that  the inequalities  $Y_i> v_i \pi_i , \ i=\overline{1, n},$ hold.

As the set of equations (\ref{bl1}) is solvable in the set $T_0$, from the Theorem \ref{vjant14} it follows  that the taxation vector\index{ taxation vector agrees with the structure of consumption}
$\pi=\{\pi_i\}_{i=1}^{n}$
agrees with the structure of consumption  given by  a vector  of levels  of satisfaction of consumers needs
$Y_i, \ i=\overline{1,n+1}. $
It is obvious that the representation holds
\begin{eqnarray*} X_i - \sum\limits_{k=1}^{n}\bar a_{ik}X_k
+ \sum\limits_{k=1}^{n+1}B_{ik} - {\cal E}_i + I_i =
 \sum\limits_{k=1}^{n+1} C_{ik}Y_k,  \quad
 \ k=\overline{1,n}. \end{eqnarray*}
Introduce notations
\begin{eqnarray*}M=V - {\cal E} +I, \quad V= \sum\limits_{i=1}^{n}\Delta_i+ \sum\limits_{j=1}^{n}\sum\limits_{m=1}^{n+1}B_{jm},
\quad {\cal E}= \sum\limits_{i=1}^{n}{\cal E}_i, \quad
I= \sum\limits_{i=1}^{n}I_i.\end{eqnarray*}
Then we have
\begin{eqnarray*} \sum\limits_{j=1}^{n}\left(C_j +N_j +\sum\limits_{m=1}^{n+1}B_{jm}\right)-\sum\limits_{i=1}^{n} \left(\pi_i\Delta_i+ \pi_i \sum\limits_{j=1}^{n}B_{ji}\right)=\end{eqnarray*} \begin{eqnarray*}=\sum\limits_{i=1}^{n}(1 - \pi_i)\Delta_i+
\sum\limits_{j=1}^{n}\sum\limits_{m=1}^{n}(1 - \pi_m)B_{jm}+\sum\limits_{j=1}^{n}B_{j,n+1} -{\cal E}+I.\end{eqnarray*}
From here we obtain
\begin{eqnarray*}
C_{ki}=\frac{\left(C_k+N_k + \sum\limits_{j=1}^{n+1} B_{kj}\right)\left(\pi_i\Delta_i+ \pi_i \sum\limits_{j=1}^{n}B_{ji}\right)}{Y_i M},
\quad  k, \ i=\overline{1,n},\end{eqnarray*}
\begin{eqnarray*}C_{k,n+1}=\left(C_k+N_k + \sum\limits_{j=1}^{n+1} B_{kj}\right) \end{eqnarray*} \begin{eqnarray*} \times\frac{\sum\limits_{i=1}^{n} (1-\pi_i)\Delta_i+ \sum\limits_{j=1}^{n}\sum\limits_{i=1}^{n}(1-\pi_i) B_{ji}+\sum\limits_{k=1}^{n} B_{k, n+1} -{\cal E}+I}{Y_{n+1}M}, \quad  k=\overline{1,n}.\end{eqnarray*}
\qed
\end{proof}

\begin{note} Just constructed solution to the set of equations
(\ref{gorl60}) is one of possible, however, very important one. It has a simple economic sense, namely, each industry consumes the goods of final consumption proportionally to net income. The construction of other possible solutions must ground on additional information about the  structure of capital investment of industries. Under Ukrainian economy conditions, where such investments is not significant, the solution constructed is admissible approximation.
\end{note}

We accept that aggregated description\index{aggregated description } from the Definition \ref{ss0s10} describes the economy in the year we call basic one.
Introduce the weighted-mean price in the $i$-th industry\index{weighted-mean price in the $i$-th industry} basing on  industry  output $X_i$
of the basic year. If the  $i$-th industry produces $m_i$ homogeneous goods\index{homogeneous goods} in quantities $x_1^i, \ldots, x_{m_{i}}^i,$ then the weighted mean price in the $i$-th industry in the basic year has the form
\begin{eqnarray}\label{bul3}
p_i^0=\frac{\sum\limits_{k=1}^{m_{i}}p_k^ix_k^i}
{\sum\limits_{k=1}^{m_{i}}x_k^i},
 \quad  i=\overline{1,n},
\end{eqnarray}
where $\sum\limits_{k=1}^{m_{i}}p_k^ix_k^i=X_i$ and notations are introduced
$\sum\limits_{k=1}^{m_{i}}x_k^i=x_i,$ \ $p_k^i$ is the price of $k$-th goods in the $i$-th industry.
After financial flow\index{financial flow} $X_{ij} $ from
the $j$-th industry into the $i$-th one, introduce aggregated matrix elements of direct inputs in natural parameters\index{aggregated matrix elements of direct inputs in natural parameters} basing on the equality
\begin{eqnarray*}X_{ij}=p_i^0a_{ij}x_j,\end{eqnarray*}
where $a_{ij}$ are aggregated matrix elements of direct inputs in natural parameters having economic sense of number of units of the $i$-th  aggregated  goods to produce one aggregated unit of the $j$-th goods.
After the matrix of financial flows for unproductive consumption\index{matrix of financial flows for unproductive consumption} $C_{ij},$
let us introduce  matrix elements  of aggregated  consumption in natural parameters\index{matrix elements  of aggregated  consumption in natural parameters} basing on the equality
\begin{eqnarray*}C_{ij}=p_i^0c_{ij},\end{eqnarray*}
where $c_{ij}$ are  matrix elements of aggregated unproductive consumption in natural parameters\index{ matrix elements of aggregated unproductive consumption in natural parameters} with economic sense of units of the $i$-th final consumption goods to be consumed by employees of the $j$-th industry.

Then the aggregated  matrix elements of direct inputs  in natural parameters\index{aggregated  matrix elements of direct inputs  in natural parameters} are  related to the aggregated matrix elements in cost  parameters\index{aggregated matrix elements in cost  parameters} as follows
\begin{eqnarray}\label{fl3}
\bar a_{ij}=\frac{X_{ij}}{X_j}=\frac{p_i^0a_{ij}}{p_j^0}.
\end{eqnarray}
This relation between aggregated quantities introduced is the same as in the case of the  economy system containing net industries only. The matrix
$\bar A(X)=||\bar a_{ij}||_{i,j=1}^{n}$ depends on vector of  gross outputs  $X.$

\begin{definition}
The description  of the economy system  being described aggregately in cost parameters
whose equilibrium state  is a solution  to the set of equations  in cost parameters
\begin{eqnarray}\label{kl16}
\bar \psi_k=\sum\limits _{i=1}^{n}
\frac{C_{ki}\pi_i\left[X_i\left(p_i -\sum\limits _{s=1}^{n} \bar a_{si}p_s\right)+ \sum\limits _{s=1}^{n}B_{si}p_s\right]}{
\sum\limits _{s=1}^{n}C_{si}p_s} \end{eqnarray}
 \begin{eqnarray*}
+ \frac{C_{k,n+1} \tilde D_{n+1}(p)}{\sum\limits _{l=1}^{n}
 C_{l,n+1}p_l},\quad k=\overline{1,n}, 
\end{eqnarray*}
for the vector of relative prices\index{vector of relative prices}
\begin{eqnarray*}p=\{p_i \}_{i=1}^{n},  \quad
p_i=\frac{\bar p_i}{\bar p_i^0}, \end{eqnarray*}
where the vector $p=\{\bar p_i\}_{i=1}^{n}$ is an equilibrium weighted mean price vector\index{equilibrium weighted mean price vector} in the year considered, $\bar p^0=\{\bar p_i^0\}_{i=1}^{n} $ is an equilibrium weighted mean price vector in the basic year,\index{ equilibrium weighted mean price vector in the basic year}
and
\begin{eqnarray*}\bar \psi_k =X_k- \sum\limits _{i=1}^{n}\bar a_{ki}X_i+ \sum\limits _{i=1}^{n+1}B_{ki} -
{\cal E}_k + I_k, \end{eqnarray*}
\begin{eqnarray*}\tilde D_{n+1}(p)= Y_{n+1}\sum\limits_{k=1}^{n}C_{k,n+1}p_k,\end{eqnarray*}
that means the equality of the demand and the supply in cost parameters,
we call associated model of  description of the  economy system.\index{associated model of  description of the  economy system}
 We call the description itself of the equilibrium state of the economy system  by the set of equations (\ref{kl16}) as non-aggregated description.\index{non-aggregated description}
\end{definition}
\begin{note}
If the  considered and basic years are the same ones, then the solution to the set of equations (\ref{kl16}) must be the vector
$p=\{p_i\}_{i=1}^{n}$ with $ \ p_i=1, \ i=\overline{1, n}.$
\end{note}
\begin{note}
The Theorem \ref{ss0s9} shows that with aggregated description of the economy one  can relate  non-aggregated description.
\end{note}
\begin{definition} We call the  economy  described aggregately  the  economy of social agreement\index{ economy of social agreement} if\\
1) every industry guarantees level of wage  corresponding to needs and law;\index{level of wage  corresponding to needs and law}\\
2) every industry is non-detrimental;\\
3) social allocation\index{social allocation} is sufficient  to guarantee adequate pension;\\
4) taxation system is such that integral taxation of  industry\index{integral taxation of  industry}  is proportional to created value in the industry\index{created value in the industry}  with the same rate of  taxation for all the industries;\index{the same rate of  taxation for all the industries}\\
5)  balance of  foreign economic relations\index{balance of  foreign economic relations} is non-negative.
\end{definition}
\begin{definition} We call the economy described aggregately the perfect competitive economy\index{perfect competitive economy of social agreement} of social agreement if it is the economy of social agreement with equal  levels of  profitability.\index{equal  levels of  profitability}
\end{definition}

\subsection{ Applications}

Suppose that in a basic year\index{basic year} we describe  the economy system $\bar E_1$  in aggregated way by cost parameters having the same notations  as in the  description of the economy system
$E_1,$ i.e., by

\noindent 1)  a vector of gross outputs\index{vector of gross outputs}  $X_1=\{X_i^1\}_{i=1}^{n}\ ;$

\noindent 2)  a matrix of  financial flows\index{matrix of  financial flows}
$||X_{ik}^1||_{i,k=1 }^{n}$ describing input structure of $n$
net industries;

\noindent 3)   an input  matrix\index{input  matrix} related to the  matrix of  financial flows and gross outputs and expressed as follows
\begin{eqnarray*}  \bar A_1=||\bar a_{ij}(X_1)||_{i,j=1}^{n},  \quad
\bar a_{ij}(X_1)=\frac{X_{ij}^1}{X_j^1}\ ;\end{eqnarray*}
4)  vectors of  export and import\index{vectors of  export and import}
${\cal E}_1=\{ {\cal E}_i^1\}_{i=1}^{n},$  $I_1=\{ I_i^1\}_{i=1}^{n}\ ;$

\noindent 5) vectors of final consumption\index{vectors of final consumption and gross accumulation and supply change}
$C_1^f=\{C_i^1\}_{i=1}^{n}$ \hfill
and gross accumulation and supply change $N_1=\{N_i^1\}_{i=1}^{n}\ ;$

\noindent 6) an added value\index{added value}  $\Delta_i^1= \Delta_i''+\Delta_i', \ i=\overline{1, n},$ of the $i$-th industry having the structure
\begin{eqnarray*}\Delta_i^1=[z_i+\gamma_i+g_i+\tau_i+\alpha_i+k_i]+ \rho_i, \quad
i=\overline{1,n}, \end{eqnarray*}
\begin{eqnarray*}  \Delta_i'' =z_i+g_i+k_i, \
\Delta_i'=\gamma_i+\tau_i+\alpha_i+ \rho_i,\quad i=\overline{1,n},\end{eqnarray*}
where $z_i$ is wage in the $i$-th industry;\index{ wage in the $i$-th industry}
$\gamma_i$ is an allocation into social funds\index{allocation into social funds} (real as well as conditional ones); $g_i$ is the net profit;\index{net profit} $\tau_i$ is a   profit  tax;\index{profit  tax}
$\alpha_i$ are the  other production taxes;\index{production taxes} $k_i$ is the consumption of  basic capital\index{consumption of  basic capital} (basic assets); $\rho_i$ is products and import tax;\index{products and import tax}

\noindent 7) a  vector of   integral taxation\index{vector of   integral taxation}  $\pi_1=\{\pi_i^1\}_{i=1}^{n}$ such that
\begin{eqnarray*}\Delta_i''=\pi_i^1\Delta_i^1,\quad   i=\overline{1, n}\ ;\end{eqnarray*}
\noindent 8) a matrix of initial goods supply\index{matrix of initial goods supply}  $||B_{ki}^1||_{k=1,i=1}^{n,\ n+1}$ at the beginning of the economy operation period.

These data form interindustry balance\index{interindustry balance}
\begin{eqnarray*}\sum\limits_{k=1}^{n}X_{ik}^{1} +C_i^1+N_i^1+ {\cal
E}_i^1 -I_i^1= X_i^{1}, \quad i=\overline{1,n},\end{eqnarray*}
\begin{eqnarray*}\sum\limits_{i=1}^{n} X_{ik}^{1} +\Delta_k^1= X_k^{1}, \quad
k=\overline{1,n}.\end{eqnarray*}

Let us  go to the   economy system $\bar E_2$ described aggregately in cost parameters with the same notations  as for the description of the  economy system $E_2$  under the additional requirement that the conditions
 $a_{ij}^1\left(x_j^1\right)x_j^1=
a_{ij}^{2}\left(x_j^2\right)x_j^{2}, \ i,j=\overline{1, n},$  hold  meaning that conditions
(\ref{dl131}) hold.

Therefore, we describe the  economy system $\bar E_2$ by\\
1) a vector of gross outputs\index{vector of gross outputs}  $X_2=\{X_i^1+\sigma_i\}_{i=1}^{n}\ ;$  \\
2) a matrix of financial flows\index{matrix of financial flows}
$||X_{ik}^2||_{i,k=1 }^{n}$ describing input structure of $n$
net industries;\\
3) an input  matrix\index{input  matrix} related to financial flows matrix and gross outputs and given as follows
\begin{eqnarray*}  \bar A_{2}=||\bar a_{ij}^{2}||_{i,j=1}^{n},  \quad
\bar a_{ij}^{2}=\frac{X_{ij}^2}{X_j^2}=\frac{X_{ij}^1}{X_j^1+ \sigma_j}= \bar a_{ij}(X_1)\beta_j,
\quad  \beta_j=\frac{X_j^1}{X^1_j+ \sigma_j}\ ;\end{eqnarray*}
4) vectors of export and import\index{vectors of export and import}
${\cal E}_1=\{ {\cal E}_i^1\}_{i=1}^{n},$  $I_1=\{ I_i^1\}_{i=1}^{n}\ ;$ \\
5)  vectors of final consumption\index{vectors of final consumption and gross accumulation and supply change}
$C_2^2=\{C_i^2\}_{i=1}^{n}$
and gross accumulation and supply change $N_2=\{N_i^2\}_{i=1}^{n}$
such that
\begin{eqnarray*}N_i^2+C_i^2=  N_i^1+C_i^1+ \sigma_i, \quad i=\overline{1, n}\ ;
\end{eqnarray*}
6) a value added\index{value added of the  $i$-th industry}  $\Delta_i^2= \Delta_i^1 + \sigma_i, \
i=\overline{1, n},$ of the $i$-th industry
where $\Delta_i^1 $ is the value added of the  $i$-th industry in $\bar E_1\ ;$

\noindent 7) the same vector of  integral taxation\index{vector of  integral taxation}  $\pi_2=\pi_1=\{\pi_i^1\}_{i=1}^{n}$ as in $\bar E_1\ ;$

\noindent 8) a  matrix of initial goods supply\index{matrix of initial goods supply} $||B_{ki}^2||_{k=1,i=1}^{n,\ n+1}$ at the beginning of the economy operation period $B_{ki}^2=B_{ki}^1, \ k=\overline{1, n}, \ i =\overline{1, n+1}\ ;$

All these data form the  interindustry balance\index{interindustry balance}
\begin{eqnarray*}\sum\limits_{k=1}^{n}X_{ik}^{1} +C_i^2+N_i^2+ {\cal
E}_i^1 -I_i^1= X_i^{1}+\sigma_i, \quad i=\overline{1,n},\end{eqnarray*}
\begin{eqnarray*}\sum\limits_{i=1}^{n} X_{ik}^{1} +\Delta_k^2= X_k^{1}+ \sigma_k, \quad
k=\overline{1,n}.\end{eqnarray*}
After these aggregated data,\index{aggregated data} we build a  matrix of unproductive consumption\index{matrix of unproductive consumption in cost parameters}
$ C_{2}=||C_{ki}^{2}||_{k=1, i=1}^{n,\ n+1}$
in cost parameters as the solution to the set of inequalities and equations (see (\ref{pgosl1}) )
\begin{eqnarray*} C_{ki}^{2} \geq 0, \quad k=\overline{1,n}, \quad
i=\overline{1,n+1}, \end{eqnarray*}
\begin{eqnarray*} \sum\limits_{i=1}^{n+1}C_{ki}^{2}Y_i^2=C_k^1+N_k^1+ \sum\limits_{i=1}^{n+1}B_{ki}^1 +\sigma_k, \quad
k=\overline{1,n}, \end{eqnarray*} \begin{eqnarray*}
Y_i^2\sum\limits_{k=1}^{n}C_{ki}^{2}=\pi_i^1\left(\Delta_i^1+\sigma_i\right) + \pi_i^1\sum\limits_{k=1}^{n}B_{ki}^1,
\quad i=\overline{1,n}, \end{eqnarray*}
\begin{eqnarray*}Y_{n+1}^2\sum\limits_{k=1}^{n}C_{k,n+1}^{2}=\sum\limits_{i=1}^{n}\left(1-\pi_i^1\right)\left(\Delta_i^1+\sigma_i\right) \end{eqnarray*}
\begin{eqnarray}\label{ss0s11}
+ \sum\limits_{i=1}^{n}\sum\limits_{s=1}^{n}\left(1 - \pi_s^1\right)B_{is}^1 + \sum\limits_{i=1}^{n}B_{i, n+1}^1 - {\cal E}_1^0 +I_1^0,
\end{eqnarray}
\begin{eqnarray*} {\cal E}_1^0= \sum\limits_{i=1}^{n}{\cal E}_i^1, \quad
I_1^0= \sum\limits_{i=1}^{n}I_i^1,\end{eqnarray*}
under the condition that the  vector of level of satisfaction of  consumers needs\index{vector of level of satisfaction of  consumers needs} is such that
\begin{eqnarray*}Y_2= \{Y_i^2\}_{i=1}^{n+1}, \quad Y_i^2=Y_i^1 > \pi_i^1v_i^1, \quad i=\overline{1, n}.\end{eqnarray*}

\begin{lemma}\label{p1}
Let the inequalities
\begin{eqnarray*}C_k^1+N_k^1+ \sum\limits_{i=1}^{n+1}B_{ki}^1 +\sigma_k> 0, \quad k=\overline{1, n},\end{eqnarray*}
\begin{eqnarray*}\sum\limits_{i=1}^{n} \left(1-\pi_i^1\right)\left(\Delta_i^1+\sigma_i\right)+ \sum\limits_{j=1}^{n}\sum\limits_{i=1}^{n}\left(1-\pi_i^1\right) B_{ji}^1+\sum\limits_{k=1}^{n} B_{k, n+1}^1- {\cal E}_1^0+ I_1^0 > 0, \end{eqnarray*}
 hold.
The solution to the set of equations and inequalities (\ref{ss0s11}) is
\begin{eqnarray*}
C_{ki}^2=\frac{\left(C_k^1+N_k^1 + \sigma_k + \sum\limits_{j=1}^{n+1} B_{kj}^1\right)\left(\pi_i^1\left(\Delta_i^1+\sigma_i\right)+ \pi_i^1 \sum\limits_{j=1}^{n}B_{ji}^1\right)}{Y_i^1 M_1},
\quad  k,i=\overline{1,n},\end{eqnarray*}
\begin{eqnarray*} C_{k,n+1}^2=\left(C_k^1+N_k^1  + \sigma_k + \sum\limits_{j=1}^{n+1} B_{kj}^1\right) \end{eqnarray*}
\begin{eqnarray*}\times\frac{\sum\limits_{i=1}^{n} \left(1-\pi_i^1\right)\left(\Delta_i^1+\sigma_i\right)+ \sum\limits_{j=1}^{n}\sum\limits_{i=1}^{n}\left(1-\pi_i^1\right) B_{ji}^1+\sum\limits_{k=1}^{n} B_{k, n+1}^1- {\cal E}_1^0+ I_1^0}{Y_{n+1}^1M_1}, \quad  k=\overline{1,n},\end{eqnarray*}
\begin{eqnarray*}M_1=V_1 - {\cal E}_1^0 +I_1^0, \quad V_1= \sum\limits_{i=1}^{n}\left(\Delta_i^1+\sigma_i\right)+ \sum\limits_{j=1}^{n}\sum\limits_{m=1}^{n+1}B_{jm}^1,\end{eqnarray*} \begin{eqnarray*}
{\cal E}_1^0= \sum\limits_{i=1}^{n}{\cal E}_i^1, \quad
I_1^0= \sum\limits_{i=1}^{n}I_i^1.\end{eqnarray*}
\end{lemma}

Finally, consider the economy system $\bar E_4$ described in an  aggregated way by

\noindent 1) a vector of  gross outputs\index{vector of  gross outputs}
\begin{eqnarray*}X(p_4)=\{X_i(p_4)\}_{i=1}^{n}, \ X_i(p_4)= p_i^4X_i^4,
\quad i=\overline{1, n}\ ;\end{eqnarray*}
2) a  matrix of financial flows \index{matrix of financial flows}
 \begin{eqnarray*}||X_{ki}(p_4)||_{k,i=1}^{n}, \quad
X_{ki}(p_4)= p_k^4\bar a_{ki}^{2}X_i^4=p_k^4\bar a_{ki}(X_1)\beta_i X_i^4,
\quad k,i=\overline{1, n}\ ;\end{eqnarray*}
\\ 3) an input  matrix\index{input  matrix}
\begin{eqnarray*} \bar A(p_4)=||\bar a_{ki}(p_4)||_{i,j=1}^{n},  \end{eqnarray*}
\begin{eqnarray*} \bar a_{ki}(p_4)=\frac{X_{ki}(p_4)}{X_i(p_4)}
=\frac{p_k^4\bar a_{ki}(X_1)\beta_i}{p_i^4}=
\frac{p_k^4X_{ki}^1\beta_i}{X_i^1p_i^4}, \quad k,i=\overline{1, n},\end{eqnarray*}
\begin{eqnarray*} \beta_i=\frac{X_i^1}{X_i^1+\sigma_i}\ ;\end{eqnarray*}
4) a matrix of  unproductive consumption\index{ matrix of  unproductive consumption}
\begin{eqnarray*} C(p_4)=||C_{ki}(p_4)||_{k,i=1}^{n, n+1},\end{eqnarray*}
\begin{eqnarray*} C_{ki}(p_4)=p_k^4C_{ki}^{2}, \quad k=\overline{1, n},
\quad i=\overline{1, n+1}, \end{eqnarray*}
and a matrix of initial goods supply\index{matrix of initial goods supply}
\begin{eqnarray*} B(p_4)=||B_{ki}(p_4)||_{k,i=1}^{n, n+1},\end{eqnarray*}
 \begin{eqnarray*} B_{ki}(p_4)=p_k^4B_{ki}^{4}, \quad B_{ki}^{4}=B_{ki}^{2} \quad k=\overline{1, n},
\quad i=\overline{1, n+1}\ ; \end{eqnarray*}
 5)  vectors of  export and import\index{vectors of  export and import}
\begin{eqnarray*}{\cal E}(p_4)=\{{\cal E}_k(p_4)\}_{k=1}^{n}, \quad
I(p_4)=\{I_k(p_4)\}_{k=1}^{n},\ \end{eqnarray*}
\begin{eqnarray*}{\cal E}_k(p_4)=p_k^4{\cal E}_k^2, \quad
 I_k(p_4)=p_k^4 I_k^2, \quad k=\overline{1, n}\ ; \end{eqnarray*}
 6)  vectors of final consumption\index{vectors of final consumption and gross accumulation and supply change} $C^f(p_4)=\{C_i(p_4)\}_{i=1}^{n}$
and gross accumulation and supply change $N(p_4)=\{N_i(p_4)\}_{i=1}^{n}$
such that
\begin{eqnarray*} N_i(p_4)+C_i(p_4)=\sum\limits_{k=1}^{n+1}C_{ik}(p_4)Y_k^4 - \sum\limits_{k=1}^{n+1}B_{ik}(p_4),
\quad i=\overline{1, n}\ ; \end{eqnarray*}
 7) a value added   of the $i$-th industry\index{value added   of the $i$-th industry}
\begin{eqnarray*}\Delta_i(p_4) = \sum\limits_{k=1}^{n}\left[C_{ki}(p_4)\frac{Y_i^4}{\pi_i^4}- B_{ki}(p_4)\right],
 \quad i=\overline{1, n}\ ;\end{eqnarray*}
 8) a taxation vector\index{taxation vector} $\pi_4=\{\pi_i^4\}_{i=1}^{n}\ ;$ \\
 9) a  vector of levels of satisfaction of   consumers needs\index{vector of levels of satisfaction of   consumers needs}
 \begin{eqnarray*}Y_4=\{Y_i^4\}_{i=1}^{n+1}, \quad  Y_k^4 > \pi_k^4v_k^2,
\quad k=\overline{1, n}. \end{eqnarray*}
Here $v_2=\{v_k^2\}_{k=1}^n$ is a non-negative vector such that the matrix $\bar C_2(v_2) - \bar B_2$
is non-negative  having no  zero  rows or columns and  $\bar A_2 + \bar C_2(v_2) - \bar B_2$ is  an indecomposable matrix. Such vector exists if the matrix $C_2=||C_{ki} ||_{k=1, i=1}^{n, \ n+1}$ is determined by the set of equations (\ref{ss0s11}) under the  conditions of the  Lemma \ref{p1}. It follows from the Theorem \ref{ss0s9}.

The  introduced quantities satisfy  the  set of balance equations
\begin{eqnarray*} X_i(p_4) - \sum\limits_{k=1}^{n} X_{ik}(p_4) - {\cal E}_i(p_4)
+ I_i(p_4) =N_i(p_4)+C_i(p_4), \quad i=\overline{1,n}, \end{eqnarray*}
\begin{eqnarray*}X_k(p_4)=\sum\limits_{i=1}^{n} X_{ik}(p_4) + \Delta_k(p_4),
\quad  k=\overline{1,n}, \end{eqnarray*}
where the vector $p_4=\{p_i^4\}_{i=1}^{n}$ is a strictly positive solution to the set of equations
\begin{eqnarray}\label{roml19}
p_i^4=\sum\limits_{k=1}^{n}p_k^4 \bar a_{ki}(X_1)\beta_i +
\frac{1}{X_i^4}\sum\limits_{k=1}^{n}\left[ C_{ki}^{2}  \frac{Y_i^4}{\pi_i^4}- B_{ki}^2\right]p_k^4, \quad
i=\overline{1,n},
\end{eqnarray}
\begin{eqnarray*}Y_i^4 > \pi_i^4 v_i^2, \quad i=\overline{1, n}, \end{eqnarray*}
and the vector $X_4=\{X_i^4\}_{i=1}^{n}$ is a strictly positive solution to the set of equations
\begin{eqnarray}\label{tyml21}
X_i^4 - \sum\limits_{k=1}^{n} \bar a_{ik}(X_1)\beta_kX_k^4 + \sum\limits_{k=1}^{n+1} B_{ik}^{2}- {\cal E}_i^4 +I_i^4 =
 \sum\limits_{k=1}^{n+1} C_{ik}^{2}Y_k^4,
\quad  i=\overline{1,n}.
\end{eqnarray}
Such strictly positive solutions of the set of equations (\ref{roml19} ) and (\ref{tyml21}) always exists if, e.g., the condition holds
\begin{eqnarray*}[E - \bar A_2]^{-1}[{\cal E}_4 - I_4 - B_2e] > 0, \quad e=\{e_i\}_{i=1}^{n+1}, \quad e_i=1, \quad i=\overline{1,n+1},\end{eqnarray*}
where $B_2e=\{\sum\limits_{k=1}^{n+1} B_{ik}^{2}\}_{i=1}^n,$  $ {\cal E}_4=\{{\cal E}_i^4\}_{i=1}^n,$
 $ I_4=\{I_i^4\}_{i=1}^n,$
and the vector $\pi_4=\{\pi_i^4\}_{i=1}^{n}$ agrees with the  structure of consumption, i.e. is such that there exists a strictly positive solution
$g_4=\{g_i^4\}_{i=1}^{n}$
to the set of equations
\begin{eqnarray}\label{euml1}
\frac{[(E- \bar A_{2})^{-1}[\bar C_2(\pi_4^{-1}\bar Y_4) -\bar B_2]g_4]_i}{X_i^4} = g_i^4,
\quad  i=\overline{1,n},
\end{eqnarray}
\begin{eqnarray*}\bar C_2(\pi_4^{-1}\bar Y_4)=\left|\left|C_{ik}^{2} \frac{Y_k^4}{\pi_k^4}\right|\right|_{i,k=1}^n, \quad  Y_k^4 > \pi_k^4v_k^2, \quad k=\overline{1,n}.\end{eqnarray*}

For subsequent study, the structure of the solution
$C_2=||C_{ki}^{2}||_{k=1, i=1}^{n,\  n+1}$
is significant. Namely, it is important to clarify  conditions of  solvability for the problem
(\ref{euml1}).
Consider the case of zero initial goods supply in the economy system $\bar E_1.$
\begin{theorem}
Let a  matrix of initial goods supply\index{ matrix of initial goods supply} $ B_{1}=||B_{ki}^{1}||_{k=1, i=1}^{n, n+1}$ at the beginning of the economy operation period
 be zero  and the inequalities hold
\begin{eqnarray*} C_k^1 +N_k^1 + \sigma_k >0, \quad  k=\overline{1,n},\end{eqnarray*}
\begin{eqnarray*}\sum\limits_{i=1}^{n} \left(1-\pi_i^1\right)\left(\Delta_i^1+\sigma_i\right)- {\cal E}_1^0+ I_1^0 > 0,\quad   0 < \pi_i^1< 1, \quad  i=\overline{1,n}. \end{eqnarray*}
If the matrix
$ C_{2}=||C_{ki}^{2}||_{k=1, i=1}^{n, n+1}$
is determined by the formulae of the Lemma \ref{p1}  where $B_{ki}^{1}=0, \ k=\overline{1,n}, \ i=\overline{1,n+1},$ then the set of equations
(\ref{euml1}) for the vector $g_4=|g_i^4\}_{i=1}^n$
has a strictly positive solution if the vector
$\pi_4=\{\pi_i^4\}_{i=1}^{n}$ solves the equation \begin{eqnarray*}\frac{1}{M_1}\sum\limits_{i=1}^{n}\frac{\pi_i^1\left(\Delta_i^1+\sigma_i\right)Y_i^4
\tilde X_i}{Y_i^1X_i^4\pi_i^4}=1.\end{eqnarray*}
Under these conditions, the set of equations
(\ref{roml19}) for the vector $p_4$
is equivalent to the set of equations
\begin{eqnarray}\label{rqml19}
p_i^4=\sum\limits_{k=1}^{n}p_k^4 \bar a_{ki}(X_1)\beta_i +
\lambda \frac{\pi_i^1\left(\Delta_i^1+\sigma_i\right)Y_i^4}{M_1\pi_i^4X_i^4Y_i^1}, \quad
\lambda >0, \quad i=\overline{1,n},
\end{eqnarray}
where the vector
 $X_4=\{X_i^4\}_{i=1}^{n}$ is a strictly positive solution to the set of equations (\ref{tyml21}) and

$\tilde X= (E- \bar A_2)^{-1}G, \quad
\tilde X=\{\tilde X_i\}_{i=1}^{n}, \quad G=\{G_i\}_{i=1}^{n},\quad  G_i= C_i^1+N_i^1+\sigma_i.$

\end{theorem}
\begin{proof}\smartqed
Denote the  matrix elements $C_{ki}^{2}$
as follows
\begin{eqnarray*}C_{ki}^{2}=G_k\gamma_i, \quad
\gamma_i= \frac{\pi_i^1\left(\Delta_i^1+\sigma_i\right)}{M_1Y_i^1}, \quad
G_k= C_k^1+N_k^1+\sigma_k,\end{eqnarray*}
and introduce the vector
 $G=\{G_i\}_{i=1}^{n}.$
Then
\begin{eqnarray*}(E- \bar A_2)^{-1}
\bar C_{2}(\pi_4^{-1}\bar Y_4)g_4=
(E- \bar A_2)^{-1}G
\sum\limits_{k=1}^{n}\frac{\gamma_k Y_k^4 g_k^4}{\pi_k^4}.\end{eqnarray*}
Denote
$(E- \bar A_2)^{-1}G=\tilde X, \
\tilde X=\{\tilde X_i\}_{i=1}^{n}.$
The set of equations (\ref{euml1}) takes the form
\begin{eqnarray}\label{masl1}
\frac{\tilde X_i}{X_i^4}
\sum\limits_{k=1}^{n}\frac{\gamma_kg_k^4Y_k^4}{\pi_k^4}= g_i^4, \quad  i=\overline{1,n}.
\end{eqnarray}
 The condition of solvability for the set of equations (\ref{masl1}) in the set of strictly positive vectors is the equality
\begin{eqnarray*}\sum\limits_{i=1}^{n}\frac{\gamma_iY_i^4\tilde X_i}{X_i^4\pi_i^4}=1,\end{eqnarray*}
or
\begin{eqnarray*}\frac{1}{M_1}\sum\limits_{i=1}^{n}\frac{\pi_i^1\left(\Delta_i^1+\sigma_i\right)Y_i^4
\tilde X_i}{Y_i^1X_i^4\pi_i^4}=1.\end{eqnarray*}
Transform the set of equations
(\ref{roml19}). In view of the  structure of  matrix elements $C_{ki}^{2},$ we have
\begin{eqnarray*}\sum\limits_{k=1}^{n}C_{ki}^{2}p_k^4=
\gamma_i \sum\limits_{k=1}^{n}G_kp_k^4,
\quad i=\overline{1,n},\end{eqnarray*}
therefore,
\begin{eqnarray}\label{masl2}
p_i^4=\sum\limits_{k=1}^{n}p_k^4 \bar a_{ki}(X_1)\beta_i +
\frac{\pi_i^1\left(\Delta_i^1+\sigma_i\right)Y_i^4}{M_1\pi_i^4X_i^4Y_i^1}
\sum\limits_{k=1}^{n}G_kp_k^4,
 \quad i=\overline{1,n}.
\end{eqnarray}
The solution to the set of equations (\ref{masl2}) is the vector
\begin{eqnarray*} p_4=\lambda (E-\bar A_2^T)^{-1}f, \quad
f= \{f_i\}_{i=1}^{n}, \quad
f_i=\frac{\pi_i^1\left(\Delta_i^1+\sigma_i\right)Y_i^4}{M_1\pi_i^4X_i^4Y_i^1},
\quad \lambda > 0,\end{eqnarray*}
where $\bar A_2^T $
denotes the matrix transposed to the matrix $\bar A_2.$
It follows from that
$\sum\limits_{k=1}^{n}G_kp_k^4 = \lambda.$
\qed
\end{proof}

\section{Ukrainian economy dynamics}

Below, we compare real states of Ukrainian economy in 1996 and 1999 with the  corresponding economy of  social agreement  and come to the conclusion about negative influence of energy and materials expensive production quantitatively measured onto Ukrainian economy development.

Suppose
\begin{eqnarray*}\tau=\frac{Y_k^4}{Y_k^1}, \quad
{\cal E}_k^4=\tau {\cal E}_k^1, \quad I_k^4=\tau I_k^1,
\quad k=\overline{1, n}. \end{eqnarray*}
Then the solution to the set of equations (\ref{tyml21}) is the vector
$X_4=\{X_i^4\}_{i=1}^{n},$ where $ X_i^4=\tau\left(X_i^1+ \sigma_i\right),\ i=\overline{1,n},\ $
because
\begin{eqnarray*}\sum\limits_{i=1}^{n+1} C_{ki}^{2}Y_i^4=\tau
\sum\limits_{i=1}^{n+1} C_{ki}^{2}Y_i^1=\tau\left(N_k^1+C_k^1+\sigma_k\right),\end{eqnarray*}
and the set of equations for the vector $X_4=\{X_i^4\}_{i=1}^{n}$ turns into the set of equations
\begin{eqnarray}\label{allatyml21}
X_i^4 - \sum\limits_{k=1}^{n} \bar a_{ik}(X_1)\beta_kX_k^4 - \tau{\cal E}_i^1 + \tau I_i^1 =
 \tau\left(N_i^1+C_i^1+\sigma_i\right),
\quad  i=\overline{1,n},
\end{eqnarray}
\begin{eqnarray*}\bar a_{ik}(X_1)\beta_k=\frac{X_{ik}^1}{X_k^1 + \sigma_k}.\end{eqnarray*}
Introduce notations
$\pi_i^1\left(\Delta_i^1+\sigma_i\right)=\delta_i^{''},$
where $\pi_i^1=\frac{\Delta_i^{''}}{\Delta_i}.$
The set of equations
(\ref{euml1})
has a strictly positive solution if the vector
$\pi_4=\{\pi_i^4\}_{i=1}^{n}$ satisfies the condition
\begin{eqnarray}\label{rhml1}
\frac{1}{M_1}\sum\limits_{i=1}^{n}\frac{\delta_i^{''}
\tilde X_i}{\left(X_i^1+\sigma_i\right)\pi_i^4}=1.
\end{eqnarray}
For subsequent study, a uniform taxation system\index{uniform taxation system} is important, i.e., the case $\pi_i^4=\pi_4, \ i=\overline{1,n}.$
In this case, for $\pi_4$ the equality
\begin{eqnarray}\label{rhml2}
\pi_4=\frac{1}{M_1}\sum\limits_{i=1}^{n}\frac{\delta_i^{''}
\tilde X_i}{X_i^1+\sigma_i}
\end{eqnarray}
holds.
The set of equations for the vector
$p_4=\{p_i^4\}_{i=1}^{n}$ takes the form
\begin{eqnarray}\label{rhml19}
p_i^4=\sum\limits_{k=1}^{n}p_k^4 \bar a_{ki}(X_1)\beta_i +
\lambda \frac{\delta_i^{''}}{M_1\pi_4\left(X_i^1+\sigma_i\right)}, \quad
\lambda >0, \quad i=\overline{1,n}.
\end{eqnarray}
Calculate the added  value of the $i$-th industry. We have
\begin{eqnarray*}\Delta_i(p_4) = \frac{Y_i^4}{\pi_i^4}\sum\limits_{k=1}^{n}C_{ki}(p_4)=\frac{Y_i^4}{\pi_i^4}\gamma_i \sum\limits_{k=1}^{n}G_kp_k^4 \end{eqnarray*} \begin{eqnarray*}=\frac{\lambda \pi_i^1\left(\Delta_i^1+\sigma_i\right) Y_i^4}{M_1Y_i^1\pi_i^4}= \frac{\lambda \tau \pi_i^1\left(\Delta_i^1+\sigma_i \right)}{M_1\pi_i^4},
 \quad i=\overline{1, n}.\end{eqnarray*}

\begin{corollary}
The economy model describing  aggregately by
\\ 1) a vector of  gross outputs\index{vector of  gross outputs }
$X(p_4)=\{\tau p_j^4\left(X_j^1+\sigma_j\right)\}_{j=1}^{n}\ ;$
\\ 2) a matrix  of financial flows\index{ matrix  of financial flows}
\begin{eqnarray*}||X_{ij}(p_4)||_{i,j}^{n}, \quad X_{ij}(p_4)=\tau p_i^4X_{ij}^1\ ,\end{eqnarray*}
input  matrix
\begin{eqnarray*}\bar A(p_4)=||a_{ij}(p_4)||_{i,j=1}^{n}, \quad
a_{ij}(p_4)=\frac{p_i^4\bar a_{ij}(X_1)\beta_j}{p_j^4},\end{eqnarray*} \begin{eqnarray*}  \beta_j=
\frac{X_j^1}{X^1_j+ \sigma_j}, \quad \bar a_{ij}(X_1)=\frac{X_{ij}^1}{X_j^1}\ ;\end{eqnarray*}
3)  vectors of export and import\index{vectors of export and import}
\begin{eqnarray*}{\cal E}(p_4)=\{\tau p_j^4{\cal E}_j^1\}_{j=1}^{n}, \quad
I(p_4)=\{\tau p_j^4 I_j^1 \}_{j=1}^{n}\ ;\end{eqnarray*}
\\ 4)  vectors of final consumption and  gross accumulation and change of supply\index{vectors of final consumption and  gross accumulation and change of supply}
\begin{eqnarray*}C^f(p_4)=\{\tau p_j^4C^2_j\}_{j=1}^{n}, \quad
N(p_4)=\{\tau p_j^4 N^2_j\}_{j=1}^{n},\end{eqnarray*}
such that
$N_j^2+C_j^2=N_j^1+C_j^1+ \sigma_j, \ j=\overline{1, n}\ ;$
\\ 5) an  added value  of
the $i$-th industry\index{added value  of
the $i$-th industry}
\begin{eqnarray*}\Delta_i^4=\frac{\lambda\tau\delta_i''}{M_1\pi_i^4},\quad \pi_i^4=\pi_4 \quad i=\overline{1, n},\quad  M_1=V_1 - {\cal E}_1^0 +I_1^0, \end{eqnarray*} \begin{eqnarray*} V_1=\sum\limits_{i=1}^{n}\left(\Delta_i^1+\sigma_i\right),\quad
{\cal E}_1^0= \sum\limits_{i=1}^{n}{\cal E}_i^1, \quad
I_1^0= \sum\limits_{i=1}^{n}I_i^1, \end{eqnarray*}
where the vector $p_4=\{p_i^4\}_{i=1}^{n}$ solves the set of equations
(\ref{rhml19})
and integral taxation level\index{integral taxation level}  $\pi_4$ is given by the formula
(\ref{rhml2}),
\begin{eqnarray*}\tilde X_j=[(E-\bar A_{2})^{-1}G]_j, \quad j=\overline{1, n},
\quad M_1=\sum\limits_{i=1}^{n}\left(\sigma_i +\Delta_i^1\right) - {\cal E}_0+I_0,
\end{eqnarray*}
is an  economy of  social agreement if\\
1) the part of newly created value\index{part of newly created value remained in the $i$-th industry} $\pi_i^4\Delta_i^4$ remained in the $i$-th industry is sufficient to guarantee adequate wages for industry employees,\index{wages for industry employees} production is profitable, renewal of fixed capital\index{renewal of fixed capital} corresponds to scientific and technological advance, the production grows;\\
2) the part of newly created value\index{part of newly created value} $\left(1 - \pi_i^4\right)\Delta_i^4$
is sufficient  to guarantee adequate pensions, to finance various social programs, to meet its debt issues.

It is perfectly competitive\index{perfectly competitive} if
\begin{eqnarray*}\Delta_i^4=\omega
\sum\limits_{k=1}^{n}X_{ki}^1, \quad
i=\overline{1, n},\end{eqnarray*}
\end{corollary}
where $\omega$ does not depend on industry index.
\begin{note}
One must take the constant $\tau >0$ within such range that at given economy development stage correspond to technological abilities to enlarge the production.
\end{note}
The achievement of standards of the  economy  of  social agreement\index{standards of the  economy  of  social agreement} with the existing wages of the basic year,\index{wages of the basic year}
with non-detrimental production,\index{non-detrimental production} and  with levels of consumption of basic  capital\index{levels of consumption of basic  capital}  as in the basic year is possible only if\\
1) industry outputs\index{industry outputs} grow according to the formula for vector of outputs\index{vector of outputs}\\
$ \{\tau p_i^4\left(X_i^1+\sigma_i\right)\}_{i=1}^{n}.$ \\
2) structures  of export and import change\index{structures  of export and import} according to formulae for  vectors of  export and import\index{vectors of  export and import}
$\{\tau p_i^4{\cal E}_i^1\}_{i=1}^{n},$
$\{\tau p_i^4I_i^1 \}_{i=1}^{n}.$ \\
3) production inputs in per cents of natural parameters\index{production inputs in per cents of natural parameters} for products consumed to produce single product unit must  reduce  according to the matrix
\begin{eqnarray}\label{allochka1}
 s=||s_{ij}||_{i,j=1}^{n}, \quad
s_{ij}=\min\left\{\frac{p_i^4\beta_j}{p_j^4}, 1\right\}-1.
\end{eqnarray}
The matrix $ s$  we call the  matrix of  reduction of inputs.\index{matrix of  reduction of inputs}
As equilibrium price vector in  built  model of the economy of social agreement\index{ model of the economy of social agreement}  is the same one as in initial model, we give reduction of production inputs  in per cents. To determine reduction of expenses  in cost parameters,\index{reduction of expenses  in cost parameters} one must multiply financial flow $X_{ij}^1$ by $s_{ij}.$

One can study economy dynamics taking into account inflation in the energy expensive Ukrainian production\index{energy expensive Ukrainian production} and National bank exchange rate policy\index{National bank exchange rate policy} by relating the real economy to a model economy of social agreement  year by year and clarify in such a manner how much the real economy achieved  or moved away from the model economy system of  social agreement.
Principles underlying in the notion of  economy of social agreement  allow wide interpretation. Two economies of  social agreement  can differ significantly one from another by levels of wages,\index{levels of wages} by profitability of industries,\index{profitability of industries}
by  balance of foreign economy relations, by levels of pensions,\index{levels of pensions} by consumption of basic  capital.\index{consumption of basic  capital} These quantities determine measures to achieve corresponding levels. When taking European average standards,\index{European average standards} we obtain one set of measures to achieve these standards, taking American standards,\index{American standards} we obtain another  set of measures to achieve them.

In this analysis, we take next standards:

\noindent 1) level of wages\index{level of wages}  in the year considered;

\noindent 2) all production industries are non-detrimental;

\noindent 3) social allocations\index{social allocations} correspond to the laws acting in the  considered year;

\noindent 4) integral levels of taxation\index{integral levels of taxation}   are neither preferential nor discriminatory;

\noindent 5) balance of foreign economic relations\index{balance of foreign economic relations} is non-negative.

Considered years are 1996 and 1999. For example, the achievement of  proposed standards in 1996 is connected with  overcoming of unprofitableness\index{unprofitableness} of coal, iron,
mechanical engineering, and light industries, with overcoming of non-uniform tax pressure\index{ non-uniform tax pressure} showing privileges for somebodies and  high levels of taxation for others, with overcoming of negative  balance of foreign economic relations.

In 1999, the achievement of proposed standards  is connected with overcoming  unprofitableness of coal, other fuel and chemical industries, with removing different integral  levels of taxation.\index{different integral  levels of taxation}

Note that high inflation rates\index{high inflation rates} from 1996 to 1999 significantly reduced levels of wages  in 1999 as compared with 1996. Industries taxation levels\index{industries taxation levels} with respect to the created  value added by industries\index{created  value added by industries} in 1999 were  significantly changed.
The most characteristic feature of this change is that the tax pressure is transferred from energy expensive production\index{energy expensive production} into industries producing consumer goods. Power industry taxation level\index{power industry taxation level} come down from 0,443 of industry value added\index{industry value added} in 1996 to 0,325 in 1999, coal industry subventions\index{coal industry subventions} raised, metallurgy taxation level\index{metallurgy taxation level} fell to a half as compared with 1996. Taxation levels for chemical industry\index{taxation levels for chemical industry} and metallurgy also were reduced. However, taxation levels of woodworking,\index{taxation levels of woodworking} building materials,\index{taxation levels of building materials} food industry\index{taxation levels of food industry} raised. Taxation levels of building, agriculture, and forestry fell.
In 1999, value added tax levels\index{value added tax levels} were high  in gas-oil, other fuel, chemical, mechanical engineering, woodworking, light, food, and fishery industries.
The highest taxation levels for products and import\index{taxation levels for products and import} were set in food, light, gas-oil, and fishery industries.

From 1996 to 1999, power production\index{power production} fell from 183 Bln kWt-h to 172,1. Oil and gas condensate production\index{oil and gas condensate production} fell from 4,1 Mln tons to 3,8, coal production\index{coal production} raised up from 70,5 Mln tons to 81,7. Commodity iron ore production\index{iron ore production} was 47,5 and 47,8 mln tons, cast iron production was 17,8 and 23 mln tons, steel production was 22,3 and 22,3 mln tons in 1996 and 1999, respectively. Mechanical engineering production\index{mechanical engineering production} as compared with
1996 decreased by 5,4 \%. Particularly, the production of coal second working combines\index{production of coal second working combines} fell from 187 to 127, mine hoists from 5 to 2,
mine loaders\index{mine loaders} from 26 to 7, mine locomotives\index{mine locomotives} from 171 to 87. Technological equipment production for light and food industries\index{technological equipment production for light and food industries} fell. Tractor production\index{tractor production} fell from 5 400 to 5 000 in 1999. Car production\index{car production} raised from 6 900 to 9 700, truck production\index{ truck production} raised from \- 4 200 to 7 800, bus production\index{bus production} from 1000 to 2000. Excavator production\index{excavator production} fell from 534 to 201,
bulldozer production\index{bulldozer production} from 4 to 3, bridge crane production\index{bridge crane production} from 48 to 43, truck cranes\index{truck cranes} from 93 to 46. Chemical and gas-oil production\index{chemical and gas-oil production} lowered by 5 \% . In particular, chemical fertilizers production\index{chemical fertilizers production} fell from 2 449 000 tons 2 319 000, chemical plant protectors production\index{chemical plant protectors production} fell from 4 600 tons to 1 800. Food production\index{food production} lowered by 4,6 \%,
agriculture production\index{agriculture production} by 16 \%, animal husbandry production\index{animal husbandry production} lowered by 12,4 \%.
Investment into building industry\index{investment into building industry} lowered by 4,3 \%, transport services\index{transport services} lowered by 18,5 \%. Retail turnover\index{retail turnover} lowered by 14,3 \%.

To achieve  the state of the  economy of  social agreement\index{ economy of  social agreement}  in each of studied years, one must reduce industry production expenses\index{industry production expenses} according to formulae (\ref{allochka1}).
One must take  magnitude of the parameter $\lambda$  such  that the added  value of certain industry  be needed.
We take the parameter $\tau$ such that in basic year\index{basic year} real gross outputs deviation from needed ones to be the smallest one. Therefore, take $\tau$ from the condition
\begin{eqnarray}\label{goal62}
\min\limits_{\tau}[\sum\limits_{i=1}^{n} X_i^1-\tau p_i^4\left(X_i^1+\sigma_i\right)]^2.
\end{eqnarray}
Form this condition, $\tau_0$ minimizing the expression (\ref{goal62}) is given by the formula
\begin{eqnarray}\label{goal63}
\tau_0 =\frac{\sum\limits_{i=1}^{n}X_i^1}
{\sum\limits_{i=1}^{n}p_i^4\left(X_i^1+\sigma_i\right)}.
\end{eqnarray}

\section{The analysis of Ukrainian economy proximity to social agreement economy}

In what direction the  Ukrainian economy developed? We accept minimum economic standards to be achieved. Integral taxation level\index{integral taxation level} $\left(1- \pi_1\right)$ in 1996 and 1999 was 0,422 and 0,391
of   GDP, respectively, i.e.  level of taxation in 1999 was lower by
3,1 \% as compared with 1996. In view of that  GDP of 1999 lowered as compared with 1996, we have certain deterioration of pension levels, wages in budget branches\index{deterioration of pension levels, wages in budget branches} others expenses. We have analyzed negative influence of some industries on another ones in 1996 in papers \cite{55, 71}. This analysis remains valid for the considered  theoretical version \cite{74,75}. In what below, we give calculations of reduction of expenses\index{reduction of expenses}  for 1996 as well as for 1999 to  achieve the  economy of social agreement.\index{economy of social agreement}

To achieve the economy of social agreement, in 1996, power production\index{power production} should reduce consumption of gas and
 black oil fuel  by 12 \% or in cost parameters by UAH 330 mln in  prices of 1996, chemistry and mechanical engineering expenses by 7 \% and 9 \% in natural parameters or by UAH 18 and 13 mln, respectively. Coal industry had negative influence from all the rest industries. The most negative influence was from energy expenses,\index{energy expenses} high industry self-expenses,\index{industry self-expenses}
metallurgy, chemistry, mechanical engineering, forestry, agriculture, transport,  materials and equipment supplying and  sale, trading expenses. Metallurgical industry is too much expensive. It should reduce expenses on electricity, gas and black oil fuel lighting, mechanical engineering products. Agriculture had negative influence from high expenses on gas and oil, chemistry, mechanical engineering,
food, and other production industries. The transport industry was highly expensive. Negative influence on it had high expenses on gas and oil, metallurgy, chemistry, mechanical engineering, and building materials.
 Housing and communal services branch were much  expensive whose expenses on gas and oil was too much high. The most energy and materials expensive industries were power production, coal, metallurgical, agriculture, transport, and
housing and communal services. All they spend too much on 3 industries, namely, gas and oil, mechanical engineering, and chemical ones.

Similar conclusions are valid concerning  negative influence of overexpenditure of the above analyzed industries  onto the economy of 1999.

It is worth to compare in natural parameters the needed reduction of expenses\index{needed reduction of expenses}  for the years considered.
In natural parameters, necessary reduction of expenses  for power production\index{reduction of expenses  for power production} raised by 8 \%  on gas and oil, by 9\% on mechanical engineering, by 10 \% on building materials as compared with 1996. The same is valid for the raising of reduction of expenses  in the metallurgy\index{reduction of expenses  in the metallurgy} to achieve  the economy of social agreement.\index{economy of social agreement}
For gas and oil they grew by 10 \%, for chemistry by 14 \%, for woodworking by 14 \%, for building materials by 10 \%, and for other industries by 12 \%. Also necessary reduction of expenses  by agriculture\index{reduction of expenses  by agriculture} raised as compared with 1996.
The same is valid for housing and communal services branch reduction of expenses on gas and oil.\index{reduction of expenses on gas and oil}

To clarify underlying reasons, analyze  matrix of raising of inputs\index{ matrix of raising of inputs}  in 1999 in  prices  and outputs of 1996
\begin{eqnarray*} \left|\left|\frac{X_{ik}^9}{\beta_i \alpha_k} -X_{ik}^6\right|\right|_{i,k=1}^{n},\end{eqnarray*}
where $\beta_i $ is a product index\index{product index} and $\alpha_k $ is an  inflation index\index{inflation index} compared with 1996 in the $i$-th and the $k$-th industries, respectively,
$X_{ik}^9,  X_{ik}^6$ are  matrices of financial flows\index{ matrices of financial flows} in 1999 and 1996.
If the $k$-th industry produces $m$ homogeneous goods,\index{homogeneous goods} then
\begin{eqnarray*}\alpha_k= \frac{\sum\limits_{i=1}^{m}t_{i}^k \bar p_i^k}
{\sum\limits_{i=1}^{m}t_{i}^k  p_i^k}, \end{eqnarray*}
where $t_i^k$ is the number of units of the $i$-th goods in the $k$-th industry produced in 1999,
$\bar p_i^k$ is the price of the $i$-th goods in the $k$-th industry in 1999,
$ p_i^k$ is the price of the $i$-th goods in the $k$-th industry in 1996.
Then
\begin{eqnarray*}\alpha_k= \frac{X_k^9}{\sum\limits_{i=1}^{m}t_{i}^k  p_i^k}, \quad
X_k^9=\sum\limits_{i=1}^{m}t_{i}^k \bar p_i^k, \end{eqnarray*}
where $X_k^9$ is the gross output of the $k$-th industry in 1999.

If $y_i^k$ is the number of units of the $i$-th goods in the $k$-th industry produced in 1996, then
\begin{eqnarray*}X_k^6=\sum\limits_{i=1}^{m}y_{i}^k p_i^k,\end{eqnarray*}
where $X_k^6$ is the gross output of the $k$-th industry in 1996.
Then product index takes the form
\begin{eqnarray*}\beta_k= \frac{\sum\limits_{i=1}^{m}t_{i}^k  p_i^k}
{\sum\limits_{i=1}^{m}y_{i}^k  p_i^k}. \end{eqnarray*}
Hence, we have
\begin{eqnarray*}\alpha_k= \frac{X_k^9}{\beta_k X_k^6}, \quad
\beta_k= \frac{X_k^9}{\alpha_k X_k^6}.\end{eqnarray*}
Therefore,
\begin{eqnarray*} \frac{X_{ik}^9}{\beta_i \alpha_k}=
\frac{\alpha_i X_{ik}^9 X_i^6}{\alpha_k X_i^9}=
\frac{\beta_k X_{ik}^9 X_k^6}{\beta_i X_k^9}. \end{eqnarray*}
The most characteristic feature of this matrix is that self-expenses raised significantly in gas and oil industry by  UAH 929,8 mln with the same output that in 1996,
in coal industry by UAH 265,9 mln, in metallurgy by UAH 185,6 mln,
in food industry by UAH 833,1 mln, in light industry by UAH 594 mln, and in agriculture by UAH 1783,6 mln.
 It is important social task to clarify reasons for this phenomenon.
\vskip 5mm
\centerline{Conclusions}
\vskip 5mm

\noindent 1) Actual taxation system is built in such a way to soften the negative influence of energy expensiveness of the Ukrainian production on  the economy transferring tax pressure on consumers. \\
2) Export stimulating policy for energy expensive production by devaluation  of national currency and transferring tax pressure on consumers to enlarge external market result in reduction of real income of the  peoples  blocking internal market development. \\
3) Lack of foresight of such policy lies in growing risks for the state to be removed from markets. \\
4) Self-expenses of energy expensive industries on product unit produced raised significantly showing implicitly raising in these industries of shadow  sector  of the economy.\\
5)  Energy inputs of the Ukrainian production raised in 1999 as compared with 1996 in\\
a) power production;\\
b) coal industry;\\
c) metallurgy;\\
d) agriculture;\\
e) transport;\\
f) housing and communal services;\\
The most overexpenditure in these industries relate to expenses on\\
a) gas and oil;\\
b) mechanical engineering;\\
c) chemistry.\\
d) Ukrainian economy in 1999 was farther from the economy of social agreement  as compared with 1996 despite that
standards of  1999 were significantly lower.

\chapter{Principles of control of   foreign economic relations and monopolistic influence }

\abstract*{ The determination of reaction of the economy system on increase of monopolistic prices is studied. The conditions for a structure matrix of production, an unproductive consumption matrix, levels of outputs, an initial goods supply matrix, and levels of taxation  under which the economy system is stable against probable change of prices for some goods group   are found.  The problem of determination of influence of prices change in  certain monopolistic industries on price levels of the rest industries and consumption levels of consumers are studied. The Theorem determining sufficient conditions and giving algorithm to build the vector of gross outputs and the vector of levels of satisfaction of consumer needs under which every industry for given vector of monopolistic  prices and a taxation system is non-detrimental is proved. In the case of exchange model with monopolists the necessary and sufficient conditions for the existence of strictly positive solution to the set of equations of the economy equilibrium with monopolists are found. }

In reality as a result of  openness of the economy system  to its environment, the equilibrium
existing during a certain operation period can break by changing of prices for some goods group.
An example of such significant influence is, in particular,  increasing of prices for  energy carriers  caused by economical as well as political reasons. The determination of reaction of the  economy system  onto such increase of prices  is an important economical and social problem. In this Chapter, the  conditions for a  structure matrix of production,\index{ structure matrix of production} an unproductive consumption matrix,\index{unproductive consumption matrix} an initial goods supply matrix,\index{initial goods supply matrix} levels of gross outputs,\index{levels of gross outputs} and levels of taxation\index{levels of taxation}  are found  under which the economy system is stable  against probable  change of prices for some goods group.

 It is very important to clarify what influence such  change of levels of prices   in a certain industries makes  on the rest levels of  prices  and to propose a protection against such change by varying levels of taxation  and increase of gross outputs of industries.

Therefore, in this Chapter, we solve the problem of determination of influence of prices  change    in certain   monopolistic industries  on price levels of the rest industries and consumption levels \cite{55, 71, 92, 106, 59, 101, 104}. We solve the problem with two restrictions, namely, prices change in monopolistic industries don't cause any industry bankruptcy and the  vector of gross outputs satisfy a certain  set of equations. These two restrictions lead to the existence of  certain dependence between levels of satisfaction of consumers needs,\index{levels of satisfaction of consumers needs}  levels of taxation of monopolistic and non-monopolistic industries,\index{ levels of taxation of monopolistic and non-monopolistic industries} and structure of foreign economic relations\index{structure of foreign economic relations}  under given  structures of consumption and production.\index{structures of consumption and production} Solution of this problem can give  answer to important practical questions, namely, what influence prices  change   in certain monopolistic industries\index{monopolistic industries} have on the rest prices in the economy system, on  levels of  consumption\index{levels of  consumption} under a given taxation system, on profitability of  non-monopolistic industries.\index{profitability of  non-monopolistic industries} We prove Theorems \ref{4taxt1}, \ref{taxt1}, and \ref{vtas1} determining conditions,  and giving  an algorithm to build the vector of gross outputs\index{vector of gross outputs} and the vector of levels of satisfaction of consumers needs\index{vector of levels of satisfaction of consumers needs} under which every industry for given vector of monopolistic prices\index{vector of monopolistic prices}  and a taxation system\index{taxation system} is non-detrimental if a structure matrix of production  does not depend on the vector of gross outputs\index{vector of gross outputs}.

In the Theorem \ref{tam1}, we establish conditions for a structure  matrix of production,\index{structure  matrix of production}  a matrix of unproductive consumption,\index{matrix of unproductive consumption} an initial goods supply matrix,\index{initial goods supply matrix}  levels of gross outputs,\index{levels of gross outputs} taxation levels,\index{taxation levels} and a  vector of monopolistic prices\index{vector of monopolistic prices} for which  the vector of levels of satisfaction of consumers needs \index{vector of levels of satisfaction of consumers needs }   remains invariant.

In the Section "Zero initial goods supply", we consider the same problem and in the same formulation. However, its solution is much simpler. Such approximation is convenient if the initial goods supply\index{initial goods supply} is small. The Theorem \ref{sort1} gives the  necessary and sufficient conditions for  a taxation vector  to agree with the structure of  consumption  under a given  vector of monopolistic prices\index{vector of monopolistic prices} and Theorems \ref{tsor2-2} and \ref{tsor5-5} contain the construction of the vector of gross outputs\index{vector of gross outputs} and the vector of levels  of satisfaction of  consumers needs \index{vector of levels  of satisfaction of  consumers needs }  for which  conditions of the  Theorem \ref{sort1} hold.
In the case of zero initial goods supply\index{zero initial goods supply}, the Theorem  \ref{tsor5-5} gives  the simple sufficient conditions for a taxation vector to agree with the  structure of consumption.

In the Section "Applications", we set out the problem of influence of  monopolistic prices\index{monopolistic prices}  on the economy system in cost parameters. As a rule, just in this manner the information about the economy system is  given.
For this, to aggregated description in cost parameters\index{aggregated description in cost parameters} we put into correspondence  the equivalent non-aggregated description\index{equivalent non-aggregated description} determined by the set of equations (\ref{2tax7}). For non-aggregated description,\index{non-aggregated description} we establish the set of equations (\ref{2tax9}) and (\ref{2tax10}) having the same form   studied in the previous Sections.

In the Section "The case of non-linear technologies", we extend results obtained in the previous Sections on the case of the presence of constant expenses.\index{constant expenses}

 In the Section "Model of the exchange with monopolists", we consider the problem of the existence of the economic equilibrium  under the condition that $m$  consumers keep fixed prices for goods they have. We prove the Theorem \ref{eq1} establishing the  necessary and sufficient conditions for the existence of a strictly positive solution to the set of equations of the economy equilibrium  with monopolists.\index{ set of equations of the economy equilibrium  with monopolists} In the Section "Monopolistic influence of external environment",\index{ influence of external environment} we show the problem of economy system interaction with environment to be reduced to the problem with internal monopolists.\index{internal monopolists}

\section{ Influence of monopolies onto price formation}
\subsection{ Statement of the Problem}

In this Section, we consider the model of open aggregated economy system \cite{55, 71, 92, 106, 59, 101, 104} described in the Section 7.3. The problem of optimization of monopolistic influence onto the economy system was considered in \cite{89, 90,  91}.
All the definitions and notations are introduced in the Section 7.3.

In economical practice, it is important to know\\
1) how does  increase of prices in certain  monopolistic industries\index{ monopolistic industries} influence on the structure of prices\index{structure of prices} of the rest industries, on the  level of welfare,\index{level of welfare} on the structure of outputs,\index{structure of outputs} on the level of  trade balance\index{level of  trade balance}  under a fixed taxation system.\index{fixed taxation system}\\
2) how does simultaneous  change of prices in some  monopolistic industries and change of  levels of taxation  influence onto levels of  profits of industries,\index{levels of  profits of industries} level of welfare, and  balance of foreign economic relations.\index{balance of foreign economic relations}

To solve these practical problems, one must develop further theory of taxation\index{theory of taxation}  that should account for formation of prices  by some  monopolistic industries.

Mathematically we formulate the problem as follows:\\
For the given technological matrix
$A\left(x^0\right)=\|a_{kj}\left(x_j^0\right)\|^{n}_{k,j=1}$ depending non-linearly on the vector of outputs
$ x^0 \in X_0,$
 the structure of unproductive consumption\index{structure of unproductive consumption}  given by the matrix
$C=\|c_{kj}\|^{n,l}_{k=1,j=1},$ and the matrix of initial goods supply\index{matrix of initial goods supply }  $B=\|b_{kj}\|^{n,l}_{k=1,j=1}$ at the beginning of the economy operation period, it is necessary to establish conditions for a strictly positive vector of  monopolistic prices\index{vector of  monopolistic prices }
$p^0=\{p_j^0\}^{m}_{j=1}$ and vector of levels  of satisfaction of consumers needs\index{vector of levels  of satisfaction of consumers needs}
$y=\{y_j\}^{l}_{j=1}$ under which for taxation vector of industries\index{taxation vector of industries }
$\pi=\{\pi_s\}^{n}_{s=1}, $  $0<\pi_s<1, \ s=\overline{1, n},$ there exists a strictly positive solution to the problem
\begin{eqnarray*} p_j^0 =
\sum\limits^m_{k=1}\left[a_{kj}\left(x^0_j\right)+\frac{y_j}{x_j^0}\left(\frac{c_{kj}}{\pi_j}-\frac{b_{kj}}{y_j}\right)\right]
p_k^0
\end{eqnarray*}
\begin{eqnarray*}  +\sum\limits^{n}_{k=m+1}\left[a_{kj}\left(x^0_j\right)+\frac{y_j}{x_j^0}\left(\frac{c_{kj}}{\pi_j}-\frac{b_{kj}}{y_j}\right)\right]p_k, \quad j=\overline{1,m},
\end{eqnarray*}
\begin{eqnarray*} p_j =
\sum\limits^m_{k=1}\left[a_{kj}\left(x^0_j\right)+\frac{y_j}{x_j^0}\left(\frac{c_{kj}}{\pi_j}-\frac{b_{kj}}{y_j}\right)\right]
p_k^0
\end{eqnarray*}
\begin{eqnarray} \label{tax1220}
+\sum\limits^{n}_{k=m+1}\left[a_{kj}\left(x^0_j\right)+\frac{y_j}{x_j^0}\left(\frac{c_{kj}}{\pi_j}-\frac{b_{kj}}{y_j}\right)\right]p_k, \quad j=\overline{m+1,n},
\end{eqnarray}

\begin{eqnarray} \label{tax1221}
x_k^0-\sum\limits^{n}_{j=1}a_{kj}\left(x_j^0\right)x_j^0+ \sum\limits_{i=1}^lb_{ki} -e_k + i_k=
\sum\limits^l_{j=1}c_{kj}y_j,
\quad k=\overline{1,n},
\end{eqnarray}
with respect to vectors
$p=\{p_{m+1}, \ldots ,p_{n}\} \in R_+^{n-m}$ and
$x^0=\{x_{1}^0, \ldots ,x_{n}^0\} \in X_0.$

\section{The agreement of taxation vector with consumption structure}

In the next Theorem, we formulate the necessary and sufficient conditions under which a taxation vector $\pi=\{\pi_j\}_{j=1}^n, \  0 < \pi_j < 1, \ j=\overline{1,n},$ agrees with the structure of consumption.\index{ agreement of taxation vector with consumption structure}
Introduce notations
 \begin{eqnarray*}  R_j\left(p^0, A\left(x^0\right), \bar B,  x^0,  p\right) = p_j^0 - \sum\limits^m_{k=1}a_{kj}\left(x_j^0\right)p_k^0
- \sum\limits^{n}_{k=m+1}a_{kj}\left(x_j^0\right) p_k \end{eqnarray*}  \begin{eqnarray*} + \frac{1}{x_j^0}\left[\sum\limits^m_{s=1}b_{sj}p_s^0+\sum\limits^n_{s=m+1}b_{sj} p_s\right], \quad j=\overline{1,m},\end{eqnarray*}
where $p=\{p_{m+1}, \ldots ,p_{n}\} \in R_+^{n-m}$ and
$x^0=\{x_{1}^0, \ldots ,x_{n}^0\} \in X_0,$
\begin{eqnarray*} D\left(x^0\right)=\left\|\frac{ y_j}{x_j^0}\left( \frac{c_{kj}}
{\pi_j} - \frac{b_{kj}}{y_j}\right) \right\|^{m}_{k,j=1},
\end{eqnarray*}
\begin{eqnarray*} \tilde{\cal A}\left(x^0\right)=\left\|a_{kj}\left(x_j^0\right)+\frac{ y_j}{x_j^0}\left( \frac{c_{kj}}
{\pi_j} - \frac{b_{kj}}{y_j}\right) \right\|^{n}_{k,j=m+1}.
\end{eqnarray*}
\begin{theorem} \label{4taxt1}
Let $\sum\limits_{k=1}^nc_{ki}>0, \ i=\overline{1,l},$ there exist a non-negative vector
$v_0=\{v_i\}_{i=1}^n, \ v_i \geq 0, \ i=\overline{1,n},$  such that $\bar C(v_0) - \bar B $ is a non-negative  matrix having no zero rows or columns and a matrix
$ A\left(x^0\right) + \bar C(v_0) - \bar B$ is indecomposable, and let the spectral radius of the matrix $A\left(x^0\right)$ be less than 1,  a strictly positive vector
$x^0=\{x_i^0\}_{i=1}^{n} \in X_0$  exist solving the set of equations (\ref{tax1221}) with rhs containing components  vector of levels of satisfaction of consumers needs   $y.$

The sufficient conditions for the taxation vector\index{taxation vector} $\pi=\{\pi_j\}_{j=1}^n$ to agree with the   structure  of consumption in the economy system\index{ structure  of consumption in the economy system} under the vector of  monopolistic prices \index{vector of  monopolistic prices }
$p^0=\{p_1^0,\ldots,p_m^0\}, \ p_i^0 > 0, \ i=\overline{1, m},$ and the vector of levels of satisfaction of consumers needs\index{ vector of levels of satisfaction of consumers needs} $y=\{y_i\}^{l}_{i=1}$ whose components satisfy inequalities $y_i > \pi_i v_i,\ i=\overline{1,n},$ are the following:

1) the spectral radius of the  matrix\index{spectral radius of the  matrix} $\tilde{\cal A}\left(x^0\right)$  is less than 1\ ;

2) there exists a strictly positive solution $\bar p=\{\bar p_i\}^{n}_{i=m+1}$ to the set of equations
\begin{eqnarray*}  p_j =
\sum\limits^m_{k=1}\left[a_{kj}\left(x_j^0\right)+  \frac{ y_j}{x_j^0}\left( \frac{c_{kj}}
{\pi_j} - \frac{b_{kj}}{y_j}\right)\right]
p_k^0  \end{eqnarray*}
\begin{eqnarray}
\label{tax2}
+ \sum\limits^{n}_{k=m+1}\left[a_{kj}\left(x_j^0\right)+ \frac{ y_j}{x_j^0}\left( \frac{c_{kj}}
{\pi_j} - \frac{b_{kj}}{y_j}\right)\right] p_k, \quad  j=\overline{m+1,n},
\end{eqnarray}
for the vector $ p=\{ p_i\}^{n}_{i=m+1}$
satisfying inequalities
\begin{eqnarray} \label{tax01}
 p_j^0 - \sum\limits^m_{k=1}a_{kj}\left(x^0\right)p_k^0
- \sum\limits^{n}_{k=m+1}a_{kj}\left(x^0\right)\bar p_k  > 0, \quad j=\overline{1,m}\ ;
\end{eqnarray}

3) there hold equalities
\begin{eqnarray} \label{tax02}
\pi_j=\frac{\sum\limits^m_{k=1}c_{kj}p_k^0 +
\sum\limits^{n}_{k=m+1}c_{kj}\bar p_k}
{R_j\left(p^0, A, \bar B, x^0, \bar p\right)}\frac{y_j}{x_j^0}, \quad j=\overline{1,m}.
\end{eqnarray}

The conditions 1) --- 3) are necessary if for a certain strictly positive vector $x^0 \in X_0 $ solving the set of equations
(\ref{tax1221}) there exists a strictly positive solution
$\bar p=\{\bar p_i\}_{i=m+1}^{n}$
to the set of equations (\ref{tax1220}), the matrix $D\left(x^0\right)$ does not contain zero columns and or the matrix $\tilde{\cal A}\left(x^0\right)$ is indecomposable  and at least for one $j, \ j=\overline{m+1, n},$
\begin{eqnarray} \label{ta01}
\sum\limits^m_{k=1}\left[a_{kj}\left(x_j^0\right)+\frac{ y_j}{x_j^0}\left( \frac{c_{kj}}
{\pi_j} - \frac{b_{kj}}{y_j}\right)\right]
p_k^0 >0,
\end{eqnarray}
or the inequalities (\ref{ta01}) hold for all
$j=\overline{m+1, n}$ and the matrix $\tilde{\cal A}\left(x^0\right)$ does not contain zero columns
\end{theorem}

\begin{proof}\smartqed \  Sufficiency. From the  conditions of the Theorem \ref{4taxt1}, it follows that there exists a strictly positive  solution to the set of equations
\begin{eqnarray*} p_j^0 =
\sum\limits^m_{k=1}\left[a_{kj}\left(x_j^0\right)+\frac{ y_j}{x_j^0}\left( \frac{c_{kj}}
{\pi_j} - \frac{b_{kj}}{y_j}\right)\right]
p_k^0 \end{eqnarray*}
\begin{eqnarray*} + \sum\limits^{n}_{k=m+1}\left[a_{kj}\left(x_j^0\right)+
\frac{ y_j}{x_j^0}\left( \frac{c_{kj}}
{\pi_j} - \frac{b_{kj}}{y_j}\right)\right]p_k, \quad j=\overline{1,m}, \end{eqnarray*}
\begin{eqnarray*} p_j =
\sum\limits^m_{k=1}\left[a_{kj}\left(x_j^0\right)+\frac{ y_j}{x_j^0}\left( \frac{c_{kj}}
{\pi_j} - \frac{b_{kj}}{y_j}\right)\right]
p_k^0  \end{eqnarray*}  \begin{eqnarray*}  + \sum\limits^{n}_{k=m+1}\left[a_{kj}\left(x_j^0\right)+
\frac{ y_j}{x_j^0}\left( \frac{c_{kj}}
{\pi_j} - \frac{b_{kj}}{y_j}\right)\right]p_k, \quad j=\overline{m+1,n},\end{eqnarray*}
for the vector $ p=\{ p_i\}_{i=m+1}^{n}$ and this solution equals to the vector
$\bar p=$ $\{\bar p_i\}_{i=m+1}^{n}.$
From here and the Theorem \ref{vjant14}, we obtain  the Proof of sufficiency because the vector $\{p_1^0, \ldots , p_m^0, \bar p_{m+1}, \ldots , \bar p_n\}$ belongs to the set $T_0.$

  Necessity. Let for the vector of monopolistic prices\index{vector of monopolistic prices}
$p^0=\{p_1^0,\ldots,p_m^0\}$ and the vector of levels of satisfaction of consumers needs
 $y=\{y_i\}^{l}_{i=1}$ whose components satisfy inequalities $y_i > \pi_i v_i,  \ i=\overline{1,n},$  a strictly positive solution
$\bar p=\{\bar p_i\}_{i=m+1}^{n}$
to the set of equations (\ref{tax1220}) exist for a certain vector $x^0$ solving the set of equations
(\ref{tax1221}).

The spectral radius  of the matrix\index{spectral radius  of the matrix}
$\tilde{\cal A}\left(x^0\right)$
is less than 1. This fact follows from that under assumptions accepted the norm of the matrix
$\tilde{\cal A}\left(x^0\right)$ or some its power
$\tilde{\cal A}^k\left(x^0\right)$ is less than 1 in the normalized space $R^{n-m}$ with the norm
\begin{eqnarray*} ||t||=\max\limits_{k \in [m+1, n]}\frac{|t_k|}{\bar p_k},\end{eqnarray*}
where $\bar p_k, \ k=\overline{m+1, n}$  are components of the solution
$\bar p=\{ \bar p_{m+1}, \ldots ,\bar p_{n}\}$
to the problem (\ref{tax1220}) that are strictly positive.
Really, let, e.g., the matrix $\tilde{\cal A}\left(x^0\right)$ be indecomposable  and at least for one $j, j=\overline{m+1,n},$   the condition (\ref{ta01}) hold.
One can write the set of equations for the vector
$\bar p=\{\bar p_i\}_{i=m+1}^{n}$ in matrix form as follows
\begin{eqnarray*} \bar p= u_0 + \tilde{\cal A}\left(x^0\right)^T\bar p,\end{eqnarray*}
 \begin{eqnarray*}   u_0=\{u_j^0\}_{j=m+1}^n, \quad u_j^0=
\sum\limits^m_{k=1}\left[a_{kj}\left(x_j^0\right)+\frac{ y_j}{x_j^0}\left( \frac{c_{kj}}
{\pi_j} - \frac{b_{kj}}{y_j}\right)\right]
p_k^0.\end{eqnarray*}
Or in the form
\begin{eqnarray*} \bar p= t_0 + [\tilde{\cal A}\left(x^0\right)^T]^n\bar p,\end{eqnarray*}
where
\begin{eqnarray*}   t_0=\{t_j^0\}_{j=m+1}^n, \quad t_0=
\sum\limits^{n-1}_{i=0}[\tilde{\cal A}\left(x^0\right)^T]^iu_0.\end{eqnarray*}
However, components of the  vector $t_0$  are strictly positive  because the matrix $\tilde{\cal A}\left(x^0\right)$ is indecomposable  and the vector $u_0$ does not equal zero and the inequalities
\begin{eqnarray*} t_i^0 < \bar p_i,\quad  i=\overline{m+1,n},\end{eqnarray*}
hold.
Then for the norm of the matrix $\tilde{\cal A}\left(x^0\right)^T$ in the above given norm the estimate
\begin{eqnarray*}  \left|\left|[\tilde{\cal A}\left(x^0\right)^T]^n\right|\right| \leq \max\limits_{m+1 \leq i \leq n}\frac{\bar p_i - t_i^0}{\bar p_i}< 1\end{eqnarray*}
holds.
For another available case the Proof is similar.

The condition 2) holds because under the conditions of the Theorem  the solution
 $\bar p$ to the set of equations (\ref{tax2}) is the vector with strictly positive components and the matrix
$D\left(x^0\right)$ has no zero columns. The condition 3) follows immediately from the condition
2) and that $\bar p$ solves the set of equations (\ref{tax1220}) with the vector $x^0.$
\qed\end{proof}

The  goal of the next two Theorems is  to establish  pair of vectors  $x^0$ and $y$ for which the taxation vector
$\pi=\{\pi_i\}_{i=1}^n, \ 0 < \pi_i < 1, \ i=\overline{1,n},$ agrees with the structure of consumption  under the vector of monopolistic prices\index{vector of monopolistic prices}  $p^0.$ First, consider the case with zero constant expenses.\index{zero constant expenses} One can reduce to this case the case when constant expenses\index{constant expenses} do not equal zero and are included into the matrix of direct expenses.\index{matrix of direct expenses} Such approximation holds if the vector of gross outputs\index{vector of gross outputs} vary insignificantly.

Introduce notations
 \begin{eqnarray*}  R_j^1\left(p^0, A, \bar B,  x,  p\right) = p_j^0 - \sum\limits^m_{k=1}a_{kj}p_k^0
- \sum\limits^{n}_{k=m+1}a_{kj} p_k \end{eqnarray*}  \begin{eqnarray*} + \frac{1}{x_j}\left[\sum\limits^m_{s=1}b_{sj}p_s^0+\sum\limits^n_{s=m+1}b_{sj} p_s\right], \quad j=\overline{1,m},\end{eqnarray*}
where $p=\{p_{m+1}, \ldots ,p_{n}\} \in R_+^{n-m}$ and
$x=\{x_{1}, \ldots ,x_{n}\} \in X_0.$

\begin{theorem}\label{taxt1}
Let $\sum\limits_{k=1}^nc_{ki}>0, \ i=\overline{1,l},$ there exist a non-negative vector
$v_0=\{v_i\}_{i=1}^n,$ $  \ v_i \geq 0, \ i=\overline{1,n},$   such  that $\bar C(v_0) - \bar B $ is a non-negative matrix having no zero rows or columns and
$ A + \bar C(v_0) - \bar B$ is an  indecomposable matrix,\index{indecomposable matrix} and let  the spectral radius of the matrix\index{spectral radius of the matrix} $A$  be less than 1,
a strictly positive vector
$\alpha=\{\alpha_i \}_{i=1}^{n},$
the vector of monopolistic prices\index{vector of monopolistic prices}  $p^0=\{p_1^0, \ldots, p_m^0 \}$ and the matrix $A$ satisfy conditions:\\
1) the spectral radius of the matrix
\begin{eqnarray*} \tilde{\cal A}=\left\|a_{kj}+\alpha_j  \frac{c_{kj}}
{\pi_j}  \right\|^{n}_{k,j=m+1},
\end{eqnarray*}
is less than 1\ ;\\
2) there exists a strictly positive solution $\bar p=\{\bar p_i\}^{n}_{i=m+1}$ to the set of equations
\begin{eqnarray*}  p_j =
\sum\limits^m_{k=1}\left[a_{kj}+  \alpha_j \frac{c_{kj}}
{\pi_j}\right]
p_k^0
\end{eqnarray*}
\begin{eqnarray}
\label{tag2}
+ \sum\limits^{n}_{k=m+1}\left[a_{kj}+ \alpha_j \frac{c_{kj}}
{\pi_j}\right] p_k, \quad  j=\overline{m+1,n},
\end{eqnarray}
relative to the vector $ p=\{ p_i\}^{n}_{i=m+1}$ satisfying inequalities
\begin{eqnarray} \label{tag01}
 p_j^0 - \sum\limits^m_{k=1}a_{kj}p_k^0
- \sum\limits^{n}_{k=m+1}a_{kj}\bar p_k > 0,
 \quad j=\overline{1,m}\ ;
\end{eqnarray}
If $x(\alpha)=\{x_k(\alpha)\}_{k=1}^n$ is a strictly positive solution to the set of equations
\begin{eqnarray} \label{tpg1}
x_k-\sum\limits^{n}_{j=1}[a_{kj}+ \alpha_j c_{kj}]x_j-e_k+i_k + \sum\limits^l_{j=1}b_{kj}=
\sum\limits^l_{j=n+1}c_{kj}y_j,
\quad k=\overline{1,n},
\end{eqnarray}
satisfying conditions $y_k(\alpha)=\alpha_k x_k(\alpha) > \pi_k v_k, \ k=\overline{1,n},$
the equalities
\begin{eqnarray} \label{ter2}
\pi_j=\frac{\sum\limits^m_{k=1}c_{kj}p_k^0 +
\sum\limits^{n}_{k=m+1}c_{kj}p_k^1}
{ R_j^1\left(p^0, A, \bar B,  x(\alpha),  p_1\right)}\alpha_j, \quad  j=\overline{1,m},
\end{eqnarray}
 hold, where $ p_1=\{ p_{m+1}^1, \ldots , p_{n}^1\}$ is a solution to the set of equations

\begin{eqnarray*} p_j =
\sum\limits^m_{k=1}\left[a_{kj}+\alpha_j \left( \frac{c_{kj}}
{\pi_j} - \frac{b_{kj}}{y_j(\alpha)}\right)\right]
p_k^0  \end{eqnarray*}
\begin{eqnarray} \label{tag1220}
  + \sum\limits^{n}_{k=m+1}\left[a_{kj}+
\alpha_j \left( \frac{c_{kj}}
{\pi_j} - \frac{b_{kj}}{y_j(\alpha)}\right)\right]p_k, \quad j=\overline{m+1,n},
\end{eqnarray}
with respect to the vector $ p=\{ p_{m+1}, \ldots , p_{n}\},$
then there exists a strictly positive solution to the set of equations
\begin{eqnarray*} p_j^0 =
\sum\limits^m_{k=1}\left[a_{kj}+\alpha_j\left(\frac{c_{kj}}{\pi_j}-\frac{b_{kj}}{y_j(\alpha)}\right)\right]
p_k^0  \end{eqnarray*}  \begin{eqnarray*}  +\sum\limits^{n}_{k=m+1}\left[a_{kj}+\alpha_j\left(\frac{c_{kj}}{\pi_j}-\frac{b_{kj}}{y_j(\alpha)}\right)\right]p_k, \quad j=\overline{1,m}, \end{eqnarray*}
\begin{eqnarray*} p_j =
\sum\limits^m_{k=1}\left[a_{kj}+\alpha_j\left(\frac{c_{kj}}{\pi_j}-\frac{b_{kj}}{y_j(\alpha)}\right)\right]
p_k^0  \end{eqnarray*}
\begin{eqnarray} \label{ter1}
+ \sum\limits^{n}_{k=m+1}\left[a_{kj}+\alpha_j\left(\frac{c_{kj}}{\pi_j}-\frac{b_{kj}}{y_j(\alpha)}\right)\right]p_k, \quad j=\overline{m+1,n},
\end{eqnarray}
for the vector $p=\{p_{m+1}, \ldots, p_n\}.$ The solution coincides with the vector $p_1.$
\end{theorem}
\begin{proof}\smartqed
As the spectral radius of the matrix $\tilde{\cal A}$ is less than 1 and the inequalities
$y_k(\alpha)=\alpha_k x_k(\alpha) > \pi_k v_k, \ k=\overline{1,n},$ hold
it follows that the spectral radius of the matrix
\begin{eqnarray*} D(\alpha)=\left|\left|a_{kj}+
\alpha_j \left( \frac{c_{kj}}
{\pi_j} - \frac{b_{kj}}{y_j(\alpha)}\right) \right|\right|_{i,j=m+1}^{n}\end{eqnarray*}
is less than 1 too. Therefore, there exists a solution to the set of equations
 (\ref{tag1220}). It is obvious, the set of inequalities holds
 \begin{eqnarray} \label{tas1}
 p_j^0 - \sum\limits^m_{k=1}a_{kj}p_k^0
- \sum\limits^{n}_{k=m+1}a_{kj} p_k^1 > 0,
 \quad j=\overline{1,m},
\end{eqnarray}
that results from the set of inequalities (\ref{tag01}) and the set of inequalities
\begin{eqnarray*} p_j^1 \leq \bar p_j,\quad j=\overline{1,n}.\end{eqnarray*}
  The last set of inequalities results from that the vector
 $\{p_j^1\}_{j=m+1}^n$ solves the set of equations (\ref{tag1220}) from which we obtain that this vector satisfy the set of inequalities
\begin{eqnarray*}  p_j^1 \leq
\sum\limits^m_{k=1}\left[a_{kj}+  \alpha_j \frac{c_{kj}}
{\pi_j}\right]
p_k^0 \end{eqnarray*}
\begin{eqnarray}
\label{tas2}
+ \sum\limits^{n}_{k=m+1}\left[a_{kj}+ \alpha_j \frac{c_{kj}}
{\pi_j}\right] p_k^1, \quad  j=\overline{m+1,n}.
\end{eqnarray}
 From the set of inequalities (\ref{tas2}) and the fact that the  spectral radius of the matrix
 $\tilde{\cal A}$ is less than 1, we obtain the needed estimate. The equalities
(\ref{ter2}) and (\ref{tag1220}) together mean the existence of a strictly positive solution to the set of equations (\ref{ter1}) for the vector $p=\{p_{m+1}, \ldots, p_n\}.$ The last results from the indecomposability  of the matrix $ A + \bar C(v_0) - \bar B$ and that the vector $\{p_1^0, \ldots, p_m^0, p_{m+1}^1, \ldots, p_{n}^1\}$ solves the set of equations
(\ref{ter1}), there hold  inequalities
 $y_k(\alpha)=\alpha_k x_k(\alpha) > \pi_k v_k, \ k=\overline{1,n},$ and the vector of monopolistic prices
$\{p_1^0, \ldots, p_m^0\}$
is strictly  positive.
It is obvious that the vector $\{p_1^0, \ldots, p_m^0, p_{m+1}^1, \ldots, p_{n}^1\}$ belongs to the set $T_0.$
\qed\end{proof}

\begin{corollary} If the vector $\alpha$ satisfies the conditions of the  Theorem \ref{taxt1},
the vector $x(\alpha)=\{x_k(\alpha)\}_{k=1}^n$ is a strictly positive solution to the set of equations
(\ref{tpg1}) and for $ i=\overline{1,n}$ to take
$y_i=\alpha_i x_i(\alpha)$
 and  the rest components
$y_i, \ i=\overline{n+1,l},$ to take such  that for the vector $x(\alpha)$ the  conditions $y_k(\alpha)=\alpha_k x_k(\alpha) > \pi_k v_k, \ k=\overline{1,n},$ were satisfied,
then the vectors
$p_1=\{ p_{m+1}^1, \ldots , p_{n}^1\}$ and
$x(\alpha)=\{x_k(\alpha)\}_{k=1}^n$
are the solutions of the set of equations
\begin{eqnarray*} p_j^0 =
\sum\limits^m_{k=1}\left[a_{kj}+
\frac{y_j}{x_j}\left(\frac{c_{kj}}{\pi_j}-\frac{b_{kj}}{y_j}\right)\right]
p_k^0  \end{eqnarray*}  \begin{eqnarray*}  +\sum\limits^{n}_{k=m+1}\left[a_{kj}+\frac{y_j}{x_j}\left(\frac{c_{kj}}{\pi_j}-\frac{b_{kj}}{y_j}\right)\right]p_k, \quad j=\overline{1,m}, \end{eqnarray*}
\begin{eqnarray*} p_j =
\sum\limits^m_{k=1}\left[a_{kj}+\frac{y_j}{x_j}\left(\frac{c_{kj}}{\pi_j}-\frac{b_{kj}}{y_j}\right)\right]
p_k^0  \end{eqnarray*}
\begin{eqnarray} \label{xter1}
+ \sum\limits^{n}_{k=m+1}\left[a_{kj}+\frac{y_j}{x_j}\left(\frac{c_{kj}}{\pi_j}-\frac{b_{kj}}{y_j}\right)\right]p_k, \quad j=\overline{m+1,n},
\end{eqnarray}
\begin{eqnarray} \label{1mag1221}
x_k-\sum\limits^{n}_{j=1}a_{kj}x_j+ \sum\limits^l_{j=1}b_{kj} -e_k+i_k=
\sum\limits^l_{j=1}c_{kj}y_j,
\quad k=\overline{1,n},
\end{eqnarray}
with such defined  vector of levels of satisfaction of  consumers needs\index{ vector of levels of satisfaction of  consumers needs}   $y.$
It means that the taxation vector\index{taxation vector} $\pi=\{\pi_i\}_{i=1}^{n}$ agrees with the structure of consumption  determined by the above vector of monopolistic prices,\index{vector of monopolistic prices} vector of levels of satisfaction of  consumers needs,\index{vector of levels of satisfaction of  consumers needs} and vector of gross outputs.\index{vector of gross outputs}
\end{corollary}

Give an algorithm to build the vector $\alpha$ satisfying  conditions of the Theorem \ref{taxt1}. Let $\alpha^0=\{\alpha_k^0\}_{k=1}^n$ be a certain  strictly positive vector. Denote  $\tilde \alpha=\{ \alpha_1, \ldots, \alpha_m, \alpha_{m+1}^0,  \ldots, \alpha_n^0\}$
an arbitrary vector whose first $m$ components satisfy inequalities
$0 \leq \alpha_i \leq \alpha_i^0, \ i=\overline{1,m}.$

Suppose the vector $\alpha^0=\{\alpha_k^0\}_{k=1}^n$ satisfy the condition:
there exists a strictly positive solution $x\left(\bar \alpha^0\right)=\{x_j\left(\bar \alpha^0\right)\}_{j=1}^n $ to the set of equations
\begin{eqnarray} \label{tas6}
 x_k - \sum\limits^{n}_{j=1}\left[a_{kj}+ \bar \alpha_j^0 c_{kj}\right]x_j-e_k+i_k + \sum\limits^l_{j=1}b_{kj}=\sum\limits^l_{j=n+1}c_{kj}y_j,
\quad k=\overline{1,n},
\end{eqnarray}
for the vector $x=\{x_j\}_{j=1}^n $
satisfying inequalities
\begin{eqnarray*} \alpha_k^0x_k\left(\bar \alpha^0\right) > \pi_k v_k, \quad  k=\overline{1,n},\end{eqnarray*}
where
 \begin{eqnarray*} \bar \alpha_0=\{\bar \alpha_1^0, \ldots, \bar \alpha_n^0\}, \quad \bar \alpha_i^0=0,\quad  i=\overline{1, m}, \quad  \bar \alpha_i^0= \alpha_i^0, \quad i=\overline{m+1, n}. \end{eqnarray*}
Take sufficiently small numbers $\varepsilon_i >0, \ i=\overline{1, m},$ such that the  inequalities
\begin{eqnarray*} \frac{\pi_i v_i + \varepsilon_i }{x_i\left(\bar \alpha^0\right)}< \alpha_i^0, \quad i=\overline{1, m},\end{eqnarray*}
hold. On  the closed set of vectors
\begin{eqnarray} \label{tak10}
H =\left\{\alpha=\{\alpha_k\}_{k=1}^m \in R_+^m,  \  \frac{\pi_i v_i + \varepsilon_i }{x_i\left(\bar \alpha^0\right)}\leq \alpha_i \leq \alpha_i^0, \ i=\overline{1, m},\right\}
\end{eqnarray}
consider the set of equations
\begin{eqnarray}
\label{tas3}
\alpha_j=f_j(\alpha_1, \ldots, \alpha_m), \quad j=\overline{1,m},
\end{eqnarray}
for the vector $\alpha=\{\alpha_i\}_{i=1}^m,$
where
\begin{eqnarray*}  f_j(\alpha_1, \ldots, \alpha_m) \end{eqnarray*}
\begin{eqnarray*} =\frac{\pi_j\left[p_j^0 - \sum\limits^m_{k=1}a_{kj}p_k^0
- \sum\limits^{n}_{k=m+1}a_{kj} p_k^1+ \frac{1}{x_j(\tilde \alpha)}\left[\sum\limits^m_{k=1}b_{kj}p_k^0+\sum\limits^n_{k=m+1}b_{kj} p_k^1\right]\right]}
{\sum\limits^m_{k=1}c_{kj}p_k^0 +
\sum\limits^{n}_{k=m+1}c_{kj}p_k^1}, \ \   j=\overline{1,m},\end{eqnarray*}
$ p_1=\{ p_{m+1}^1, \ldots , p_{n}^1\}$ solves the set of equations
\begin{eqnarray*} p_j =
\sum\limits^m_{k=1}\left[a_{kj}+\alpha_j^0 \left( \frac{c_{kj}}
{\pi_j} - \frac{b_{kj}}{y_j(\tilde \alpha)}\right)\right]
p_k^0  \end{eqnarray*}
\begin{eqnarray} \label{tas4}
  + \sum\limits^{n}_{k=m+1}\left[a_{kj}+
\alpha_j^0 \left( \frac{c_{kj}}
{\pi_j} - \frac{b_{kj}}{y_j(\tilde \alpha)}\right)\right]p_k, \quad j=\overline{m+1,n},
\end{eqnarray}
relative to the vector $ p=\{ p_{m+1}, \ldots , p_{n}\},$
 $y_j(\tilde \alpha)=\alpha_j^0x_j(\tilde \alpha), \ j=\overline{m+1, n},$ and
$x(\tilde \alpha)=\{x_j(\tilde \alpha)\}_{j=1}^n $ solves the set of equations
\begin{eqnarray} \label{tas5}
 x_k-\sum\limits^{m}_{j=1}[a_{kj}+ \alpha_j c_{kj}]x_j - \sum\limits^{n}_{j=m+1}\left[a_{kj}+ \alpha_j^0 c_{kj}\right]x_j-e_k+i_k
\end{eqnarray}
\begin{eqnarray*} + \sum\limits^l_{j=1}b_{kj}=\sum\limits^l_{j=n+1}c_{kj}y_j,
\quad k=\overline{1,n},
\end{eqnarray*}
for the vector $x=\{x_j\}_{j=1}^n. $

Note that for components $p_j^1,\  j=\overline{1,n},$ of the solution to the set of equations (\ref{tas4}) the following  inequalities
\begin{eqnarray*}   p_j^0(0)\leq  p_j^1 \leq \bar p_j\left( \alpha^0\right), \quad j=\overline{m+1,n},\end{eqnarray*}
hold,  where $p^0(0)=\{p_j^0(0)\}_{j=m+1}^n$ solves the set of equations
\begin{eqnarray} \label{tas8}
p_j =\sum\limits^m_{k=1}a_{kj}
p_k^0 +
 \sum\limits^{n}_{k=m+1}a_{kj} p_k, \quad  j=\overline{m+1,n},
\end{eqnarray}
and $\bar p\left(\alpha^0\right) =\{\bar p_j\left( \alpha^0\right)\}_{j=m+1}^n$ solves the set of equations
\begin{eqnarray*} p_j =
\sum\limits^m_{k=1}\left[a_{kj}+\alpha_j^0 \left( \frac{c_{kj}}
{\pi_j} - \frac{b_{kj}}{y_j\left( \alpha^0\right)}\right)\right]
p_k^0  \end{eqnarray*}
\begin{eqnarray} \label{tas9}
  + \sum\limits^{n}_{k=m+1}\left[a_{kj}+
\alpha_j^0 \left( \frac{c_{kj}}
{\pi_j} - \frac{b_{kj}}{y_j\left( \alpha^0\right)}\right)\right] p_k, \quad j=\overline{m+1,n},
\end{eqnarray}
for the vector $ p=\{ p_{m+1}, \ldots , p_{n}\},$
where $y_j\left(\alpha^0\right)=\alpha_j^0x_j\left(\alpha^0\right),\ j=\overline{m+1, n}$ and vector $x\left(\alpha^0\right)=\{x_j\left(\alpha^0\right)\}_{j=1}^n$ is a solution to the set of equations (\ref{tas5})
under conditions that $\alpha_j= \alpha_j^0, \ j=\overline{1,m}.$
\begin{theorem}\label{vtas1}
Let $\sum\limits_{k=1}^nc_{ki}>0, \ i=\overline{1,l},$ there exist  a non-negative vector
$v_0=\{v_i\}_{i=1}^n, \ v_i \geq 0, \ i=\overline{1,n},$ such that $\bar C(v_0) - \bar B $ is a non-negative matrix having no zero rows or columns and
$ A + \bar C(v_0) - \bar B$ is an  indecomposable matrix, and let the spectral radius of the matrix $A$ be less than 1,
a strictly positive vector
$\alpha^0=\{\alpha_i^0 \}_{i=1}^{n},$
the vector of  monopolistic prices  $p^0=\{p_1^0, \ldots, p_m^0 \}$ and the matrix $A$ satisfy conditions:\\
1) the  spectral radius of matrices
\begin{eqnarray*} \left\|a_{kj}+\alpha_j^0  \frac{c_{kj}}
{\pi_j}  \right\|^{n}_{k,j=1}, \quad \left\|a_{kj}+\alpha_j^0  \frac{c_{kj}}
{\pi_j}  \right\|^{n}_{k,j=m+1}
\end{eqnarray*}
is less than 1\ ;\\
2) there exists a strictly positive solution $\bar p=\{\bar p_i\}^{n}_{i=m+1}$ to the set of equations
\begin{eqnarray}
\label{allatag2}
  p_j =
\sum\limits^m_{k=1}\left[a_{kj}+  \alpha_j^0 \frac{c_{kj}}
{\pi_j}\right]
p_k^0+ \sum\limits^{n}_{k=m+1}\left[a_{kj}+ \alpha_j^0\frac{c_{kj}}
{\pi_j}\right] p_k, \quad  j=\overline{m+1,n},
\end{eqnarray}
for the vector $ p=\{ p_i\}^{n}_{i=m+1}$ satisfying the inequalities
\begin{eqnarray} \label{allatag01}
 p_j^0 - \sum\limits^m_{k=1}a_{kj}p_k^0
- \sum\limits^{n}_{k=m+1}a_{kj}\bar p_k > 0,
 \quad j=\overline{1,m}\ ;
\end{eqnarray}
3) there exists a strictly positive solution $x\left(\bar \alpha^0\right)=\{x_j\left(\bar \alpha^0\right)\}_{j=1}^n $ to the set of equations (\ref{tas6}) satisfying the inequalities
\begin{eqnarray*} \alpha_k^0x_k\left(\bar \alpha^0\right) > \pi_k v_k, \quad  k=\overline{1,n},\end{eqnarray*}
where
 \begin{eqnarray*} \bar \alpha_0=\{\bar \alpha_1^0, \ldots, \bar \alpha_n^0\}, \quad \bar \alpha_i^0=0,\quad  i=\overline{1, m}, \quad  \bar \alpha_i^0= \alpha_i^0, \quad i=\overline{m+1, n}\ ; \end{eqnarray*}
4) the inequalities
\begin{eqnarray*} \frac{\pi_j\left[p_j^0 - \sum\limits^m_{k=1}a_{kj}p_k^0
- \sum\limits^{n}_{k=m+1}a_{kj} \bar p_k\right]}
{\sum\limits^m_{k=1}c_{kj}p_k^0 +
\sum\limits^{n}_{k=m+1}c_{kj}\bar p_k} \geq \frac{\pi_j v_j + \varepsilon_j }{x_j\left(\bar \alpha^0\right)},
 \quad j=\overline{1, m},
\end{eqnarray*}
\begin{eqnarray*}  \frac{\pi_j \left[p_j^0+ \frac{1}{x_j\left(\bar \alpha^0\right)} [\sum\limits^m_{k=1}b_{kj}p_k^0+\sum\limits^n_{k=m+1}b_{kj} \bar p_k]\right]}
{\sum\limits^m_{k=1}c_{kj}p_k^0 +
\sum\limits^{n}_{k=m+1}c_{kj}p_k^0(0)} \leq \alpha_j^0,
 \quad j=\overline{1, m},
\end{eqnarray*}
\begin{eqnarray*} \max\limits_{1\leq j \leq m} \pi_j\sum\limits_{s=1}^m d_{js}\left(p^0, A, B, C, \alpha^0\right)<1,
\end{eqnarray*}
hold, where
\begin{eqnarray*} d_{js}\left(p^0,   A, \bar  B, C, \alpha^0\right)=\frac{\sum\limits^{n}_{k=m+1}a_{kj} \left [\left[E - \tilde{\cal A}\left( \alpha^0\right)^T\right]^{-1}\varphi^s\left(\alpha^0\right)\right]_k }{\sum\limits^m_{k=1}c_{kj}p_k^0 +
\sum\limits^{n}_{k=m+1}c_{kj}p_k^0(0)} \end{eqnarray*}

\begin{eqnarray*} +\frac{ x_s\left(\alpha^0\right)\left[\left[E - (A+ \alpha^0 C)\right]^{-1}C\right]_{js} \left[\sum\limits^m_{k=1}b_{kj}p_k^0+\sum\limits^n_{k=m+1}b_{kj} \bar p_k\left( \alpha^0\right)\right]}
{x_j\left(\bar \alpha^0\right)^2\left[\sum\limits^m_{k=1}c_{kj}p_k^0 +
\sum\limits^{n}_{k=m+1}c_{kj}p_k^0(0)\right]} \end{eqnarray*}

\begin{eqnarray*} +\frac{1}{x_j\left(\bar \alpha^0\right)}\frac{ \sum\limits^n_{k=m+1}b_{kj}\left[\left[E - \tilde{\cal A}\left(\alpha^0\right)^T\right]^{-1}\varphi^s\left(\alpha^0\right)\right]_k }{\sum\limits^m_{k=1}c_{kj}p_k^0 +
\sum\limits^{n}_{k=m+1}c_{kj}p_k^0(0)} \end{eqnarray*}

\begin{eqnarray*} +\frac{\left[p_j^0 - \sum\limits^m_{k=1}a_{kj}p_k^0
- \sum\limits^{n}_{k=m+1}a_{kj} p_k^0(0)\right]}
{\left[\sum\limits^m_{k=1}c_{kj}p_k^0 +
\sum\limits^{n}_{k=m+1}c_{kj}p_k^0(0)\right]^2} \end{eqnarray*}  \begin{eqnarray*}  \times \sum\limits^{n}_{k=m+1}c_{kj}\left[\left[E - \tilde{\cal A}\left( \alpha^0\right)^T\right]^{-1}\varphi^s\left(\alpha^0\right)\right]_k \end{eqnarray*}

\begin{eqnarray*} + \frac{1}{x_j\left(\bar \alpha^0\right)}\frac{\left[\sum\limits^m_{k=1}b_{kj}p_k^0+\sum\limits^n_{k=m+1}b_{kj} \bar p_k\left(\alpha^0\right)\right]}
{\left[\sum\limits^m_{k=1}c_{kj}p_k^0 +
\sum\limits^{n}_{k=m+1}c_{kj}p_k^0(0)\right]^2} \end{eqnarray*}  \begin{eqnarray*} \times\sum\limits^{n}_{k=m+1}c_{kj}\left[\left[E - \tilde{\cal A}\left(\alpha^0\right)^T\right]^{-1}\varphi^s\left(\alpha^0\right)\right]_k, \quad j, s=\overline{1, m},\end{eqnarray*}

\begin{eqnarray*} \varphi^s\left(\alpha^0\right)=\{\varphi^s_j\left(\alpha^0\right)\}_{j=m+1}^n,\end{eqnarray*}
\begin{eqnarray*} \varphi^s_j\left(\alpha^0\right) =\frac{x_s\left(\alpha^0\right)}{x_j\left(\bar \alpha^0\right)^2}\left[\left[E - \left(A+ \alpha^0 C\right)\right]^{-1}C\right]_{js} \left[\sum\limits^{m}_{k=1}b_{kj}p_k^0+\sum\limits^{n}_{k=m+1}b_{kj} \bar p_k\left( \alpha^0\right)\right].\end{eqnarray*}
Then there exists a vector $ \beta^0=\{ \beta^0_i\}_{i=1}^n$  such  that \begin{eqnarray*}  \beta^0_i= \alpha^0_i, \quad i=\overline{m+1, n}, \quad  \frac{\pi_iv_i +\varepsilon_i}{x_i\left(\bar \alpha^0\right)}\leq   \beta^0_i  \leq \alpha^0_i, \quad i=\overline{1, m},\end{eqnarray*}
and this vector solves the set of equations (\ref{tas3}).
The solution built satisfies  the same conditions as the vector $\alpha$ that figures in the  conditions of the  Theorem \ref{taxt1}.
\end{theorem}
\begin{proof}\smartqed  Consider on a closed set $H$ of vectors $\alpha=\{\alpha_k\}_{k=1}^m,$ where
\begin{eqnarray} \label{tas7}
H =\left \{\alpha=\{\alpha_k\}_{k=1}^m \in R_+^m,  \  \frac{\pi_i v_i + \varepsilon_i }{x_i\left(\bar \alpha^0\right)}\leq \alpha_i \leq \alpha_i^0, \ i=\overline{1, m}\right\},
\end{eqnarray}
the set of equations (\ref{tas3}). It is obvious that for any vector
\begin{eqnarray*} \tilde \alpha=\{\alpha_1, \ldots, \alpha_m, \alpha_m^0, \ldots, \alpha_n^0\}, \quad \alpha_i^0 >0, \quad i=\overline{m+1, n},\end{eqnarray*}
 whose first $m$ components satisfy the set of inequalities (\ref{tas7})
there exists a solution to the set of equations (\ref{tas5}) satisfying inequalities
\begin{eqnarray*} x_i(\tilde \alpha) \alpha_i^0 > \pi_i v_i, \quad  i=\overline{1,n}, \end{eqnarray*}
because the inequalities $x_i(\tilde \alpha) \geq x_i\left(\bar \alpha^0\right), \ i=\overline{1,n},$  hold.
Prove the existence of the solution to the set of equations (\ref{tas3}).
The inequalities
\begin{eqnarray*} \frac{\pi_j\left[p_j^0 - \sum\limits^m_{k=1}a_{kj}p_k^0
- \sum\limits^{n}_{k=m+1}a_{kj} \bar p_k\right]}
{\sum\limits^m_{k=1}c_{kj}p_k^0 +
\sum\limits^{n}_{k=m+1}c_{kj}\bar p_k} \leq  f_j(\alpha_1, \ldots, \alpha_m)
\end{eqnarray*}
\begin{eqnarray*} \leq  \frac{\pi_j\left[p_j^0+ \frac{1}{x_j\left(\bar \alpha^0\right)} \left[\sum\limits^m_{k=1}b_{kj}p_k^0+\sum\limits^n_{k=m+1}b_{kj} \bar p_k\right]\right]}
{\sum\limits^m_{k=1}c_{kj}p_k^0 +
\sum\limits^{n}_{k=m+1}c_{kj}p_k^0(0)}, \quad j=\overline{1, m},\end{eqnarray*}
hold.
From these inequalities and the  conditions of the Theorem, it follows that the map
\begin{eqnarray*} f(\alpha_1, \ldots, \alpha_m)=\{f_j(\alpha_1, \ldots, \alpha_m)\}_{j=1}^m\end{eqnarray*}
maps the set $H$ into itself. To prove contractibility  of this map,
estimate solutions to the sets of equations (\ref{tas4}) and (\ref{tas5}).

Estimate derivatives of the solution to the set of equations
(\ref{tas4}). For this, find the set of equations the vector
\begin{eqnarray*} \frac{\partial p_1}{\partial \alpha_s}=\left\{\frac{\partial p_j^1}{\partial \alpha_s}\right\}_{j=m+1}^n, \quad s=\overline{1,m},\end{eqnarray*}
 satisfies.

 For $s=\overline{1,m}$ we have
\begin{eqnarray*} \frac{\partial p_j^1}{\partial \alpha_s}= - \sum\limits^{m}_{k=1}b_{kj}p_k^0\frac{\partial}{\partial \alpha_s}\left[\frac{1}{x_j(\tilde \alpha)}\right] - \sum\limits^{n}_{k=m+1}b_{kj}p_k^1\frac{\partial}{\partial \alpha_s}\left[\frac{1}{x_j(\tilde \alpha)}\right] \end{eqnarray*}
\begin{eqnarray} \label{tas10}
+\sum\limits^{n}_{k=m+1}\left[a_{kj}+
\alpha_j^0 \left( \frac{c_{kj}}
{\pi_j} - \frac{b_{kj}}{y_j(\tilde \alpha)}\right)\right]\frac{\partial p_k^1}{\partial \alpha_s},  \quad j=\overline{m+1,n}.
\end{eqnarray}
From the fact that $x(\tilde \alpha) $ satisfy (\ref{tas5})  the vector
\begin{eqnarray*} \frac{\partial x(\tilde \alpha)}{\partial \alpha_s} = \left\{\frac{\partial x_k(\tilde \alpha)}{\partial \alpha_s}\right\}_{k=m+1}^n, \quad s=\overline{1,m},\end{eqnarray*}
satisfies the set of equations
\begin{eqnarray*} \frac{\partial x_k(\tilde \alpha)}{\partial \alpha_s}-\sum\limits^{m}_{j=1}[a_{kj}+ \alpha_j c_{kj}]\frac{\partial x_j(\tilde \alpha)}{\partial \alpha_s}  \end{eqnarray*}
\begin{eqnarray} \label{tas11}
- \sum\limits^{n}_{j=m+
1}[a_{kj}+ \alpha_j^0 c_{kj}]\frac{\partial x_j(\tilde \alpha)}{\partial \alpha_s}=
c_{ks}x_s(\tilde \alpha),
\quad k=\overline{1,n}.
\end{eqnarray}
From (\ref{tas11}) we obtain
\begin{eqnarray} \label{tas12}
\frac{\partial x_k(\tilde \alpha)}{\partial \alpha_s}=x_s(\tilde \alpha)\left[\left[E - (A+ \tilde \alpha C)\right]^{-1}C\right]_{ks},
\end{eqnarray}
where
$\tilde \alpha =\{\alpha_1, \ldots, \alpha_m, \alpha_{m+1}^0, \ldots,
\alpha_{n}^0\},$ and  $\left[\left[E - (A+ \tilde \alpha C)\right]^{-1}C\right]_{ks}$ is a matrix element of the matrix
 $\left[E - (A+ \tilde \alpha C)\right]^{-1}C,$
\begin{eqnarray*}  A+ \tilde \alpha C=||a_{kj}+ \tilde \alpha_j c_{kj}||_{k, j=1}^n, \quad \tilde \alpha_j=\alpha_j, \quad j=\overline{1,m}, \quad \tilde \alpha_j=\alpha_j^0, \quad j=\overline{m+1,n}.\end{eqnarray*}
Therefore, the estimate
\begin{eqnarray} \label{tas13}
\left|\frac{\partial x_k(\tilde \alpha)}{\partial \alpha_s}\right|\leq x_s\left(\alpha^0\right)\left[\left[E - \left(A+  \alpha^0 C\right)\right]^{-1}C\right]_{ks}
\end{eqnarray}
holds.
As the equality
\begin{eqnarray*}  \frac{\partial }{\partial \alpha_s}\left[\frac{1}{x_k(\tilde \alpha)}\right]=- \frac{1}{x_k(\tilde \alpha)^2}\frac{\partial x_k(\tilde \alpha)}{\partial \alpha_s} \end{eqnarray*}
 holds we have
\begin{eqnarray*}  \frac{\partial }{\partial \alpha_s}\left[\frac{1}{x_k(\tilde \alpha)}\right]=-
\frac{x_s(\tilde \alpha)}{x_k(\tilde \alpha)^2}\left[\left[E - (A+ \tilde \alpha C)\right]^{-1}C\right]_{ks}.\end{eqnarray*}
Write the set of equations (\ref{tas10}) in the form
\begin{eqnarray*} \frac{\partial p_j^1}{\partial \alpha_s}=\frac{x_s(\tilde \alpha)}{x_j(\tilde \alpha)^2}  \left[\left[E - (A+ \tilde \alpha C)\right]^{-1}C\right]_{js}\sum\limits^{m}_{k=1}b_{kj}p_k^0 \end{eqnarray*}
\begin{eqnarray*} +\frac{x_s(\tilde \alpha)}{x_j(\tilde \alpha)^2}\left[\left[E - (A+ \tilde \alpha C)\right]^{-1}C\right]_{js} \sum\limits^{n}_{k=m+1}b_{kj}p_k^1 \end{eqnarray*}
\begin{eqnarray} \label{!tas14}
+\sum\limits^{n}_{k=m+1}\left[a_{kj}+
\alpha_j^0 \left( \frac{c_{kj}}
{\pi_j} - \frac{b_{kj}}{y_j(\tilde \alpha)}\right)\right]\frac{\partial p_k^1}{\partial \alpha_s}.
\end{eqnarray}
From the set of equations (\ref{!tas14} ) for the vector
\begin{eqnarray*} \frac{\partial p_1}{\partial \alpha_s}= \left\{\frac{\partial p_j^1}{\partial\alpha_s}\right\}_{j=m+1}^n\end{eqnarray*}
we obtain the set of inequalities
\begin{eqnarray*} \frac{\partial p_j^1}{\partial \alpha_s}\leq \frac{x_s\left(\alpha^0\right)}{x_j\left(\bar \alpha^0\right)^2} \left[\left[E - (A+ \alpha^0 C)\right]^{-1}C\right]_{js}\sum\limits^{m}_{k=1}b_{kj}p_k^0 \end{eqnarray*}
\begin{eqnarray*} +\frac{x_s\left(\alpha^0\right)}{x_j\left(\bar \alpha^0\right)^2}\left[\left[E - \left(A+ \alpha^0 C\right)\right]^{-1}C\right]_{js} \sum\limits^{n}_{k=m+1}b_{kj} \bar p_k\left(\alpha^0\right) \end{eqnarray*}
\begin{eqnarray} \label{tas14}
+\sum\limits^{n}_{k=m+1}\left[a_{kj}+
\alpha_j^0 \left( \frac{c_{kj}}
{\pi_j} - \frac{b_{kj}}{y_j\left(\alpha^0\right)}\right)\right]\frac{\partial p_k^1}{\partial \alpha_s}.
\end{eqnarray}
Introduce the vector
\begin{eqnarray*} \varphi^s\left(\alpha^0\right)=\{\varphi^s_j\left(\alpha^0\right)\}_{j=m+1}^n,\end{eqnarray*}
where
\begin{eqnarray*} \varphi^s_j\left(\alpha^0\right)=\frac{x_s\left(\alpha^0\right)}{x_j\left(\bar \alpha^0\right)^2}\left[\left[E - \left(A+ \alpha^0 C\right)\right]^{-1}C\right]_{js} \left[\sum\limits^{m}_{k=1}b_{kj}p_k^0+\sum\limits^{n}_{k=m+1}b_{kj} \bar p_k\left( \alpha^0\right)\right],\end{eqnarray*}
and the matrix
\begin{eqnarray*}  \tilde{\cal A}\left( \alpha^0\right)=\left|\left|a_{kj}+\alpha_j^0\left(\frac{c_{kj}}{\pi_j} -\frac{b_{kj}}{y_j\left(\alpha^0\right)}\right)\right|\right|_{k,j=m+1}^n.\end{eqnarray*}
From the set of inequalities (\ref{tas14}), the estimate
\begin{eqnarray*} 0 \leq \frac{\partial p_j^1}{\partial\alpha_s} \leq \left[\left[E - \tilde{\cal A}\left( \alpha^0\right)^T\right]^{-1}\varphi^s\left(\alpha^0\right)\right]_j, \quad j=\overline{m+1,n}, \quad  s=\overline{1,m},\end{eqnarray*}
 follows.
Here $ \tilde{\cal A}\left( \alpha^0\right)^T $ is the matrix transposed to the matrix $\tilde{\cal A}\left(\alpha^0\right).$

Find conditions under which the map
\begin{eqnarray*} f(\alpha_1, \ldots, \alpha_m)=\{f_j\left(\alpha_1, \ldots, \alpha_m\right)\}_{j=1}^m\end{eqnarray*}
is a contraction on the set considered.
The equality
\begin{eqnarray*} f_j\left(\alpha_1^{''}, \ldots, \alpha_m^{''}\right) - f_j\left(\alpha_1^{'}, \ldots, \alpha_m^{'}\right) \end{eqnarray*}
\begin{eqnarray*} =\int\limits_{0}^{1}\sum\limits_{s=1}^m\left(\alpha_s^{''}- \alpha_s^{'}\right)\frac{\partial}{\partial \alpha_s}f_j\left(\alpha_1^{'}+t \left(\alpha_1^{''}-\alpha_1^{'}\right), \ldots,\alpha_m^{'}+t \left(\alpha_m^{''}- \alpha_m^{'}\right)\right)dt\end{eqnarray*}
holds.
From the last equality, it follows that to prove the map $f(\alpha_1, \ldots, \alpha_m)$ to be a contraction one must estimate
$\frac{\partial}{\partial \alpha_s}f_j(\alpha_1, \ldots,\alpha_m)$ on the set considered.
There holds the equality
\begin{eqnarray*} \frac{\partial}{\partial \alpha_s}f_j(\alpha_1, \ldots,\alpha_m) \end{eqnarray*}
\begin{eqnarray*} =\pi_j\frac{\left[- \sum\limits^{n}_{k=m+1}a_{kj}\frac{\partial p_k^1}{\partial \alpha_s} - \frac{1}{x_j(\tilde \alpha)^2}\frac{\partial x_j(\tilde \alpha)}{\partial \alpha_s} \left[\sum\limits^m_{k=1}b_{kj}p_k^0+\sum\limits^n_{k=m+1}b_{kj} p_k^1\right]\right]}
{\sum\limits^m_{k=1}c_{kj}p_k^0 +
\sum\limits^{n}_{k=m+1}c_{kj}p_k^1} \end{eqnarray*}
\begin{eqnarray*} +\frac{\pi_j}{x_j(\tilde \alpha)}\frac{ \sum\limits^n_{k=m+1}b_{kj}\frac{\partial p_k^1}{\partial \alpha_s}}{\sum\limits^m_{k=1}c_{kj}p_k^0 +
\sum\limits^{n}_{k=m+1}c_{kj}p_k^1} \end{eqnarray*}
\begin{eqnarray*} -\frac{\pi_j\left[p_j^0 - \sum\limits^m_{k=1}a_{kj}p_k^0
- \sum\limits^{n}_{k=m+1}a_{kj} p_k^1+ \frac{1}{x_j(\tilde \alpha)}\left[\sum\limits^m_{k=1}b_{kj}p_k^0+\sum\limits^n_{k=m+1}b_{kj} p_k^1\right]\right]}
{\left[\sum\limits^m_{k=1}c_{kj}p_k^0 +
\sum\limits^{n}_{k=m+1}c_{kj}p_k^1\right]^2} \end{eqnarray*}
\begin{eqnarray*}  \times \sum\limits^{n}_{k=m+1}c_{kj}\frac{\partial p_k^1}{\partial \alpha_s}.\end{eqnarray*}
In view of estimates obtained, the inequality
\begin{eqnarray*}
\left|\frac{\partial}{\partial \alpha_s}f_j(\alpha_1, \ldots,\alpha_m)\right| \leq \pi_j\frac{\sum\limits^{n}_{k=m+1}a_{kj} \left[\left[E - \tilde{\cal A}\left( \alpha^0\right)^T\right]^{-1}\varphi^s\left(\alpha^0\right)\right]_k }{\sum\limits^m_{k=1}c_{kj}p_k^0 +
\sum\limits^{n}_{k=m+1}c_{kj}p_k^0(0)} 
\end{eqnarray*}
\begin{eqnarray*} +\pi_j\frac{ x_s\left(\alpha^0\right)\left[\left[E - \left(A+ \alpha^0 C\right)\right]^{-1}C\right]_{js} \left[\sum\limits^m_{k=1}b_{kj}p_k^0+\sum\limits^n_{k=m+1}b_{kj} \bar p_k\left( \alpha^0\right)\right]}
{x_j\left(\bar \alpha^0\right)^2\left[\sum\limits^m_{k=1}c_{kj}p_k^0 +
\sum\limits^{n}_{k=m+1}c_{kj}p_k^0(0)\right]} 
\end{eqnarray*}
\begin{eqnarray*} +\frac{\pi_j}{x_j\left(\bar \alpha^0\right)}\frac{ \sum\limits^n_{k=m+1}b_{kj}\left[\left[E - \tilde{\cal A}\left( \alpha^0\right)^T\right]^{-1}\varphi^s\left(\alpha^0\right)\right]_k }{\sum\limits^m_{k=1}c_{kj}p_k^0 +
\sum\limits^{n}_{k=m+1}c_{kj}p_k^0(0)} \end{eqnarray*}
\begin{eqnarray*} +\frac{\pi_j\left[p_j^0 - \sum\limits^m_{k=1}a_{kj}p_k^0
- \sum\limits^{n}_{k=m+1}a_{kj} p_k^0(0)\right]}
{\left[\sum\limits^m_{k=1}c_{kj}p_k^0 +
\sum\limits^{n}_{k=m+1}c_{kj}p_k^0(0)\right]^2} \end{eqnarray*}  \begin{eqnarray*}  \times \sum\limits^{n}_{k=m+1}c_{kj}\left[\left[E - \tilde{\cal A}\left( \alpha^0\right)^T\right]^{-1}\varphi^s\left(\alpha^0\right)\right]_k 
\end{eqnarray*}
\begin{eqnarray*} + \frac{\pi_j}{x_j\left(\bar \alpha^0\right)}\frac{\left[\sum\limits^m_{k=1}b_{kj}p_k^0+\sum\limits^n_{k=m+1}b_{kj} \bar p_k\left(\alpha^0\right)\right]}
{\left[\sum\limits^m_{k=1}c_{kj}p_k^0 +
\sum\limits^{n}_{k=m+1}c_{kj}p_k^0(0)\right]^2} 
\end{eqnarray*}  
\begin{eqnarray*} \times\sum\limits^{n}_{k=m+1}c_{kj}\left[\left[E - \tilde{\cal A}\left(\alpha^0\right)^T\right]^{-1}\varphi^s\left(\alpha^0\right)\right]_k  \end{eqnarray*}
\begin{eqnarray*} =\pi_jd_{js}\left(p^0, A, \bar B,\bar C,  \alpha^0\right)\end{eqnarray*}
holds.
With notations introduced, the last estimate turns into
\begin{eqnarray*} \left|\frac{\partial}{\partial \alpha_s}f_j(\alpha_1, \ldots,\alpha_m)\right|\leq \pi_jd_{js}\left(p^0, A,\bar B,\bar C,  \alpha^0\right).\end{eqnarray*}
Introduce on the  considered set the norm $||\alpha||=\max\limits_{1 \leq i \leq m}|\alpha_i|.$
In this norm, we estimate this map $f(\alpha_1, \ldots, \alpha_m)=\{f_i(\alpha_1, \ldots, \alpha_m)\}_{i=1}^m$ as follows \begin{eqnarray*}  \left|f\left(\alpha_1^{''}, \ldots, \alpha_m^{''}\right)- f\left(\alpha_1^{'}, \ldots, \alpha_m^{'}\right)\right|  \end{eqnarray*}  \begin{eqnarray*}  \leq \left| \alpha^{''} - \alpha^{'}\right| \max\limits_{1\leq j \leq m} \pi_j\sum\limits_{s=1}^m d_{js}\left(p^0, A, \bar B,\bar C,  \alpha^0\right).\end{eqnarray*}
Therefore, it is a contraction map according to the  condition of the  Theorem.
From the conditions of the  Theorem  it follows that there exists a unique solution to the set of equations (\ref{tas3}) satisfying the  conditions of the Theorem \ref{taxt1}.
\qed\end{proof}

\section{Guaranteeing given levels of consumers needs satisfaction}

The problem of determination of  conditions under which monopolistic prices\index{monopolistic prices} increase in the economy system does not deteriorate welfare\index{deteriorate welfare} is important. From the mathematical point of view, it means: what must be  levels of  gross outputs,\index{levels of  gross outputs}
technological coefficients,\index{technological coefficients}  vectors of export and import\index{vectors of export and import}   for levels of satisfaction of  consumers needs\index{levels of satisfaction of  consumers needs}   do not deteriorate.
One can, e.g., correct levels of  taxation of monopolistic industries in the way that  levels of consumption of consumers  do not deteriorate. In this Section, we solve this problem.

Let the vector $\alpha^0=\{\alpha_j^0\}_{j=1}^{n} $ be such that the spectral radius of the matrix $||a_{kj}+\alpha_j^0c_{kj}||_{k,j=1}^{n} $ is less than 1.
Consider the set of equations
\begin{eqnarray} \label{tar1222}
x_k-\sum\limits^{n}_{j=1}\left[a_{kj}+\alpha_jc_{kj}\right]x_j
+\sum\limits^l_{j=1}b_{kj}-e_k+i_k=
\sum\limits^l_{j=n+1}c_{kj}y_j,
\quad k=\overline{1,n},
\end{eqnarray}
relative to  the vector $ x= \{ x_i \}_{i=1}^{n},$ where the vector $y=\{y_i\}_{i=1}^l$ is a strictly positive vector of levels of satisfaction of consumers needs   and
 $ x(\alpha)= \{ x_i( \alpha_1,\ldots, \alpha_{n}) \}_{i=1}^{n}$
solves the set of equations (\ref{tar1222})
for the vectors
 $\alpha=\{\alpha_j\}_{j=1}^{n} $ belonging to the set
 \begin{eqnarray*}  T=\{\alpha=\{\alpha_i\}_{i=1}^{n},  \
0 \leq \alpha_i\leq \alpha_i^0,
\ i=\overline{1, n}\},\end{eqnarray*}   the vector
$x^0=\{x_j^0\}_{j=1}^{n}$ solves the set of equations (\ref{tar1222})
for the vector
 $\alpha=\{\alpha_j\}_{j=1}^{n} $ with zero components.
Consider the set of equations
\begin{eqnarray*}  p_j =
\sum\limits^m_{k=1}\left[a_{kj}+  \alpha_j\left( \frac{c_{kj}}
{\pi_j} - \frac{b_{kj}}{y_j}\right)\right]
p_k^0  \end{eqnarray*}
\begin{eqnarray}
\label{tam9}
+ \sum\limits^{n}_{k=m+1}\left[a_{kj}+ \alpha_j\left( \frac{c_{kj}}
{\pi_j} - \frac{b_{kj}}{y_j}\right)\right] p_k, \quad  j=\overline{m+1,n}\ ;
\end{eqnarray}
for the vector $ p=\{ p_j\}_{j=m+1}^n,$
where the vector $ \bar \alpha \in T_1,$
\begin{eqnarray*} T_1=\{\bar \alpha=\{\alpha_i\}_{i=m+1}^{n},  \
0 \leq \alpha_i\leq \alpha_i^0,
\ i=\overline{m+1, n}\}.\end{eqnarray*}
We denote a solution to the set of equations (\ref{tam9}) for arbitrary vector $\bar \alpha=\{\alpha_i\}_{i=m+1}^{n}$ from the set $T_1$ by $ p(\bar \alpha)=\{ p_j(\bar \alpha)\}_{j=m+1}^n.$

So, $ p\left(\bar \alpha_0\right)=\{ p_j\left(\bar \alpha_0\right)\}_{j=m+1}^n$ solves the set of equations (\ref{tam9}) for the vector $\bar \alpha_0=\{\alpha_i^0\}_{i=m+1}^{n},$ and
$ p(0)=\{ p_j(0)\}_{j=m+1}^n$ solves the set of equations (\ref{tam9}) for the vector  $\alpha=\{\alpha_i\}_{i=m+1}^{n}, \ \alpha_i=0, \ i=\overline{m+1, n}.$

\begin{theorem}\label{tam1}
Let $\sum\limits_{k=1}^nc_{ki}>0, \ i=\overline{1,l},$ there  exist a non-negative vector
$v_0=\{v_i\}_{i=1}^n, \ v_i \geq 0, \ i=\overline{1,n},$  such that $\bar C(v_0) - \bar B $ is a non-negative matrix having no zero rows or columns and
$ A + \bar C(v_0) - \bar B$ is an  indecomposable matrix,\index{ indecomposable matrix} and let the  spectral radius of the  matrix\index{ spectral radius of the  matrix} $A$ be less than 1.
And also let a strictly positive vector
$\alpha^0=\{\alpha_i^0 \}_{i=1}^{n}$
and a vector of  monopolistic prices\index{vector of  monopolistic prices}  $p^0=\{p_1^0, \ldots, p_m^0 \}$ satisfy conditions:
the spectral radius of the matrix $||a_{kj}+\alpha_j^0c_{kj}||_{k,j=1}^n$ is less than 1,
there exists a strictly positive solution $x^0=\{x_j^0\}_{j=1}^n $ to the set of equations (\ref{tar1222})
for the vector $\alpha=0.$

If inequalities
\begin{eqnarray*} \frac{y_j}{x_j^0} \leq \alpha_j^0, \quad j=\overline{1,n}, \end{eqnarray*}
\begin{eqnarray*} \max\limits_{j=\overline{1,n}}\frac{y_j}{\left[x_j^0\right]^2}
\sum\limits_{s=1}^n\left[\left[E -\left(A+\alpha^0C\right)\right]^{-1}C\right]_{js}x_s\left(\alpha^0\right)
<1, \end{eqnarray*}
hold, then there exists the unique solution $  \alpha_1=\{ \alpha_j^1\}_{j=1}^n $ to the set of equations
\begin{eqnarray} \label{tax022}
\alpha_j=\frac{ y_j}
{ x_j(\alpha_1,\ldots,\alpha_{n})}, \quad j=\overline{1,n},
\end{eqnarray}
for the vector $  \alpha=\{ \alpha_j\}_{j=1}^n $ in the set $T.$

If, in addition,
\begin{eqnarray} \label{tam3}
 p_j^0 - \sum\limits^m_{k=1}a_{kj}p_k^0
- \sum\limits^{n}_{k=m+1}a_{kj} p_k\left(\bar \alpha^0\right)  > 0, \quad j=\overline{1,m},
\end{eqnarray}
\begin{eqnarray*} y_j >
\frac{v_j  \alpha_j^0\left[\sum\limits^m_{k=1}c_{kj}p_k^0 + \sum\limits^{n}_{k=m+1}c_{kj} p_k(\bar \alpha^0)\right]}
{M_j\left(p^0, A, \bar B, x\left(\alpha^0\right),  p(0),  p\left(\bar \alpha^0\right)\right)}, \quad j=\overline{1,m},\end{eqnarray*}
\begin{eqnarray} \label{tam2}
\quad y_j > \pi_j v_j, \quad j=\overline{m+1,n},
\end{eqnarray}
where
\begin{eqnarray*} M_j\left(p^0, A, \bar B, x\left(\alpha^0\right),  p(0), p\left(\bar \alpha^0\right)\right)= p_j^0 - \sum\limits^m_{k=1}a_{kj}p_k^0
- \sum\limits^{n}_{k=m+1}a_{kj} p_k\left(\bar \alpha^0\right) \end{eqnarray*}  \begin{eqnarray*} + \frac{1}{x_j\left(\alpha^0\right)}\left[\sum\limits^m_{s=1}b_{sj}p_s^0+\sum\limits^n_{s=m+1}b_{sj}  p_s(0)\right],\end{eqnarray*}
 the spectral radius of the matrix
\begin{eqnarray*} \tilde{\cal A}\left(\bar \alpha^0\right)=\left\|a_{kj}+ \alpha^0_j\left( \frac{c_{kj}}
{\pi_j} - \frac{b_{kj}}{y_j}\right) \right\|^{n}_{k,j=m+1}
\end{eqnarray*}
is less than 1, the solution $ p(0)=\{ p_j(0)\}_{j=m+1}^n$ to the set of equations (\ref{tam9}) for zero vector $\alpha$ is strictly positive, then there exists a strictly positive solution to the set of equations
\begin{eqnarray*} p_j^0 =
\sum\limits^m_{k=1}\left[a_{kj}+
\frac{y_j}{x_j( \alpha_1)}\left(\frac{c_{kj}}{\pi_j( \alpha_1)}-\frac{b_{kj}}{y_j}\right)\right]
p_k^0  \end{eqnarray*} 
 \begin{eqnarray*}  +\sum\limits^{n}_{k=m+1}\left[a_{kj}+\frac{y_j}{x_j( \alpha_1)}\left(\frac{c_{kj}}{\pi_j(\alpha_1)}-\frac{b_{kj}}{y_j}\right)\right]p_k, \quad j=\overline{1,m}, \end{eqnarray*}
\begin{eqnarray*} p_j =
\sum\limits^m_{k=1}\left[a_{kj}+\frac{y_j}{x_j( \alpha_1)}\left(\frac{c_{kj}}{\pi_j}-\frac{b_{kj}}{y_j}\right)\right]
p_k^0  \end{eqnarray*}
\begin{eqnarray} \label{tam5}
+ \sum\limits^{n}_{k=m+1}\left[a_{kj}+\frac{y_j}{x_j( \alpha_1)}\left(\frac{c_{kj}}{\pi_j}-\frac{b_{kj}}{y_j}\right)\right]p_k, \quad j=\overline{m+1,n},
\end{eqnarray}
for the vector $p=\{p_{i}\}_{i=m+1}^n,$
if to put
\begin{eqnarray}\label{tam7}
\pi_j(\alpha_1) =\frac{\alpha_j^1\left[\sum\limits^m_{k=1}c_{kj}p_k^0 +
\sum\limits^{n}_{k=m+1}c_{kj} p_k(\bar  \alpha_1)\right]}
{V_j\left(p^0, A,  \bar B,  p(\bar \alpha_1), x( \alpha_1)\right)}, \quad j=\overline{1,m},\end{eqnarray}
 \begin{eqnarray*} V_j\left(p^0, A, \bar B,  p(\bar \alpha_1), x( \alpha_1)\right) =p_j^0- \sum\limits^m_{k=1}a_{kj}p_k^0
- \sum\limits^{n}_{k=m+1}a_{kj} p_k(\bar \alpha_1) \end{eqnarray*}  \begin{eqnarray*} + \frac{1}{x_j( \alpha_1)}\left[\sum\limits^m_{s=1}b_{sj}p_s^0+\sum\limits^n_{s=m+1}b_{sj} p_s(\bar \alpha_1)\right],\end{eqnarray*}
and the vector
 $ x(\alpha_1)=\{ x_k(\alpha_1)\}_{k=1}^n$
solves the set of equations
\begin{eqnarray} \label{tam6}
x_k-\sum\limits^{n}_{j=1}a_{kj}x_j+ \sum\limits^l_{j=1}b_{kj} -e_k+i_k=
\sum\limits^l_{j=1}c_{kj}y_j,
\quad k=\overline{1,n},
\end{eqnarray}
relative to the vector $ x=\{ x_k\}_{k=1}^n,$
where
$ p(\bar \alpha_1)=\{ p_i(\bar \alpha_1)\}^{n}_{i=m+1}$ is a solution of the set of equations (\ref{tam9}) for the vector $\bar \alpha_1=\{ \alpha_i^1\}_{i=m+1}^n$ whose components are built after the unique solution $ \alpha_1=\{ \alpha_i^1\}_{i=1}^n$ to the set of equations (\ref{tax022}).
\end{theorem}
\begin{proof}\smartqed  Let
$x(\alpha_1,\ldots,\alpha_{n})=
\{x_j(\alpha_1,\ldots,\alpha_{n})\}_{j=1}^{n}$ be a solution of  the set of equations
(\ref{tar1222}) for the vector $ \alpha=\{ \alpha_1, \ldots,  \alpha_n\} \in T.$ The non-linear map
\begin{eqnarray*}  f(\alpha_1,\ldots,\alpha_{n})=\{ f_j(\alpha_1,\ldots,\alpha_{n}) \}_{j=1}^{n},\end{eqnarray*}
where
\begin{eqnarray*} f_j(\alpha_1,\ldots,\alpha_{n})=\frac{ y_j}
{ x_j(\alpha_1,\ldots,\alpha_{n})}, \qquad j=\overline{1,n},\end{eqnarray*}
is a monotonously decreasing map of the set $T$ into itself. The map
\begin{eqnarray*} F(\alpha_1,\ldots,\alpha_{n})=f^2(\alpha_1,\ldots,\alpha_{n}) \end{eqnarray*}
\begin{eqnarray*} =\{f_j(f_1(\alpha_1,\ldots,\alpha_{n}),\ldots,
f_{n}(\alpha_1,\ldots,\alpha_{n}))\}_{j=1}^{n}\end{eqnarray*}
is a monotonously non-decreasing map of the set $T$ into itself, therefore the following vector inequalities
\begin{eqnarray*} F^k(0,\ldots,0)\leq F^{k+1}(0,\ldots,0),\end{eqnarray*}
\begin{eqnarray*} F^k\left(\alpha_1^0,\ldots,\alpha_{n}^0\right)\geq
F^{k+1}\left(\alpha_1^0,\ldots,\alpha_{n}^0\right),\end{eqnarray*}
hold that one must understand componentwise.
By $F^m(\alpha_1,\ldots,\alpha_{n})$, we denote the $m$-th power of the map
$F(\alpha_1,\ldots,\alpha_{n}).$
In addition,
\begin{eqnarray*} F^k(0,\ldots,0)\le
F^k\left(\alpha_1^0,\ldots,\alpha_{n}^0\right).\end{eqnarray*}
Therefore, the sequence $F^k(0,\ldots,0)$ is  monotone non-decreasing  and
the sequence $F^k\left(\alpha_1^0,\ldots,\alpha_{n}^0\right)$ is monotone non-increasing.
Denote
\begin{eqnarray*} \lim\limits_{k \to \infty}F^k(0,\ldots,0)=F^\infty(0,\ldots,0), \quad
\lim\limits_{k \to \infty}F^k\left(\alpha_1^0,\ldots,\alpha_{n}^0\right)
=F^\infty\left(\alpha_1^0,\ldots,\alpha_{n}^0\right).\end{eqnarray*}
If the equality
\begin{eqnarray}
\label{tax023}
F^\infty(0,\ldots,0)=
F^\infty\left(\alpha_1^0,\ldots,\alpha_{n}^0\right)
\end{eqnarray}
holds, then the vector $\bar \alpha^1=F^\infty(0,\ldots,0)$ solves the problem (\ref{tax022}).

Find sufficient conditions under which  the equality (\ref{tax023}) holds.
There hold the equalities
\begin{eqnarray*} x_j\left(\alpha_1^{''}, \ldots, \alpha_n^{''}\right) - x_j\left(\alpha_1^{'}, \ldots, \alpha_n^{'}\right) \end{eqnarray*}
\begin{eqnarray*} =\int\limits_{0}^{1}\sum\limits_{s=1}^m\left(\alpha_s^{''}- \alpha_s^{'}\right)\frac{\partial}{\partial \alpha_s}x_j\left(\alpha_1^{'}+t \left(\alpha_1^{''}-\alpha_1^{'}\right), \ldots,\alpha_n^{'}+t \left(\alpha_n^{''}- \alpha_n^{'}\right)\right)dt,\end{eqnarray*}
\begin{eqnarray} \label{tam11}
\frac{\partial x_k( \alpha)}{\partial \alpha_s} - \sum\limits^{n}_{j=1}\left[a_{kj}+ \alpha_j c_{kj}\right]\frac{\partial x_j( \alpha)}{\partial \alpha_s}=
c_{ks}x_s( \alpha),
\quad k=\overline{1,n}.
\end{eqnarray}
From here
\begin{eqnarray} \label{tam12}
\frac{\partial x_k(\alpha)}{\partial \alpha_s}=x_s( \alpha)\left[\left[E - (A+  \alpha\bar  C)\right]^{-1}\bar C\right]_{ks},
\end{eqnarray}
where
$ \alpha =\{\alpha_1, \ldots, \alpha_n\}\in T,$ and
$ A+ \alpha \bar C=||a_{kj}+ \alpha_j c_{kj}||_{k, j=1}^n.$
Therefore, the estimate
\begin{eqnarray} \label{tam13}
\left|\frac{\partial x_k(\alpha)}{\partial \alpha_s}\right|\leq x_s\left(\alpha^0\right)\left[\left[E - \left(A+  \alpha^0 \bar C\right)\right]^{-1}\bar C\right]_{ks}
\end{eqnarray}
and inequalities
\begin{eqnarray*} \max\limits_{j \in [1,n]}\left|f_j(\alpha_1^{''},\ldots,\alpha_{n}^{''}) -
f_j\left(\alpha_1^{'},\ldots,\alpha_{n}^{'}\right)\right|  \end{eqnarray*}  \begin{eqnarray*} =
\max\limits_{j \in [1,n]}
\left|\frac{ y_j}{ x_j(\alpha_1^{''},\ldots,\alpha_{n}^{''})} -
\frac{ y_j}{ x_j\left(\alpha_1^{'},\ldots,\alpha_{n}^{'}\right)}\right| \end{eqnarray*}
\begin{eqnarray*} \leq
\max\limits_{j \in [1,n]}
\frac{y_j\left|x_j(\alpha_1^{''},\ldots,\alpha_{n}^{''}) -
x_j\left(\alpha_1^{'},\ldots,\alpha_{n}^{'}\right)\right|}
{\left[x_j^0\right]^2} \end{eqnarray*}
\begin{eqnarray*} \leq \max\limits_{j \in [1,n]}
\frac{y_j}{\left[x_j^0\right]^2}
\sum\limits_{s=1}^n\left[\left[E - \left(A+  \alpha^0 \bar C\right)\right]^{-1}\bar C\right]_{js}x_s\left(\alpha^0\right)
\max\limits_{s \in [1,n]}\left|\alpha_s^{''} - \alpha_s^{'}\right|\end{eqnarray*}
hold.

Therefore, if
\begin{eqnarray*} \max\limits_{j \in [1,n]}
\frac{y_j}{\left[x_j^0\right]^2}
\sum\limits_{s=1}^n\left[\left[E - \left(A+  \alpha^0 \bar C\right)\right]^{-1}\bar C\right]_{js}x_s\left(\alpha^0\right)
<1,\end{eqnarray*}
then the solution to the set of equations
(\ref{tax022}) is unique and the equality (\ref{tax023}) holds.
From the inequalities
\begin{eqnarray} \label{tam15}
y_j >
\frac{v_j  \alpha_j^0\left[\sum\limits^m_{k=1}c_{kj}p_k^0 + \sum\limits^{n}_{k=m+1}c_{kj} p_k\left(\bar \alpha^0\right)\right]}
{M_j\left(p^0, A, \bar B, x\left(\alpha^0\right),  p(0),  p\left(\bar \alpha^0\right)\right)},
\quad j=\overline{1,m},
\end{eqnarray}
the inequalities
\begin{eqnarray*}   y_j > \pi_j (\alpha_1)v_j, \quad j=\overline{1,m},\end{eqnarray*}
follow.

Under the  fulfilment of additional conditions, there exists  a solution to the sets of equations (\ref{tam5}) and (\ref{tam6}).
\qed\end{proof}
\begin{note}
From the Theorem \ref{tam1}, it follows that the vector \begin{eqnarray*} \{p_1^0, \ldots, p_m^0, p_{m+1}(\bar \alpha_1), \ldots, p_{n}(\bar \alpha_1)\}\end{eqnarray*}  is a strictly positive solution to the set of equations (\ref{tam5}) in the set $T_0,$ therefore, the taxation vector given by the formula \begin{eqnarray*} \{\pi_1 (\alpha_1), \ldots, \pi_m (\alpha_1),\pi_{m+1},  \ldots, \pi_{n}\}\end{eqnarray*}  agrees with the  structure of consumption.
\end{note}

\section{Zero initial goods supply}

The case of zero initial goods supply\index{zero initial goods supply} is important  in view of  simplicity of results and available applications to real economy systems when one can neglect initial goods supply at the beginning of   the economy operation  period.
As in the case of non-zero goods supply,
 for given technological matrix\index{technological matrix}
$A\left(x^0\right)=\|a_{kj}\left(x_j^0\right)\|^{n}_{k,j=1}$ non-linearly depending on the strictly positive vector of  gross outputs\index{vector of  gross outputs }
$ x^0 \in X,$
the structure of  unproductive consumption\index{structure of  unproductive consumption} given by the matrix
$C=\|c_{kj}\|^{n,l}_{k=1,j=1},$ one must establish conditions for a strictly positive  vector of monopolistic prices\index{ vector of monopolistic prices}
$p^0=\{p_j^0\}^{m}_{j=1}$ and a vector of levels of satisfaction of  consumers needs\index{vector of levels of satisfaction of  consumers needs}
$y=\{y_j\}^{l}_{j=1}$ under which for industries taxation vector\index{industries taxation vector}
$\pi=\{\pi_s\}^{n}_{s=1}, $  $0<\pi_i<1, \ i=\overline{1, n},$ there exists a strictly positive solution to the problem

\begin{eqnarray*} p_j^0 =
\sum\limits^m_{k=1}\left(a_{kj}\left(x_j^0\right)+\frac{y_j}{\pi_jx_j^0}c_{kj}\right)
p_k^0   \end{eqnarray*}  \begin{eqnarray*} + \sum\limits^{n}_{k=m+1}\left(a_{kj}\left(x_j^0\right)+
\frac{y_j}{\pi_jx_j^0}c_{kj}\right)p_k, \quad j=\overline{1,m}, \end{eqnarray*}

\begin{eqnarray*}  p_j =
\sum\limits^m_{k=1}\left(a_{kj}\left(x_j^0\right)+\frac{y_j}{\pi_jx_j^0}c_{kj}\right)
p_k^0  \end{eqnarray*}
\begin{eqnarray} \label{sor1220}
+\sum\limits^{n}_{k=m+1}\left(a_{kj}\left(x_j^0\right)+
\frac{y_j}{\pi_jx_j^0}c_{kj}\right)p_k, \quad j=\overline{m+1,n},
\end{eqnarray}

\begin{eqnarray} \label{sor1221}
x_k^0-\sum\limits^{n}_{j=1}a_{kj}\left(x_j^0\right)x_j^0-e_k+i_k=
\sum\limits^l_{j=1}c_{kj}y_j,
\quad k=\overline{1,n},
\end{eqnarray}
relative to vectors
$p=\{p_{m+1}, \ldots ,p_{n}\} \in R_+^{n-m}$ and
$x^0=\{x_{1}^0, \ldots ,x_{n}^0\} \in X.$

\begin{theorem} \label{sort1}
Let $\sum\limits_{s=1}^nc_{si}>0, \ i=\overline{1,l},$
the matrix $\bar  C$ have no zero rows, the matrix
$(A\left(x^0\right) + \tilde C), \ x^0 \in X,$ be   indecomposable and let the vector
$x^0=\{x_i^0\}_{i=1}^{n} \in X$ be a strictly positive solution to the set of equations (\ref{sor1221}) with rhs containing components $y_i, \ i =\overline{1,l},$ of  levels of satisfaction  of  consumers needs.\index{levels of satisfaction  of  consumers needs}
The taxation vector\index{taxation vector agrees with the structure of consumption  in the economy system in the presence of monopolists} $\pi$  agrees with the structure of consumption  in the economy system in the presence of monopolists with the  strictly positive vector of monopolistic prices \index{ vector of monopolistic prices }
$p^0=\{p_1^0,\ldots,p_m^0\}$ and vector of levels of satisfaction of consumers needs   $y=\{y_i\}^l_{i=1}$ if\\
1) the spectral radius of the matrix\index{spectral radius of the matrix}
\begin{eqnarray*} \tilde{\cal A}\left(x^0\right)=\left\|a_{kj}\left(x_j^0\right)+\frac{y_jc_{kj}}
{\pi_jx_j^0}\right\|^{n}_{k,j=m+1},
\end{eqnarray*}
is less than 1\ ;\\
2) there exists a strictly positive solution $\bar p=\{\bar p_i\}^{n}_{i=m+1}$ to the set of equations
\begin{eqnarray*}  p_j =
\sum\limits^m_{k=1}\left(a_{kj}\left(x_j^0\right)+\frac{y_j}{\pi_jx_j^0}c_{kj}\right)
p_k^0  \end{eqnarray*}
\begin{eqnarray} \label{sor2}
+ \sum\limits^{n}_{k=m+1}\left(a_{kj}\left(x_j^0\right)+
\frac{y_j}{\pi_jx_j^0}c_{kj}\right) p_k, \quad  j=\overline{m+1,n},
\end{eqnarray}
for the vector $ p=\{ p_i\}^{n}_{i=m+1}$ such that the inequalities hold
\begin{eqnarray}
\label{sor01}
p_j^0 - \sum\limits^m_{k=1}a_{kj}\left(x_j^0\right)p_k^0
- \sum\limits^{n}_{k=m+1}a_{kj}\left(x_j^0\right)\bar p_k > 0, \quad j=\overline{1,m}\ ;
\end{eqnarray}
3) the equalities hold
\begin{eqnarray}
\label{sor02}
\pi_j=\frac{\sum\limits^m_{k=1}c_{kj}p_k^0 +
\sum\limits^{n}_{k=m+1}c_{kj}\bar p_k}
{p_j^0 - \sum\limits^m_{k=1}a_{kj}\left(x_j^0\right)p_k^0
- \sum\limits^{n}_{k=m+1}a_{kj}\left(x_j^0\right)\bar p_k}\frac{y_j}{x_j^0}, \quad
j=\overline{1,m}.
\end{eqnarray}
Conditions 1) --- 3) are necessary  if for the strictly positive vector of gross outputs\index{vector of gross outputs}  $x^0 \in X_0 $ solving the set of equations
(\ref{sor1221}), there exists a strictly positive solution
$ p=\{ p_i\}_{i=m+1}^{n}$
to the set of equations (\ref{sor1220}) under the vector of gross outputs  $x^0$ and, in addition,
or the matrix $\tilde{\cal A}\left(x^0\right)$ is indecomposable  and at least for one $j=\overline{m+1, n}$
\begin{eqnarray} \label{ra01}
\sum\limits^m_{k=1}\left(a_{kj}\left(x_j^0\right)+\frac{y_j}{\pi_jx_j^0}c_{kj}\right)
p_k^0 >0,
\end{eqnarray}
or the inequalities (\ref{ra01}) hold for all
$j=\overline{m+1, n}$ and the matrix $\tilde{\cal A}\left(x^0\right)$ has no zero columns.
\end{theorem}

\begin{proof}\smartqed   Sufficiency. From the conditions of the  Theorem, under the strictly positive vector of gross outputs  $x^0$ there exists a strictly positive solution to the set of equations (\ref{sor1220})
for the vector
$ p=\{ p_i\}_{i=m+1}^{n}$ and this solution equals to the vector $\bar p=\{\bar p_i\}_{i=m+1}^{n}$ solving the set of equations (\ref{sor2}).
From the latter and indecomposability of the matrix $A\left(x^0\right) +\bar C,$  it follows that there exists a strictly positive solution to the conjugate problem
\begin{eqnarray*} \sum\limits^{n}_{k=1}\left(a_{jk}\left(x_k^0\right)+\frac{y_k}{\pi_kx_k^0}c_{jk}\right)
z_k=z_j, \quad j=\overline{1,n},\end{eqnarray*}
for the vector $z=\{z_i\}_{i=1}^n.$
This solution $z=\{z_j\}_{j=1}^{n}$ satisfies the set of equations
\begin{eqnarray*} \sum\limits^{n}_{s=1}\left(E-A\left(x^0\right)\right)^{-1}_{js}\sum\limits^{n}_{k=1}c_{sk}
\frac{y_kz_k}{\pi_kx_k^0}=z_j, \quad j=\overline{1,n}, \end{eqnarray*}
and the latter is equivalent to the existence of a strictly positive vector $g=\{g_i\}_{i=1}^{n}$  such  that $x^0=\{x_i^0\}_{i=1}^{n} \in X$ solves the set of equations
\begin{eqnarray} \label{sor0002}
\frac{\left[\left(E-A\left( x^0\right)\right)^{-1}\bar
C(\pi^{-1}\bar y) g\right]_i}{g_i}=x_i^0, \quad i=\overline{1,n}.
\end{eqnarray}
Here $y=\{y_i\}_{i=1}^l$ is the vector of levels of satisfaction of  consumers needs,\index{vector of levels of satisfaction of  consumers needs}      $x^0 \in X$ is the vector of gross outputs\index{vector of gross outputs} corresponding to the vector $y,$ the  strictly positive vector $g=\{g_i\}_{i=1}^n $ is the vector with components $g_i=z_i/x_i^0, \ i=\overline{1,n}.$

  Necessity. Let for the vector of monopolistic prices
$p^0=\{p_1^0,\ldots,p_m^0\}$ and the vector of levels of satisfaction of  consumers needs $y=\{y_i\}^l_{i=1}$  a strictly positive solution
$\bar p=\{\bar p_i\}_{i=m+1}^{n}$
to the set of equations (\ref{sor1220}) exist under the strictly positive vector of gross outputs  $x^0.$
The spectral radius of the matrix
$\tilde{\cal A}\left(x^0\right)$
is less than 1. The latter follows from the fact  that under assumptions made norms of the matrix
$\tilde{\cal A}\left(x^0\right)$ or a certain  its power
$\tilde{\cal A}^k\left(x^0\right)$ are less than 1 in the norm
\begin{eqnarray*} ||t||=\max\limits_{k \in [m+1, n]}\frac{|t_k|}{\bar p_k}\end{eqnarray*}
in the space \ $R^{n-m},$ \ where \ $\bar p_k, \ k=\overline{m+1, n},$ \
are components of the solution \ $\bar p=\{\bar p_j\}_{j=m+1}^{n}$
to the problem (\ref{sor1220}) that are strictly positive.
The Proof of this fact is the same one as in the case of non-zero initial goods supply.
\qed\end{proof}

Consider  technologies  determined by matrix $A(x)=||a_{ij}||_{i,j=1}^{n},$ where
$A(x)$ is a non-negative matrix independent of $x \in X.$

The aim of two subsequent Theorems is to establish pair of vectors  $x$ and $y$ for which the taxation vector $\pi$  agrees with the  structure  of consumption under the vector of monopolistic prices\index{taxation vector  agrees with the  structure  of consumption under the vector of monopolistic prices}  $p^0.$
\begin{theorem} \label{tsor2-2}
Let $\sum\limits_{k=1}^nc_{ki}>0, \ i=\overline{1,l},$ the matrix $A+\bar C$ be indecomposable, the taxation vector\index{taxation vector} $\pi=\{\pi_s\}^{n}_{s=1} $ be such that $0<\pi_i<1, \ i=\overline{1, n},$ and let   a strictly positive vector
$\alpha=\{\alpha_i \}_{i=1}^{n},$
and a strictly positive vector of  monopolistic prices\index{vector of  monopolistic prices}  $p^0=\{p_1^0, \ldots, p_m^0,\}$ satisfy conditions: \\
1) the spectral radius of the matrix\index{spectral radius of the matrix}
\begin{eqnarray*} \tilde{\cal A}=\left\|a_{kj}+\frac{ \alpha_jc_{kj}}
{\pi_j}\right\|^{n}_{k,j=m+1}
\end{eqnarray*}
is less than 1\ ;\\
2) there exists a strictly positive solution $\bar p=\{\bar p_i\}^{n}_{i=m+1}$ to the set of equations
\begin{eqnarray}
\label{qag2}
 p_j =
\sum\limits^m_{k=1}\left(a_{kj}+\frac{\alpha_j}{\pi_j}c_{kj}\right)
p_k^0 + \sum\limits^{n}_{k=m+1}\left(a_{kj}+
\frac{\alpha_j}{\pi_j}c_{kj}\right) p_k, \quad j=\overline{m+1,n}\ ;
\end{eqnarray}
for the vector $ p=\{ p_i\}^{n}_{i=m+1}$ such that the inequalities
\begin{eqnarray}
\label{qag01}
p_j^0 - \sum\limits^m_{k=1}a_{kj}p_k^0
- \sum\limits^{n}_{k=m+1}a_{kj}\bar p_k > 0, \quad j=\overline{1,m},
\end{eqnarray}
hold\ ;\\
3) the equalities
\begin{eqnarray}
\label{qag03}
\pi_j=\frac{\sum\limits^m_{k=1}c_{kj}p_k^0 +
\sum\limits^{n}_{k=m+1}c_{kj}\bar p_k}
{p_j^0 - \sum\limits^m_{k=1}a_{kj}p_k^0
- \sum\limits^{n}_{k=m+1}a_{kj}\bar p_k}\alpha_j, \quad
j=\overline{1,m},
\end{eqnarray}
hold.

If the matrix $\bar C$ has no zero rows and  components of the vector
\begin{eqnarray} \label{sag1221}
u=(E- A)^{-1}[e - i]
\end{eqnarray}
 are strictly positive,
then the sets of equations
\begin{eqnarray*} p_j^0 =
\sum\limits^m_{k=1}\left(a_{kj}+\frac{\alpha_j}{\pi_j}c_{kj}\right)
p_k^0 + \sum\limits^{n}_{k=m+1}\left(a_{kj}+
\frac{\alpha_j}{\pi_j}c_{kj}\right)p_k, \quad j=\overline{1,m},\end{eqnarray*}
\begin{eqnarray} \label{qag1220}
p_j =
\sum\limits^m_{k=1}\left (a_{kj}+\frac{\alpha_j}{\pi_j}c_{kj}\right )
p_k^0 + \sum\limits^{n}_{k=m+1}\left (a_{kj}+
\frac{\alpha_j}{\pi_j}c_{kj}\right )p_k, \quad j=\overline{m+1,n},
\end{eqnarray}

\begin{eqnarray} \label{qag1221}
x_k-\sum\limits^{n}_{j=1}a_{kj}x_j-e_k+i_k=
\sum\limits^l_{j=1}c_{kj}y_j,
\quad k=\overline{1,n},
\end{eqnarray}
relative to vectors
$p=\{p_j\}_{j=m+1}^{n}$ and $x=\{x_j\}_{j=1}^{n}$
have strictly positive solutions
$\bar p=\{\bar p_{m+1}, \ldots ,\bar p_{n}\}$ and
$\bar x=\{ \bar x_{1}, \ldots , \bar x_{n}\}.$
The vector of levels of satisfaction of consumers needs \index{vector of levels of satisfaction of consumers needs }  $y=\{y_i\}_{i=1}^l$ under which the taxation vector $\pi $
agrees with the structure  of consumption\index{taxation vector
agrees with the structure  of consumption} is determined by vectors $\bar x$ and
$\alpha$ through the formula $y_i=\alpha_i \bar x_i,\ i=\overline{1,n},$
the rest components of $y_i, \ i=\overline{n+1, l}$ are determined by conditions of the economy system state.
\end{theorem}
\begin{proof}\smartqed  If vectors
$p^0=\{p_1^0, \ldots, p_m^0\},$
$\alpha=\{\alpha_i \}_{i=1}^{n}$ satisfy the  conditions of the  Theorem \ref{tsor2-2}, then there exists a strictly positive solution to the problem (\ref{qag1220}) for the vector
$ p=\{ p_{m+1}, \ldots , p_{n}\}.$ It remains to prove that for such a  vector $\alpha=\{\alpha_i \}_{i=1}^{n}$ there exists a strictly positive solution to the problem (\ref{qag1221}).
We can write the problem (\ref{qag1221}) as follows
\begin{eqnarray} \label{sor1222}
x_k-\sum\limits^{n}_{j=1}[a_{kj}+\alpha_jc_{kj}]x_j
-e_k+i_k=
\sum\limits^l_{j=n+1}c_{kj}y_j,
\quad k=\overline{1,n},
\end{eqnarray}
if to put $y_i=\alpha_ix_i, \ i=\overline{1,n}.$

To prove the Theorem, let us  prove that the spectral radius of the matrix\index{spectral radius of the matrix}
\begin{eqnarray*} A+\alpha \bar C=||a_{ij}+\alpha_jc_{ij} ||_{i,j=1}^{n}\end{eqnarray*}
 is less than 1.
Let us use that, under the conditions of the  Theorem,  the problem
\begin{eqnarray*} \sum\limits^{n}_{k=1}\left[a_{kj}+
\frac{c_{kj}\alpha_j}{\pi_j}\right]p_k=p_j, \quad j=\overline{1,n},\end{eqnarray*}
has a strictly positive solution $\tilde p^0=\{\tilde p_j^0\}_{j=1}^{n},$  \ $ \tilde p_j^0>0,\
j=\overline{1,n}.$ Introduce into $R^{n}$ the norm
\begin{eqnarray*} ||p||=\max\limits_{i}\frac{|p_i|}{\tilde p_i^0}.\end{eqnarray*}  Then
\begin{eqnarray*} ||(A+\alpha\bar C)^T|| \leq
\max\limits_{j}\frac{\sum\limits^{n}_{k=1}\left[a_{kj}+c_{kj}\alpha_j\right]\tilde p_k^0}{\tilde p_j^0}
< \max\limits_{j}\frac{\sum\limits^{n}_{k=1}\left[a_{kj}+
\frac{c_{kj}\alpha_j}{\tilde \pi_k}\right]\tilde p_k^0}{\tilde p_j^0}= 1,\end{eqnarray*}
because $ 0< \pi_k<1, \quad k=\overline{1,n}.$
\qed\end{proof}

In the next Theorem, we clarify conditions for a strictly positive vector $\alpha=\{\alpha_i\}_{i=1}^n $ under which the  sets of equations (\ref{qag1220}) and (\ref{qag1221}) have strictly positive solutions if matrix  elements of the matrix  $||c_{ij}||_{i=1, j=1}^{n, l}$ have a special form
$c_{ij}=u_iv_j$, \ $i=\overline{1,n}$, \ $j=\overline{1,l}.$

\begin{theorem}\label{tsor5-5}
Let matrices $A=||a_{ij}||_{i,j=1}^{n}$ and
$\bar C=||c_{ij}||_{i,j=1}^{n}$ satisfy the  conditions of the  Theorem \ref{tsor2-2} and
 $c_{ij}=u_iv_j, \ i=\overline{1,n}, \ j=\overline{1,n+1}.$
Assume that
\begin{eqnarray*} (E- A)^{-1}(e-i)>0.\end{eqnarray*}
A strictly positive vector $\alpha=\{\alpha_j\}_{j=1}^{n}$ satisfies the set of equations
\begin{eqnarray*}
\nonumber
\lambda d_j(\alpha_1,\ldots,\alpha_{n})= p_j^0, \quad j=\overline{1,m},
\end{eqnarray*}
\begin{eqnarray*}
\lambda d_j(\alpha_1,\ldots,\alpha_{n})= p_j, \quad j=\overline{m+1,n},
\end{eqnarray*}
\begin{eqnarray}\label{2tax10'}
\sum\limits^{n}_{k=1}u_kd_k(\alpha_1,\ldots,\alpha_{n})=1,
\end{eqnarray}
where $\lambda>0$, \ $d(\alpha_1,\ldots,\alpha_{n})=\{d_k(\alpha_1,\ldots,\alpha_{n})\}_{k=1}^n=
(E-A^T)^{-1}t,$
\begin{eqnarray*} t=\{t_k\}_{k=1}^{n}, \quad t_k=\frac{\alpha_kv_k}{\pi_k},\end{eqnarray*}
if and only if the vector $p=\{p_i\}_{i=m+1}^n$ solves the set of equations (\ref{qag1220})
for the vector of monopolistic prices\index{ vector of monopolistic prices}
$p^0=\{p^0_k\}_{k=1}^{m},$ the vector of levels of  satisfaction of consumers needs\index{vector of levels of  satisfaction of consumers needs}   $y=\{y_i\}_{i=1}^l,$
and there exists a strictly positive solution to the set of equations (\ref{qag1221}).
\end{theorem}
\begin{proof}\smartqed  Necessity. Let    solutions to the problems
(\ref{qag1220}) and  (\ref{qag1221}) exist for a certain strictly positive vector
$\alpha=\{\alpha_k\}_{k=1}^{n}$ relative to vectors
$ p=\{ p_k\}_{k=m+1}^{n}$ and $x=\{x_k\}_{k=1}^{n}.$
 Introduce the vector
$\hat p=\{\hat p_k\}_{k=1}^{n},$ where
\begin{eqnarray*} \hat p_k=p_k^0, \quad k=\overline{1,m}, \quad
\hat p_k= p_k, \quad k=\overline{m+1,n}.\end{eqnarray*}
Then the vector $\hat p$ solves the set of equations
\begin{eqnarray}
\label{2tax11}
p_j =
\sum\limits^{n}_{k=1} a_{kj}p_k +
\frac{\alpha_jv_j}{\pi_j}\sum\limits^{n}_{k=1}u_kp_k, \quad
j=\overline{1,n},
\end{eqnarray}
for the vector $ p=\{ p_k\}_{k=1}^{n},$
or solves the set of equations
\begin{eqnarray}
\label{2tax12}
p_j = d_j(\alpha_1,\ldots,\alpha_{n})\sum\limits^{n}_{k=1}u_kp_k,
\quad  j=\overline{1,n},
\end{eqnarray}
\begin{eqnarray*} d_j(\alpha_1,\ldots,\alpha_{n})=[(E-A^T)^{-1}t]_j,
\quad
t=\left\{\frac{\alpha_kv_k}{\pi_k}\right\}_{k=1}^{n}.
\end{eqnarray*}
As the set of equations (\ref{2tax12}) has a solution if and only if
\begin{eqnarray*} \sum\limits^{n}_{k=1}u_kd_k(\alpha_1,\ldots,\alpha_{n})=1,\end{eqnarray*}
we obtain that the vector $\alpha=\{\alpha_j\}_{j=1}^{n}$
solves the set of equations
\begin{eqnarray*} \hat p_j=\lambda d_j(\alpha_1,\ldots,\alpha_{n}),
\quad j=\overline{1,n},\end{eqnarray*}
\begin{eqnarray*} \sum\limits^{n}_{k=1}u_kd_k(\alpha_1,\ldots,\alpha_{n})=1,\end{eqnarray*}
where $\lambda$ is a certain  positive number. From here, we obtain the necessity.

 Sufficiency.
Consider the vector
$\hat p=\{\hat p_k\}_{k=1}^{n}$, where
\begin{eqnarray*} \hat p_k=p_k^0, \quad k=\overline{1,m}, \quad
\hat p_k=p_k, \quad k=\overline{m+1,n},\end{eqnarray*}
and components $ p_k, \ k=\overline{m+1,n},$
are arbitrary strictly positive.
 Let for the considered vector $\hat p$  a strictly positive vector
$\alpha=\{\alpha_j\}_{j=1}^{n}$ exist solving the set of equations
(\ref{2tax10'}) for certain $\lambda>0.$
Then the vector $\hat p$ solves the set of equations (\ref{2tax11}) and, therefore, solves the set of equations (\ref{qag1220}).

Consider the set of equations
\begin{eqnarray} \label{2tax14}
x_k-\sum\limits^{n}_{j=1}( a_{kj} + c_{kj}\alpha_j)x_j -
e_k+i_k= \sum\limits_{i=n+1}^l c_{ki}y_i,
\quad k=\overline{1,n},
\end{eqnarray}
for the vector $x=\{x_i\}_{i=1}^n.$
As in the Proof of the Theorem \ref{tsor2-2}, we establish that the spectral radius of the matrix
$\|a_{kj} + c_{kj}\alpha_j\|_{k,j=1}^{n}$ is less than 1.
Therefore, there exists a strictly positive solution to the set of equations (\ref{2tax14}).
\qed\end{proof}

\begin{corollary} If after vector \ $\alpha$ \ satisfying the conditions of the Theorem
\ref{tsor2-2}  to take the vector \ $\bar x=\{ \bar x_{1}, \ldots , \bar x_{n}\},$ \ solving the set of equations (\ref{sor1222}) and for  $i=\overline{1,n}$ to put
$y_i=\alpha_i\bar x_i$
 and the rest components
$y_i, \ i=\overline{n+1,l}$ to take from  conditions of  the economy state,
then vectors
$\bar p=\{\bar p_{m+1}, \ldots ,\bar p_{n}\}$ and
$\bar x=\{ \bar x_{1}, \ldots , \bar x_{n}\}$
solve the set of equations
\begin{eqnarray*} p_j^0 =
\sum\limits^m_{k=1}\left (a_{kj}+\frac{y_j}{\pi_jx_j}c_{kj}\right )
p_k^0 + \sum\limits^{n}_{k=m+1}\left (a_{kj}+
\frac{y_j}{\pi_jx_j}c_{kj}\right )p_k, \quad j=\overline{1,m},\end{eqnarray*}
\begin{eqnarray*} p_j =
\sum\limits^m_{k=1}\left (a_{kj}+\frac{y_j}{\pi_jx_j}c_{kj}\right )
p_k^0  \end{eqnarray*}
\begin{eqnarray} \label{mag1220}
+\sum\limits^{n}_{k=m+1}\left (a_{kj}+
\frac{y_j}{\pi_jx_j}c_{kj}\right )p_k, \quad j=\overline{m+1,n},
\end{eqnarray}

\begin{eqnarray} \label{mag1221}
x_k-\sum\limits^{n}_{j=1}a_{kj}x_j-e_k+i_k=
\sum\limits^l_{j=1}c_{kj}y_j,
\quad k=\overline{1,n},
\end{eqnarray}
with such defined vector of levels  of satisfaction of consumers needs \index{vector of levels  of satisfaction of consumers needs }  $y.$
It means that the taxation vector $\pi=\{\pi_i\}_{i=1}^{n}$ agrees with the structure of consumption \index{taxation vector  agrees with the structure of consumption } determined by mentioned vectors of monopolistic prices,\index{vectors of monopolistic prices}
levels  of satisfaction of consumers needs,\index{levels  of satisfaction of consumers needs} and gross outputs.\index{gross outputs}
\end{corollary}

Let us show the  sufficient conditions and an algorithm to build  a strictly positive vector $\alpha=\{\alpha_i \}_{i=1}^{n}$ satisfying the  conditions of the Theorem \ref{tsor2-2}.
\begin{theorem} \label{tsor2-3}
Let $\sum\limits_{k=1}^nc_{ki}>0, \ i=\overline{1,l},$ the matrix $A$ be productive  and the matrix $A+\bar C$ be indecomposable, the taxation vector\index{taxation vector} $\pi=\{\pi_s\}^{n}_{s=1} $ be such that $0<\pi_i<1, \ i=\overline{1, n},$
the condition (\ref{sag1221}) hold,
and let   the matrix
$\bar C$ have no zero rows, the strictly positive  vector of monopolistic prices\index{vector of monopolistic prices}
$p^0=\{p_1^0, \ldots, p_m^0\}$ and the matrix $A$
be such that there exists a strictly positive solution $\bar p=\{\bar p_i\}_{i=m+1}^n$ to the problem
\begin{eqnarray}
\label{sor1223}
 p_j =
\sum\limits^m_{k=1}a_{kj}p_k^0 + \sum\limits^{n}_{k=m+1}a_{kj}
 p_k, \quad j=\overline{m+1,n},
\end{eqnarray}
for the vector $ p=\{ p_i\}_{i=m+1}^n$
satisfying conditions
\begin{eqnarray}
\label{sor1224}
p_j^0 - \sum\limits^m_{k=1}a_{kj}p_k^0
- \sum\limits^{n}_{k=m+1}a_{kj}\bar p_k > 0, \quad j=\overline{1,m}.
\end{eqnarray}

Under these conditions there exists a strictly positive vector
$\alpha=\{\alpha_i \}_{i=1}^{n}$ satisfying conditions:\\
1) the spectral radius of the matrix\index{spectral radius of the matrix}
\begin{eqnarray*} \tilde{\cal A}=\left\|a_{kj}+\frac{ \alpha_j}{\pi_j}c_{kj}
\right\|^{n}_{k,j=m+1}
\end{eqnarray*}
is less than 1\ ;\\
2) the strictly positive solution $\tilde p=\{\tilde p_i\}_{i=m+1}^n$ to the problem
\begin{eqnarray}
\label{sor1226}
 p_j =
\sum\limits^m_{k=1}\left (a_{kj}+\frac{\alpha_j}{\pi_j}c_{kj}\right )
p_k^0 + \sum\limits^{n}_{k=m+1}\left (a_{kj}+
\frac{\alpha_j}{\pi_j}c_{kj}\right) p_k, \quad j=\overline{m+1,n},
\end{eqnarray}
for the vector $ p=\{ p_i\}_{i=m+1}^n$
satisfies  the set of inequalities
\begin{eqnarray} \label{sor1227}
p_j^0 - \sum\limits^m_{k=1}a_{kj}p_k^0
- \sum\limits^{n}_{k=m+1}a_{kj}\tilde p_k > 0, \quad j=\overline{1,m}\ ;
\end{eqnarray}
3) components $\alpha_j$ for $j=\overline{1,m}$ are determined by formulae
\begin{eqnarray} \label{sor1228}
\alpha_j=\pi_j\frac{p_j^0 - \sum\limits^m_{k=1}a_{kj}p_k^0
- \sum\limits^{n}_{k=m+1}a_{kj}\tilde p_k}
{\sum\limits^m_{k=1}c_{kj}p_k^0 +
\sum\limits^{n}_{k=m+1}c_{kj}\tilde p_k}, \quad j=\overline{1,m}.
\end{eqnarray}
\end{theorem}
\begin{proof}\smartqed  Let a solution to the problem
(\ref{sor1223}) satisfy  the condition (\ref{sor1224}).
Then a vector
$\alpha=\{\alpha_i \}_{i=1}^{n}$ exists such that the set of equations
(\ref{qag1220}) has a solution. Really, take strictly positive components
$ \alpha_i, \ i=\overline{m+1,n},$ of the vector
$\alpha$ such that the spectral radius of the matrix
\begin{eqnarray*} \tilde{\cal A}=\left\|a_{kj}+\frac{ \alpha_j}{\pi_j}c_{kj}
\right\|^{n}_{k,j=m+1}  \end{eqnarray*}
is less than 1 and, additionally, the vector $ \tilde p=\{\tilde p_i\}_{m+1}^{n}$
solving the set of equations
(\ref{sor1226}) satisfies the set of inequalities
(\ref{sor1227}). One can reach the latter taking rather small components
$\alpha_j>0, \ j=\overline{m+1,n}.$
We take the rest components
$\alpha_j>0, \ j=\overline{1,m},$ such that the equalities
(\ref{sor1228}) hold. From the latter it follows that there exists a strictly positive solution to the set of equations
\begin{eqnarray*} p_j^0 =
\sum\limits^m_{k=1}\left(a_{kj}+\frac{\alpha_j}{\pi_j}c_{kj}\right )
p_k^0 + \sum\limits^{n}_{k=m+1}\left(a_{kj}+
\frac{\alpha_j}{\pi_j}c_{kj}\right )p_k, \quad j=\overline{1,m},\end{eqnarray*}
\begin{eqnarray} \label{sor1229}
p_j =
\sum\limits^m_{k=1}\left(a_{kj}+\frac{\alpha_j}{\pi_j}c_{kj}\right )
p_k^0 + \sum\limits^{n}_{k=m+1}\left(a_{kj}+
\frac{\alpha_j}{\pi_j}c_{kj}\right )p_k, \quad j=\overline{m+1,n},
\end{eqnarray}
\begin{eqnarray} \label{mga1}
x_k-\sum\limits^{n}_{j=1}a_{kj}x_j-e_k+i_k=
\sum\limits^l_{j=1}c_{kj}y_j,
\quad k=\overline{1,n}.
\end{eqnarray}
\qed\end{proof}

\section{Applications}

Let us show how one can solve the problem of influence of  change of prices by  monopolistic  industries\index{monopolistic  industries}  onto the rest  prices in the  economy system in the case of aggregated description of the economy.\index{aggregated description of the economy}

As earlier,  we assume that the  economy system production is described by technological maps built after the  productive matrix\index{technological maps built after the  productive matrix}
$A=\|a_{kj}\|^{n}_{k,j=1},$
a structure  of  unproductive consumption is given by the matrix
$C=\|c_{kj}\|^{n,l}_{k=1,j=1}$ determined by fields of information evaluation by consumers,\index{ fields of information evaluation by consumers} and   vectors of property of consumers\index{vectors of property of consumers} that they have at the beginning of the economy operation period is given  by columns of the matrix of the initial goods supply\index{matrix of the initial goods supply}  $B_0=\|b_{kj}^0\|^{n,l}_{k=1,j=1}.$

Let, within a certain basic  period of the  economy operation, the equilibrium price vector\index{equilibrium price vector}
$\bar p=\{\bar p_k\}_{k=1}^{n}$ and the gross outputs vector\index{gross outputs vector}
$x^0=\{x^0_k\}_{k=1}^{n}$ with  vector of levels of satisfaction of consumers needs\index{vector of levels of satisfaction of consumers needs}   $y^0=\{y^0_k\}_{k=1}^{l}$ satisfy the sets of equations
\begin{eqnarray} \label{2tax1}
\bar p_j = \sum\limits^{n}_{k=1}\left[a_{kj}+\frac{y_j^0}{x_j^0}\left(\frac{c_{kj}}{\pi_j}-\frac{b_{kj}^0}{y_j^0}\right)\right]\bar p_k, \quad j=\overline{1,n},
\end{eqnarray}

\begin{eqnarray} \label{2tax2}
x_k^0-\sum\limits^{n}_{j=1}a_{kj}x_j^0+ \sum\limits_{i=1}^lb_{ki}^0 -e_k^0 + i_k^0=
\sum\limits^l_{j=1}c_{kj}y_j^0,
\quad k=\overline{1,n},
\end{eqnarray}
for vectors
$\bar p=\{\bar p_{1}, \ldots ,\bar p_{n}\} \in R_+^{n}$ and
$x^0=\{x_{1}^0, \ldots ,x_{n}^0\} \in X_0,$
where $e^0=\{e^0_k\}_{k=1}^{n}$,
$i^0=\{i^0_k\}_{k=1}^{n}$ are  vectors of  export and import.

Assume that, within the next period of the  economy operation, a certain change of prices  happen in certain $m, \ m < n,$  industries that we  consider by monopolists, and the existing  equilibrium price vector $\{\bar p_k\}_{k=1}^{m}$ in the previous period of the economy operation
turns into the equilibrium price vector $\{p^0_k\}_{k=1}^{m}$ in the next period of the economy operation.  The question arises how the economy system should respond on such change of  prices in monopolistic industries?\index{ monopolistic industries}

Within the next period,  the  production in  the economy system is described  by technological maps built after productive matrix\index{ technological maps built after productive matrix}
$A=\|a_{kj}\|^{n}_{k,j=1},$
 a structure  of  unproductive consumption\index{ structure  of  unproductive consumption} is given by the matrix
$C=\|c_{kj}\|^{n,l}_{k=1,j=1}$ determined by
fields of information evaluation by consumers,\index{fields of information evaluation by consumers}
 and  vectors of property of consumers\index{vectors of property of consumers} that they  have at the beginning of the next economy operation period is given by columns of the matrix of initial goods supply\index{matrix of initial goods supply}  $B=\|b_{kj}\|^{n,l}_{k=1,j=1}.$
Establish conditions for the vector of monopolistic prices\index{vector of monopolistic prices}
$p^0=\{p_j^0\}^{m}_{j=1}$ and the vector of levels of satisfaction of consumers needs\index{vector of levels of satisfaction of consumers needs}
$y=\{y_j\}^{l}_{j=1}$ under which for industries taxation vector\index{industries taxation vector}
$\pi=\{\pi_s\}^{n}_{s=1}, $  $0<\pi_i<1, \ i=\overline{1, n},$ there exists a strictly positive solution to the problem

\begin{eqnarray*} p_j^0 =
\sum\limits^m_{k=1}\left[a_{kj}+\frac{y_j}{x_j}\left(\frac{c_{kj}}{\pi_j}-\frac{b_{kj}}{y_j}\right)\right]
p_k^0   \end{eqnarray*}  \begin{eqnarray*}  +\sum\limits^{n}_{k=m+1}\left[a_{kj}+\frac{y_j}{x_j}\left(\frac{c_{kj}}{\pi_j}-\frac{b_{kj}}{y_j}\right)\right]p_k, \quad j=\overline{1,m}, \end{eqnarray*}
\begin{eqnarray*} p_j =
\sum\limits^m_{k=1}\left[a_{kj}+\frac{y_j}{x_j}\left(\frac{c_{kj}}{\pi_j}-\frac{b_{kj}}{y_j}\right)\right]
p_k^0  \end{eqnarray*}
\begin{eqnarray} \label{2tax3}
+ \sum\limits^{n}_{k=m+1}\left[a_{kj}+\frac{y_j}{x_j}\left(\frac{c_{kj}}{\pi_j}-\frac{b_{kj}}{y_j}\right)\right]p_k, \quad j=\overline{m+1,n},
\end{eqnarray}

\begin{eqnarray} \label{2tax4}
x_k-\sum\limits^{n}_{j=1}a_{kj}x_j+ \sum\limits_{i=1}^lb_{ki} -e_k + i_k=
\sum\limits^l_{j=1}c_{kj}y_j,
\quad k=\overline{1,n},
\end{eqnarray}
relative to vectors
$p=\{p_{m+1}, \ldots ,p_{n}\} \in R_+^{n-m}$ and
$x=\{x_{1}, \ldots ,x_{n}\} \in X_0,$
where $x=\{x_k\}_{k=1}^{n}$ is the  vector of  gross outputs,\index{vector of  gross outputs} $e=\{e_k\}_{k=1}^{n}$ and
$i=\{i_k\}_{k=1}^{n}$ are  vectors of export and import,\index{vectors of export and import}
$y=\{y_k\}_{k=1}^{l}$ is a vector of levels of satisfaction of consumers needs,\index{vector of levels of satisfaction of consumers needs} and
$p=\{p_k\}_{k=m+1}^{n}$ is an equilibrium price vector of non-monopolistic industries\index{equilibrium price vector of non-monopolistic industries}  in  the economy system.

For practical application of the results obtained, one must go to cost parameters description. Let
\begin{eqnarray*} \tilde p^0_j =\frac{p^0_j}{\bar p_j}, \quad j=\overline{1,m},
\quad \tilde p_j =\frac{p_j}{\bar p_j}, \quad j=\overline{m+1,n},\end{eqnarray*}
where $\bar p=\{\bar p_k\}_{k=1}^{n}$ is an equilibrium price vector within the basic period of the economy operation\index{equilibrium price vector within the basic period of the economy operation} and $p^1=\{p^0_1,\ldots,p^0_m,p_{m+1},\ldots,p_{n}\}$  is an equilibrium price vector
in the next period of the economy operation\index{equilibrium price vector
in the next period of the economy operation}
 satisfying   the set of equations (\ref{2tax3}).

In cost parameters, the set of equations (\ref{2tax3}), (\ref{2tax4}) takes the form

\begin{eqnarray*} \tilde  p_j^0 =
\sum\limits^m_{k=1}\left[\bar a_{kj}+\frac{Y_j}{X_j^1}\left(\frac{C_{kj}}{\pi_j}-\frac{B_{kj}}{Y_j}\right)\right]
\tilde p_k^0  \end{eqnarray*}  \begin{eqnarray*}  +\sum\limits^{n}_{k=m+1}\left[\bar a_{kj}+\frac{Y_j}{X_j^1}\left(\frac{C_{kj}}{\pi_j}-\frac{B_{kj}}{Y_j}\right)\right]\tilde p_k, \quad j=\overline{1,m}, \end{eqnarray*}
\begin{eqnarray*} \tilde p_j =
\sum\limits^m_{k=1}\left[\bar a_{kj}+\frac{Y_j}{X_j^1}\left(\frac{C_{kj}}{\pi_j}-\frac{B_{kj}}{Y_j}\right)\right]
\tilde p_k^0  \end{eqnarray*}
\begin{eqnarray} \label{2tax5}
+ \sum\limits^{n}_{k=m+1}\left[\bar a_{kj}+\frac{Y_j}{X_j^1}\left(\frac{C_{kj}}{\pi_j}-\frac{B_{kj}}{Y_j}\right)\right]\tilde p_k, \quad j=\overline{m+1,n},
\end{eqnarray}

\begin{eqnarray} \label{2tax6}
X_k^1-\sum\limits^{n}_{j=1}\bar a_{kj}X_j^1+ \sum\limits_{i=1}^lB_{ki} - {\cal E}^1_k + I_k=
\sum\limits^l_{j=1}C_{kj}Y_j,
\quad k=\overline{1,n},
\end{eqnarray}
where notations are introduced
\begin{eqnarray*} \bar a_{kj} =\frac{\bar p_k a_{kj}}{\bar p_j}, \quad
   C_{ks} =\bar p_k c_{ks}, \quad
 B_{ks} =\bar p_k b_{ks}, \quad X^1_j =x_j\bar p_j, \quad
 {\cal E}^1_j =e_j\bar p_j,  \end{eqnarray*}
\begin{eqnarray*}    I^1_j =i_j\bar p_j, \quad Y_s =y_s, \quad k,j=\overline{1,n},
   \quad s=\overline{1,l}.\end{eqnarray*}

In the basic year,\index{basic year} we describe the economy system aggregately, i.e., suppose that  information about the economy system  is given by\\
1) a vector of gross outputs\index{vector of gross outputs}  $X=\{X_i\}_{i=1}^{n}, \ X_i >0, \ i=\overline{1,n}\ ;$  \\
2) a matrix of financial flows\index{matrix of financial flows}
$||X_{ik}||_{i,k=1 }^{n}$ describing the structure of inputs of $n$
net industries;\\
3) a structure matrix of inputs\index{structure matrix of inputs}  related to financial flows and gross outputs and given by the formula
\begin{eqnarray*}   \bar A=||\bar a_{ij}||_{i,j=1}^{n},  \quad
\bar a_{ij}=\frac{X_{ij}}{X_j}\ ;\end{eqnarray*}
4) vectors of export and import\index{vectors of export and import}
${\cal E}=\{ {\cal E}_i\}_{i=1}^{n},$  $I=\{ I_i\}_{i=1}^{n}\ ;$ \\
5) vectors of final consumption\index{vectors of final consumption and gross accumulation and  change of supply}
$C=\{C_i\}_{i=1}^{n}$
and gross accumulation and  change of supply $N=\{N_i\}_{i=1}^{n}\ ;$ \\
6) a  value added of the $i$-th industry\index{value added of the $i$-th industry}
$\Delta_i, \ i=\overline{1, n}\ ;$ \\
7) a matrix of initial goods supply\index{matrix of initial goods supply}  $ B=||B_{ki}||_{k=1, i=1}^{n, n+1}$ at the beginning of the economy operation period; \\
8) a vector of integral taxation\index{vector of integral taxation of industries}  $\pi=\{\pi_i\}_{i=1}^{n}$ of industries;\\
All these data form interindustry balance\index{interindustry balance}
\begin{eqnarray*} \sum\limits_{k=1}^{n}X_{ik} +C_i+N_i+ {\cal
E}_i -I_i= X_i, \quad i=\overline{1,n},\end{eqnarray*}
\begin{eqnarray*} \sum\limits_{i=1}^{n} X_{ik} +\Delta_k= X_k, \quad
k=\overline{1,n}.\end{eqnarray*}

Relate aggregated description of the economy\index{aggregated description of the economy} to non-aggregated.
Introduce a non-negative rectangular matrix
$ C=||C_{ki}||_{k=1, i=1}^{n, n+1}$
describing unproductive consumption\index{unproductive consumption} of $(n+1)$ consumers,
where matrix elements $C_{ki}$ satisfy the set of equations
\begin{eqnarray*}  \sum\limits_{i=1}^{n+1}C_{ki}Y_i=C_k+N_k+ \sum\limits_{i=1}^{n+1}B_{ki}, \quad
k=\overline{1,n}, \end{eqnarray*}  \begin{eqnarray*}
Y_i\sum\limits_{k=1}^{n}C_{ki}=\pi_i\Delta_i + \pi_i\sum\limits_{k=1}^{n}B_{ki},
\quad i=\overline{1,n}, \end{eqnarray*}
\begin{eqnarray} \label{2tax7}
Y_{n+1}\sum\limits_{k=1}^{n}C_{k,n+1}=\sum\limits_{i=1}^{n}(1-\pi_i)\Delta_i+
\sum\limits_{i=1}^{n}\sum\limits_{s=1}^{n}(1 - \pi_s)B_{is} + \sum\limits_{i=1}^{n}B_{i, n+1} - {\cal E} +I,
\end{eqnarray}
where ${\cal E}= \sum\limits_{i=1}^{n}{\cal E}_i, \
I= \sum\limits_{i=1}^{n}I_i,$ and
$Y= \{Y_i\}_{i=1}^{n+1}$ is a certain vector with strictly positive components.

Further we consider the solution to the set of equations (\ref{2tax7}) that can be expressed as follows
\begin{eqnarray*}
C_{ki}=\frac{\left(C_k+N_k + \sum\limits_{j=1}^{n+1} B_{kj}\right)\left(\pi_i\Delta_i+ \pi_i \sum\limits_{j=1}^{n}B_{ji}\right)}{Y_i M},
\quad  k,i=\overline{1,n},\end{eqnarray*}
\begin{eqnarray*} C_{k,n+1}=\left(C_k+N_k + \sum\limits_{j=1}^{n+1} B_{kj}\right) \end{eqnarray*}  \begin{eqnarray*}  \times\frac{\sum\limits_{i=1}^{n} \left(1-\pi_i\right)\Delta_i+ \sum\limits_{j=1}^{n}\sum\limits_{i=1}^{n}\left(1-\pi_i\right) B_{ji}+\sum\limits_{k=1}^{n} B_{k, n+1} -{\cal E}+I}{Y_{n+1}M}, \quad  k=\overline{1,n},\end{eqnarray*}  
where notations are introduced
\begin{eqnarray*} M=V - {\cal E} +I, \quad V= \sum\limits_{i=1}^{n}\Delta_i+ \sum\limits_{j=1}^{n}\sum\limits_{m=1}^{n+1}B_{jm},
\quad {\cal E}= \sum\limits_{i=1}^{n}{\cal E}_i, \quad
I= \sum\limits_{i=1}^{n}I_i.\end{eqnarray*}
Introduce notations
\begin{eqnarray} \label{2tax8}
C_{jk}^0=C_{jk}Y^0_k, \quad B_{jk}^0=B_{jk}, \quad j=\overline{1,n},
\quad k=\overline{1,n+1},
\end{eqnarray}
and accept plausible hypothesis that we can describe change of monopolistic prices  under aggregated description of the economy system by the set of equations (\ref{2tax5}), (\ref{2tax6}) with matrix elements
$C_{jk}^0$ given by the expression (\ref{2tax8}) and matrix elements
$\bar a_{kj}$ given by the expression
\begin{eqnarray*} \bar a_{kj}=\frac{X_{kj}}{X_j},\quad k,j=\overline{1,n},\end{eqnarray*}
where $X_{kj}$ and $X_j$ correspond to the basic year. Under these conditions, we write the set of equations (\ref{2tax5}) --- (\ref{2tax6}) in the form

\begin{eqnarray*} \tilde p_j^0 = \sum\limits^m_{k=1}\left[\bar a_{kj}+\frac{\tau_j}{X^1_j}\left(
\frac{C^0_{kj}}{\pi_j} - \frac{B^0_{kj}}{\tau_j}\right)\right]\tilde p_k^0 \end{eqnarray*}
 \begin{eqnarray*} + \sum\limits^{n}_{k=m+1}\left[\bar a_{kj}+\frac{\tau_j}{X^1_j}\left(
\frac{C^0_{kj}}{\pi_j} - \frac{B^0_{kj}}{\tau_j}\right)\right]\tilde p_k, \quad j=\overline{1,m},\end{eqnarray*}

\begin{eqnarray*} \tilde p_j =
\sum\limits^m_{k=1}\left[\bar a_{kj}+\frac{\tau_j}{X^1_j}\left(
\frac{C^0_{kj}}{\pi_j} - \frac{B^0_{kj}}{\tau_j}\right)\right]
\tilde p_k^0  \end{eqnarray*}
\begin{eqnarray} \label{2tax9}
+\sum\limits^{n}_{k=m+1}\left[\bar a_{kj}+\frac{\tau_j}{X^1_j}\left(
\frac{C^0_{kj}}{\pi_j} - \frac{B^0_{kj}}{\tau_j}\right)\right]\tilde p_k, \quad j=\overline{m+1,n},
\end{eqnarray}
\begin{eqnarray} \label{2tax10}
X^1_k-\sum\limits^{n}_{j=1}\bar a_{kj}X^1_j+ \sum\limits^{n+1}_{j=1}B_{kj}^0-{\cal E}^1_k+I^1_k=
\sum\limits^{n+1}_{j=1}C^0_{kj}\tau_j,
\quad k=\overline{1,n},
\end{eqnarray}
where we denote
\begin{eqnarray*} \tau_k=\frac{Y_k}{Y^0_k}, \quad k=\overline{1,n}.\end{eqnarray*}

\begin{note} If
\begin{eqnarray*}  {\cal E}^1_j ={\cal E}_j, \quad I^1_j =I_j, \quad \tau_s =1,
\quad j=\overline{1,n}, \quad s=\overline{1,n+1},\end{eqnarray*}
then the set of equations (\ref{2tax9}) --- (\ref{2tax10}) is solvable  and
\begin{eqnarray*}  X^1_j =X_j, \quad j=\overline{1,n}, \quad \tilde p_k^0=1,  \quad k=\overline{1,m}, \quad \tilde p_s=1, \quad s=\overline{m,n}.\end{eqnarray*}
\end{note}
Therefore, the set of equations (\ref{2tax9}) --- (\ref{2tax10}) describes the deviation
from the economy characteristics  of the basic year that
  happen in the economy system in consequence of change of  prices by monopolists.

The set of equations (\ref{2tax9}) --- (\ref{2tax10}) has the same form as the set of equations  (\ref{xter1}) --- (\ref{1mag1221}) if to do corresponding identifications.
Therefore, to study it, one can apply the Theorem \ref{vtas1}.

\section{Non-linear technologies}

\subsection{The agreement of taxation vector with consumption structure within non-linear technologies}

Below, we consider  production technologies depending  non-linearly on  vector of gross outputs\index{production technologies depending  non-linearly on  vector of gross outputs} if the economy system has constant production inputs,\index{constant production inputs} i.e., matrix elements of technological  matrix is given by the formula
\begin{eqnarray*} A(x)=||a_{ij}(x_j)||_{i,j=1}^{n}, \quad a_{ij}(x_j)=a_{ij}+r_{ij}/x_j,\end{eqnarray*}
  where non-negative matrix
$R=||r_{ij}||_{i,j=1}^{n}$ describes constant production inputs.
Consider the set of equations
\begin{eqnarray*} x_k-\sum\limits^{n}_{j=1}\left[a_{kj}+\alpha_jc_{kj}\right]x_j
-\sum\limits^{n}_{j=1}r_{kj} +  \end{eqnarray*}
\begin{eqnarray} \label{dar1222}
 + \sum\limits^{l}_{j=1}b_{kj}-e_k+i_k=\sum\limits^l_{j=n+1}c_{kj}y_j,
\quad k=\overline{1,n},
\end{eqnarray}
for the  vector of  gross outputs\index{ vector of  gross outputs} $x=\{x_i\}_{i=1}^n,$
and let  the vector
\begin{eqnarray*}  x( \alpha_1,\ldots, \alpha_{n})
= \{ x_i( \alpha_1,\ldots, \alpha_{n}) \}_{i=1}^{n}\end{eqnarray*}
solve the set of equations (\ref{dar1222})
for the vector
 $\alpha=\{\alpha_j\}_{j=1}^{n} $ belonging to the set
\begin{eqnarray*} L=\{ \alpha=\{\alpha_j\}_{j=1}^{n}, \
0 \leq \alpha_i \leq \alpha_i^0, \ i=\overline{1, n}\},\end{eqnarray*}  and the vector
$x^0=\{x_j^0\}_{j=1}^{n}$ be a strictly positive solution to the set of equations (\ref{dar1222})
for the vector
 $\alpha=\{\alpha_j\}_{j=1}^{n} $ with zero components. The conditions of the  existence  of a strictly positive solution to the set of equations (\ref{dar1222}) for the vector
 $\alpha=\{\alpha_j\}_{j=1}^{n} $ with zero components we give later. Suppose that the matrix
$A\left(x^0\right)$ is productive.
Let a  strictly positive vector $\alpha^0=\{\alpha_j^0\}_{j=1}^{n} $ be such that the spectral radius of the matrix\index{spectral radius of the matrix} $||a_{kj}\left(x_j^0\right)+\alpha_j^0c_{kj}||_{k,j=1}^{n} $ is less than 1.
 Introduce the set
\begin{eqnarray*} X_0=\{x=\{x_i\}_{i=1}^{n} \in R_+^{n},\ x_i \geq x_i^0,\
i=\overline{1,n} \},\end{eqnarray*}
on which the matrix of inputs $A(x)$ is given, and notations
 \begin{eqnarray*}  R_j^2\left(p^0, A, \bar B,  x,  p\right) = p_j^0 - \sum\limits^m_{k=1}a_{kj}(x_j)p_k^0
- \sum\limits^{n}_{k=m+1}a_{kj}(x_j) p_k \end{eqnarray*}  \begin{eqnarray*} + \frac{1}{x_j}\left[\sum\limits^m_{k=1}b_{kj}p_k^0+\sum\limits^n_{k=m+1}b_{kj} p_k\right], \quad j=\overline{1,m},\end{eqnarray*}
where $p=\{p_{m+1}, \ldots ,p_{n}\} \in R_+^{n-m},$ and
$x=\{x_{1}, \ldots ,x_{n}\} \in X_0.$

\begin{theorem}\label{tak1}
Let $\sum\limits_{k=1}^nc_{ki}>0, \ i=\overline{1,l},$  there exist a non-negative vector
$v_0=\{v_i\}_{i=1}^n, \ v_i \geq 0, \ i=\overline{1,n},$  such that $\bar C(v_0) - \bar B $ is a non-negative matrix having no zero rows or columns and a  matrix  $ A\left(x^0\right) + \bar C(v_0) - \bar B$ is indecomposable, and let  the spectral radius of the matrix $A\left(x^0\right)$ be less than 1,
 a non-negative vector
$\alpha^0=\{\alpha_i^0 \}_{i=1}^{n},$
the  vector of  monopolistic prices\index{vector of  monopolistic prices} $p^0=\{p_1^0, \ldots, p_m^0,\}$ and the matrix $A\left(x^0\right)$ satisfy conditions: \\
1) the spectral radius of the matrix
\begin{eqnarray*} \tilde{\cal A}\left(x^0\right)=\left\|a_{kj}\left(x_j^0\right)+\alpha_j^0  \frac{c_{kj}}
{\pi_j}  \right\|^{n}_{k,j=m+1},
\end{eqnarray*}
is less than 1\ ;\\
2) there exists a strictly positive solution $\bar p=\{\bar p_i\}^{n}_{i=m+1}$ to the set of equations
\begin{eqnarray*}  p_j =
\sum\limits^m_{k=1}\left[a_{kj}\left(x_j^0\right)+  \alpha_j^0 \frac{c_{kj}}
{\pi_j}\right]
p_k^0  \end{eqnarray*}
\begin{eqnarray}
\label{tak3}
+ \sum\limits^{n}_{k=m+1}\left[a_{kj}\left(x_j^0\right)+ \alpha_j^0 \frac{c_{kj}}
{\pi_j}\right] p_k, \quad  j=\overline{m+1,n},
\end{eqnarray}
for the vector $ p=\{ p_i\}^{n}_{i=m+1}$
such that the inequalities
\begin{eqnarray} \label{tak2}
 p_j^0 - \sum\limits^m_{k=1}a_{kj}\left(x_j^0\right)p_k^0
- \sum\limits^{n}_{k=m+1}a_{kj}\left(x_j^0\right)\bar p_k > 0,
 \quad j=\overline{1,m},
\end{eqnarray}
hold.
If there exists a vector $\alpha=\{\alpha_i \}_{i=1}^{n} \in L$ such that a  strictly positive solution $x(\alpha)=\{x_k(\alpha)\}_{k=1}^n$ to the set of equations (\ref{dar1222})
satisfies conditions $y_k(\alpha)=\alpha_k x_k(\alpha) > \pi_k v_k, \ k=\overline{1,n},$
the equalities
\begin{eqnarray} \label{tak5}
\pi_j=\frac{\sum\limits^m_{k=1}c_{kj}p_k^0 +
\sum\limits^{n}_{k=m+1}c_{kj}p_k^1}
{ R_j^2\left(p^0, A, \bar B,  x(\alpha),  p_1\right)}\alpha_j, \quad  j=\overline{1,m},
\end{eqnarray}
hold, where $ p_1=\{ p_{m+1}^1, \ldots , p_{n}^1\}$ solves the set of equations

\begin{eqnarray*} p_j =
\sum\limits^m_{k=1}\left[a_{kj}(x_j(\alpha))+\alpha_j \left( \frac{c_{kj}}
{\pi_j} - \frac{b_{kj}}{y_j(\alpha)}\right)\right]
p_k^0  \end{eqnarray*}
\begin{eqnarray} \label{tak6}
  + \sum\limits^{n}_{k=m+1}\left[a_{kj}(x_j(\alpha))+
\alpha_j \left( \frac{c_{kj}}
{\pi_j} - \frac{b_{kj}}{y_j(\alpha)}\right)\right]p_k, \quad j=\overline{m+1,n},
\end{eqnarray}
relative to the vector $ p=\{ p_{m+1}, \ldots , p_{n}\},$
then there exists a solution to the problem
\begin{eqnarray*} p_j^0 =
\sum\limits^m_{k=1}\left[a_{kj}(x_j(\alpha))+\alpha_j\left(\frac{c_{kj}}{\pi_j}-\frac{b_{kj}}{y_j(\alpha)}\right)\right]
p_k^0  \end{eqnarray*}  \begin{eqnarray*}  +\sum\limits^{n}_{k=m+1}\left[a_{kj}(x_j(\alpha))+\alpha_j\left(\frac{c_{kj}}{\pi_j}-\frac{b_{kj}}{y_j(\alpha)}\right)\right]p_k, \quad j=\overline{1,m}, \end{eqnarray*}
\begin{eqnarray*} p_j =
\sum\limits^m_{k=1}\left[a_{kj}(x_j(\alpha))+\alpha_j\left(\frac{c_{kj}}{\pi_j}-\frac{b_{kj}}{y_j(\alpha)}\right)\right]
p_k^0  \end{eqnarray*}
\begin{eqnarray} \label{tak7}
+ \sum\limits^{n}_{k=m+1}\left[a_{kj}(x_j(\alpha))+\alpha_j\left(\frac{c_{kj}}{\pi_j}-\frac{b_{kj}}{y_j(\alpha)}\right)\right]p_k, \quad j=\overline{m+1,n},
\end{eqnarray}
for the vector $p=\{p_{m+1}, \ldots, p_n\}.$ This solution coincides with the vector $p_1.$
\end{theorem}
\begin{proof}\smartqed
Obviously, for any vector $\alpha \in L$ the set of inequalities $x(\alpha) \geq x^0$ holds.
A solution to the set of equations (\ref{tak6}) exists as a result of the  conditions of the Theorem.
In view of the set of inequalities
\begin{eqnarray*} y_k(\alpha)=\alpha_k x_k(\alpha) > \pi_k v_k, \quad  k=\overline{1,n},\end{eqnarray*}
  the vector $p_1$ solving the set of equations (\ref{tak6}) satisfies  the set of inequalities
\begin{eqnarray*} p_j^1 \leq
\sum\limits^m_{k=1}\left[a_{kj}\left(x_j^0\right)+\alpha_j^0 \frac{c_{kj}}
{\pi_j}\right]
p_k^0   \end{eqnarray*}
\begin{eqnarray} \label{tak8}
  + \sum\limits^{n}_{k=m+1}\left[a_{kj}\left(x_j^0\right)+
\alpha_j^0 \frac{c_{kj}}
{\pi_j}\right]p_k^1, \quad j=\overline{m+1,n}.
\end{eqnarray}
From which and the fact that the spectral radius of the matrix $\tilde{\cal A}\left(x^0\right)$ is less than 1, it follows that the set of inequalities
$ p_j^1 \leq \bar p_j, \ j=\overline{m+1, n},$ holds, therefore, the  inequalities

\begin{eqnarray*} p_j^0 - \sum\limits^m_{k=1}a_{kj}(x_j(\alpha))p_k^0 - \sum\limits^n_{k=m+1}a_{kj}(x_j(\alpha))p_k^1   \end{eqnarray*}  \begin{eqnarray*}  \geq p_j^0 - \sum\limits^m_{k=1}a_{kj}\left(x_j^0\right)p_k^0 - \sum\limits^n_{k=m+1}a_{kj}\left(x_j^0\right)\bar p_k > 0, \quad  j=\overline{1,m}, \end{eqnarray*}
hold.

Hence, $ \pi_j >0, \ j=\overline{1,m}.$  The  validity  of the equalities
(\ref{tak5}) and the fact that the vector $p_1=\{p_i^1\}_{i=m+1}^n$ solves the set of equations (\ref{tak6}) together mean that the vector $p_1$ solves the set of equations (\ref{tak7}).
\qed\end{proof}

Build the vector $\alpha $ presented in the  conditions of the  Theorem \ref{tak1}.
Find conditions for the validity of the  set of  equalities (\ref{tak5}).
 Let $\alpha_0=\{\alpha_k^0\}_{k=1}^n$ be the same strictly positive vector as before.
Denote by $\tilde \alpha=\{ \alpha_1, \ldots, \alpha_m, \alpha_{m+1}^0,  \ldots, \alpha_n^0\}$
vectors whose first $m$ components satisfy the set of inequalities
$0 \leq \alpha_i \leq \alpha_i^0, \ i=\overline{1,m}.$

Suppose the condition holds:
there exists a strictly positive solution $x(\bar \alpha_0)=\{x_j(\bar \alpha_0)\}_{j=1}^n $ to the set of equations
\begin{eqnarray*} x_k(\bar \alpha_0) - \sum\limits^{n}_{j=1}\left[a_{kj}+ \bar \alpha_j^0 c_{kj}\right]x_j(\bar \alpha_0)-e_k+i_k  \end{eqnarray*}
\begin{eqnarray} \label{tak11}
- \sum\limits^n_{j=1}r_{kj}+ \sum\limits^l_{j=1}b_{kj}=\sum\limits^l_{j=n+1}c_{kj}y_j,
\quad k=\overline{1,n},
\end{eqnarray}
satisfying the inequalities
\begin{eqnarray*} \alpha_k^0x_k(\bar \alpha_0) > \pi_k v_k, \quad  k=\overline{1,n},\end{eqnarray*}
where
 \begin{eqnarray*} \bar \alpha_0=\{\bar \alpha_1^0, \ldots, \bar \alpha_n^0\}, \quad \bar \alpha_i^0=0,\quad  i=\overline{1, m}, \quad  \bar \alpha_i^0= \alpha_i^0, \quad i=\overline{m+1, n}. \end{eqnarray*}
Take  rather small numbers $\varepsilon_i >0, \ i=\overline{1, m},$ such that the inequalities
\begin{eqnarray*} \frac{\pi_i v_i + \varepsilon_i }{x_i(\bar \alpha_0)}< \alpha_i^0, \quad i=\overline{1, m},\end{eqnarray*}
hold.
On the closed set of  vectors
\begin{eqnarray} \label{tak12}
H =\left\{\alpha=\{\alpha_k\}_{k=1}^m \in R_+^m,  \  \frac{\pi_i v_i + \varepsilon_i }{x_i(\bar \alpha_0)}\leq \alpha_i \leq \alpha_i^0, \ i=\overline{1, m}\right\},
\end{eqnarray}
consider the set of equations
\begin{eqnarray}
\label{tak13}
\alpha_j=f_j(\alpha_1, \ldots, \alpha_m), \quad j=\overline{1,m},
\end{eqnarray}
relative to the vector $\alpha=\{\alpha_i\}_{i=1}^m,$
where
\begin{eqnarray*}  f_j(\alpha_1, \ldots, \alpha_m) \end{eqnarray*}
\begin{eqnarray*} =\frac
{ R_j^2\left(p^0, A, \bar B,  x(\tilde \alpha),  p_1\right)}{\sum\limits^m_{k=1}c_{kj}p_k^0 +
\sum\limits^{n}_{k=m+1}c_{kj}p_k^1}\pi_j, \quad  j=\overline{1,m},\end{eqnarray*}
 $ p_1=\{ p_{m+1}^1, \ldots , p_{n}^1\}$ solves the set of equations
\begin{eqnarray*} p_j =
\sum\limits^m_{k=1}\left[a_{kj}(x_j(\tilde \alpha))+\alpha_j^0 \left( \frac{c_{kj}}
{\pi_j} - \frac{b_{kj}}{y_j(\tilde \alpha)}\right)\right]
p_k^0  \end{eqnarray*}
\begin{eqnarray} \label{tak14}
  + \sum\limits^{n}_{k=m+1}\left[a_{kj}(x_j(\tilde \alpha))+
\alpha_j^0 \left( \frac{c_{kj}}
{\pi_j} - \frac{b_{kj}}{y_j(\tilde \alpha)}\right)\right]p_k, \quad j=\overline{m+1,n},
\end{eqnarray}
for the vector $ p=\{ p_{m+1}, \ldots , p_{n}\},$
 $y_j(\tilde \alpha)=\alpha_j^0x_j(\tilde \alpha), \ j=\overline{m+1, n},$ and
$x(\tilde \alpha)=\{x_j(\tilde \alpha)\}_{j=1}^n $ is a solution of the set of equations
\begin{eqnarray*} x_k(\tilde \alpha)-\sum\limits^{m}_{j=1}\left[a_{kj}+ \alpha_j c_{kj}\right]x_j(\tilde \alpha) - \sum\limits^{n}_{j=m+1}\left[a_{kj}+ \alpha_j^0 c_{kj}\right]x_j(\tilde \alpha)-e_k+i_k  \end{eqnarray*}
\begin{eqnarray} \label{tak15}
- \sum\limits^n_{j=1}r_{kj}+\sum\limits^l_{j=1}b_{kj}=\sum\limits^l_{j=n+1}c_{kj}y_j,
\quad k=\overline{1,n}.
\end{eqnarray}

Let us note that for components of the solution to the set of equations (\ref{tak14}) the next inequalities
\begin{eqnarray*}   p_j^0(0)\leq  p_j^1 \leq \bar p_j( \alpha_0) \leq \bar p_j, \quad j=\overline{m+1,n},\end{eqnarray*}
hold, where $p^0(0)=\{p_j^0(0)\}_{j=m+1}^n$ solves the set of equations
\begin{eqnarray} \label{tak17}
p_j =\sum\limits^m_{k=1}a_{kj}
p_k^0 +
 \sum\limits^{n}_{k=m+1}a_{kj} p_k, \quad  j=\overline{m+1,n},
\end{eqnarray}
$\bar p(\alpha_0) =\{\bar p_j( \alpha_0)\}_{j=m+1}^n$ solves the set of equations
\begin{eqnarray*}  p_j =
\sum\limits^m_{k=1}\left[a_{kj}\left(x_j^0\right)+\alpha_j^0 \left( \frac{c_{kj}}
{\pi_j} - \frac{b_{kj}}{y_j( \alpha_0)}\right)\right]
p_k^0  \end{eqnarray*}
\begin{eqnarray} \label{tak18}
  + \sum\limits^{n}_{k=m+1}\left[a_{kj}\left(x_j^0\right)+
\alpha_j^0 \left( \frac{c_{kj}}
{\pi_j} - \frac{b_{kj}}{y_j( \alpha_0)}\right)\right] p_k, \quad j=\overline{m+1,n},
\end{eqnarray}
for the vector $p=\{p_j)\}_{j=m+1}^n$
 and $\bar p =\{\bar p_j\}_{j=m+1}^n$ solves the set of equations (\ref{taalla1}).

\begin{theorem}\label{vtak1}
Let $\sum\limits_{k=1}^nc_{ki}>0, \ i=\overline{1,l},$ there exist  a non-negative vector
$v_0=\{v_i\}_{i=1}^n, \ v_i \geq 0, \ i=\overline{1,n},$  such  that $\bar C(v_0) - \bar B $ is a non-negative matrix having no zero rows or columns and
$ A\left(x^0\right) + \bar C(v_0) - \bar B$ is an indecomposable matrix, and let  the spectral radius of the matrix\index{spectral radius of the matrix} $A$ be less than 1,
a strictly positive vector
$\alpha_0=\{\alpha_i^0 \}_{i=1}^{n},$
the vector of  monopolistic prices\index{vector of  monopolistic prices}  $p^0=\{p_1^0, \ldots, p_m^0\}$ and the matrix $ A\left(x^0\right)$ satisfy conditions: \\
1) the spectral radius of matrices\index{spectral radius of matrices}
\begin{eqnarray*} \left\|a_{kj}\left(x_j^0\right)+\alpha_j^0  \frac{c_{kj}}
{\pi_j}  \right\|^{n}_{k,j=1}, \quad \left\|a_{kj}\left(x_j^0\right)+\alpha_j^0  \frac{c_{kj}}
{\pi_j}  \right\|^{n}_{k,j=m+1},
\end{eqnarray*}
is less than 1\ ;\\
2) there exists a strictly positive solution $x(\bar \alpha_0)=\{x_j(\bar \alpha_0)\}_{j=1}^n $ to the set of equations (\ref{tak15})
satisfying the inequalities
\begin{eqnarray*} \alpha_k^0x_k(\bar \alpha_0) > \pi_k v_k, \quad  k=\overline{1,n},\end{eqnarray*}
where
 \begin{eqnarray*} \bar \alpha_0=\{\bar \alpha_1^0, \ldots, \bar \alpha_n^0\}, \quad \bar \alpha_i^0=0,\quad  i=\overline{1, m}, \quad  \bar \alpha_i^0= \alpha_i^0, \quad i=\overline{m+1, n}\ ; \end{eqnarray*}
3) there exists a strictly positive solution $\bar p=\{\bar p_i\}^{n}_{i=m+1}$ to the set of equations
\begin{eqnarray*}  p_j =
\sum\limits^m_{k=1}\left[a_{kj}\left(x_j^0\right)+  \alpha_j^0 \frac{c_{kj}}
{\pi_j}\right]
p_k^0 \end{eqnarray*}
\begin{eqnarray}
\label{taalla1}
+ \sum\limits^{n}_{k=m+1}\left[a_{kj}\left(x_j^0\right)+ \alpha_j^0 \frac{c_{kj}}
{\pi_j}\right] p_k, \quad  j=\overline{m+1,n},
\end{eqnarray}
for the vector $ p=\{ p_i\}^{n}_{i=m+1}$
such that the inequalities
\begin{eqnarray} \label{taalla2}
 p_j^0 - \sum\limits^m_{k=1}a_{kj}\left(x_j^0\right)p_k^0
- \sum\limits^{n}_{k=m+1}a_{kj}\left(x_j^0\right)\bar p_k > 0,
 \quad j=\overline{1,m},
\end{eqnarray}
hold;\\
4) the inequalities
\begin{eqnarray*} \frac{\pi_j\left[p_j^0 - \sum\limits^m_{k=1}a_{kj}\left(x_j^0\right)p_k^0
- \sum\limits^{n}_{k=m+1}a_{kj}\left(x_j^0\right) \bar p_k\right]}
{\sum\limits^m_{k=1}c_{kj}p_k^0 +
\sum\limits^{n}_{k=m+1}c_{kj}\bar p_k} \geq \frac{\pi_j v_j + \varepsilon_j }{x_j\left(\bar \alpha^0\right)}, \quad j=\overline{1,m},\end{eqnarray*}
\begin{eqnarray*}  \frac{\pi_j \left[p_j^0+ \frac{1}{x_j\left(\bar \alpha^0\right)} \left[\sum\limits^m_{k=1}b_{kj}p_k^0+\sum\limits^n_{k=m+1}b_{kj} \bar p_k\right]\right]}
{\sum\limits^m_{k=1}c_{kj}p_k^0 +
\sum\limits^{n}_{k=m+1}c_{kj}p_k^0(0)} \leq \alpha_j^0, \quad j=\overline{1,m},\end{eqnarray*}
 \begin{eqnarray*} \max\limits_{1\leq j \leq m} \pi_j\sum\limits_{s=1}^m d_{js}^1\left(p^0, A, B, C, \alpha_0\right)<1,\end{eqnarray*}
hold, where
\begin{eqnarray*}  d_{js}^1\left(p^0,  A, \bar  B, C, \alpha_0\right)=\frac{ x_s(\alpha_0)\left[\left[E - (A+ \alpha_0 C)\right]^{-1}C\right]_{js} \sum\limits^m_{k=1}|b_{kj}- r_{kj}|p_k^0}
{x_j(\bar \alpha_0)^2\left[\sum\limits^m_{k=1}c_{kj}p_k^0 +
\sum\limits^{n}_{k=m+1}c_{kj}p_k^0(0)\right]} \end{eqnarray*}

\begin{eqnarray*} +\frac{ x_s(\alpha_0)\left[\left[E - (A+ \alpha_0 C)\right]^{-1}C\right]_{js} \sum\limits^n_{k=m+1}|b_{kj}- r_{kj}| \bar p_k( \alpha_0)}
{x_j(\bar \alpha_0)^2\left[\sum\limits^m_{k=1}c_{kj}p_k^0 +
\sum\limits^{n}_{k=m+1}c_{kj}p_k^0(0)\right]} \end{eqnarray*}

\begin{eqnarray*} +\frac{\sum\limits^{n}_{k=m+1}a_{kj}(x_j(\bar \alpha_0)) \left[\left[E - \tilde{\cal A}( \alpha_0)^T\right]^{-1}\varphi^s(\alpha_0)\right]_k }{\sum\limits^m_{k=1}c_{kj}p_k^0 +
\sum\limits^{n}_{k=m+1}c_{kj}p_k^0(0)} \end{eqnarray*}

\begin{eqnarray*} +\frac{ \sum\limits^n_{k=m+1}b_{kj}\left[\left[E - \tilde{\cal A}( \alpha_0)^T\right]^{-1}\varphi^s(\alpha_0)\right]_k }{x_j(\bar \alpha_0)\left[\sum\limits^m_{k=1}c_{kj}p_k^0 +
\sum\limits^{n}_{k=m+1}c_{kj}p_k^0(0)\right]} \end{eqnarray*}

\begin{eqnarray*} + \frac{1}{x_j(\bar \alpha_0)}\frac{\left[\sum\limits^m_{k=1}b_{kj} p_k^0+\sum\limits^n_{k=m+1}b_{kj}\left[\left[E - \tilde{\cal A}( \alpha_0)^T\right]^{-1}\varphi^s(\alpha_0)\right]_k     \right]}
{\left[\sum\limits^m_{k=1}c_{kj}p_k^0 +
\sum\limits^{n}_{k=m+1}c_{kj}p_k^0(0)\right]^2} \end{eqnarray*}  \begin{eqnarray*} \times\sum\limits^{n}_{k=m+1}c_{kj}\left[\left[E - \tilde{\cal A}(\alpha_0)^T\right]^{-1}\varphi^s(\alpha_0)\right]_k  \end{eqnarray*}

\begin{eqnarray*} +\frac{p_j^0 }
{\left[\sum\limits^m_{k=1}c_{kj}p_k^0 +
\sum\limits^{n}_{k=m+1}c_{kj}p_k^0(0)\right]^2} \sum\limits^{n}_{k=m+1}c_{kj}\left[\left[E - \tilde{\cal A}( \alpha_0)^T\right]^{-1}\varphi^s(\alpha_0)\right]_k.\end{eqnarray*}
Then there exists  a  vector $\beta_0=\{ \beta^0_i\}_{i=1}^n$ such that
\begin{eqnarray*}  \beta^0_i= \alpha^0_i, \quad i=\overline{m+1, n}, \quad  \frac{\pi_iv_i +\varepsilon_i}{x_i(\bar \alpha_0)}\leq  \beta^0_i \leq \alpha^0_i, \quad i=\overline{1, m},\end{eqnarray*}
and this vector solves the set of equations (\ref{tak13}). The  built solution satisfies the same conditions as the vector $\alpha$ presented in the  conditions of the Theorem \ref{tak1}.
\end{theorem}
\begin{proof}\smartqed   Consider on the closed set $H$ of vectors $\alpha=\{\alpha_k\}_{k=1}^m,$ where
\begin{eqnarray} \label{ktas7}
H =\left\{\alpha=\{\alpha_k\}_{k=1}^m \in R_+^m,  \  \frac{\pi_i v_i + \varepsilon_i }{x_i(\bar \alpha_0)}\leq \alpha_i \leq \alpha_i^0, \ i=\overline{1, m}\right\},
\end{eqnarray}
the set of equations (\ref{tak13}).
It is obvious that for any vector
\begin{eqnarray*} \tilde \alpha=\{\alpha_1, \ldots, \alpha_m, \alpha_m^0, \ldots, \alpha_n^0\}, \quad \alpha_i^0 >0, \quad i=\overline{m+1, n},\end{eqnarray*}
 whose first $m$ components belong to the set $H$
there exists a solution to the set of equations (\ref{tak15})  satisfying the set of inequalities
\begin{eqnarray*} x_i(\tilde \alpha) \alpha_i^0 > \pi_i v_i, \quad  i=\overline{1,n}, \end{eqnarray*}
because the set of inequalities $x_i(\tilde \alpha) \geq x_i(\bar \alpha_0), \ i=\overline{1,n},$ holds.
Prove the existence of a solution to the set of equations (\ref{tak13}).

The inequalities
\begin{eqnarray*} \frac{\pi_j\left[p_j^0 - \sum\limits^m_{k=1}a_{kj}\left(x_j^0\right)p_k^0
- \sum\limits^{n}_{k=m+1}a_{kj}\left(x_j^0\right) \bar p_k\right]}
{\sum\limits^m_{k=1}c_{kj}p_k^0 +
\sum\limits^{n}_{k=m+1}c_{kj}\bar p_k} \leq  f_j(\alpha_1, \ldots, \alpha_m)
\end{eqnarray*}
\begin{eqnarray*} \leq  \frac{\pi_j\left[p_j^0+ \frac{1}{x_j\left(\bar \alpha^0\right)} \left[\sum\limits^m_{k=1}b_{kj}p_k^0+\sum\limits^n_{k=m+1}b_{kj} \bar p_k\right]\right]}
{\sum\limits^m_{k=1}c_{kj}p_k^0 +
\sum\limits^{n}_{k=m+1}c_{kj}p_k^0(0)}, \quad j=\overline{1,m},\end{eqnarray*}
 hold.
From these inequalities and the Theorem conditions, it follows that the map
\begin{eqnarray*} f(\alpha_1, \ldots, \alpha_m)=\{f_j(\alpha_1, \ldots, \alpha_m)\}_{j=1}^m\end{eqnarray*}
maps the set $H$ into itself, where
\begin{eqnarray*} f_j(\alpha_1, \ldots, \alpha_m)=\frac
{ R_j^2\left(p^0, A, \bar B,  x(\tilde \alpha),  p_1\right)}{\sum\limits^m_{k=1}c_{kj}p_k^0 +
\sum\limits^{n}_{k=m+1}c_{kj}p_k^1}\pi_j, \quad  j=\overline{1,m}.\end{eqnarray*}

To prove   the  map $ f(\alpha_1, \ldots, \alpha_m)$ is contractive  on the set $H,$ let us estimate the solution to the sets of equations (\ref{tak14}) and (\ref{tak15}).

Find estimate for derivatives of the solution to the set of equations
(\ref{tak14}). For this, find the set of equations satisfied by the vector
\begin{eqnarray*} \frac{\partial p_1}{\partial \alpha_s}=\left\{\frac{\partial p_j^1}{\partial \alpha_s}\right\}_{j=m+1}^n, \quad s=\overline{1,m}.\end{eqnarray*}
 For $s=\overline{1,m}$ we have
\begin{eqnarray*} \frac{\partial p_j^1}{\partial \alpha_s}= - \sum\limits^{m}_{k=1}(b_{kj}- r_{kj})p_k^0\frac{\partial}{\partial \alpha_s}\left[\frac{1}{x_j(\tilde \alpha)}\right] - \sum\limits^{n}_{k=m+1}(b_{kj}- r_{kj})p_k^1\frac{\partial}{\partial \alpha_s}\left[\frac{1}{x_j(\tilde \alpha)}\right] \end{eqnarray*}
\begin{eqnarray} \label{tak19}
+\sum\limits^{n}_{k=m+1}\left[a_{kj}(x_j(\tilde \alpha))+
\alpha_j^0 \left( \frac{c_{kj}}
{\pi_j} - \frac{b_{kj}}{y_j(\tilde \alpha)}\right)\right]\frac{\partial p_k^1}{\partial \alpha_s},  \quad j=\overline{m+1,n}.
\end{eqnarray}
The vector
\begin{eqnarray*} \frac{\partial x(\tilde \alpha)}{\partial \alpha_s} = \left\{\frac{\partial x_k(\tilde \alpha)}{\partial \alpha_s}\right\}_{k=m+1}^n, \quad s=\overline{1,m},\end{eqnarray*}
satisfies the set of equations
\begin{eqnarray*} \frac{\partial x_k(\tilde \alpha)}{\partial \alpha_s}-\sum\limits^{m}_{j=1}\left[a_{kj}+ \alpha_j c_{kj}\right]\frac{\partial x_j(\tilde \alpha)}{\partial \alpha_s}  \end{eqnarray*}
\begin{eqnarray} \label{tak20}
- \sum\limits^{n}_{j=m+
1}\left[a_{kj}+ \alpha_j^0 c_{kj}\right]\frac{\partial x_j(\tilde \alpha)}{\partial \alpha_s}=
c_{ks}x_s(\tilde \alpha),
\quad k=\overline{1,n}, \quad s=\overline{1,m}.
\end{eqnarray}
From here
\begin{eqnarray} \label{tak21}
\frac{\partial x_k(\tilde \alpha)}{\partial \alpha_s}=x_s(\tilde \alpha)\left[\left[E - (A+ \tilde \alpha C)\right]^{-1}C\right]_{ks},
\quad k=\overline{1,n}, \quad s=\overline{1,m},
\end{eqnarray}
where
$\tilde \alpha =\{\alpha_1, \ldots, \alpha_m, \alpha_{m+1}^0, \ldots,
\alpha_{n}^0\},$
and $\left[\left[E - (A+ \tilde \alpha C)\right]^{-1}C\right]_{ks}$ is the matrix element of the matrix
 $\left[E - (A+ \tilde \alpha C)\right]^{-1}C,$
\begin{eqnarray*}  A+ \tilde \alpha C=||a_{kj}+ \tilde \alpha_j c_{kj}||_{k, j=1}^n, \quad \tilde \alpha_j=\alpha_j, \quad j=\overline{1,m}, \quad \tilde \alpha_j=\alpha_j^0, \quad j=\overline{m+1,n}.\end{eqnarray*}
Therefore, the estimate
\begin{eqnarray} \label{tak22}
\left|\frac{\partial x_k(\tilde \alpha)}{\partial \alpha_s}\right|\leq x_s(\alpha_0)\left[\left[E - (A+  \alpha_0 C)\right]^{-1}C\right]_{ks},
\quad k=\overline{1,n}, \quad s=\overline{1,m},
\end{eqnarray}
holds.
As the equality
\begin{eqnarray*}  \frac{\partial }{\partial \alpha_s}\left[\frac{1}{x_k(\tilde \alpha)}\right]=- \frac{1}{x_k(\tilde \alpha)^2}\frac{\partial x_k(\tilde \alpha)}{\partial \alpha_s},
\quad k=\overline{1,n}, \quad s=\overline{1,m},\end{eqnarray*}
holds, we have
\begin{eqnarray*}  \frac{\partial }{\partial \alpha_s}\left[\frac{1}{x_k(\tilde \alpha)}\right]=-
\frac{x_s(\tilde \alpha)}{x_k(\tilde \alpha)^2}\left[\left[E - (A+ \tilde \alpha C)\right]^{-1}C\right]_{ks}.\end{eqnarray*}
We rewrite the set of equations (\ref{tak19}) as follows

\begin{eqnarray*} \frac{\partial p_j^1}{\partial \alpha_s}=\frac{x_s(\tilde \alpha)}{x_j(\tilde \alpha)^2}  \left[\left[E - (A+ \tilde \alpha C)\right]^{-1}C\right]_{js}\sum\limits^{m}_{k=1}(b_{kj}- r_{kj})p_k^0 \end{eqnarray*}
\begin{eqnarray*} +\frac{x_s(\tilde \alpha)}{x_j(\tilde \alpha)^2}\left[\left[E - (A+ \tilde \alpha C)\right]^{-1}C\right]_{js} \sum\limits^{n}_{k=m+1}(b_{kj}- r_{kj})p_k^1 \end{eqnarray*}
\begin{eqnarray} \label{tak23}
+\sum\limits^{n}_{k=m+1}\left[a_{kj}(x_j(\tilde \alpha))+
\alpha_j^0 \left( \frac{c_{kj}}
{\pi_j} - \frac{b_{kj}}{y_j(\tilde \alpha)}\right)\right]\frac{\partial p_k^1}{\partial \alpha_s},
\quad j=\overline{1,n}.
\end{eqnarray}
From the set of equations (\ref{tak23}) for the vector
\begin{eqnarray*} \frac{\partial p_1}{\partial \alpha_s}= \left\{\frac{\partial p_j^1}{\partial\alpha_s}\right\}_{j=m+1}^n\end{eqnarray*}
we obtain the set of inequalities
\begin{eqnarray*} \left|\frac{\partial p_j^1}{\partial \alpha_s}\right|\leq \frac{x_s(\alpha_0)}{x_j(\bar \alpha_0)^2} \left[\left[E - (A+ \alpha_0 C)\right]^{-1}C\right]_{js}\sum\limits^{m}_{k=1}|b_{kj}- r_{kj}|p_k^0 \end{eqnarray*}
\begin{eqnarray*} +\frac{x_s(\alpha_0)}{x_j(\bar \alpha_0)^2}\left[\left[E - (A+ \alpha_0 C)\right]^{-1}C\right]_{js} \sum\limits^{n}_{k=m+1}|b_{kj}- r_{kj}| \bar p_k(\alpha_0) \end{eqnarray*}
\begin{eqnarray} \label{tak24}
+\sum\limits^{n}_{k=m+1}\left[a_{kj}\left(x_j^0\right)+
\alpha_j^0 \left( \frac{c_{kj}}
{\pi_j} - \frac{b_{kj}}{y_j(\alpha_0)}\right)\right]\left|\frac{\partial p_k^1}{\partial \alpha_s}\right|,
\quad j=\overline{1,n}.
\end{eqnarray}
Introduce the vector
\begin{eqnarray*} \varphi^s(\alpha_0)=\{\varphi^s_j(\alpha_0)\}_{j=m+1}^n,\end{eqnarray*}
where
\begin{eqnarray*} \varphi^s_j(\alpha_0)=\frac{x_s(\alpha_0)}{x_j(\bar \alpha_0)^2}\left[\left[E - (A+ \alpha_0 C)\right]^{-1}C\right]_{js} \end{eqnarray*}  \begin{eqnarray*} \times \left[\sum\limits^{m}_{k=1}|b_{kj}- r_{kj}|p_k^0+\sum\limits^{n}_{k=m+1}|b_{kj}- r_{kj}| \bar p_k(\alpha_0)\right],
\quad j=\overline{1,n}, \quad s=\overline{1,m},\end{eqnarray*}
and the matrix
\begin{eqnarray*}  \tilde{\cal A}( \alpha_0)=\left|\left|a_{kj}\left(x_j^0\right)+\alpha_j^0\left(\frac{c_{kj}}{\pi_j} -\frac{b_{kj}}{y_j(\alpha_0)}\right)\right|\right|_{k,j=m+1}^n.\end{eqnarray*}
The estimate
\begin{eqnarray*} 0 \leq \left|\frac{\partial p_j^1}{\partial\alpha_s}\right| \leq \left[\left[E - \tilde{\cal A}( \alpha_0)^T\right]^{-1}\varphi^s(\alpha_0)\right]_j, \quad j=\overline{m+1,n}, \quad  s=\overline{1,m},\end{eqnarray*}
 holds.
Here $ \tilde{\cal A}( \alpha_0)^T $ is the matrix transposed to the matrix $\tilde{\cal A}(\alpha_0).$
Find conditions for which the map
\begin{eqnarray*} f(\alpha_1, \ldots, \alpha_m)=\{f_j(\alpha_1, \ldots, \alpha_m)\}_{j=1}^m\end{eqnarray*}
is a contraction on the considered set $H.$
As
\begin{eqnarray*} f_j\left(\alpha_1^{''}, \ldots, \alpha_m^{''}\right) - f_j\left(\alpha_1^{'}, \ldots, \alpha_m^{'}\right) \end{eqnarray*}
\begin{eqnarray*} =\int\limits_{0}^{1}\sum\limits_{s=1}^m\left(\alpha_s^{''}- \alpha_s^{'}\right)\frac{\partial}{\partial \alpha_s}f_j\left(\alpha_1^{'}+t \left(\alpha_1^{''}-\alpha_1^{'}\right), \ldots,\alpha_m^{'}+t \left(\alpha_m^{''}- \alpha_m^{'}\right)\right)dt,\end{eqnarray*}
it is worth to estimate on the considered set $H$
\begin{eqnarray*} \frac{\partial f_j(\alpha_1, \ldots,\alpha_m)}{\partial \alpha_s}=\frac{\partial R_j^2\left(p^0, A, \bar B, x(\tilde \alpha), p_1\right)}{\partial \alpha_s}  \frac{\pi_j}{\sum\limits^m_{k=1}c_{kj}p_k^0 +
\sum\limits^{n}_{k=m+1}c_{kj}p_k^1}    \end{eqnarray*}  \begin{eqnarray*}  - \frac{ \pi_jR_j^2\left(p^0, A, \bar B, x(\tilde \alpha), p_1\right)\sum\limits^{n}_{k=m+1}c_{kj}\frac{\partial p_k^1}{\partial \alpha_s}}
{\left[\sum\limits^m_{k=1}c_{kj}p_k^0 +
\sum\limits^{n}_{k=m+1}c_{kj}p_k^1\right]^2}.\end{eqnarray*}
The equality
\begin{eqnarray*}  \frac{\partial R_j^2\left(p^0, A, \bar B, x(\tilde \alpha), p_1\right)}{\partial \alpha_s} \end{eqnarray*}
\begin{eqnarray*} =  \frac{\partial}{\partial \alpha_s}\left[\frac{1}{ x_j(\tilde \alpha)}\right]\left[ \sum\limits^m_{s=1}(b_{sj}- r_{sj}) p_s^0+\sum\limits^n_{s=m+1}(b_{sj}- r_{sj}) p_s^1\right]  \end{eqnarray*}
\begin{eqnarray*}  - \sum\limits^{n}_{k=m+1}a_{kj}(x(\tilde \alpha))\frac{\partial p_k^1}{\partial \alpha_s}+\frac{1}{ x_j(\tilde \alpha)}\sum\limits^{n}_{k=m+1}b_{kj}\frac{\partial p_k^1}{\partial \alpha_s} \end{eqnarray*}
\begin{eqnarray*} = - \frac{\partial x_j(\tilde \alpha)}{\partial \alpha_s}
\frac{\left[ \sum\limits^m_{s=1}(b_{sj}- r_{sj}) p_s^0+\sum\limits^n_{s=m+1}(b_{sj}- r_{sj}) p_s^1\right]}{ x_j(\tilde \alpha)^2}  \end{eqnarray*}
\begin{eqnarray*}  - \sum\limits^{n}_{k=m+1}a_{kj}(x(\tilde \alpha))\frac{\partial p_k^1}{\partial \alpha_s}+\frac{1}{ x_j(\tilde \alpha)}\sum\limits^{n}_{k=m+1}b_{kj}\frac{\partial p_k^1}{\partial \alpha_s} \end{eqnarray*}
holds.
However,
\begin{eqnarray*}  R_j^2\left(p^0, A, \bar B,  x(\tilde \alpha),  p_1\right) = p_j^0 - \sum\limits^m_{k=1}a_{kj}(x_j(\tilde \alpha))p_k^0
- \sum\limits^{n}_{k=m+1}a_{kj}(x_j(\tilde \alpha)) p_k^1 \end{eqnarray*}  \begin{eqnarray*} + \frac{1}{x_j(\tilde \alpha)}\left[\sum\limits^m_{k=1}b_{kj}p_k^0+\sum\limits^n_{k=m+1}b_{kj} p_k^1\right]  \end{eqnarray*}
\begin{eqnarray*}  \geq p_j^0 - \sum\limits^m_{k=1}a_{kj}\left(x_j^0\right)p_k^0
- \sum\limits^{n}_{k=m+1}a_{kj}\left(x_j^0\right)\bar  p_k > 0.\end{eqnarray*}
Therefore,
\begin{eqnarray*}  R_j^2\left(p^0, A, \bar B,  x(\tilde \alpha),  p_1\right) \leq p_j^0 + \frac{1}{x_j(\tilde \alpha)}\left[\sum\limits^m_{k=1}b_{kj}p_k^0+\sum\limits^n_{k=m+1}b_{kj} p_k^1\right].\end{eqnarray*}

With estimates obtained, we have
\begin{eqnarray*} \left|\frac{\partial}{\partial \alpha_s}f_j(\alpha_1, \ldots,\alpha_m)\right|  \end{eqnarray*}
\begin{eqnarray*} \leq \frac{\pi_j x_s(\alpha_0)\left[\left[E - (A+ \alpha_0 C)\right]^{-1}C\right]_{js} \sum\limits^m_{k=1}|b_{kj}- r_{kj}|p_k^0}
{x_j(\bar \alpha_0)^2\left[\sum\limits^m_{k=1}c_{kj}p_k^0 +
\sum\limits^{n}_{k=m+1}c_{kj}p_k^0(0)\right]} \end{eqnarray*}

\begin{eqnarray*} +\frac{\pi_j x_s(\alpha_0)\left[\left[E - (A+ \alpha_0 C)\right]^{-1}C\right]_{js} \sum\limits^n_{k=m+1}|b_{kj}- r_{kj}| \bar p_k( \alpha_0)}
{x_j(\bar \alpha_0)^2\left[\sum\limits^m_{k=1}c_{kj}p_k^0 +
\sum\limits^{n}_{k=m+1}c_{kj}p_k^0(0)\right]} \end{eqnarray*}

\begin{eqnarray*} +\frac{\pi_j\sum\limits^{n}_{k=m+1}a_{kj}(x_j(\bar \alpha_0)) \left[\left[E - \tilde{\cal A}( \alpha_0)^T\right]^{-1}\varphi^s(\alpha_0)\right]_k }{\sum\limits^m_{k=1}c_{kj}p_k^0 +
\sum\limits^{n}_{k=m+1}c_{kj}p_k^0(0)} \end{eqnarray*}

\begin{eqnarray*} +\frac{\pi_j \sum\limits^n_{k=m+1}b_{kj}\left[\left[E - \tilde{\cal A}( \alpha_0)^T\right]^{-1}\varphi^s(\alpha_0)\right]_k }{x_j(\bar \alpha_0)\left[\sum\limits^m_{k=1}c_{kj}p_k^0 +
\sum\limits^{n}_{k=m+1}c_{kj}p_k^0(0)\right]} \end{eqnarray*}

\begin{eqnarray*} + \frac{\pi_j}{x_j(\bar \alpha_0)}\frac{\left[\sum\limits^m_{k=1}b_{kj} p_k^0+\sum\limits^n_{k=m+1}b_{kj}\left[\left[E - \tilde{\cal A}( \alpha_0)^T\right]^{-1}\varphi^s(\alpha_0)\right]_k     \right]}
{\left[\sum\limits^m_{k=1}c_{kj}p_k^0 +
\sum\limits^{n}_{k=m+1}c_{kj}p_k^0(0)\right]^2} \end{eqnarray*}  \begin{eqnarray*} \times\sum\limits^{n}_{k=m+1}c_{kj}\left[\left[E - \tilde{\cal A}(\alpha_0)^T\right]^{-1}\varphi^s(\alpha_0)\right]_k  \end{eqnarray*}

\begin{eqnarray*} +\frac{\pi_j p_j^0 }
{\left[\sum\limits^m_{k=1}c_{kj}p_k^0 +
\sum\limits^{n}_{k=m+1}c_{kj}p_k^0(0)\right]^2} \sum\limits^{n}_{k=m+1}c_{kj}\left[\left[E - \tilde{\cal A}( \alpha_0)^T\right]^{-1}\varphi^s(\alpha_0)\right]_k.\end{eqnarray*}
We rewrite the last estimate as follows
\begin{eqnarray*} \left|\frac{\partial}{\partial \alpha_s}f_j(\alpha_1, \ldots,\alpha_m)\right|\leq \pi_jd_{js}^1\left(p^0, A,\bar B,\bar C,  \alpha_0\right).\end{eqnarray*}
Introduce on the considered set of vectors $\alpha$ the norm $||\alpha||=\max\limits_{1 \leq i \leq m}|\alpha_i|.$
In this norm, for the map $f(\alpha_1, \ldots, \alpha_m)=\{f_i(\alpha_1, \ldots, \alpha_m)\}_{i=1}^m$ the  estimate
\begin{eqnarray*}  \left|\left|f\left(\alpha_1^{''}, \ldots, \alpha_m^{''}\right)- f\left(\alpha_1^{'}, \ldots, \alpha_m^{'}\right)\right|\right|  \end{eqnarray*}  \begin{eqnarray*}  \leq \left|\left| \alpha^{''} - \alpha^{'}\right|\right| \max\limits_{1\leq j \leq m} \pi_j\sum\limits_{s=1}^m d_{js}^1\left(p^0, A, \bar B,\bar C,  \alpha_0\right)\end{eqnarray*}
holds and, therefore, it is a contraction map.
From the  conditions of the Theorem, it follows that there exists a unique solution to the set of equations (\ref{tak13}) satisfying the  conditions of the  Theorem \ref{vtak1}.
\qed\end{proof}

\subsection{The satisfaction of fixed consumers needs levels under non-linear technologies}

In this Subsection, we clarify conditions under which  increase of monopolistic prices\index{monopolistic prices} does not lead to lowering levels  of satisfaction of consumers needs.

 Suppose a  matrix
$A$ of direct inputs\index{ matrix
 of direct inputs} is productive.
Consider the set of equations
\begin{eqnarray*} x_k-\sum\limits^{n}_{j=1}[a_{kj}+\alpha_jc_{kj}]x_j   \end{eqnarray*}
\begin{eqnarray} \label{1tar1222}
-\sum\limits^n_{j=1}r_{kj}
+\sum\limits^l_{j=1}b_{kj} -e_k+i_k =\sum\limits^l_{j=n+1}c_{kj}y_j,
\quad k=\overline{1,n},
\end{eqnarray}
relative to the vector of gross outputs\index{vector of gross outputs}  $x=\{x_i\}_{i=1}^n.$

Denote by $ x(\alpha)= \{ x_i( \alpha_1,\ldots, \alpha_{n}) \}_{i=1}^{n}$
the solution to the set of equations (\ref{1tar1222})
for the vector
 $\alpha=\{\alpha_j\}_{j=1}^{n} $ belonging to the set \begin{eqnarray*}  T=\left\{\alpha=\{\alpha_i\}_{i=1}^{n},  \
0 \leq \alpha_i\leq \alpha_i^0,
\ i=\overline{1, n}\right\}.\end{eqnarray*}
Suppose a strictly positive solution
$x^0=\{x_j^0\}_{j=1}^{n}$  to the set of equations (\ref{1tar1222}) exists
for the vector
 $\alpha=\{\alpha_j\}_{j=1}^{n} $ with zero components  such  that the matrix
$A\left(x^0\right)$ is productive.
Take a  strictly positive vector $\alpha^0=\{\alpha_j^0\}_{j=1}^{n} $  such  that the spectral radius of the matrix $\left|\left|a_{kj}\left(x_j^0\right)+\alpha_j^0c_{kj}\right|\right|_{k,j=1}^{n} $ is less than 1.

Introduce the set
\begin{eqnarray*} X=\{x=\{x_i\}_{i=1}^{n} \in R_+^{n},\ x_i \geq x_i^0,\
i=\overline{1,n} \}\end{eqnarray*}
and consider the set of equations
\begin{eqnarray*}  p_j =
\sum\limits^m_{k=1}\left[a_{kj}(x_j(\alpha))+  \alpha_j\left( \frac{c_{kj}}
{\pi_j} - \frac{b_{kj}}{y_j}\right)\right]
p_k^0  \end{eqnarray*}
\begin{eqnarray}
\label{1tam9}
+ \sum\limits^{n}_{k=m+1}\left[a_{kj}(x_j(\alpha))+ \alpha_j\left( \frac{c_{kj}}
{\pi_j} - \frac{b_{kj}}{y_j}\right)\right] p_k, \quad  j=\overline{m+1,n},
\end{eqnarray}
for the vector $ p=\{ p_j\}_{j=m+1}^n,$
where the vector $ \alpha \in T.$
We denote the solution to the set of equations (\ref{1tam9}) for arbitrary vector $\alpha=\{\alpha_i\}_{i=1}^{n}$ from the set $T$ by $\bar p(\alpha)=\{\bar p_j(\alpha)\}_{j=m+1}^n.$
\begin{theorem}\label{1tam1}
Let $\sum\limits_{k=1}^nc_{ki}>0, \ i=\overline{1,l},$ there  exist  a  non-negative vector
$v_0=\{v_i\}_{i=1}^n, \ v_i \geq 0, \ i=\overline{1,n},$ such that $\bar C(v_0) - \bar B $ is a non-negative matrix having no zero rows or columns and
$ A\left(x^0\right) + \bar C(v_0) - \bar B$ is an  indecomposable matrix, and let the spectral radius of the matrix\index{ spectral radius of the matrix} $A\left(x^0\right)$ be less than 1.
And also let a strictly positive vector
$\alpha^0=\{\alpha_i^0 \}_{i=1}^{n}$
and a  vector of  monopolistic prices\index{vector of  monopolistic prices} $p^0=\{p_1^0, \ldots, p_m^0\}$ satisfy conditions:
there exists a strictly positive solution $x^0=\{x_j^0\}_{j=1}^n $ to the set of equations (\ref{1tar1222})
for the vector $\alpha=0$ such  that the matrix $ A\left(x^0\right)$ is productive  and
the spectral radius of the matrix $||a_{kj}\left(x_j^0\right)+\alpha_j^0c_{kj}||_{k,j=1}^n$ is less than 1.

If the inequalities
\begin{eqnarray*} \frac{y_j}{x_j^0} \leq \alpha_j^0, \quad j=\overline{1,n}, \end{eqnarray*}
\begin{eqnarray*} \max\limits_{j=\overline{1,n}}\frac{y_j}{\left[x_j^0\right]^2}
\sum\limits_{s=1}^n\left[\left[E -\left(A+\alpha^0C\right)\right]^{-1}C\right]_{js}x_s\left(\alpha^0\right)
<1, \end{eqnarray*}
\begin{eqnarray*} y_j >
\frac{v_j  \alpha_j^0\left[\sum\limits^m_{k=1}c_{kj}p_k^0 + \sum\limits^{n}_{k=m+1}c_{kj}\bar p_k\left(\alpha^0\right)\right]}
{M_j\left(p^0, A, \bar B, x(\alpha^0), \bar p(0), \bar p(\alpha^0)\right)}, \quad j=\overline{1,m},\end{eqnarray*}
\begin{eqnarray} \label{1tam2}
\quad y_j > \pi_j v_j, \quad j=\overline{m+1,n},
\end{eqnarray}
hold, where
\begin{eqnarray*} M_j\left(p^0, A, \bar B, x\left(\alpha^0\right),\bar p\left(\alpha^0\right)\right), \bar p(0))
\end{eqnarray*} 
\begin{eqnarray*} 
= p_j^0 - \sum\limits^m_{k=1}a_{kj}\left(x_j^0\right)p_k^0
- \sum\limits^{n}_{k=m+1}a_{kj}\left(x_j^0\right)\bar p_k\left(\alpha^0\right) \end{eqnarray*}  \begin{eqnarray*} + \frac{1}{x_j\left(\alpha^0\right)}\left[\sum\limits^m_{s=1}b_{sj}p_s^0+\sum\limits^n_{s=m+1}b_{sj} \bar p_s(0)\right],\end{eqnarray*}
$y_i, \ i=\overline{1,n}$ are consumer-producers needs satisfaction levels,\index{consumer-producers needs satisfaction levels}
then there  exists a  unique solution $ \bar \alpha=\{\bar \alpha_j\}_{j=1}^n $ to the set of equations
\begin{eqnarray} \label{1tax022}
\alpha_j=\frac{ y_j}
{ x_j(\alpha_1,\ldots,\alpha_{n})}, \quad j=\overline{1,n},
\end{eqnarray}
for the vector $ \alpha=\{ \alpha_j\}_{j=1}^n, $
where
$x(\alpha_1,\ldots,\alpha_{n})=
\{x_j(\alpha_1,\ldots,\alpha_{n})\}_{j=1}^{n}$ is a solution of  the set of equations
(\ref{1tar1222}).

Under conditions that
 the spectral radius of the matrix\index{spectral radius of the matrix}
\begin{eqnarray*} \tilde{\cal A}\left(\alpha^0\right)=\left\|a_{kj}\left(x_j^0\right)+ \alpha^0_j\left( \frac{c_{kj}}
{\pi_j} - \frac{b_{kj}}{y_j}\right) \right\|^{n}_{k,j=m+1}
\end{eqnarray*}
is less than 1,
 the inequalities
\begin{eqnarray} \label{1tam3}
 p_j^0 - \sum\limits^m_{k=1}a_{kj}\left(x_j^0\right)p_k^0
- \sum\limits^{n}_{k=m+1}a_{kj}\left(x_j^0\right)\bar p_k\left(\alpha^0\right)  > 0, \quad j=\overline{1,m},
\end{eqnarray}
hold, where $\bar p\left( \alpha^0\right)=\{\bar p_i\left(\alpha^0\right)\}^{n}_{i=m+1}$ is a solution of  the set of equations
(\ref{1tam9}) for the vector $\alpha^0=\{ \alpha_i^0\}_{i=1}^n,$
 there exists a solution to the set of equations
\begin{eqnarray*} p_j^0 =
\sum\limits^m_{k=1}\left[a_{kj}(x_j)+
\frac{y_j}{x_j}\left(\frac{c_{kj}}{\pi_j(\bar \alpha)}-\frac{b_{kj}}{y_j}\right)\right]
p_k^0   \end{eqnarray*}  \begin{eqnarray*}  +\sum\limits^{n}_{k=m+1}\left[a_{kj}(x_j)+\frac{y_j}{x_j}\left(\frac{c_{kj}}{\pi_j(\bar \alpha)}-\frac{b_{kj}}{y_j}\right)\right]p_k, \quad j=\overline{1,m}, \end{eqnarray*}
\begin{eqnarray*} p_j =
\sum\limits^m_{k=1}\left[a_{kj}(x_j)+\frac{y_j}{x_j}\left(\frac{c_{kj}}{\pi_j}-\frac{b_{kj}}{y_j}\right)\right]
p_k^0  \end{eqnarray*}
\begin{eqnarray} \label{1tam5}
+ \sum\limits^{n}_{k=m+1}\left[a_{kj}(x_j)+\frac{y_j}{x_j}\left(\frac{c_{kj}}{\pi_j}-\frac{b_{kj}}{y_j}\right)\right]p_k, \quad j=\overline{m+1,n},
\end{eqnarray}
\begin{eqnarray} \label{1tam6}
x_k-\sum\limits^{n}_{j=1}a_{kj}x_j- \sum\limits^n_{j=1}r_{kj}+ \sum\limits^l_{j=1}b_{kj} -e_k+i_k=
\sum\limits^l_{j=1}c_{kj}y_j,
\quad k=\overline{1,n},
\end{eqnarray}
for vectors $p=\{p_i\}_{i=m+1}^n$ and $ \ x=\{x_i\}_{i=1}^n,$
if to put
\begin{eqnarray} \label{1tam7}
\pi_j(\bar \alpha)
 =\frac{\bar \alpha_j\left[\sum\limits^m_{k=1}c_{kj}p_k^0 +
\sum\limits^{n}_{k=m+1}c_{kj}\bar p_k(\bar \alpha)\right]}
{V_j\left(p^0, A,  \bar B, \bar p(\bar \alpha), x(\bar \alpha)\right)}, \quad j=\overline{1,m},\end{eqnarray}
\begin{eqnarray*} V_j\left(p^0, A, \bar B, \bar p(\bar \alpha), x(\bar \alpha)\right) \end{eqnarray*}  \begin{eqnarray*} =p_j^0- \sum\limits^m_{k=1}a_{kj}(x_j(\bar \alpha))p_k^0
- \sum\limits^{n}_{k=m+1}a_{kj}(x_j(\bar \alpha))\bar p_k(\bar \alpha) \end{eqnarray*}  \begin{eqnarray*} + \frac{1}{x_j(\bar \alpha)}\left[\sum\limits^m_{s=1}b_{sj}p_s^0+\sum\limits^n_{s=m+1}b_{sj}\bar p_s(\bar \alpha)\right],\end{eqnarray*}
where
$\bar p(\bar \alpha)=\{\bar p_i(\bar \alpha)\}^{n}_{i=m+1}$ solves the set of equations (\ref{1tam9}) for the vector $\bar \alpha=\{\bar \alpha_i\}_{i=1}^n$ that is unique solution to the set of equations (\ref{1tax022}).
\end{theorem}
\begin{proof}\smartqed  Let
$x(\alpha_1,\ldots,\alpha_{n})=
\{x_j(\alpha_1,\ldots,\alpha_{n})\}_{j=1}^{n}$ be a solution of  the problem
(\ref{1tar1222}). The non-linear map
\begin{eqnarray*}  f(\alpha_1,\ldots,\alpha_{n})=\{ f_j(\alpha_1,\ldots,\alpha_{n}) \}_{j=1}^{n},\end{eqnarray*}
where
\begin{eqnarray*} f_j(\alpha_1,\ldots,\alpha_{n})=\frac{ y_j}
{ x_j(\alpha_1,\ldots,\alpha_{n})}, \qquad j=\overline{1,n},\end{eqnarray*}
is a monotonously decreasing map\index{monotonously decreasing map} of the set $T$
into itself and the map
\begin{eqnarray*} F(\alpha_1,\ldots,\alpha_{n})=f^2(\alpha_1,\ldots,\alpha_{n})
\end{eqnarray*}
\begin{eqnarray*} =\{f_j(f_1(\alpha_1,\ldots,\alpha_{n}),\ldots,
f_{n}(\alpha_1,\ldots,\alpha_{n}))\}_{j=1}^{n}\end{eqnarray*}
is \hfill a \hfill monotonously \hfill non-decreasing \hfill map\index{ monotonously  non-decreasing  map} \hfill of \hfill the \hfill set \hfill $T$ \hfill into \hfill itself. \hfill
Denote \hfill by \\ $F^m(\alpha_1,\ldots,\alpha_{n})$ the $m$-th power of the map
$F(\alpha_1,\ldots,\alpha_{n}).$
The next vector inequalities
\begin{eqnarray*} F^k(0,\ldots,0)\leq F^{k+1}(0,\ldots,0),\end{eqnarray*}
\begin{eqnarray*} F^k(\alpha_1^0,\ldots,\alpha_{n}^0)\geq
F^{k+1}(\alpha_1^0,\ldots,\alpha_{n}^0),\end{eqnarray*}
hold that one must  understand componentwise.
Additionally,
\begin{eqnarray*} F^k(0,\ldots,0)\le
F^k(\alpha_1^0,\ldots,\alpha_{n}^0).\end{eqnarray*}
Therefore, the sequence $F^k(0,\ldots,0)$ is monotonously non-decreasing  and the sequence $F^k(\alpha_1^0,\ldots,\alpha_{n}^0)$ is monotonously non-increasing.
Denote
\begin{eqnarray*} \lim\limits_{k \to \infty}F^k(0,\ldots,0)=F^\infty(0,\ldots,0), \quad
\lim\limits_{k \to \infty}F^k(\alpha_1^0,\ldots,\alpha_{n}^0)
=F^\infty(\alpha_1^0,\ldots,\alpha_{n}^0).\end{eqnarray*}
If the equality
\begin{eqnarray}
\label{1tax023}
F^\infty(0,\ldots,0)=
F^\infty(\beta_1^0,\ldots,\beta_{n}^0)
\end{eqnarray}
 holds, then $\bar \alpha=F^\infty(0,\ldots,0)$ solves the problem (\ref{1tax022}).

Find the sufficient conditions for this to be the case.

There holds the set of equalities
\begin{eqnarray*} x_j\left(\alpha_1^{''}, \ldots, \alpha_n^{''}\right) - x_j\left(\alpha_1^{'}, \ldots, \alpha_n^{'}\right) \end{eqnarray*}
\begin{eqnarray*} =\int\limits_{0}^{1}\sum\limits_{s=1}^m\left(\alpha_s^{''}- \alpha_s^{'}\right)\frac{\partial}{\partial \alpha_s}x_j\left(\alpha_1^{'}+t \left(\alpha_1^{''}-\alpha_1^{'}\right), \ldots,\alpha_n^{'}+t \left(\alpha_n^{''}- \alpha_n^{'}\right)\right)dt,\end{eqnarray*}
\begin{eqnarray*} j=\overline{1,n},\end{eqnarray*}
\begin{eqnarray} \label{1tam11}
\frac{\partial x_k( \alpha)}{\partial \alpha_s}-\sum\limits^{n}_{j=1}\left[a_{kj}+ \alpha_j c_{kj}\right]\frac{\partial x_j( \alpha)}{\partial \alpha_s}=
c_{ks}x_s( \alpha),
\quad k=\overline{1,n}.
\end{eqnarray}
From here,
\begin{eqnarray} \label{1tam12}
\frac{\partial x_k(\alpha)}{\partial \alpha_s}=x_s( \alpha)\left[\left[E - (A+  \alpha\bar  C)\right]^{-1}\bar C\right]_{ks},
\end{eqnarray}
where
$ \alpha =\{\alpha_1, \ldots, \alpha_n\},$ and
$ A+ \alpha \bar C=||a_{kj}+ \alpha_j c_{kj}||_{k, j=1}^n.$
Therefore, the estimate
\begin{eqnarray} \label{1tam13}
\left|\frac{\partial x_k(\alpha)}{\partial \alpha_s}\right|\leq x_s\left(\alpha^0\right)\left[\left[E - \left(A+  \alpha^0 \bar C\right)\right]^{-1}\bar C\right]_{ks}
\end{eqnarray}
and  the inequalities
\begin{eqnarray*} \max\limits_{j \in [1,n]}\left|f_j(\alpha_1^{''},\ldots,\alpha_{n}^{''}) -
f_j\left(\alpha_1^{'},\ldots,\alpha_{n}^{'}\right)\right|
 \end{eqnarray*}
 \begin{eqnarray*} =
\max\limits_{j \in [1,n]}
\left|\frac{ y_j}{ x_j(\alpha_1^{''},\ldots,\alpha_{n}^{''})} -
\frac{ y_j}{ x_j\left(\alpha_1^{'},\ldots,\alpha_{n}^{'}\right)}\right| \end{eqnarray*}
\begin{eqnarray*} \le
\max\limits_{j \in [1,n]}
\frac{y_j\left|x_j\left(\alpha_1^{''},\ldots,\alpha_{n}^{''}\right) -
x_j\left(\alpha_1^{'},\ldots,\alpha_{n}^{'}\right)\right|}
{\left[x_j^0\right]^2} \end{eqnarray*}
\begin{eqnarray*} \le\max\limits_{j \in [1,n]}
\frac{y_j}{\left[x_j^0\right]^2}
\sum\limits_{s=1}^n\left[\left[E - \left(A+  \alpha^0 \bar C\right)\right]^{-1}\bar C\right]_{js}x_s\left(\alpha^0\right)
\max\limits_{s \in [1,n]}\left|\alpha_s^{''} - \alpha_s^{'}\right| \end{eqnarray*}
hold. Therefore, if
\begin{eqnarray*} \max\limits_{j \in [1,n]}
\frac{y_j}{\left[x_j^0\right]^2}
\sum\limits_{s=1}^n\left[\left[E - \left(A+  \alpha^0 \bar C\right)\right]^{-1}\bar C\right]_{js}x_s\left(\alpha^0\right)
<1,\end{eqnarray*}
then the solution to the set of equations
(\ref{1tax022}) is unique  and the equality (\ref{1tax023}) holds.
From the inequalities
\begin{eqnarray} \label{1tam15}
y_j >
\frac{v_j  \alpha_j^0\left[\sum\limits^m_{k=1}c_{kj}p_k^0 + \sum\limits^{n}_{k=m+1}c_{kj}\bar p_k\left(\alpha^0\right)\right]}
{M_j\left(p^0, A, \bar B, x\left(\alpha^0\right), \bar p(0), \bar p\left(\alpha^0\right)\right)},
\quad j=\overline{1,m},
\end{eqnarray}
the inequalities
\begin{eqnarray*}   y_j > \pi_j (\bar \alpha)v_j, \quad j=\overline{1,m},\end{eqnarray*}
follow. Really, as
\begin{eqnarray} \label{2tam7}
\pi_j(\bar \alpha)
 =\frac{\bar \alpha_j\left[\sum\limits^m_{k=1}c_{kj}p_k^0 +
\sum\limits^{n}_{k=m+1}c_{kj}\bar p_k(\bar \alpha)\right]}
{V_j\left(p^0, A,  \bar B, \bar p(\bar \alpha), x(\bar \alpha)\right)}, \quad j=\overline{1,m},\end{eqnarray}
\begin{eqnarray*} V_j\left(p^0, A, \bar B, \bar p(\bar \alpha), x(\bar \alpha)\right) \end{eqnarray*}  \begin{eqnarray*} =p_j^0- \sum\limits^m_{k=1}a_{kj}(x_j(\bar \alpha))p_k^0
- \sum\limits^{n}_{k=m+1}a_{kj}(x_j(\bar \alpha))\bar p_k(\bar \alpha)
\end{eqnarray*}
\begin{eqnarray*} + \frac{1}{x_j(\bar \alpha)}\left[\sum\limits^m_{s=1}b_{sj}p_s^0+\sum\limits^n_{s=m+1}b_{sj}\bar p_s(\bar \alpha)\right],\end{eqnarray*}
where
$\bar p(\bar \alpha)=\{\bar p_i(\bar \alpha)\}^{n}_{i=m+1}$ is a solution of  the set of equations (\ref{1tam9}) for the vector $\bar \alpha=\{\bar \alpha_i\}_{i=1}^n$ that is a unique solution to the set of equations (\ref{1tax022}),
we have the estimates
\begin{eqnarray*} V_j\left(p^0, A, \bar B, \bar p(\bar \alpha), x(\bar \alpha)\right) \end{eqnarray*}
\begin{eqnarray*}  \geq p_j^0- \sum\limits^m_{k=1}a_{kj}\left(x_j^0\right)p_k^0
- \sum\limits^{n}_{k=m+1}a_{kj}\left(x_j^0\right)\bar p_k\left(\alpha^0\right) \end{eqnarray*}  \begin{eqnarray*} + \frac{1}{x_j\left(\alpha^0\right)}\left[\sum\limits^m_{s=1}b_{sj}p_s^0+\sum\limits^n_{s=m+1}b_{sj}\bar p_s(0)\right],\end{eqnarray*}
\begin{eqnarray*} \bar \alpha_j\left[\sum\limits^m_{k=1}c_{kj}p_k^0 +
\sum\limits^{n}_{k=m+1}c_{kj}\bar p_k(\bar \alpha)\right] \end{eqnarray*}
\begin{eqnarray*} \leq  \alpha_j^0\left[\sum\limits^m_{k=1}c_{kj}p_k^0 +
\sum\limits^{n}_{k=m+1}c_{kj}\bar p_k\left(\alpha^0\right)\right]. \end{eqnarray*}
From these estimates, we obtain the set of inequalities
\begin{eqnarray*}  \pi_j(\bar \alpha)\leq \frac{\alpha_j^0\left[\sum\limits^m_{k=1}c_{kj}p_k^0 + \sum\limits^{n}_{k=m+1}c_{kj}\bar p_k\left(\alpha^0\right)\right]}
{M_j\left(p^0, A, \bar B, x\left(\alpha^0\right), \bar p(0), \bar p\left(\alpha^0\right)\right)}, \quad j=\overline{1,m},\end{eqnarray*}
that together with inequalities (\ref{1tam2}) leads to the set of inequalities
\begin{eqnarray*} v_j \pi_j(\bar \alpha)< y_j, \quad j=\overline{1,m}.\end{eqnarray*}
It is obvious that the vector
$x(\bar \alpha_1,\ldots,\bar \alpha_{n})=
\{x_j(\bar \alpha_1,\ldots,\bar \alpha_{n})\}_{j=1}^{n},$ solving the problem
(\ref{1tar1222}), where $\bar \alpha=\{\bar \alpha_1,\ldots,\bar \alpha_{n}\}$ is a unique solution to the set of equations (\ref{1tax022}), solves the set of equations (\ref{1tam6}) too.
If one inserts into the set of equations (\ref{1tam5}) the vector $x(\bar \alpha_1,\ldots,\bar \alpha_{n}),$
solving the problem
(\ref{1tar1222}), where $\bar \alpha=\{\bar \alpha_1,\ldots,\bar \alpha_{n}\}$ is a unique solution to the set of equations (\ref{1tax022}), then there exists also unique solution to the set of equations (\ref{1tam5}) for the vector $p=\{p_j\}_{j=m+1}^n.$ This vector satisfies  the rest equations of this set because equalities (\ref{1tam7}) hold.
\qed\end{proof}

\subsection{Exchange model with monopolists}

Study of the influence of external economic agents\index{external economic agents} onto  formation of prices is rather important in view of formation of domestic policy\index{domestic policy}  as well as foreign one \cite{59,77, 55, 71, 92}. Change of prices by monopolists   can result in significant price disbalances\index{price imbalances} that in turn can have serious economical consequences. Another factor that can significantly influence onto  formation of prices\index{formation of prices} within    the  economy system is the presence of unequal foreign economical relations.\index{unequal foreign economical relations} Price formation model accounting for all these factors can be applied to analyze formation of prices  on regional or industry level\index{formation of prices  on regional or industry level} and equilibrium industries outputs\index{equilibrium industries outputs} if one takes an industry or a region to be economy system and all the rest to be external economic environment.\index{external economic environment} Consider an economy system containing $m$ monopolists keeping fixed prices or added values for goods produced.\index{added values for goods produced}

Suppose the economy system contains $n$ consumers everyone of which has goods vector $B_i^0=\{b_i \delta_{ik}\}_{k=1}^n, \ b_i>0, \ i=\overline{1,n}.$ We describe consumers choice\index{consumers choice} by fields of information evaluation  by consumers\index{fields of information evaluation  by consumers}  that are columns of a  certain matrix $A.$ We treat this model as exchange one and call the  structure matrix $A$ the exchange matrix. We give expression for the matrix $A$  in terms of technological matrix\index{technological matrix} and matrix of unproductive consumption\index{matrix of unproductive consumption}  in the presence of the production within the next Subsection. Therefore, the economy system has $m$
monopolists keeping prices at the level given by the strictly positive vector
 $p^{0}=\{p_1^0,\dots ,
p_m^0\}$ or added values in the presence of the production. The rest industries have a certain  vector of goods  $b^0=\{b_{m+1}^0 ,\dots
,b_n^0 \}$ to sell. The problem is to find volume of  monopolistic sales\index{volume of  monopolistic sales}  under the condition they can guarantee the  needed  volume  of sales and equilibrium prices of the rest non-monopolistic goods.\index{equilibrium prices of the rest non-monopolistic goods}

For conditions considered, we give a demand vector\index{demand vector of the $i$-th insatiable consumer} of the $i$-th insatiable consumer in the form
\begin{eqnarray} \label{r5l1}
\gamma _i={\left\{\gamma _{ik}\left(p^0 ,p\right)\right\} }_{k=1}^n,
\end{eqnarray}
where
\begin{eqnarray*} {\gamma _{is}\left(p_0 ,p\right)}={a_{si}p_s^0\over\sum\limits
_{l=1}^ma_{li}p_l^0 +\sum\limits _{l=m+1}^na_{li}p_l} ,\quad
s=\overline {1,m},\end{eqnarray*}
\begin{eqnarray*} \gamma _{is}\left(p_0 ,p\right)={a_{si}p_s\over\sum\limits
_{l=1}^ma_{li}p_l^0 +\sum\limits _{l=m+1}^na_{li}p_l} ,\quad
s=\overline {m+1,n},\end{eqnarray*}
\begin{eqnarray*} p^0 =\{p_1^0 ,\dots ,p_m^0 \},\quad p=\{p_{m+1},\dots
,p_n\}, \quad A=||a_{ij}||_{i,j=1}^n.\end{eqnarray*}
Write the condition of the economy equilibrium  in such  economy system introducing additional notations. Introduce the vector
\begin{eqnarray*} \{ b,p\}
=\{ b_1,\dots ,b_m,p_{m+1},\dots ,p_n\},\end{eqnarray*}  where $b_i,\ i=\overline
{1,m}, $ is volume of sales  of the $i$-th goods by
the $i$-th monopolist\index{volume of sales  of the $i$-th goods by
the $i$-th monopolist} and $p_i,\ i=\overline{m+1,n}$ is the price of the $i$-th goods of  non-monopolist.

 The demand on the $k$-th goods for demand vectors introduced is
\begin{eqnarray} \label{r5l2}
 \Phi _k^A(\{
b,p\})=\sum\limits _{i=1}^m\frac{a_{ki}b_ip_i^0}{\sum\limits
_{l=1}^ma_{li}p_l^0 +\sum\limits _{l=m+1}^na_{li}p_l}
\end{eqnarray}
\begin{eqnarray*} +\sum\limits _{i=m+1}^n{a_{ki}b_i^0
p_i\over\sum\limits _{l=1}^ma_{li}p_l^0 +\sum\limits
_{l=m+1}^na_{li}p_l},\quad k=\overline {1,n}.\end{eqnarray*}
Introduce the society demand vector
$\Phi ={\{\Phi _k^A(\{ b,p\} )\}
}_{k=1}^n. $

\begin{theorem} If matrix elements of the matrix  $A$  are strictly positive, then for all strictly positive vector of goods
$b^{0}=\{b_i^0\}_{i=m+1}^n$ there exists corresponding price vector
$\bar p^{0}=\{\bar p_i^0\}_{i=m+1}^n,$ $\bar
 p_i^0\geq 0,\ i=\overline{m+1,n},$ and a strictly positive vector of goods sales\index{vector of goods sales by monopolists} $\bar b^0=\{\bar b_i^0\}_{i=1}^m,$
~~ $\bar b_i^0>0,\quad i=\overline {1,m},$ by monopolists for which the economy system is in the Walras  equilibrium state,
i.e., the inequalities hold
 \begin{eqnarray*} \Phi _k^A\left(\{\bar b^0 ,\bar p^0\} \right)\leq \bar
b_k^0 ,\quad  k=\overline {1,m},\end{eqnarray*}
\begin{eqnarray} \label{r5l4}
 \Phi _k^A\left(\{\bar b^0 ,\bar
p^0\}\right)\leq b_k^0 ,\quad  k=\overline {m+1,n},
\end{eqnarray}
where
\begin{eqnarray*} \{\bar b^0 ,\bar p^0\} =\{\bar b_1^0 ,\dots ,\bar
b_m^0 ,\bar p_{m+1}^0 ,\dots ,\bar p_n^0\}.\end{eqnarray*}
\end{theorem}
\begin{proof}\smartqed  Let $B_1$  be  a certain  positive number. Consider on the set
\begin{eqnarray*} B=\left\{\{b,p\} \in
R_+^n,\quad \sum\limits _{i=1}^mb_ip_i^0 +\sum\limits
_{i=m+1}^nb_i^0 p_i=B_1\right\} \end{eqnarray*}
the sequence of operators
$\Gamma ^{\varepsilon _k}(\{ b,p\}
)={\{\Gamma _s^{\varepsilon _k}(\{ b,p\} )\} }_{s=1}^n, $ where
\begin{eqnarray*}  \Gamma _s^{\varepsilon _k}(\{b,p\} )=\frac{1}{
p_s^0 }\left\{ \sum\limits _{i=1}^m {(a_{si}p_s^0
+{\varepsilon _k/n})b_ip_i^0\over\sum\limits
_{l=1}^ma_{li}p_l^0 +\sum\limits
_{l=m+1}^na_{li}p_l+\varepsilon _k}\right. \end{eqnarray*}
\begin{eqnarray*} +\left.\sum\limits _{i=m+1}^n\frac{(a_{si}p_s^0+\varepsilon _k/n)
b_i^0 p_i}{\sum\limits _{l=1}^ma_{li}p_l^0
+\sum\limits _{l=m+1}^na_{li}p_l+\varepsilon _k} \right\},
\quad  s=\overline {1,m},\end{eqnarray*}
\begin{eqnarray*} \Gamma_s^{\varepsilon _k}(\{ b,p\} )=\frac{1}{
b_s^0}\left\{\sum\limits _{i=1}^m{(a_{si}p_s +
\varepsilon _k/n)b_ip_i^0\over\sum\limits
_{l=1}^ma_{li}p_l^0 +\sum\limits
_{l=m+1}^na_{li}{p_l+\varepsilon _k}}\right.  \end{eqnarray*}
\begin{eqnarray} \label{r5l5}
 +  \left.\sum\limits
_{i=m+1}^n\frac{(a_{si}p_s+\varepsilon _k/n )b_i^0
p_i}{\sum\limits _{l=1}^ma_{li}p_l^0 +\sum\limits
_{l=m+1}^na_{li}{p_l+\varepsilon _k}} \right\},\quad  s=\overline{m+1,n},
\end{eqnarray}
\begin{eqnarray*} 1>\varepsilon
_k>0,\quad\varepsilon _k\rightarrow 0.\end{eqnarray*}

For every fixed $1\leq k<\infty $ the operator $\Gamma
 ^{\varepsilon _k}(\{ b,p\} )$ maps $B$ into itself and is a continuous map on $B.$ As the set $B$ is convex and closed,
 by the Brouwer Theorem \cite{150},\index{ Brouwer Theorem} there exists a fixed point of this map $\{ b^{(k)}, p^{(k)}\} $ such that
\begin{eqnarray*} \Gamma _s^{\varepsilon _k}\left(\left\{ b^{(k)},p^{(k)}\right\}
\right)=b_s^{(k)}, \quad s=\overline{1,m},\end{eqnarray*}
\begin{eqnarray} \label{r5l6}
\Gamma _s^{\varepsilon _k}\left(\left\{ b^{(k)},p^{(k)}\right\} \right)=p_s^{(k)},
\quad s=\overline{m+1,n}.
\end{eqnarray}
 From the last equalities and the  assumptions of the Theorem, the inequalities follow \begin{eqnarray*} b_s^{(k)}\geq
{\min\limits _{i\in [1,n]}} a_{si}{B_1\over K}>0, \quad
s=\overline{1,m},\end{eqnarray*}  \begin{eqnarray} \label{r5l7} p_s^{(k)}\geq
{\varepsilon _k\over nKb_s^0} B_1>0,\quad s=\overline{m+1,n},
\end{eqnarray}  where \begin{eqnarray*} K= \max\limits _i\left(\sum\limits
_{l=1}^ma_{li}p_l^0 +\sum\limits _{l=m+1}^na_{li}{1\over
b_l^0 }B_1+1\right) .\end{eqnarray*}  Taking into account these inequalities and the set of equations (\ref{r5l6}), we obtain the set of inequalities
\begin{eqnarray*} \sum\limits _{i=1}^m{a_{si}b_i^{(k)}p_i^0\over\sum\limits
_{l=1}^ma_{li}p_l^0 +\sum\limits
_{l=m+1}^na_{li}p_l^{(k)}+\varepsilon _k} \end{eqnarray*}
\begin{eqnarray*} +\sum\limits _{i=m+1}^n\frac{a_{si}b_i^0
p_i^{(k)}}{\sum\limits _{l=1}^ma_{li}p_l^0+ \sum\limits
_{l=m+1}^na_{li}p_l^{(k)}+\varepsilon _k}\leq b_s^{(k)},\quad
s=\overline {1,m}, \end{eqnarray*}
\begin{eqnarray*} \sum\limits _{i=1}^m{a_{si}b_i^{(k)}p_i^0\over\sum\limits
_{l=1}^ma_{li}p_l^0 +\sum\limits
_{l=m+1}^na_{li}p_l^{(k)}+\varepsilon _k}+\sum\limits
_{i=m+1}^n\frac{a_{si}b_i^0 p_i^{(k)}}{\sum\limits
_{l=1}^ma_{li}p_l^0 +\sum\limits
_{l=m+1}^na_{li}p_l^{(k)}+\varepsilon _k}\leq b_s^0, \end{eqnarray*}
\centerline{$s=\overline{m+1,n}$.}
 As the sequence $d^{(k)}=\{ b^{(k)},
p^{(k)}\}$ belongs to the set $B, $ due to  simple reasons of compactness, there  exists  a subsequence
$d^{(k_p)}$ converging to
\begin{eqnarray*}
\{\bar b^{0}, \bar p^{0}\}=\{\bar b_1^0 ,\dots ,\bar
b_m^0 ,\bar p_{m+1}^0 ,\dots , \bar p_n^0\} \in B.
 \end{eqnarray*}
Thus, the vector $\{\bar b^{0}, \bar p^{0}\}$ is constructed and with
(\ref{r5l7}) we have
 \begin{eqnarray*} \bar b_s^0\geq {\min\limits_{i \in
[1,n]}a_{si}B_1\over K}>0,\quad s=\overline {1,m}.\end{eqnarray*}  It means that in (\ref{r5l4}) first $m$ inequalities are equalities because inequalities hold
$p_s^0 >0,\ s=\overline {1,m}.$ \qed\end{proof}

One can remove the requirement of strict positivity for the matrix $A$.

{\bf Corollary.} If \begin{eqnarray*}  \sum\limits_{l=1}^ma_{li}p_l^0 >0,
\quad \forall i=\overline{1,n},\end{eqnarray*}  then Theorem 1 is valid only without guarantee that the volume of  monopolistic sales  is strictly positive.

The next aim is to describe the necessary and sufficient conditions for strict positivity of solutions to the problem (\ref{r5l4}).

Let $\varepsilon $ be a diagonal matrix with elements $\varepsilon_{ii}=\varepsilon_{i}>0,
\ i=\overline{1,n},$ such that there exists positive inverse matrix to the matrix $E-A\varepsilon$ where $E$ is unit matrix and matrix   $\varepsilon $ are such that the spectral radius of the matrix\index{spectral radius of the matrix} $A\varepsilon$ is less than 1.
Denote matrix elements of the matrix  ${(E-A\varepsilon )}^{-1}$  by
$\alpha _{ki}^{-1}(\varepsilon ),\  k, i = \overline{1,n}.$
Let
\begin{eqnarray} \label{r5l10}
d_{si}(\varepsilon)=\sum\limits
 _{k=1}^n\alpha _{sk}^{-1} (\varepsilon )a_{ki},\quad
D(\varepsilon )=||d_{si}(\varepsilon ) ||_{s,i=1}^n.
 \end{eqnarray}
Note that if $A$ is an  indecomposable matrix,\index{indecomposable matrix} then $d_{si}(\varepsilon)>0,\ s,i=\overline{1,n}.$
Introduce the vector
 $ \{\delta^0,  b^0\}=\{\delta _{1}^0,\dots ,\delta _m^0, b_{m+1}^0,\dots
b_n^0\},$ where $\delta _i^0> 0, \ i=\overline{1,m}.$
 Consider the set of equations \begin{eqnarray*} \sum\limits
_{i=1}^m{d_{si}(\varepsilon )b_i\delta _i^0\over\sum\limits
_{l=1}^md_{li}(\varepsilon )\delta _l^0 +\sum\limits
_{l=m+1}^nd_{li}(\varepsilon )\delta _l} \end{eqnarray*}
\begin{eqnarray} \label{r5l11}
+\sum\limits _{i=m+1}^n{d_{si}(\varepsilon )b_i^0\delta
_i\over\sum\limits _{l=1}^md_{li}(\varepsilon )\delta _l^0
+\sum\limits _{l=m+1}^nd_{li}(\varepsilon )\delta _l}=b_s,\quad
s=\overline{1,m},
\end{eqnarray}
\begin{eqnarray*} \sum\limits _{i=1}^m \frac {d_{si}(\varepsilon )b_i\delta _i^0}{
  \sum\limits _{l=1}^m d_{li} (\varepsilon )\delta _l^0 +
  \sum\limits _{l=m+1}^n d_{li}(\varepsilon )\delta _l} \end{eqnarray*}  \begin{eqnarray*} +
  \sum\limits _{i=m+1}^n{d_{si}(\varepsilon )b_i^0\delta _i\over
  \sum\limits _{l=1}^md_{li}(\varepsilon )\delta _l^0
 +\sum\limits _{l=m+1}^nd_{li}(\varepsilon )\delta _l}=b_s^0 ,\quad
s=\overline {m+1,n},\end{eqnarray*}  with respect to the vector  $ \{b,\delta \}=\{b_1,\dots
b_m,\delta _{m+1},\dots ,\delta _n\}.$
\begin{theorem}\label{var1}
 Let $A$ be an indecomposable matrix,\index{ indecomposable matrix} the spectral radius of the matrix\index{spectral radius of the matrix} $A\varepsilon $ be less than 1 and let $D(\varepsilon )$ be a matrix built after the matrix $A$ by the formula (\ref{r5l10}). If for a certain vector
 \begin{eqnarray*} \{\delta^{0},b^{0}\}=\{\delta
_1^0 ,\dots ,\delta _m^0, b_{m+1}^0 ,\dots
,b_m^0 \}, \ \ \delta _i^0\geq 0, \ \ i=\overline
{1,m},\ \ b_i^0 >0,\ \ i=\overline {m+1,n}, \end{eqnarray*}
there exists a solution to the set of equations (\ref{r5l11}) relative to the vector $ \{b,\delta \}$ given by the vector
\begin{eqnarray*} \{\bar b^{0},\bar \delta^{0}\}= \left(\bar b_1^0 ,\dots, \bar
b_m^0 ,\bar\delta _{m+1}^0 ,\dots ,\bar\delta _n^0
\right) \in R_+^n,\end{eqnarray*}  then there exists a strictly positive solution $\{\bar b_1^0 ,\dots ,\bar b_m^0 ,\bar p_{m+1}^0 ,\dots ,\bar p_n^0 \}$= $\{\bar b^{0}, \bar p^{0}\}$ to the problem
\begin{eqnarray} \label{r5l12}
\Phi _k^A(p,b)=b_k,
\quad  k=\overline {1,m},\quad  \Phi _k^A(p,b)=b_k^0, \quad k=\overline
{m+1,n},
\end{eqnarray}
where
$\{ p^{0}, \bar p^{0}\}=\{p_1^0 ,\dots
,p_m^0 ,\bar p_{m+1}^0 , \dots ,\bar p_n^0 \}$
satisfies  the set of equations
\begin{eqnarray*} p_i^0 -\varepsilon _i\sum\limits
_{l=1}^ma_{li}p_l^0 -\varepsilon _i\sum\limits
_{l=m+1}^na_{li}\bar p_l^0 =\delta _i^0 ,\quad  i=\overline
{1,m},\end{eqnarray*}
\begin{eqnarray} \label{r5l13}
\bar p_i^0 -\varepsilon _i\sum\limits
_{l=1}^ma_{li}p_l^0 -\varepsilon _i\sum\limits
_{l=m+1}^na_{li}\bar p_l^0 =\bar\delta _i^0 ,\quad  i=\overline
{m+1,n}.
\end{eqnarray}
The map $\Phi_k^A(p,b)$ is built after the  vector of  goods supply\index{ vector of  goods supply}
$b^{0}=\{b_{m+1}^0 ,\dots ,b_n^0\} $ and
the vector of  monopolistic prices\index{vector of  monopolistic prices} $p^{0}=\{p_1^0 ,\dots ,p_m^0\} $ by the formula
 (\ref{r5l2}).  \end{theorem}
\begin{proof}\smartqed  We denote by $A^T$ the matrix transposed to the matrix $A$. From the existence of the inverse matrix
${(E-A \varepsilon)}^{-1}$, the fact that the solution to the problem (\ref{r5l13}) is given by the formula
\begin{eqnarray*} \{p^0,\bar p^0 \}={\left(E- \varepsilon A^T\right)}^{-1}
\{\delta^0, \bar \delta^0\} ,\end{eqnarray*}  where
\begin{eqnarray*} \{\delta^0, \bar
\delta^0\}=\{\delta _1^0 ,\dots ,\delta _m^0
,\bar\delta _{m+1}^0 ,\dots ,\bar\delta _n^0 \},
\{p^0,\bar p^0 \}=\{p_1^0 ,\dots ,p_m^0 ,\bar
p_{m+1}^0 ,\dots ,\bar p_n^0 \},\end{eqnarray*}  and also from the structure of the matrix $D(\varepsilon )$
and that the vector $ \{\bar b^0,\bar \delta^0 \}$ solves the set of equations
(\ref{r5l11}), it follows that \begin{eqnarray*} \{\bar b^{0},
\bar p^{0}\}=\{\bar b_1^0 ,\dots ,\bar b_m^0 ,~~\bar
p_{m+1}^0 ,\dots ,\bar p_n^0 \}\end{eqnarray*}  solves the problem
(\ref{r5l12}).
\qed\end{proof}

\begin{theorem}\label{var2}
Let a diagonal matrix\index{diagonal matrix} $\varepsilon $ be such that the spectral radius of the matrix\index{spectral radius of the matrix } $A \varepsilon $ be less than 1, the matrix $A$ be indecomposable, then to any non-zero vector
$\{\delta^{0},\bar \delta^{0}\}=(\delta_1^{0}
,\dots,\delta_m^{0} ,\bar\delta _{m+1}^{0} ,\dots
,\bar\delta _n^{0} ) \in R_+^n,$   that is rhs of the set of equations (\ref{r5l13}) there  corresponds  one and only one up to constant factor a strictly positive vector
\begin{eqnarray} \label{r6l1}
\{\bar b^0, b^0\}=\{\bar b_1^0,\dots ,\bar b_m^0,
b_{m+1}^0 ,\dots ,b_n^0\}
\end{eqnarray}
such that the problem (\ref{r5l12}) is solvable in the set of strictly positive vectors under the  vector  of  monopolistic prices\index{vector  of  monopolistic prices} $p^0=\{p_1^0 ,\ldots , p_m^0\},$ whose components are first $m$ components of the vector
\begin{eqnarray} \label{r6l2}
\{p^0, \bar p^0\}=\{p_1^0 ,\dots , p_m^0,  \bar
p_{m+1}^0 ,\ldots ,\bar p_n^0 \}
\end{eqnarray}
solving the problem (\ref{r5l13}) with  vector of  goods supply $b^{0}=\{b_{m+1}^0 ,\dots ,b_n^0\}.$
The solution to the problem (\ref{r5l12}) can be written as follows
\begin{eqnarray} \label{r6l3}
\{\bar b^0, \bar p^0\}=\{\bar b_1^0,\dots ,\bar
b_m^0, \bar p_{m+1}^0 ,\ldots ,\bar p_n^0 \},
\end{eqnarray}
where first $m$ components of the vector $\{\bar b^0, \bar p^0\} $
are first $m$ components of the vector $\{\bar b^0, b^0\}$
from (\ref{r6l1}) and the rest $n-m$ components of this vector are corresponding components of the vector $\{ p^0, \bar
p^0\}.$ The vector (\ref{r6l1}) solves the problem
 (\ref{r5l11}) for the vector
\begin{eqnarray*} \{\delta^0,
\delta\}=\{\delta_1^0,\dots,\delta_m^0, \delta
_{m+1} ,\dots ,\delta _n
\}=\{\delta_1^0,\dots,\delta_m^0, \bar\delta _{m+1}^0
,\dots ,\bar\delta _n^0 \}=\{\delta^0,
\bar \delta^0 \}.\end{eqnarray*}
\end{theorem}
\begin{proof}\smartqed
Establish the existence and uniqueness up to constant factor of a strictly positive solution to the problem (\ref{r5l11}) for the vector
$\{b, b^0\}$ under arbitrary non-zero vector $\{\delta^0, \delta \}=
\{\delta_1^0,\dots,\delta_m^0, \delta _{m+1} ,\dots
,\delta _n \}.$ First consider the case of the strictly positive vector
$\{\delta^0, \delta \}.$ Introduce in $R^n$ the linear operator given by the matrix
$H^{\varepsilon}\left(\{\delta^0, \delta
\}\right)=||H_{ki}^{\varepsilon}\left(\{\delta^0, \delta
\}\right)||_{k,i=1}^n$ with matrix elements \begin{eqnarray*}
H_{si}^{\varepsilon}\left(\{\delta^0, \delta
\}\right)={d_{si}(\varepsilon )\delta _i^0\over\sum\limits
_{l=1}^md_{li}(\varepsilon )\delta _l^0 +\sum\limits
_{l=m+1}^nd_{li}(\varepsilon )\delta _l}, \quad  s=\overline{1,n},\quad
i=\overline{1,m}, \end{eqnarray*}
\begin{eqnarray*}  H_{si}^{\varepsilon}\left(\{\delta^0,
\delta \}\right)=\frac{d_{si}(\varepsilon )\delta_i}{\sum\limits
_{l=1}^md_{li}(\varepsilon )\delta _l^0 +\sum\limits
_{l=m+1}^nd_{li}(\varepsilon )\delta _l}, \quad s=\overline{1,n},\quad
i=\overline{m+1,n}. \end{eqnarray*}
Then the set of equations (\ref{r5l11})
is rewritten as  follows \begin{eqnarray*} b_k=\sum\limits_{i=1}^m
H_{ki}^{\varepsilon}\left(\{\delta^0, \delta \}\right)b_i
+\sum\limits_{i=m+1}^n H_{ki}^{\varepsilon}\left(\{\delta^0, \delta
\}\right)b_i^0, \quad k=\overline{1,m} ,\end{eqnarray*}
\begin{eqnarray*} b_k^0=\sum\limits_{i=1}^m
H_{ki}^{\varepsilon}\left(\{\delta^0, \delta \}\right)b_i
+\sum\limits_{i=m+1}^n H_{ki}^{\varepsilon}\left(\{\delta^0, \delta
\}\right)b_i^0, \quad k=\overline{m+1,n} .\end{eqnarray*}
Or in operator form
\begin{eqnarray*} \{b, b^0\}= H^{\varepsilon}\left(\{\delta^0, \delta
\}\right)\{b, b^0\}, \quad \{b, b^0\}=\{b_1, \ldots, b_m,
b_{m+1}^0, \ldots, b_n^0\}.\end{eqnarray*}
Consider the closed bounded convex set\index{closed bounded convex set}
\begin{eqnarray*} B=\left\{\{b, b_0\}=
\{ b_1,\dots , b_m,
b_{m+1}^0 ,\dots ,b_n^0\} \in R_+^n, \ \sum\limits_{i=1}^m
\delta_i^0 b_i+\sum\limits_{i=m+1}^n
\delta_i b_i^0=C\right\},\end{eqnarray*}
where fixed number $C>0.$
The continuous operator $H^{\varepsilon}\left(\{\delta^0, \delta \}\right)$ maps
$B$ into itself. By the Brouwer Theorem \cite{150}\index{Brouwer Theorem} there exists fixed point\index{fixed point} of this map. As $d_{ki}(\varepsilon)>0,  \
k,i=\overline{1,n},$ fixed point
$\{\bar b^0, b^0\}=\{\bar b_1^0, \ldots, \bar
b_m^0, b_{m+1}^0, \ldots, b_n^0\}$ of this map is strictly positive  because
\begin{eqnarray*}  \bar b_k^0 \geq C \min\limits_{k,i}
\bar H_{ki}^{\varepsilon}\left(\{\delta^0, \delta \}\right), \
k=\overline{1,m}, \ b_k^0 \geq C \min\limits_{k,i}
\bar H_{ki}^{\varepsilon}\left(\{\delta^0, \delta \}\right),
\quad  k=\overline{m+1,n},\end{eqnarray*}
where
\begin{eqnarray*} \bar H_{ki}^{\varepsilon}\left(\{\delta^0, \delta \}\right)=\frac{
H_{ki}^{\varepsilon}\left(\{\delta^0, \delta \}\right)}{\delta_i^0},
\quad k=\overline{1,n}, \quad i=\overline{1,m},    \end{eqnarray*}
\begin{eqnarray*} \bar H_{ki}^{\varepsilon}\left(\{\delta^0, \delta \}\right)=\frac{
H_{ki}^{\varepsilon}\left(\{\delta^0, \delta \}\right)}{\delta_i},
\quad k=\overline{1,n}, \quad i=\overline{m+1,n}.\end{eqnarray*}
Build this solution. Introduce the matrix $K=||
H_{ki}^{\varepsilon}\left(\{\delta^0, \delta \}\right)||_{k,i=1}^{n-1},$
build the vector-column \begin{eqnarray*} u=\left\{H_{1 n}^{\varepsilon}\left(\{\delta^0,
\delta \}\right), \ldots, H_{(n-1),  n}^{\varepsilon}\left(\{\delta^0, \delta
\}\right)\right\}\end{eqnarray*}  and the vector-column \begin{eqnarray*} v=\left\{\frac{\bar b_1^0}{b_n^0},
\ldots, \frac{\bar b_m^0}{b_n^0}, \frac{
b_{m+1}^0}{b_n^0} , \ldots, \frac{
b_{n-1}^0}{b_n^0}\right\}\end{eqnarray*}
after components of the fixed point of operator $H^{\varepsilon}\left(\{\delta^0, \delta \}\right).$

Then the vector $v$ is strictly positive  and satisfies  the set of equations
\begin{eqnarray} \label{d6l4}
v=Kv+u.
\end{eqnarray}
Show that the spectral radius of the operator\index{spectral radius of the operator} $K$ is strictly less than 1. Introduce the norm into the space $R^{n-1}$ by the formula
$|x|=\max\limits_{i}\frac{|x_i|}{v_i}.$ Then the estimate holds
\begin{eqnarray*} |Kx| \leq \max\limits_{i}\frac{[Kv]_i}{v_i}|x| =
\max\limits_{i}\left(1-\frac{u_i}{v_i}\right)|x|.\end{eqnarray*}
 As $u_i< v_i,\  i=\overline{1,n-1},$ we have
$\max\limits_{i}(1-\frac{|u_i|}{v_i})<1,$ and, therefore,
$|K|<1.$

Introduce the vector
\begin{eqnarray*} t=\{t_1, \ldots, t_{n-1}\}=\sum\limits_{n=0}^{\infty}K^nu.\end{eqnarray*}
This series converges because the spectral radius of the operator $K$
is less than 1, as it is just proven. Then the vector that is a fixed point of the operator\index{fixed point of the operator} $H^{\varepsilon}\left(\{\delta^0,
\delta \}\right)$ can be given as follows \begin{eqnarray*} \{\bar b^0,
b^0\}=\lambda\{t_1, \ldots, t_m, t_{m+1},\ldots, t_{n-1}, 1\},
\quad  \lambda>0. \end{eqnarray*}
We consider only non-negative $\lambda.$
Let now the vector $\{\delta^0, \delta \}$
be not strictly positive. Denote by $I$ the set of indices for which  components of this vector are non-zero  and assume this set is non-empty. Let $J$ be the set of indices of zero components. Consider the set of equations \begin{eqnarray*} \sum\limits_{i \in
I}H_{ki}^{\varepsilon}\left(\{\delta^0, \delta \}\right)x_i=x_k , \quad k
\in I. \end{eqnarray*}  This set of equations  has the same structure as just considered  set of equations. Then according to proven, there exists a strictly positive vector solving this set of equations.
 Build after this vector the vector $\phi_{0}=\{\phi_1,
\ldots, \phi_n \},$ putting $\phi_i=x_i, \ i \in I, \quad
\phi_i=0, \ i \in J,$ where $x_i$ is the $i$-th component of the strictly positive solution to above considered set of equations.
Then the vector \begin{eqnarray*} \{\bar b^0,
b^0\}=H^{\varepsilon}\left(\{\delta^0, \delta \}\right)\phi_0\end{eqnarray*}
is a strictly positive solution to the problem (\ref{r5l11}).
We finish the Proof referring onto the Theorem \ref{var1} if for the vector $\{\delta^0, \delta \}$
to take the vector $\{\delta^0, \bar \delta^0 \}.$
\qed\end{proof}

\begin{theorem}\label{eq1} Let a matrix $A$ be indecomposable. The necessary and sufficient conditions for the existence of a strictly positive solution to the set of equations
(\ref{r5l12}) for the strictly positive vector of    monopolistic prices\index{vector of    monopolistic prices} $p^0=\{p_1^0
,\dots , p_m^0\}$
are the existence of a diagonal matrix $\varepsilon^0$
with strictly positive diagonal matrix elements and the  spectral radius of the matrix\index{spectral radius of the matrix} $A\varepsilon^0 $
less than 1 and a strictly positive vector $\delta^0=\{\delta_1^0,\dots,\delta_m^0\}, \ 0<  \delta_i^0 < p_i^0, \ i=\overline{1,m},$
 such that for  components of goods supply vector\index{components of goods supply vector}
$b^0=\{ b_{m+1}^0 ,\ldots , b_n^0 \}$ the representation holds
\begin{eqnarray} \label{r6l4} b_s^0=\sum\limits
_{i=1}^m{d_{si}\left(\varepsilon^0\right)\bar b_i^0\delta
_i^0\over\sum\limits _{l=1}^md_{li}\left(\varepsilon^0 \right)\delta
_l^0}, \quad s=\overline{m+1,n}, \end{eqnarray}  where the vector $
\bar b_0=\{\bar b_i^0 \}_{i=1}^m$ is a  strictly positive solution to the set of equations \begin{eqnarray} \label{r6l11} \bar
b_s^0=\sum\limits _{i=1}^m{d_{si}\left(\varepsilon^0\right)\bar
b_i^0\delta _i^0\over\sum\limits
_{l=1}^md_{li}\left(\varepsilon^0 \right)\delta _l^0}, \quad
s=\overline{1,m}.
\end{eqnarray}
\end{theorem}
\begin{proof}\smartqed
  Necessity. Let there exist a strictly positive solution
\begin{eqnarray*}
\{ \bar b^0,  \bar p^0\}=\{\bar b_1^0, \ldots,  \bar
b_m^0, \bar p_{m+1}^0, \ldots,  \bar
p_n^0 \}  \end{eqnarray*}
 to the problem (\ref{r5l12}). Consider the vector $\{p^0, \bar p^0\}=\{ p_1^0, \ldots,
p_m^0, \bar p_{m+1}^0, \ldots,  \bar
p_n^0 \}.$ Show that a diagonal matrix\index{ diagonal matrix}
$\varepsilon_0$
exists such that the vector $\{p^0, \bar p^0\}$ solves the set of equations
\begin{eqnarray*} p_i^0 -\varepsilon_i^0\sum\limits _{l=1}^ma_{li}p_l^0
-\varepsilon_i^0\sum\limits _{l=m+1}^na_{li}\bar p_l^0
=\delta _i^0 ,\quad  i=\overline {1,m},\end{eqnarray*}
\begin{eqnarray} \label{r6l6}
\bar p_i^0
-\varepsilon_i^0\sum\limits _{l=1}^ma_{li}p_l^0
-\varepsilon_i^0\sum\limits _{l=m+1}^na_{li}\bar p_l^0
=0 ,\quad  i=\overline {m+1,n}.
\end{eqnarray}
Take a certain strictly positive vector $\delta^0=\{\delta_1^0,\dots,\delta_m^0\}$ whose components satisfy  inequalities $ 0<  \delta_i^0 < p_i^0, \ i=\overline{1,m}.$
Determine components of a matrix $\varepsilon_0$  from the set of equations
\begin{eqnarray*} \delta_i^0=p_i^0-\varepsilon_i^0\left(\sum\limits_{l=1}^ma_{li}p_l^0+
\sum\limits_{l=m+1}^na_{li}\bar p_l^0\right), \quad
i=\overline{1,m},\end{eqnarray*}
\begin{eqnarray*} \bar
p_i^0=\varepsilon_i^0\left(\sum\limits_{l=1}^ma_{li}p_l^0+
\sum\limits_{l=m+1}^na_{li}\bar p_l^0\right), \quad
i=\overline{m+1,n}.\end{eqnarray*}
From indecomposability of the matrix $A$, it has no zero columns, therefore
\begin{eqnarray*} \sum\limits_{l=1}^ma_{li}p_l^0+
\sum\limits_{l=m+1}^na_{li}\bar p_l^0 > 0, \quad
i=\overline{1,n}.\end{eqnarray*}
Put
\begin{eqnarray*} \varepsilon_i^0=\frac{\bar p_i^0 - \delta _i^0}{\sum\limits_{l=1}^ma_{li}p_l^0+
\sum\limits_{l=m+1}^na_{li}\bar p_l^0}, \quad
i=\overline{1,m},\end{eqnarray*}
\begin{eqnarray*} \varepsilon_i^0=\frac{\bar p_i^0}{\sum\limits_{l=1}^ma_{li}p_l^0+
\sum\limits_{l=m+1}^na_{li}\bar p_l^0}, \quad
i=\overline{m+1,n}.\end{eqnarray*}
Show that the matrix
$\varepsilon_{0}=||\delta_{ij}\varepsilon_i^{0}||_{i,j=1}^n,$~
where $\varepsilon_i^0, \ i=\overline{1,n},$ are just built numbers, is such that the spectral radius of the matrix\index{spectral radius of the matrix}
$A\varepsilon_{0}$ is less than 1. Strictly positive vector-column $\{p^{0}, \bar p^{0} \}$ satisfies  the set of equations \begin{eqnarray*} \{p^{0}, \bar p^{0}\}
=\varepsilon_{0} A^T \{p^{0}, \bar p^{0} \}
+\{\delta^{0},0\},\end{eqnarray*}  where the vector-column
$\{\delta^{0},0\}=\{\delta_1^{0}, \ldots,
\delta_m^{0}, 0, \ldots , 0\}.$ Therefore, the vector-column
$\{p^0, \bar p^0\}$ satisfies also the set of equations
\begin{eqnarray} \label{r6l10} \{p^{0}, \bar
p^{0} \}=\left(\varepsilon_{0} A^T\right)^n \{p^{0}, \bar
p^{0} \}+\sum\limits_{k=0}^{n-1}\left(\varepsilon_{0} A^T\right)^k
\{\delta^{0}, 0\}.
\end{eqnarray}
 From the indecomposability of $A$, it follows the indecomposability of $\varepsilon_{0}A^T,$ therefore, the vector
$\sum\limits_{k=0}^{n-1}\left(\varepsilon_0 A^T\right)^k
\{\delta^0,0\}$ is strictly positive.
It is obvious that components of the vector
$\left(\varepsilon_{0} A^T\right)^n \{p^{0}, \bar p^{0} \}$ are strictly positive  because the matrix $(\varepsilon_{0} A^T)^n$ has no zero rows or columns and the vector $\{p^{0}, \bar p^{0} \}$ is strictly positive.
From here, with the same arguments as in the Proof of the Theorem \ref{var2},
it follows that the spectral radius of the matrix\index{spectral radius of the matrix} $(\varepsilon_{0}
A^T)^n $ is strictly less than 1. From here, the spectral radius of the matrix
$A\varepsilon_{0} $ is less than 1.
To obtain the representation for the  vector of  goods supply the Theorem states, it is sufficient to remember that by the Theorem \ref{var2} the vector
$\{\bar b^0, b^0\}=\{\bar b_1^0, \ldots,  \bar
b_m^0, b_{m+1}^0, \ldots, b_n^0\}$ corresponding to rhs of the set of equations (\ref{r6l6}) and guaranteeing the existence of a strictly positive solution to the problem (\ref{r5l12})
must satisfy the set of equations (\ref{r5l11}). However,
(\ref{r6l4}) and (\ref{r6l11}) mean just that.

 Sufficiency. Let such matrix $\varepsilon_0$
exist that representations (\ref{r6l4}) and (\ref{r6l11}) hold,
then by the Theorem \ref{var1}, there exists a strictly positive solution to the set of equations (\ref{r5l12}). \qed\end{proof}

The next Theorem gives the algorithm to build strictly positive solutions to the set of equations (\ref{r5l12}).
\begin{theorem}\label{eq3}
 Let the matrix $A$ be indecomposable  and  a diagonal matrix $\varepsilon^0$ with strictly positive elements be such that the spectral  radius of the matrix $A\varepsilon^0 $ be less than 1, the set of equations
\begin{eqnarray*} \bar p_i^0=\varepsilon_i^0\left(\sum\limits_{l=1}^ma_{li}p_l^0+
\sum\limits_{l=m+1}^na_{li}\bar p_l^0\right), \quad
i=\overline{m+1,n},\end{eqnarray*}
for the vector $\bar p^0 = \{\bar p_i^0\}_{i=m+1}^n$
has a strictly positive solution satisfying conditions
\begin{eqnarray*} p_i^0-\varepsilon_i^0\left(\sum\limits_{l=1}^ma_{li}p_l^0+
\sum\limits_{l=m+1}^na_{li}\bar p_l^0\right) > 0, \quad
i=\overline{1,m}.\end{eqnarray*}
The sufficient condition of the existence of a strictly positive solution to the set of equations
(\ref{r5l12}) for a strictly positive vector of  monopolistic prices\index{vector of  monopolistic prices}  $p^0=\{p_1^0,\dots , p_m^0\}, \ p_i^0>0, \ i=\overline{1,m}, $
and a non-monopolistic vector of  initial goods supply\index{non-monopolistic vector of  initial goods supply}   $b^0=\{ b_{m+1}^0 ,\ldots , b_n^0 \}$ is the validity of the representation
\begin{eqnarray} \label{eq4}
 b_s^0=\sum\limits
_{i=1}^m{d_{si}\left(\varepsilon^0\right)\bar b_i^0\delta
_i^0\over\sum\limits _{l=1}^md_{li}\left(\varepsilon^0 \right)\delta
_l^0}, \quad s=\overline{m+1,n},
\end{eqnarray}
where the vector $
\bar b_0=\{\bar b_i^0 \}_{i=1}^m$ is a strictly positive solution to the set of equations
\begin{eqnarray} \label{eq5} \bar
b_s^0=\sum\limits _{i=1}^m{d_{si}\left(\varepsilon^0\right)\bar
b_i^0\delta _i^0\over\sum\limits
_{l=1}^md_{li}\left(\varepsilon^0 \right)\delta _l^0}, \quad
s=\overline{1,m},
\end{eqnarray}
for the vector $\delta^0=\{\delta_1^0,\dots,\delta_m^0\}$
with components
\begin{eqnarray*} \delta_i^0=p_i^0-\varepsilon_i^0\left(\sum\limits_{l=1}^ma_{li}p_l^0+
\sum\limits_{l=m+1}^na_{li}\bar p_l^0\right), \quad
i=\overline{1,m}.\end{eqnarray*}
The solution to the set of equations (\ref{r5l12}) is the vector $\{\bar b_0,\bar p^0\},$ where $\bar b_0=\{\bar b_i^0 \}_{i=1}^m,$  $\bar p^0 = \{\bar p_i^0\}_{i=m+1}^n.$
\end{theorem}

Give alternative the  necessary and sufficient conditions for the existence of a strictly positive solution to the set of equations
(\ref{r5l12}).

\begin{theorem} Let $A$ be an  indecomposable matrix,\index{ indecomposable matrix} there exists
$A^{-1},$ then between vectors
$\{b,b^0\}=\{b_1,\ldots,b_m,b_{m+1}^0,\ldots,
b_n^0\}$ belonging to the interior of positive cone created by vector-columns  of the  matrix\index{interior of positive cone created by vector-columns  of the  matrix} $A$  and to the set
\begin{eqnarray*} B_2=\left\{\{b, b_0\}, ~b_i>0, ~i=\overline{1,m},
~b_i^0>0 , ~i=\overline{m+1,n}, ~\sum\limits_{i=1}^mb_i+
 \sum\limits_{i=m+1}^mb_i^0=1\right\}\end{eqnarray*}
and  vectors $\{p^0,p\}=\{p_1^0, \ldots ,p_m^0,
p_{m+1},\ldots, p_n\} $ belonging to the interior of the set
\begin{eqnarray*} \bar P= \left\{\{p^0, p\}, p_i^0
>0,~i=\overline{1,m},~p_i>0,
~i=\overline{m+1,n},~\sum\limits_{i=1}^mp_i^0+\sum\limits_{i=m+1}^mp_i=1
\right\}\end{eqnarray*}
one-to-one correspondence exists.
 \end{theorem}
\begin{proof}\smartqed  Let the vector $\{b, b^0\}$ belong to the interior of the cone created by columns of the matrix $A.$
Then there exists a  unique vector
$y=\{y_i\}_{i=1}^n ,$ $ y_i>0,$ $
i=\overline{1,n},$ such that
\begin{eqnarray} \label{r5l14}
b_k=\sum\limits_{i=1}^n a_{ki} y_i ,\quad k=\overline{1,m},\quad
b_k^0=\sum\limits_{i=1}^n a_{ki} y_i ,\quad  k=\overline{m+1,n},
\end{eqnarray}
where  $ b_k>0,\ k=\overline{1,m},\ b_k^0>0, \
k=\overline{m+1,n},$ because $A$ is an indecomposable matrix\index{indecomposable matrix} and the unique vector $y$ is strictly positive. Put
\begin{eqnarray*} z_i=\frac{y_i}{\sum\limits_{l=1}^n
a_{il} y_l} ,\quad i=\overline{1,n}.\end{eqnarray*}
The vector $z=\{ z_i\}_{i=1}^n$
is determined by the vector $y=\{y_i\} _{i=1}^n$ unambiguously. The set of equalities (\ref{r5l14}) can be given as follows
\begin{eqnarray*}
b_k&=&\sum\limits_{i=1}^m a_{ki}b_iz_i+\sum\limits_{i=m+1}^n a_{ki}b_i
^0 z_i ,\quad k=\overline{1,m},
\end{eqnarray*}
\begin{eqnarray}\label{r5l15}
b_k^0&=&\sum\limits_{i=1}^m a_{ki}b_iz_i+\sum\limits_{i=m+1}^n a_{ki}b_i
^0 z_i ,\quad k=\overline{m+1,n}.
\end{eqnarray}
 As $a_{ij}\geq 0,$  $z_i>0,$  $i=\overline{1,n},$ the vector
$\{ p_1^0 ,\ldots ,p_m^0,p_{m+1},\ldots ,p_n\}$ determined from the set of equations
\begin{eqnarray}\nonumber \frac{p_i^0}{\sum\limits_{l=1}^n a_{li}p_l^0 +
\sum\limits_{l=m+1}^na_{li}p_l}&=&z_i ,\quad i=\overline{1,m} ,
\\\label{r5l16}&&\\\nonumber \frac{ p_i}{\sum\limits_{l=1}^m
a_{li}p_l^0 + \sum\limits_{l=m+1}^na_{li}p_l}&=&z_i ,\quad
i=\overline{m+1,n},
\end{eqnarray}
exists because the problem (\ref{r5l16})
is solvable due to the  solvability of conjugate problem\index{solvability of conjugate problem}
(\ref{r5l15}). Therefore, a strictly positive solution  to the set of equations (\ref{r5l16}) exists determined up to constant $\lambda>0.$
Take $\lambda$ from the condition
\begin{eqnarray*} \sum\limits_{i=1}^m p_i^0 +
\sum\limits_{i=m+1}^n p_i=1.\end{eqnarray*}
In view of uniqueness of $\lambda$, to every vector $\{ b,b^0\} $ satisfying the conditions of the   Theorem a  single vector $\{ p^0, p\}\in \bar P$ corresponds.

Inversely, let
$\{ p_1^0,\ldots, p_m^0, p_{m+1},\ldots
,p_n\} $ $\in$ $ \bar P$ be the vector having all the components positive. Build the vector $z=\{z_i\}_{i=1}^n$ with the formula (\ref{r5l16}). Then a strictly positive solution
$\{b_1,\ldots,b_m, b_{m+1}^0,\ldots, b_n^0\}$ to the problem
(\ref{r5l15})   exists that is  determined up to constant $\lambda>0.$ Take such $\lambda > 0$ that the condition $\sum\limits_{i=1}^m b_i + \sum\limits_{i=m+1}^n b_i^0
=1 $ holds. \qed\end{proof}

Give now alternative necessary and sufficient conditions of the existence for positive solutions.

We build after the matrix $D(\varepsilon)=||d_{ij}(\varepsilon)||_{i,j=1}^n $
rectangular matrix\index{rectangular matrix} $\hat
D(\varepsilon)=||d_{ij}(\varepsilon) ||_{i=m+1,j=1}^n. $ Assume that the vector $b^0=\{b_{m+1}^0, \ldots,b_n^0 \}$ belongs to the interior of the cone created by columns of the matrix\index{interior of the cone created by columns of the matrix}
 $\hat D(\varepsilon). $ If the rank of the  matrix $\hat D(\varepsilon) $  equals
$r\leq n-m$, then there exists manyfold of positive solutions to the set of equations
\begin{eqnarray} \label{r5l17}
\sum\limits_{j=1}^n d_{ij}(\varepsilon)y_j =b_i^0.
\end{eqnarray}
We have described these solutions in the Chapter 6 (see, for example, the Theorems \ref{jant39}, \ref{allupotka4}).
Denote this manyfold by $ \Gamma(\varepsilon). $

After the vector $b^0$, we build the set of vectors  in $R^n_+$
\begin{eqnarray*} \left(b,b^0\right)=\{b_1,\ldots,b_m,b_{m+1}^0,\ldots,b_n^0\} \end{eqnarray*}
supposing
\begin{eqnarray} \label{r5l19}
\sum\limits_{j=1}^n d_{ij}(\varepsilon)y_j =b_i,\quad i=\overline{1,m},
\quad y \in \Gamma (\varepsilon).
\end{eqnarray}
We denote  aggregate of vectors $\left(b,b^0\right)$ for
$y$ $\in$ $\Gamma(\varepsilon)$ by $F(\varepsilon).$
The dimension of  manyfold of  vectors\index{dimension of  manyfold of  vectors} $F(\varepsilon)$ depends on the dimension of  manyfold\index{dimension of  manyfold} $\Gamma(\varepsilon).$
Build the image of this manyfold by the next algorithm.
To every vector
$\{b,b^0\} \in F(\varepsilon)$ built after the vector $y=\{y_i\}_{i=1}^n$ we put into  correspondence  the vector $z=\{z_i\}_{i=1}^n,$ where
\begin{eqnarray} \label{r5l20} z_i=\frac{y_i}{\sum\limits_{l=1}^n
d_{il}(\varepsilon ) y_l},\quad i=\overline{1,n}.  \end{eqnarray}
Determine the vector
$\{\delta_1^0,\ldots,\delta_m^0,\delta_{m+1},
\ldots,\delta_n\}$ from the set of equations
\begin{eqnarray*}
\frac{ \delta_i^0}{\sum\limits_{l=1}^n d_{li}(\varepsilon)\delta_l^0 +
\sum\limits_{l=m+1}^nd_{li}(\varepsilon)\delta_l}&=&z_i ,
\quad i=\overline{1,m},
\end{eqnarray*}
 \begin{eqnarray}\label{r5l21}
\frac{ \delta_i}{\sum\limits_{l=1}^m d_{li}(\varepsilon)\delta_l^0 +
\sum\limits_{l=m+1}^nd_{li}(\varepsilon)\delta_l}&=&z_i ,\quad i=\overline{m+1,n}.
\end{eqnarray}
Finally, solve the set of equations
\begin{eqnarray}\nonumber
p_i^0-\varepsilon_i \left\{\sum\limits_{l=1}^ma_{li}p_l^0+
\sum\limits_{l=m+1}^na_{li}p_l\right\}&=&\delta_i^0 ,\quad i=\overline{1,m}
 ,\\\label{r5l22}&&\\\nonumber
p_i-\varepsilon_i \left\{\sum\limits_{l=1}^ma_{li}p_l^0+
\sum\limits_{l=m+1}^na_{li}p_l\right\}&=&\delta_i ,\quad i=\overline{m+1,n},
\end{eqnarray}
for the vector $\{p_1^0,\ldots, p_m^0,
p_{m+1},\ldots, p_n\} =\{p^0,p\}.$ Now build the needed  map
$G$. To every vector $\{b, b^0\}$ $\in$
$F(\varepsilon)$ we put into  correspondence the vector $G(\{b,
 b_0\})=\{p^0,p\}$ solving the set of equations
(\ref{r5l22}). We normalize this solution by the condition \begin{eqnarray*} \sum\limits_{i=1}^m p_i^0 b_i+\sum\limits_{i=m+1}^n
p_ib_i^0=B_1\end{eqnarray*}
 because the solution to the set of equations
(\ref{r5l21}) is determined up to positive constant.
  Put \begin{eqnarray*}
G(F(\varepsilon))=\bigcup\limits_{\{b,b^0\} \in
F(\varepsilon)} G\left(\{b,b^0\}\right) ,\quad
G=\bigcup\limits_{\varepsilon>0} G(F(\varepsilon)).\end{eqnarray*}  The union in the last formula is made over those matrices
 $\varepsilon>0 $ for which the vector
$b^0$ belongs to the interior of the cone created by vector-columns of the matrix\index{ interior of the cone created by vector-columns of the matrix}
 $\hat D( \varepsilon).$
\begin{theorem}
Let a matrix $A$ be indecomposable, the inverse matrix $A^{-1}$ exist, the vector
 $b^0=\{b_{m+1}^0,\ldots,b_n^0\}$ belong to the interior of positive cone created by vector-columns of the matrix
 $\hat D(\varepsilon)$ for a certain  matrix $\varepsilon$  such that the spectral radius of the matrix $A
\varepsilon$ is less than 1. The set of equations
\begin{eqnarray} \label{r5l23}
\Phi _k^A(\{ b,p\})= b_k, \quad k=\overline {1,m}, \quad \Phi _k^A(\{ b,p\} )= b_k^0
,\quad  k=\overline {m+1,n},
\end{eqnarray}  has a solution in the set of strictly positive vectors  if and only if the set
$T\bigcap G$ is non-empty, where \begin{eqnarray*} T=\{\{p_1^0,
\ldots,p_m^0,p_{m+1},\ldots,p_n\}, p_i>0, \
i=\overline{m+1,n}\}.\end{eqnarray*}
\end{theorem} \begin{proof}\smartqed   Necessity. Let  a strictly positive solution $\{\bar b,\bar p^0\}$
to the problem (\ref{r5l23}) exist. Then the vector $\{\bar b, b^0\} $
belongs to the interior of the cone created by vector-columns of the matrix\index{interior of the cone created by vector-columns of the matrix}
  $A.$  From the inequalities
 \begin{eqnarray*} p_i^0 >0, \quad   i=\overline{1,m},
 \quad \bar p_i^0>0,\quad  i=\overline{m+1,n},\end{eqnarray*}
 where
 $\{\bar b, \bar p^0\}$ solves the set of equations
(\ref{r5l23}), it follows the existence of the  set of positive numbers
 $\varepsilon_i$,  $i=\overline
{1,n},$ and a strictly positive vector
$\{\delta_1^0,\ldots,\delta_m^0, \bar \delta_{m+1}^0
,\ldots ,\bar \delta_n^0\}$ such  that  $\{p_1^0
 ,\ldots,  p_m^0,  \bar p_{m+1}^0, \ldots,\bar
 p_n^0\}$ satisfies  the set of equations (\ref{r5l13}). From
(\ref{r5l13}) and (\ref{r5l23}), it follows that the vector
$\{\delta_1^0,\ldots,\delta_m^0, \bar \delta_{m+1}^0
,\ldots ,\bar \delta_n^0\}$ satisfies the set of equations
 (\ref{r5l11}). It means that the vector $b^0$ belongs to the interior of the cone created by vector-columns of the matrix $\hat
D(\varepsilon)$ for the matrix $\varepsilon$ built after mentioned set $\varepsilon_i$.
Hence, $T\bigcap G$ is non-empty set.
Sufficiency is obvious from the construction proposed before this Theorem.
\qed\end{proof}

\subsection{Monopolistic influence of external environment}

In this Subsection, we show that one can reduce external influence onto economy system\index{external influence onto economy system} to the problem with internal monopolists.

Consider an economy system interacting with its economy environment. Such interaction shows itself through the trade with foreign economic agents.\index{trade with foreign economic agents}
Suppose the  prices of  goods by which the economy system trades with its foreign economic agents are known. They can be prices of the world market or agreement ones. Prices within the economy system are formed by the demand and the supply.

As the model can be applied to real prognosis equilibrium prices within the economy system,
suppose the Leontieff matrix determines technological map.
Suppose that the economy system consumes $n$ kinds of goods producing $(n - t) $ of them within the economy system and importing $t$ kinds of goods.
It is convenient to suppose that goods imported into the economy  system\index{goods imported into the economy  system} are numbered from $1$ to $t$ and goods produced within the economy system\index{goods produced within the economy system} are numbered from $t+1$ to $n.$
Therefore, we suppose  the economy system has $n - t$ inside industries each of which produces a single kind of goods. We describe the economy system production by technological matrix\index{technological matrix}
${||a_{ki}||}_{k=1,i=t+1}^{n,n}.$
Assume that the economy system has $l$ consumers each of which has goods vector $b_i=\{b_{ki}\}_{k=1}^n$ and $b_{ki}=0, \ k=\overline{1,t},\  i=\overline{1,l}.$
The last means that domestic economic agents\index{domestic economic agents} have no goods imported into the system.
Let, further, $\pi= \{\pi_i\}_{i={t+1}}^n$ be a  vector of industries taxation.\index{vector of industries taxation}
 We give the  $j$-th, $j=\overline {t+1,n},$ industry income by the formula
\begin{eqnarray} \label{r5l24}
D_j(p)=\pi_jx_j\left[\left(p_j-\sum\limits _{s=1}^na_{sj}p_s\right)+ <b_i,p>\right],
\end{eqnarray}
where $p=(p_1,\dots ,p_n)$ is the price vector; $p_i,~i=\overline
{1,t},$ is the price for the $i$-th goods imported into the economy system;\index{price for goods imported into the economy system} $p_i,~i=\overline {t+1,n},$ is the price of
the $i$-th goods produced within the economy  system;\index{price of
the $i$-th goods produced within the economy  system} $x_j$
is the $j$-th industry output.
It is convenient to suppose that foreign economic agents\index{foreign economic agents} buy goods produced by industries numbered from $t+1$ to $s< n.$
Therefore, if vectors  $C_i=\{c_{ki}\}_{k=1}^n, \ i=\overline{1,l}$ are fields of information evaluation  by $l$ consumers,\index{ fields of information evaluation  by  consumers} we suppose that first $t$ fields of information evaluation by consumers describe choice  of foreign agents\index{choice  of foreign agents} and, according to assumptions made $c_{ki}=0$  for $ \ k=\overline{1,t}, \ k=\overline{s+1,n}, \ i=\overline{1,t}.$ We suppose every consumer is insatiable.

Assume conditions of the external trade balance\index{external trade balance}  hold.
To write conveniently trade balance conditions,\index{trade balance conditions} we treat components of fields of information evaluation  by consumer-importers\index{fields of information evaluation  by consumer-importers} as: $c_{mj}$ is the number of units of the $m$-th goods,
$m=\overline{t+1,s},$ exchanged with one unit of
the $j$-th goods, $j=\overline{1,t}.$

Within such treatment of components of fields of information evaluation  by consumer-importers\index{ fields of information evaluation  by consumer-importers }  we write conditions of  trade balance\index{conditions of  trade balance}  as follows
\begin{eqnarray} \label{r5l26}
\sum\limits _{m=t+1}^sc_{mj}p_m=p_j,\quad j=\overline {1,t}.
\end{eqnarray}
If $d=\{d_i\}_{i=1}^n$ is a vector of import,\index{vector of import} then it is obvious that $d_i=0, \ i=\overline {t+1,n}.$

The condition that  the demand is equal to the supply is the set of equations of the economic equilibrium
\begin{eqnarray*} \sum\limits _{j=t+1}^l\frac{c_{kj}D_j(p)}{\sum\limits
_{s=1}^nc_{sj}p_s}=d_k - \sum\limits _{i=t+1}^na_{ki}x_i, \quad 1\leq k\leq t, \end{eqnarray*}
\begin{eqnarray} \label{r5l27}
\sum\limits _{j=1}^t\frac{c_{kj}d_jp_j}{\sum\limits
_{m=t+1}^sc_{mj}p_m}
\end{eqnarray}
\begin{eqnarray*} +\sum\limits
_{j=t+1}^l\frac{c_{kj}D_j(p)}{\sum\limits
_{s=1}^nc_{sj}p_s}=x_k-\sum\limits _{i=t+1}^na_{ki}x_i + \sum\limits_{i=1}^l b_{ki},\quad
t+1\leq k\leq s, \end{eqnarray*}
\begin{eqnarray*} \sum\limits _{j=t+1}^l\frac{c_{kj}D_j(p)}{\sum\limits
_{s=1}^nc_{sj}p_s}=x_k-\sum\limits _{i=t+1}^na_{ki}x_i + \sum\limits_{i=1}^l b_{ki}, \quad
s+1\leq k\leq n.\end{eqnarray*}

If the conditions of the  trade balance\index{ trade balance}  hold, the relations (\ref{r5l26}) are valid.
Using this, we can write (\ref{r5l27}) as follows
\begin{eqnarray} \label{r5l28}
\sum\limits _{j=t+1}^n\frac{c_{kj}^*D_j(p)}{\sum\limits
_{s=1}^nc_{sj}p_s}=x_k-\sum\limits _{i=t+1}^na_{ki}x_i
\end{eqnarray}
\begin{eqnarray*} -\sum\limits _{j=1}^tc_{kj}\sum\limits _{m=t+1}^nx_ma_{jm}+ \sum\limits_{i=1}^l b_{ki},\quad
1+t\leq k\leq s, \end{eqnarray*}
\begin{eqnarray*} \sum\limits _{j=t+1}^n\frac{c_{kj}D_j(p)}{\sum\limits
_{s=1}^nc_{sj}p_s}=x_k-\sum\limits _{i=t+1}^na_{ki}x_i+ \sum\limits_{i=1}^l b_{ki},\quad
s+1\leq k\leq n,\end{eqnarray*}
where
\begin{eqnarray*} c_{kj}^*=\sum\limits _{l=1}^tc_{kl}c_{lj}+c_{kj},\quad
k=\overline {t+1,s},~~j=\overline {t+1,n}. \end{eqnarray*}
Here we insert first $t$ equations (\ref{r5l27}) into the middle set of equations (\ref{r5l27}).
Note that
\begin{eqnarray*} \sum\limits _{l=1}^nc_{lj}p_l=\sum\limits
_{l=1}^tc_{lj}\sum\limits _{m=t+1}^sc_{ml}p_m+\sum\limits
_{l=t+1}^sc_{lj}p_l+\sum\limits _{l=s+1}^nc_{lj}p_l \end{eqnarray*}
\begin{eqnarray*} =\sum\limits _{m=t+1}^s\left(\sum\limits
_{l=1}^tc_{ml}c_{lj}+c_{mj}\right)p_m+\sum\limits _{m=s+1}^nc_{mj}p_m. \end{eqnarray*}
Therefore,
\begin{eqnarray*} \sum\limits _{l=1}^nc_{lj}p_l=\sum\limits
_{m=t+1}^sc_{mj}^*p_m+\sum\limits _{m=s+1}^nc_{mj}p_m.\end{eqnarray*}
Introduce a  matrix of unproductive consumption $\bar C=||\bar c_{kj}||_{k,j=t+1}^n, $
where \begin{eqnarray*} \bar c_{kj}=c_{kj}^*=\sum\limits
_{l=1}^tc_{kl}c_{lj}+c_{kj},\quad k=\overline
{t+1,s},~~~j=\overline {t+1,n}, \end{eqnarray*}
\begin{eqnarray*} \bar c_{kj}=c_{kj},\quad k=\overline {s+1,n},~~~j=\overline
{t+1,l}. \end{eqnarray*}
Build after $\bar C$
the demand vector \begin{eqnarray*} \gamma _j(p)=\left\{ \frac{c_{t+1,j}^*
p_{t+1}}{\sum\limits _{l=t+1}^sc_{lj}^* p_l+\sum\limits
_{l=s+1}^nc_{lj}p_l},\dots ,{c_{s,j}^*p_s\over\sum\limits
_{l=t+1}^sc_{lj}^*p_l+\sum\limits _{l=s+1}^nc_{lj}p_l},\right.
\end{eqnarray*}
\begin{eqnarray*} \left.
\frac{c_{s+1,j}p_{s+1}}{
\sum\limits_{l=t+1}^sc_{lj}^*p_l+\sum\limits_{l=s+1}^nc_{lj}p_l},\dots ,
{c_{nj}p_n\over
\sum\limits_{l=t+1}^sc_{lj}^*p_l+\sum\limits _{l=s+1}^nc_{lj}p_l}\right\} \end{eqnarray*}
\begin{eqnarray*} ={\{ \gamma _{kj}^* (p)\} }_{k=t+1}^n,\quad j=\overline
{t+1,n},\end{eqnarray*}
\begin{eqnarray*} \gamma _{kj}^*(p)={c_{k,j}^*p_k\over\sum\limits
_{l=t+1}^sc_{lj}^*p_l+\sum\limits _{l=s+1}^nc_{lj}p_l},\quad
k=\overline {t+1,s},~~j=\overline {t+1,n},\end{eqnarray*}
\begin{eqnarray*} \gamma _{kj}^*(p)={c_{kj}p_k\over\sum\limits
_{l=t+1}^sc_{lj}^*p_l+\sum\limits _{l=s+1}^nc_{lj}p_l},\quad
k=\overline {s+1,n},~~~j=\overline {t+1,l}. \end{eqnarray*}
From here, it is easy to see that
\begin{eqnarray} \label{r5l29}
\frac{1}{p_k}\sum\limits _{j=t+1}^l\gamma
_{kj}^*(p)D_j(p)=\sum\limits
_{j=t+1}^l\frac{c_{kj}^*D_j(p)}{\sum\limits
_{m=t+1}^sc_{mj}^*p_m+\sum\limits _{m=s+1}^nc_{mj}p_m}
\end{eqnarray}
\begin{eqnarray*} =\sum\limits _{j=t+1}^l\frac{c_{kj}^*D_j(p)}{\sum\limits
_{m=1}^nc_{mj}p_m},\quad t+1\leq k\leq s, \end{eqnarray*}
\begin{eqnarray*} \frac{1}{ p_k}\sum\limits _{j=t+1}^l\gamma
_{kj}^*(p)D_j(p)=\sum\limits
_{j=t+1}^l\frac{c_{kj}D_j(p)}{\sum\limits
_{m=t+1}^sc_{mj}^*p_m+\sum\limits
_{m=s+1}^lc_{mj}p_m} \end{eqnarray*}  \begin{eqnarray*} =\sum\limits
_{j=t+1}^l\frac{c_{kj}D_j(p)}{\sum\limits
_{m=1}^nc_{mj}p_m},\quad k=\overline{s+1,n}.\end{eqnarray*}
Now transform
 \begin{eqnarray*} D_j(p)=\pi_j\left[x_j\left(p_j-\sum\limits _{l=1}^na_{lj}p_l\right)+ <b_j, p>\right], \quad
j=\overline {t+1,n},\end{eqnarray*}  \begin{eqnarray*} \sum\limits _{l=1}^na_{lj}p_l=\sum\limits
_{l=1}^ta_{lj}p_l+\sum\limits _{l=t+1}^sa_{lj}p_l+\sum\limits
_{l=s+1}^na_{lj}p_l \end{eqnarray*}  \begin{eqnarray*} =\sum\limits _{l=1}^ta_{lj}\sum\limits
_{m=t+1}^sc_{ml}p_m+\sum\limits _{m=t+1}^sa_{mj}p_m+\sum\limits
_{m=s+1}^na_{mj}p_m  \end{eqnarray*}
\begin{eqnarray*} =\sum\limits _{m=t+1}^s\left(\sum\limits
_{l=1}^tc_{ml}a_{lj}+a_{mj}\right)p_m+\sum\limits _{m=s+1}^na_{mj}p_m. \end{eqnarray*}
Introduce the matrix
\begin{eqnarray*} a_{mj}^*=\sum\limits _{l=1}^tc_{ml}a_{lj}+a_{mj}, \quad  m=\overline
{t+1,s}, \quad j=\overline {t+1,n}, \end{eqnarray*}
\begin{eqnarray*}  a_{mj}^*=a_{mj}, \quad m=\overline {s+1,n}, \quad j=\overline
{t+1,n}.\end{eqnarray*}
Then
\begin{eqnarray*} \sum\limits _{l=1}^na_{lj}p_l=\sum\limits
_{m=t+1}^na_{mj}^*p_m,\quad j=\overline {t+1,n}.\end{eqnarray*}
Input vector for such matrix is
\begin{eqnarray*} \sum\limits _{j=t+1}^n a_{kj}^* x_j=\sum\limits
_{m=t+1}^n\sum\limits _{j=1}^tc_{kj}a_{jm}x_m+\sum\limits
_{j=t+1}^na_{kj}x_j,\quad k=\overline {t+1,s},\end{eqnarray*}
\begin{eqnarray*} \sum\limits _{j=t+1}^n a_{kj}^* x_j=\sum\limits
_{j=t+1}^na_{kj}x_j,\quad k=\overline {s+1,n},\end{eqnarray*}
that equals to rhs of
(\ref{r5l28}). Therefore, we can write (\ref{r5l28}) with
 (\ref{r5l29}) as follows
\begin{eqnarray} \label{mav1}
\sum\limits _{j=t+1}^l\frac{\bar c_{kj}D_j(p)}{\sum\limits
_{l=t+1}^n\bar c_{lj}p_l}=x_k-\sum\limits
_{j=t+1}^na_{kj}^*x_j+ \sum\limits_{i=1}^l b_{ki},\quad k=\overline {t+1,n},
\end{eqnarray}
where
\begin{eqnarray*}
a_{kj}^*=\sum\limits _{l=1}^tc_{kl}a_{lj}+a_{kj}, \quad k=\overline
{t+1,s}, \quad  j=\overline {t+1,n},\end{eqnarray*}  \begin{eqnarray*}
a_{kj}^*=a_{kj}, \quad k=\overline {s+1,n}, \quad j=\overline
{t+1,n},\end{eqnarray*}
\begin{eqnarray*} \bar c_{kj}=\sum\limits _{l=1}^tc_{kl}c_{lj}+c_{kj},\quad  k=\overline
{t+1,s}, \quad  j=\overline {t+1,n},\end{eqnarray*}
  \begin{eqnarray*}
\bar c_{kj}=c_{kj}, \quad k=\overline {s+1,n}, \quad j=\overline
{t+1,n},\end{eqnarray*}
\begin{eqnarray*} D_j(p)=\pi_j\left[x_j\left(p_j-\sum\limits _{k=t+1}^na_{kj}^*p_k\right)+ <b_j,p>\right]. \end{eqnarray*}

If   the  conditions of  trade balance\index{conditions of  trade balance} (\ref{r5l26}) account for with known rhs and one  takes available solution to the set of trade balance equations\index{set of trade balance equations } (\ref{r5l26}) for vector of domestic prices\index{vector of domestic prices}  $\{p_{t+1}, \ldots, p_{s}\},$ one can suppose this vector known and fixed  in the set of equations (\ref{mav1}) of the economy equilibrium.

Therefore, one must further solve the problem formulated as follows.

Economy \hfill system \hfill with \hfill production \hfill has
\hfill  $m$ \hfill monopolists \hfill that \hfill dictate \hfill prices \\
$\{p_1^0,\dots ,p_m^0\},$ the rest prices $\{p_{m+1},\dots ,p_n\}$ are formed within  the economy system by the demand and the supply. We say monopolists dictate prices\index{monopolists dictate prices} in presence of production   if they get needed added value. A monopolist is interested not in prices but in added value he gets.
Of course, monopolistic prices\index{monopolistic prices} are determined by these added values and non-monopolistic equilibrium prices\index{non-monopolistic equilibrium prices} unambiguously. If $C={||c_{ik}||}_{i,k=1}^{n,l} $ is a matrix built after  fields  of information evaluation by consumers\index{fields  of information evaluation by consumers} and $A={||a_{ik}||}_{i=1,k=1}^n$ is  a technological matrix,\index{technological matrix} the problem is that to describe vectors of gross outputs\index{vector of gross outputs}  $x=\{x_1,\dots ,x_n\},$ for which the set of equations
\begin{eqnarray} \label{r5l30} \sum\limits
_{j=1}^l\frac{c_{kj}D_j(p)}{\sum\limits
_{l=1}^nc_{lj}p_l}=x_k-\sum\limits _{i=1}^na_{ki}x_i+ \sum\limits_{i=1}^l b_{ki},\quad
k=\overline {1,n},
\end{eqnarray}
relative to  the vector $\{p_{m+1},\dots ,p_n\}$ has economically available solutions,
where
\begin{eqnarray*} D_j(p)=\pi_j \left[x_j\left(p_j^0-\sum\limits _{s=1}^ma_{sj}p_s^0 - \sum\limits _{s=m+1}^na_{sj}p_s\right)+
< b_j, p>\right], \quad j=\overline{1, m},\end{eqnarray*}
\begin{eqnarray*} D_j(p)=\pi_j\left[ x_j\left(p_j-\sum\limits _{s=1}^ma_{sj}p_s^0 - \sum\limits _{s=m+1}^na_{sj}p_s\right) + < b_j, p>\right], \quad  j=\overline{m+1, n}.\end{eqnarray*}
For the rest $j=\overline{n+1, l}, \ D_j(p)$ is determined by a taxation system.\index{taxation system}
  However, we considered such problem in the Subsection 1 of this Section.

Consider partial case of this problem supposing that $b_i=0, \  i=\overline{1, l},$ take  vector of taxation\index{vector of taxation } being  unit  and introduce notations
$\{\sigma^0, x^0\}=\{\sigma_1^0,\dots,\sigma_m^0,x_{m+1},\dots ,x_n\}$,
where $\sigma_j^0=p_j-\sum\limits _{l=1}^na_{lj}p_l, \
j=\overline {1,m},$ are added values for monopolistic goods\index{added values for monopolistic goods} and
$x^0=\{x_{m+1},\dots ,x_n\}$ is non-monopolistic output vector.

Introduce a matrix $[E-A]^{-1}C=D=||d_{ij}||_{i,j=1}^n$ and
 created added values\index{ created added values} 
 \begin{eqnarray*}\sigma _j=\left(p_j-\sum\limits _{l=1}^na_{lj}p_l\right), \quad j=\overline
{m+1, n},  \end{eqnarray*} 
Then the vector
$\{x_1,\dots ,x_m,~\sigma_{m+1},\dots ,\sigma_n \}$
obeys the set of equations
\begin{eqnarray*} \sum\limits _{j=1}^m
{d_{kj}x_j\sigma _j^0\over\sum\limits_{l=1}^m d_{lj}
\sigma_l^0
+\sum\limits_{l=m+1}^n d_{lj}\sigma _l} \end{eqnarray*}
\begin{eqnarray} \label{r5l31}
+\sum\limits _{j=m+1}^n {d_{kj}x_j\sigma
_j\over\sum\limits_{l=1}^m d_{lj}\sigma
_l^0+\sum\limits_{l=m+1}^n d_{lj}\sigma _l}=x_k,
\quad k=\overline {1, n}.
\end{eqnarray}
Suppose that added values of  monopolistic goods  $\sigma _j^0 ,$ \
$j=\overline {1,m}, $ are known.
The problem is to find the vector
$(x_1,\dots ,x_m, \sigma_{m+1},\dots ,\sigma _n)$
from the set of equations (\ref{r5l31})
for known vector $\left(\sigma _1^0 ,\dots ,\sigma
_m^0 , x_{m+1} ,\dots ,x_n \right).$

We solved this problem in the previous Subsection. Note that one can reduce the consideration of an open economy system\index{open economy system} to closed one with domestic monopolists\index{domestic monopolists} and effective  matrix  of unproductive consumption\index{effective  matrix  of unproductive consumption} and effective technological matrix.\index{effective technological matrix} Every industry profitability\index{industry profitability} depends on whether the effective technological matrix is productive one or not and whether the final consumption vector agrees with the consumption structure for industries-consumers,\index{final consumption vector agrees with the consumption structure for industries-consumers} i.e., whether it belongs to the interior of positive cone created by columns of effective industries unproductive consumption matrix\index{interior of positive cone created by columns of effective industries unproductive consumption matrix} or not.

Therefore, efficacy  of the economy system interaction with environment\index{efficacy  of the economy system interaction with environment} is determined by effective technological matrix\index{effective technological matrix} and unproductive consumption matrix.\index{unproductive consumption matrix} If the effective technological matrix is productive  and the final consumption vector\index{final consumption vector} belongs to the interior of positive cone created by columns of the effective unproductive consumption matrix,\index{cone created by columns of the effective unproductive consumption matrix} then the economy system can operate in open way guaranteeing zero trade balance with the external environment.\index{zero trade balance with the external environment}

No one of economy systems is perfectly competitive.\index{ perfectly competitive} Under uncertainty conditions, the firm with advanced technologies\index{firm with advanced technologies} can influence onto formation of prices     in the direction of their decrease as well as in the direction of their increase. In view of its monopolistic state, the firm can increase the part of the profit in value created. As it was shown earlier, keeping certain level of trade balance by industries-exporters also leads to monopolism in price formation. For example, to increase exporters competition, the economy system can decrease levels of taxation  for exporters products\index{levels of taxation  for exporters products} and
devaluate national currency\index{devaluate national currency} that is price discrimination\index{price discrimination} for those industries that work for exporters. The devaluation of national currency\index{devaluation of national currency} to stimulate export for certain industries and transfer of  levels of  taxation\index{levels of  taxation} onto industries producing consumers goods, low level of wages\index{low level of wages} can block domestic market development\index{domestic market development} and domestic investment.\index{domestic investment} Therefore, study of monopolistic formation of  prices\index{monopolistic formation of  prices }  and  influence of  levels of taxation onto the rest prices, income levels of firms\index{income levels of firms} and consumers and onto  the dynamics of economy development\index{dynamics of economy development}  is central topic of mathematical economics.

\backmatter

\include{index}
\printindex

\end{document}